\documentclass[preprint2,longabstract]{aastex}

\newcommand{\solar}{$_{\odot}$}
\newcommand{\tco}{$^{12}$CO}
\newcommand{\ttco}{$^{13}$CO}
\newcommand{\ceto}{C$^{18}$O}
\newcommand{\hcop}{HCO$^+$}
\newcommand{\httco}{H$^{13}$CO$^+$}
\newcommand{\nnh}{N$_2$H$^+$}
\newcommand{\httcn}{H$^{13}$CN}
\newcommand{\hctn}{HC$_3$N}
\newcommand{\joz}{$J$=1$\rightarrow$0}
\newcommand{\jto}{$J$=2$\rightarrow$1}

\newcommand{\kms}{\,km\,s$^{-1}$}
\newcommand{\degree}{$^{\circ}$\hspace{-1mm}}

\def\lapp{\ifmmode\stackrel{<}{_{\sim}}\else$\stackrel{<}{_{\sim}}$\fi}
\def\gapp{\ifmmode\stackrel{>}{_{\sim}}\else$\stackrel{>}{_{\sim}}$\fi}

\shorttitle{CHaMP I.  Mopra \hcop\ Maps}
\shortauthors{Barnes et al.}

\begin{document}


\title{The Galactic Census of High- and Medium-mass Protostars. \\
    I.  Catalogues and First Results from Mopra \hcop\ Maps}


\author{Peter J.Barnes\altaffilmark{1}, Yoshinori Yonekura\altaffilmark{2}, Yasuo Fukui\altaffilmark{3}, 
Andrew T.Miller\altaffilmark{4}, Martin M\"uhleg-\\
ger\altaffilmark{4,5}, 
Lawrence C.Agars\altaffilmark{4}, Yosuke Miyamoto\altaffilmark{3}, Naoko Furukawa\altaffilmark{3}, George Papadopoulos\altaffilmark{4,6}, Scott L.Jones\altaffilmark{4,7}, 
Audra K.Hernandez\altaffilmark{1}, Stefan N.O'Dougherty\altaffilmark{1}, and Jonathan C.Tan\altaffilmark{1}}
\email{peterb@astro.ufl.edu}


\altaffiltext{1}{Astronomy Department, University of Florida, Gainesville, FL 32611, USA}
\altaffiltext{2}{Center for Astronomy, Ibaraki University, 2-1-1 Bunkyo, Mito, Ibaraki 310-8512, Japan}
\altaffiltext{3}{Department of Astrophysics, Nagoya University, Furo-cho, Chikusa-ku, Nagoya 464-8602, Japan}
\altaffiltext{4}{School of Physics A28, University of Sydney, NSW 2006, Australia}
\altaffiltext{5}{Max Planck Institut f\"ur Extraterrestrische Physik, Giessenbachstrasse 1, 85748 Garching, Germany}
\altaffiltext{6}{School of Mathematics and Statistics F07, University of Sydney, NSW 2006, Australia}
\altaffiltext{7}{School of Electrical Engineering \& Telecommunications, University of NSW, Sydney NSW 2052, Australia.}


\begin{abstract}
The {\em Census of High- and Medium-mass Protostars} (CHaMP) is the first large-scale, unbiased, uniform mapping survey at sub-parsec scale resolution of 90\,GHz line emission from massive molecular clumps in the Milky Way.  We present the first Mopra (ATNF) maps of the CHaMP survey region ($300^{\circ}$$>$$l$$>$$280^{\circ}$) in the \hcop \joz\ line, which is usually thought to trace gas at densities up to 10$^{11}$m$^{-3}$.  In this paper we introduce the survey and its strategy, describe the observational and data reduction procedures, and give a complete catalogue of moment maps of the \hcop \joz\ emission from the ensemble of 301 massive molecular clumps.  From these maps we also derive the physical parameters of the clumps, using standard molecular spectral-line analysis techniques.  This analysis yields the following range of properties: integrated line intensity 1--30\,K\kms, peak line brightness 1--7\,K, linewidth 1--10\kms, integrated line luminosity 0.5--200\,K\kms\,pc$^2$, FWHM size 0.2--2.5\,pc, mean projected axial ratio 2, optical depth 0.08--2, total surface density 30--3000\,M\solar\,pc$^{-2}$, number density 0.2--30$\times$10$^9$\,m$^{-3}$, mass 15--8000\,M\solar, virial parameter 1--55, and total gas pressure 0.3--700\,pPa.  We find that the CHaMP clumps do not obey a Larson-type size-linewidth relation. 
Among the clumps, there exists a large population of subthermally excited, weakly-emitting (but easily detectable) dense molecular clumps, confirming the prediction of \cite{ncs08}.  These weakly-emitting clumps comprise 95\% of all massive clumps by number, and 87\% of the molecular mass, in this portion of the Galaxy; their properties are distinct from the brighter massive star-forming regions that are more typically studied.  If the clumps evolve by slow contraction, the 95\% of fainter clumps may represent a long-lived stage of pressure-confined, gravitationally stable massive clump evolution, while the CHaMP clump population may not engage in vigorous massive star formation until the last 5\% of their lifetimes.  The brighter sources are smaller, denser, more highly pressurised, and closer to gravitational instability than the less bright sources.  Our data suggest that massive clumps approach critical Bonnor-Ebert like states at constant density, while others' suggest that lower-mass clumps reach such states at constant pressure.  Evidence of global gravitational collapse of massive clumps is rare, suggesting this phase lasts $<$1\% of the clumps' lifetime.
\end{abstract}



\begin{minipage}[h]{120mm}{
\rule{0mm}{196mm}}
\end{minipage}

\section{Introduction\label{intro}}

\begin{figure*}[ht]
\vspace{-16mm}
\centerline{\rotatebox{90}{\includegraphics[width=48.6mm]{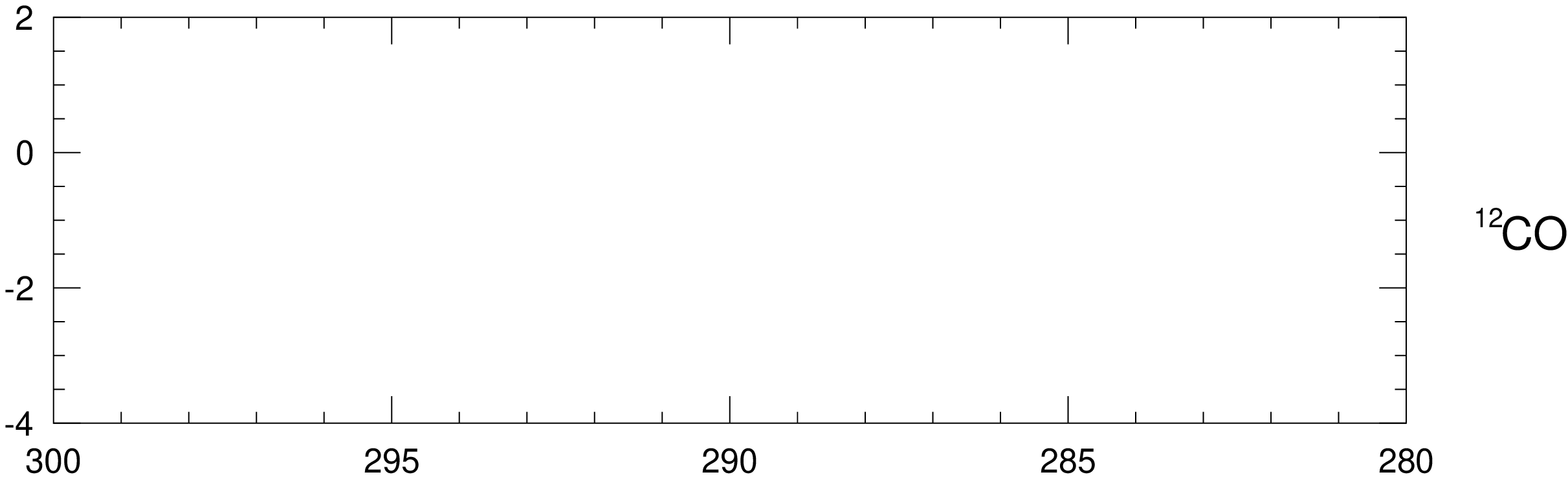}}}
\vspace{-14mm}
\centerline{\rotatebox{90}{\includegraphics[width=52.2mm]{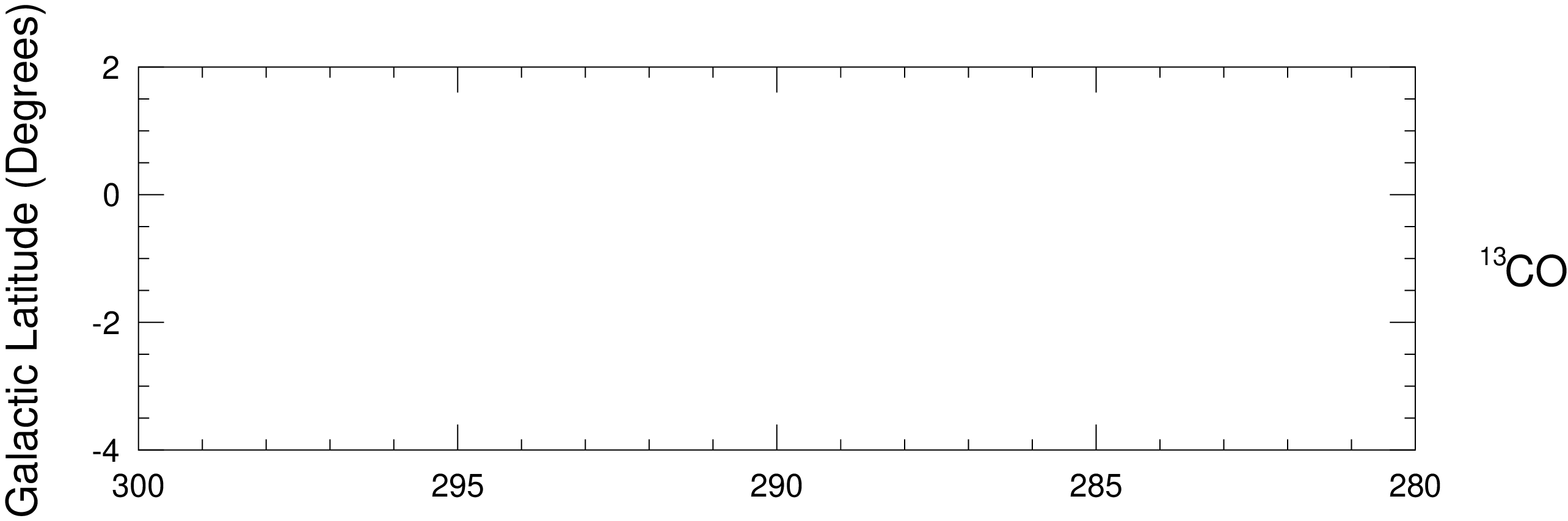}}}
\vspace{-10mm}
\centerline{\rotatebox{90}{\includegraphics[width=50.4mm]{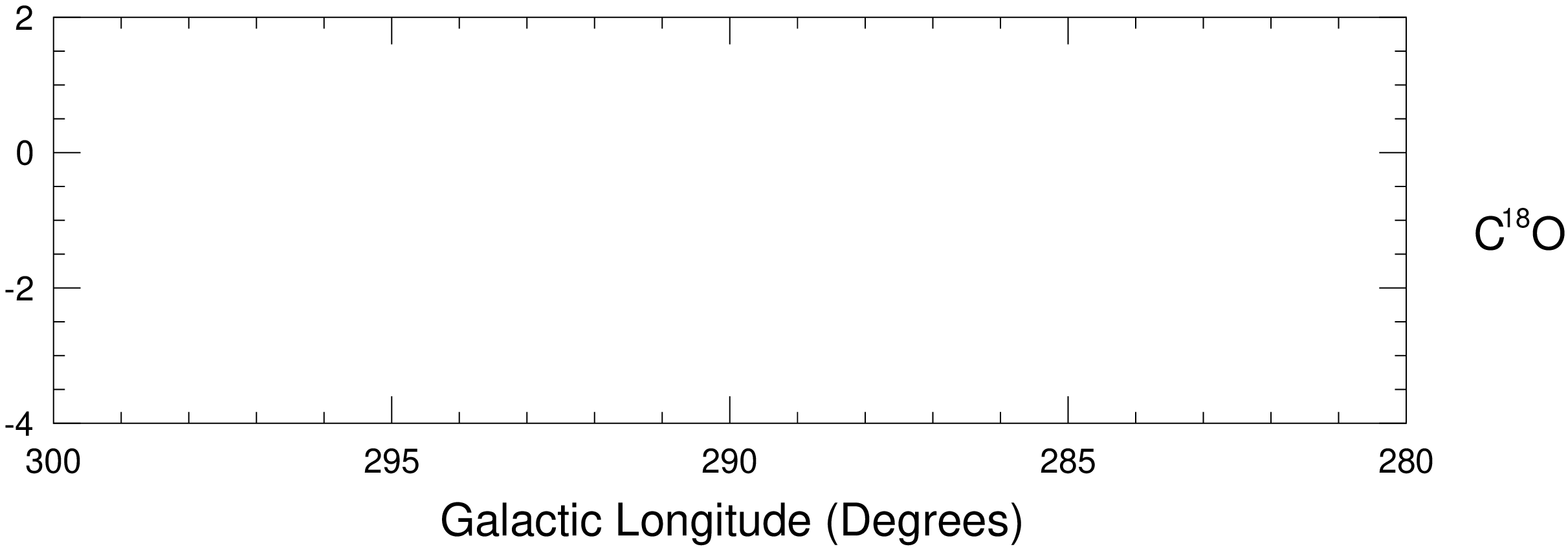}}}
\vspace{-124mm}
\centerline{\hspace{10.5mm}\rotatebox{0}{\includegraphics[width=11.16cm]{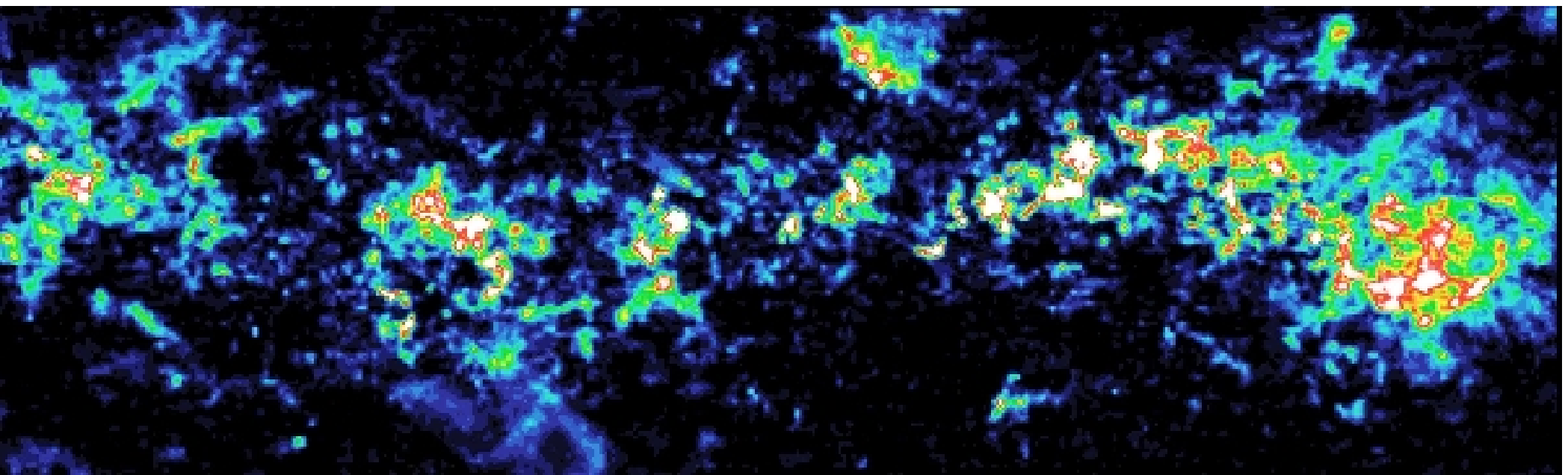}}}
\vspace{5mm}
\centerline{\hspace{10.5mm}\rotatebox{0}{\includegraphics[width=11.16cm]{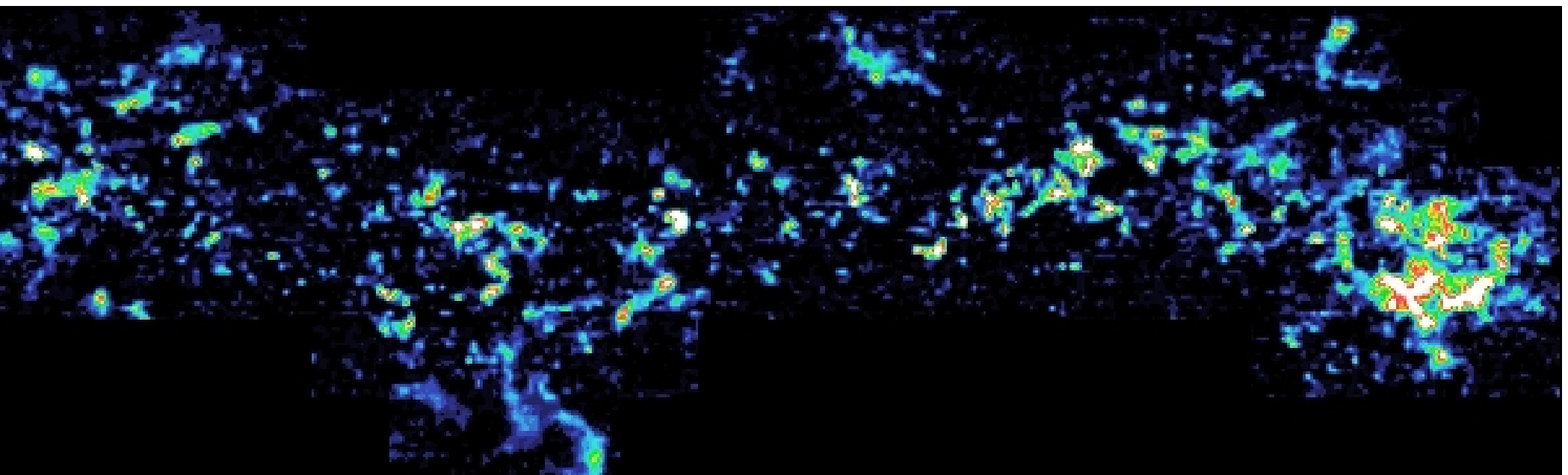}}}
\vspace{5mm}
\centerline{\hspace{10.5mm}\rotatebox{0}{\includegraphics[width=11.16cm]{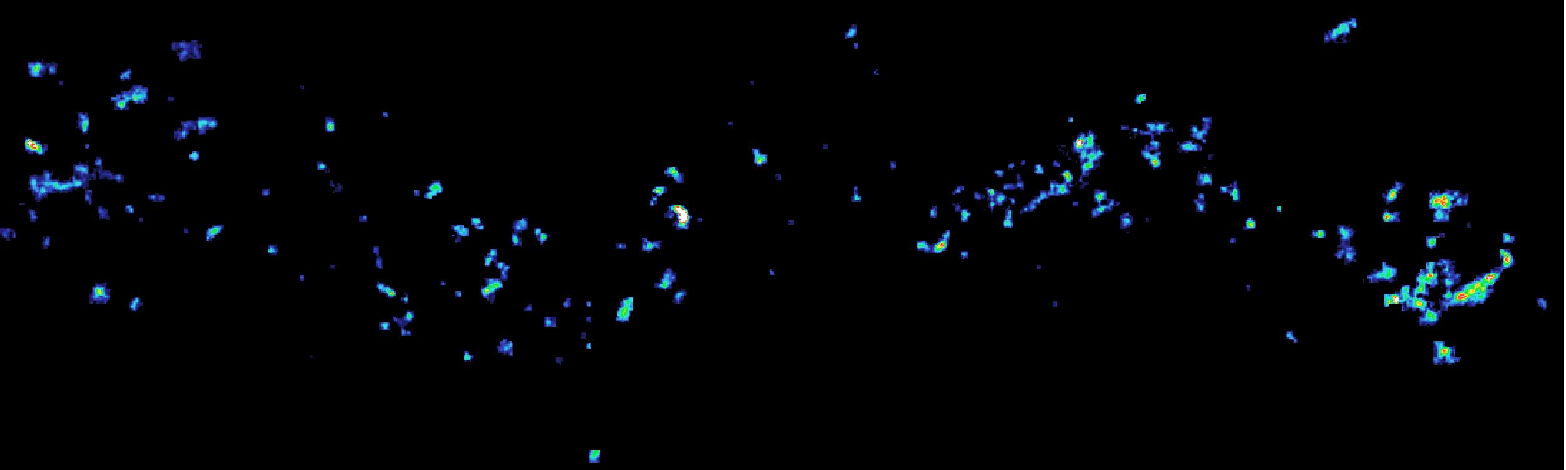}}}
\vspace{8mm}
\caption{{\small Nanten integrated intensity images of CO isotopologues.  The sequence \tco$\rightarrow$\ttco$\rightarrow$\ceto\ samples progressively denser molecular gas environments 
for the same line brightness.  Thus \tco\ maps can serve as finder charts for \ttco\ emission, and \ttco\ for \ceto.  Active star formation only occurs within the \ceto\ emission, despite the \tco\ being very widespread.  Therefore the Nanten \ceto, and similar \hcop\ maps, can serve as finder charts for most, if not all, star formation within this window.}
\label{nanten}}
\vspace{-3mm}
\end{figure*}

Several recent studies have shown that massive stars and the majority of all stars, perhaps including the Sun, have formed together in star clusters \cite[e.g.][]{ll03, dtp05,g09}, yet the question of {\em how} massive stars and star clusters form remains largely open despite much effort, both observationally and theoretically, over the past two decades \citep{bcm07}.  Their formation can be viewed as a single astrophysical process occurring in giant molecular clouds (GMCs), where parsec-scale {\it clumps} transform into star clusters via turbulent fragmentation into {\it cores}, which form individual stars \cite[e.g.][]{pn02, mt03}.  This process underpins all theories of galaxy evolution, and in particular the global process of gas conversion into stars that is empirically described by the Kennicutt-Schmidt relations \citep{k98, lwb08, T10}.  Massive stars in particular also play a dominant role in the regulation of the interstellar medium (ISM) via their energy, momentum, and metallicity feedback.

\begin{figure*}[ht]
\vspace{-15mm}
\centerline{(a)\includegraphics[angle=0,scale=0.83]{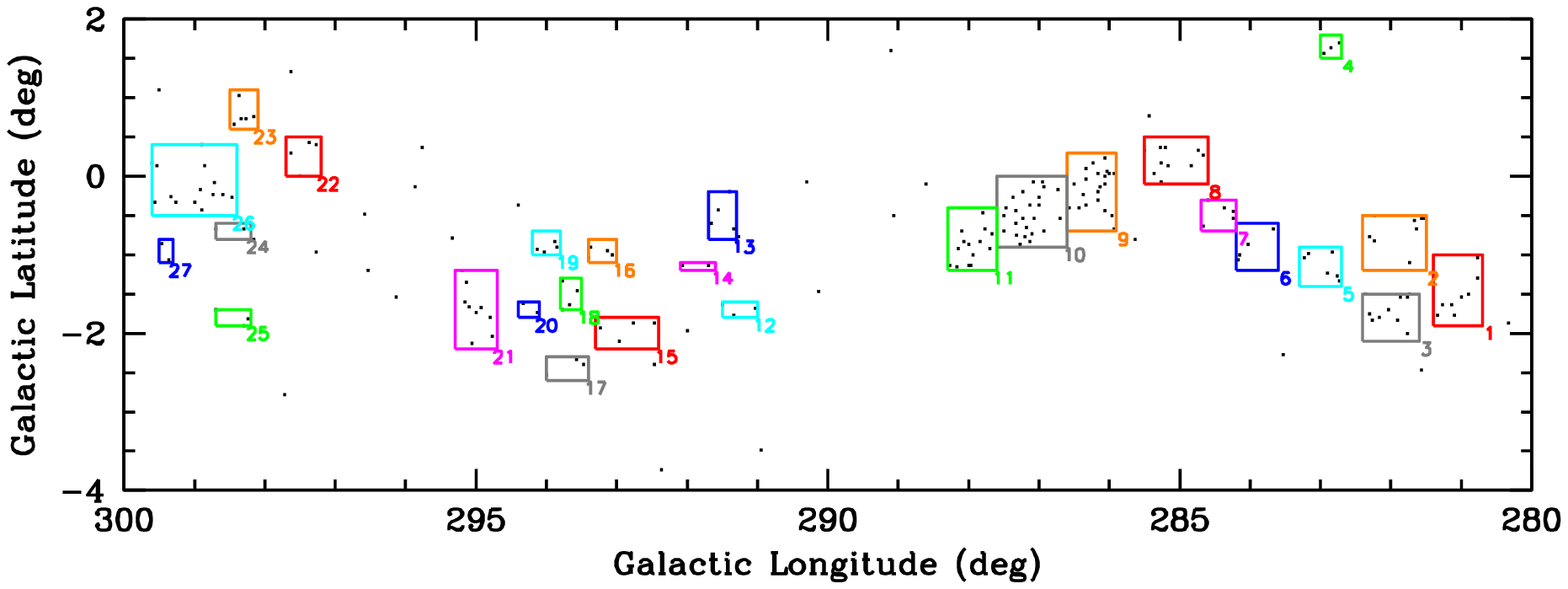}}
\centerline{(b)\includegraphics[angle=0,scale=0.83]{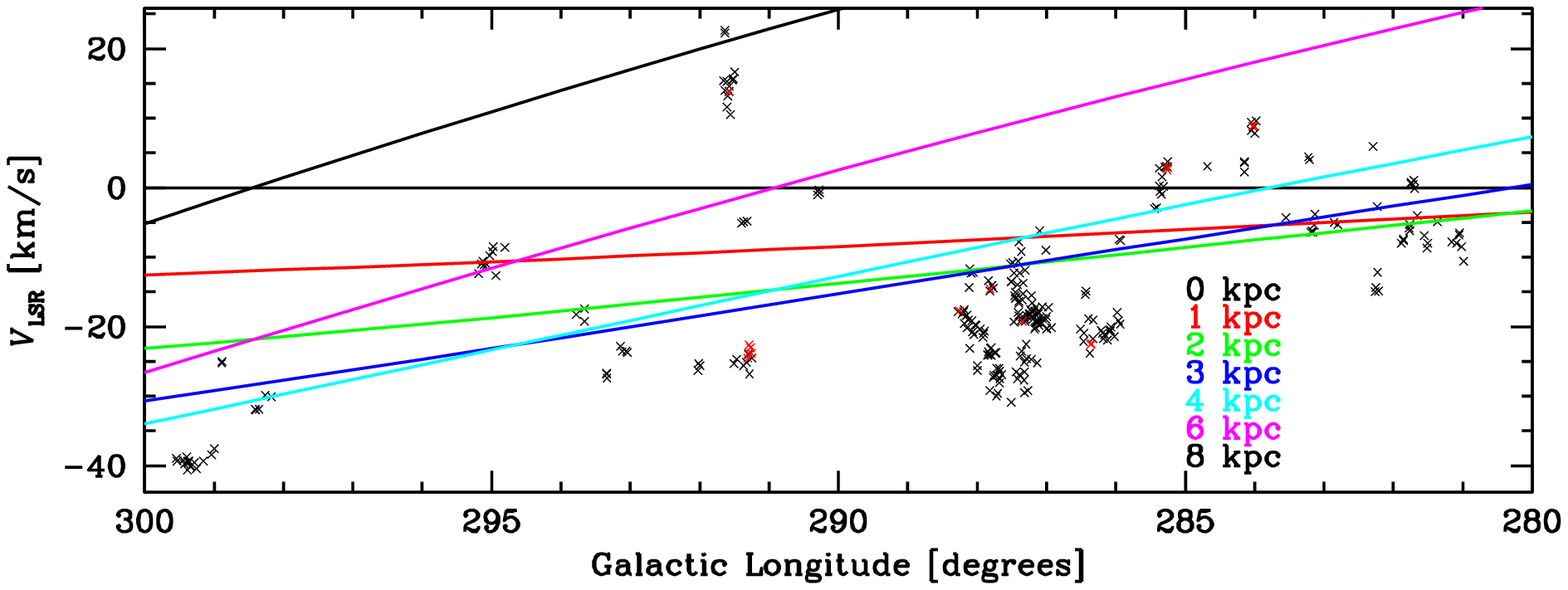}}
\vspace{-1mm}
\caption{\small (a) Diagram showing the location of all sources (dots) and ``Regions'' (coloured and numbered boxes) listed in Table \ref{NMC}; see the text for further description.  (b) Longitude-velocity diagram of our Mopra \hcop\ clumps from Table \ref{sources}, overlaid by distance contours based on the \cite{r09} algorithm.
\label{regions}}
\vspace{-2mm}
\end{figure*}

Despite this importance, our understanding of the massive star and star cluster formation process remains very primitive.  We still lack systematic data on the kinematics and physical conditions in the {\rm prestellar} dense gas.  Although numerous studies exist, most are constrained in key ways.  Detailed, unbiased studies at mm wavelengths usually include a relatively small number of sources \cite[e.g.][]{sm04, pzc07, h09}.  Other studies choose sources for mm-line study based on their emission at other\newline wavelengths \cite[e.g.][]{bsm02, s02, f05, r06, L07, w10}, creating possible statistical biases to the derived gas properties that may be difficult to quantify.  Yet others observe multiple mm lines but only at single points in each source (Klaassen \& Wilson 2007; Chen et al. 2010), thus losing any information on spatial variability of the mm emission within sources.

Partly because of these observational limitations, there is little clear consensus on even the basic formation mechanism, whether 
as massive gas cores collapsing via relatively well-ordered accretion disks \cite[e.g.][]{mt03}, 
competitive accretion of ambient cluster gas 
\citep{BBV03}, 
or more radical theories \citep{bz05, kk08}.  Even among the more conventional core models, a vast range of parameters, such as formation timescale or accretion rate, are debated.  For star clusters, is the overall formation timescale a few \citep{Elm07} or many \citep{TKM06} free fall times?  The influence of feedback is uncertain in setting both the stellar initial mass function (IMF), including its upper limit, and the efficiency of star formation in clusters.  
We are also unsure if the IMF is universal or variable \citep{hg08}.

To address many of these problems, the Galactic {\em Census of High- and Medium-mass Protostars} (CHaMP) was developed \citep{BYM06}.  By compiling a {\em complete, unbiased, uniform, sensitive, and multi-wavelength} survey, CHaMP's objectives are to systematically obtain the properties of the more massive dense gas clumps giving rise to higher-mass star formation, characterise the properties of the stars and clusters emerging from these clouds, identify all the important evolutionary stages, and through a demographic analysis enabled by our statistical approach to this problem, derive the timescales of these stages for the first time.  CHaMP is also intended to provide a valuable legacy in the southern Milky Way during 

\begin{deluxetable}{ccccccc|ccccccc}
\tabletypesize{\scriptsize}
\tablecaption{Nanten Master Catalogue\label{NMC}}
\tablewidth{0pt}
\tablehead{
\colhead{BYF} & \colhead{$l$} & \colhead{$b$} & \colhead{ID$^b$} & \colhead{$T_{peak}$} & \colhead{$V_{LSR}$} & \colhead{Region$^c$} & \colhead{BYF} & \colhead{$l$} & \colhead{$b$} & \colhead{ID$^b$} & \colhead{$T_{peak}$} & \colhead{$V_{LSR}$} & \colhead{Region$^c$} \\
\colhead{no.$^a$} & \colhead{deg} & \colhead{deg} & & \colhead{K} & \colhead{\kms} & & \colhead{no.$^a$} & \colhead{deg} & \colhead{deg} & & \colhead{K} & \colhead{\kms} & \vspace{-3mm} \\
}
\startdata
  1  & 280.3333 & -1.8667 &\ceto& 0.27 & -5 & -- &  56* & 285.3667 &  0.0333 &\ceto& 0.28 & -3  & 8 \\ 
  2* & 280.7667 & -1.3000 & CF	& 0.55 & -8 & 1 &   57* & 285.4333 &  0.7667 & CF  & 0.41 & -3  & --\\ 
  3* & 280.7667 & -1.0333 & CF	& 0.44 & -5 & 1 &   58  & 285.5000 &  0.3333 &\ceto& 0.29 & -22 & 8 \\ 
  4* & 280.9000 & -1.5000 & CF	& 0.48 & -9 & 1 &   59  & 285.6333 & -0.8000 &\ceto& 0.35 &  9  & --\\ 
  5* & 281.0000 & -1.5333 & CF	& 0.55 & -8 & 1 &   60* & 285.9333 & -0.6667 & CF  & 0.68 & -7  & 9\vspace{2mm} \\ 
  6* & 281.1000 & -1.7667 & CF	& 0.45 & -7 & 1 &   61* & 285.9333 &  0.0333 &\ceto& 0.35 & -18 & 9 \\ 
  7* & 281.1333 & -1.6333 & CF	& 0.48 & -7 & 1 &   62* & 285.9667 & -0.5000 & CF  & 0.52 & -16 & 9 \\ 
  8* & 281.2667 & -1.6333 & CF	& 0.44 & -6 & 1 &   63* & 286.0000 &  0.0333 &\hcop& 0.09 &-2.5 & 9 \\ 
  9* & 281.3333 & -1.7667 & CF	& 0.70 & -6 & 1 &   64* & 286.0333 &  0.0667 &\hcop& 0.12 &14.5 & 9 \\ 
 10* & 281.5333 & -0.5333 & CF	& 0.42 & -3 & 2a &  65  & 286.0667 & -0.4333 &\hcop& 0.14 & -19 & 9\vspace{2mm} \\ 
 11* & 281.5667 & -2.4667 & CF	& 0.60 & -7 & -- &  66* & 286.0667 & -0.1000 & CF  & 0.50 & -21 & 9 \\ 
 12* & 281.5667 & -0.5333 & CF	& 0.56 &  3 & 2a &  67* & 286.0667 &  0.0000 & CF  & 0.58 & -21 & 9 \\ 
 13* & 281.6333 & -0.6667 & CF	& 0.40 & -4 & 2a &  68* & 286.0667 &  0.2333 & CF  & 0.48 & -20 & 9 \\ 
 14* & 281.6667 & -0.5667 & CF	& 0.69 & -4 & 2a &  69* & 286.1333 & -0.1333 &\hcop& 0.43 & -22 & 9 \\ 
 15* & 281.7333 & -1.5000 & CF	& 0.42 & -13 & 3 &  70* & 286.1667 & -0.2000 &\hcop& 0.39 & -21 & 9\vspace{2mm} \\ 
 16* & 281.7333 & -1.1000 & CF	& 0.50 & -1 & 2b &  71* & 286.1667 & -0.3000 &\ceto& 0.23 & -24 & 9 \\ 
 17* & 281.7667 & -2.0000 & CF	& 0.48 & -6 & 2b &  72* & 286.1667 &  0.0333 &\ceto& 0.38 & -22 & 9 \\ 
 18* & 281.7667 & -1.5333 & CF	& 0.45 &  2 & 2b &  73* & 286.2333 &  0.1667 & CF  & 1.14 & -20 & 9 \\ 
 19* & 281.8667 & -1.5333 & CF	& 0.55 & -6 & 2b &  74  & 286.3333 &  0.1000 &\hcop& 0.17 &-17.5& 9 \\ 
 20* & 281.9000 & -1.8333 & CF	& 0.53 & -10 & 3 &  75  & 286.3333 & -0.0333 &\ceto& 0.18 & -21 & 9\vspace{2mm} \\ 
 21* & 282.0333 & -1.7000 & CF	& 0.42 & -10 & 3 &  76* & 286.3333 & -0.3667 &\hcop& 0.20 & -19 & 9 \\ 
 22* & 282.1667 & -1.8000 & CF	& 0.57 & -11 & 3 &  77* & 286.3667 & -0.2333 & CF  & 0.62 & -24 & 9 \\ 
 23* & 282.2333 & -0.5000 & CF	& 0.76 & -2 & 2c &  78* & 286.4333 & -0.4000 & CF  & 0.56 & -13 & 9 \\ 
 24* & 282.2333 & -0.8200 &\hcop& 0.23 &-11.5&2c &  79* & 286.5000 & -0.1000 &\ceto& 0.29 & -22 & 9 \\ 
 25* & 282.2667 & -1.8333 & CF	& 0.47 & -14 & 3 &  80* & 286.5667 & -0.4000 &\ceto& 0.26 & -18 & 9\vspace{2mm} \\ 
 26* & 282.3000 & -1.7500 &\ceto& 0.35 & -13 & 3 &  81  & 286.7000 & -0.5333 &\ceto& 0.24 & -16 & 10 \\ 
 27* & 282.3000 & -0.7667 & CF	& 0.88 &  6 & 2c &  82  & 286.7333 & -0.1667 &\ceto& 0.29 &  16 & 10 \\ 
 28  & 282.3667 & -1.5000 &\ceto& 0.39 & -13 & 3 &  83* & 286.9333 & -0.7000 &\hcop& 0.25 & -18 & 10 \\ 
 29  & 282.7333 &  1.7000 &\ceto& 0.31 & -20 & 4 &  84* & 286.9333 & -0.1333 &\hcop& 0.15 &  19 & 10 \\ 
 30  & 282.7333 & -1.3333 &\hcop& 0.19 &-13.5& 5 &  85* & 286.9500 & -0.0667 &\hcop& 0.26 &-20.5& 10\vspace{2mm} \\ 
 31  & 282.7667 & -1.2667 &\ceto& 0.31 & -6 & 5 &   86* & 287.0000 & -0.2667 &\hcop& 0.23 &-18.5& 10 \\ 
 32* & 282.8333 & -0.9667 & CF	& 0.41 & -6 & 5 &   87* & 287.0000 & -0.3667 &\hcop& 0.36 & -20 & 10 \\ 
 33  & 282.8500 &  1.6333 &\ceto& 0.36 & -22 & 4 &  88* & 287.0667 & -0.5333 &\hcop& 0.23 & -19 & 10 \\ 
 34  & 282.9000 & -1.2333 &\ceto& 0.31 & -5 & 5 &   89* & 287.0833 & -0.0667 &\hcop& 0.20 & -21 & 10 \\ 
 35  & 282.9500 &  1.5667 &\ceto& 0.35 & -23 & 4 &  90* & 287.1000 & -0.7333 &\ceto& 0.35 & -19 & 10\vspace{2mm} \\ 
 36* & 283.1667 & -0.9800 &\ceto& 0.37 & -7 & 5 &   91* & 287.1333 & -0.8333 &\hcop& 0.38 & -20 & 10 \\ 
 37* & 283.2333 & -1.0333 &\ceto& 0.20 & -10 & 5 &  92* & 287.1333 & -0.3667 &\ceto& 0.25 & -21 & 10 \\ 
 38* & 283.5333 & -2.2667 &\ceto& 0.39 & -5 & -- &  93* & 287.1833 & -0.6500 &\hcop& 0.26 & -21 & 10 \\ 
 39  & 283.6667 & -0.6667 &\ceto& 0.34 & 12 & 6 &   94* & 287.2000 & -0.7667 &\hcop& 0.33 &-15.5& 10 \\ 
 40* & 284.0333 & -0.8667 & CF	& 0.52 &  9 & 6 &   95* & 287.2333 & -0.5333 & CF  & 0.41 & -17 & 10\vspace{2mm} \\ 
 41* & 284.1500 & -1.0000 &\hcop& 0.19 & 1.5 & 6 &  96* & 287.2333 & -0.2000 &\ceto& 0.30 & -18 & 10 \\ 
 42* & 284.1667 & -1.0667 &\hcop& 0.19 & 3.5 & 6 &  97* & 287.2667 & -0.8667 &\hcop& 0.35 & -15 & 10 \\ 
 43  & 284.2333 & -0.5333 &\ceto& 0.26 & 17 & 7 &   98* & 287.3167 & -0.7500 &\hcop& 0.20 &-28.5& 10 \\ 
 44  & 284.2333 & -0.4500 &\ceto& 0.33 & 16 & 7 &   99* & 287.3333 & -0.6000 &\hcop& 0.57 & -24 & 10 \\ 
 45  & 284.3667 & -0.4000 &\ceto& 0.26 &  4 & 7 &  100* & 287.3333 & -0.4333 &\hcop& 0.38 &-19.5& 10\vspace{2mm} \\ 
 46  & 284.6000 & -0.3000 &\ceto& 0.28 & 12 & 7 &  101* & 287.3667 & -0.2667 &\hcop& 0.16 & -15 & 10 \\ 
 47* & 284.6667 & -0.6333 & CF	& 0.43 &  4 & 7 &  102* & 287.4667 & -0.4000 &\hcop& 0.19 &-15.5& 10 \\ 
 48  & 284.6667 &  0.2667 &\ceto& 0.25 & 12 & 8 &  103* & 287.5000 & -0.5000 &\ceto& 0.32 & -14 & 10 \\ 
 49  & 284.7333 &  0.3333 &\ceto& 0.27 & 12 & 8 &  104* & 287.5000 & -0.7000 &\hcop& 0.22 &-26.5& 10 \\ 
 50* & 284.8333 &  0.1333 & CF	& 0.44 & -12 & 8 & 105* & 287.6667 & -0.7333 & CF  & 0.42 & -26 & 11\vspace{2mm} \\ 
 51* & 285.1667 &  0.1333 &\hcop& 0.13 & -4 & 8 &  106* & 287.7000 & -0.9167 &\hcop& 0.21 & -28 & 11 \\ 
 52  & 285.2000 &  0.3667 &\ceto& 0.33 & -22 & 8 & 107* & 287.7667 & -0.6667 &\hcop& 0.37 & -27 & 11 \\ 
 53  & 285.2667 &  0.1667 &\ceto& 0.23 & -2 & 8 &  108* & 287.8000 & -0.4667 &\hcop& 0.18 &-22.5& 11 \\ 
 54* & 285.2667 & -0.0667 & CF	& 0.57 & 3 & 8 &   109* & 287.8333 & -0.8333 &\hcop& 0.41 &-14.5& 11 \\ 
 55  & 285.2800 &  0.3667 &\ceto& 0.31 & -21 & 8 & 110* & 287.9333 & -1.0000 &\ceto& 0.32 & -20 & 11\vspace{2mm} \\ 
 \\
 \\
 111* & 287.9667 & -1.1333 & CF  & 0.69 & -21 & 11 & 161* & 294.8000 & -1.8000 &\hcop& 0.41 &-8.5 & 21 \\ 
 112* & 288.0000 & -1.1333 & CF  & 0.51 & -19 & 11 & 162* & 294.9300 & -1.6700 &\hcop& 0.26 & -13 & 21 \\ 
 113* & 288.0000 & -0.8667 &\hcop& 0.17 &-26.5& 11 & 163* & 295.0000 & -1.7333 & CF  & 0.48 & -9  & 21 \\ 
 114* & 288.0667 & -0.8333 &\hcop& 0.21 &-20.5& 11 & 164  & 295.0600 & -2.1300 &\ceto& 0.29 & -18 & 21 \\ 
 115* & 288.1000 & -0.7000 & CF  & 0.46 & -12 & 11 & 165* & 295.1000 & -1.6600 &\hcop& 0.45 & -11 & 21\vspace{2mm} \\ 
 116* & 288.1500 & -0.9167 &\hcop& 0.19 & -20 & 11 & 166  & 295.1333 & -1.3500 &\ceto& 0.34 & -17 & 21 \\ 
 117* & 288.1667 & -1.1500 &\ceto& 0.34 & -18 & 11 & 167* & 295.1600 & -1.6000 &\hcop& 0.43 &-10.5& 21 \\ 
 118* & 288.2667 & -1.1333 & CF  & 0.42 & -18 & 11 & 168  & 295.2000 & -1.2000 &\ceto& 0.27 & -30 & 21 \\ 
 119  & 288.6000 & -0.1000 &\ceto& 0.29 & -33 & -- & 169  & 295.3333 & -0.7833 &\ceto& 0.22 & -1  & -- \\ 
 120  & 289.0667 & -0.5000 &\ceto& 0.37 & -20 & -- & 170  & 295.7667 &  0.3667 & CF  & 0.53 & -25 & --\vspace{2mm} \\ 
 121  & 289.1000 &  1.6000 &\ceto& 0.40 & -28 & -- & 171  & 295.8667 & -0.1333 & CF  & 0.42 & -27 & -- \\ 
 122  & 290.1333 & -1.4667 &\ceto& 0.26 &  9  & -- & 172  & 296.1333 & -1.5333 &\ceto& 0.22 &  18 & -- \\ 
 123* & 290.3000 & -0.0667 & CF  & 0.52 & -1  & -- & 173  & 296.5333 & -1.2000 & CF  & 0.52 & -27 & -- \\ 
 124  & 290.9439 & -3.4880 &\tco & 3.60 &-4.5 & -- & 174  & 296.5833 & -0.4833 &\ceto& 0.17 &  10 & -- \\ 
 125  & 291.0333 & -1.6805 &\ttco& 0.68 &-4.75& 12 & 175  & 297.2667 & -0.9667 & CF  & 0.63 & -35 & --\vspace{2mm} \\ 
 126* & 291.2667 & -0.7667 & CF  & 2.25 & -24 & 13 & 176  & 297.2667 &  0.4000 &\ceto& 0.32 & -37 & 22 \\ 
 127* & 291.3333 & -1.7667 & CF  & 0.61 & -4  & 12 & 177  & 297.3667 &  0.4333 & CF  & 0.47 & -37 & 22 \\ 
 128* & 291.3333 & -0.6667 & CF  & 0.87 & -26 & 13 & 178  & 297.5000 &  0.0000 &\ceto& 0.20 &  15 & 22 \\ 
 129* & 291.4000 & -0.2000 & CF  & 0.81 & -6  & 13 & 179  & 297.6333 &  0.3000 &\ceto& 0.31 & -35 & 22 \\ 
 130* & 291.5000 & -1.6333 & CF  & 0.59 & -25 & 12 & 180  & 297.6333 &  1.3333 &\ceto& 0.25 & -30 & --\vspace{2mm} \\ 
 131* & 291.5600 & -0.4300 &\hcop& 0.72 & 14.5& 13 & 181  & 297.7225 & -2.7772 &\tco & 5.20 & -4.5& -- \\ 
 132* & 291.6600 & -0.6000 &\hcop& 0.31 & 11.5& 13 & 182  & 298.1600 & -0.8000 &\hcop& 0.18 & 23.5& 24 \\ 
 133  & 291.7000 & -1.1333 &\ceto& 0.32 & -25 & 14 & 183* & 298.1600 &  0.7600 &\ceto& 0.32 & -31 & 23 \\ 
 134* & 292.0000 & -1.9667 & CF  & 0.53 & -26 & 15 & 184  & 298.2333 & -1.8167 &\ceto& 0.31 & -29 & 25 \\ 
 135  & 292.0667 & -1.1333 &\ceto& 0.48 & -20 & 14 & 185* & 298.2667 &  0.7333 & CF  & 0.65 & -31 & 23\vspace{2mm} \\ 
 136  & 292.3667 & -3.7333 & CF  & 1.13 & -5  & -- & 186  & 298.3000 & -1.9000 &\ceto& 0.32 & -29 & 25 \\ 
 137  & 292.4600 & -2.4000 &\ceto& 0.24 & -25 & 15 & 187  & 298.3000 & -0.6667 &\ceto& 0.37 & -34 & 24 \\ 
 138  & 292.4667 & -1.8667 &\ceto& 0.33 & -28 & 15 & 188* & 298.3333 &  0.7333 & CF  & 0.53 & -32 & 23 \\ 
 139  & 292.7667 & -1.8667 &\ceto& 0.33 & -25 & 15 & 189  & 298.3667 &  1.0333 &\ceto& 0.29 & -27 & 23 \\ 
 140  & 292.9667 & -2.1000 & CF  & 0.43 & -24 & 15 & 190* & 298.4333 &  0.6667 & CF  & 0.58 & -33 & 23\vspace{2mm} \\ 
 141* & 293.0600 & -1.0000 &\ceto& 0.39 & -23 & 16 & 191  & 298.4667 & -0.2667 &\ceto& 0.33 & -38 & 26 \\ 
 142* & 293.1333 & -0.9500 &\ceto& 0.39 & -24 & 16 & 192  & 298.6000 & -0.2333 &\ceto& 0.29 & -37 & 26 \\ 
 143  & 293.2333 & -1.9333 &\ceto& 0.25 & -24 & 15 & 193  & 298.7000 & -1.7000 & CF  & 0.64 & -30 & 25 \\ 
 144* & 293.3667 & -0.9000 & CF  & 0.42 & -29 & 16 & 194  & 298.7000 & -0.6667 &\ceto& 0.29 & -40 & 24 \\ 
 145  & 293.4000 & -1.0333 & CF  & 0.46 & -24 & 16 & 195  & 298.7167 & -0.0833 &\hcop& 0.11 & -16 & 26\vspace{2mm} \\ 
 146  & 293.4667 & -2.4000 &\ceto& 0.31 & -27 & 17 & 196  & 298.7333 & -0.2333 &\ceto& 0.25 & -37 & 26 \\ 
 147  & 293.5600 & -1.4600 &\ceto& 0.28 & -27 & 18 & 197  & 298.8600 &  0.1333 &\ceto& 0.18 &  21 & 26 \\ 
 148  & 293.5667 & -2.3333 &\ceto& 0.27 & -27 & 17 & 198  & 298.9000 & -0.4300 &\hcop& 0.27 &  31 & 26 \\ 
 149* & 293.6667 & -1.6333 & CF  & 0.48 & -17 & 18 & 199* & 298.9000 &  0.4000 & CF  & 0.62 & -25 & 26 \\ 
 150* & 293.7667 & -1.7000 & CF  & 0.58 & -18 & 18 & 200  & 298.9167 & -0.1667 &\ceto& 0.33 & -20 & 26\vspace{2mm} \\ 
 151  & 293.7667 & -1.3333 & CF  & 0.43 & -29 & 18 & 201* & 299.0000 & -0.3333 & CF  & 0.42 & -38 & 26 \\ 
 152  & 293.8500 & -0.9000 &\ceto& 0.25 & -25 & 19 & 202* & 299.2667 & -0.3333 & CF  & 0.56 & -40 & 26 \\ 
 153  & 293.8833 & -0.8333 &\ceto& 0.28 &  34 & 19 & 203* & 299.3300 & -0.2600 &\ceto& 0.35 & -40 & 26 \\ 
 154  & 294.0000 & -2.5300 &\ceto& 0.34 & -11 & 17 & 204  & 299.3667 & -1.0667 &\ceto& 0.33 & -35 & 27 \\ 
 155  & 294.0333 & -0.9667 & CF  & 0.41 & -27 & 19 & 205  & 299.4600 & -0.8600 &\hcop& 0.17 & -35 & 27\vspace{2mm} \\ 
 156  & 294.1333 & -1.7333 &\ceto& 0.24 & -25 & 20 & 206  & 299.5000 &  1.1000 & CF  & 0.68 & -33 & -- \\ 
 157  & 294.1333 & -0.9333 &\ceto& 0.39 & -26 & 19 & 207  & 299.5333 &  0.1333 & CF  & 0.61 & -4  & 26 \\ 
 158  & 294.3333 & -1.6200 &\ceto& 0.23 & -15 & 20 & 208* & 299.5667 & -0.3333 & CF  & 0.48 & -39 & 26 \\ 
 159  & 294.4000 & -0.3667 & CF  & 0.59 & -26 & -- & 209  & 299.6000 &  0.1667 & CF  & 0.68 & -8  & 26 \\ 
 160  & 294.7667 & -2.0333 & CF  & 0.45 & -14 & 21 \\ 

\enddata
\vspace{-2mm}
\tablenotetext{a}{An asterisk indicates a Nanten clump that was mapped at Mopra (see Table \ref{sources}).}
\tablenotetext{b}{Method of clump identification: ``CF'' is from the CLUMPFIND program operating on the Nanten \ceto\ data cube (see Paper II); ``\ceto'' or ``\hcop'' is from visual inspection of the respective Nanten data cubes.}
\tablenotetext{c}{Grouping of sources by Region is for convenience of mapping only; see Fig.\,\ref{regions} and \S\ref{maps}.  Some of these sources were too isolated to be conveniently grouped into regions.}
\end{deluxetable}

\hspace{-4mm}the SOFIA-ALMA era for future studies of star formation and Galactic structure.

\section{The CHaMP Survey}
\subsection{Survey Strategy}

CHaMP began with complete surveys of a large portion of the southern Galactic Plane in four molecular spectral lines in the 3-millimetre waveband 
made with the 4m Nanten telescope (see Fig.\,\ref{nanten}).  These lines are the \joz\ transitions of \tco, \ttco, \ceto, and \hcop\ at 115.271, 110.201, 109.781, and 89.189\,GHz, respectively; the area mapped is a $20^{\circ}$$\times$$6^{\circ}$ section of the Milky Way in Vela, Carina, and Centaurus, $300^{\circ}$$>$$l$$>$$280^{\circ}$ and $-4^{\circ}$$<$$b$$<$$+2^{\circ}$.  The results of these observations are presented in other papers \cite[][hereafter Paper II]{YAK05,YBF11}.  These lines were chosen to bootstrap from the standard GMC tracers \tco\ and \ttco, which are sensitive to molecular densities $\sim$10$^8$\,m$^{-3}$, to the higher-density tracers \ceto\ and \hcop, sampling gas at densities up to $\sim$10$^9$ and 10$^{11}$\,m$^{-3}$, resp.  The objective was not merely to map some of the dense gas in GMCs, but to do so systematically across the entire region, mapping (as far as our flux limit allowed) {\bf all} the dense gas in our window.

In order to do this efficiently, each line in the sequence above was used as a ``finder chart'' for the next line.  Thus, \ttco\ was only observed where the \tco\ integrated intensity was above 5\,K\kms, and \ceto\ \& \hcop\ were only mapped where \ttco\ was brighter than 2\,K\kms.  This means that the time required to integrate down to high S/N levels in each line did not go up dramatically as the line brightness dropped, since the sky coverage required to do so simultaneously became much less.  Thanks to the finder chart approach, which minimises the amount of time spent integrating on blank sky in the weaker lines, we developed a highly efficient way to obtain the complete dense-gas coverage we sought over a very large area of sky, and at the same time to effectively ``peel the onion layers'' of density in GMCs.  

\subsection{The Nanten Master Catalogue \label{master}}

The Nanten Master Catalogue (hereafter NMC) was constructed in the following way.  We performed a CLUMPFIND analysis on the Nanten \ceto\ data cube in order to compile a uniformly-selected sample of clouds for further analysis.  This will be described in some detail in Paper II, and was aimed at identifying only the most clearly-defined clumps in the data.  For follow-up mapping campaigns, however, a more inclusive approach was needed in order to meet our goal of completeness.  Using the CLUMPFIND list as a starting point, we then visually inspected the Nanten data \ceto\ and \hcop\ cubes to find in ($l$,$b$,$v$) space all local maxima of emission in either line, above a minimum cutoff level of 0.15\,K\kms.  This gave our NMC list of 209 clumps (see Table \ref{NMC}).  Many of the sources are organised for convenience into smaller ``Regions'', which are also given in Table \ref{NMC} and shown approximately in Figure \ref{regions}a.  See Paper II for more details on the Nanten mapping and CLUMPFIND procedures.

\subsection{Mopra Observations}
A higher-resolution follow-up campaign was then begun to map these clumps in a number of 3mm molecular transitions with the 22m-diameter Mopra dish of the Australia Telescope National Facility\footnote{The Mopra telescope is part of the Australia Telescope which is funded by the Commonwealth of Australia for operation as a National Facility managed by CSIRO.  The University of New South Wales Digital Filter Bank used for the observations with the Mopra telescope was provided with support from the Australian Research Council.}, 
at much higher sensitivity and angular resolution than with Nanten.  Over the period described below, we observed the brightest 121 of these clumps, with an effective brightness limit of 0.25\,K\kms\ in the Nanten data.  We therefore estimate that, over the areas mapped with Mopra, our completeness limit on the Nanten scale is essentially 100\% above this level.  On the Mopra scale the completeness is more complex; we discuss this in \S\ref{ensemble}.  

In this way, the NMC and the Mopra maps form a unique resource for a true census of all massive star formation phenomenology, from $\mu$m to cm wavelengths.  Therefore, while making use of existing archives from IRAS, MSX, 2MASS, and Spitzer-IRAC, we have also begun more sensitive and higher-resolution observations of these regions in the near-IR using the IRIS2 camera of the Australian Astronomical Telescope, in the mid-IR using T-ReCS at Gemini-South, and at 3mm using the Australia Telescope Compact Array.  These results will be reported in future papers.

\begin{deluxetable}{lccccc}
\tabletypesize{\footnotesize}
\tablecaption{Mopra Observing History\label{obshist}}
\tablewidth{0pt}
\tablehead{
\colhead{Dates} & \colhead{Map Mode} & \colhead{Receiver} & \colhead{Backend Mode} & \colhead{Clump Area Coverage} & \colhead{Species} \vspace{-3mm} \\
}
\startdata
  2004Jul21--25 & OTF & SIS & MPCOR 32MHz/31kHz & 11$\times$ 5$'$ single/dual-raster & \nnh$\!$/\ceto\vspace{1mm} \\
  $\!\!\!\!\! \left. \begin{array}{l}
  \textit{\rm 2005Jul13--23} \\
  \textit{\rm 2005Sep30--Oct02}
  \end{array} \right\}$ & OTF & MMIC & MPCOR 32MHz/31kHz & $\left\{ \begin{array}{c}
  \textit{\rm 51$\times$ 5$'$ single-raster} \\
  \textit{\rm 7$\times$ 5$'$ single-raster}
  \end{array} \right.$ & $\begin{array}{c}
  \textit{\rm \hcop} \\
  \textit{\rm \nnh$\!$/\httco$\!$/\hctn/CS}\vspace{1mm}
  \end{array}$ \\
  $\!\!\!\!\! \left. \begin{array}{l}
  \textit{\rm 2006Jun04--11} \\
  \textit{\rm 2006Jul04--12} \\
  \textit{\rm 2006Jul24--Aug01} \\
  \textit{\rm 2006Oct23--29} 
  \end{array} \right\}$ & OTF & MMIC & MOPS zoom 138MHz/34kHz & 120$\times$ 3$'$--7$'$ dual-raster & $\left\{ \begin{array}{c}
  \textit{\rm Setup 1$^a$} \\
  \textit{\rm Setup 2$^a$} \\
  \textit{\rm Setup 2$^a$} \\
  \textit{\rm Setup 2$^a$}\vspace{1mm}
  \end{array} \right.$ \\
  2007Aug26--Sep20 & OTF & MMIC & MOPS zoom 138MHz/34kHz & 120$\times$ 3$'$--7$'$ dual-raster & Setup 3$^a$\vspace{1mm} \\
\enddata
\vspace{-2mm}
\tablenotetext{a}{ See Table \ref{setups} for MOPS spectral-line setups.}
\vspace{-5mm}
\end{deluxetable}

Observations with the Mopra antenna were conducted over the period 2004--07, during which time a number of significant upgrades were completed on the telescope.  The Mopra antenna's performance at the beginning of this period has been described by \citet{LPW05}.  Since that study, an on-the-fly (OTF) mapping capability has been implemented in the control software (in 2004; T. Wong 2005, unpublished), new 3mm MMIC receivers were installed (in 2005) which were at least as sensitive as the previous SIS mixers and much more efficient to operate, and the MOPS wideband digital filterbank was commissioned \cite[in 2006;][]{WMF06}.  This latter innovation especially, when combined with the Nanten maps as finder charts, makes an ambitious survey like CHaMP possible.  Because of these changes, we describe each season's data-taking in turn.  Table \ref{obshist} summarises this observational progress.
 
\subsubsection{2004 and OTF mapping}
In 2004 we began CHaMP with a pilot survey of 11 Nanten clumps in the \joz\ lines of \ceto\ and \nnh\ at 109.781 and 93.177\,GHz, respectively.  The clumps were arbitrarily chosen from among the westernmost Nanten clumps.

\begin{figure}[ht]
\vspace{-6mm}
\centerline{\includegraphics[angle=0,scale=0.43]{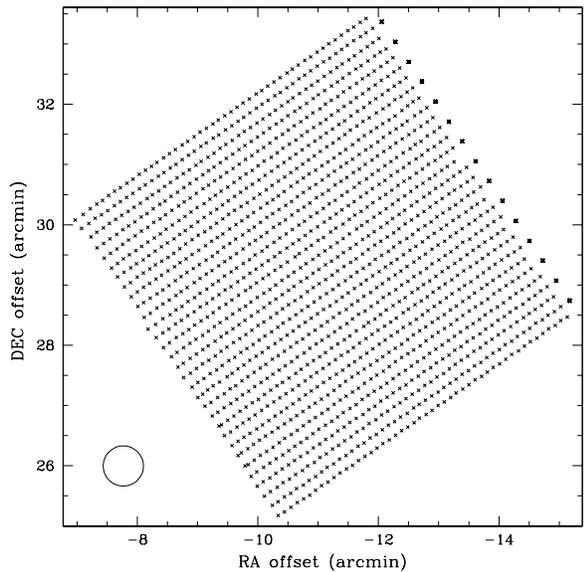}}
\vspace{-5mm}
\caption{\small Sample raster pattern for a Mopra OTF map of one of the CHaMP sources (BYF\,36).  The reference position is located at (0,0) in these coordinates.  The position angle of this map corresponds to the orientation of the Galactic Plane at this location.  Another map in the orthogonal raster direction was usually made of the same area.  Also shown in the lower left corner is the effective smoothed Mopra beam in the final maps (after resampling with Gridzilla).
\label{raster}}
\vspace{-2mm}
\end{figure}

\hspace{-3mm}
The 3mm receiver at this time was a dual-channel SIS mixer which has been described by \cite{msp97}.  In order to check the pointing on standard SiO maser sources, one of the channels was re-tuned every hour or two to 86\,GHz, and then back to the observing frequency when the pointing was completed.  Tuning was achieved by manually adjusting the bias voltage(s) and other parameters for the mixer, using a custom tuning program and standard electronic diagnostics, such as measuring the sideband rejection ratio, whether the receiver was locked to the tuning frequency, etc.  However this regimen was awkward and time-consuming, sometimes taking up to 30\,min to find a satisfactory tuning, and the overhead for each pointing measurement ranged from 15--20\,min (usually) up to 45\,min on some occasions.  For this reason the pointing was not checked as often as would have been ideal, sometimes between each OTF map (see below) but usually only every two maps.  Typical pointing corrections were $\sim$10--15$''$ in these cases.

Mopra's backend at this time was a 1024-channel autocorrelator, MPCOR, with a selectable bandwidth; we chose a 32\,MHz (31\,kHz resolution) configuration, which gave respective velocity resolutions 0.085 and 0.101\kms\ for the above lines.  The observing dates for this season were July 21--25, during which time the system temperature ranged from 160--320\,K while conditions were stable enough for mapping.

Mopra's OTF mapping mode has been described by T. Wong (2005, unpublished) and summarised by \citet{BYR10}; here we give a fuller description since it is a key aspect of the survey.  The telescope is driven in a raster pattern across the sky (which can be in any of the $l$, $b$, $\alpha$, or $\delta$ directions, but for CHaMP was usually $l$ or $b$) at a rate such that the data dump interval (2\,s) from the spectrometer to mass storage is consistent with Nyquist- (or better) sampling of the sky, given the telescope beam and observing frequency.  (A diagram illustrating the sampling pattern is given in Figure \ref{raster}).  At 90 GHz this drive rate across the sky equates to approximately $6''$\,s$^{-1}$.  Each raster row is then offset by a similar amount from the previous row (i.e., $12''$ at 90\,GHz), until a square map with a size chosen by the user is built up.  The user also selects whether a reference position (which can be specified in either relative or absolute coordinates) is observed at the beginning of each row, or only once every 2 rows.  Additionally, the user can choose from which corner of the square map the raster pattern is begun, i.e. the NE, NW, SE, or SW (in the respective coordinate system being used).  Finally the frequency of hot-cold load measurements of $T_{\rm sys}$ needs to be specified; this is typically every 10--30\,min, depending on the stability of sky conditions.  [In the 2007 season, however, a noise-diode calibration system was introduced into the data stream, effectively giving continuous $T_{\rm sys}$ measurements and making separate hot-cold load scans somewhat redundant.]  Skydip measurements of the atmospheric opacity were not found to be necessary.  At the beginning of each day a calibration spectrum of a known source such as Orion-KL confirmed the long-term gain stability of the system.

In this way a typical 5$'\times$5$'$ map, in the given raster direction, can be built up over a period of about 70\,min at 90\,GHz, of which about 45--50\,min are spent on-source, the rest of the time being consumed by reference spectra and telescope slews.  Thus the OTF mapping efficiency, defined as (time on-source/clock time), is quite high, around 70\%.  In order to minimise rastering artifacts, however, a second map is usually made of the same field, but in an orthogonal rastering direction.  Including time ($\sim$20\,min) for pointing checks between each map, such a 5$'\times$5$'$ field is ``complete'' in about 2.6\,hr. Further rasters can be made  of the same field, and this not only improves the S/N in the usual way, but under variable sky conditions will also minimise noise variations across a map, which might otherwise give erratic sensitivity coverage of the user's field.  After just 2 raster maps, however, the noise variations are usually acceptable ($\lapp$20\%) in all but the worst conditions.  For all our Mopra OTF maps, reference positions were chosen from emission-free areas of the Nanten maps, usually within a degree or less of the OTF map being constructed.

\begin{figure*}[ht]
\vspace{-15mm}
\centerline{\includegraphics[angle=0,scale=0.85]{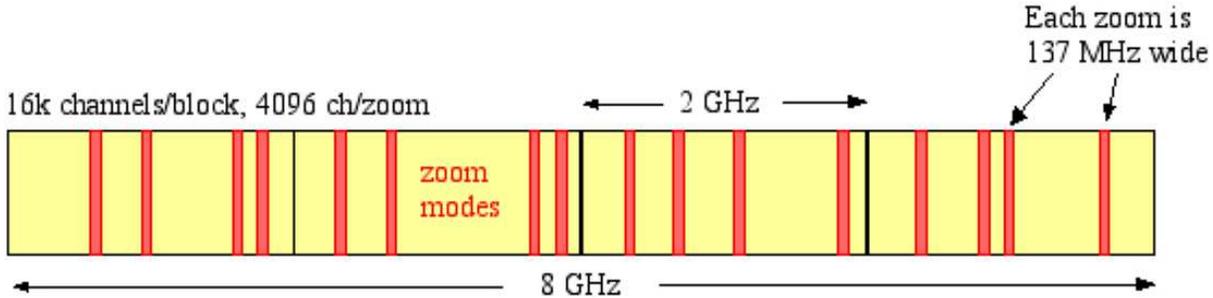}}
\vspace{-2mm}
\caption{{\small Schematic diagram (not to scale) of Mopra's MOPS digital filterbank, illustrating the flexibility available from 2007 onwards.}
\label{mops}}
\vspace{-2mm}
\end{figure*}

\subsubsection{2005 and the MMIC receivers \label{rx}}
After the pilot work described above, in 2005 we embarked upon CHaMP proper, after the installation of a new 3mm dual-channel MMIC receiver.  This frontend was as sensitive as the SIS system ($T_{\rm sys}$$<$200\,K at 90\,GHz in good conditions), but it was much more efficient to operate, since tuning was now automatically controlled in software and trivial to change.  This meant that pointing checks could easily be performed every hour or so between each OTF map, typically taking less than 10\,min to complete, and the typical corrections needed decreased to $\sim$7$''$.

On July 13--23 we made a single-raster 5$'$ OTF map in \hcop \joz\ for each of the first 50 sources in the NMC, effectively covering longitudes $288^{\circ}$$>$$l$$>$$280^{\circ}$ in our CHaMP window.  Some additional observations of the 3mm lines of \nnh, \httco, \hctn, and CS were also made of some clumps during Sept 30 -- Oct 2.  However, it was apparent that in many cases the 5$'$ fields were insufficient to contain all the \hcop\ emission, although there were some cases also where the \hcop\ was quite weak, e.g.\,for some of the Nanten \ceto\ peak positions.  In stark contrast, during the Galactic Ring Survey it was found that the CS $J$=2$\rightarrow$1 emission was {\em extremely} localised compared to the \ttco: only $\sim$2\% of the survey area had CS emission above 0.5\,K\kms (McQuinn et al. 2002).  In 2005 we found that \hcop\ was bright ($>$2\,K\kms) in $\sim$20\% of our maps, easily detectable ($>$0.5\,K\kms) in another $\sim$50\%, and weaker in the remainder.  This confirms the great efficacy of finding dense gas clumps with our finder chart strategy.  We revisit the \hcop\ brightness distribution in \S\ref{ensemble}.

\subsubsection{2006--07 and the MOPS digital filterbank}
The success in 2005 meant that we needed somewhat more spatial coverage to complete our maps in the western half of the CHaMP window, and a somewhat broader mapping strategy for the eastern half.  More significantly, it also became obvious from the sample maps of other tracers that no one tracer would give us an unbiased look at all dense molecular gas; the imperfect correlations between (e.g.) \ceto\ and \hcop\ visible in the Nanten data became even more glaring once we went to Mopra's higher resolution.  The commissioning of the MOPS digital filterbank in the early part of the 2006 observing season largely circumvented this issue, and enabled us to enlarge the scope of CHaMP at no extra cost in observing time.  This revolutionary 128k-channel backend can instantaneously accept up to 8\,GHz from the receiver, engineering a paradigm shift in how spectral line surveys can be conducted.

MOPS can be employed in either ``broadband'' or ``zoom'' mode.  With the former, the full 8.2\,GHz available bandwidth is observed with 32,768 270-kHz-wide channels in each polarisation, corresponding to a velocity resolution of 0.90\kms\ at 90\,GHz.  In contrast the latter, shown schematically in Figure \ref{mops}, allows up to eight (in 2006) or sixteen (from 2007 onwards) independently selectable 138-MHz-wide ``zoom IFs'' to be observed simultaneously from within the filterbank's 8\,GHz total instantaneous bandwidth, up to four of these per 2.2\,GHz block of the filterbank.  Although there is a limit to {\em how many} zoom modes can be observed within each block, the {\em positioning} of the zoom modes in frequency is completely flexible within the block, in steps of 69\,MHz (i.e.,\,half the zoom width).  Thus any 4 of 32 possible centre frequencies can be chosen across each 2.2\,GHz, giving arbitrary coverage of each block.  This ``interleaved'' zoom mode, which allowed better simultaneous centering of the chosen spectral lines within each zoom band, was also available from 2007 on.  Each zoom mode is then assigned 4096 channels in each of two orthogonal polarisations, resulting in a spectral resolution of 34\,kHz, or 0.11\kms\ at 90 GHz.  See Table \ref{setups} for a summary of the spectral coverage in these two seasons.

\begin{deluxetable}{llllccc}
\tabletypesize{\small}
\tablecaption{MOPS filterbank setups for 2006 \& 2007\label{setups}}
\tablewidth{0pt}
\tablehead{
\colhead{Species} & \colhead{Transition} & \colhead{Frequency} & \colhead{Utility} & \multicolumn{3}{c}{IF Zoom Number$^a$} \\
 &  & \colhead{(GHz)} &  & \colhead{Setup 1} & \colhead{Setup 2} & \colhead{Setup 3}\vspace{-3mm} \\
}
\startdata
  NH$_2$D & $J_{K+K-}$=1$_{11}$$\rightarrow$1$_{01}$ & 85.925-8 (6hf) & coldest dense gas	& & 8 & 16 \\
  SiO			& \jto\ v=1				& 86.243			& maser				& & 7 & 15 \\
  \httcn		& \joz				& 86.339-44 (3hf)	& Class I tracer			& 8 & 6 & 14 \\
  $\!\!\!\!\! \left\{ \begin{array}{l}
  \textit{\rm \httco} \\
  \textit{\rm HCO} \\
  \textit{\rm SiO}
  \end{array} \right.$ & $ \left. \begin{array}{l}
  \textit{\rm \joz} \\
  \textit{\rm $J_{K+K-}$=1$_{01}$$\rightarrow$0$_{00}$} \\
  \textit{\rm \jto}
  \end{array} \right.$ & $ \left. \begin{array}{l}
  \textit{\rm 86.754} \\
  \textit{\rm 86.777,806 (2hf)} \\
  \textit{\rm 86.847}
  \end{array} \right.$ & $ \left. \begin{array}{l}
  \textit{\rm densest gas} \\
  \textit{\rm PDR interface} \\
  \textit{\rm outflows}
  \end{array} \right\}$ & & 5 & 13\vspace{2mm} \\
  \hline\vspace{-2mm} \\
  HNCO & $J_{K+K-}$=4$_{04}$$\rightarrow$3$_{03}$ & 87.925		& chemistry	& & & 12 \\
  HCN		& \joz						& 88.630-4 (3hf)	& Class I tracer	& & 4 & 11 \\
  CH$_3$OH & $J_{K+K-}$=15$_{3,12}$$\rightarrow$14$_{4,11}$\,A & 88.940 & hot core/maser	& & & 10 \\
  $\!\!\!\!\! \left\{ \begin{array}{l}
  \textit{\rm \hcop} \\
  \textit{\rm H$^+$}
  \end{array} \right.$ & $ \left. \begin{array}{l}
  \textit{\rm \joz} \\
  \textit{\rm 59$\alpha$}
  \end{array} \right.$ & $ \left. \begin{array}{l}
  \textit{\rm 89.189} \\
  \textit{\rm 89.247}
  \end{array} \right.$ & $ \left. \begin{array}{l}
  \textit{\rm infall, outflow} \\
  \textit{\rm HII regions}
  \end{array} \right\}$ & 6 & 3 & 9\vspace{2mm} \\
  \hline\vspace{-2mm} \\
  CH$_3$CH$_2$CN & $J_{K+K-}$=10$_{91}$$\rightarrow$9$_{90}$ & 89.549 & organic chemistry	& & & 8 \\
  HNC		& \joz				& 90.664			& chemistry			& & & 7 \\
  \hctn		& $J$=10$\rightarrow$9	& 90.979			& prestellar gas		& & & 6 \\
  CH$_3$OCH$_3$ & $J_{K+K-}$=3$_{22}$$\rightarrow$3$_{13}$ & 91.474-9 (4cpts) & organic chemistry & & & 5\vspace{2mm} \\
  \hline\vspace{-2mm} \\
  CH$_2$DOH & $J_{K+K-}$=4$_{13}$$\rightarrow$4$_{04}$ & 91.587	& cold to hot gas & & & 4 \\
  CH$_3$CN	& $J$=5$\rightarrow$4	& 91.959-87 (K-lad)	& thermometer			& & 2 & 3 \\
  $^{13}$CS	& \jto					& 92.494			& dense gas, infall		& & & 2 \\
  $\!\!\!\!\! \left\{ \begin{array}{l}
  \textit{\rm \nnh} \\
  \textit{\rm CH$_3$OH} 
  \end{array} \right.$ & $ \left. \begin{array}{l}
  \textit{\rm \joz} \\
  \textit{\rm $J_{K+K-}$=1$_{01}$$\rightarrow$2$_{12}$\,E} 
  \end{array} \right.$ & $ \left. \begin{array}{l}
  \textit{\rm 93.171-6 (7hf)} \\
  \textit{\rm 93.197}
  \end{array} \right.$ & $ \left. \begin{array}{l}
  \textit{\rm cold dense gas} \\
  \textit{\rm hot core/maser}
  \end{array} \right\}$ & 2 & 1 & 1 \\
\enddata
\vspace{-2mm}
\tablenotetext{a }{See Table \ref{obshist} for dates on which these setups were used.}
\vspace{-18.5mm}
\end{deluxetable}

In 2006--07 maps were made by coadding OTF fields which abut each other on the sky to cover larger areas.  Usually the individual OTF fields were 5$'$$\times$5$'$, but they ranged in size from 3$'$$\times$3$'$ to 7$'$$\times$7$'$ where necessary to better conform to the desired coverage, given the extent of emission seen in the Nanten maps.  The reference positions used for sky-subtraction during all mapping were at locations which show no emission in the Nanten CO maps.  Most map areas were scanned at least once in each of $l$ and $b$ in order to minimise rastering artifacts and noise variations, although some areas were only scanned once due to time limitations.  

In summary, for the 2004--07 austral winter seasons we mapped the brightest 121 Nanten clumps to yield 301 Mopra \hcop\ clumps (see below), but 
\begin{minipage}[t]{120mm}{
\rule{0mm}{11mm}}
\end{minipage}
simultaneously covering many other spectral lines in the 85--93\,GHz range, among them the \joz\ transitions of \hcop, HCN, \nnh, \httco, and \httcn.  At these frequencies, Mopra has a half-power beamwidth of 36$''$, an inner error beam which extends to $\sim$80$''$, and a coupling efficiency of 0.65--0.62 (at 85--93\,GHz, resp.) to sources of size a few arcmin \citep{LPW05}.


\begin{figure}[ht]
\vspace{-5mm}
\centerline{(a)\hspace{-5mm}\includegraphics[angle=0,scale=0.48]{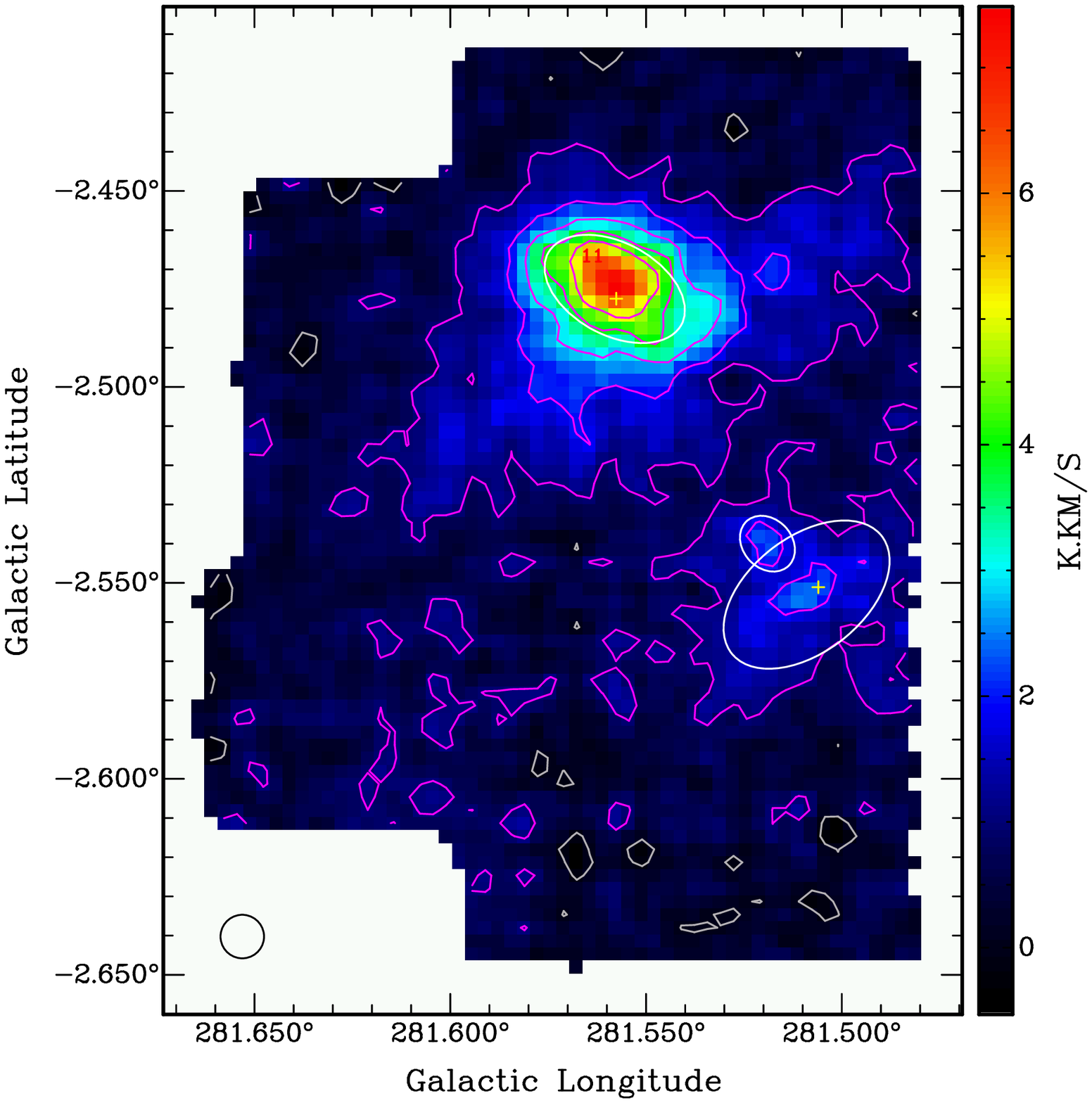}}
\vspace{2mm}
\centerline{(b)\hspace{2mm}\includegraphics[angle=-90,scale=0.28]{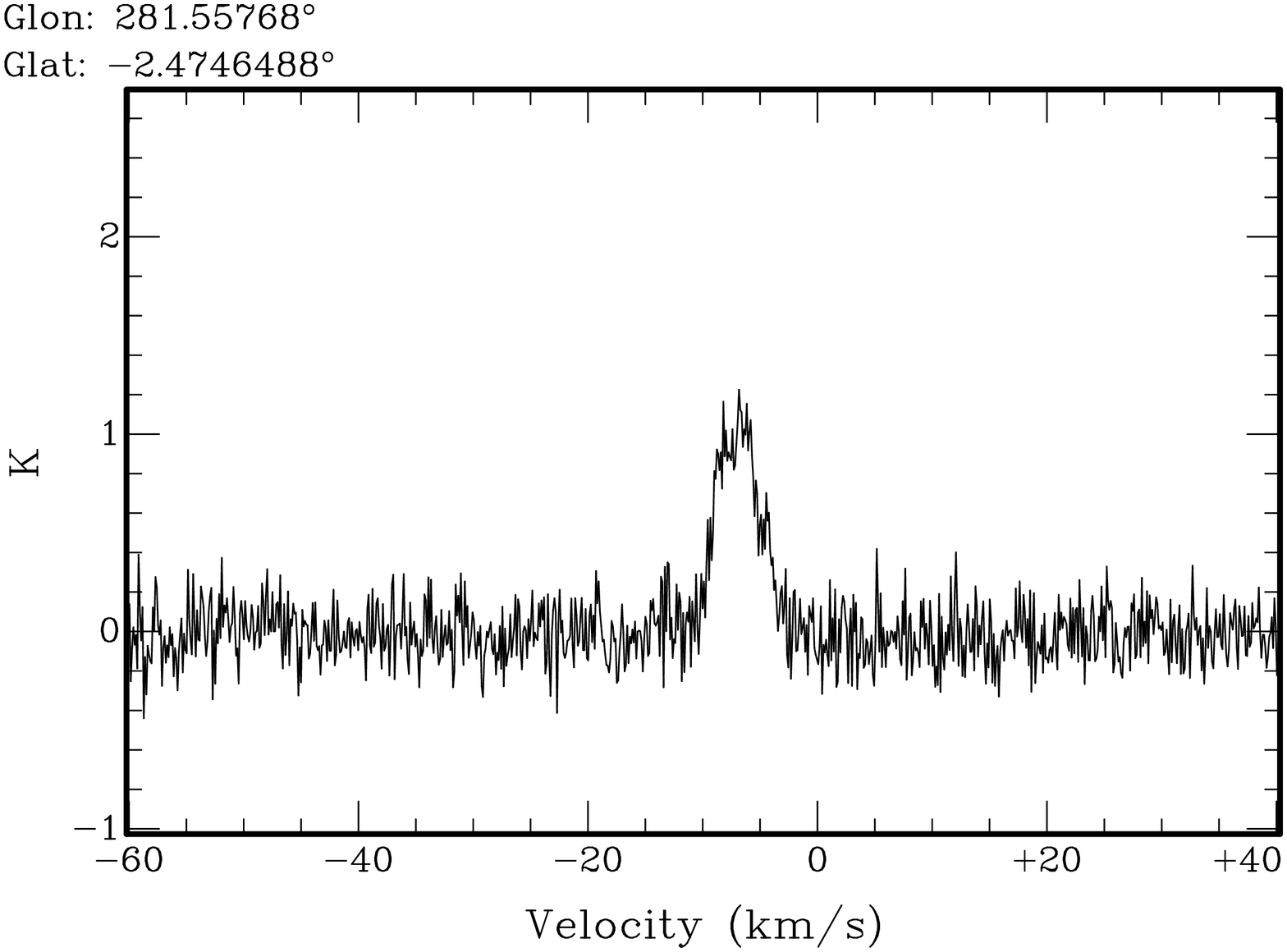}\hspace{6mm}}
\caption{\small (a) Mopra integrated intensity \hcop\ \joz\ map of BYF\,11 from the Nanten Master Catalogue, on the $T_R^*$ scale as given by the colourbar.  The integration is over the range --12 to --3\,\kms\ or 80 channels, yielding an rms noise level 0.238\,K\kms: hence the contiguous low-level emission above $\sim$0.5\,K\kms\ is real.  To aid in the identification of significant features, contour levels spaced every 4$\sigma$ ( = 0.952\,K\kms) are overlaid.  Emission peaks in ($l$,$b$,$v$) space, fitted by the gaussians listed in Table \ref{sources}, are also shown at their half-power widths.  The smoothed Mopra HPBW (40$''$) is shown for reference in the lower-left corner.  At a distance of 3.2\,kpc (see \S\ref{distances}), the scale is 40$''$ = 0.621\,pc or 1\,pc = 64$''$\hspace{-1mm}.5. 
(b) Sample spectrum from the peak of the map in (a).  Note that this spectrum is on the $T_A^*$ scale.
\label{sample0}}
\end{figure}

\subsection{Data Reduction and Processing}

As the flagship species in our filterbank setup (it is usually the brightest and most widespread), we report here only the \hcop\ results from the observing campaign described above; we leave for later papers the presentation of results of other observed species.

The raw OTF data from each season were processed with the Livedata-Gridzilla package \citep{BSD01} by bandpass division and baseline subtraction. The 2s-long OTF samples 
\begin{minipage}[t]{120mm}{
\rule{0mm}{11mm}}
\end{minipage}
were then regridded onto a regular grid of 12$''$ pixels, where the samples were weighted by $T_{\rm sys}^{-2}$, before averaging them into each gridded pixel.  Weighting by the rms$^{-2}$ of the spectra was not an option provided by Gridzilla; however as described above, since 2007 the continuously-measured $T_{\rm sys}$ has effectively given the same information for each 2\,s sample.  For all Mopra maps shown here, the effective telescope HPBW has been smoothed at the gridding stage to 40$''$ from the intrinsic 36$''$, in order to further improve the S/N.  The resulting spectral line data cubes have low but, due to variations in weather and coverage, somewhat variable rms noise levels, typically ranging from 0.2\,K up to 0.5\,K (with rare extremes up to 0.8\,K) on the $T_R^*$ scale, per 0.11\kms\ channel; the mean$\pm$SD across all maps is 0.31$\pm$0.09\,K per channel.  Although the pointing (checked on the SiO maser source R Carinae every hour or two) was typically good to 10$''$ or better ($<$1 pixel on the scale of our maps), because of the simultaneity of the spectral line mapping afforded by MOPS, the spatial registration of features between these lines is perfect.

\section{Mopra \hcop\ Maps}
\subsection{Catalogue of Integrated Intensity Maps by Region \label{maps}}

We show in the electronic edition of the Journal (\S\ref{mom0maps}) all \hcop \joz\ integrated intensity (i.e.,\,zeroth-moment) maps from our processed data cubes; an example is also shown in Figure \ref{sample0}.  In all cases the maps presented here are on the $T_R^*$ scale, where we have used a conversion $\eta_c$ = 0.64 at 89\,GHz from $T_A^*$ to $T_R^*$ \citep{LPW05}, where $T_R^*$ = $T_A^*/\eta_c$.  As mentioned in \S\ref{master} above, the sources from the NMC are organised into ``Regions'' for convenience; these are simply areas no larger than $\sim$1\degree\ \ square around groups of sources at all velocities, and do not necessarily indicate a physical association.

\notetoeditor{Figures a--d should appear 2x2 on each page} 
\begin{figure*}[ht]
(a){\includegraphics[angle=-90,scale=0.45]{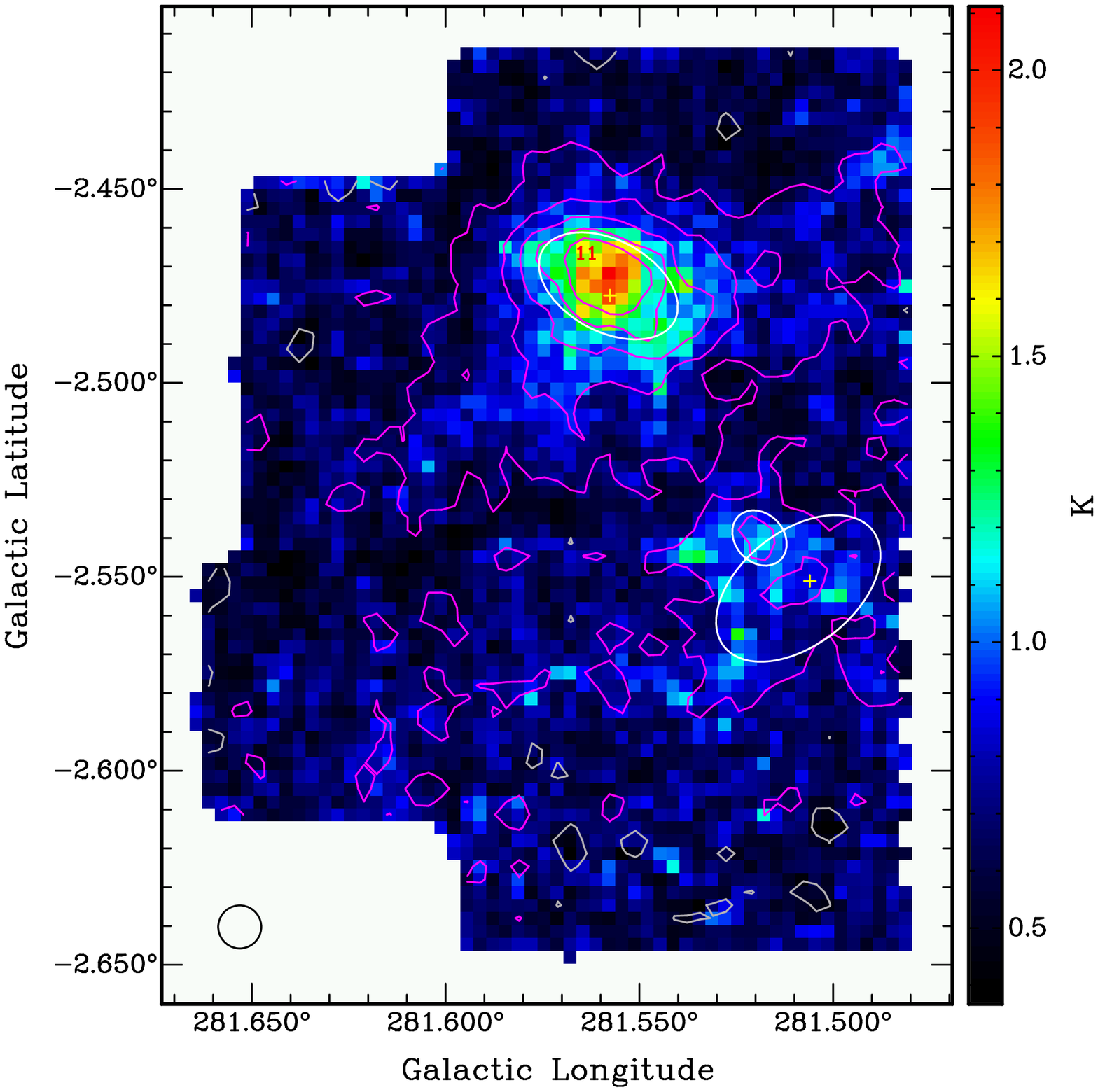}}
(b){\includegraphics[angle=-90,scale=0.45]{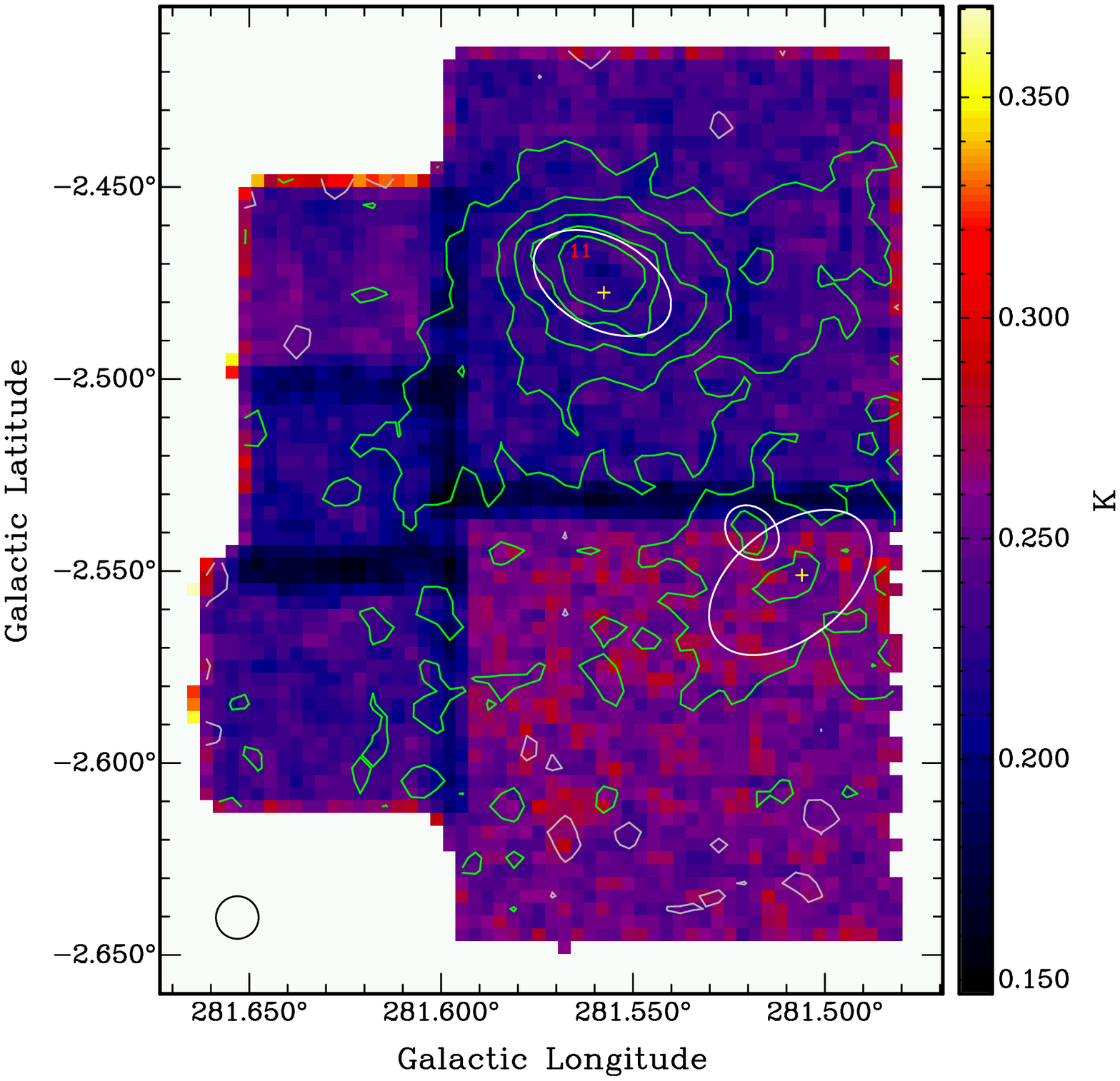}}
(c){\includegraphics[angle=0,scale=0.45]{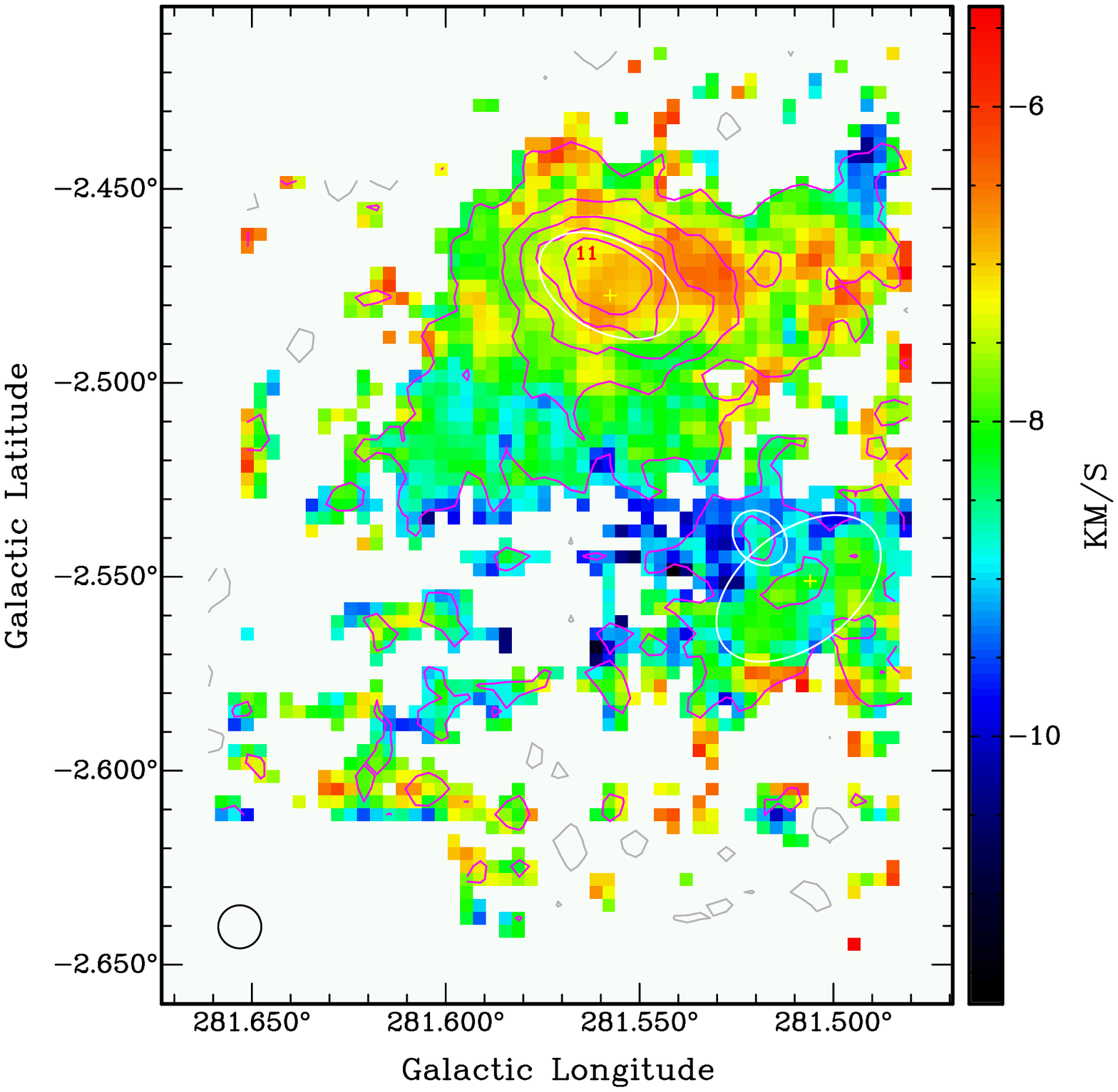}}
(d){\includegraphics[angle=0,scale=0.45]{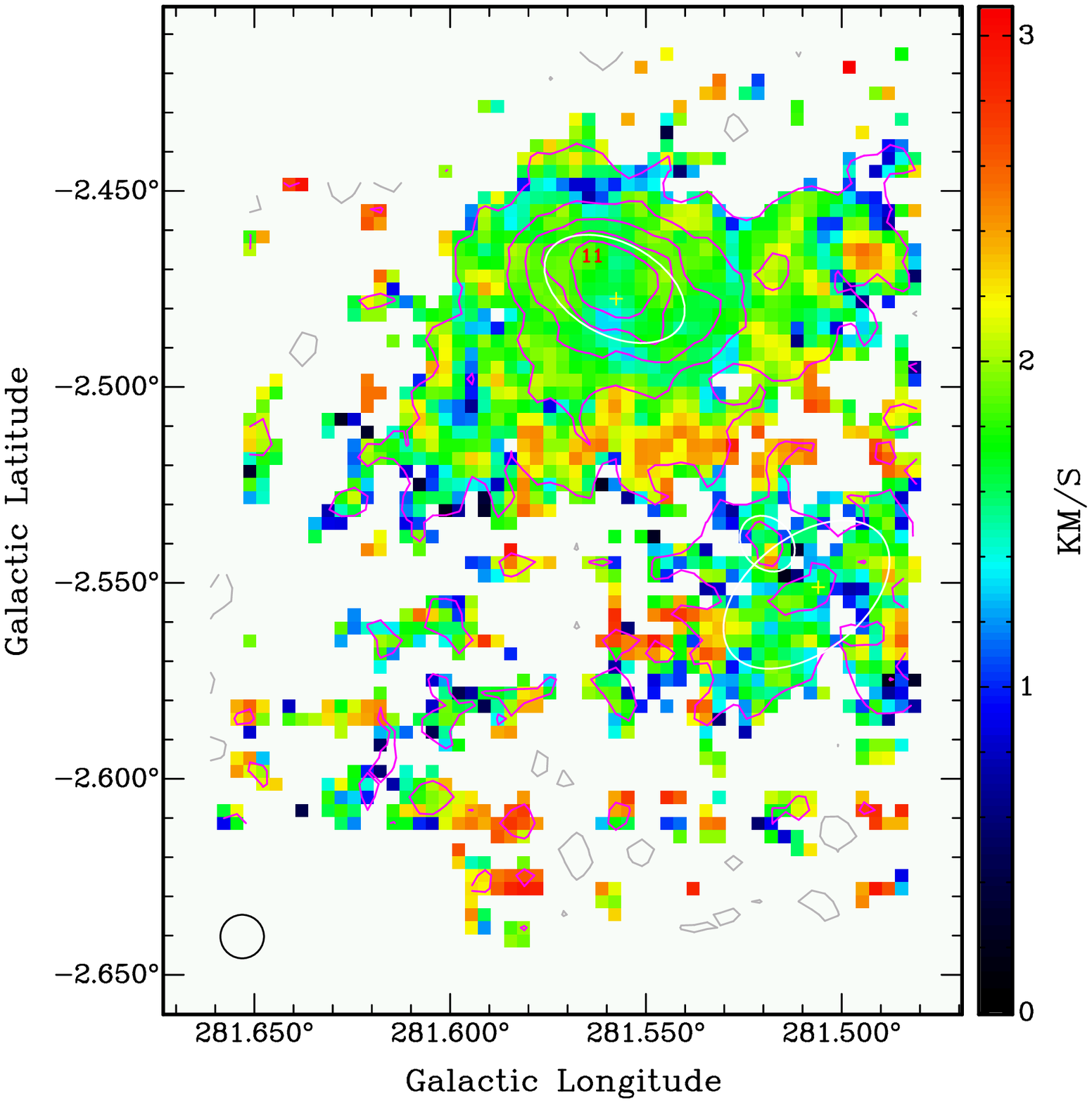}}
\caption{\small Higher-moment Mopra \hcop \joz\ images for BYF\,11, with contours of integrated intensity at 4$\sigma$ ( = 0.952\,K\kms) intervals.  All moments were calculated over the same velocity range as in Fig.\,\ref{sample0}.  The fitted gaussians and 40$''$ smoothed telescope beam are shown as in Fig.\,\ref{sample0}; at a source distance of 3.2\,kpc, 40$''$ corresponds to a linear scale of 0.62\,pc.  ($a$) Peak \hcop\ line temperature $T_{\rm peak}$.  ($b$) rms noise level $\sigma$ over line-free channels.  ($c$)  Intensity-weighted mean velocity field $V_{\rm LSR}$ (first moment).  ($d$) Velocity dispersion $\sigma_V$ (second moment).
\label{samplehigh}}
\end{figure*}

In each figure the brightness scale (shown by the colour bar) is linear and chosen to show the full range of emission features; it varies (by a factor of $>$10) from Region to Region.  Such a display, however, makes it difficult for the viewer to gauge the relative strength of features between maps.  Therefore to facilitate comparisons, we have also overlaid contours at 2- to 5$\sigma$-intervals on each map (where $\sigma$ = the rms noise level over line-free channels in that map).  While the noise level does also vary somewhat from map to map (as described above), the contours give the reader a much more intuitive way to judge what features should be viewed with suspicion vs. those that are reliably above the noise.

Sources are identified by generally requiring a combination of peak integrated intensity $>$5$\sigma$ in the zeroth-moment map and peak temperature $>$3$\sigma$ in the data cube.  For the integrated intensities (and indeed all moment maps), we limited the velocity intervals of integration to only those ranges over which emission can be reliably discerned in the full data cubes, i.e.\,contiguous areas above 2$\sigma$.  This is described further in the next section.  We also show in each figure the smoothed Mopra HPBW (40$''$) which, as captioned beneath each figure, also gives a physical size scale at the given distance to the source (see \S\ref{distances}).  Secondary peaks near a brighter source were generally considered to be separate sources if their peak position fell outside the brighter source's half-power contour; otherwise the secondary peak was deemed to be part of the brighter source, unless both had high S/N and seemed to be a well-formed double peak.

We also attempted various automated clump-finding algorithms but, in our complex maps with variable noise levels, these proved less than satisfactory.  Our heuristic approach, although difficult to reproduce algorithmically, nevertheless ensures a high reliability for source identification.  We estimate our detection rate for clumps brighter than 5$\sigma$ {\em at the position of each source} in the integrated intensity maps is at least 95\%, and likely close to 100\%.  Similarly, we estimate our false positive rate is likely less than 5\%, even in cases where the map noise varies with position due to known incomplete coverage or instances of poor weather.  After allowing for these complications, one can see in these maps a wide range of emission morphologies (from compact to fairly extended, compared to the beam) and brightnesses (from the detection threshold to S/N ratios up to 100).  Altogether we identified 301 ``Mopra clumps'' from the 121 Nanten clumps mapped, reinforcing the high detection rate described in \S\ref{rx}.

\subsection{Catalogue of Higher-Moment Maps by Region \label{moments}}

In the online edition of the Journal (\S\ref{highmom}) we next show all the higher-moment maps for each Region, with an example in Figure \ref{samplehigh}.  For each Region or source map in \S\ref{mom0maps}, we give 4 panels in \S\ref{highmom} with the same contours overlaid from \S\ref{mom0maps}.  The panels are (a) the peak temperature ($T_{\rm peak}$) map, 
(b) the rms ($\sigma$) map computed over line-free channels, (c) the intensity-weighted mean LSR-velocity ($V_{\rm LSR}$) of the emission (first moment), and (d) the velocity dispersion ($\sigma_V=\Delta V_{\rm FWHM}/\sqrt{8\,{\rm ln}2}$) in the emission line (second moment).

As mentioned in the last section, each Region's data cube has been integrated over a different range of velocities to derive the various moment maps.  The velocity ranges chosen were those where contiguous emission above 2$\sigma$ was detected.  Considerable care was taken to ensure that each emission feature in ($l$,$b$,$v$) was separately identified and, where necessary, appropriately integrated.  In many maps, this was relatively straightforward since the 3D emission patterns were fairly isolated in ($l$,$b$,$v$).  In some Regions, however, multiple spectral features at different velocities overlay each other spatially, and in such cases have been separately integrated.  The most complex case was that of BYF\,99 in Region 10, requiring 18 separate ($l$,$b$,$v$) components to be fully decomposed due to its very complex variation of velocity and linewidth with position.  For each separate Mopra clump, however, all moment maps were integrated over the same velocity range to derive the parameters listed in Table \ref{sources} for that clump.  In a few such cases, more moment-velocity ranges were computed for measurement of clump properties than are shown in the Figures (see Table \ref{sources} for all of these).  For the Figures that are shown, however, the integration ranges are given in the caption to each.  In addition, to guide their visual inspection, each of the higher-moment maps for a given Region or feature has been overlaid in \S\ref{highmom} by the same contours of integrated intensity as in \S\ref{mom0maps}, as given in each caption.


\subsection{The Mopra Source Catalogue \label{mopcat}}

For each source in these maps that could be separated in 3 dimensions ($l$,$b$,$v$) as described in \S\ref{maps}, we have measured standard line parameters which we present in Table \ref{sources}.  We also provide online versions of Tables \ref{NMC}, \ref{sources}, and \ref{physpar}, and all the moment maps described above and appearing in the Appendices, at our website {\bf www.astro.ufl.edu/champ}.  The coordinates and new designations in columns 1--3 of Table \ref{sources} are now from the Mopra \hcop\ maps as shown in Figures \ref{reg1}--\ref{reg26b}.  Here we have added letter suffixes to the NMC catalogue numbers to indicate how the higher-resolution Mopra beam breaks up the emission in the Nanten maps into smaller sub-clumps.  These coordinates are {\bf defined} by the peak positions of the Mopra integrated intensity (or zeroth moment), which values we give in column 4.  Moment values in subsequent columns of this table, measured from Figures \ref{momR1}--\ref{momR26b}, are given only at the coordinate values of column 4, even if (for example) the peak temperature at that coordinate is not a local maximum.  The velocity ranges over which all the moments in a given map were computed are given in column 5 (these are the same ranges as given in the respective figure captions).  We then respectively give in columns 

\begin{deluxetable}{ccccccccccccc}
\tabletypesize{\footnotesize}
\tablecaption{Mopra \hcop \joz\ Observed Parameters\label{sources}}
\tablewidth{0pt}
\tablehead{
\colhead{BYF} & \colhead{$l$} & \colhead{$b$} & \colhead{$\int T_R^* dV$} & \colhead{V range} & \colhead{$T_p$} & \colhead{$V_{\rm LSR}$} & \colhead{$\sigma_V$} & \colhead{$\theta_{maj}$} & \colhead{$\theta_{min}$} & \colhead{PA} & \colhead{$d_{\rm cl}^a$} & \colhead{$d_{\rm kin}^a$} \\
\colhead{no.} & \colhead{deg} & \colhead{deg} & \colhead{K\,\kms} & \colhead{\kms} & \colhead{K} & \colhead{\kms} & \colhead{\kms} & \colhead{arcsec} & \colhead{arcsec} & \colhead{deg} & \colhead{kpc} & \colhead{kpc}\vspace{-3mm} \\ %
}
\startdata
  2  &     N   &     N   &   -- (8)  & --14.0,--1.8  &  -- (6)  &     --      &   --     &    --    &    --    &   --   & 3.2(5) &   --/--   \\
  3  &     N   &     N   &   -- (7)  & --14.0,--1.8  &  -- (5)  &     --      &   --     &    --    &    --    &   --   & 3.2(5) &   --/--   \\
  4  &     N   &     N   &   -- (6)  & --14.0,--1.8  &  -- (5)  &     --      &   --     &    --    &    --    &   --   & 3.2(5) &   --/--   \\
  5a & 280.988 & --1.543 &  5.85(32) & --14.0,--1.8  & 1.60(23) & --10.6 (3)  & 1.9 (3)  &  206(12) &   82(13) & 117(3) & 3.2(5) & 1.60/1.60 \\
  5b & 281.018 & --1.587 &  3.36(32) & --14.0,--1.8  & 1.0 (4)  &  --8.5(14)  & 3.1 (3)  &   64(14) &   $<$40  & 115(5) & 3.2(5) & 1.61/1.61 \\
  5c & 281.048 & --1.503 &  2.63(32) & --14.0,--1.8  & 1.03(30) &  --6.5(13)  & 3.0 (7)  &  121(13) &   66(14) & 132(5) & 3.2(5) & 1.61/1.61 \\
  5d & 281.051 & --1.537 &  2.32(32) & --14.0,--1.8  & 0.93(20) &  --6.8 (3)  & 2.6 (4)  &  183(12) &  122(13) &  28(5) & 3.2(5) & 1.61/1.61 \\
  6  &     N   &     N   &   -- (6)  & --14.0,--1.8  &  -- (5)  &     --      &   --     &    --    &    --    &   --   & 3.2(5) &   --/--   \\
\enddata
\vspace{-6mm}
\tablecomments{An uncertainty of 1.23$\pm$0.10 is shown as 1.23(10).}
\vspace{-1mm}
\tablenotetext{a}{Distances in column 12 are from the literature \citep{gcb88,YAK05,sbb06} using various classical optical or kinematic methods (see Paper II); distances in column 13 are kinematic (in the format near/far) using \citet{r09}.}
\vspace{-5.5mm}
\end{deluxetable}

\begin{figure*}[ht]
\vspace{-15mm}
(a)\hspace{-4mm}\includegraphics[angle=0,scale=.42]{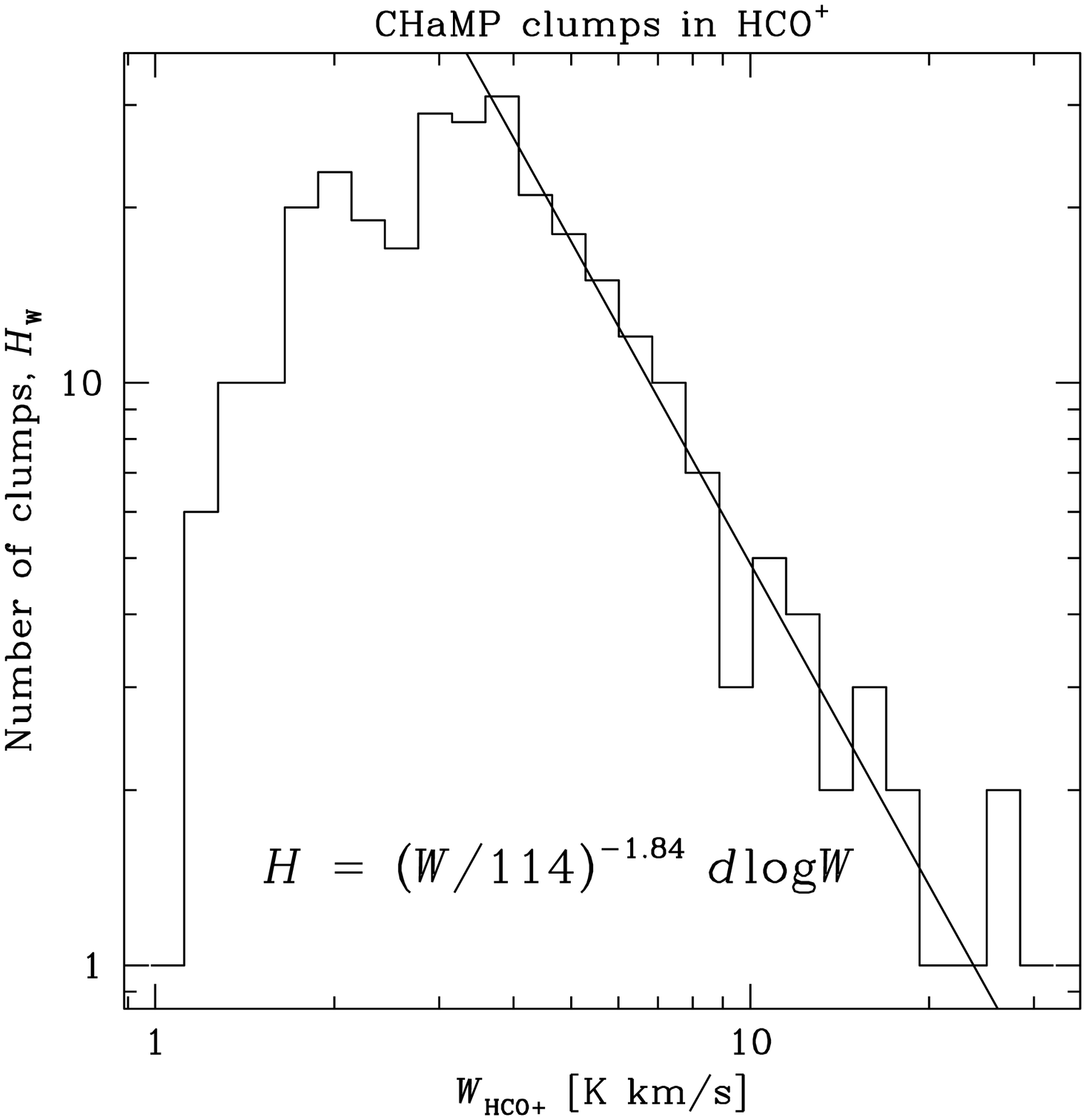}
(b)\hspace{-4mm}\includegraphics[angle=0,scale=.42]{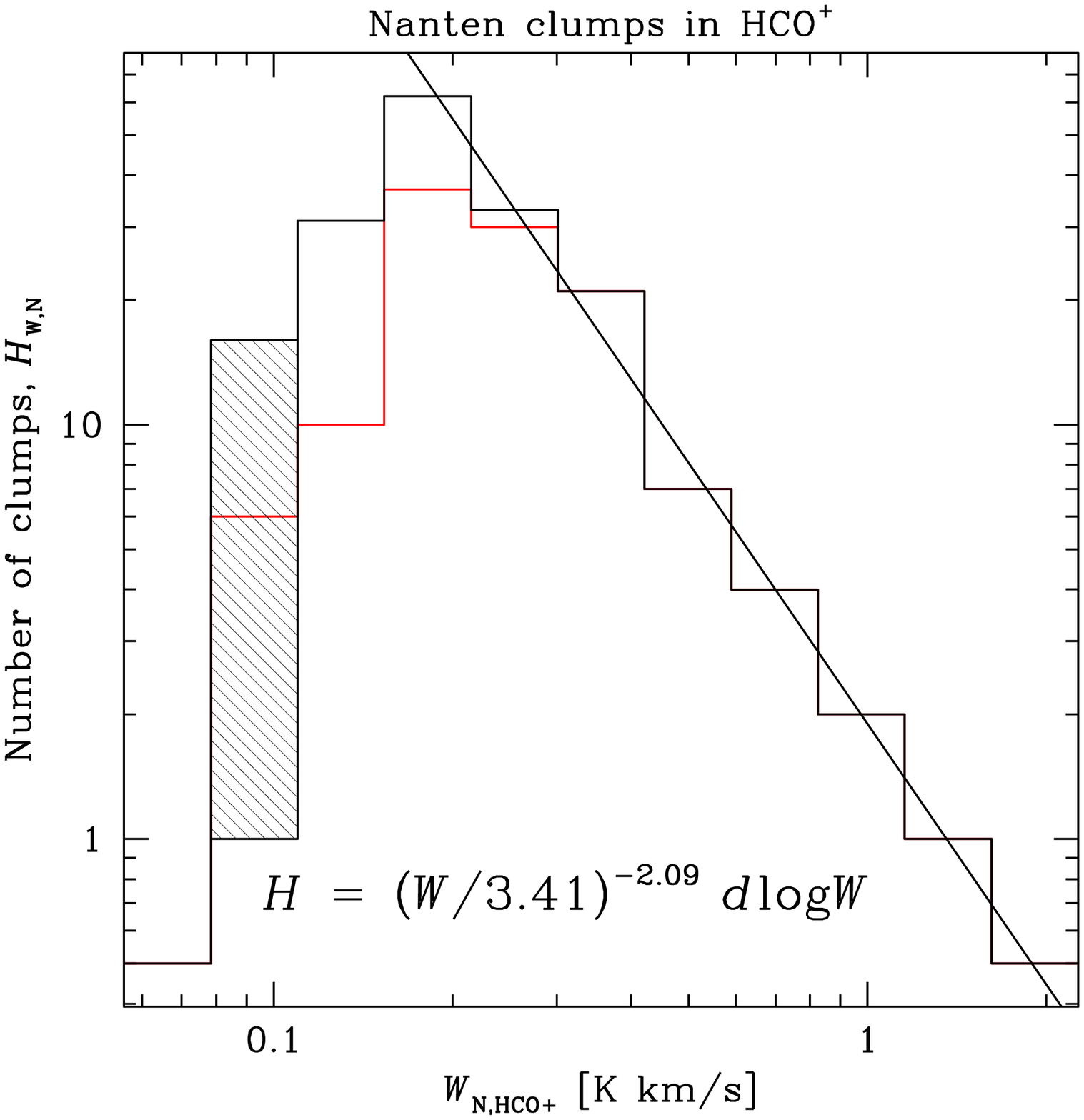}
\vspace{-5mm}
\caption{\small (a) Mopra \hcop\ source probability density function (PDF), as defined in the text.  This is just one realisation of the PDF for 27 histogram bins, therefore the fitted power and scale factor are just examples.  The mean fitted power and scale over many realisations are $W_0$ = 109$\pm$7\,K\kms\ and $p$ = 1.86$\pm$0.05.  (b) One realisation of the Nanten \hcop\ source PDF in black, and the subset of these mapped at Mopra in red.  The shaded part of the bin at $W_N$=0.093\,K\kms\ shows upper limits in \hcop\ for some of the Nanten clumps; the bins at $W_N$=0.13 and 0.18\,K\kms\ are modelled in Fig.\,\ref{theta}.  See that figure and the text for more details.
\label{sourcefn}}
\vspace{-2mm}
\end{figure*}

\hspace{-4mm}6--8 the peak $T_R^*$ (= $T_p$, see \S\ref{ensemble}) in that range $\pm$ the noise level at that position from the rms map; the intensity-weighted mean velocity (or first moment) $V_{\rm LSR}$; and the velocity dispersion (or second moment) $\sigma_V$.  The uncertainties listed in columns 7 \& 8 are from the average variation in the respective quantities in the 8 pixels (36$''$) surrounding the given coordinate.  In the integrated intensity maps, we also measured the angular scale of the emission.  This is approximated by 2D gaussians, parametrised by FWHMs and PAs along the estimated major and minor axes of the emission (suitably deconvolved from the 40$''$ resolution of the maps) around the peak positions.  These are given in columns 9--11, even where the emission pattern is not particularly gaussian; it should also be noted that these angular extents are not always symmetric about the peak positions.  Finally we give distances in the last two columns.  Column 12 lists those adopted by \cite{gcb88} (which are a combination of optical and kinematic distances for the GMC complexes of the Columbia-CfA survey in Carina), except where superseded by \cite{YAK05} and \cite{sbb06} for the $\eta$ Car GMC (Regions 9--11) and NGC\,3603 (Region 13c), respectively.  Since some of these entries are a little dated, in column 13 we also list kinematic distance solutions using the more recent algorithm of \citet[][see \S\ref{distances}]{r09}.

In a few sources no \hcop\ emission could be discerned in the Mopra data cubes at or near the nominal NMC coordinates; this is indicated in Table \ref{sources} by ``N'' in columns 2 \& 3, and only noise levels are given in columns 4 \& 6.  In other cases, multiple subcomponents of a Nanten clump could be seen in the Mopra data, but they were not completely separated at the half-power level and so unambiguous sizes were difficult to determine.  In those cases limits or no values are given in one or both size columns (9 \& 10).

It may be useful here to illustrate the source identification procedure with some brief examples, e.g., why are some sources separated into multiple components, while others fit by a single gaussian?  BYF\,40 and 99 are very bright and have complex but easily-measured velocity and linewidth variations with position: the derived decompositions were extracted from the data cubes component by component.  BYF\,26 and 97, on the other hand, are ``simpler'': the S/N is low for BYF \,26, meaning the derived decomposition was necessarily simple in order to avoid over-interpretation; and the structures in BYF\,97 are low-contrast, meaning they all had to be ascribed to a single gaussian within the source half-power contour.

\section{Analysis of Clump Properties}
\vspace{-1mm}
\subsection{Distance Determinations \label{distances}}

In this and the following sections we derive some elementary physical parameters for each clump, and present the results in Table \ref{physpar}.  For purposes of uniformity and archival utility, the physical parameters for each Mopra clump are given in Table \ref{physpar} mainly in either natural (i.e., M\solar, pc) or SI units.  In subsequent plots illustrating some of these parameters, we also show cgs axes where convenient.

In column 2 of Table \ref{physpar} we indicate which distance (from Table \ref{sources}) was used for each clump.  
Most clumps can be associated with sources at previously determined distances.  For a few clumps which lack a classical distance indicator, we have taken the near kinematic distance using the algorithm of \citet{r09}, in which $R_0$ = 8.4\,kpc and $\Theta_0$ = 254\kms\ are adopted as best fits to recent maser observations of sources over a wide range of Galactic longitudes.  In the CHaMP window, however, a large fraction (perhaps 75\%) of sources have $V_{\rm LSR}$ more negative than allowed by any rotation curve at any distance; most of the rest have $V_{\rm LSR}$ that puts them beyond the solar circle: see the longitude-velocity diagram in Figure \ref{regions}b.  In those cases which lack a classical distance, we simply take the tangent-point distance for that longitude, $R_0$cos$(l)$, as giving the best distance estimate.  The source of the distance estimate for each source is also noted in column 2 of Table \ref{physpar}.  
For the most part, we note from Table \ref{sources} that the classical and kinematic distances agree reasonably well, except for Regions 1--3 where \cite[as noted by][]{gcb88} velocity crowding makes distances quite uncertain between local and tangent positions.

\subsection{Clump Distribution Functions and Completeness of the Sample \label{ensemble}}

With such a large dataset, we plan to explore a number of applications (and anticipate other workers will make further use of the data) to address current and future questions in our understanding of massive star and star cluster formation.  However much of this would be beyond the scope of the present work.  We limit ourselves here to a few elementary analyses to illustrate the power of CHaMP's demographic approach, and leave more advanced studies for later papers.

As a first step we construct the ``\hcop\ source probability density function'' (PDF), $H_W$.  We define this observable function to be the number of \hcop\ clumps as a function of $W_{\rm HCO^+}$ = $\int T_R^* dV$, and this is plotted in Figure \ref{sourcefn}a as a histogram, binned in equal intervals of log\,$W$.  Interestingly, between a lower limit $W$=4\,K\kms\ (typically $>$10$\times$ our noise threshhold) and an upper limit $W$=12\,K\kms, $H_W$ seems to obey a power law,
\begin{mathletters}
\begin{equation} 
	H_W = (W/W_0)^{-p}~d{\rm log}W  .
\end{equation}
From a least-squares fit between these limits, 
we find values $W_0$ = 109$\pm$7\,K\kms\ and $p$ = 1.86$\pm$0.05; an example is shown in the figure.  The quoted statistical uncertainties are for different realisations of the figure for different numbers of histogram bins.

\begin{figure*}[ht]
\vspace{-15mm}
(a)\hspace{-4mm}\includegraphics[angle=0,scale=.42]{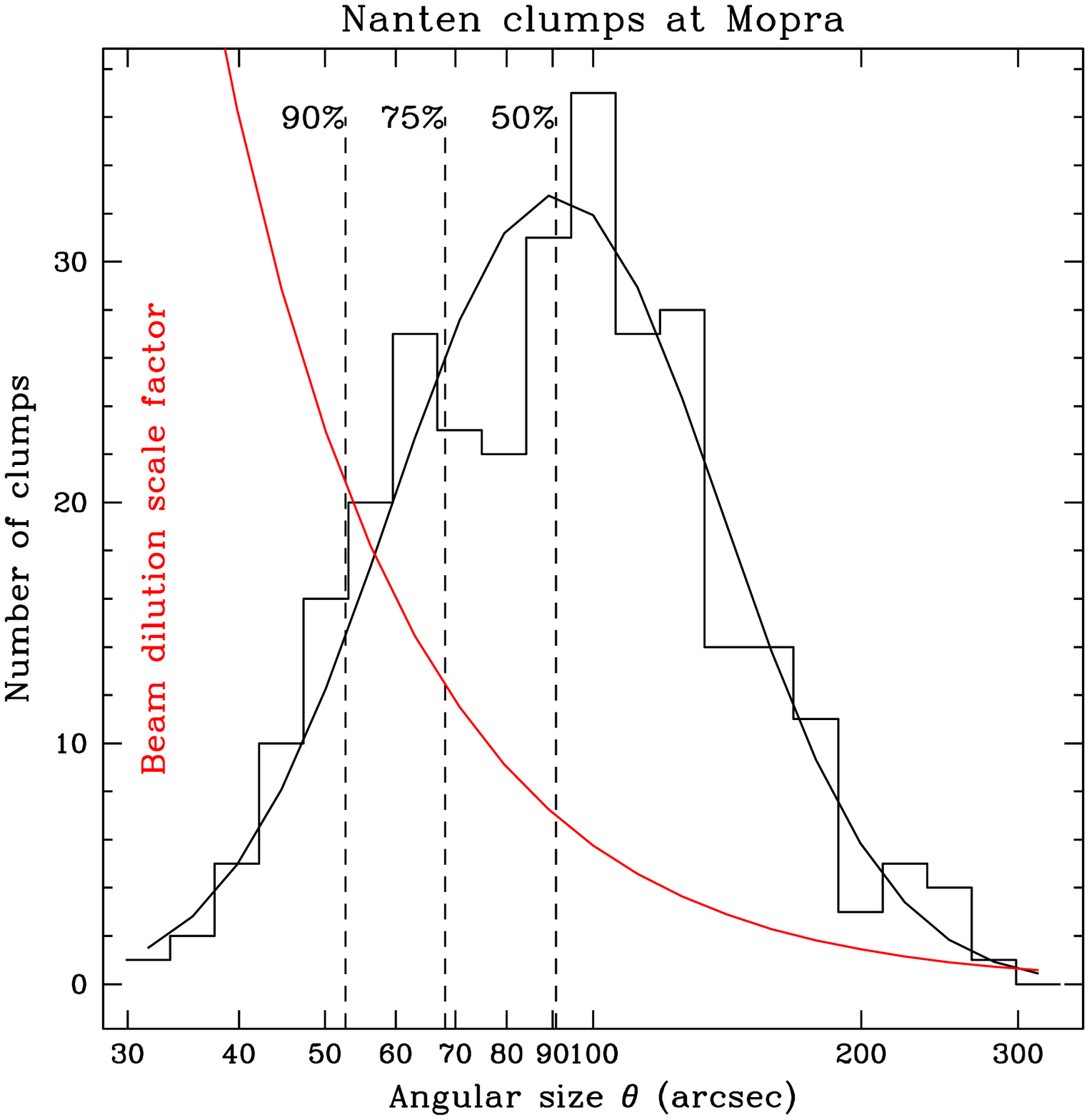}
(b)\hspace{-4mm}\includegraphics[angle=0,scale=.42]{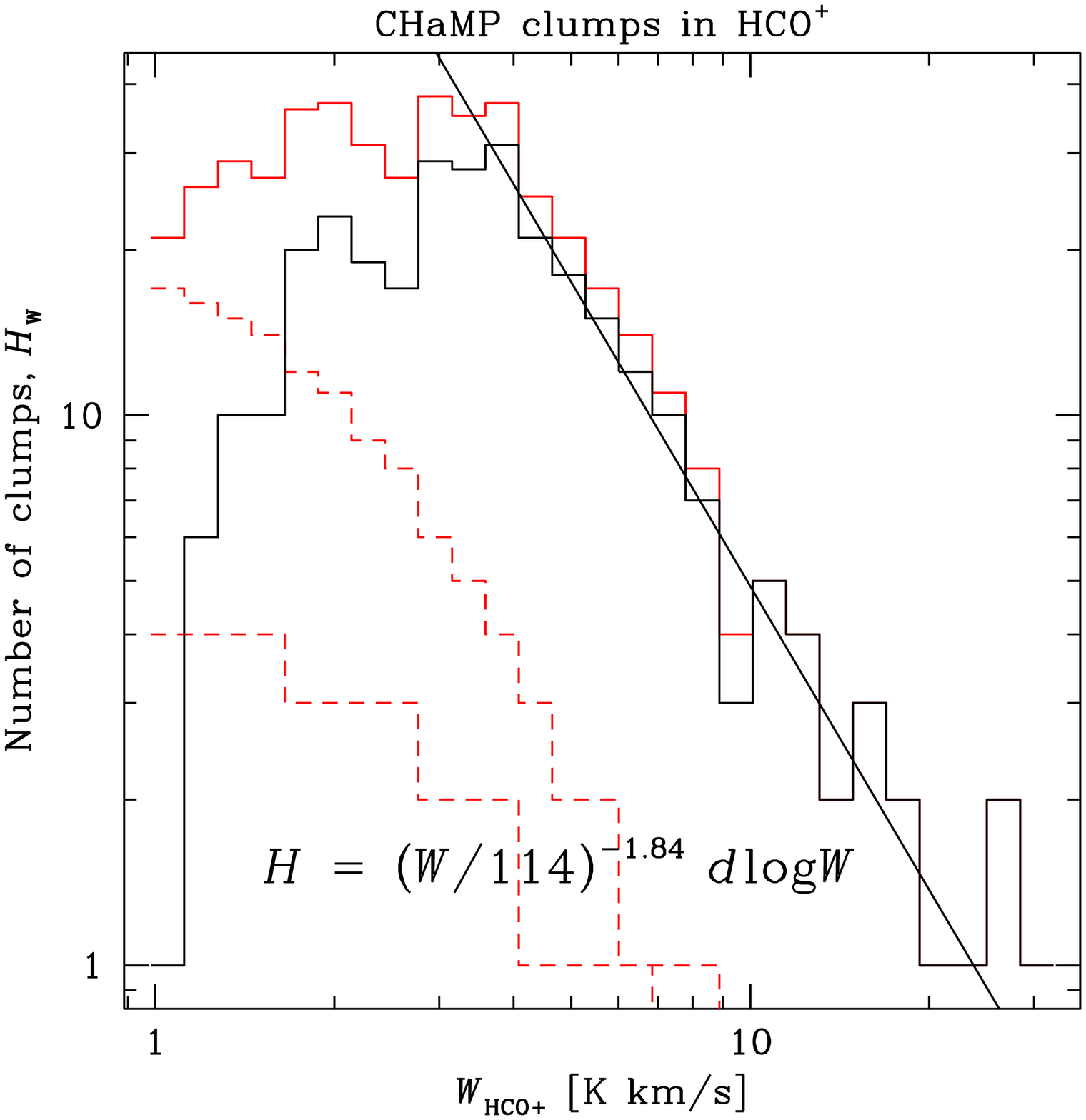}
\vspace{-5mm}
\caption{\small (a) Distribution of Mopra clump deconvolved angular sizes (histogram), overlayed by a gaussian fit. 
Sizes above which the distribution is 90\%, 75\%, and 50\% complete (53$''$, 68$''$, and 91$''$, resp.) are shown as dotted vertical lines.  Also overlayed (in red) on the same vertical scale is the factor by which a compact Nanten source's brightness would be increased in the smaller Mopra beam.  (b) Source PDF in black as in Fig.\,\ref{sourcefn}a, overlaid in red by the PDF corrected for the unmapped faint Nanten clumps, assuming a size distribution as in panel a of this Figure.  The dashed red histograms are predicted Mopra clump counts for the unmapped faint Nanten populations from each of two bins in Fig.\,\ref{sourcefn}b, while the solid red histogram includes all Mopra clumps, both observed and predicted.
\label{theta}}
\vspace{-2mm}
\end{figure*}

Above $W$=12\,K\kms\ lie the brighter sources in the tail of this distribution; we show below (\S\S\ref{ensemble}--\ref{compare}) that these ``bright-tail'' sources are distinct in several respects from the much more numerous weaker clumps.  Indeed, the identification of a large population of weaker clumps confirms a prediction by \cite{ncs08} (see \S\ref{compare}).

However, this power law may not truly represent the complete clump population.  Below $W$=4\,K\kms\ one might think our survey becomes incomplete, but the 3$\sigma$ threshhold for $W$ lies near 1\,K\kms\ (this will vary slightly depending on the velocity range over which $W$ is integrated).  Therefore all clumps with a peak $W$$>$2\,K\kms\ will have their half-power size well-determined, and we are confident we have measured all such clumps within our Mopra maps.  Incompleteness should only enter our statistics where relatively compact sources have been weakly detected in the Nanten beam, putting such sources into the weaker half of the NMC, i.e.\,those not mapped with Mopra.  But since we {\em have} mapped $>$50\% of all Nanten clumps at Mopra, we suppose a similar completeness at the 2\,K\kms\ level for our Mopra data.  This is much higher than that suggested by extrapolating the fitted power law to the same level, i.e.,\,a completeness of only $\sim$20\% at 2\,K\kms.  This suggests that the power-law distribution may turn over from the steep slope fitted to the brighter clumps.

To see whether Figure \ref{sourcefn}a's suggestion of a significant flattening of $H_W$ for $W\leqslant$ 4\,K\kms\ is real, we need to examine the Mopra data completeness more carefully.  The Nanten brightness cutoff of 0.25\,K\kms\ within its 200$''$ beam corresponds to a point-source brightness of 6.2\,K\kms\ in the Mopra maps.  (Such sources would have been easily detectable with Mopra if they had been mapped.)  Because the Nanten maps are $\sim$20\% undersampled in each dimension, however, the true point-source sensitivity in the Nanten maps may be up to $\sim$40\% worse, or 9.0\,K\kms\ in the Mopra maps.  However, at least for the 301 brighter clumps measured at Mopra, most are actually well-resolved (taking the geometric means of the sizes in columns 9 \& 10 of Table \ref{sources}), and very few are point-like (only 6 of 301 have deconvolved sizes $\leqslant$40$''$).  Thus, we model the Mopra clump size distribution as a gaussian in log($\theta$); see Figure \ref{theta}a.  From this we find that the deconvolved minimum sizes for 90\%, 75\%, and 50\% of the mapped Mopra clumps are 53$''$, 68$''$, and 91$''$, respectively. 
If we assume that any population of fainter clumps has a similar size distribution to this, then this gives effective minimum beam-filling factors of such clumps in the Nanten ``beam'' of 0.048, 0.081, and 0.14, respectively.  Such clumps would appear brighter in Mopra maps by a factor indicated by the red curve in Figure \ref{theta}a. 
Now, all but one of the Nanten \hcop\ sources brighter than 0.25\,K\kms\ were mapped at Mopra (121 Nanten sources in total).  At fainter levels, between $W_N$=0.11\,K\kms\ (roughly the Nanten 3$\sigma$ sensitivity limit) and 0.25\,K\kms, there are 44 more Nanten clumps; another 15 Nanten clumps were not detected in \hcop\ above 3$\sigma$, and 29 more were not mapped in \hcop\ at Nanten (giving 88 total Nanten clumps not mapped at Mopra).  Considering only the 44 fainter detected clumps, at a ``typical'' brightness $W_N$=0.15\,K\kms\ we predict that 90\%, 75\%, and 50\% of these clumps would have equivalent maximum brightnesses of 3.1, 1.9, and 1.1\,K\kms, respectively, when mapped at Mopra. 
This is close to the range of $H_W$ that we are considering.  In particular, it should be clear there will be relatively few missing Mopra sources with $W$$>$4\,K\kms.  

But we have not yet established the reality of the apparent turnover in Figure \ref{sourcefn}a.  To estimate the effect of these fainter, unmapped Nanten sources on our Mopra \hcop\ source PDF, we model the Nanten \hcop\ source PDF by
\begin{equation} 
	H_{W,N} = (W/W_{N0})^{-n}~d{\rm log}W
\end{equation}
\end{mathletters}
\hspace{-1mm}similarly to eq.\,(1a), and this is shown in Figure \ref{sourcefn}b.  Between $W_N$=0.15 and 1.0, the fit gives $n$ = 2.00$\pm$0.10 and $W_{N0}$ = 3.7$\pm$0.4\,K\kms.  We suppose that this PDF continues to lower $W_N$ (at least down to 0.10\,K\kms) and that the sharp cutoff below $W_N$ = 0.20\,K\kms\ is due only to Nanten's sensitivity limit.  We now scale up the number of Nanten clumps in the $W_N$=0.13 and 0.18\,K\kms\ bins of Figure \ref{sourcefn}b to the fitted line, from 10 and 37 clumps, to 116 and 59 clumps, respectively.  For simplicity we assume these fainter Nanten clumps all have $W_N$ at the middle value of each bin.  If we then apply the Mopra size model to the Nanten clumps in each bin, we project 264 and 55 (resp.)\,{\bf more} Mopra clumps, with a brightness and size distribution as above, would have been found.  Adding these hypothetical sources to the observed Mopra source PDF gives us the ``corrected'' Mopra source PDF in Figure \ref{theta}b.

We see now that the turnover in Figure \ref{theta}b is softer than in Figure \ref{sourcefn}a, but is still present.  We could continue to model even fainter Nanten clump populations in the same way, but we find that the turnover is still present in the Mopra source PDF, although the turnover becomes progressively more gentle as fainter sources are included.  We note, however, that the position of this turnover ($W$=4\,K\kms) doesn't change as fainter Nanten sources are modelled.  Moreover, this model depends on the Nanten PDF continuing down to very low levels with the same power law $n$=--2.0, but we don't have sufficient data to extrapolate this law so far.  Therefore we cannot rule out the possibility of the turnover in the Mopra source PDF being real, although the value of the slope below $W$=4\,K\kms\ is quite uncertain.  Ideally, completing the Mopra survey to cover the fainter half of the Nanten clumps would give a strong test of the form of the source PDF at lower brightness levels.

If confirmed, this turnover means that there is a ``characteristic'' brightness $W_c$ = 4\,K\kms, above which the clump population scales as a power-law, but below which perhaps either the clump formation or line emission mechanism might change.  For typical excitation temperatures, this $W_c$ corresponds to an \hcop\ column density around 10$^{17}$m$^{-2}$, or a mass column around 0.4\,kg\,m$^{-2}$ = 190\,M\solar\,pc$^{-2}$ for a standard \hcop\ abundance $X_{{\rm HCO}^+}$ = 10$^{-9}$ (see \S\ref{tausig}).

From the arguments above, we estimate our completeness at $W$=4\,K\kms\ is $\sim$90\%, while we estimate minimum 75\% and 60\% completeness at 3 and 2\,K\kms, respectively, even if the Nanten source PDF power law continues down to very low levels.  
If there is even a modest break in the Nanten power law, these completeness levels would be somewhat higher.  The modelling of the faint Nanten clump population in Figure \ref{theta}b also suggests a slight steepening of the source PDF power law above $W_c$, to perhaps $p$$\sim$2.

\begin{figure}[ht]
\vspace{-3mm}
\hspace{-5mm}\includegraphics[angle=0,scale=.42]{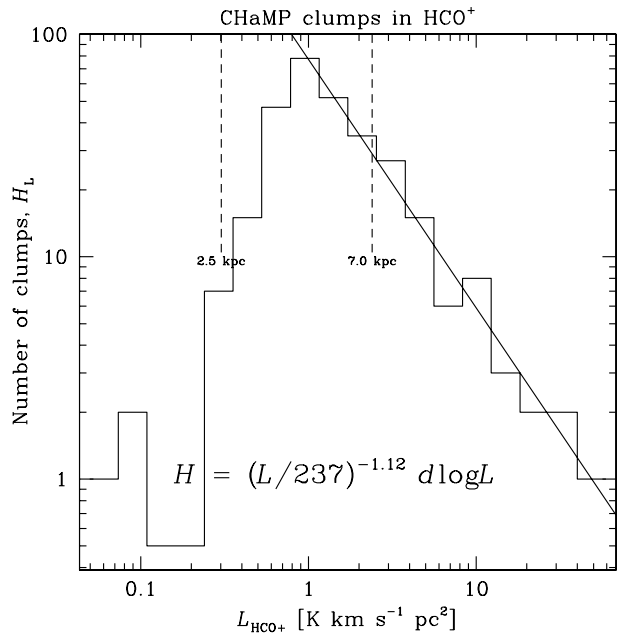}
\vspace{-6.5mm}
\caption{\small One realisation of the Mopra \hcop\ luminosity PDF for 18 histogram bins; mean values of the power and scale factors over many realisations are $L_0$ = 230$\pm$17\,K\kms\,pc$^2$ and $q$ = 1.11$\pm$0.02.  The two dotted vertical lines show our sensitivity limits at the two indicated distances.
\label{lumfn}}
\vspace{-2mm}
\end{figure}

Similarly to $H_W$, we also calculate the \hcop\ peak luminosity PDF $H_L$, defined as the number of \hcop\ clumps as a function of the peak line luminosity $L_{\rm HCO^+}$ = $W_{\rm HCO^+} d^2\theta_{\rm HPBW}^2$ in equal-sized bins of log$L$; this is plotted in Figure \ref{lumfn}.  Here we also have $H_L$ obeying a power law,
\begin{equation} 
	H_L = (L/L_0)^{-q}~d{\rm log}L  ,
\end{equation}
with a least-squares fit over the range 1.0\,K\kms\,pc$^2$ $\leqslant L \leqslant$ 34\,K\kms\,pc$^2$ yielding $L_0$ = 230 $\pm$ 17\,K\kms\,pc$^2$ and $q$ = 1.11$\pm$0.02, as shown in the figure.  Once again we see a flattening of the fitted power law below a characteristic peak line luminosity $L_c$ = 1.0\,K\kms\,pc$^2$.  This $L_c$ is commensurate with the $W_c$ above and the preponderance of clumps at the distance of the $\eta$ Car complex near 2.5\,kpc.  The skewing of $H_L$ compared to $H_W$, i.e. the smaller fraction of clumps below $L_c$ compared to the fraction below $W_c$ and the flattening of the power law from $p$ = --2.0 to $q$ = --1.2, is also consistent with our clump distance distribution.

This line luminosity PDF is strikingly different to recent determinations of similar functions for massive clumps.  For example, \cite{w10} mapped CS and HCN transitions in a sample of 50 clumps which were selected for their evidence of massive star formation activity, namely H$_2$O maser emission; many also contain compact or ultracompact HII regions.  They found their luminosity functions peaked between 10 and 100\,K\kms\,pc$^2$, with little evidence of a power-law distribution.  This is likely due to their small sample size, and the fact that their source selection was strongly biased towards the most luminous massive star-forming regions in the Galaxy.  We claim that our relatively unbiased source selection yields an $H_L$ in Figure \ref{lumfn} that is more likely to represent the true distribution of the population of massive clumps.

From our data we can also examine the Clump Mass Function and PDF, but we defer this discussion to \S\ref{MLF}.

Besides these PDFs, the Mopra \hcop\ clumps have three intrinsic observable parameters from which all others are derived.  These are the gaussian-equivalent mean projected radius to half-power $R_{\rm HP}$ (in pc), the peak brightness temperature $T_p$ (K), and the velocity dispersion $\sigma_V$ (\kms; equivalently, the linewidth).  The radius is derived simply from the distance as in the last section and the deconvolved angular size at each clump's half-power level (the geometric mean of columns 9 \& 10 in Table \ref{sources}); this is also listed in Table \ref{physpar}, column 6.  Although we use $W_{\rm HCO^+}$ to define the location of clump peak and linewidth measurements in Table \ref{sources}, we do this mainly for S/N reasons, and since it is essentially a combination of two of the above parameters, we don't consider $W$ to be an intrinsic parameter by itself.

\begin{figure}[ht]
\vspace{-20mm}
(a)\includegraphics[angle=0,scale=.33]{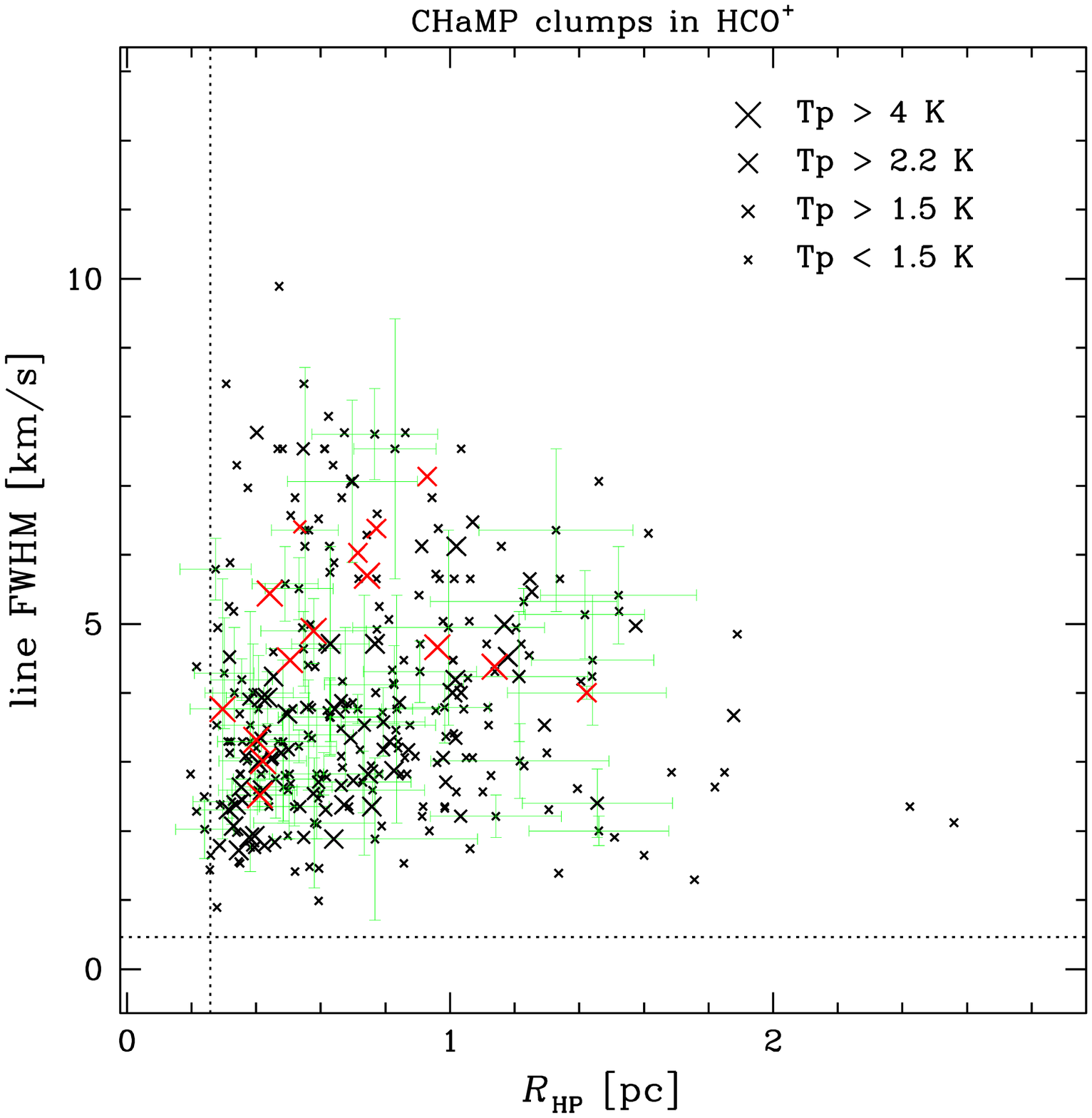}
(b)\includegraphics[angle=0,scale=.33]{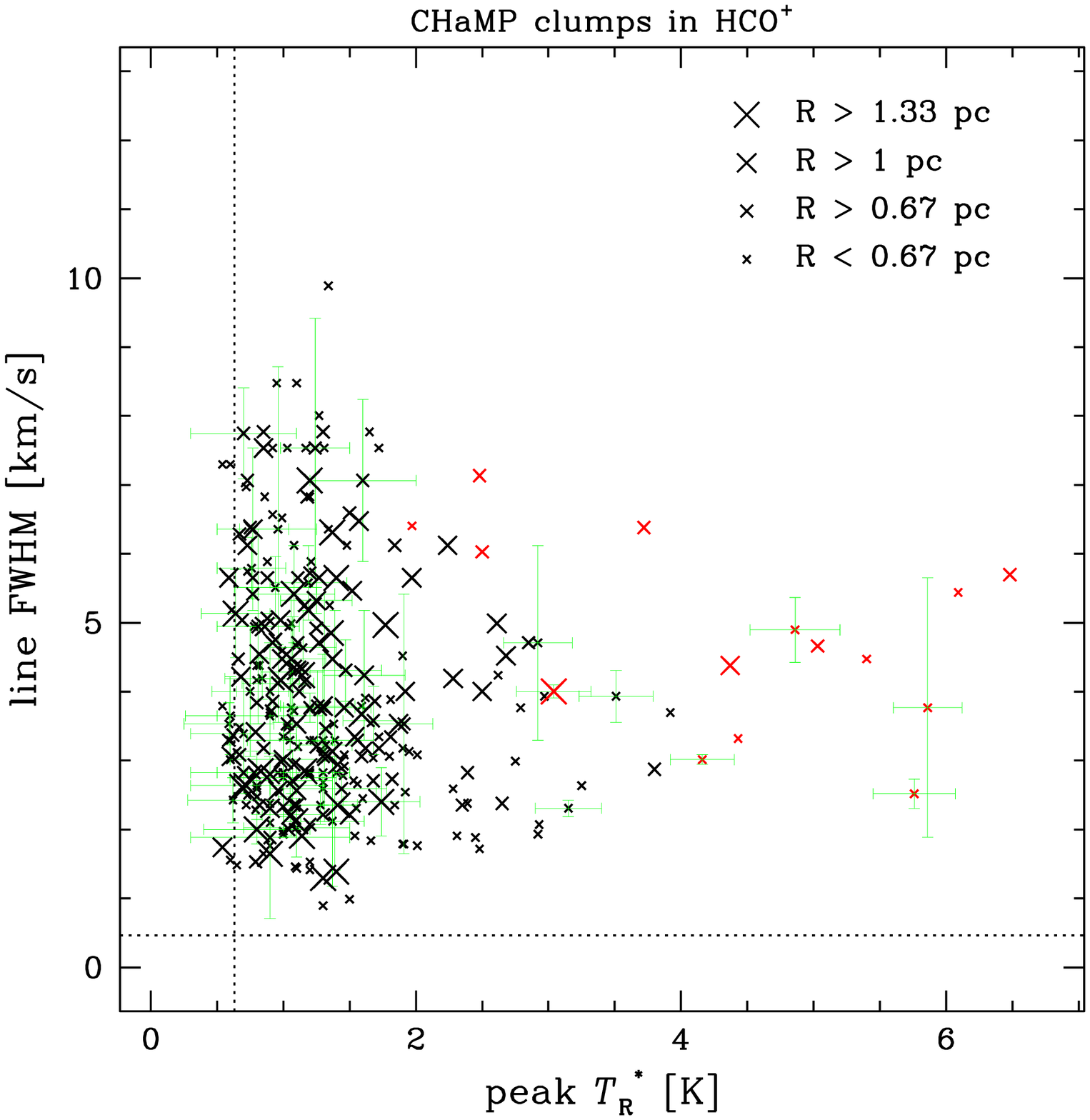}
(c)\includegraphics[angle=0,scale=.33]{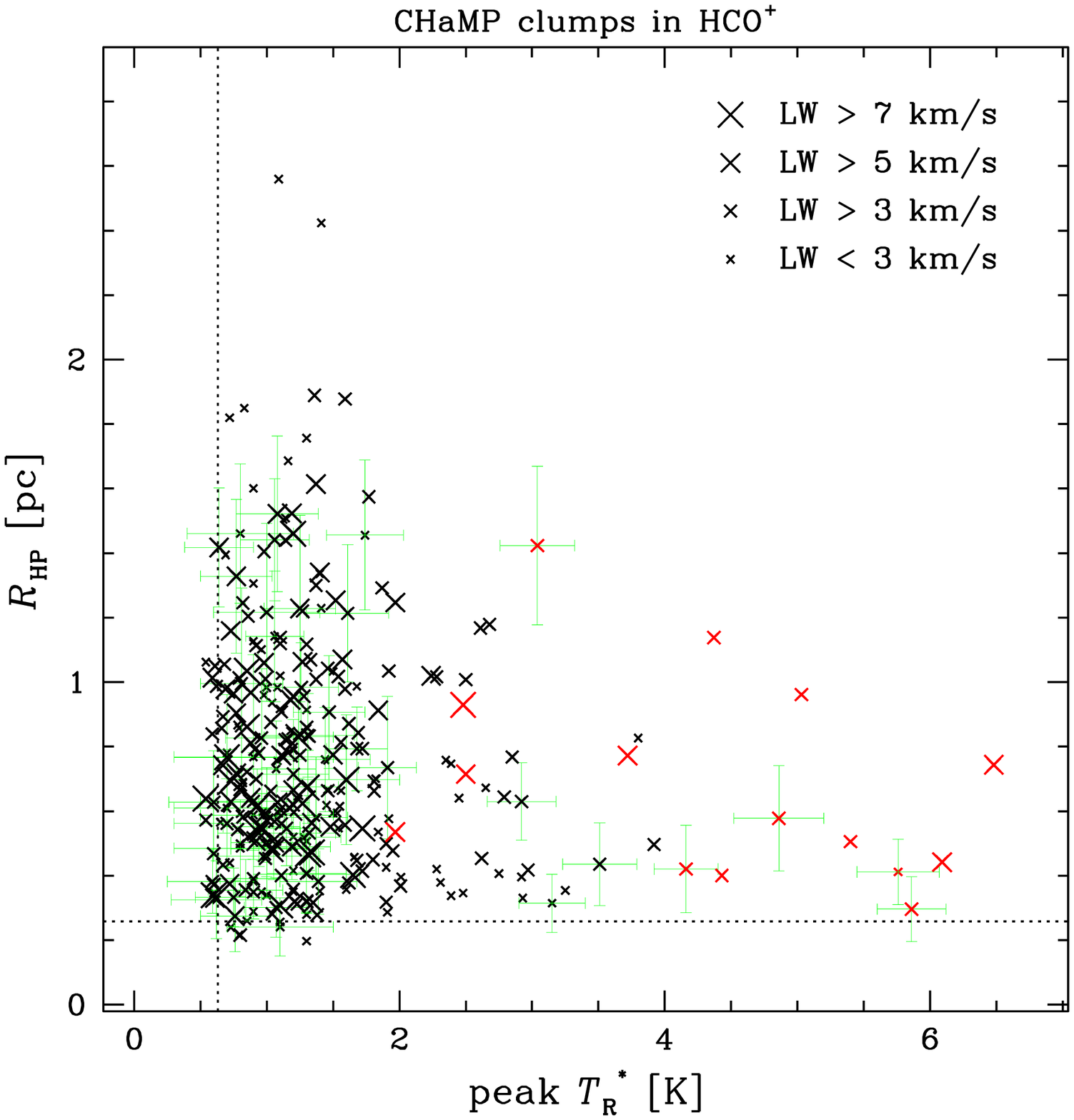}
\vspace{-2mm}
\caption{\small Observables of the Mopra \hcop\ maps.  (a) Size vs. linewidth; (b) linewidth vs. peak temperature; and (c) peak temperature vs. size for \hcop\ clumps.  In each panel the symbol size corresponds roughly to the magnitude of (a) the peak temperature, (b) the clump radius, or (c) the linewidth.  Uncertainties for 1 in 5 points are shown as green error bars and mean sensitivities as dotted lines in all panels.  
The red symbols show the brightest clumps ($W$$\geqslant$12\,K\kms) from the source function of Fig.\,\ref{sourcefn}a. 
\label{obspars}}
\end{figure}

In Figure \ref{obspars} we plot the relationships among these three parameters.  The first panel shows that the clump size seems not to be particularly correlated with the linewidth, i.e.\,that there is {\bf no} strong \citet{L81} relation of the form $\Delta V \propto R^{s}$: we obtain $s$ = 0.12$\pm$0.05.  This is unlike the situation for low-mass cores \citep{g98}, but similar to previous work on massive clumps \citep{bm92, cm95, pje97}.  For example, \cite{cm95} find $s$ = 0.21$\pm$0.03.

However, the clump size and linewidth both show an interesting relationship to the peak temperature.  Consider first Figure \ref{obspars}b: at low brightness ($<$3.3\,K), the clumps' linewidths range from quite narrow to quite broad, 1--10\kms\ (the mean $\pm$ SD = 3.9$\pm$1.7\kms). 
As the clump brightness rises, however, the linewidths become more tightly constrained, to values around 3--6\kms\ (mean $\pm$ SD = 4.0$\pm$1.2\kms) for all clumps $>$3.3\,K.  Obviously these bright clumps are the same ones that comprise the bright tail of the \hcop\ source PDF.  This relationship {\em among} our Mopra clumps is closely mimicked by the pixel variations {\em within} each clump.  Thus, for a given clump, in the fainter pixels the linewidths vary by a large amount, but become more and more constrained as one looks towards the bright peaks in each map.  This is a common feature of the Mopra clumps at all brightnesses and linewidths. 

Similarly, Figure \ref{obspars}c shows that peak temperature and clump size are related.  At low brightness ($T_p$$<$2.5\,K) we find clumps of any radius from 0.2 to 2.5\,pc; the median $\pm$ SIQR size is 0.68$\pm$0.25\,pc.  However, for clumps brighter than 2.5\,K their sizes are more constrained, 0.3--1.0\,pc, with a median $\pm$ SIQR = 0.58$\pm$0.18\,pc.  This suggests that whatever mechanism is responsible for producing bright \hcop\ emission in these clumps, its range is limited to about a parsec.

\begin{figure}[ht]
\vspace{-2mm}
\hspace{-4mm}\includegraphics[angle=0,scale=.42]{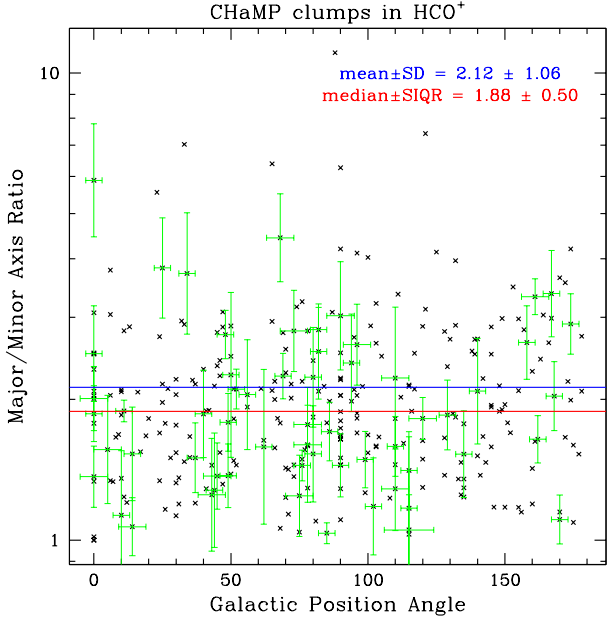}
\vspace{-8mm}
\caption{\small Major/minor axis ratio vs.\,position angle (in Galactic coordinates) for CHaMP clumps.  The typical axial ratio is around 2 as indicated, while the PAs appear to be randomly distributed on the sky.  Error bars for 1 in 5 points are shown in green. 
\label{shapes}}
\end{figure}

Since we fit the clumps by elliptical gaussians, we can also analyse the clump shapes, presented in Figure \ref{shapes}.  The mean projected aspect ratio (defined as the ratio of each clump's major to mi-

\begin{deluxetable}{cccccccccccccc}
\tabletypesize{\footnotesize}
\tablecaption{Mopra Clump Physical Parameters (for $T_{ex}$=10\,K)\label{physpar}}
\tablewidth{0pt}
\tablehead{
\colhead{BYF} & \colhead{$d_{\rm final}^a$} & \colhead{$\tau_p$} & \colhead{$N_p$} & \colhead{$\Sigma_p$} & \colhead{$R$} & \colhead{$n_{\rm col}$} & \colhead{$\rho_{\rm col}$} & \colhead{$M_{\rm col}$} & \colhead{$P_{\rm gas}$} & \colhead{$L_{\rm HCO+}$} & \colhead{$\alpha_{\rm vir}$} & \colhead{$R_{\rm Jeans}$} & \colhead{$M_{\rm BE}$} \\ 
\colhead{no.} & \colhead{kpc} & & \colhead{m$^{-2}$} & \colhead{M\solar\,pc$^{-2}$} & \colhead{pc} & \colhead{m$^{-3}$} & \colhead{M\solar\,pc$^{-3}$} & \colhead{M\solar} & \colhead{pPa} & \colhead{\tiny K\kms\,pc$^2$} & & \colhead{pc} & \colhead{M\solar} \vspace{-3mm} \\ 
}
\startdata
%
   5a &    3.2c &    0.208 &   1.06e+17 &   195.8 &   1.01 &   1.51e+09 &    85.7 &    1022 &    21.15 &    30.54 &     6.11 &   0.31 &   4879 \\ 
   5b &    3.2c &    0.086 &   7.15e+16 &   131.8 &   0.34 &   2.29e+09 &   129.9 &     136 &    84.84 &    3.458 &    54.39 &   0.25 &  17211 \\ 
   5c &    3.2c &    0.106 &    8.5e+16 &   156.7 &   0.69 &   1.68e+09 &    95.7 &     420 &    58.52 &    7.054 &    26.54 &   0.30 &  18178 \\ 
   5d &    3.2c &    0.106 &   7.37e+16 &   135.8 &   1.16 &   9.33e+08 &    53.0 &     891 &    24.40 &    15.22 &    14.71 &   0.40 &  15895 \\ 
   7a &    3.2c &    0.112 &    8.1e+16 &   149.2 &   1.33 &   8.86e+08 &    50.4 &    1307 &    25.00 &     22.7 &    12.49 &   0.41 &  18258 \\ 
   7b &    3.2c &    0.085 &   5.44e+16 &   100.3 &   1.01 &    7.8e+08 &    44.4 &     512 &    17.41 &     12.2 &    19.24 &   0.43 &  13667 \\ 
    9 &    3.2c &    0.167 &   7.62e+16 &   140.4 &   0.77 &   1.39e+09 &    78.8 &     445 &    15.61 &    8.219 &     8.75 &   0.33 &   3645 \\ 
  11a &    3.2c &    0.264 &   1.16e+17 &   214.1 &   0.84 &   1.96e+09 &   111.2 &     793 &    20.52 &    27.68 &     4.94 &   0.27 &   2754 \\
\enddata
\vspace{-2mm}
\tablenotetext{a}{Distances taken from Table \ref{sources} as follows: $c$ = literature-based value, $n$ = near kinematic, $f$ = far kinematic, $t$ = tangent point.}
\vspace{-6mm}
\end{deluxetable}

\hspace{-4mm}nor axis) is around 2, while the position angles 
seem to be randomly aligned on the sky.  Although there is an excess of clumps oriented at PAs of 0$^{\circ}$ and 90$^{\circ}$, we don't consider this significant since we attribute it to rastering artifacts, which affect the measured orientations for the $\sim$10\% lowest-brightness clumps in our sample.  However, the axial ratio is significantly different from 1: the median $\pm$ SIQR = 1.88$\pm$0.50, i.e., half our clumps have ratios between 1.38 and 2.38.  This is intriguingly similar to the aspect ratios of embedded stellar clusters: \cite{g09} found that, for a sample of 39 nearby ($d$$<$1\,kpc) young clusters, their median$\pm$SIQR aspect ratio is 1.82$\pm$0.31, and our aspect ratio distribution resembles their Figure 9b very closely, including the skewness.  A two-sided KS test cannot reject the hypothesis that the two axial ratio distributions are consistent with each other (probability 51\%).  This suggests that, structurally at least, our \hcop\ clumps may be closely related to the clusters that form within them.  In contrast, \citet{bcc06} found that their massive dust clumps were quite symmetric, with both mean and median axial ratios of 1.04.  This suggests that the dust emission may not trace the cluster-forming gas as accurately as the \hcop.  We also find it significant that our mean \hcop\ axial ratio is not very much larger than 2.  In particular, we find very few extremely filamentary objects 
(only 7 clumps with axial ratios $>$5, the other 294 clumps have ratios $<$4.5): our most elongated clump is BYF\,202a, with an axial ratio $\sim$11.  A few collections of clumps, which look more filamentary at the Nanten resolution, are broken up by Mopra's smaller beam, but these comprise a small fraction of our clump population.  Thus, filamentary structures in \hcop\ are rare at this resolution.  

\subsection{Excitation Temperatures, Optical Depths, \& Column Densities \label{tausig}}

For linear molecules with a simple rotational $J$-ladder, the calculation of physical parameters of molecular clouds usually starts with a determination of the transition's optical depth $\tau$ and excitation temperature $T_{\rm ex}$.  Often one uses two lines to solve simultaneously for these two parameters using a form of the radiative transfer equation (i.e., in the Rayleigh-Jeans limit)
\begin{equation} 
	T_{\rm mb} = (T_{\rm ex}-T_{\rm bg})(1-e^{-\tau})\hspace{1mm} ,
\end{equation}
where $T_{\rm bg}$=2.72\,K and $T_{\rm mb}$ is the brightness temperature of the emission within our telescope beam.  However, here we only have the \hcop \joz\ line available.  In order to make some simple estimates of $\tau$ and other physical quantities, we will assume two different values for $T_{\rm ex}$, and then examine how this affects our results.  The two assumed values are $T_{\rm ex}$ = 10\,K, typical of dark, low-mass clouds \citep{bm89}, and also $T_{\rm ex}$ = $T_d$ = 30\,K, typical for massive clumps observed in dust continuum emission 
\citep{fbg04,fbb05}; we further justify these choices below.  Then the peak optical depth $\tau_p$ for each clump is obtained simply from the peak brightness $T_p$ (which we use for $T_{\rm mb}$ hereafter) with
\begin{equation} 
	\tau_p = -{\rm ln}[1-T_p/(T_{\rm ex}-T_{\rm bg})]\hspace{1mm} ,
\end{equation}
and this is given in column 3 of Table \ref{physpar} for the case $T_{\rm ex}$=10\,K.  

To determine the excitation temperature more precisely, a number of methods are available in principle.  Because of its very high optical depth and the likelihood that its emission fills arcminute-scale beams, the \tco\ \joz\ line can be used as a proxy for the gas kinetic temperature in molecular clouds.  However we are still collecting such information at the Mopra resolution.  Alternatively we can use the clumps' spectral energy distributions (SEDs) to derive a bolometric temperature \citep{MAC98} or effective temperature at some wavelength.  For example, \citet{fbg04} derived $T_d$ = 32$\pm$5\,K for a sample of massive clumps in the southern Milky Way from 1.2mm continuum dust emission (they also found it necessary to fit two temperature components to the SEDs of most of their sources, with an average warm component $T_d \sim$ 140\,K).  \citet{bcc06} obtained a similar result ($T_d$=30\,K) for the dust emission from their clumps.  Other methods include temperatures from ratios of different $J$ lines of linear molecules, e.g. \citet{fbb05} found that the dense gas traced by C$^{17}$O in the same clumps as \citet{bcc06} was largely subthermally excited, with $T_{\rm ex}$=8--10\,K.  ``Thermometer'' molecules like CH$_3$CN with $K$-ladders in their symmetric-top spectra, isotopologue analysis, and other alternatives are also used.  For CHaMP we have obtained some such data, specifically the \httco \joz\ line, but this line tends to be quite weak in most clumps.  Where detectable, we will present an isotopologue analysis in a later paper.  Here we adopt values of $T_{\rm kin}$ = $T_{\rm ex}$ for our clumps as above, based on the Benson\newline \& Myers (1989), \citet{fbg04}, \citet{fbb05}, and \citet{bcc06} results.

Since we are assuming values for $T_{\rm ex}$, the optical depths so computed will only be estimates.  For those clumps with higher excitation in reality (e.g., near embedded sources), the true optical depths will be lower than estimated by eq.\,(4); conversely, in clumps with low excitation (e.g., in starless, dark clumps), the true optical depths will be higher.  
Nevertheless, we present this simplified analysis here for uniformity, while the collection and reduction of other data on $T_{\rm ex}$ is still ongoing.

\begin{figure*}[ht]
\vspace{-21mm}
(a)\hspace{-4mm}\includegraphics[angle=0,scale=.42]{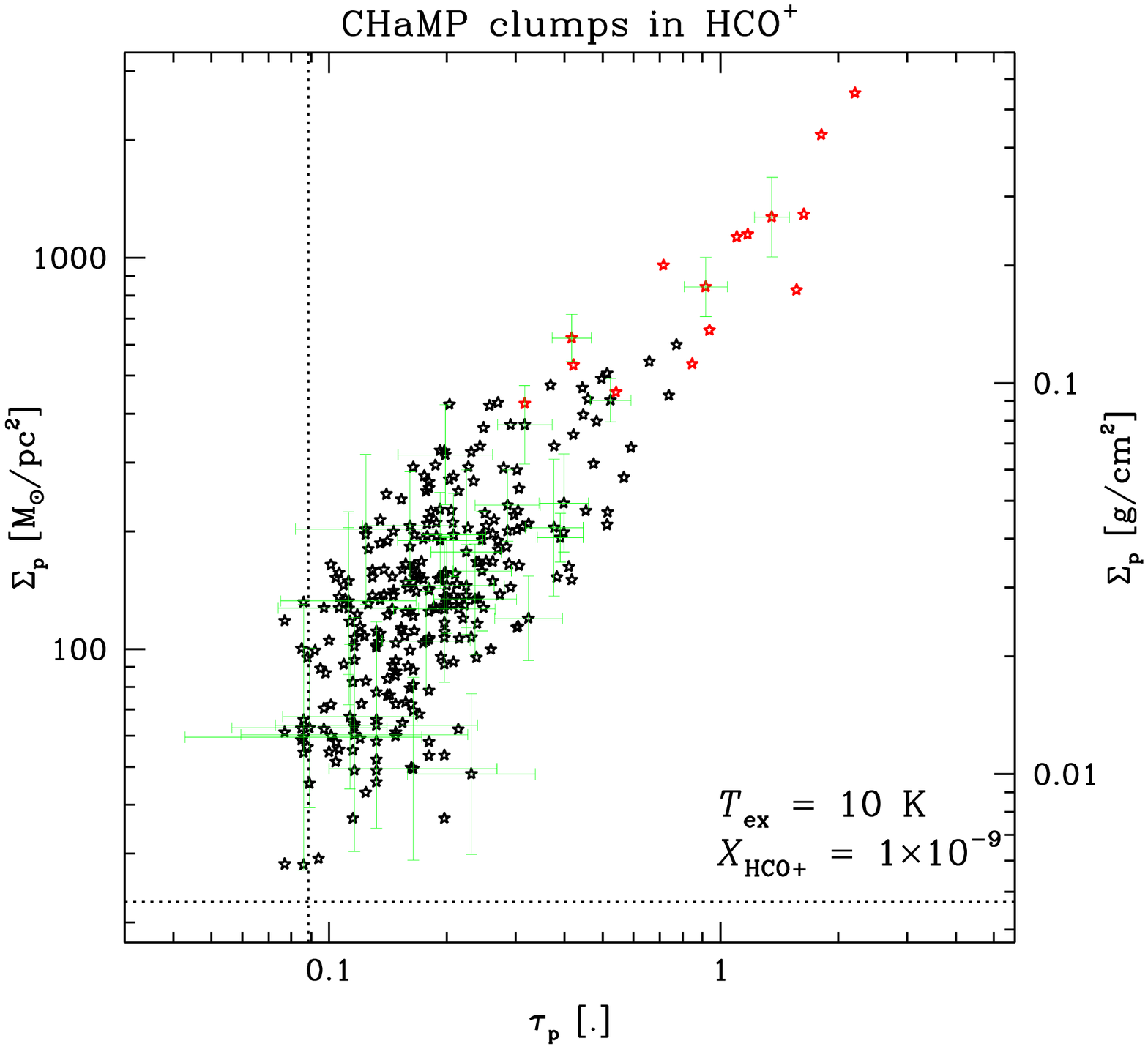}
(b)\hspace{-4mm}\includegraphics[angle=0,scale=.42]{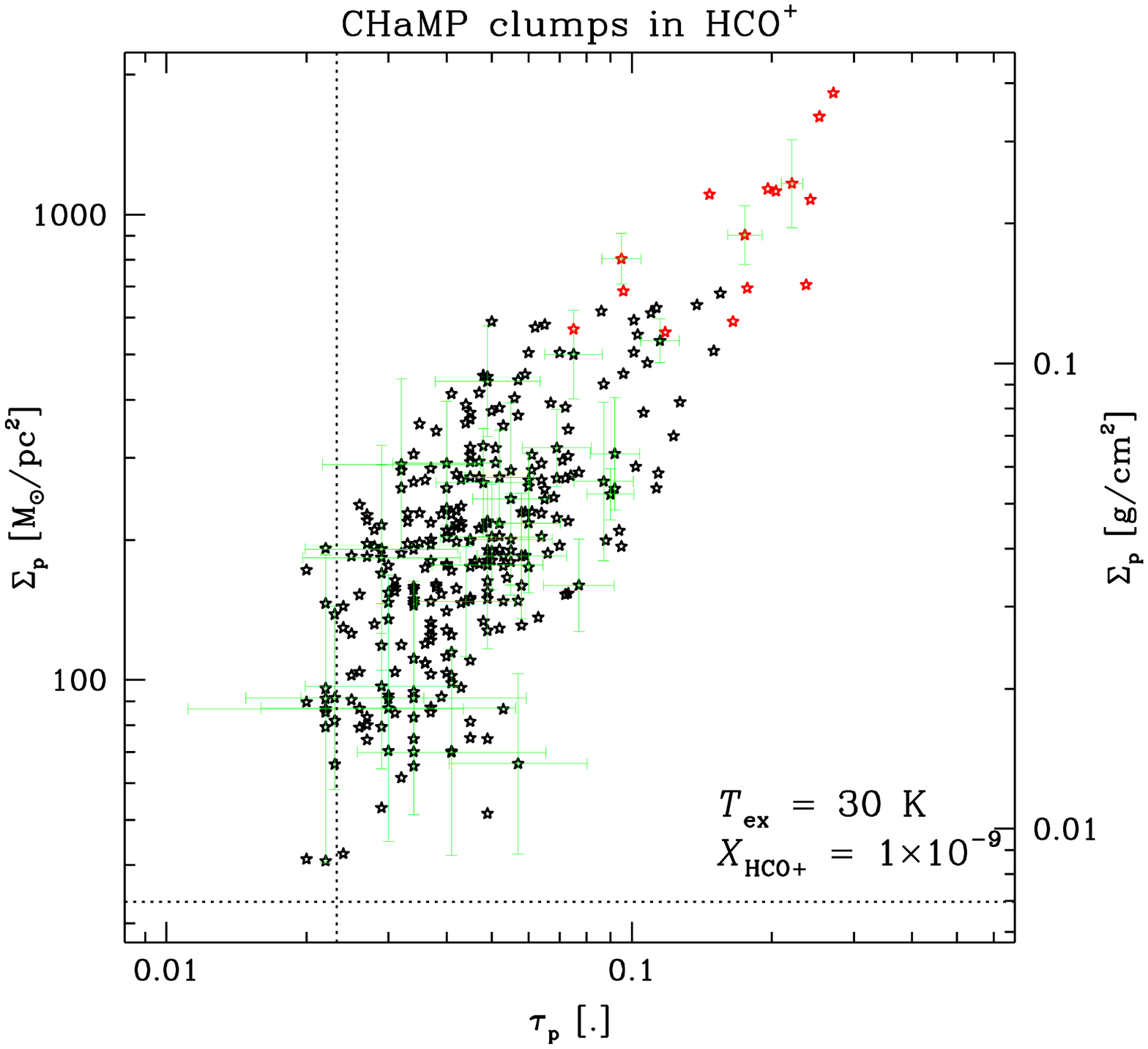}
\vspace{-6mm}
\caption{\small Mass surface density vs. \hcop \joz\ optical depth for CHaMP clumps, assuming two different $T_{\rm ex}$ values in panels $a$ and $b$.  We provide two different scales for the surface density: the natural scale from Table \ref{physpar} on the left side of each panel, and a cgs scale for convenience on the right.  Uncertainties for 10\% of the points, which are representative of all the points, are shown as green error bars, while the dotted lines indicate mean sensitivity limits for each axis: 2$\sigma$ for the optical depth in the peak channel, 5$\sigma$ for the velocity-integrated surface density.  Some points are below the mean $\tau_p$ sensitivity limit since the noise level varies from map to map.  The red symbols show those clumps from the high-brightness tail ($W\geqslant$ 12\,K\kms) of the source PDF from Fig.\,\ref{sourcefn}a.  For a given linewidth and excitation temperature, $\Sigma$$\propto$$\tau$; since we assume $T_{\rm ex}$, the scatter in each panel is due entirely to the linewidth.
\label{tauZig}}
\vspace{-2mm}
\end{figure*}

Here it is worth mentioning the special case of BYF\,73, which has been examined in detail by \cite{BYR10}.  The highly self-absorbed \hcop\ line profiles in this source do not allow a correct derivation of the \hcop\ optical depth from eq.\,(4): the correct maximum $\tau$=6.1 comes instead from a full radiative transfer treatment: see \citet{BYR10} for details.  It may be that some other clumps, e.g. a few of the bright sources, may also show higher optical depths once the \httco\ data are analysed, and this would change their derived physical parameters in a similar way to BYF\,73 (which we indicate below).  However we suggest changes to other sources' parameters will likely be less extreme than the case of BYF\,73, which we therefore view as providing a ``worst case'' for the corrections that may be needed for the analysis following.

First, we compute the column density without assumptions about whether the optical depth is large or small, and without approximations to the stimulated emission correction \cite[e.g.,][]{RW06}.  Assuming LTE applies we obtain a column density for {\em each beam-averaged line of sight}
\begin{mathletters}
\begin{equation} 
	N({\rm HCO^+}) = \frac{3h}{8\pi^3\mu_D^2J_u}\, \frac{ Q(T_{\rm ex})e^{E_u / kT_{\rm ex}}}{(1-e^{-h\nu/kT_{\rm ex}})} \int \tau dV\hspace{1mm} ,
\end{equation}
where $Q$ is the partition function for \hcop\ at the excitation temperature $T_{\rm ex}$, $E_u$ is the energy level of the upper state $J_u$ of the transition, $\mu_D$ is the molecule's electric dipole moment, and the line optical depth $\tau$ 
is integrated over the velocity range from Table \ref{sources}.  Because \hcop\ is a linear molecule, its partition function is straightforward to calculate \citep{RW06}.  At $T_{\rm ex}$ = 10 or 30\,K, $Q$ = 5.02 or 14.4 respectively, giving 
\begin{eqnarray} 
	\hspace{-4mm}N_p\hspace{-2mm} & = & \hspace{-2mm}(1.07~{\rm or}~6.02)\times10^{17} {\rm m}^{-2} \tau_p \int \phi\,dV, 
\end{eqnarray}
\end{mathletters}
\hspace{-1mm}where the velocity is in \kms\ and we take gaussian line profiles $\phi(V)$ for the integral at the positional and velocity peak of the emission in each clump: $\int \tau dV$ = $\tau_p \int \phi dV$ = $\sqrt{2\pi}\tau_p\sigma_V$.  This can also be expressed as a mass surface density
\begin{equation} 
	\Sigma_p = (N_p/X)(\mu_{\rm mol} m_{\rm H})\hspace{1mm} . 
\end{equation}
Here we use an abundance $X = 10^{-9}$ for \hcop\ relative to H$_2$, which is our best estimate from a number of studies.  For example, it is an upper limit from recent models of massive core chemistry \cite[e.g.,][]{GWH08}, which show $X_{{\rm HCO}^+}$ is a strong function of time, and is not necessarily the main charge carrier in such regions.  Thus $X_{{\rm HCO}^+}$ in massive cores may be lower than our adopted value, which is perhaps more typical of low-mass cores \cite[e.g.,][]{LWW90,CWZ02, LES03}.  On the other hand, \citet{ZCP09} obtain $X_{{\rm HCO}^+}\sim$ 2.3--12$\times$10$^{-9}$ from observations of a sample of massive clumps, although they cautioned that their values are probably overestimates.  This means that parameters derived here that depend on $X$ are quite uncertain.  

\begin{figure}[ht]
\vspace{-2mm}
\hspace{-4mm}\includegraphics[angle=0,scale=.42]{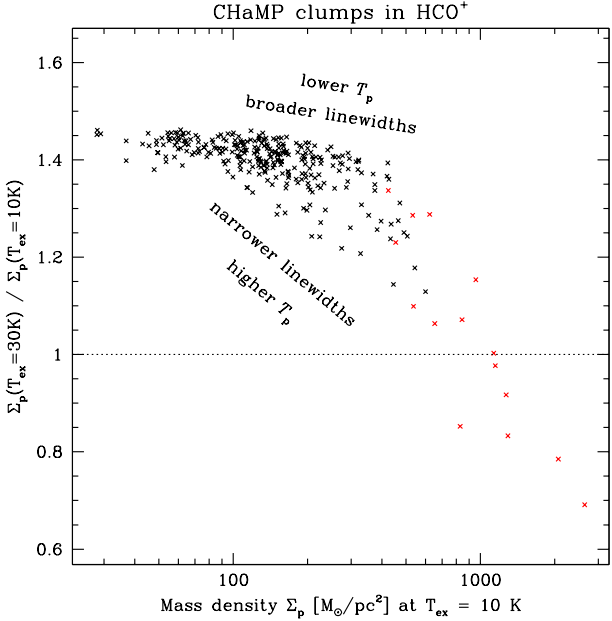}
\vspace{-8mm}
\caption{\small Ratio of column densities calculated at different assumed excitation temperatures from Fig.\,\ref{tauZig}.  The red symbols show those clumps from the high-brightness tail ($W\geqslant$ 12\,K\kms) of the source PDF from Fig.\,\ref{sourcefn}a.  The horizontal dotted line indicates a ratio of unity, i.e. where the column densities at the two assumed $T_{\rm ex}$s are equal.
\label{colrat}}
\vspace{-2mm}
\end{figure}

These peak column densities are listed (for $T_{\rm ex}$=10\,K) in columns 4 \& 5 of Table \ref{physpar}, and are shown with the optical depths in Figure \ref{tauZig} (for both $T_{\rm ex}$=10\,K and 30\,K).  Even with the uncertainties mentioned so far, it is clear that the high-brightness tail of the clump source function (the red points in Fig.\,\ref{tauZig}) forms a distinct sub-population in our clump sample, with significantly higher optical depth and column density than the rest of the sample.  For example, at 10\,K the mean $\pm$ SD of log($\Sigma_p$) for the whole clump population is 2.17$\pm$0.31 (in natural units as per Fig.\,\ref{tauZig}a), while for the 15 bright-tail sources it is 2.95$\pm$0.23.  Thus, there are approximately triple the number of sources in the bright tail compared to that expected from a pure a log-normal distribution.  The corrected point for BYF\,73 lies beyond the range of points plotted in Figure \ref{tauZig}, at ($\tau$, $N$) = (6.1, 1.70$\times$10$^4$\,M\solar\,pc$^{-2}$).

Note that we use the peak optical depths and column densities here.  We prefer this parameter over ``average'' values since with the latter, one needs to define the area or depth over which one is averaging.  In the literature of molecular cloud observations, this is often done with reference to a noise-multiple or half-power level, yet this is either not intrinsic to, or may not represent the emission from, the entire cloud.  A peak value with an assumed gaussian profile is a better approach in our view, since it allows a finite integration for the whole cloud of the relevant quantity.  If desired, one can simply use half the peak value of the relevant quantity as a representative average value.

Figure \ref{tauZig} also shows the effects of a different assumed $T_{\rm ex}$ in these calculations.  The optical depths at the lower $T_{\rm ex}$ are at least a factor of 3 higher than with the higher $T_{\rm ex}$, more so for the brighter peaks since the $T_p$/$T_{\rm ex}$ to $\tau_p$ conversion is nonlinear.  Because of this anticorrelation between $\tau$ and $T_{\rm ex}$ for a fixed observed brightness, the column densities from eqs.\,(5) \& (6) increase only weakly with $T_{\rm ex}$ --- see Figure \ref{colrat} for an illustration.  Thus, at low optical depth (most of the weaker peaks) the column densities are $<$50\% larger for the higher $T_{\rm ex}$ compared to the lower $T_{\rm ex}$ values; at high optical depth (bright peaks) the column densities are actually less at higher $T_{\rm ex}$ than at low $T_{\rm ex}$, again because of the nonlinear conversion from $T_p$/$T_{\rm ex}$ to $\tau_p$.  We conclude that the column densities (and densities \& masses, see below) are not very sensitive to the assumed $T_{\rm ex}$ in this range.


\begin{figure*}[ht]
\vspace{-16mm}
(a)\hspace{-4mm}\includegraphics[angle=0,scale=.42]{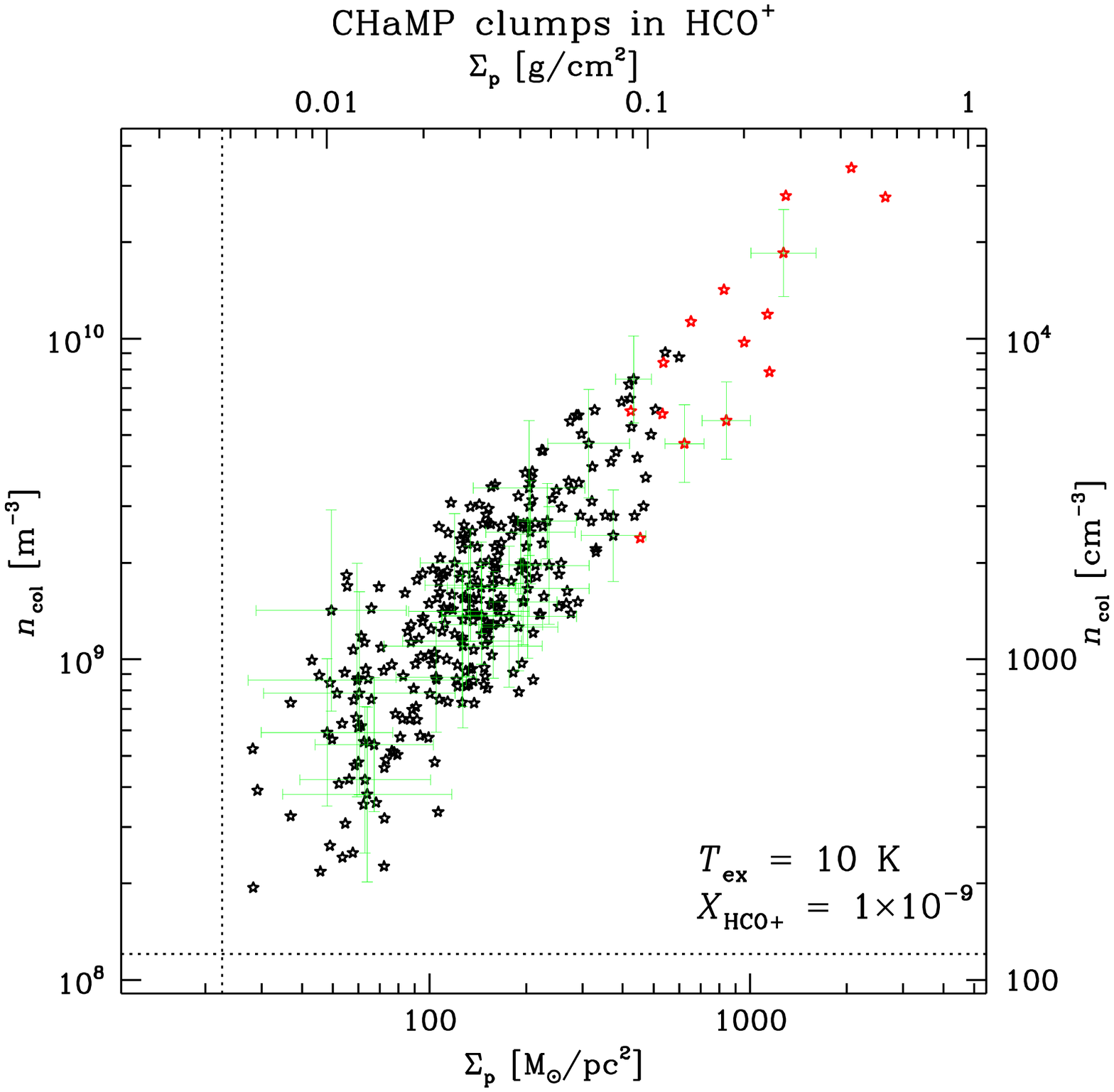}
(b)\hspace{-4mm}\includegraphics[angle=0,scale=.42]{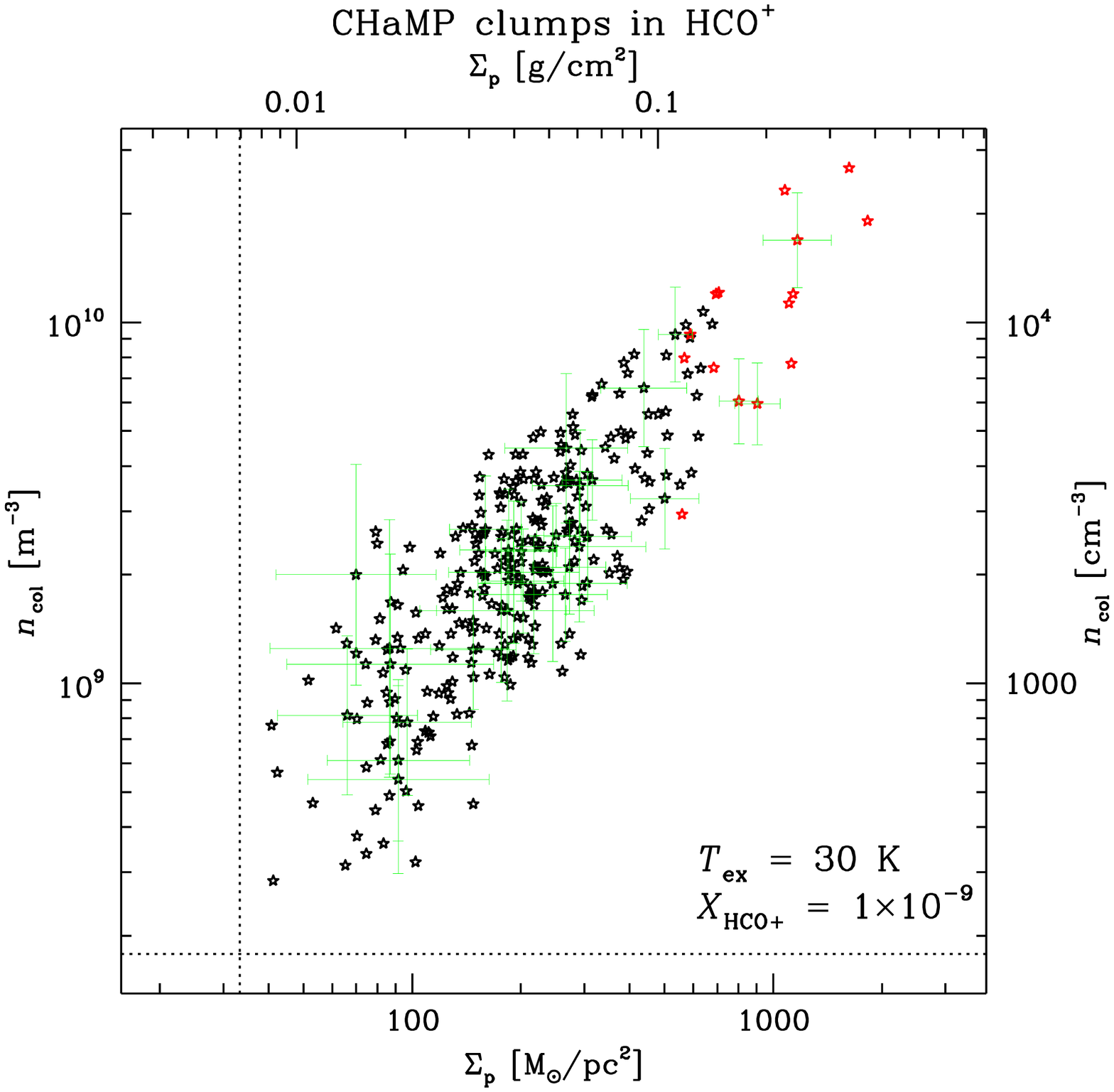}
\vspace{-6mm}
\caption{\small Volume density vs. \hcop \joz\ mass surface density for CHaMP clumps at two different assumed values for $T_{\rm ex}$.  For each panel and axis, we provide a natural and cgs scale, as in Fig.\,\ref{tauZig}.  Uncertainties, 5$\sigma$ sensitivities, and the high-brightness tail of the source PDF are also shown as in Fig.\,\ref{tauZig}.  Since $n$$\propto$$(N/R)$, the scatter in these plots is due entirely to the clump radius.  Increasing the $T_{\rm ex}$ (from panel $a$ to $b$) shifts the weaker emitters from Fig.\,\ref{sourcefn}$a$ (black points) towards slightly higher $\Sigma$ and $n$, but compresses slightly the high-brightness tail (red points).
\label{Zigdens}}
\vspace{-2mm}
\end{figure*}

\subsection{Clump Volume Densities, Masses, and Internal Pressures \label{mass}}

The gas volume density where we see \hcop\ emission is often assumed in the literature to be near the \joz\ transition's critical density $n_{\rm cr}$ of 3$\times$10$^{11}$\,m$^{-3}$ \citep{BC90}, or higher.  However, significant molecular emission can occur below a rotational line's critical density \citep{E99}.  Therefore it is important to determine whether the \hcop\ emission we see is thermalised to the local gas (i.e., H$_2$) temperature, or sub-thermally excited.  Combining the column density (eq.\,5) with the size measurement from \S\ref{distances} (hereafter we use {\bf un}deconvolved sizes in order to obtain correct beam-averaged and integrated quantities) and assuming that the physical depth of the source is comparable to the projected size gives a density estimate at the emission peak
\begin{eqnarray} 
	n_{\rm col} & = & \sqrt{\frac{\ln2}{\pi}} \frac{N_p}{RX} \\
	& = & \frac{(4.08~{\rm or}~23)\times10^{9}\,{\rm m}^{-3}} {(R/{\rm pc})(X/10^{-9})} \int \tau dV \nonumber 
\end{eqnarray}
at the assumed $T_{\rm ex}$.  Note that, since we are assuming a gaussian density profile along the line of sight, eq.(7) is effectively an average density through the clump along this line of peak emission.  An equivalent mass density is obtained from $\rho_{\rm col}$ = $\mu_{\rm mol} m_{H} n_{\rm col}$, where $\mu_{\rm mol}$ = 2.30 is the mean molecular mass in the gas.  These densities are given in columns 7 \& 8 of Table \ref{physpar} and compared with the column density in Figure \ref{Zigdens}.  From this we obtain the somewhat surprising result that {\em the \hcop \joz\ emission from nearly all of our clumps apparently arises from gas that is well below the critical density for this line}, averaged over the physical size of our telescope beam.  Thus, unless the beam-averaged optical depths have been greatly underestimated, or the filling-factor of the emission in our beam is $\ll$1, our Mopra \hcop\ clumps may be mostly subthermally excited.  The only exception to this rule is BYF\,73, for which the density (3.5$\times$10$^{11}$\,m$^{-3}$) is actually close to the \joz\ critical value for thermalisation.

This point is further illustrated by a number of mass calculations.  If we simply assume the clump volume contains gas at the \hcop \joz\ line's critical density, the clump mass is given by
\begin{eqnarray} 
	M & = & f\mu_{\rm mol} m_{H} n_{\rm cr} (\pi/\ln2)^{3/2} R^{3} \\
	 & = & 5.3\times 10^4\,{\rm M}_{\odot} f \left(\frac{n_{\rm cr}}{10^{11}\,{\rm m}^{-3}}\right)\left(\frac{R}{\rm pc}\right)^3 \nonumber 
\end{eqnarray}
where $f$ is the emission beam-filling factor, $n_{cr}$ is the critical density for the \joz\ transition, and we have used the volume for a 3D gaussian.  However, this assumes no emission comes from gas below the transition's critical density, and that $f$ for the dense gas is known and $\sim$constant across the cloud.  Over some of this emission, e.g. especially at the outer parts of a cloud, the dense gas giving rise to the emission may not fill the beam, in which case $f$ would be spatially variable and possibly $\ll$1.  Moreover, \citet{E99} has shown that many dense-gas tracers like \hcop\ can easily be excited to emit below their critical density, as mentioned above.  Nevertheless, if we compare the mass estimate from eq.\,(8) to those below, we may infer values for the required $f$.

\begin{figure*}[ht]
\vspace{-16mm}
(a)\hspace{-4mm}\includegraphics[angle=0,scale=.42]{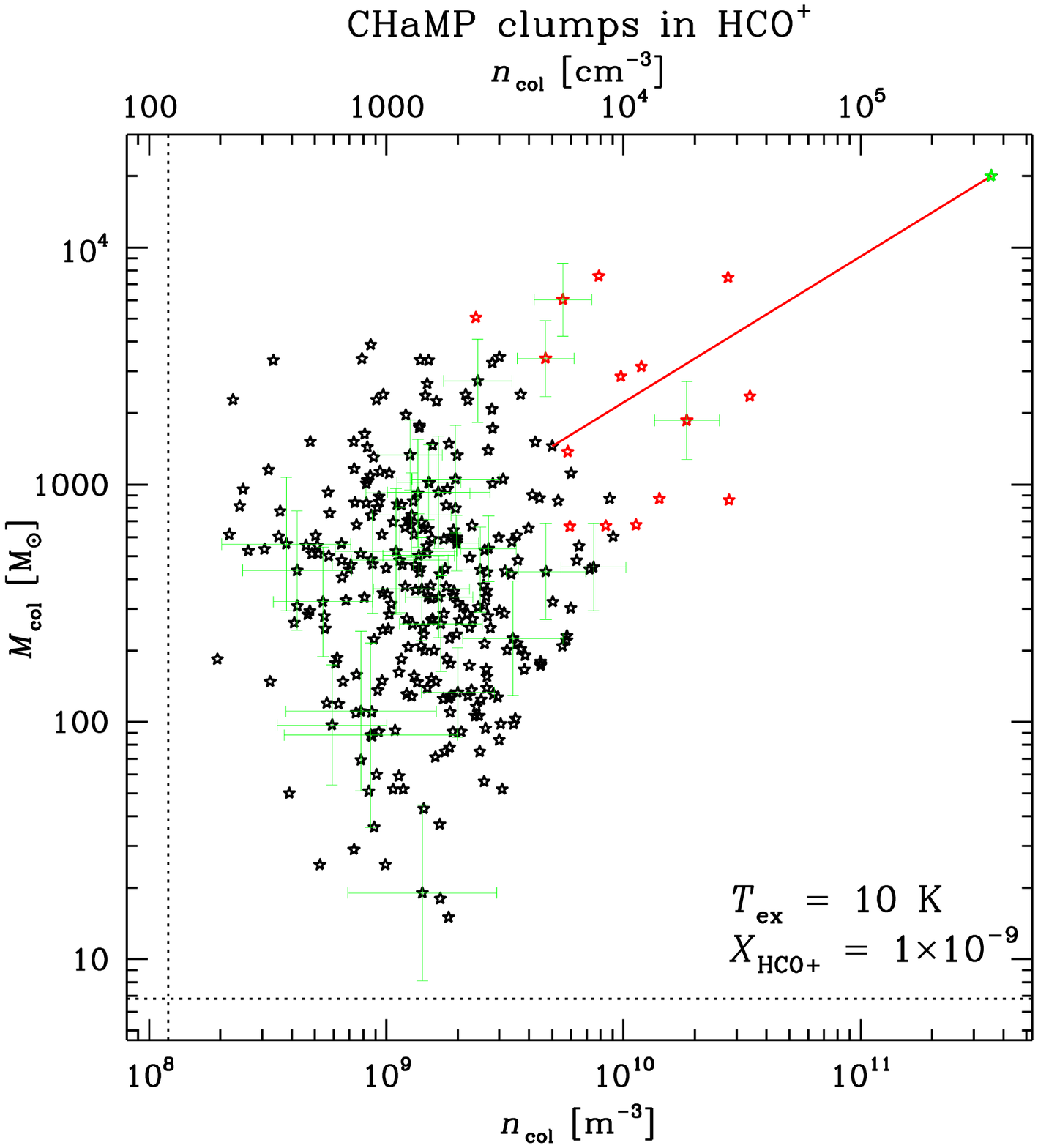}
(b)\hspace{-4mm}\includegraphics[angle=0,scale=.42]{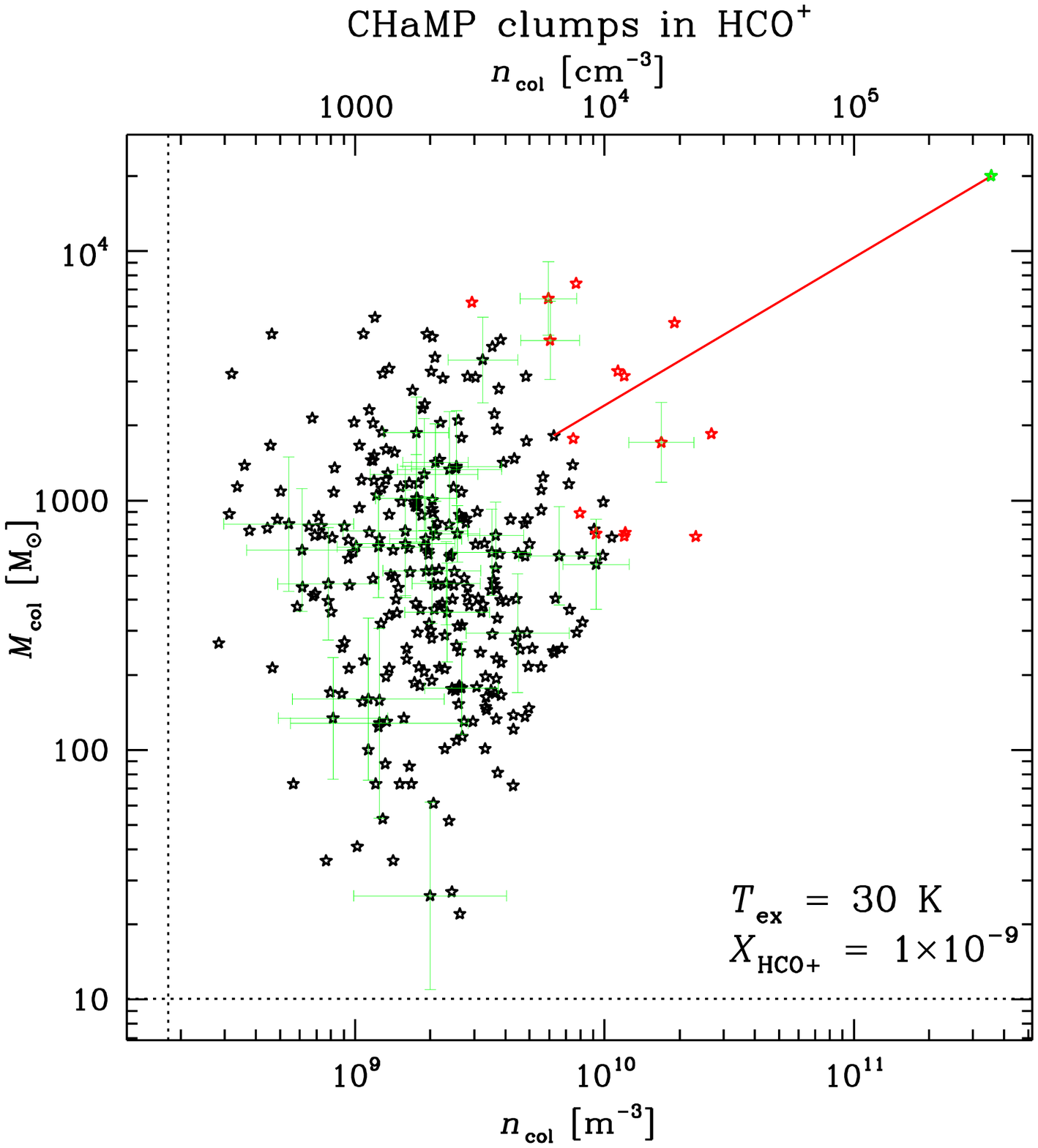}
\vspace{-5mm}
\caption{\small Mass from column density vs.\,volume density for Mopra \hcop\ clumps.  
The red line connects the values for BYF\,73 in Table \ref{physpar} to corrected points (in green) from a radiative transfer treatment by \citet{BYR10}.  Other details, including 5$\sigma$ sensitivity limits, are as in Fig.\,\ref{Zigdens}.  This figure gives perhaps the clearest indication of the physical difference between the brighter (red points) and weaker (black points) \hcop\ clumps from Fig.\,\ref{sourcefn}a.
\label{densMass}}
\vspace{-1mm}
\end{figure*}

A better mass estimate comes from the \hcop\ column density, eq.\,(5).  Integrating this over the emission region yields a total cloud mass
\begin{eqnarray} 
	M_{\rm col} & = & \frac{N}{X}(\mu_{\rm mol} m_H) \frac{\pi R^2}{\ln2} \\
	 & = & (2.24~{\rm or}~12.6)\times10^3\,{\rm M}_{\odot} \frac{(R/{\rm pc})^2}{(X/10^{-9})} \int \tau dV \nonumber
\end{eqnarray} 
for the two assumed excitation temperatures, and where the former appears in column 9 of Table \ref{physpar}.  (For BYF\,73 we estimate its LTE mass to be around 2.0$\times$10$^4$\,M\solar, significantly higher than the other clumps.)  We show the variation of our clumps' volume density with this mass in Figure \ref{densMass}.  This plot makes plain the physical difference between the weaker and brighter sources, and that the latter are unlikely to be simply the tail of a log-normal distribution for the former.  For example, in Fig.\,\ref{densMass}a the mean $\pm$ SD of log($n_{\rm col}$) for all clumps is 9.20$\pm$0.37; for the weaker sources (black points) this is 9.16$\pm$0.32, but is 10.01$\pm$0.32 for the bright clumps (red points).

Comparing the values from eqs.\,(8) \& (9), we find that our clumps are typically 1--3 orders of magnitude less massive than would be expected if the emission was coming from gas near the \joz\ transition's critical density, which is the same result as in Figure \ref{Zigdens}.  This effectively means that either the filling factor in eq.\,(8) is typically 0.001--0.1, that the optical depths are severely underestimated, or that the gas sampled by the \hcop\ emission is subthermally excited.  Although interferometric observations of \hcop\ would reveal the true filling factor, we discount the likelihood of a very low value since a majority of the clumps are well-resolved in the Mopra beam, and we see little evidence for highly clumpy substructure in those cases.  This suggests that eq.\,(8) is not likely to be applicable to our clumps.  Additionally, spot checks of \httco\ maps (currently being reduced) reveal that this isotopologue is usually quite weak (where it is detected at all, typically \lapp0.2\,K), confirming the requirement for low optical depth.  We conclude that subthermal excitation due to low density is the most reasonable explanation for our clump emission properties.

\begin{figure*}[ht]
\vspace{-16mm}
(a)\hspace{-4mm}\includegraphics[angle=0,scale=.42]{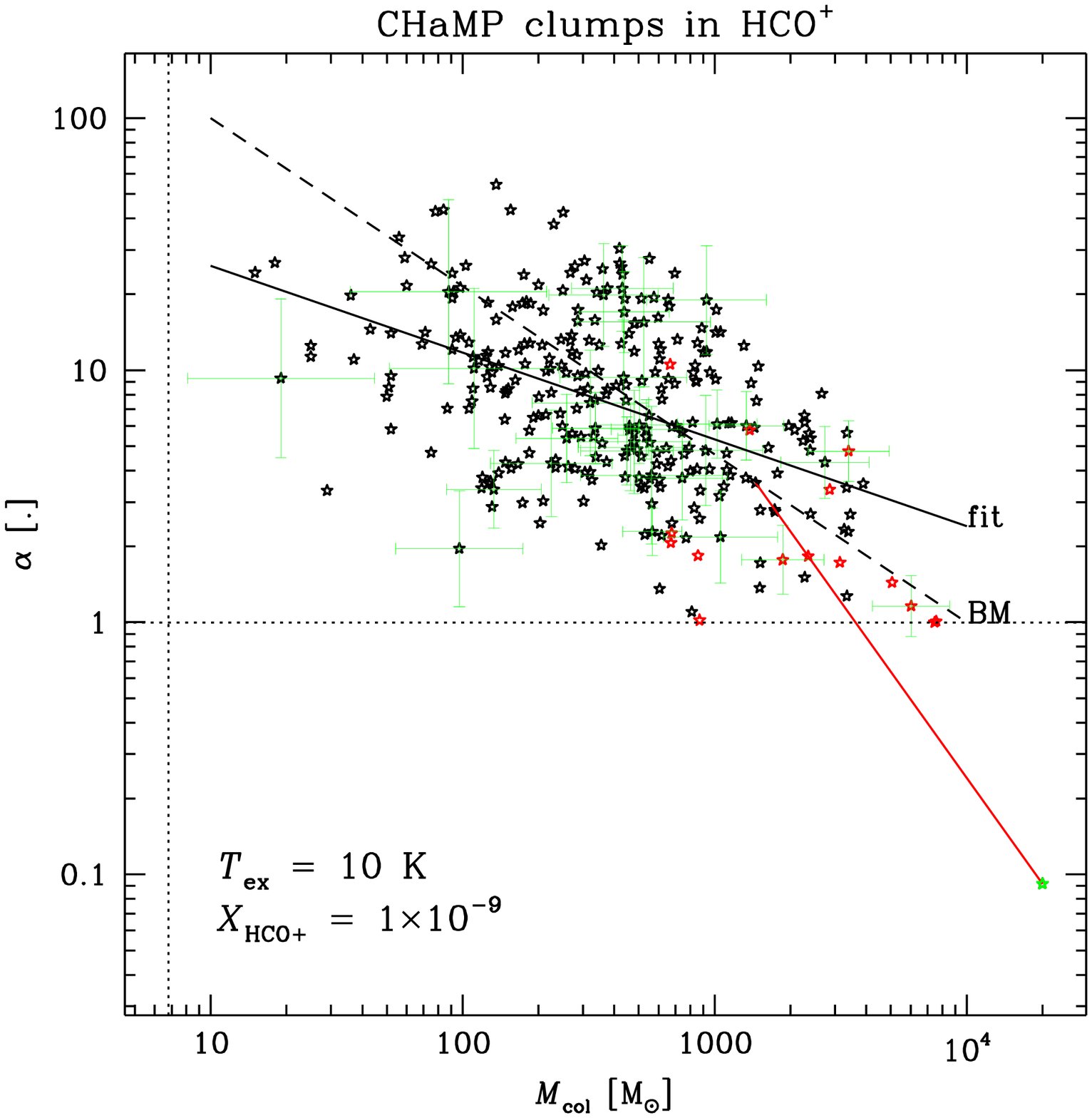}
(b)\hspace{-4mm}\includegraphics[angle=0,scale=.42]{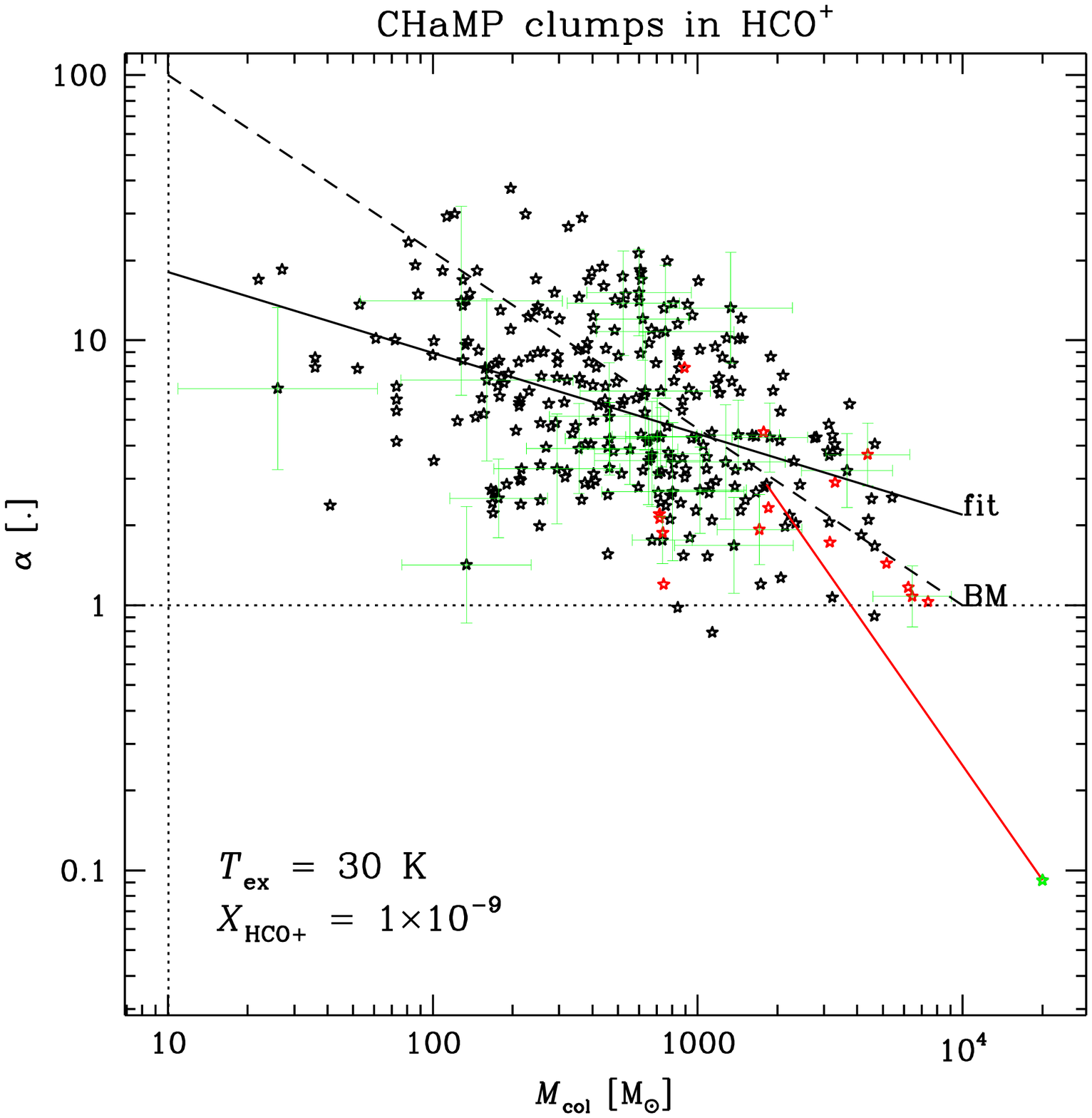}
\vspace{-6mm}
\caption{\small \cite{bm92} $\alpha$ parameter vs.\,mass from column density for Mopra \hcop\ clumps.  The solid line in each panel (labelled ``fit'') is the least-squares fit to the clump data, while the dashed line (labelled ``BM'') is the theoretical line from \cite{bm92}.  The dotted lines show the 5$\sigma$ mass sensitivity and the $\alpha$=1 limit for gravitationally supported clouds.  Other details are as in Fig.\,\ref{densMass}. 
\label{alpha}}
\vspace{-2mm}
\end{figure*}

We now consider the virial mass according to \cite{bm92}, using the velocity dispersion (Table \ref{sources}) and clump radius (Table \ref{physpar}): 
\begin{eqnarray} 
	M_{\rm vir} & = & 5\sigma_V^2 R/G \\
	 & = & 1160\,{\rm M}_{\odot} \left(\frac{\sigma_V}{{\rm km s}^{-1}}\right)^2 \left(\frac{R}{\rm pc}\right) . \nonumber 
\end{eqnarray}
An interesting result is seen when comparing this mass to the LTE mass $M_{\rm col}$ (eq.\,9), 
through the virial $\alpha$ parameter, where $\alpha$=$M_{\rm vir}$/$M_{\rm col}$, as shown in Figure \ref{alpha}.  Most of our Mopra \hcop\ clumps have $M_{\rm col}$ $<$ $M_{\rm vir}$, i.e. $\alpha$$>$1 (typically by about an order of magnitude), indicating they are probably stable against gravitational forces.  This also means that virial masses, as usually calculated in the literature, are likely to be overestimates by the same factor for most dense cloud masses.  However, the clumps in the bright tail of the source function (the red points) lie closer to the line of equal masses ($\alpha$=1) on average, showing they are only a factor of $\sim$3 away from being gravitationally dominated, in the virial sense.  Once we obtain better estimates for $\tau$ or $T_{\rm ex}$ for these points, we may find some of them actually lie on or below the critical line.  As before, BYF\,73 is a special case: we have also plotted corrected values in green for this source from \citet{BYR10} in Figure \ref{alpha}, and connected them with a red line to the respective uncorrected values from Table \ref{physpar}.  This shows that BYF\,73 appears unique in our sample, lying well below the virial equilibrium limit, and suggests it should be strongly dominated by gravitational forces.  As \cite{BYR10} found, this is exactly correct, since BYF\,73 appears to be in a state of global gravitational collapse.

\cite{bm92} summarised the results of four other GMC clump surveys in such a diagram.  Compared to the clumps in the Ophiuchus, Rosette, Orion B, and Cepheus OB3 clouds, the CHaMP clumps' $\alpha$-$M_{\rm col}$ power-law relation has a substantially flatter index (the solid lines with slope --0.34$\pm$0.04 and --0.31$\pm$0.04 at $T_{\rm ex}$ = 10\,K and 30\,K in Fig.\,\ref{alpha}a and b resp., compared to slopes from --0.50 to --0.68 in the above four GMCs) and shows roughly twice the scatter ($\sim$1 vs.\,0.5 orders of magnitude in $\alpha$).  \cite{bm92} argued that, for $\alpha$$\gg$1 at least, the expected index would be close to --$\frac{2}{3}$, indicated in Figure \ref{alpha} by the dashed line.  Thus, our $\alpha$-$M$ relation seems to be much weaker than in these other cloud samples.

However, $\alpha$ is strongly dependent on the linewidth, and given our clumps' large linewidth range, this is the single biggest contributor to the large scatter in $\alpha$ at any given mass.  Indeed, Figure \ref{alpha} can almost be contoured by lines of constant linewidth with slope --0.5 and increasing to higher $\alpha$ and $M$.  
Similarly, the mass is a strong function of clump size; thus, the smaller clumps are all to the left (at the high-$\alpha$ end) in Figure \ref{alpha}, while the larger clumps have the lower $\alpha$s on the right.  Our low value of the power-law index for the whole CHaMP clump population is then seen to be at least partially due to clumps with small linewidths being somewhat lower-brightness (and hence lower-column and lower-mass) than average, and clumps with large linewidths being somewhat brighter than average, or at least including more bright clumps within their number.  This combination of trends then shifts our fit away from the \cite{bm92} expectation.

\begin{figure}[ht]
\vspace{-7mm}
\hspace{-5mm}\includegraphics[angle=0,scale=.42]{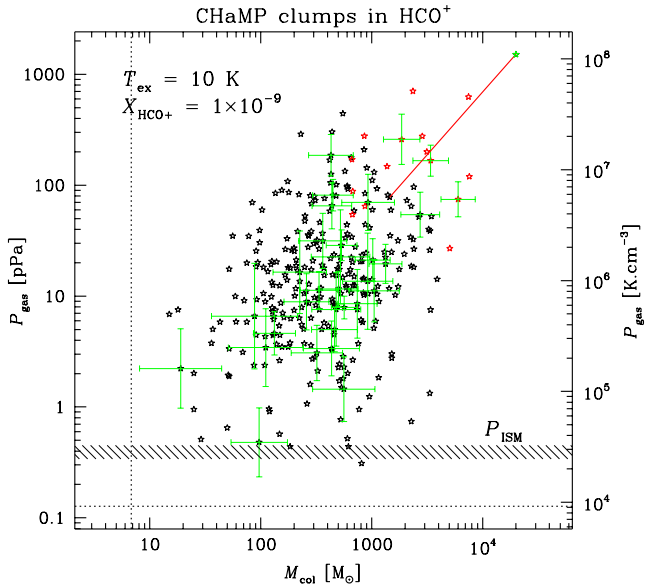}
\vspace{-7mm}
\caption{\small Total internal gas pressure vs.\,mass from column density for Mopra \hcop\ clumps.  Increasing $T_{\rm ex}$ by a factor of 3 changes $P$ and $M$ by relatively small amounts.  The horizontal shaded region shows the level of the general ISM pressure from Boulares \& Cox (1990).  Other details are as in Fig.\,\ref{densMass}.
\label{press}}
\vspace{-2mm}
\end{figure}

\cite{bm92} argued that large-$\alpha$ (i.e., lower-mass) clumps are likely to be pressure-confined, while \cite{ll08} showed that pressure-confinement by the overlying GMC was also likely to be the case for a large number of low-mass thermal cores in the Pipe nebula.  We may calculate the total internal pressure of the CHaMP clumps from
\begin{eqnarray} 
	P_{\rm gas} & = & \rho_{\rm col}c_{\rm tot}^2 = \rho_{\rm col} (c_{\rm iso}^2 + \sigma_V^2)  ,
\end{eqnarray}
where 
$c_{\rm iso}$ is the one-dimensional isothermal sound speed in the (hydrogen-dominated) gas at the given $T_{\rm ex}$, and $\sigma_V$ is from Table \ref{sources}; this is shown in Figure \ref{press}.  We note that virtually all of our clumps are high-pressure objects compared to the general ISM value $\sim$0.4\,pPa 
\citep{bc90}, which seems to be a strong lower bound to the points in Fig.\,\ref{press}, but the range of pressures is commensurate with that in other cluster-forming clouds such as $\rho$ Oph and Orion B \cite[][and references therein]{jb06}.  It seems that massive clumps are well-defined objects that are likely to be pressure-confined by their embedding GMCs, as suggested by \cite{bm92} and shown for low-mass cores by \cite{ll08}.  However, unlike the results of these studies, $P$ is not $\sim$constant or a slowly-rising function of $M$; instead we see at least a linear rise of $P$ with $M$, although the scatter in $P$ for a given $M$ is very large, about 3 orders of magnitude (due mostly to the large range of linewidths in our sample).  Notably, the CHaMP bright-tail sources are all at the high-pressure end of Figure \ref{press}.  Thus, while our bright-tail sources span the upper half of the range of all clump masses in Figures \ref{alpha} \& \ref{press}, as a group they are smaller, denser, have higher internal pressure, and are closer to being gravitationally dominated than the rest of the CHaMP clumps.  Additionally, except for BYF\,73 there is a dearth of clumps in the strongly supercritical ($\alpha$$<$1; Fig.\,\ref{alpha}) regime, although future corrections to the optical depth of the bright-tail sources may somewhat fill in this region.

\begin{figure}[ht]
\vspace{-2mm}
\hspace{-3mm}\includegraphics[angle=0,scale=.42]{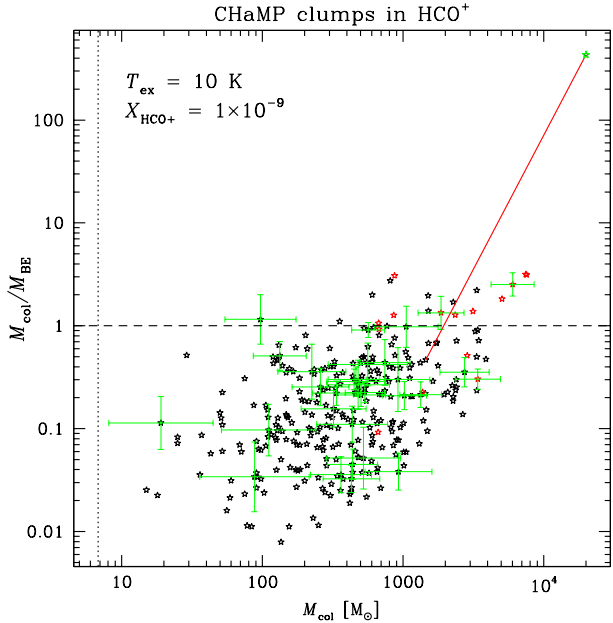}
\vspace{-7mm}
\caption{\small Ratio of (clump mass to Bonnor-Ebert critical mass) vs.\,mass from column density for Mopra \hcop\ clumps, following \cite{ll08}.  Increasing $T_{\rm ex}$ by a factor of 3 raises both the mass and mass ratio by a small factor.  The dotted line shows the 5$\sigma$ mass sensitivity, while the dashed line shows unit mass ratio for critically supported clouds.  Other details are as in Fig.\,\ref{densMass}.
\label{BE}}
\vspace{-2mm}
\end{figure}

To illustrate this further we look at the clumps' gravitational stability.  The critical limit for a non-magnetic, pressurised, gravitating cloud is given by the Bonnor-Ebert  mass,
\vspace{-2mm}
\begin{mathletters}
\begin{eqnarray} 
	M_{\rm BE} & = & 1.182 \frac{c_{\rm tot}^4}{(G^3 P_{\rm ext})^{1/2}} \\
			   & = & 1.858 \frac{c_{\rm tot}^3}{(G^3 \rho_{\rm col})^{1/2}}
\end{eqnarray}
\end{mathletters}
\hspace{-4mm}
\cite[called the Jeans mass by][]{bm92}, where we have used $\rho_{\rm col}$ = 2.47$\rho_{\rm ext}$ from the critical Bonnor-Ebert solution.  When \cite{ll08} plotted $M_{\rm core}$/$M_{\rm BE}$ for their low-mass, thermal cores as a function of core mass, they found a clear trend of rising mass ratio which crossed unity for core masses 2--3\,M\solar.  In Figure \ref{BE} we see a similar trend of rising mass ratio, but with a much larger scatter (due to our highly non-thermal linewidths $\Delta$$V$) and crossing unity at a much higher mass around 1000\,M\solar.  In fact, this crossing point rises rapidly with the linewidth: for smaller $\Delta$$V$$\sim$2\kms, the crossing point is $\sim$400\,M\solar.  For larger $\Delta$$V$$\sim$5\kms, it is $\sim$3000\,M\solar.  For the low-mass cores of \cite{ll08} with total $\Delta$$V$$\sim$0.6\kms, the crossing point is 2--3\,M\solar.  Of course this is roughly consistent with eq.\,(12) by construction, but intriguingly the power-law index $\beta$ for $M_{\rm cross}\propto$ $\Delta$$V$$^{\beta}$ between 3\,M\solar and 400\,M\solar is closer to 4, while for the CHaMP clumps (i.e., between 400 and 3000\,M\solar) $\beta$ is closer to 2.3.  This suggests that {\em lower-mass cores or clumps reach critical BE-like states at constant pressure, while more massive clumps reach such states at constant density}.  As for the majority of the clumps in Figure \ref{BE}, they may be in equilibrium states with respect to gravity, mirroring the result in Figure \ref{alpha}.  Few clumps rise above a mass ratio of 1 indicating instability to collapse (with BYF\,73 being the obvious exception).  The bright sources (red points) are generally closer to criticality as before.  We discuss these findings further in \S\ref{timescales}.

\subsection{The Clump Mass Function\label{MLF}}

\begin{figure*}[ht]
\vspace{-16mm}
(a)\hspace{-4mm}\includegraphics[angle=0,scale=.42]{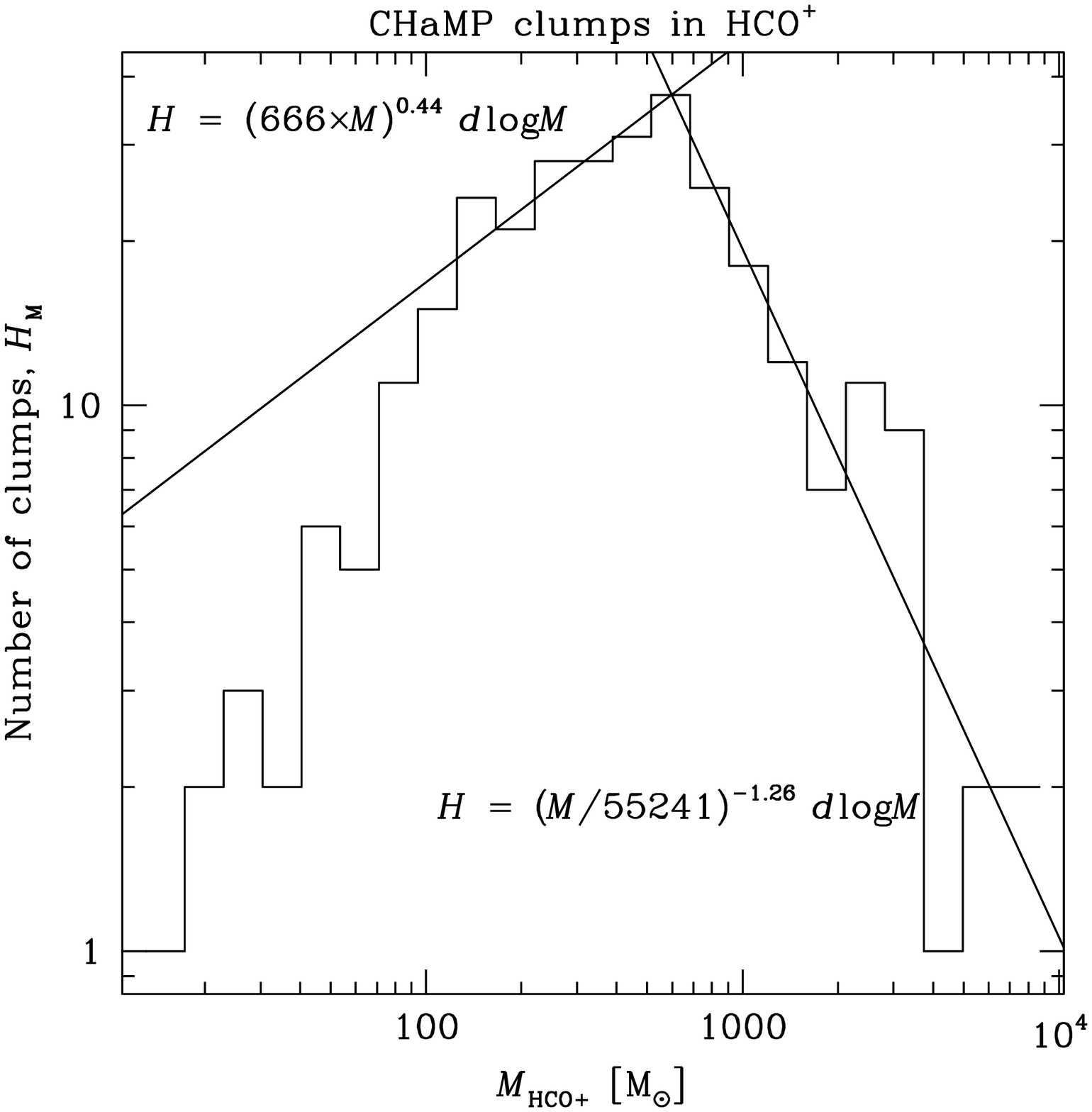}
(b)\hspace{-4mm}\includegraphics[angle=0,scale=.42]{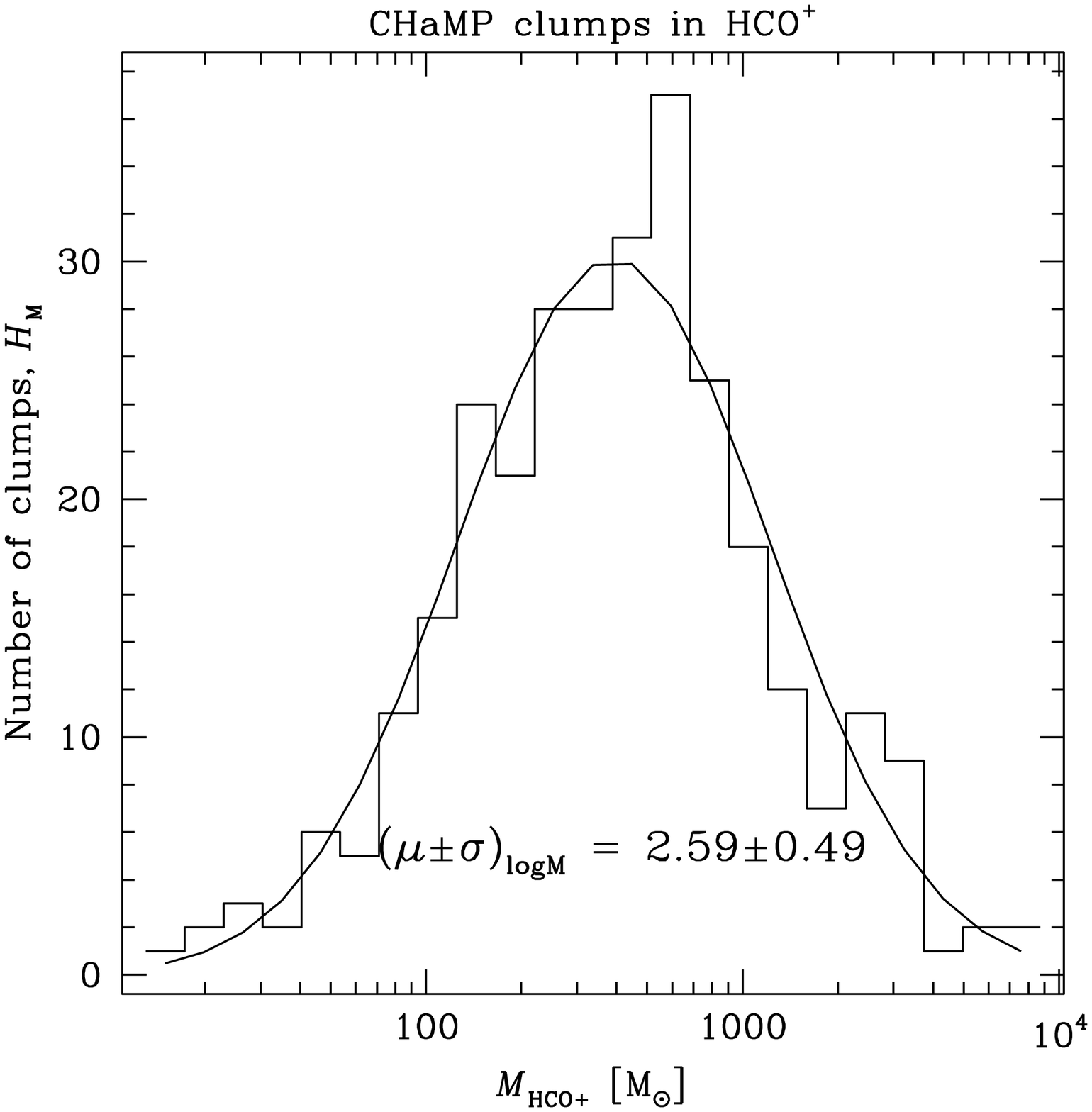}
\vspace{-6mm}
\caption{\small Mopra \hcop\ mass PDF, as defined in the text, for $T_{\rm ex}$ = 10\,K.  (a) One realisation of power-law fits for 23 histogram bins.  (b) Equivalent gaussian fit. 
\label{massfn}}
\vspace{-1mm}
\end{figure*}

Besides the source and luminosity PDFs presented in \S\ref{ensemble}, we can also construct the clump mass PDF 
\begin{equation} 
	H_M = (M/M_0)^{-r}~d{\rm log}M  ,
\end{equation}
in a similar way (Fig.\,\ref{massfn}), although this will change slightly with the assumed $T_{\rm ex}$.  Here we do not find a single power-law behaviour as found for the other two functions, rather we have two distinct power laws, one each over a lower- and a higher-mass range.  For $T_{\rm ex}$ = 10\,K, fitting to clumps with mass $M$$>$600\,M\solar\ gives a fit with slope $r$=--1.25$\pm$0.04 over a range of realisations, while a fit to clumps in the range 100\,M\solar$<$$M$$<$600\,M\solar\ similarly gives $r$=+0.37$\pm$0.05.  However for $T_{\rm ex}$ = 30\,K, splitting the mass ranges at 120 and 750\,M\solar gives fits $r$=--1.19$\pm$0.04 and $r$=+0.39$\pm$0.07.  Once again we find that these results are not very sensitive to the assumed $T_{\rm ex}$. 

We can understand the change in the form of the PDF from Figure \ref{sourcefn} to Figure \ref{massfn} as follows.  Dimensionally, the clump mass goes as the peak line luminosity times the projected area.  Except for the bright-tail sources, the bulk of the clumps have a size that is relatively independent of the brightness (Fig.\,\ref{obspars}b).  Thus the distinct form of the luminosity PDF is essentially erased when multiplied by the square of a somewhat random factor that has a dynamic range of 6.  Indeed, the mass PDF appears to resemble a normal distribution in log(mass), with a mean$\pm$SD = 2.59$\pm$0.49 in the log at 10\,K, i.e., a mean mass of 390\,M\solar (Fig.\,\ref{massfn}$b$).

\begin{figure}[h]
\vspace{-2mm}
\hspace{-4mm}\includegraphics[angle=0,scale=.42]{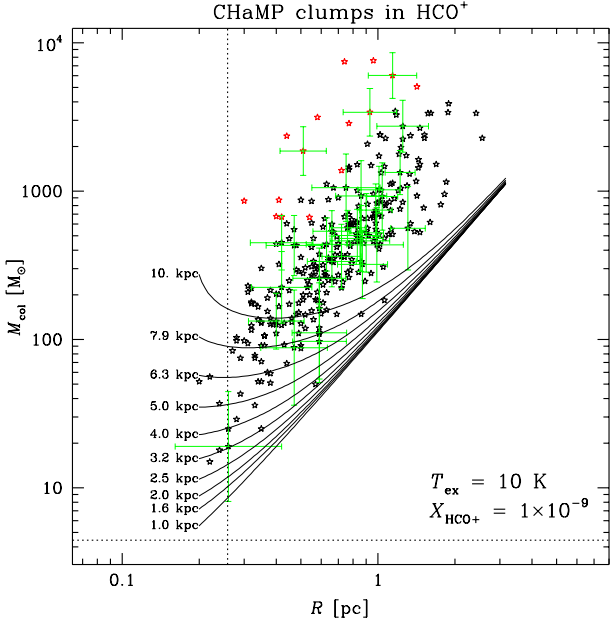}
\vspace{-7mm}
\caption{\small Mass vs.\,radius for Mopra \hcop\ clumps (with 5$\sigma$ mass sensitivity and 2$\sigma$ radius sensitivity at a mean clump distance of 3.2\,kpc; other details as in Fig.\,\ref{tauZig}).  The points are overlaid by simulations of clump detectability at our mean 3$\sigma$ sensitivity limit of $W$ = 0.90\,K\kms.  The convergence of curves in the upper right to a slope of 2 reflects the equivalent column density sensitivity for well-resolved clumps, whereas the upturn in the curves at small $R$, especially at the larger distances, reflects our angular resolution limit.  Each curve, calculated for a linewidth of 3\kms, is labelled by the clump distance where $W$ would be at this limit; above each curve is the locus of detectable clumps at this distance, below each curve are undetectable clumps.  Increasing  $T_{\rm ex}$ by a factor of 3 raises both the points and the simulated curves by a small factor in mass.  See the text for more details.
\label{sims}}
\vspace{-2mm}
\end{figure}

The lower-mass range in Figure \ref{massfn}$a$ also represents clumps which are incompletely sampled at our largest distances.  
This is shown more clearly in Figure \ref{sims}, where we overlay simulations of our clump detection limits on a mass-radius plot of our measured clumps.  Here we see that, as the mass drops further below $\sim$100\,M\solar, clumps can only be detected at smaller and smaller distances, leading to a completeness which drops with mass.  
Similar curves can also be drawn for various linewidths at a fixed distance, but for constant $W$, varying $\Delta$$V$ changes the curves very little, since the simulated $T_p$ drops as $\Delta$$V$ rises.  The main point to note in Figure \ref{sims}, however, is that the distribution in ($R$,$M$) seems well-sampled above $\sim$100\,M\solar\ at {\em all} distances, since there is a clear gap between the detected massive clumps and our sensitivity limit.  In contrast, although all our clumps are nearer than 7\,kpc, below $\sim$60\,M\solar\ the true distribution is likely undersampled.  This underscores the point made in \S\ref{ensemble} that, while the slope of the PDFs below the turnover is quite uncertain, the existence of the turnover at characteristic values of $W$, $L$, or $M$, and its value in mass of $M_c$$\sim$600\,M\solar, seems fairly reliable.

On the other hand, the higher-mass range most likely forms a complete clump sample.  In this case, we can then relate the mass PDF to the Clump Mass Function (CMF),
\begin{equation} 
	\frac{dH_M}{dM} = \frac{H_{M}}{M~d{\rm log}\,M} \propto M^{-\gamma}  ,
\end{equation}
which means that $\gamma$=$r$--1, and this is illustrated in Figure \ref{CMF}.  This value for $\gamma$ is very similar to that in several other CMF studies \cite[e.g.][and references therein]{ll03,r06,all07,bcc06}, and to the Salpeter IMF for stars.  In addition, the overall distribution Figure \ref{CMF} is reminiscent of the dense core mass function (DCMF) derived by \cite{all07}, in that the high-mass objects obey a power law above some critical mass, but break to a much flatter power law below the critical mass.  In our case though, the critical mass in the CMF is about 300 times that of the DCMF at 2\,M\solar.  As \cite{bcc06}, \cite{all07}, and \cite{ll08} point out in comparing the CMF/DCMF and IMF, such mimicry between the shapes of these mass functions suggests that similar fragmentation processes could be at work on the different scales, apart from~a uniform efficiency factor for the formation of dense cores from massive clumps.  However as mentioned above, our lower-mass range likely suffers from decreasing completeness towards lower masses, so the exact slope in this range 
is uncertain.  While it is clear from Figure \ref{sims} that our completeness level drops below 60\,M\solar, it is not clear that any missing clumps will be of sufficient number to meaningfully change the existence of the mass function turnover.  In order to confirm this suggestion, one would need (for example) to map the fainter 88 Nanten clumps not mapped at Mopra, thereby improving our statistics for clumps in the lower half of our mass range.  Integrating more deeply in the existing fields, thereby lowering our column density limit, would also be helpful.  Obtaining the actual $T_{\rm ex}$ for each clump will reduce the uncertainty in the CMF slope for our higher-mass range as well; higher-resolution observations of our bright clumps would help in this regard.

\subsection{Comparison with Models and Other Massive Clump Samples \label{compare}}

\cite{kth07} and \cite{ncs08} modelled the radiative transfer of molecular line emission in a clumpy density distribution representing populations of clouds in galaxies.  In such large systems, observed indicators of star formation rate (SFR), such as the bolometric infrared luminosity $L_{\rm IR}$, scale as the underlying Kennicutt-Schmidt law, i.e.\,as a power $N$ of the volume-averaged mean density $n_{\rm gas}$ of the molecular mass in the galaxy; typically $N$=1.4--1.6 \citep{gs4a,gs4b}.  Tracers of dense molecular gas such as CS, HCN \citep{w10}, or \hcop\ are also often used to trace star formation activity, both in the Milky Way and external galaxies.  However as \cite{krt07} point out, the exact SFR--$L_{\rm IR}$ relation in individual young clusters is subject to much more uncertainty, due mainly to the variable contribution by still-forming stars to $L_{\rm IR}$ when the cluster age is $\lapp$3\,Myr.  Nevertheless, many studies \cite[e.g.,][and references therein]{w10} find that $L_{\rm IR}$ and $L_{\rm mol}$ (the molecular line luminosity from one of the aforementioned dense gas tracers) are well-correlated from cluster to galactic scales, and therefore the relation of either of these quantities to $n_{\rm gas}$ is of great interest.

\cite{ncs08} suppose that if the~SFR (which they take in their case to be indicated by $L_{\rm IR}$) $\propto$ $L_{\rm mol}^a$, 
and if $L_{\rm mol}$ $\propto$ $n_{\rm gas}^b$, 
then one expects $N$=$a$.$b$.  \cite{ncs08} predict that, in a galaxy-wide population of clouds with a mean density below the HCN \joz\ critical density, $a$$\sim$1 and $b$$\sim$1.5, while \cite{kth07} obtain a similar result for \hcop, but with $a$ rising to 1.5 and $b$ dropping to 1 as the median density rises (\hcop\ has a slightly lower critical density than HCN and so probes clouds with a wider range of properties).  Strikingly, \cite{ncs08}'s model also predicted a very large fraction of the molecular line emission may be coming from a ``vast population'' of subthermally-excited clumps, which we strongly confirm in our data.

Because of our large sample of Galactic clumps, we are in a better position than previous dense gas studies of the Milky Way to examine some of these predictions, in particular the $L_{\rm mol}$--$n_{\rm gas}$ relation.  Here we compare our sample properties with results from some recent surveys that approach CHaMP's comprehensiveness.  \cite{w10} have recently completed a similar molecular line study to CHaMP, though smaller in sample size (50) and selected towards massive star-forming regions traced by water masers.  A comparison of our results with this work nicely illustrates how this selection can strongly affect the derived properties of such clump samples. 

\begin{figure*}[ht]
\vspace{-16mm}
(a)\hspace{-5mm}\includegraphics[angle=0,scale=.42]{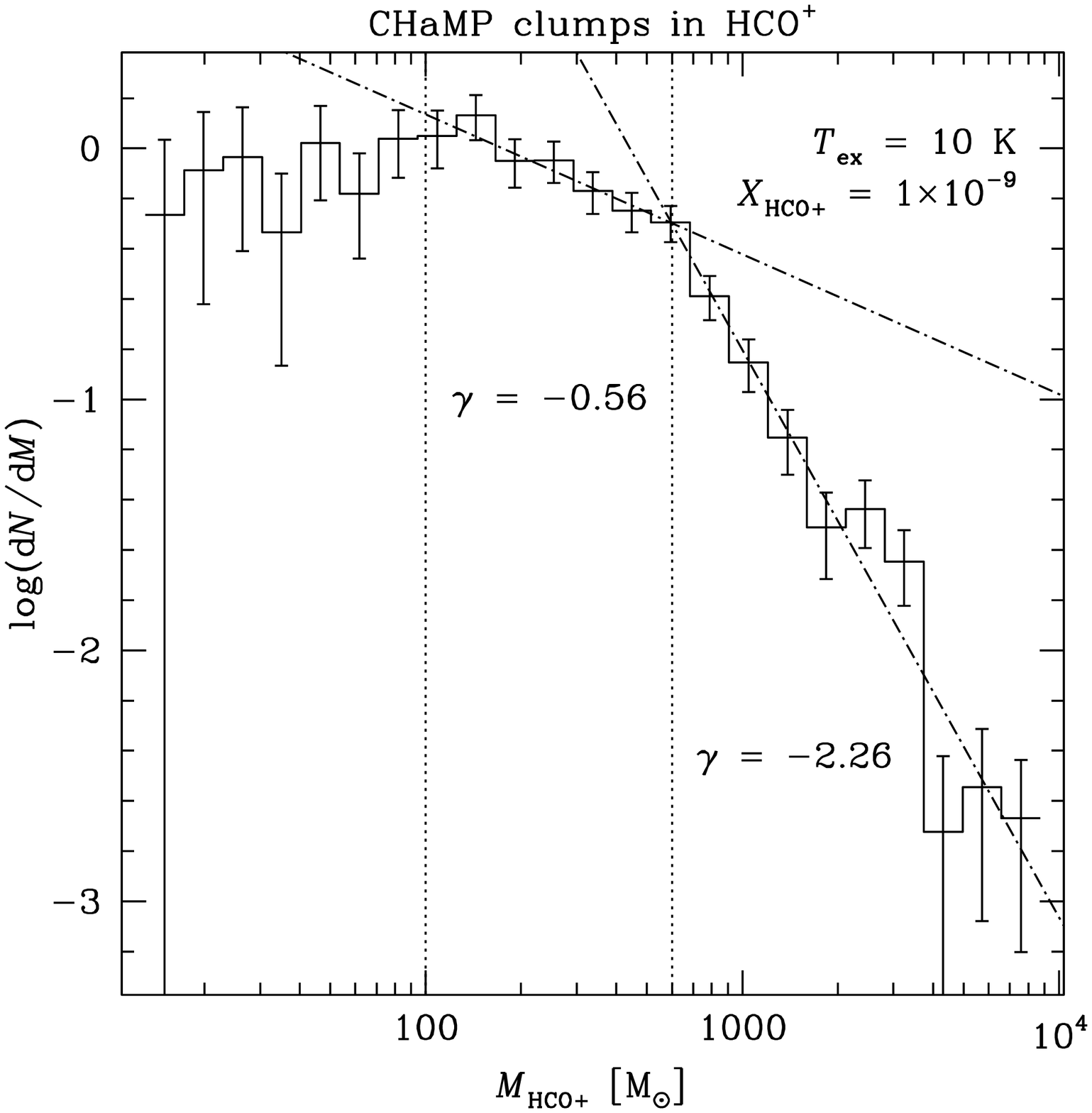} 
(b)\hspace{-5mm}\includegraphics[angle=0,scale=.42]{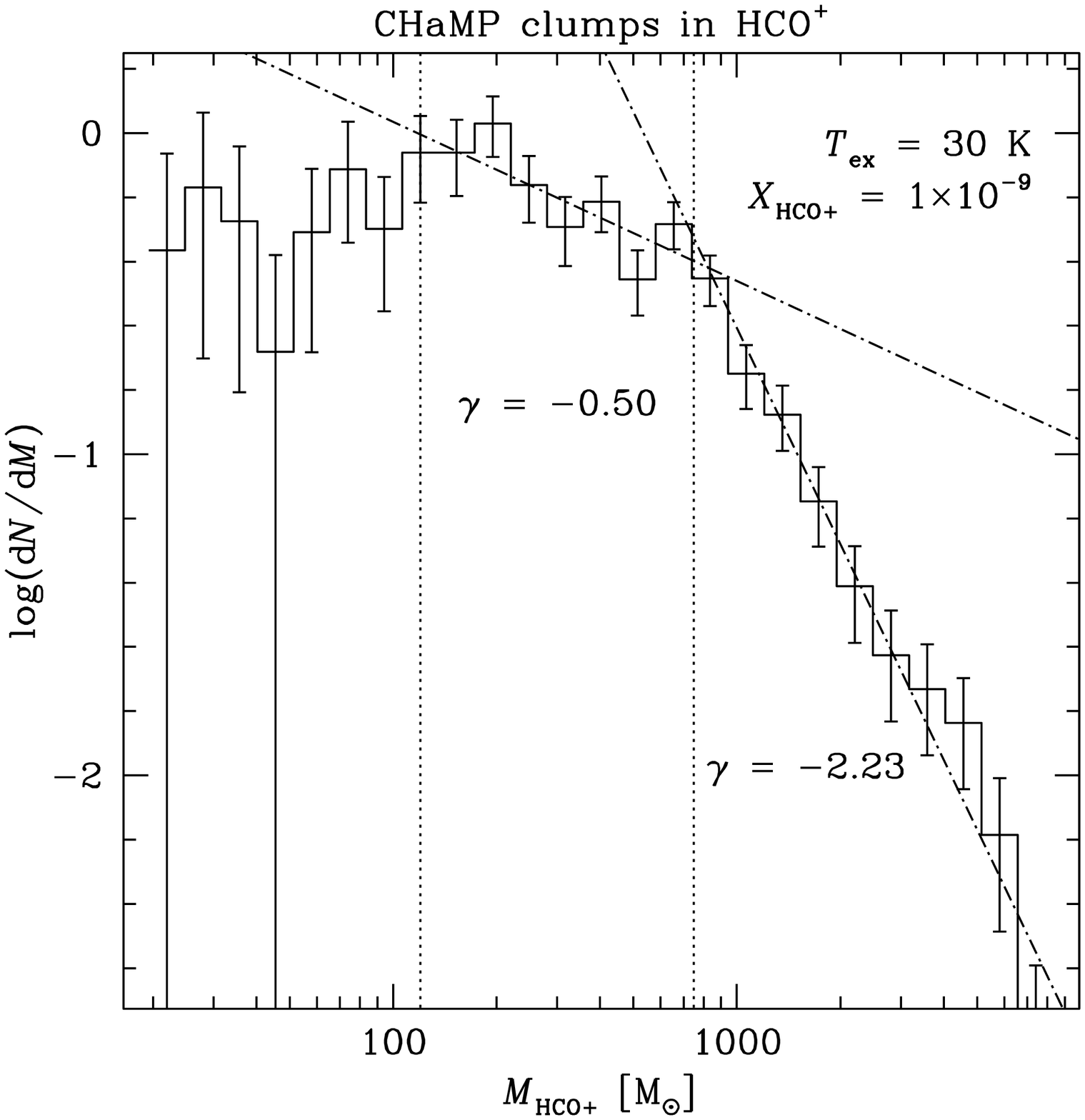} 
\vspace{-6mm}
\caption{\small Mopra \hcop\ Clump Mass Function (CMF) after \cite{bcc06}, and plotted for ($a$) $T_{\rm ex}$ = 10\,K, and ($b$) $T_{\rm ex}$ = 30\,K.  Dotted vertical lines show the mass ranges over which power-law fits (shown by dot-dashed lines) were obtained.  Each power-law fit is labelled by the fitted slope $\gamma$ for this realisation (23 and 25 histogram bins for panels $a$ and $b$, resp.).  At 10\,K (panel $a$) the mean $\pm$ SD values for many realisations are $\gamma$ = --0.63$\pm$0.05 over the lower-mass range and --2.25$\pm$0.04 for the higher masses; at 30\,K (panel $b$) $\gamma$ = --0.61$\pm$0.07 and --2.19$\pm$0.04, resp.
\label{CMF}}
\vspace{-1mm}
\end{figure*}

\begin{figure}[ht]
\vspace{-1mm}
\hspace{-3mm}\includegraphics[angle=0,scale=.45]{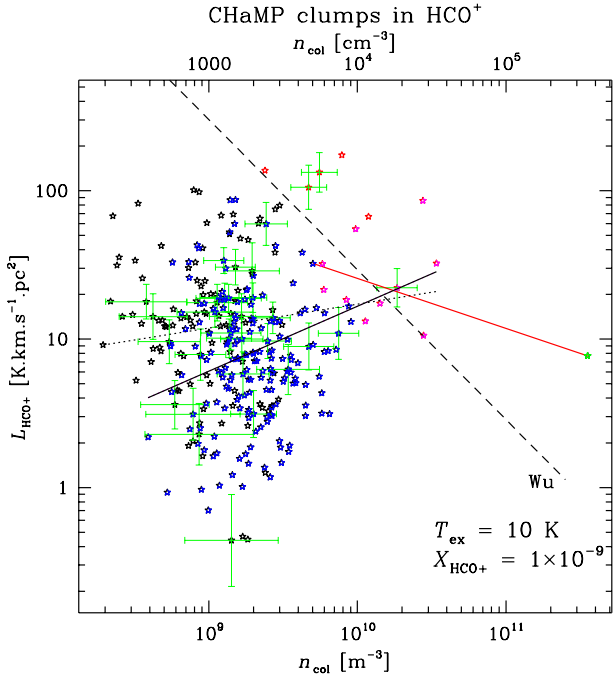}
\vspace{-8mm}
\caption{\small 
Mopra \hcop\ integrated line luminosity vs.\,\,density plotted for $T_{\rm ex}$ = 10\,K, with weaker and brighter clumps from Fig.\,\ref{sourcefn} in black and red as before, except for 191 clumps with distance 2.4\,kpc $\leq d\leq$ 2.5\,kpc, which are shown in blue and magenta, resp.  The correction for BYF\,73 is shown as a red line.  We show two fits to these points: (i) all points were binned in equal intervals of log $n$ and these bins equally weighted in a least-squares fit, giving a power-law index 0.16$\pm$0.12 (dotted line), and (ii) a similar fit to the blue and magenta points, giving a power-law index 0.44$\pm$0.24 (solid line).  
We also show, as a dashed line labelled ``Wu'', the trend of \cite{w10} on the same scale, from a robust fit to a sample of massive clumps from selected luminous Galactic star-forming regions, with a power-law index --1.01$\pm$0.02.  This is actually an average of \cite{w10}'s robust linear fits for $L_{\rm CS2-1}$ and $L_{\rm HCN1-0}$ vs.\,$n$, which differ from this average by less than 0.05 in the log over this plot.  See text for further discussion.
\label{LN}}
\vspace{-1mm}
\end{figure}

In Figure \ref{LN} we show the distribution of the CHaMP clumps' \hcop\ integrated line luminosity with density, and give two fits to these data (discussed further below).  Both fits show that, while the value of the index $b$ is quite uncertain, it is clearly positive, i.e., the clumps' \hcop\ line luminosity increases with the clumps' gas density, although possibly with a low value for the index $b$.  \cite{w10} found, in contrast, that their massive clumps' HCN and CS line luminosities, as well as total $L_{\rm IR}$, are {\em anticorrelated} with $n$ (i.e., $b$ around --1, also shown in Figure \ref{LN}).  They attributed this disparity to their selection of luminous star-forming regions, with more luminous ones tending to excite emission from larger areas, making their derived mean densities lower.  At face value, this may be reasonable: we see in Figure \ref{LN} that the upper envelope of our brightest sources straddles the trend seen by \cite{w10}, while all the weaker clumps are well below this trend.  
This supports the idea that this bright sub-sample is qualitatively different from the majority of weakly-emitting clumps, and leads us further to suggest that \cite{w10}'s sample may be just the upper envelope of the whole clump population which, apart from a normalisation, may follow the \cite{ncs08} trend after all.  Thus, sample selection may make a critical difference to the derived $L_{\rm mol}$--$n_{\rm gas}$ relation.  We suggest that because CHaMP selects a clump population, the trend we see here is more representative of all clouds at all evolutionary stages, rather than a law that describes only the most luminous clouds in a galaxy that may, in fact, represent only a single evolutionary stage \cite[H$_2$O maser-containing clouds in the case of][]{w10}.

However, there is a further difference between the CHaMP and \cite{w10} results that may be significant.  Here we derive clump masses based on column density, and depending only on the assumed $T_{\rm ex}$ (which has only a weak influence on the results, \S\ref{tausig}) and the \hcop\ abundance.  In contrast, \cite{w10} calculate the virial masses for their HCN and CS data, and derive densities based on this mass.  At least in the case of \hcop, Figure \ref{alpha} shows that the virial mass is typically 3--30 times larger than the actual cloud mass.  Therefore it may be that, if the clouds traced by HCN and CS are similar, \cite{w10}'s masses, and hence densities, could be overestimated.  Moreover, this overestimate might not be uniform: substituting densities for our \hcop\ clumps based on our virial masses into Figure \ref{LN}, the lower-luminosity sources, having larger $\alpha$, are most shifted to higher densities, while the higher-luminosity sources shift relatively little.  Although there is a large scatter, the resulting distribution (not shown) evenly fills the space between the unshifted points plotted in Figure \ref{LN} and the \cite{w10} trend, while the brighter sources follow the \cite{w10} HCN and CS results more closely ($b=-0.71\pm0.24$).

Instead, we prefer to use the density based on the calculated column density, as displayed in Figure \ref{LN}.  Fitting these data with a single law gives an index $b$ near zero (not shown); however we maintain that this is mostly due to the distribution of our sample in $L_{\rm mol}$--$n_{\rm gas}$ space, which is highly concentrated towards the centre of this diagram and distorts the least-squares fit to the higher-density points.  Binning the data in log($n$) and weighting the bins uniformly, we obtain $b$ = 0.16$\pm$0.12 (shown as a dotted line), which is still quite shallow.  Notably, the more distant clouds are all to the left of the plot, while nearer clouds are all to the right.  Indeed, some of the structure in this plot must be attributed to the fact that most clumps are associated with only a few complexes at a small number of quite different distances.  Restricting the points to subsamples within smaller distance ranges makes the $L_{\rm mol}$--$n_{\rm gas}$ correlation plainer.  An example is shown for the $\eta$ Carinae (BYF\,60--118) and other complexes with fairly well determined distances in the range 2.4--2.5\,kpc which, being nearer, are complete to a lower value of $L_{\rm mol}$ than the rest of the sample.  This yields the fit $b$ = 0.44$\pm$0.24 
and is shown as a solid line in Figure \ref{LN}.  Indeed, the trend seems significantly steeper than this to the eye, but as described above the centrally concentrated distribution of points skews the least-squares fitting.  As a consequence, though, we see that even our large sample of clumps probably doesn't give an adequate sampling of $L_{\rm mol}$--$n_{\rm gas}$ space.  Therefore, despite our best fits giving values for $b$ possibly less than the $\sim$1--1.5 predicted by Krumholz \& Thomp
\clearpage

\noindent son (2007) and \cite{ncs08}, we consider that the stronger statement $b$$>$0 for the CHaMP \hcop\ clumps does give conditional support to these models.  


It will be important to measure $L_{\rm IR}$ in the clumps to investigate how the extragalactic relations between $L_{\rm IR}$ and $L_{\rm mol}$ are established: we are now compiling bolometric data on our clumps and will examine this issue in a future CHaMP paper (Ma et al., in prep).  Likewise, once data from other species included in our survey (such as HCN, HNC, or \nnh, see Table \ref{setups}) are analysed, we can investigate whether there is evidence for larger indices $b$ in the $L_{\rm mol}$--$n_{\rm gas}$ relation for molecules with higher critical densities.  Additionally, we are also surveying these clumps in the near IR for their current star formation activity, and so will eventually be able to examine the relation, modelled by \cite{krt07}, between the true SFR and $L_{\rm IR}$ for a very large sample of Galactic clusters. 
For now, we believe we can reconcile the apparent disparity between the \cite{w10} $L_{\rm mol}$--$n_{\rm gas}$ relation and its theoretical counterparts. 
This emphasises again the advantage of our unbiased, large-population approach.

Among other studies, \cite{r06} mapped the 1.2mm continuum emission from 38 Infrared Dark Clouds (IRDCs) and compared this to Galactic Ring Survey \ttco\ emission to derive properties of 140 dark ``cores'' within the IRDCs.  Their IRDC population has a median mass (10$^3$\,M\solar) and mass range (10$^2$--10$^4$\,M\solar) somewhat larger than the CHaMP \hcop\ clumps, while the ``cores'' are more similar in mass to our clumps.  The IRDCs and their cores similarly seem to bracket our clumps' sizes and densities, and the IRDC core mass function has $\gamma\sim -2$ compared to our high-mass slope near --2.2 (Fig.\,\ref{CMF}a), while their dust temperatures seem somewhat higher than our excitation temperatures (which are apparently subthermal).  \cite{r06} find 
a very large number of such clouds that could conceivably represent a population that accounts for all the star formation in the Milky Way.  They proposed that IRDCs form an earlier stage of evolution for massive star and cluster formation than hot cores, but  did not quantify the timescales in that picture.

While the absorption of IR emission against the bright Galactic background would seem to afford a fairly reliable method of detecting massive dark clouds, \cite{r06} themselves point out that such a method only works where a bright background exists; equivalent clouds with a dark background are undetectable with this technique.  This introduces a bias towards nearby objects, and those at smaller absolute longitudes (eg $|l|$$<$50\degr) which have a higher chance of lying in front of a bright background.  \cite{r06} estimate their detection rate across the Galaxy to be $\frac{1}{3}$ due to this bias, and although it is unclear if this introduces a selection effect on the {\em properties} of the molecular clouds, the detection rate is otherwise hard to quantify: it may well be smaller.  It seems clear that our method of finding all line emission within a given window is a necessarily less biased technique.

\cite{bcc06} also mapped 1.2mm emission from 235 massive clumps selected on their IRAS colours, similar to the study of \cite{fbg04} of a somewhat higher-luminosity sample.  \cite{bcc06}'s median clump size is somewhat smaller than ours, 0.4 vs 0.67\,pc, although this may be partially due to the smaller SIMBA beam compared to Mopra.  In contrast, sizes of clumps from a related study \citep{fbb05} are very similar to our CHaMP \hcop\ sizes.  However \cite{bcc06} found most of their clumps to be symmetric, unlike our results and those of \cite{g09}.  \cite{fbb05} also found that their clumps' excitation temperatures were 8--10\,K based on multi-line C$^{17}$O measurements, while the dust temperatures derived from SED fitting were similar to the \cite{fbg04} result (30\,K).  This supports the conclusion here that the dense gas is mostly subthermally excited.  The \cite{bcc06} median clump mass is 102\,M\solar, median volume density 4$\times$10$^{10}$m$^{-3}$, and median surface density 1.4\,kg\,m$^{-2}$, compared to our results of 325\,M\solar, 1.7$\times$10$^{9}$m$^{-3}$, and 0.29\,kg\,m$^{-2}$.  Thus the Mopra \hcop\ clumps are larger and more massive, and of lower column density than the \cite{bcc06} sample, but our clumps are less massive than the \cite{fbg04} and similar high-luminosity samples.  This means that, even when selecting for massive star-forming regions based on mid- or far-IR colours or flux densities, care must be taken when intercomparing surveys to allow for possibly subtle differences in sample properties.

\subsection{Clump Sub-populations, Evolution, \& Timescales \label{timescales}}

As described earlier (\S\S\ref{intro}, \ref{ensemble}, \ref{mass}), one of CHaMP's primary science goals was to obtain, by demographic analysis of an unbiased sample, constraints on the lifetimes of different stages of clump evolution, where they could be discerned.  From the analysis presented above, we propose that for the dense clumps traced by \hcop, we can identify two main sub-populations within this sample: a small subset of clumps in the ``high-brightness tail'' of the source PDF $H_W$, which are smaller, denser, closer to being gravitationally-dominated, and somewhat more massive; and the remaining large majority of ``weakly-emitting'' (but easily-detected) clumps.  The latter, larger group have a fairly uniform set of properties compared to the brighter sources.

This source PDF is quite significant.  As far as we are aware, this is the first time that a weakly-emitting subpopulation among all dense clumps has been identified.  The high-brightness tail of this distribution is, in contrast, similar to the kinds of clumps that have been studied in several other surveys \cite[such as][]{w10}.  If the clumps sampled here uniformly represent, as intended, the entire population of dense molecular clumps in a range of evolutionary states, e.g. from ones forming out of a GMC envelope to others actively making massive stars or clusters from denser cores within them, then the high-brightness tail of the distribution, comprising only 5\% of our sample by number (13\% by mass), represents a state that only lasts 5\% of the overall lifetime of the clumps, if they evolve at constant number (e.g. by becoming denser), and only 13\% of the clump lifetime if they evolve at constant mass (e.g. by mergers).  These fractions would be even lower if allowance is made for our survey incompleteness at masses below 100\,M\solar.  Thus, if the typical clump lifetime, from formation in a GMC environment to disruption by an embedded massive cluster, is 50\,Myr (for example), then the bright phase is implied to only last 2.5 or 6\,Myr in these respective scenarios.  Indeed, this is very suggestive: studies of typical young embedded clusters \cite[e.g.][]{dr10} consistently give ages on the order of 1--5\,Myr.  To confirm this number, we would need to establish that the weakly-emitting clump population does not show (e.g.) an abundance of Brackett-$\gamma$ emission indicating the presence of already-formed massive young stars.  If this is indeed the case, we will be able to reconcile a long-standing discrepancy between different estimates of molecular cloud lifetimes \citep{kss09,bsd10}.

The calculations in \S\S\ref{ensemble}--\ref{mass}, in particular the volume density estimates, strongly suggest our clump population is, for the most part, sub-thermally excited, unless (for example) the beam filling-factor of the \hcop\ emission is $\ll$1.  In this case we would need very filamentary cloud structures, of size $\lapp$0.05\,pc, to produce an effective beam-dilution of up to two orders of magnitude.  While this is possible, there is little evidence in the literature, or our data, to support this idea.  
This is not to say that massive clumps cannot have small-scale structure, since such structure is ultimately necessary to form individual stars.  However for our emission to come mostly from thermally-excited gas, {\em most} of the gas mass would have to reside in very filamentary or clumpy, high-density structures, which are then beam-diluted to produce maps of apparently low optical depth.

Instead, it is well-known that molecules like \hcop\ can be detected from regions with density well below their critical density, if their abundance or column density is high enough.  In this case, they are simply sub-thermally excited \citep{E99}.  We posit that the weakly-emitting clump population is in this state, and as such may comprise a pre-massive star formation sample.  Such a population of subthermal massive clumps is, in fact, predicted by models such as \cite{ncs08}.  In this state, their internal dynamics are still disorganised, perhaps indicating their formation conditions, since they have not yet become dominated by the consequences of massive star or cluster formation in their interiors.  Thus, they have no particular size-linewidth relation, and their \hcop\ brightness is low, but detectable.  Such clumps may yet have ongoing low-mass star formation within their confines, since at the typical distances to these clumps ($\sim$2--7\,kpc) such low-luminosity effects would be hard to detect.

The brighter clumps in the tail of the \hcop\ source function then represent, in this scenario, those massive clumps which are being actively altered by the formation of an embedded massive star or cluster, although such effects are typically limited to a volume \lapp1\,pc in radius.  They have higher optical depths in \hcop\ and higher columns of gas, as well as higher temperatures, densities, and pressures, but have a smaller range of linewidths: the \hcop\ is now closer to being thermally excited.  This idea is strongly supported by Figures \ref{densMass}, \ref{alpha}, \ref{press}, and \ref{BE}: these bright-tail clumps are the closest to being unstable against gravity, compared to the weakly-emitting clumps.  
For \hcop, the action of such massive star-cluster formation seems to produce a consistent degree of heating and/or dynamical stirring, producing brighter \hcop\ lines of width $\sim$4--5\kms\ and clumps of size $\sim$1\,pc.  We suggest this linewidth may not be random: it corresponds to a thermal linewidth of H$_2$ gas at a temperature of 1300--2000\,K, typical of shocks in (for example) Herbig-Haro objects and similar locations \cite[e.g.][]{CGN06}.

\section{Conclusions\label{conclude}}

We report the first Mopra-ATNF results of the CHaMP project, a large-scale, uniform, and unbiased census of higher-mass star formation in a large portion of the Galactic Plane.  We present several catalogues as part of this project, including a complete set of integrated intensity and higher moment maps, of the \hcop \joz\ emission from 301 massive molecular clumps in this window.  Our results include:
\vspace{-2mm}
\begin{itemize}
\item We detect a large population of ``weakly-emitting'' (but easily-detectable) massive clumps, comprising 95\% of our sample.  The ``high-brightness tail'' of this distribution, with properties similar to some popularly-studied massive star-forming regions, make up only 5\% of our sample.
\vspace{-2mm}
\item The clump source and peak luminosity PDFs show power-law distributions above characteristic levels $W_c$=4\,K\kms\ ($\sim$10$\times$ our noise threshhold) and $L_c$=1.0\,K\kms\,pc$^2$, while the clump mass function shows two power-law regimes with a break point near  $M_c$=600\,M\solar; the mass function also resembles a normal distribution in log(mass), with a mean$\pm$SD = 2.59$\pm$0.49, or a mean mass near 390\,M\solar, although this may be skewed somewhat to higher masses due to completeness limits in our survey.
\vspace{-2mm}
\item The power law of the higher-mass clumps is similar to the Salpeter value for the stellar IMF; the two power laws taken together resemble similar patterns seen for stars and low-mass dense cores, indicating (i) there is a characteristic scale to structure (i.e., clump, core, or star) formation in the dense ISM, and (ii) the process is self-similar over many decades in mass, except for an efficiency factor in converting clumps to cores to stars.
\vspace{-3mm}
\vspace{-3.3mm}
\item The observed properties of the clump population are: radius 0.2--2.5\,pc, peak temperature 1--7\,K, FWHM linewidth 1--10\kms, and mean axial ratio of around 2.  The clumps show no linewidth-size relation of the classic \citet{L81} variety.  However both size and linewidth seem more narrowly distributed for the brighter clumps.
\vspace{-2.3mm}
\item The derived physical properties of the clump population are: integrated line luminosity 0.5--200\,K\kms\,pc$^2$, optical depth 0.08--2 (assuming $T_{\rm ex}$=10\,K), mass surface density 30--3000\,M\solar\,pc$^{-2}$ (assuming $X_{\rm HCO+}$=10$^{-9}$), number density (0.2--30)$\times$10$^9$\,m$^{-3}$, LTE mass 15--8000\,M\solar, virial 
parameter 1--55, and total gas pressure 0.3--700\,pPa.
\vspace{-2.3mm}
\item In order to explain all these clump properties, most of the \hcop\ emission must come from subthermally excited clouds, rather than from structures having a small beam-filling factor.
\vspace{-2.3mm}
\item The \hcop\ clump population does not seem to follow the exact \citet{bm92} $\alpha$-$M$ relation; instead we see a shallower dependence of $\alpha$ on $M$, mostly due to the wide range of linewidths of our clumps.  Most clumps may be in pressure-confined virial equilibrium with their surroundings.
\vspace{-2.3mm}
\item Comparison with populations of low-mass cores suggests that lower-mass cores or clumps reach critical Bonnor-Ebert-like states at constant pressure, while more massive clumps reach such states at constant density.  Few clumps exceed the criterion for gravitational instability.
\end{itemize}
\vspace{-2.3mm}
These properties are markedly different to those of massive clumps in other, more biased, studies.  In particular most of our clumps seem to be sub-thermally excited and not gravitationally bound; only the brightest $\sim$5\% of our clumps are similar in their properties to other massive cloud studies, which we attribute to bias due to various selection criteria in these studies.

One of the main objectives of CHaMP is to use the unbiased sampling of a complete population of massive molecular clumps to infer lifetimes for the various phases seen.  This suggests that a weakly-emitting (in \hcop) phase of massive clump evolution lasts $\sim$95\% of the lifetime of a clump, during which few massive stars or star clusters form.  When the clump at last becomes bright in traditional dense-gas tracers such as \hcop, we claim that massive stars and clusters are already forming/have formed, disturbing the gas from its initial conditions.  The transition between these two phases, where gravity triggers a global collapse of the clump and a cluster first begins to form, appears to be very short compared to the clump lifetime, $<$1\%, since we only see one such example in our population.  For a GMC/clump lifetime of perhaps 50\,Myr, the pre-massive, collapse, and cluster phases then correspond to lifetimes 47, 0.2, and 2.5\,Myr, respectively.

We have used these data to synthesise a new observational picture of the overall evolution of massive clumps which consistently explains a number of observed and theoretically predicted phenomena for the first time.  We look forward to further observational and theoretical tests of these ideas.

\acknowledgments

The concept of CHaMP as a systematic survey was inspired by the example of Phil Myers and coworkers and their systematic studies of low-mass star formation.  
But CHaMP would have remained only a concept if the Nanten surveys had not been available to bootstrap from, and if the Mopra dish had not been able to leverage this opportunity into new science.  We are indeed fortunate that the efforts of many talented people have enabled Mopra's many recent upgrades and enhancements.  In particular, we thank Warwick Wilson and his team in the ATNF receiver group for an outstanding job in engineering the MOPS and MMIC upgrades.  Mopra's OTF capability owes a great deal to Tony Wong, who also provided the foundation for our data analysis scripts.  We also thank Ned Ladd and Erik Muller for their valuable contributions to Mopra's development.  Besides the normal TAC process from 2004--07, CHaMP was also granted ``UNSW time'' on Mopra in 2006, for which we thank Michael Burton.  

Finally we thank the anonymous referee for a thorough reading of the manuscript and many helpful comments and suggestions which improved the paper, and Paola Caselli for additional comments on the manuscript.  PJB gratefully acknowledges support through NSF grant AST-0645412 to JCT at the University of Florida.  


{\it Facilities:} \facility{Mopra (MOPS)}



\appendix
\section{\hcop\ Integrated Intensity Maps \label{mom0maps}}

\begin{figure*}[ht]
\centerline{\includegraphics[angle=-90,scale=0.50]{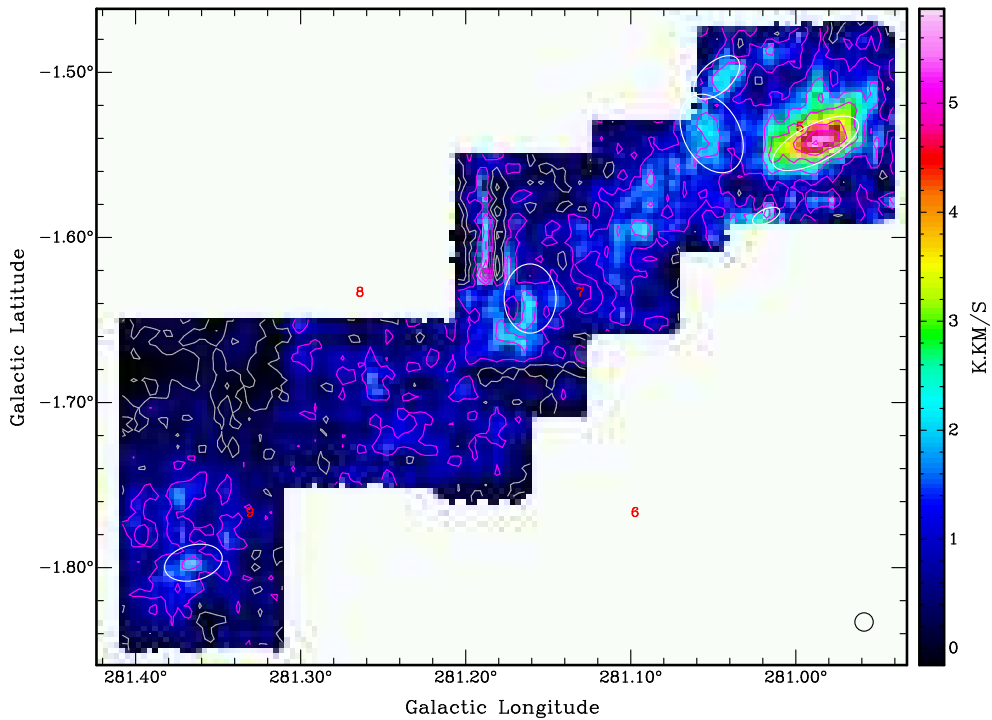}}
\caption{\small Mopra integrated intensity \hcop \joz\ map of Region 1 sources BYF\,5--9.  The integration is over the range --14.0 to --1.8\,\kms\ or 109 channels, yielding an average rms noise level 0.318\,K\kms, and contour levels spaced every 3$\sigma$ are overlaid.  At a distance of 3.2\,kpc, the smoothed Mopra beam (lower right corner) scales to 40$''$ = 0.621\,pc or 1\,pc = 64$''$\hspace{-1mm}.5. 
\label{reg1}}
\end{figure*}

\begin{figure}[htp]
\centerline{\includegraphics[angle=0,scale=0.37]{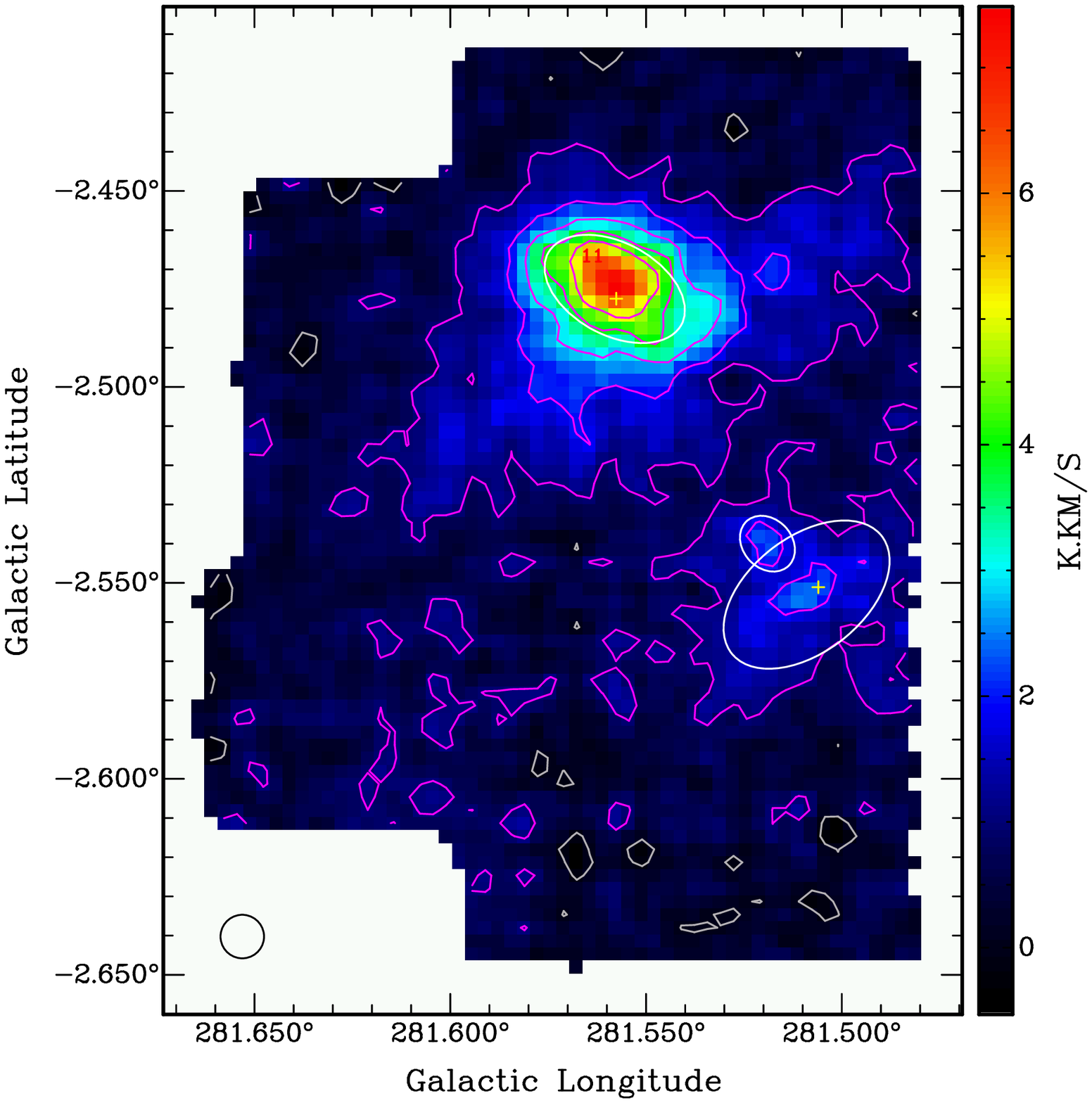}}
\caption{\small Same as Fig.\,\ref{reg1}, but for isolated source BYF\,11.  The integration here is over the range --12.0 to --3.0\,\kms\ or 80 channels, yielding an average rms noise level 0.238\,K\kms, and contour levels spaced every 4$\sigma$ are overlaid.  At a distance of 3.2\,kpc, the smoothed Mopra beam (lower left corner) scales to 40$''$ = 0.621\,pc or 1\,pc = 64$''$\hspace{-1mm}.5.
\label{byf11}}
\end{figure}

\begin{figure}[htp]
\centerline{\includegraphics[angle=-90,scale=0.25]{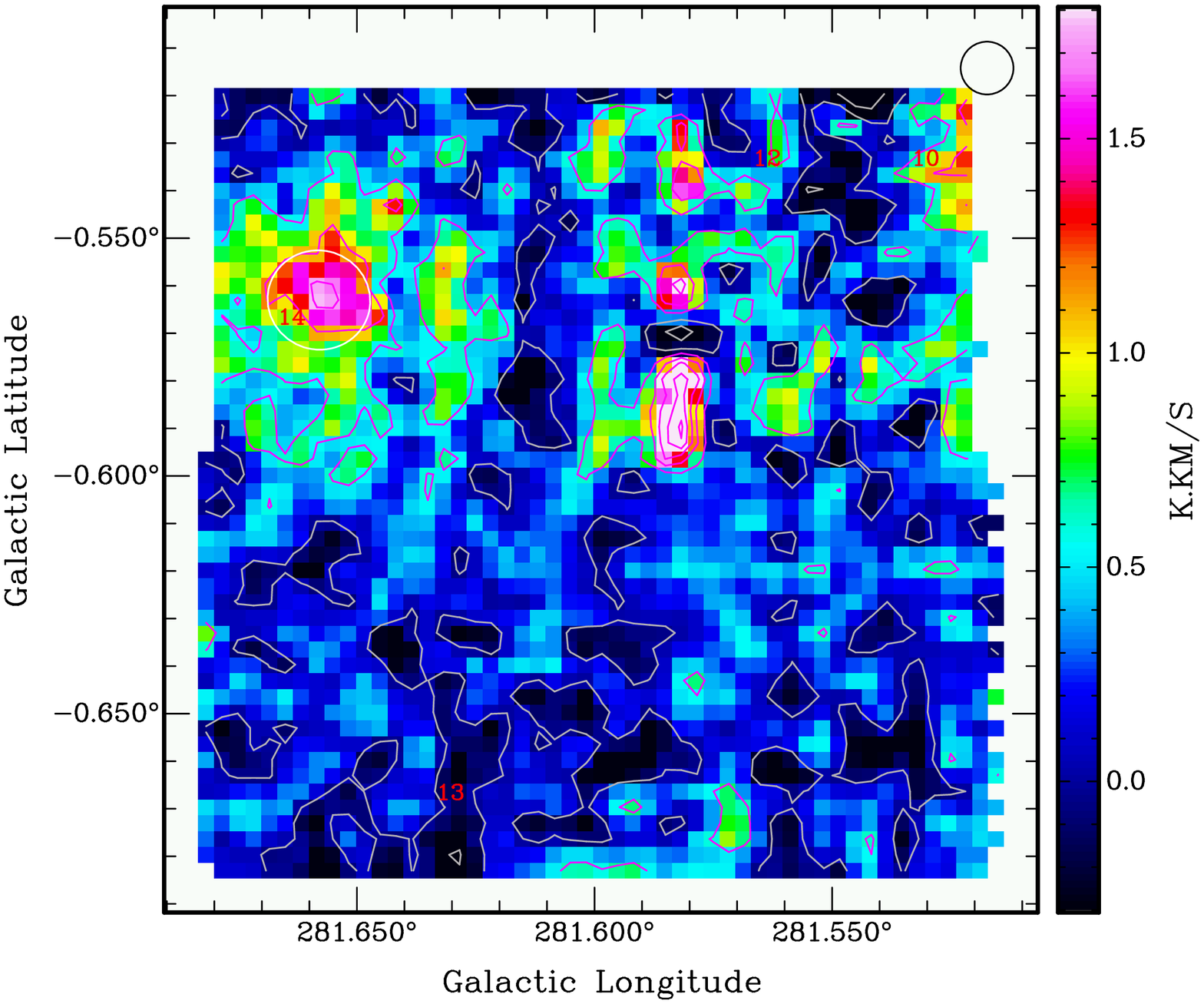}}
\caption{\small Same as Fig.\,\ref{reg1}, but for Region 2a sources BYF\,10 and 12--14.  The integration here is over the range --4.8 to --2.5\,\kms\ or 23 channels, yielding an average rms noise level 0.279\,K\kms, and contour levels spaced every 2$\sigma$ are overlaid.  At a distance of 3.2\,kpc, the smoothed Mopra beam (upper right corner) scales to 40$''$ = 0.621\,pc, or 1\,pc = 64$''$\hspace{-1mm}.5.
\label{reg2a}}
\end{figure}

\begin{figure*}[htp]
\centerline{(a)\includegraphics[angle=0,scale=0.3]{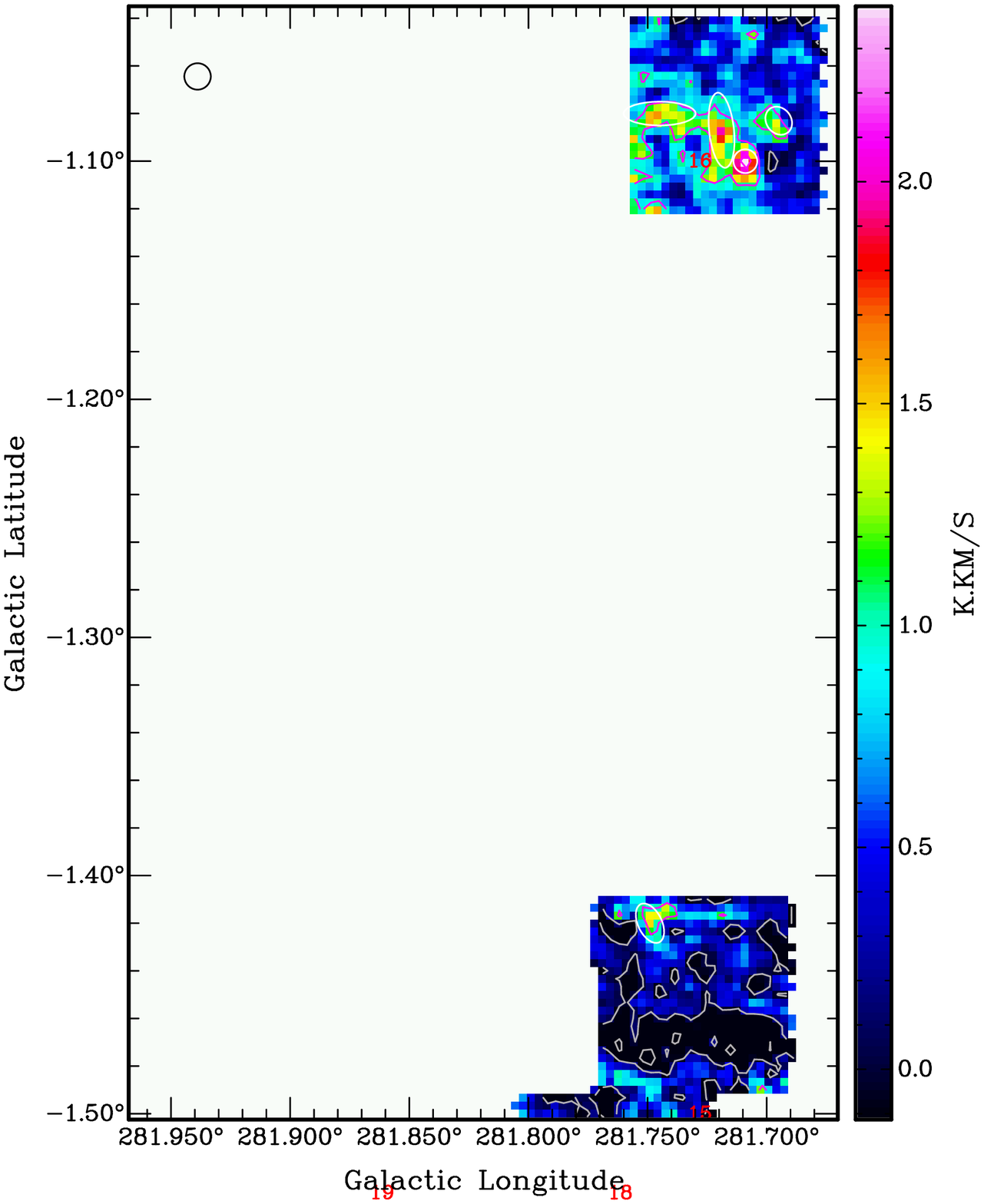}
		(b)\includegraphics[angle=0,scale=0.3]{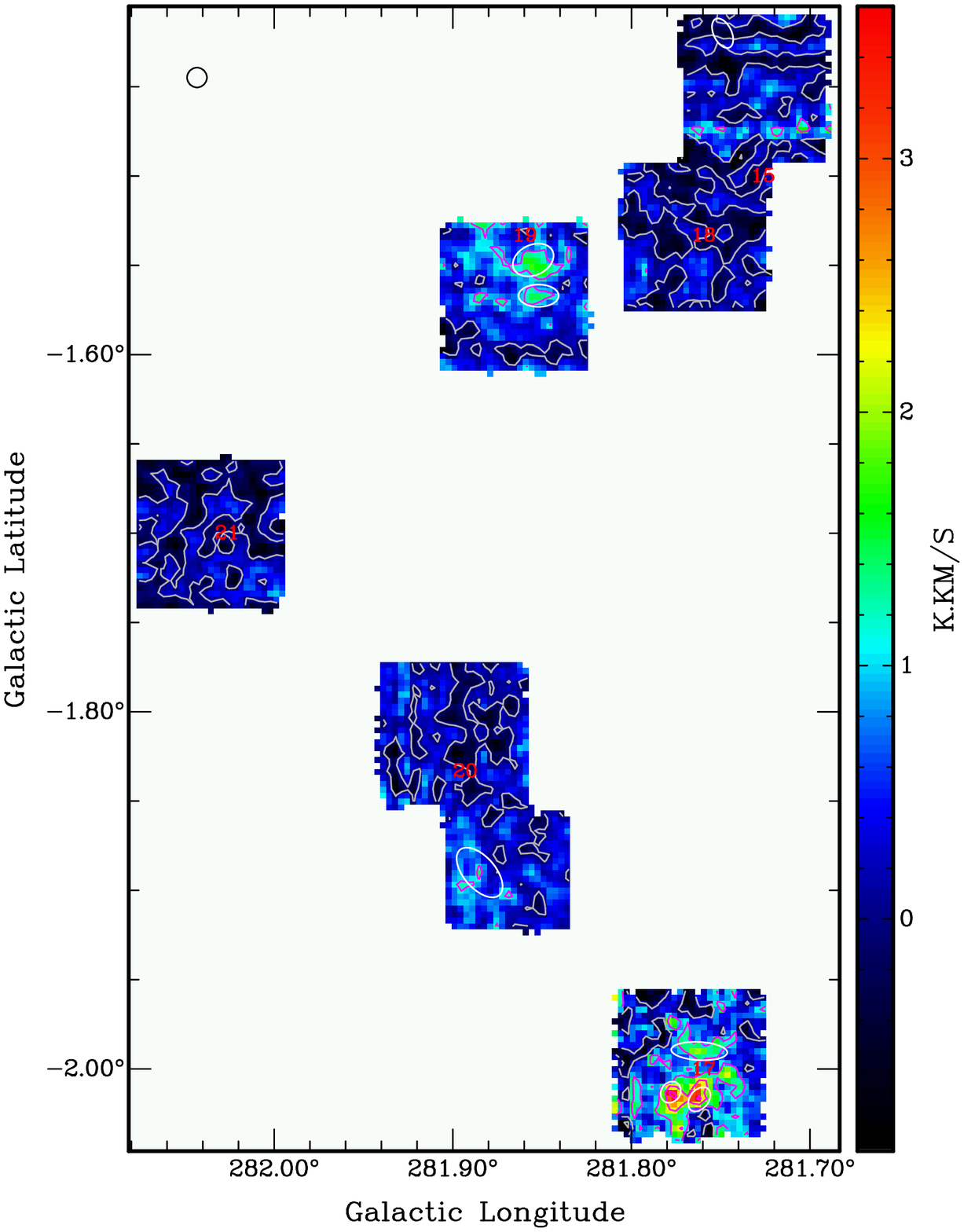}}
\caption{\small Same as Fig.\,\ref{reg1}, but for Region 2b \& 3 sources ($a$) BYF\,15 and 16, and ($b$) BYF\,15 and 17--21.  The respective integrations are over the ranges --2.0 to +2.5\,\kms\ (44 channels) and --8.3 to --3.0\,\kms\ (51 channels), however the various clumps are at different $V_{\rm LSR}$ and are imaged at higher S/N when integrated over more restricted velocity ranges, as in Table \ref{sources}.  Here the maps have average respective rms noise levels of 0.347\,K\kms and 0.374\,K\kms, and contour levels spaced every 3$\sigma$ (for both) are overlaid.  At a distance of 3.2\,kpc, the smoothed Mopra beam (upper left corner) scales to 40$''$ = 0.621\,pc, or 1\,pc = 64$''$\hspace{-1mm}.5.
\label{reg2b3a}}
\end{figure*}

\begin{figure*}[htp]
\centerline{\includegraphics[angle=-90,scale=0.2]{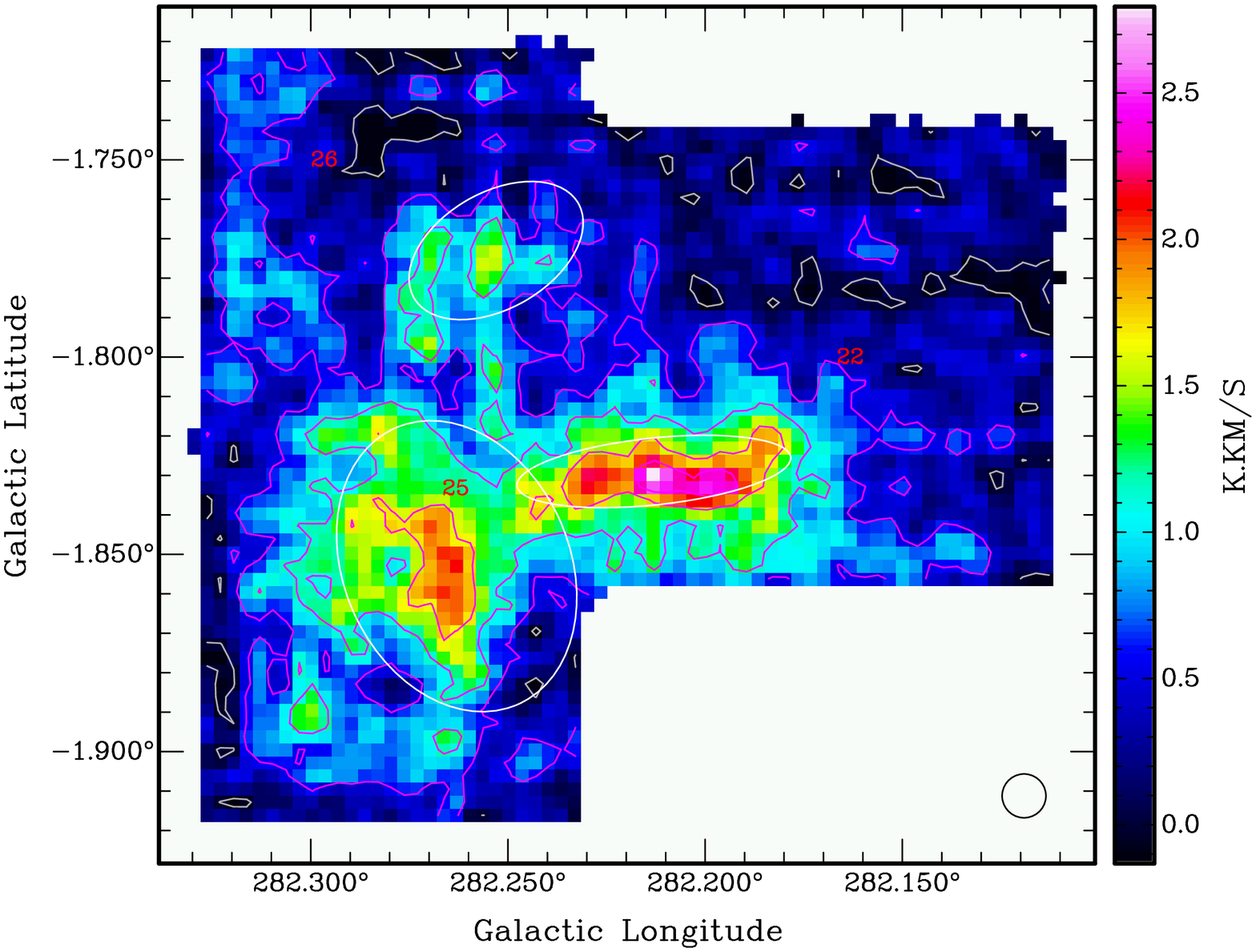}}
\caption{\small Same as Fig.\,\ref{reg1}, but for Region 3 sources BYF\,22,25,26.  The integration here is over the range --17.2 to --12.2\,\kms\ or 45 channels, yielding an average rms noise level 0.186\,K\kms, and contour levels spaced every 3$\sigma$ are overlaid.  At a distance of 3.2\,kpc, the smoothed Mopra beam (lower right corner) scales to 40$''$ = 0.621\,pc, or 1\,pc = 64$''$\hspace{-1mm}.5.
\label{reg3b}}
\end{figure*}

\begin{figure*}[htp]
\centerline{
(a)\includegraphics[angle=0,scale=0.27]{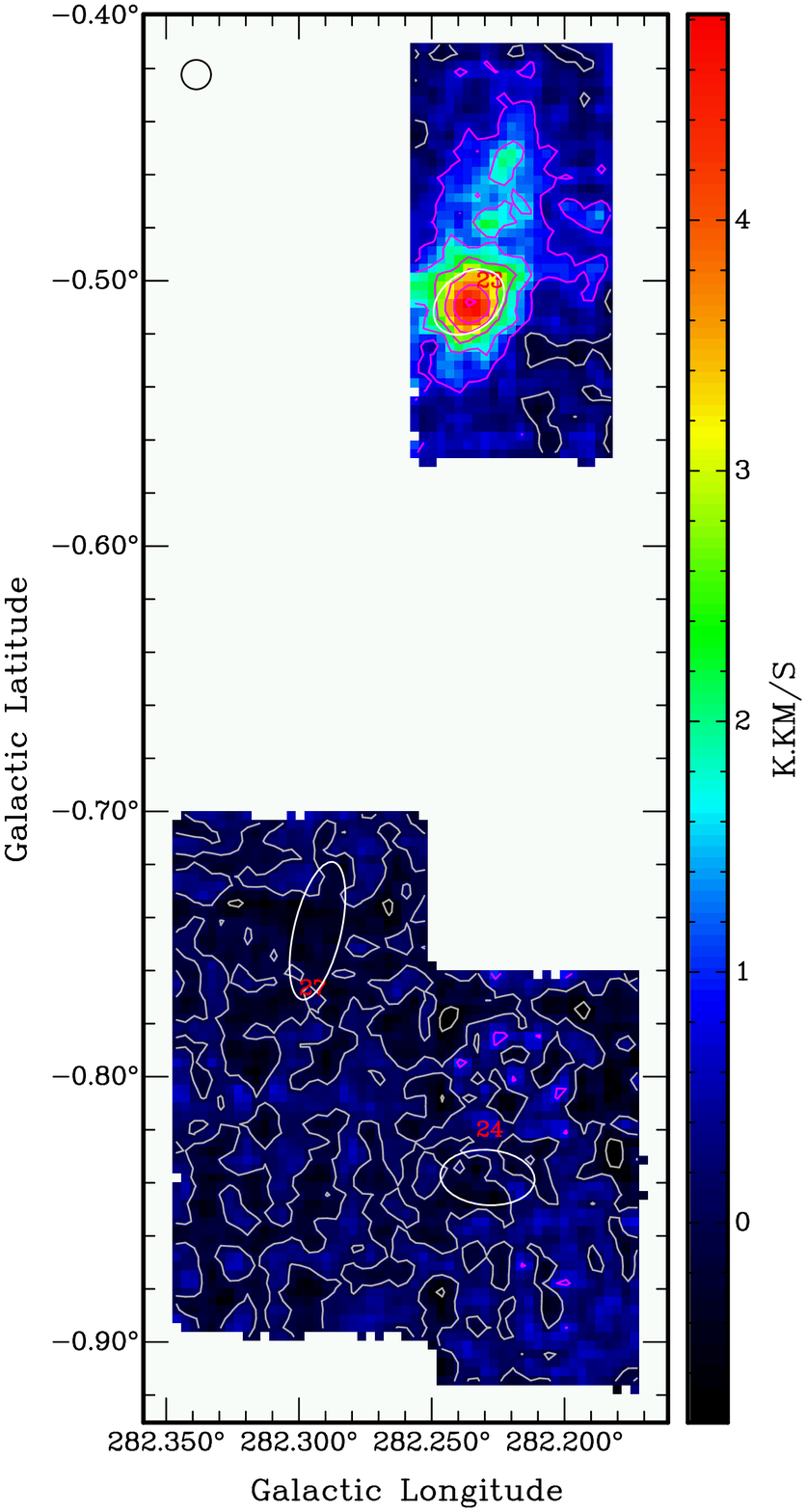}
(b)\includegraphics[angle=0,scale=0.27]{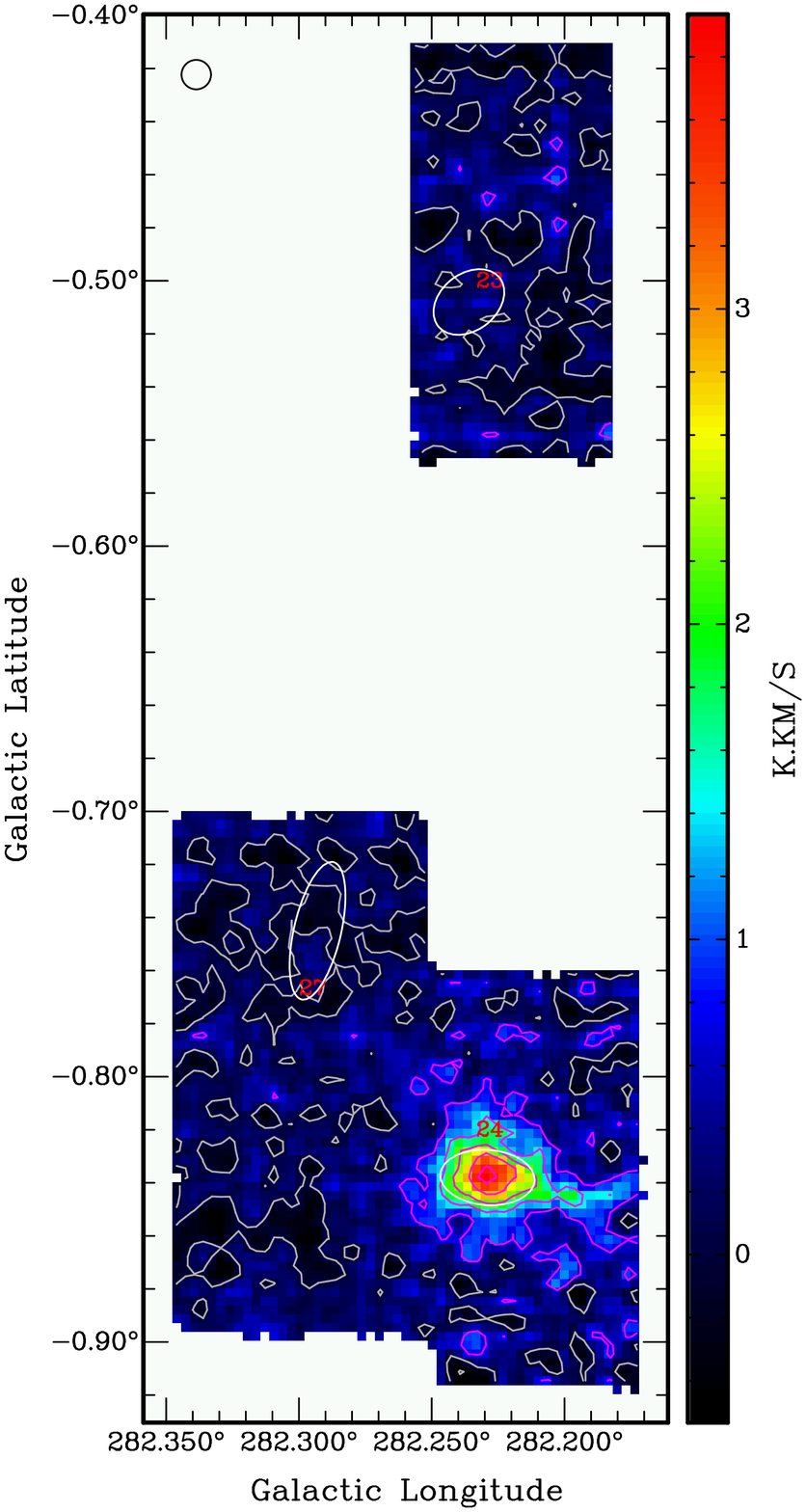}
(c)\includegraphics[angle=0,scale=0.27]{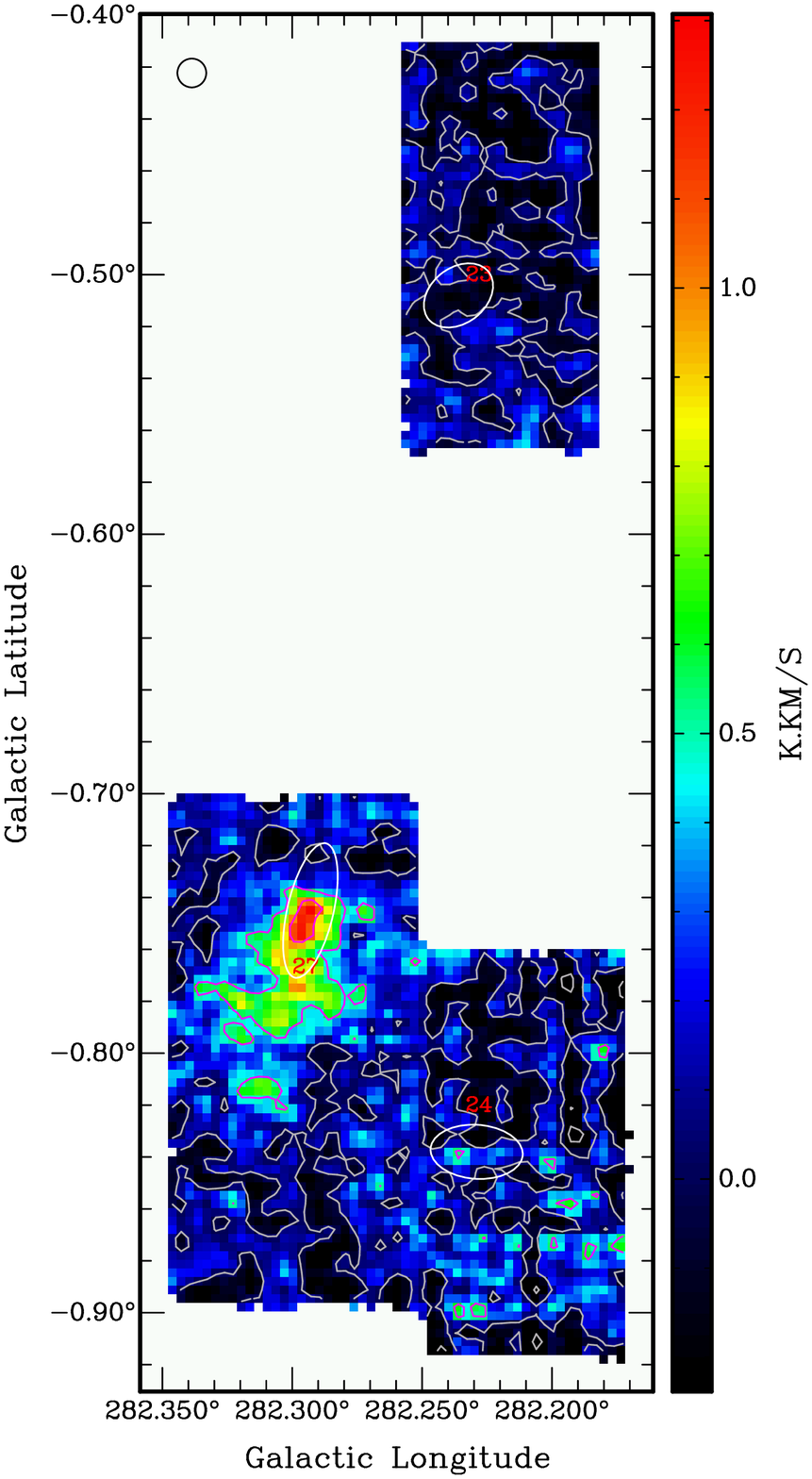}}
\caption{\small Same as Fig.\,\ref{reg1}, but for Region 2c sources ($a$) BYF\,23, ($b$) 24, and ($c$) 27.  The respective integrations are over the ranges --5.0 to +0.3\,\kms\ (48 channels), --14.6 to --10.0\,\kms\ (42 channels), and +4.8 to +7.0\,\kms\ (21 channels), yielding average rms noise levels of 0.260\,K\kms, 0.243\,K\kms, and 0.169\,K\kms; contour levels spaced every 3$\sigma$ (for all) are overlaid.  At a distance of 3.2\,kpc, the smoothed Mopra beam (upper left corner) scales to 40$''$ = 0.621\,pc, or 1\,pc = 64$''$\hspace{-1mm}.5.
\label{reg2c}}
\end{figure*}

\begin{figure*}[htp]
\centerline{(a)\includegraphics[angle=-90,scale=0.2]{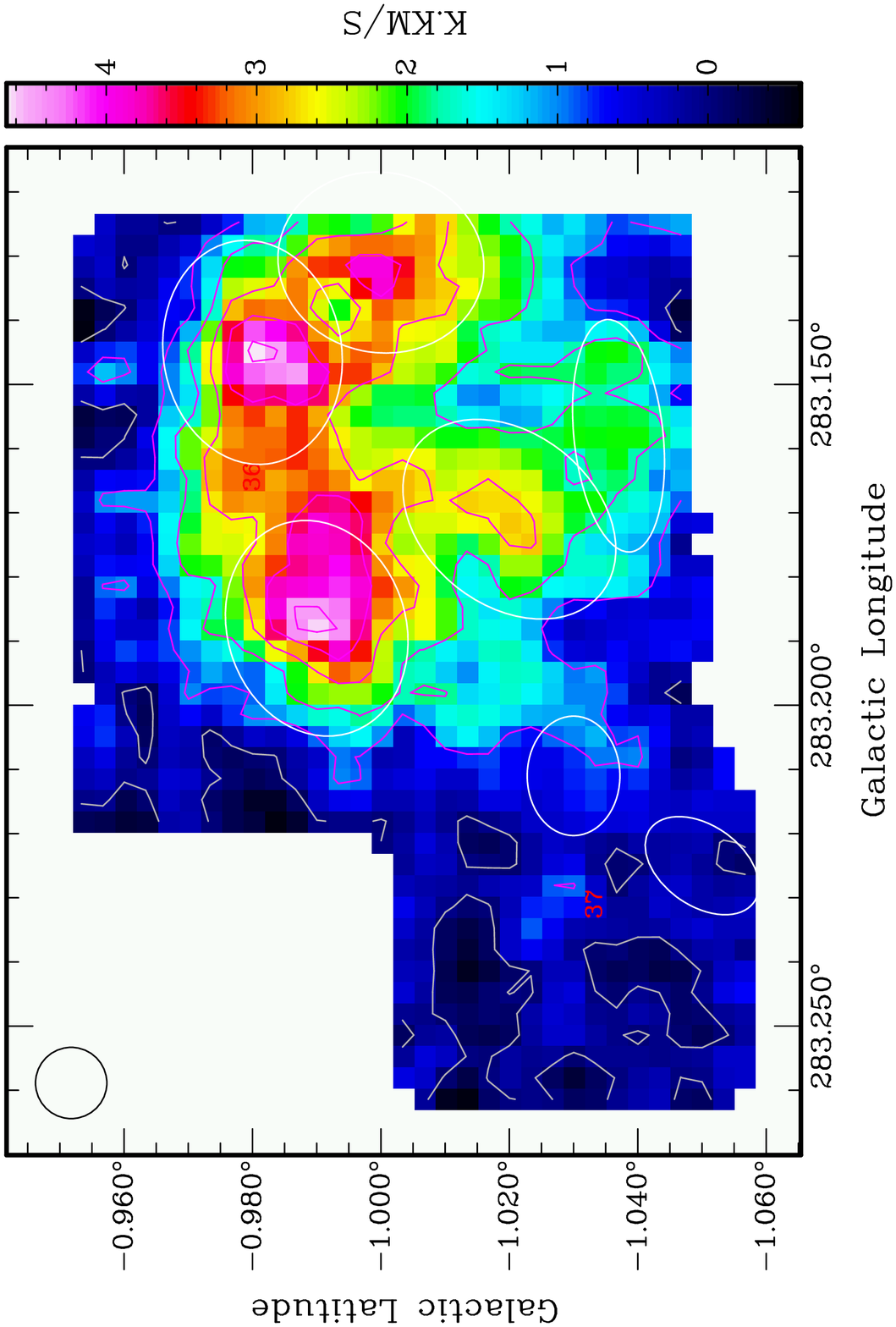}
		(b)\includegraphics[angle=-90,scale=0.2]{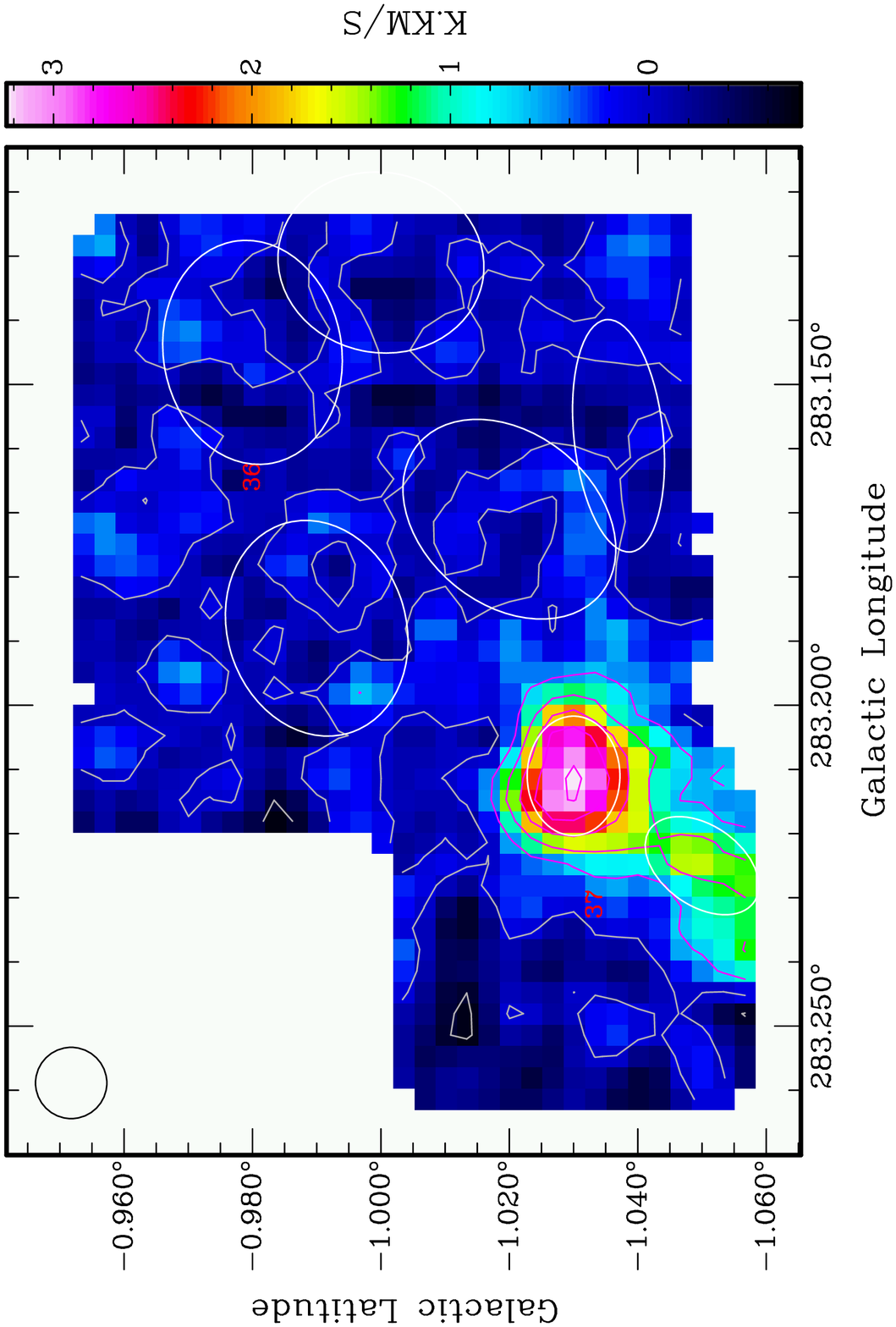}}
\caption{\small Same as Fig.\,\ref{reg1}, but for Region 5 sources ($a$) BYF\,36 and ($b$) 37.  The respective integrations are over the ranges --9.1 to --1.2\,\kms\ (70 channels) and +2.1 to +5.85\,\kms\ (35 channels), yielding average rms noise levels of 0.297\,K\kms and 0.208\,K\kms; contour levels spaced every 3$\sigma$ (for both) are overlaid.  At a distance of 3.2\,kpc, the smoothed Mopra beam (upper left corner) scales to 40$''$ = 0.621\,pc, or 1\,pc = 64$''$\hspace{-1mm}.5.
\label{reg5}}
\end{figure*}

\begin{figure*}[htp]
\centerline{\includegraphics[angle=-90,scale=0.25]{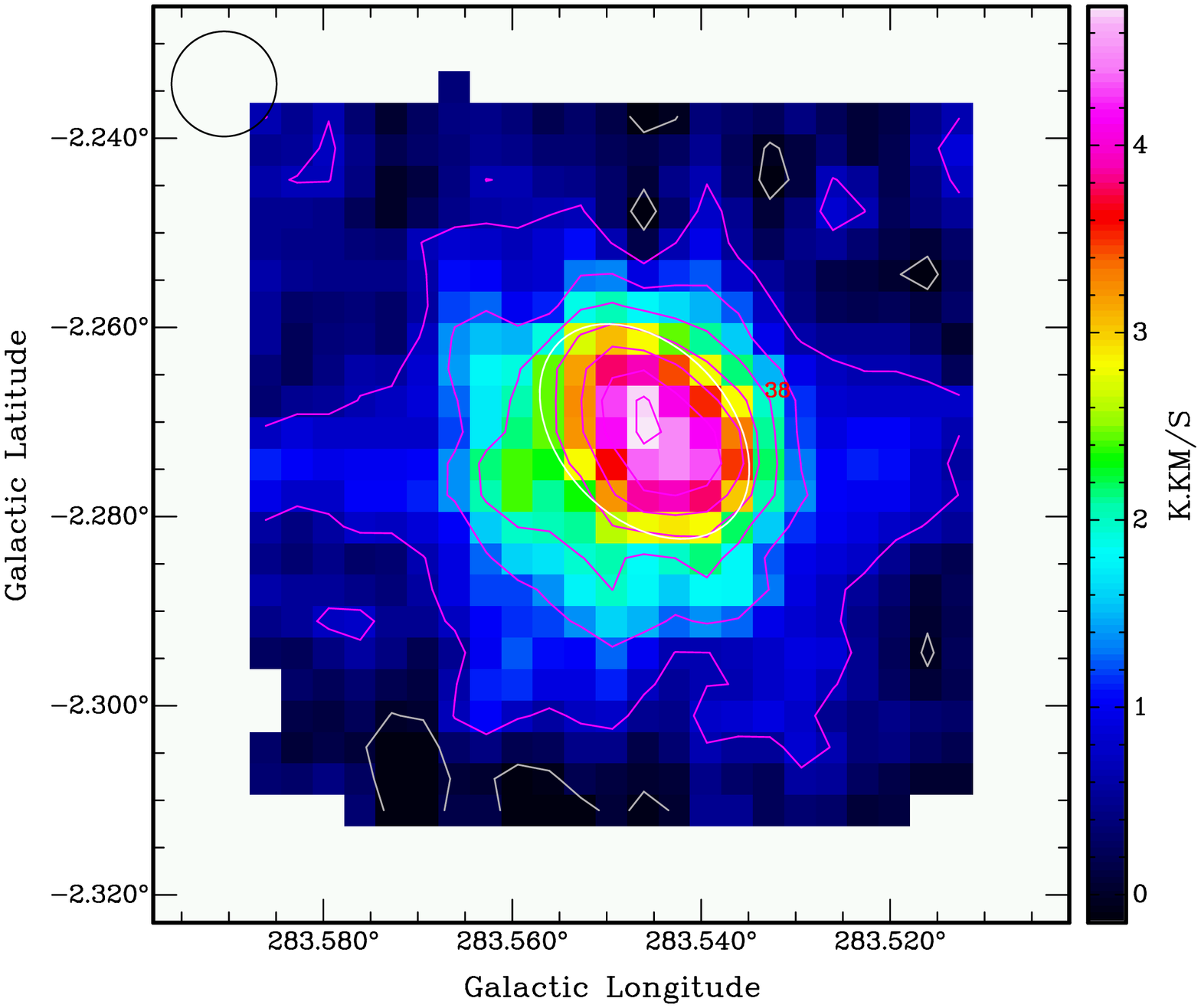}}
\caption{\small Same as Fig.\,\ref{reg1}, but for isolated source BYF\,38.  The integration here is over the range --6.5 to --2.7\,\kms\ or 34 channels, yielding an average rms noise level 0.219\,K\kms, and contour levels spaced every 3$\sigma$ are overlaid.  At a distance of 2.0\,kpc, the smoothed Mopra beam (upper left corner) scales to 40$''$ = 0.388\,pc or 1\,pc = 103$''$\hspace{-1mm}.1.
\label{by38}}
\end{figure*}

\begin{figure*}[htp]
\centerline{(a)\includegraphics[angle=0,scale=0.33]{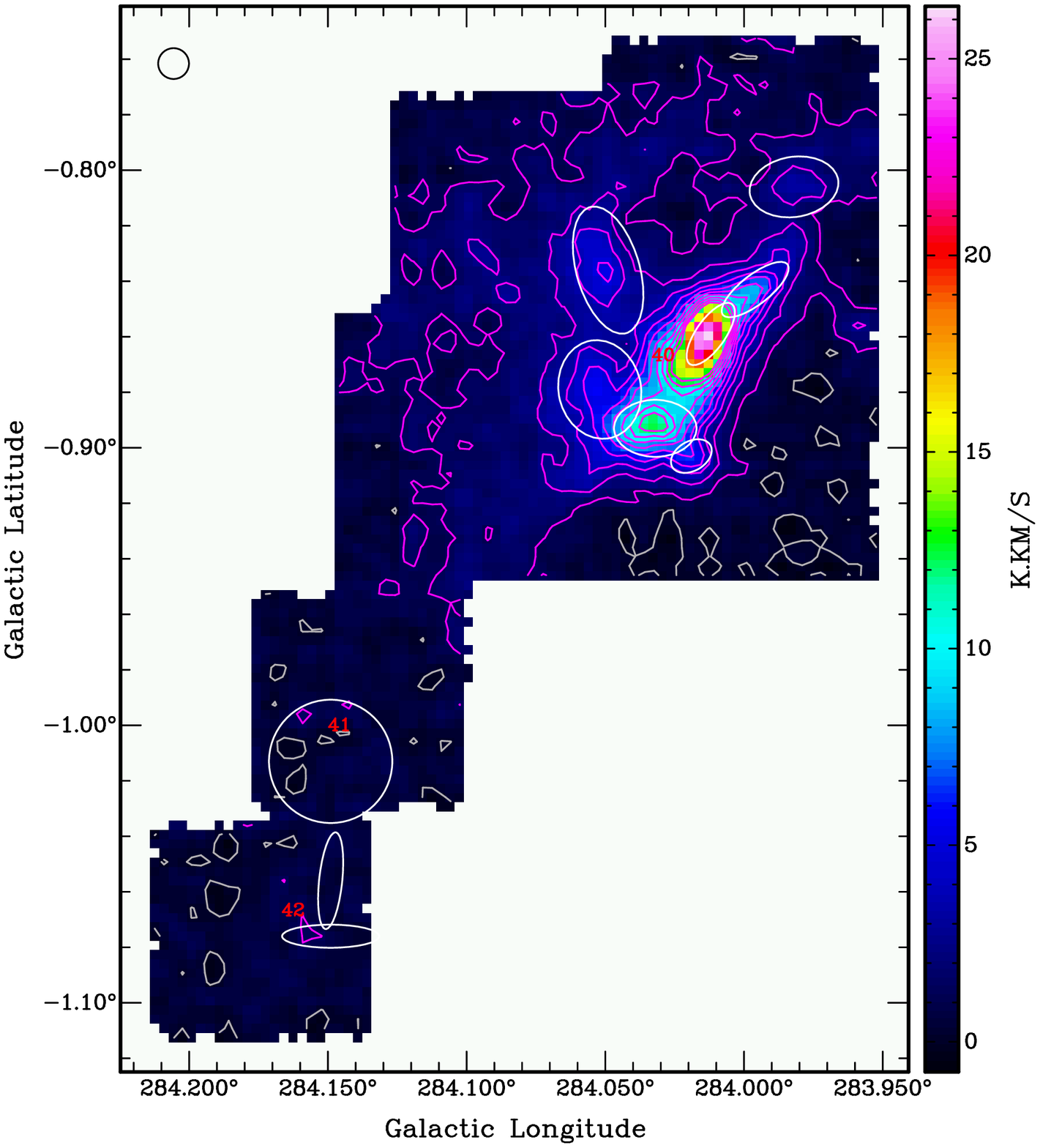}
		(b)\includegraphics[angle=0,scale=0.33]{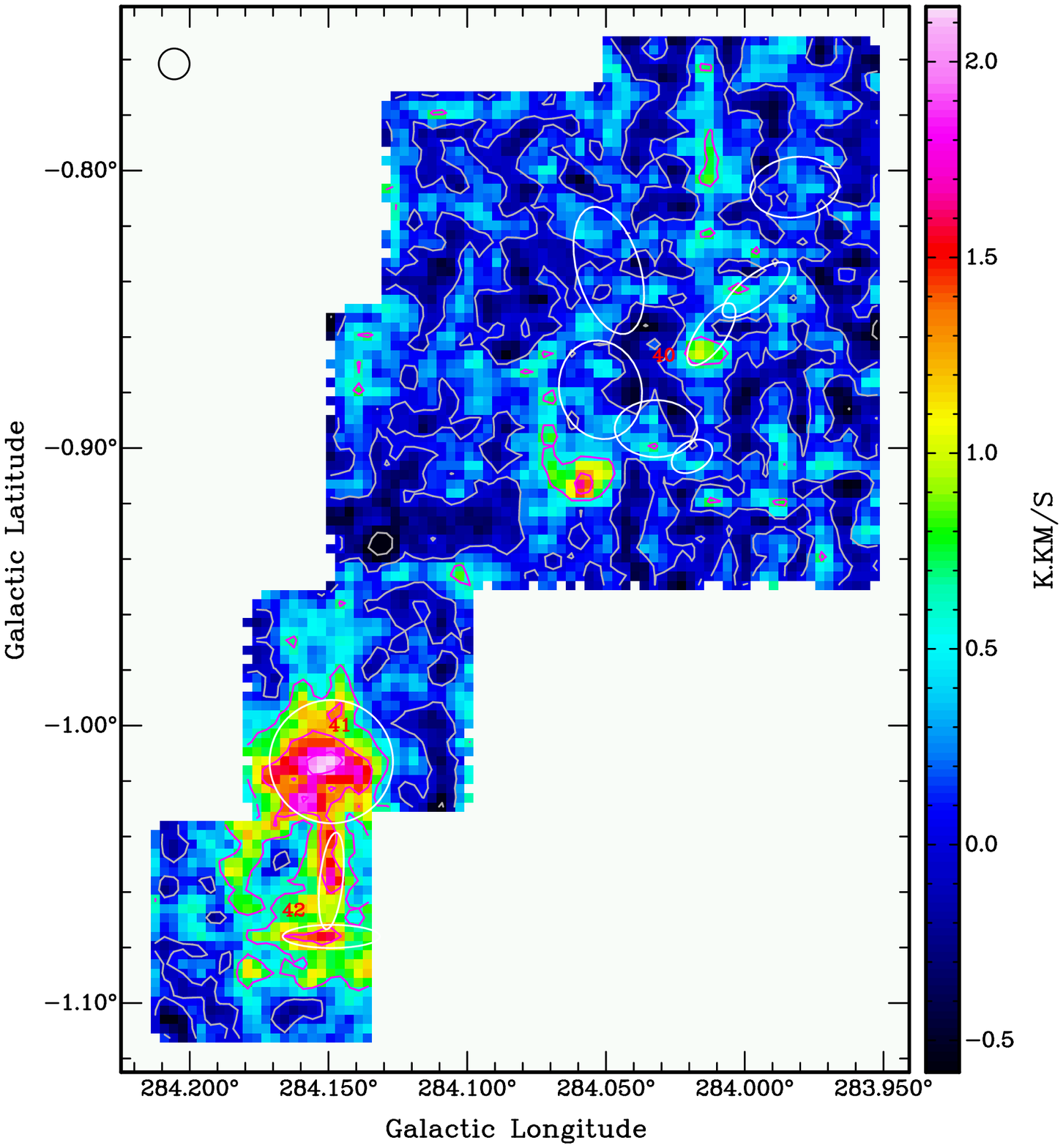}}
\caption{\small Same as Fig.\,\ref{reg1}, but for Region 6 sources ($a$) BYF\,40 and ($b$) 41 \& 42.  The respective integrations are over the ranges +4.5 to +13.9\,\kms\ (84 channels) and +0.7 to +5.0\,\kms\ (39 channels), yielding average rms noise levels of 0.312\,K\kms and 0.213\,K\kms; contour levels spaced every 4$\sigma$ and 3$\sigma$ (respectively) are overlaid.  At a distance of 6.6\,kpc, the smoothed Mopra beam (upper left corner) scales to 40$''$ = 1.280\,pc, or 1\,pc = 31$''$\hspace{-1mm}.3.
\label{reg6}}
\end{figure*}

\begin{figure*}[htp]
\centerline{\includegraphics[angle=-90,scale=0.2]{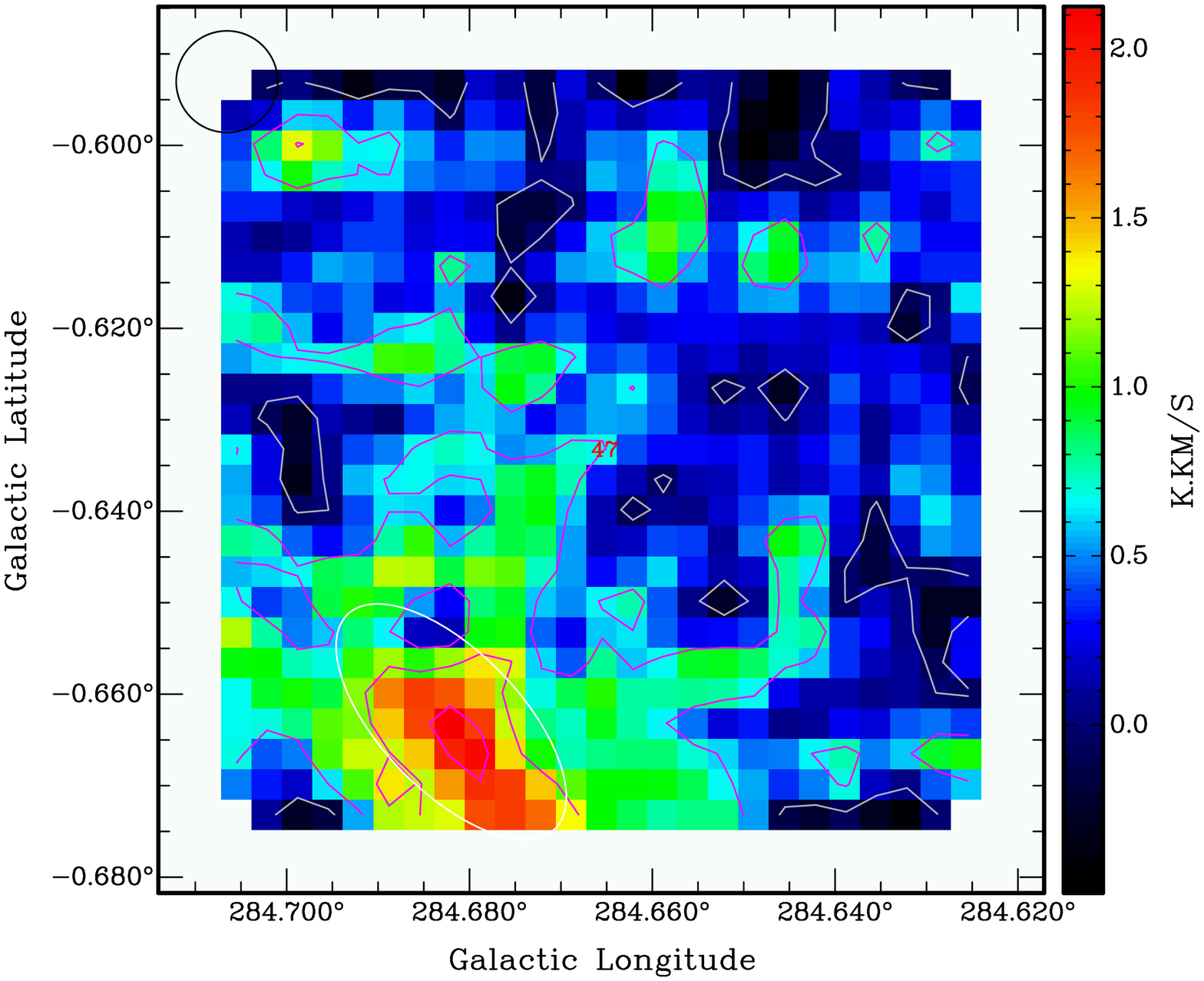}}
\caption{\small Same as Fig.\,\ref{reg1}, but for Region 7 source BYF\,47.  The integration here is over the range +1.0 to +7.0\,\kms\ or 58 channels, yielding an average rms noise level 0.318\,K\kms, and contour levels spaced every 2$\sigma$ are overlaid.  At a distance of 5.3\,kpc, the smoothed Mopra beam (upper left corner) scales to 40$''$ = 1.028\,pc or 1\,pc = 38$''$\hspace{-1mm}.9.
\label{reg7}}
\end{figure*}

\begin{figure*}[ht]
\centerline{\includegraphics[angle=-90,scale=0.4]{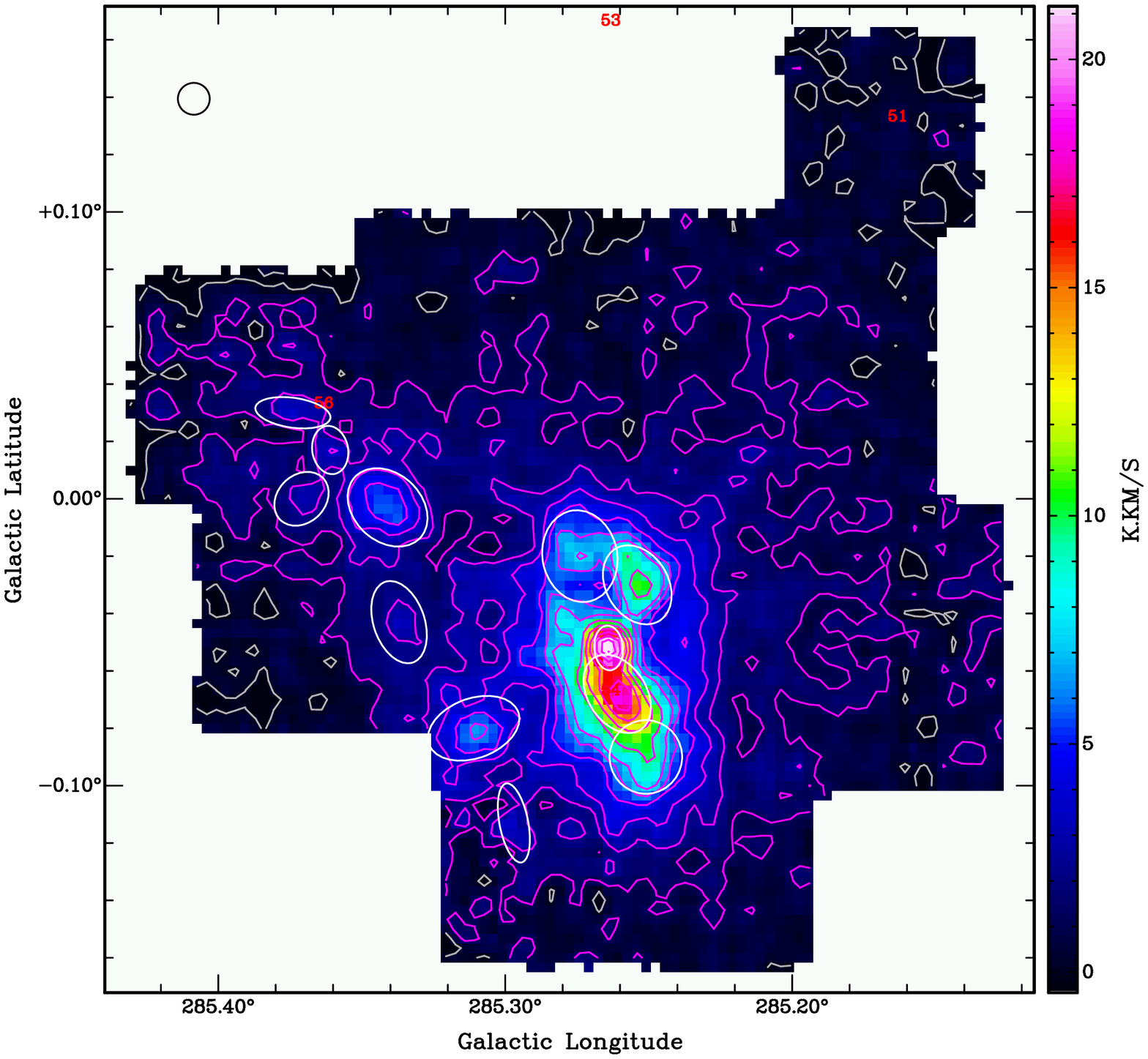}}
\caption{\small Same as Fig.\,\ref{reg1}, but for Region 8 sources BYF\,51--56.  The integration here is over the range --3.7 to +7.8\,\kms\ or 103 channels, yielding an average rms noise level 0.401\,K\kms, and contour levels spaced every 3$\sigma$ up to 18$\sigma$, and then every 6$\sigma$, are overlaid.  At a distance of 5.3\,kpc, the smoothed Mopra beam (upper left corner) scales to 40$''$ = 1.028\,pc or 1\,pc = 38$''$\hspace{-1mm}.9.
\label{reg8}}
\end{figure*}

\begin{figure*}[ht]
\centerline{\includegraphics[angle=-90,scale=0.74]{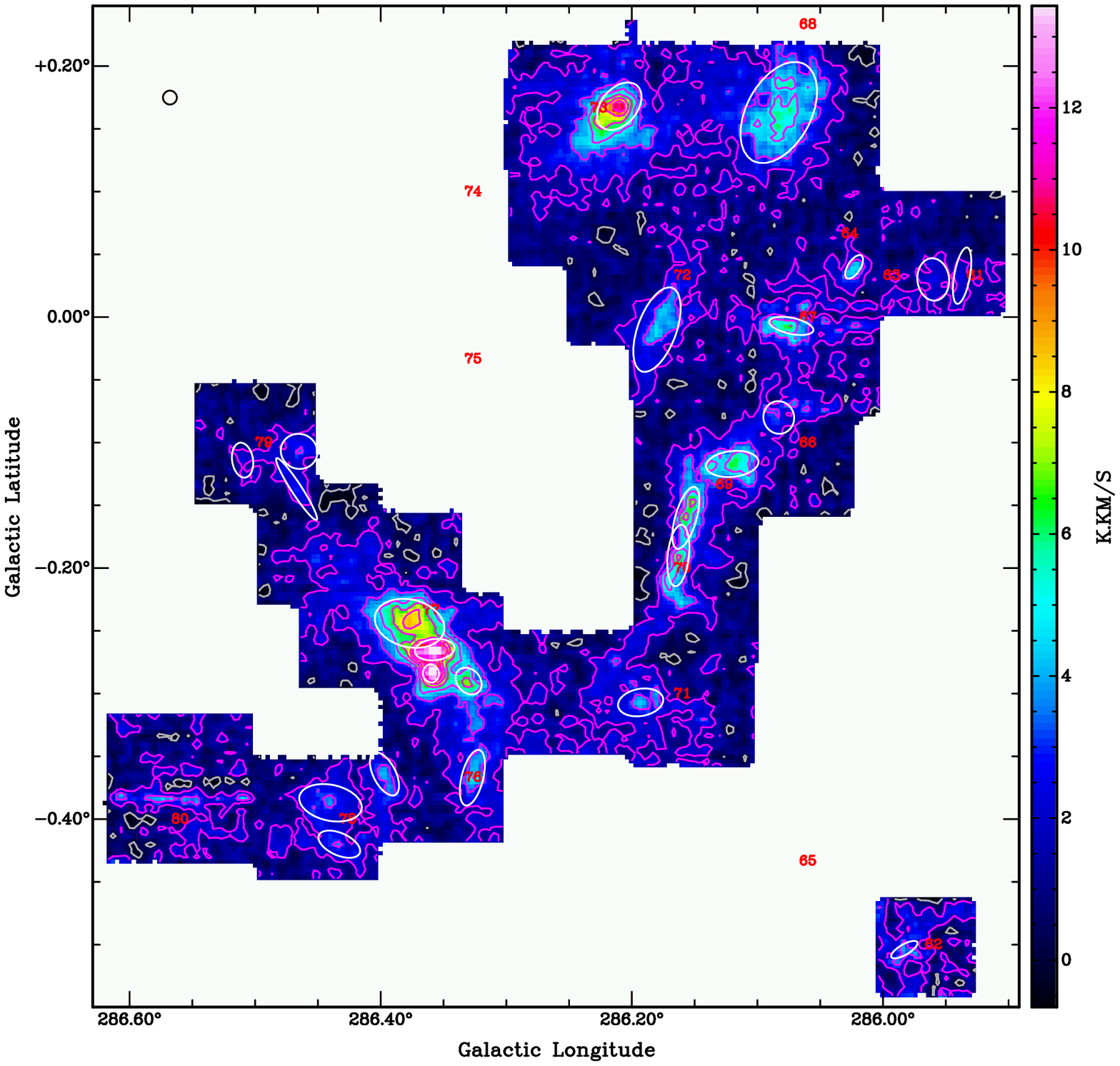}} 
\caption{\small Same as Fig.\,\ref{reg1}, but for Region 9 sources BYF\,61--80.  The integration here is over the range --27.0 to --11.0\,\kms\ or 143 channels, however the various clumps are at different $V_{\rm LSR}$ and are imaged at higher S/N when integrated over more restricted velocity ranges, as in Table \ref{sources}.  Here the map has an average rms noise level 0.404\,K\kms, and contour levels spaced every 4$\sigma$ are overlaid.  At a distance of 2.5\,kpc, the smoothed Mopra beam (upper left corner) scales to 40$''$ = 0.485\,pc or 1\,pc = 82$''$\hspace{-1mm}.5.
\label{reg9}}
\end{figure*}

\begin{figure*}[ht]
\centerline{\includegraphics[angle=0,scale=0.62]{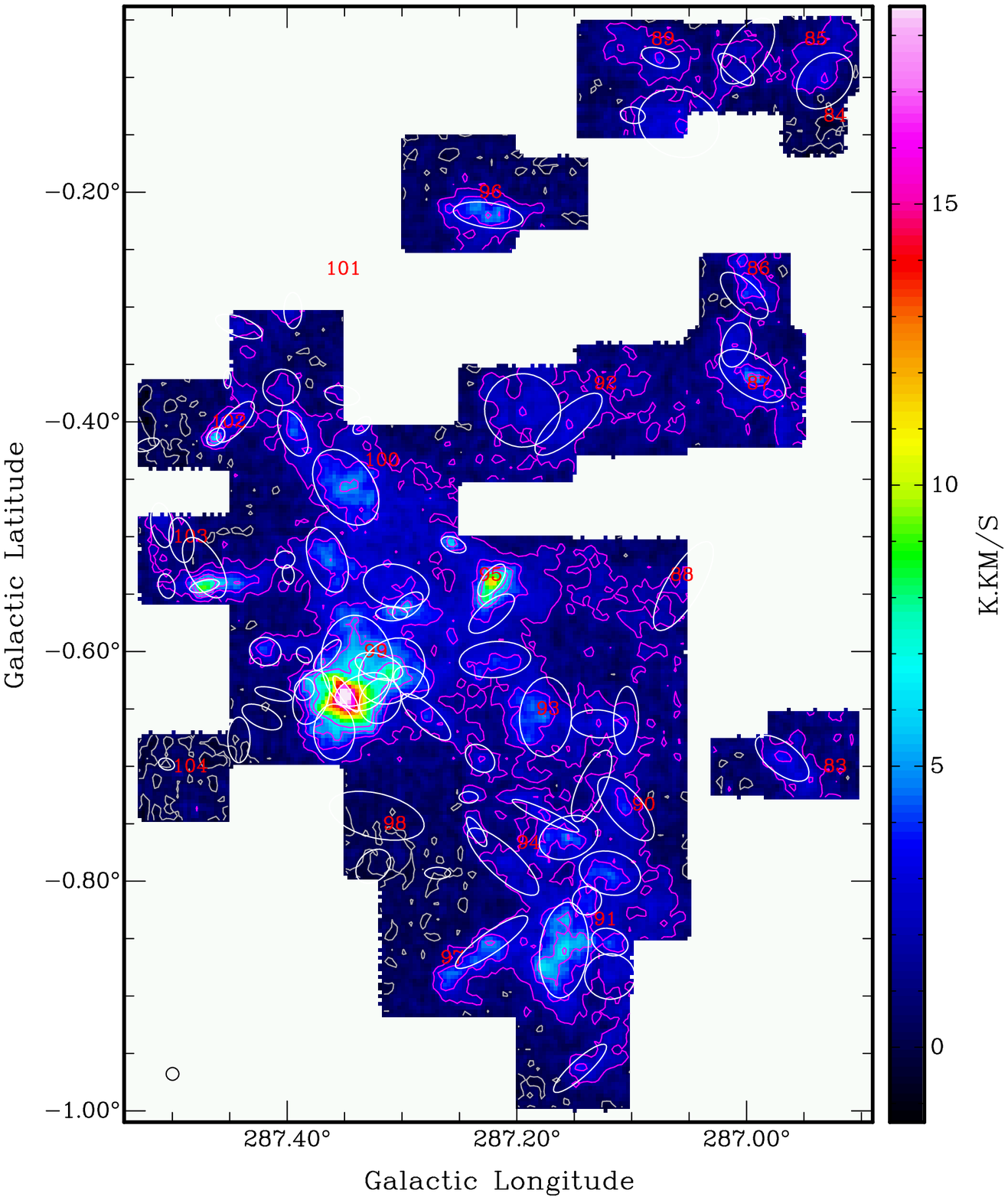}} 
\caption{\small Same as Fig.\,\ref{reg1}, but for Region 10 sources BYF\,83--104.  The integration here is over the range --23.5 to --12.9\,\kms\ or 95 channels, however the various clumps are at different $V_{\rm LSR}$ and are imaged at higher S/N when integrated over more restricted velocity ranges, as in Table \ref{sources}.  Here the map has an average rms noise level 0.321\,K\kms, and contour levels spaced every 5$\sigma$ are overlaid.  At a distance of 2.5\,kpc, the smoothed Mopra beam (lower left corner) scales to 40$''$ = 0.485\,pc or 1\,pc = 82$''$\hspace{-1mm}.5.
\label{reg10}}
\end{figure*}

\begin{figure*}[ht]
\centerline{\includegraphics[angle=-90,scale=0.74]{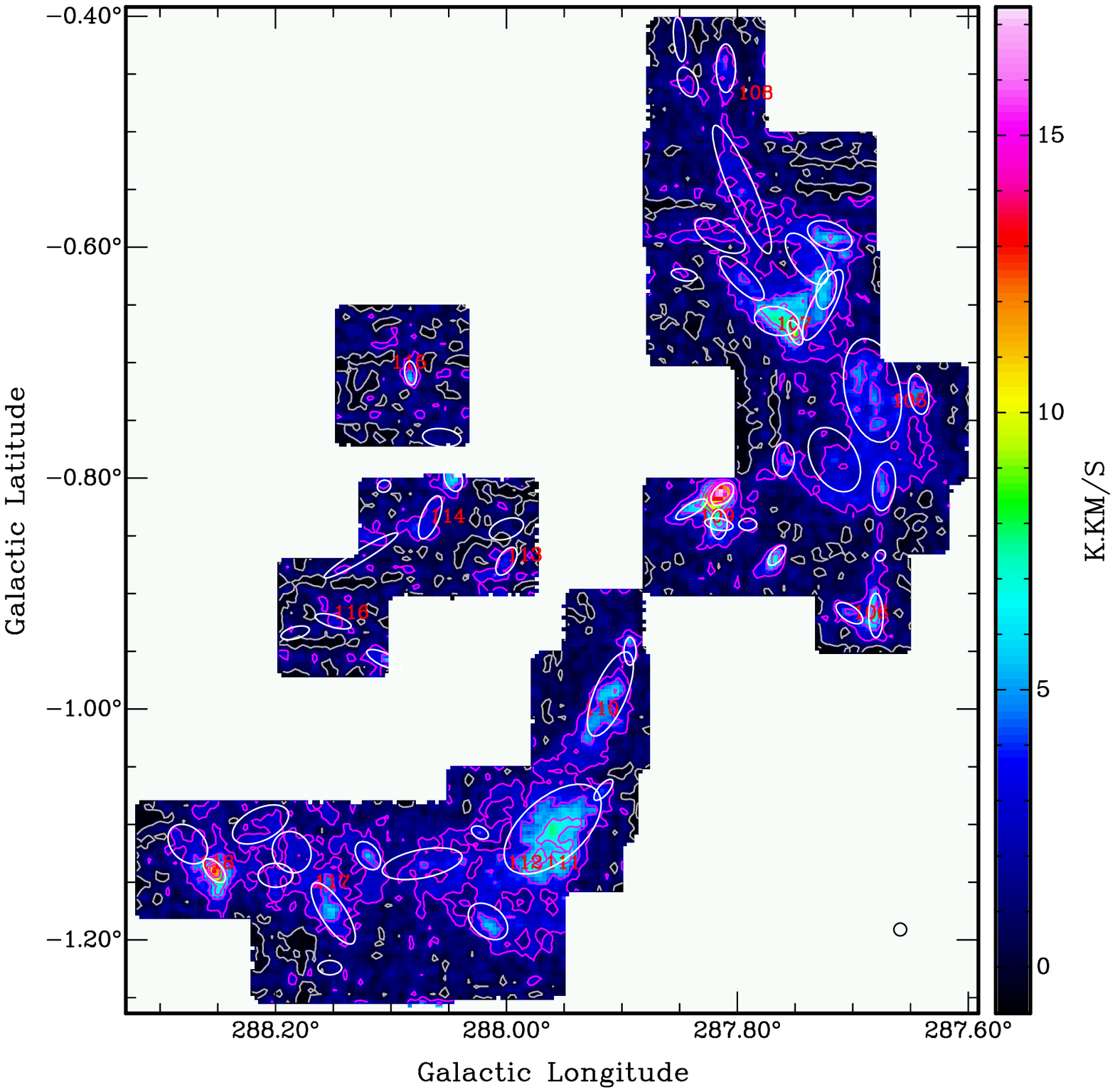}} 
\caption{\small Same as Fig.\,\ref{reg1}, but for Region 11 sources BYF\,105--118.  The integration here is over the range --33.0 to --9.5\,\kms\ or 209 channels, however the various clumps are at different $V_{\rm LSR}$ and are imaged at higher S/N when integrated over more restricted velocity ranges, as in Table \ref{sources}.  Here the map has an average rms noise level 0.511\,K\kms, and contour levels spaced every 4$\sigma$ are overlaid.  At a distance of 2.5\,kpc, the smoothed Mopra beam (lower right corner) scales to 40$''$ = 0.485\,pc or 1\,pc = 82$''$\hspace{-1mm}.5.
\label{reg11}}
\end{figure*}

\begin{figure*}[ht]
\centerline{\includegraphics[angle=-90,scale=0.25]{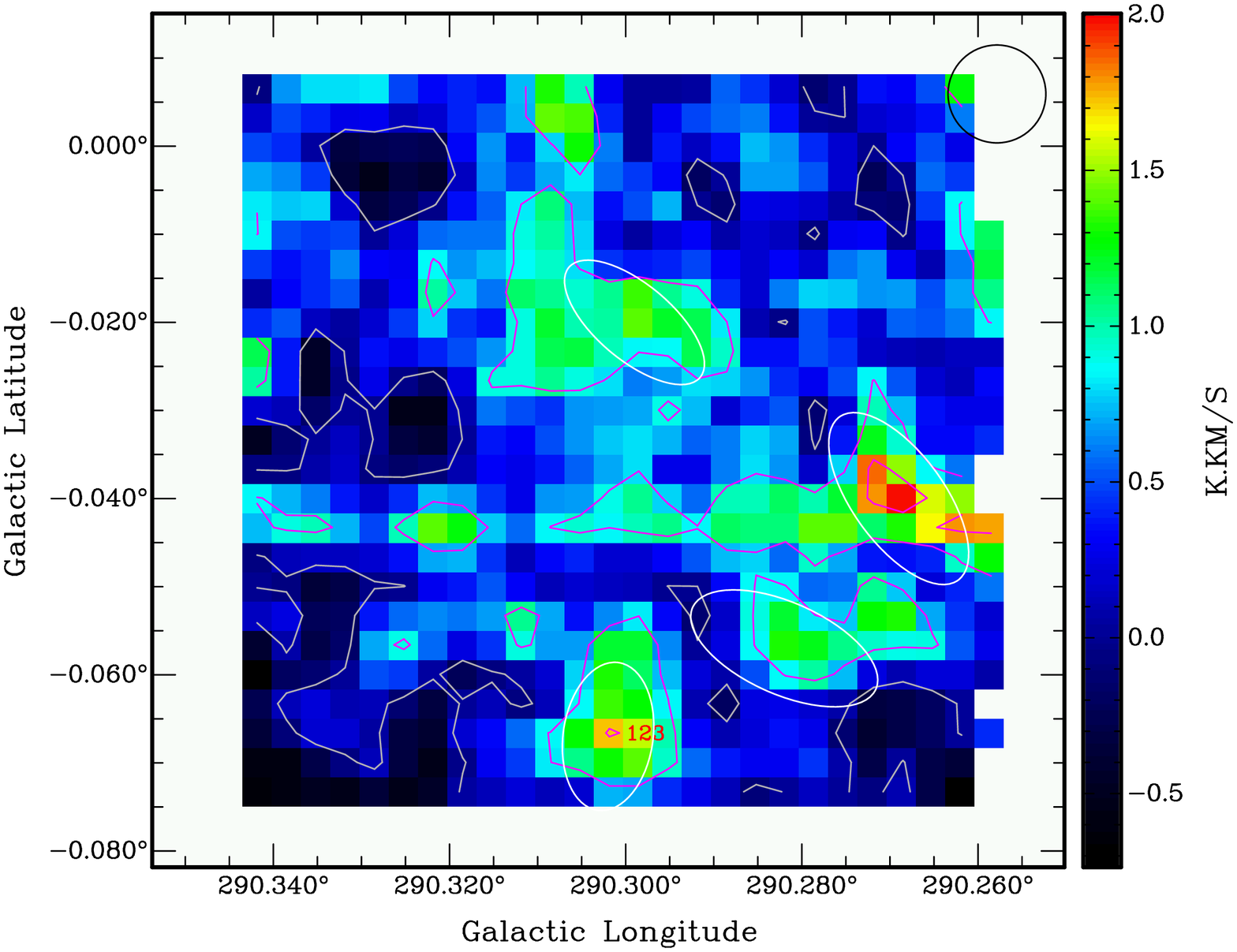}}
\caption{\small Same as Fig.\,\ref{reg1}, but for isolated source BYF\,123.  The integration here is over the range --3.0 to +1.5\,\kms\ or 44 channels, yielding an average rms noise level 0.418\,K\kms, and contour levels spaced every 2$\sigma$ are overlaid.  At a distance of 6.8\,kpc, the smoothed Mopra beam (upper right corner) scales to 40$''$ = 1.32\,pc or 1\,pc = 30$''$\hspace{-1mm}.3.
\label{byf123}}
\end{figure*}

\begin{figure*}[ht]
\centerline{(a)\includegraphics[angle=-90,scale=0.35]{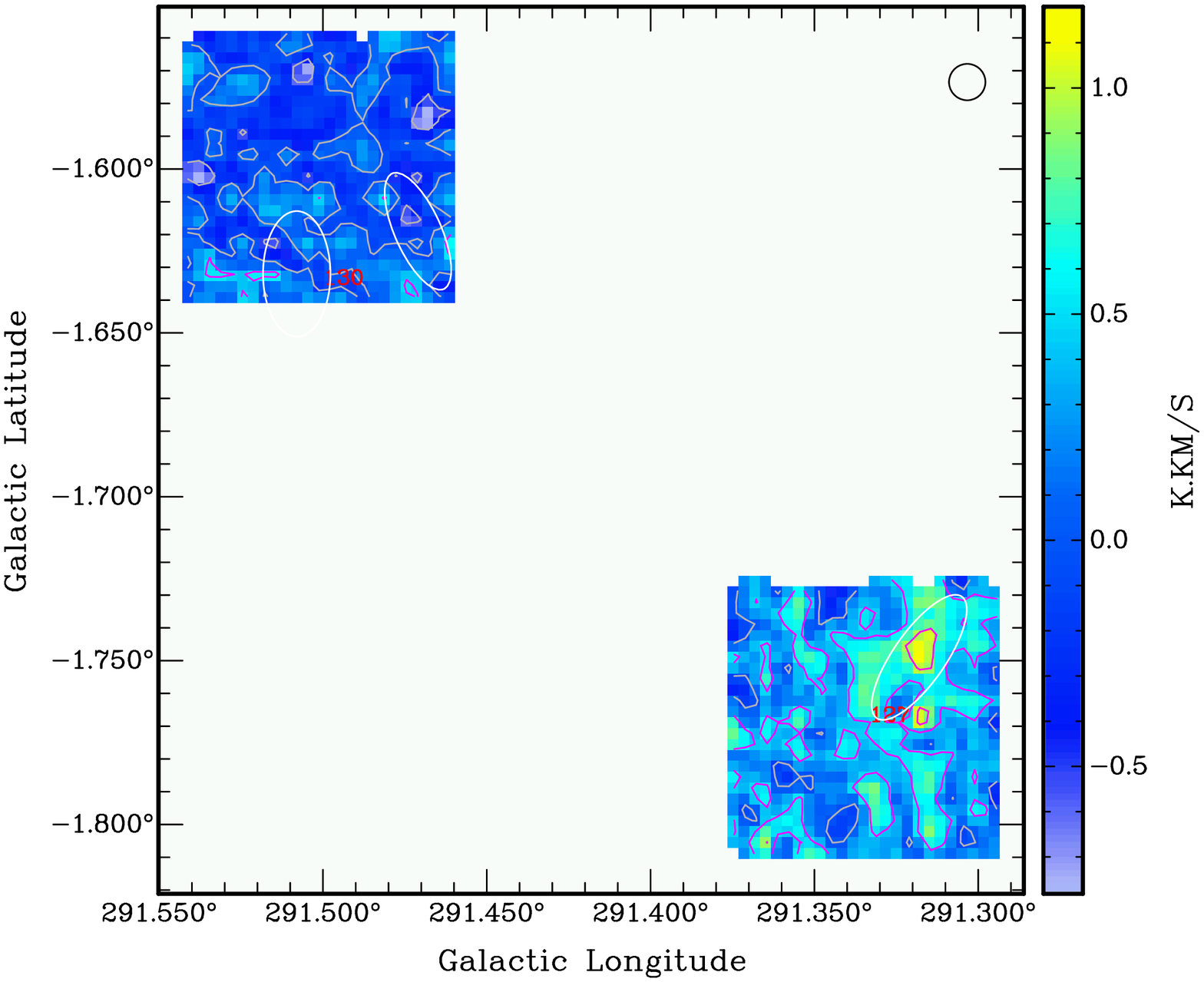}
		(b)\includegraphics[angle=-90,scale=0.35]{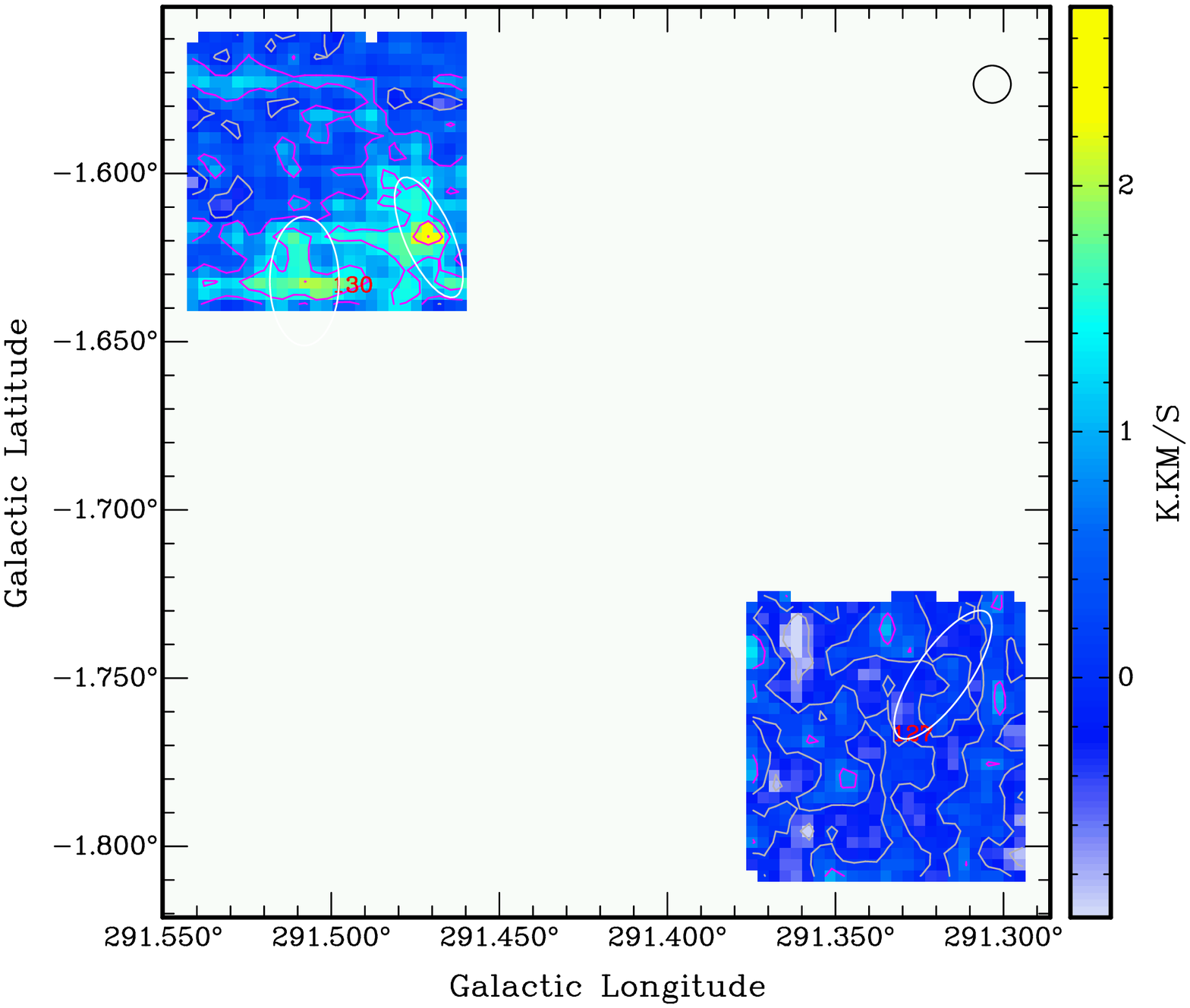}}
\caption{\small Same as Fig.\,\ref{reg1}, but for Region 12 sources ($a$) BYF\,127 and ($b$) BYF\,130.  The respective integrations are over the ranges --6.1 to --3.8\,\kms\ (23 channels) and --27 to --22\,\kms\ (48 channels), yielding average rms noise levels of 0.230\,K\kms and 0.338\,K\kms; contour levels spaced every 2$\sigma$ (for both) are overlaid.  At distances of 1.1 and 2.4\,kpc, the smoothed Mopra beam (lower left corner) scales to 40$''$ = 0.213 and 0.465\,pc, or 1\,pc = 187$''$\hspace{-1mm}.5 and 85$''$\hspace{-1mm}.9, for BYF\,127 and BYF\,130, respectively.
\label{reg12}}
\end{figure*}

\begin{figure*}[htp]
\centerline{\includegraphics[angle=-90,scale=0.37]{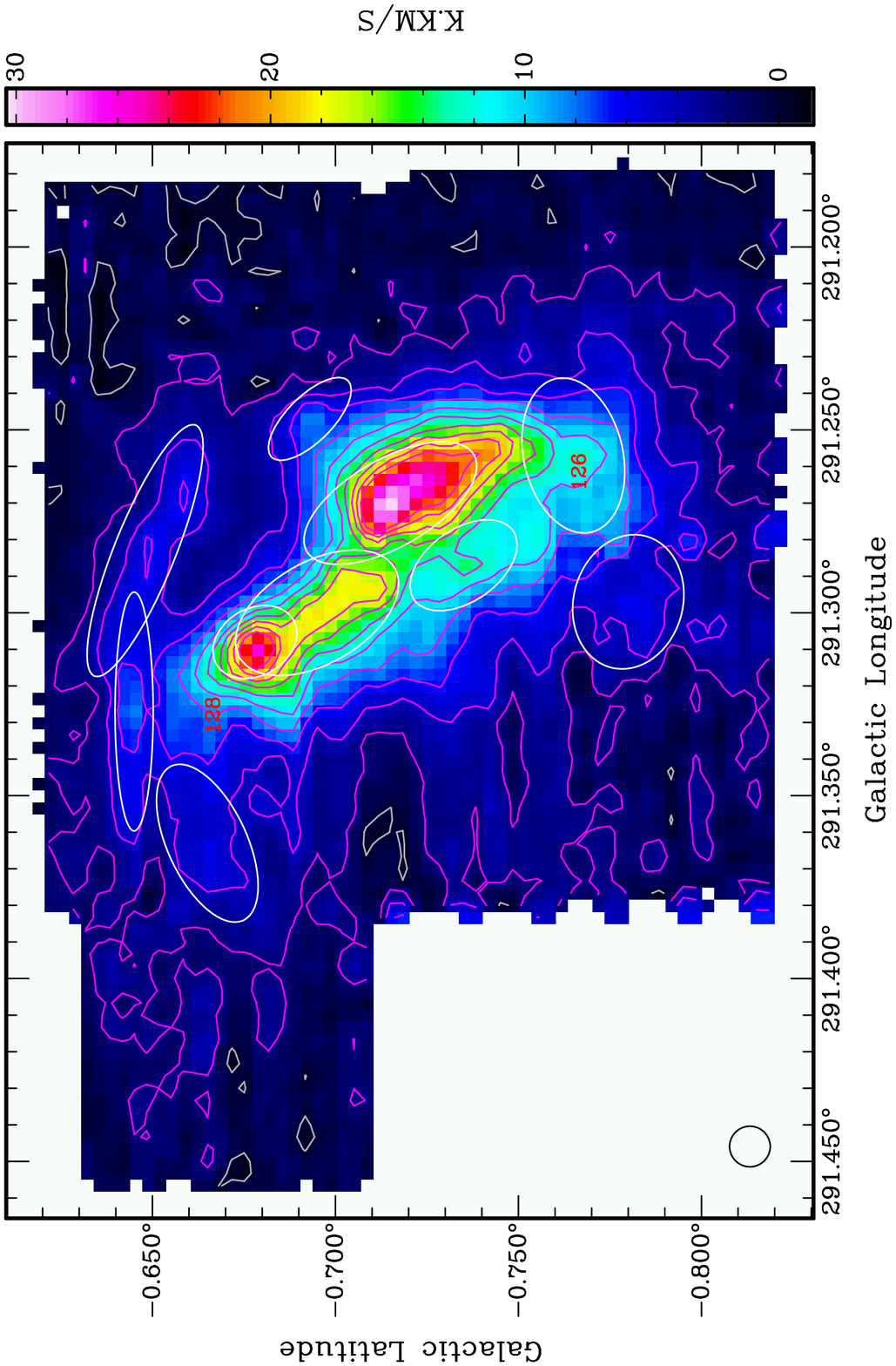}} 
\caption{\small Same as Fig.\,\ref{reg1}, but for Region 13 sources BYF\,126 and 128.  The integration here is over the range --31.0 to --15.0\,\kms\ or 142 channels, yielding an average rms noise level 0.424\,K\kms, and contour levels spaced every 5$\sigma$ are overlaid.  At a distance of 2.4\,kpc, the smoothed Mopra beam (lower left corner) scales to 40$''$ = 0.465\,pc or 1\,pc = 85$''$\hspace{-1mm}.9.
\label{reg13a}}
\end{figure*}

\begin{figure*}[htp]
\centerline{\includegraphics[angle=-90,scale=0.25]{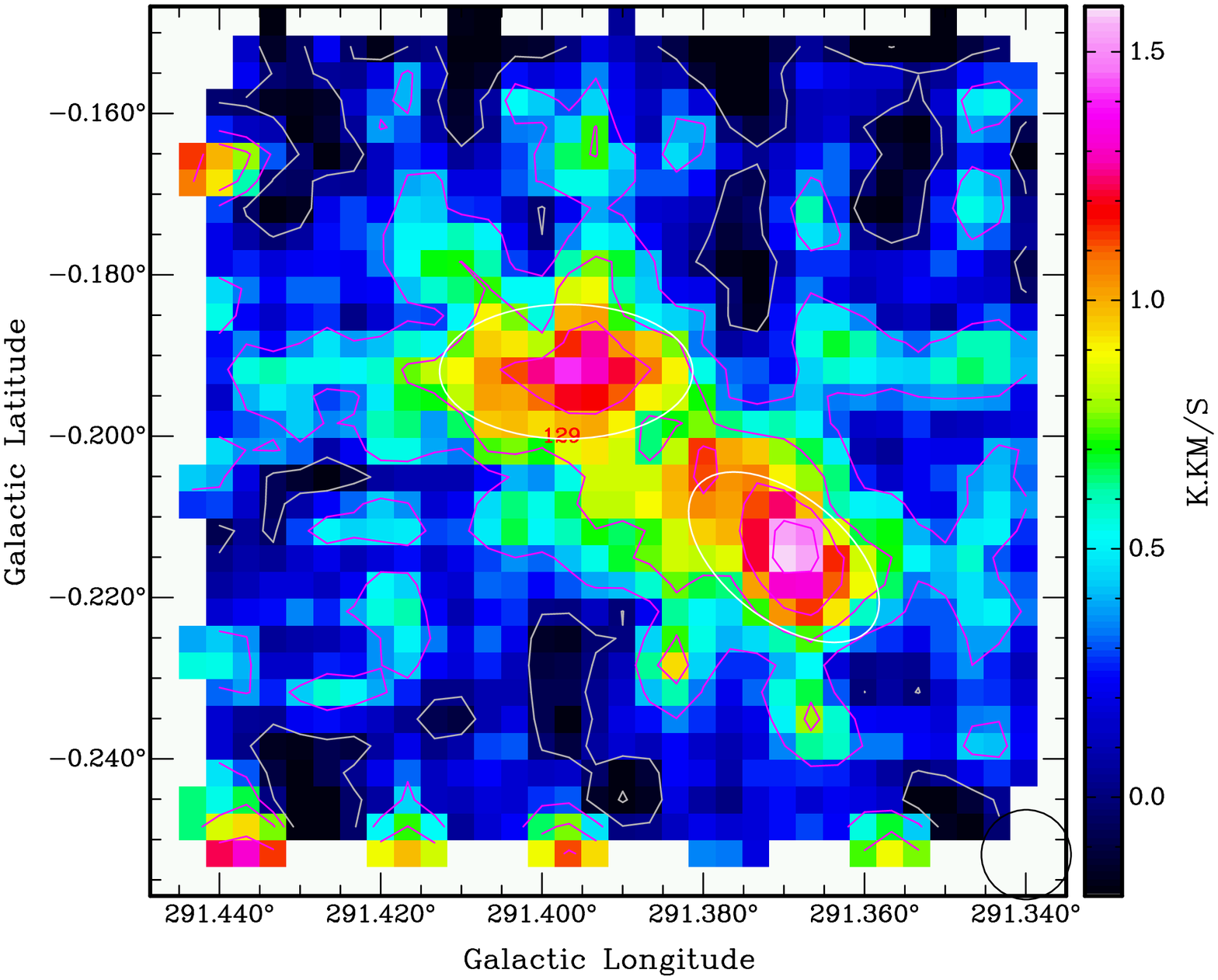}}
\caption{\small Same as Fig.\,\ref{reg1}, but for Region 13 source BYF\,129.  The integration here is over the range --7.0 to --3.0\,\kms\ or 36 channels, yielding an average rms noise level 0.179\,K\kms, and contour levels spaced every 2$\sigma$ are overlaid.  At a distance of 1.2\,kpc, the smoothed Mopra beam (lower right corner) scales to 40$''$ = 0.233\,pc or 1\,pc = 171$''$\hspace{-1mm}.9.
\label{reg13b}}
\end{figure*}

\begin{figure*}[ht]
\centerline{\includegraphics[angle=-90,scale=0.65]{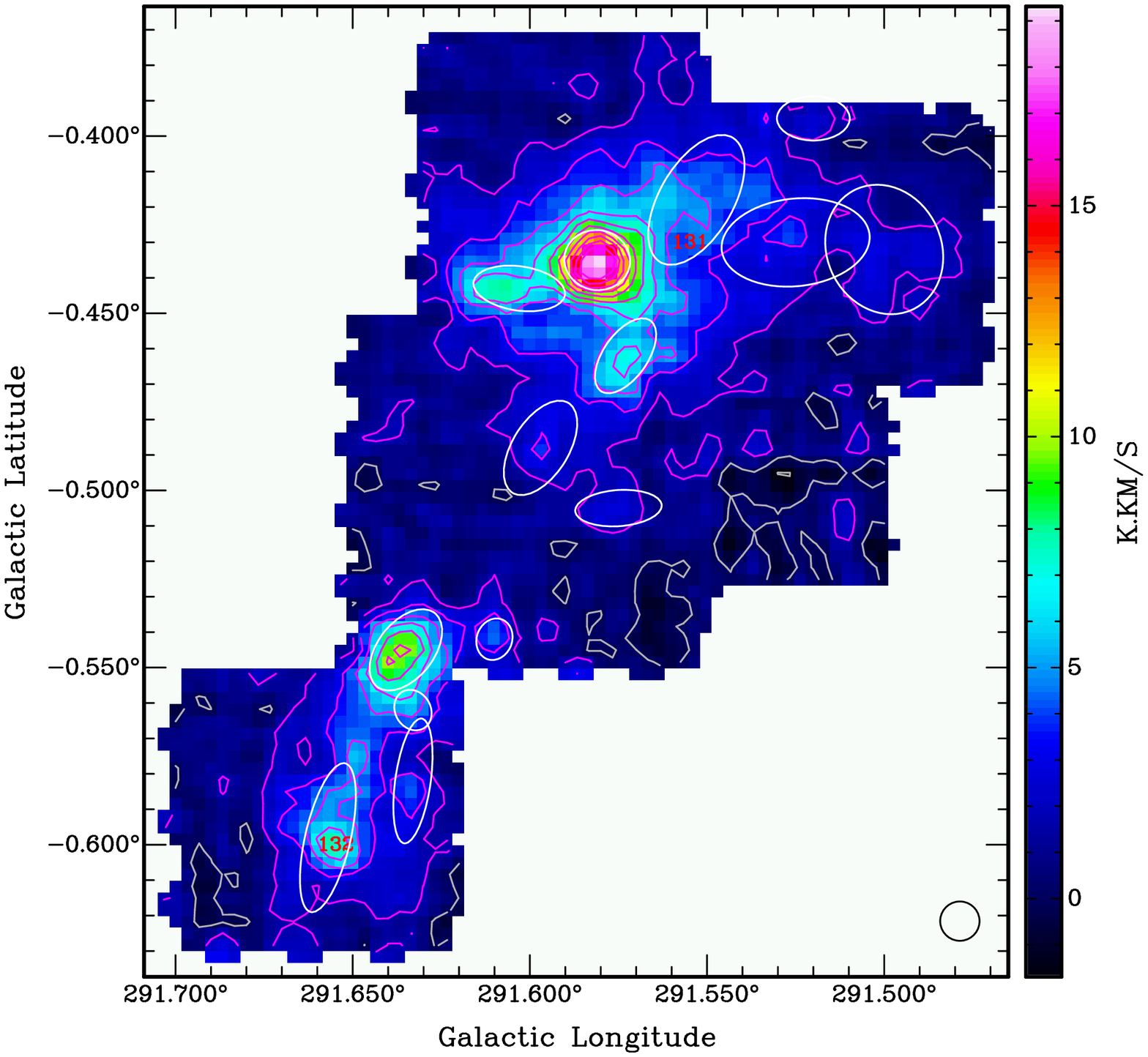}}
\caption{\small Same as Fig.\,\ref{reg1}, but for Region 13 sources BYF\,131 and 132.  The integration here is over the range +6.0 to +25.0\,\kms\ or 169 channels, however the various clumps are at different $V_{\rm LSR}$ and are imaged at higher S/N when integrated over more restricted velocity ranges, as in Table \ref{sources}.  Here the map has an average rms noise level 0.400\,K\kms, and contour levels spaced every 4$\sigma$ are overlaid.  At a distance of 6.0\,kpc, the smoothed Mopra beam (lower right corner) scales to 40$''$ = 1.16\,pc or 1\,pc = 34$''$\hspace{-1mm}.4.
\label{reg13c}}
\end{figure*}

\begin{figure*}[ht]
\centerline{\includegraphics[angle=-90,scale=0.25]{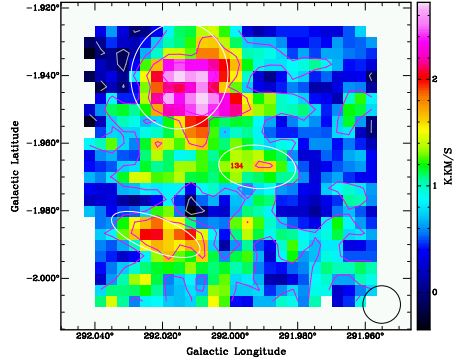}}
\caption{\small Same as Fig.\,\ref{reg1}, but for Region 15 source BYF\,134.  The integration here is over the range --28.7 to --23.0\,\kms\ or 55 channels, yielding an average rms noise level 0.383\,K\kms, and contour levels spaced every 2$\sigma$ are overlaid.  At a distance of 2.4\,kpc, the smoothed Mopra beam (lower right corner) scales to 40$''$ = 0.465\,pc or 1\,pc = 85$''$\hspace{-1mm}.9.
\label{reg15}}
\end{figure*}

\begin{figure*}[ht]
\centerline{(a)\includegraphics[angle=-90,scale=0.35]{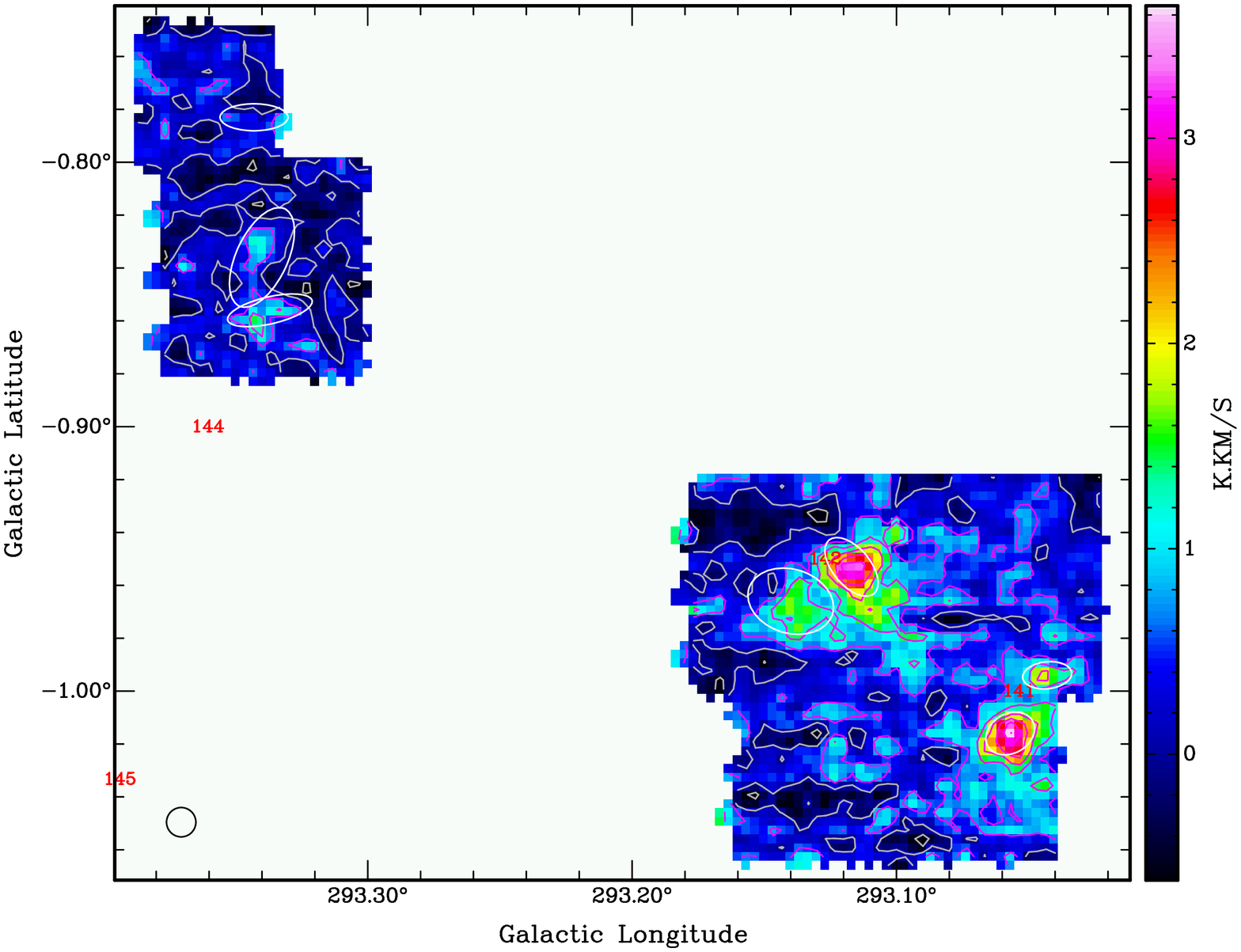}
		(b)\includegraphics[angle=-90,scale=0.35]{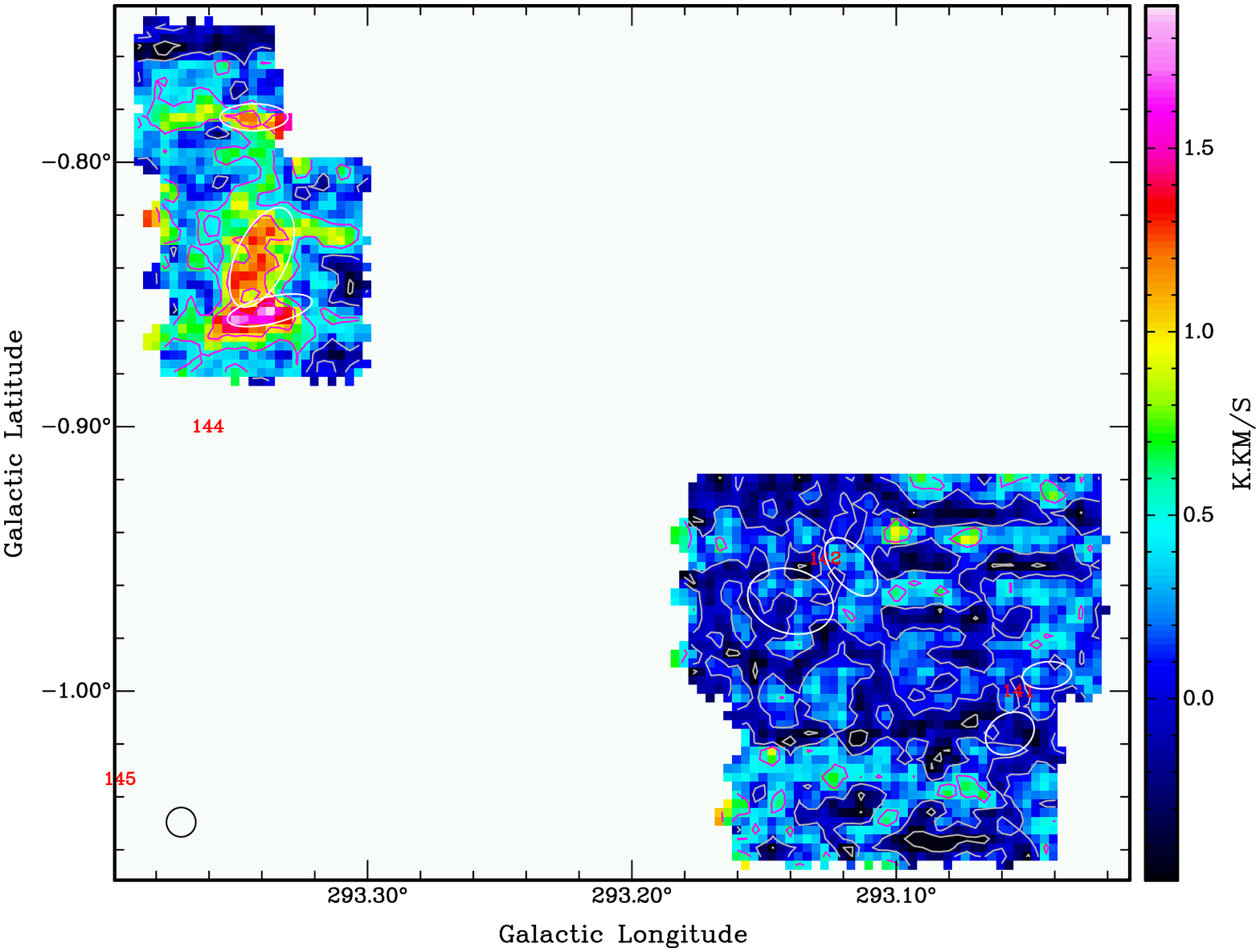}}
\caption{\small Same as Fig.\,\ref{reg1}, but for Region 16 sources ($a$) BYF\,141,142 and ($b$) BYF\,144.  The respective integrations are over the ranges --26.0 to --21.0\,\kms\ (45 channels) and --28.7 to --25.4\,\kms\ (31 channels), yielding average rms noise levels of 0.303\,K\kms and 0.250\,K\kms; contour levels spaced every 2$\sigma$ are overlaid.  At a distance of 2.4\,kpc, the smoothed Mopra beam (lower left corner) scales to 40$''$ = 0.465\,pc, or 1\,pc = 85$''$\hspace{-1mm}.9, for all of these sources.
\label{reg16}}
\end{figure*}

\begin{figure*}[ht]
\centerline{\includegraphics[angle=-90,scale=0.30]{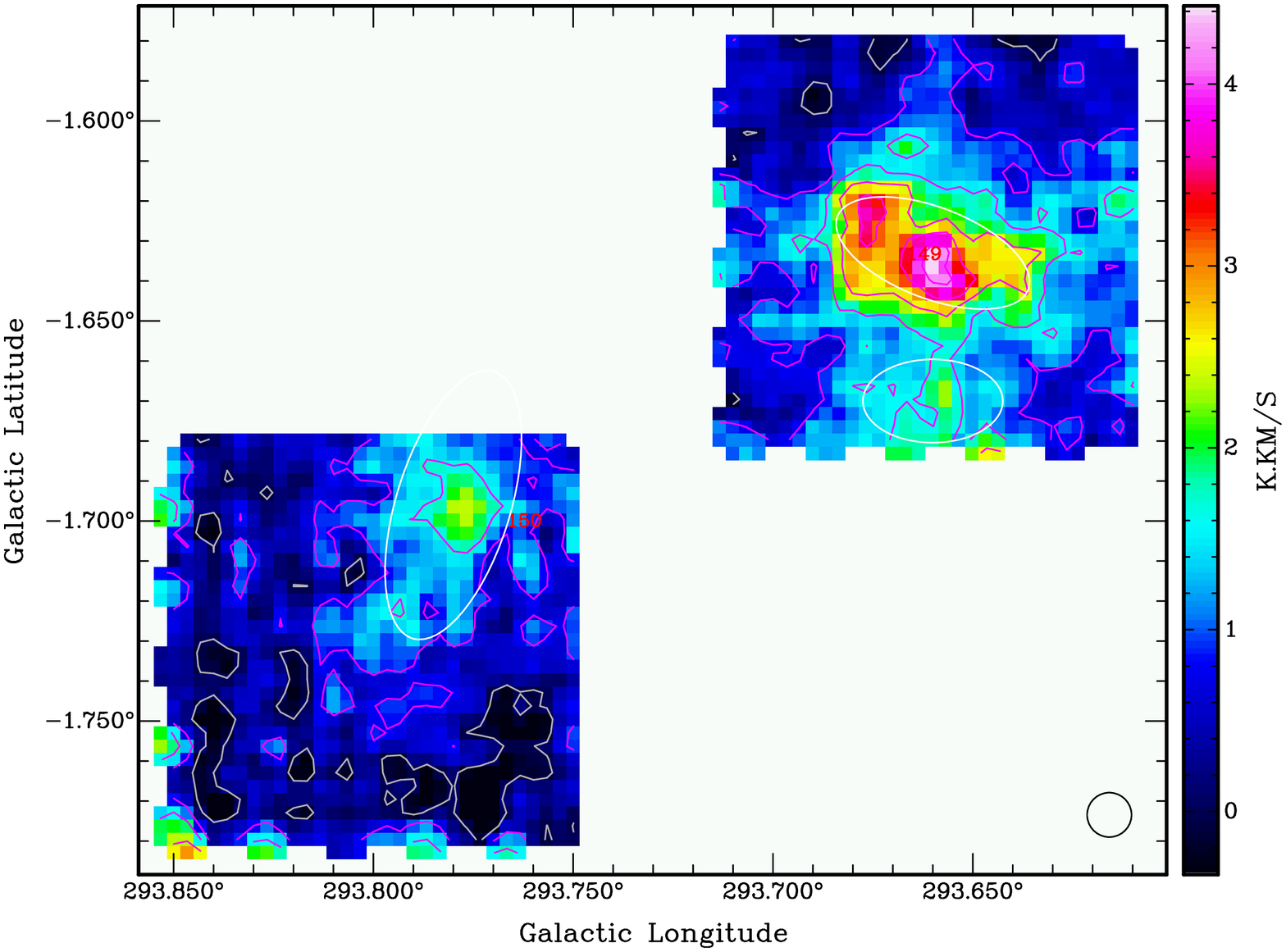}}
\caption{\small Same as Fig.\,\ref{reg1}, but for Region 18 sources BYF\,149 and 150.  The integration here is over the range --22.0 to --14.5\,\kms\ or 67 channels, yielding an average rms noise level 0.270\,K\kms, and contour levels spaced every 3$\sigma$ are overlaid.  At a distance of 2.4\,kpc, the smoothed Mopra beam (lower right corner) scales to 40$''$ = 0.465\,pc or 1\,pc = 85$''$\hspace{-1mm}.9.
\label{reg18}}
\end{figure*}

\begin{figure*}[ht]
\centerline{\includegraphics[angle=-90,scale=0.60]{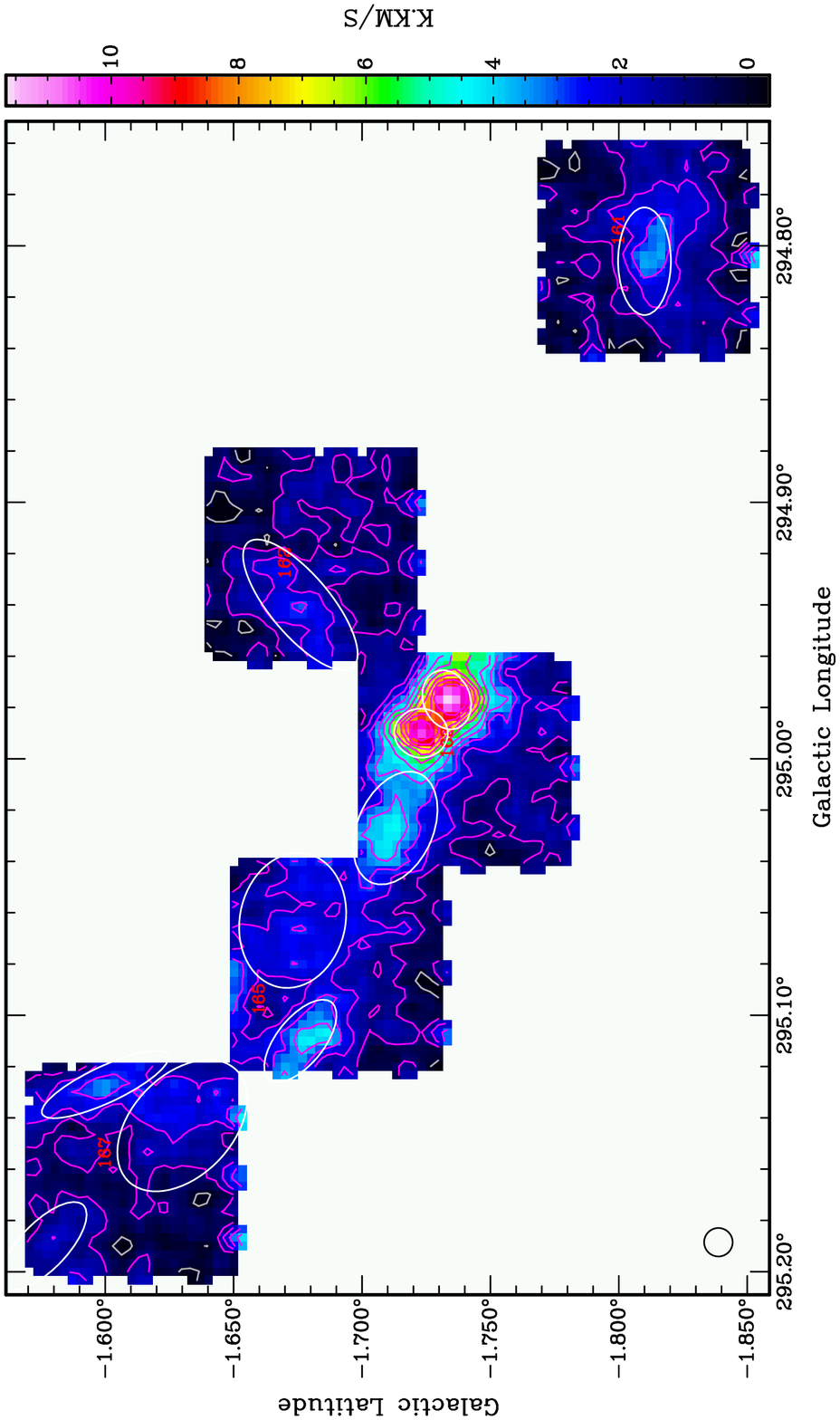}}
\caption{\small Same as Fig.\,\ref{reg1}, but for Region 21 sources BYF\,161--167.  The integration here is over the range --16.1 to --4.0\,\kms\ or 107 channels, however the various clumps are at different $V_{\rm LSR}$ and are imaged at higher S/N when integrated over more restricted velocity ranges, as in Table \ref{sources}.  Here the map has an average rms noise level 0.308\,K\kms, and contour levels spaced every 3$\sigma$ are overlaid.  At a distance of 2.4\,kpc, the smoothed Mopra beam (lower left corner) scales to 40$''$ = 0.465\,pc or 1\,pc = 85$''$\hspace{-1mm}.9.
\label{reg21}}
\end{figure*}

\begin{figure*}[ht]
\centerline{\includegraphics[angle=-90,scale=0.50]{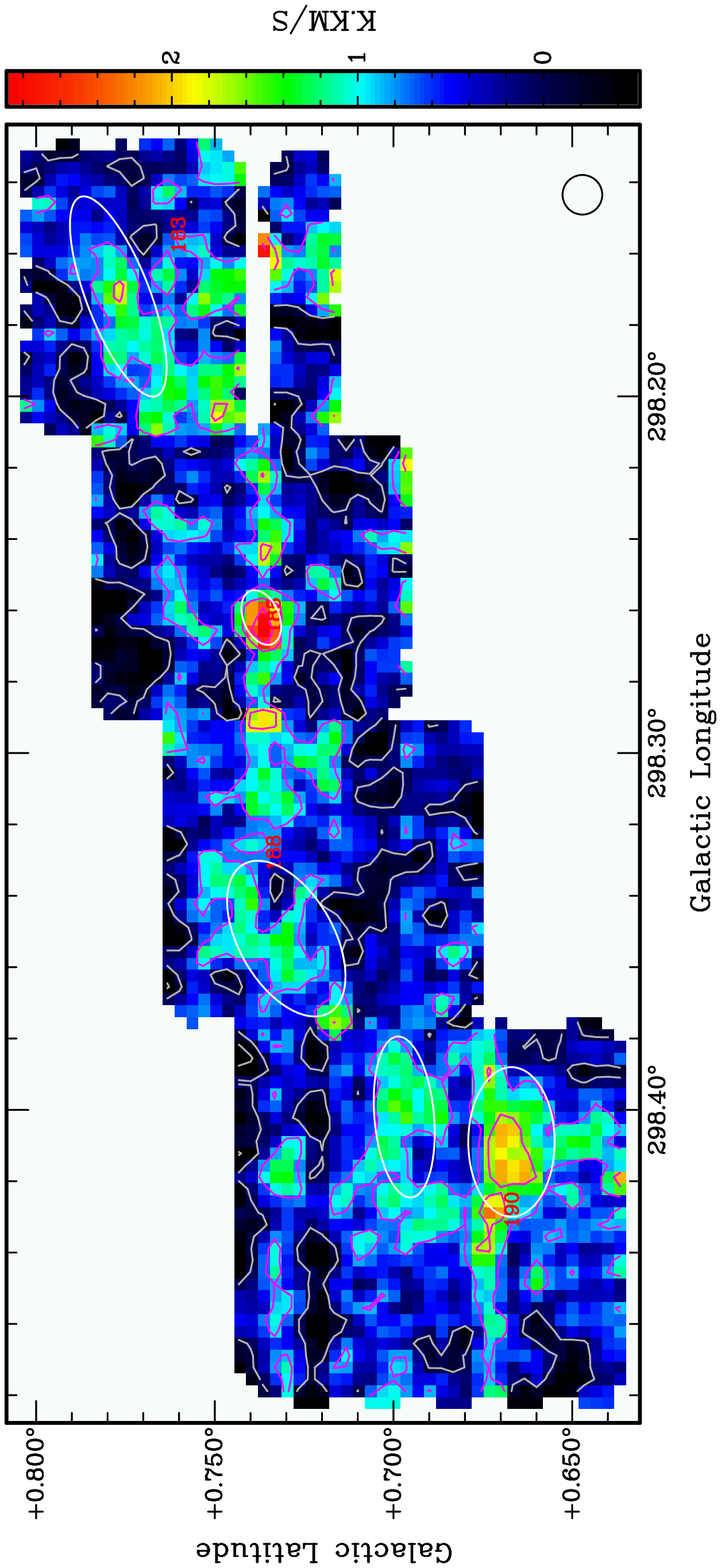}}
\caption{\small Same as Fig.\,\ref{reg1}, but for Region 23 sources BYF\,183--190.  The integration here is over the range --34.2 to --28.8\,\kms\ or 49 channels, however the various clumps are at different $V_{\rm LSR}$ and are imaged at higher S/N when integrated over more restricted velocity ranges, as in Table \ref{sources}.  Here the map has an average rms noise level 0.414\,K\kms, and contour levels spaced every 2$\sigma$ are overlaid.  At a distance of 4.7\,kpc, the smoothed Mopra beam (lower right corner) scales to 40$''$ = 0.911\,pc or 1\,pc = 43$''$\hspace{-1mm}.9.
\label{reg23}}
\end{figure*}

\begin{figure*}[ht]
\centerline{\includegraphics[angle=0,scale=0.30]{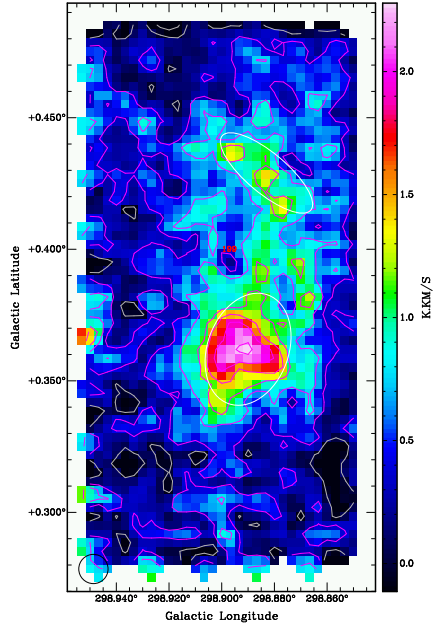}}
\caption{\small Same as Fig.\,\ref{reg1}, but for Region 26 source BYF\,199.  The integration here is over the range --27.05 to --23.4\,\kms\ or 33 channels, yielding an average rms noise level 0.184\,K\kms, and contour levels spaced every 2$\sigma$ are overlaid.  At a distance of 4.7\,kpc, the smoothed Mopra beam (lower left corner) scales to 40$''$ = 0.911\,pc or 1\,pc = 43$''$\hspace{-1mm}.9.
\label{reg26a}}
\end{figure*}

\begin{figure*}[ht]
\centerline{\includegraphics[angle=-90,scale=0.60]{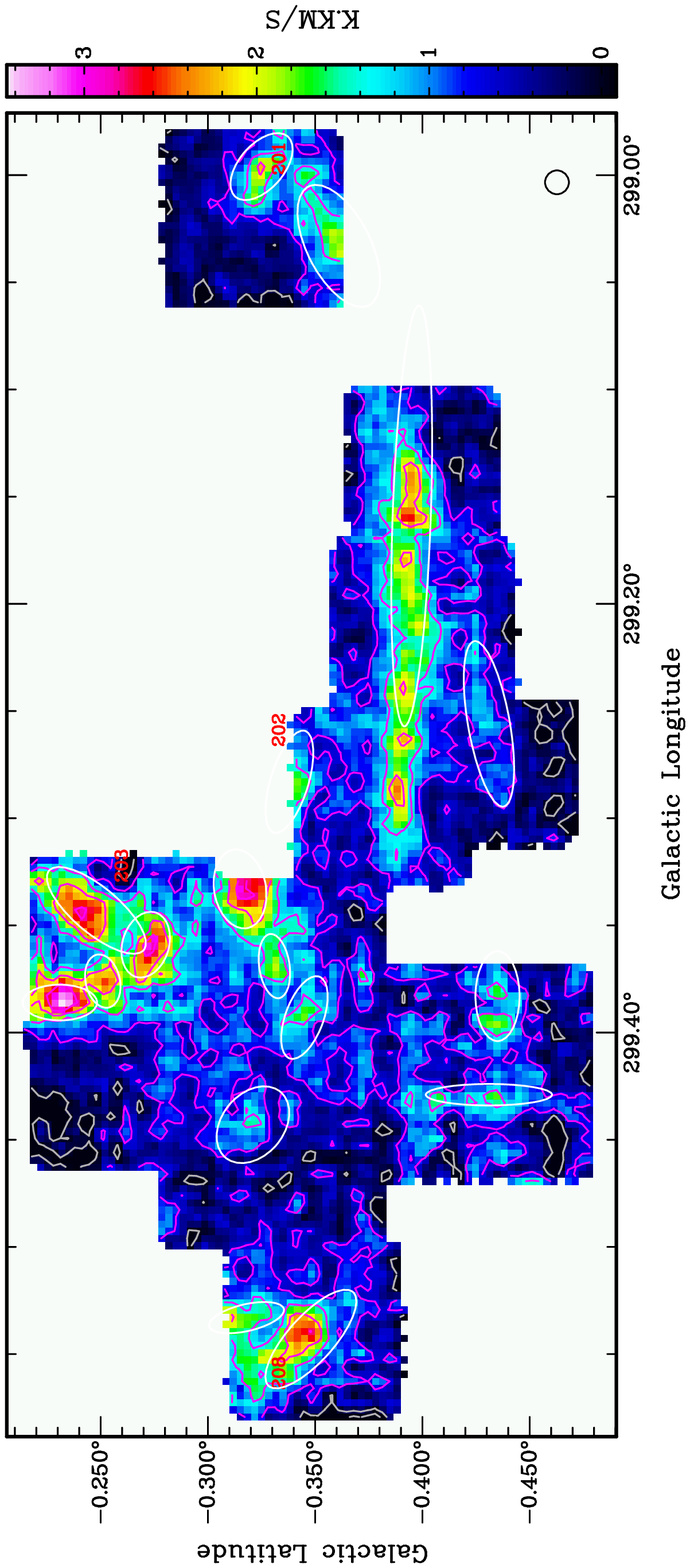}}
\caption{\small Same as Fig.\,\ref{reg1}, but for Region 26 sources BYF\,201--208.  The integration here is over the range --41.65 to --36.0\,\kms\ or 51 channels, yielding an average rms noise level 0.230\,K\kms, and contour levels spaced every 3$\sigma$ are overlaid.  At a distance of 4.7\,kpc, the smoothed Mopra beam (lower right corner) scales to 40$''$ = 0.911\,pc or 1\,pc = 43$''$\hspace{-1mm}.9.
\label{reg26b}}
\end{figure*}

The Mopra catalogue of \hcop\ moment-0 maps of sources from the NMC (Table \ref{NMC}) is presented in Figures \ref{reg1}--\ref{reg26b} on the $T_R^*$ scale as given by each colourbar; they are also available for download as individual PDF and FITS files from {\bf www.astro.ufl.edu/champ/}.  Details such as velocity integration ranges, contouring, noise levels, and distance scales are included in the caption to each figure.  Generally, however, {\em contiguous} low-level emission above $\sim$2$\sigma$ tends to be real, although of course the detailed emission structure is reliable only above $\sim$4$\sigma$.  To aid in the identification of significant features, contour levels spaced every few $\sigma$ are overlaid (this varies from figure to figure; in some cases the noise level across each map also varies --- see \S\ref{highmom} for the rms maps).  For the moment-0 maps, negative contours (and zero) are grey, while positive contours are magenta.  Also shown in each figure are white ellipses for the fitted 2D gaussians of each Mopra clump, and the smoothed Mopra HPBW (40$''$) as a black circle in one corner; the latter can be used as a physical scale indicator in each case.

\vspace{10mm}

\clearpage


\clearpage


\clearpage


\clearpage


\clearpage


\clearpage

\newpage

\newpage

\newpage


\clearpage


\clearpage

\newpage


\newpage


\newpage


\clearpage 

\section{\hcop\ Higher-moment Maps \label{highmom}}
\notetoeditor{Figures a--d should appear 2x2 on each page}

Higher-moment Mopra \hcop \joz\ images for sources from \S\ref{mom0maps}.  The same contours of integrated intensity from the moment-0 maps are overlaid here: contours are green in the rms map and magenta in the others for positive contours, and grey in all maps for 0 and any negative contours.  All moments were calculated over the same velocity ranges as in \S\ref{mom0maps}.  The fitted gaussian ellipses for the Mopra clumps and smoothed Mopra beam are shown as before.  ($a$) Peak \hcop\ line temperature $T_{\rm peak}$.  ($b$) rms noise level over line-free channels.  ($c$)  Intensity-weighted mean velocity field $V_{\rm LSR}$ (first moment).  ($d$) Velocity dispersion $\sigma_{V}$ (second moment).  The first and second moments are only computed where the integrated intensity maps are more than a few (usually 3) times the rms noise.

\begin{figure*}[ht]
\centerline{(a){\includegraphics[angle=-90,scale=0.30]{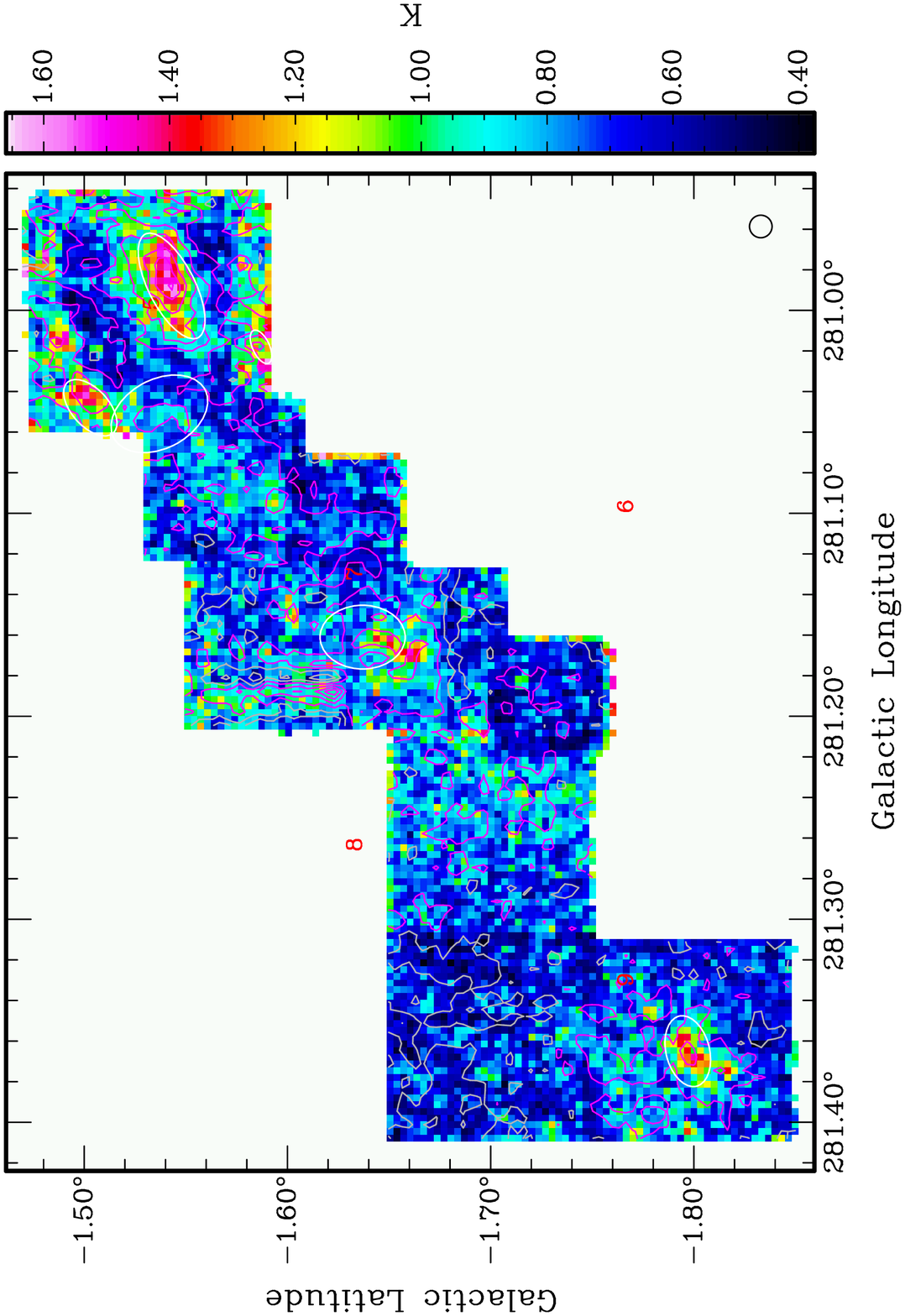}}
		(b){\includegraphics[angle=-90,scale=0.30]{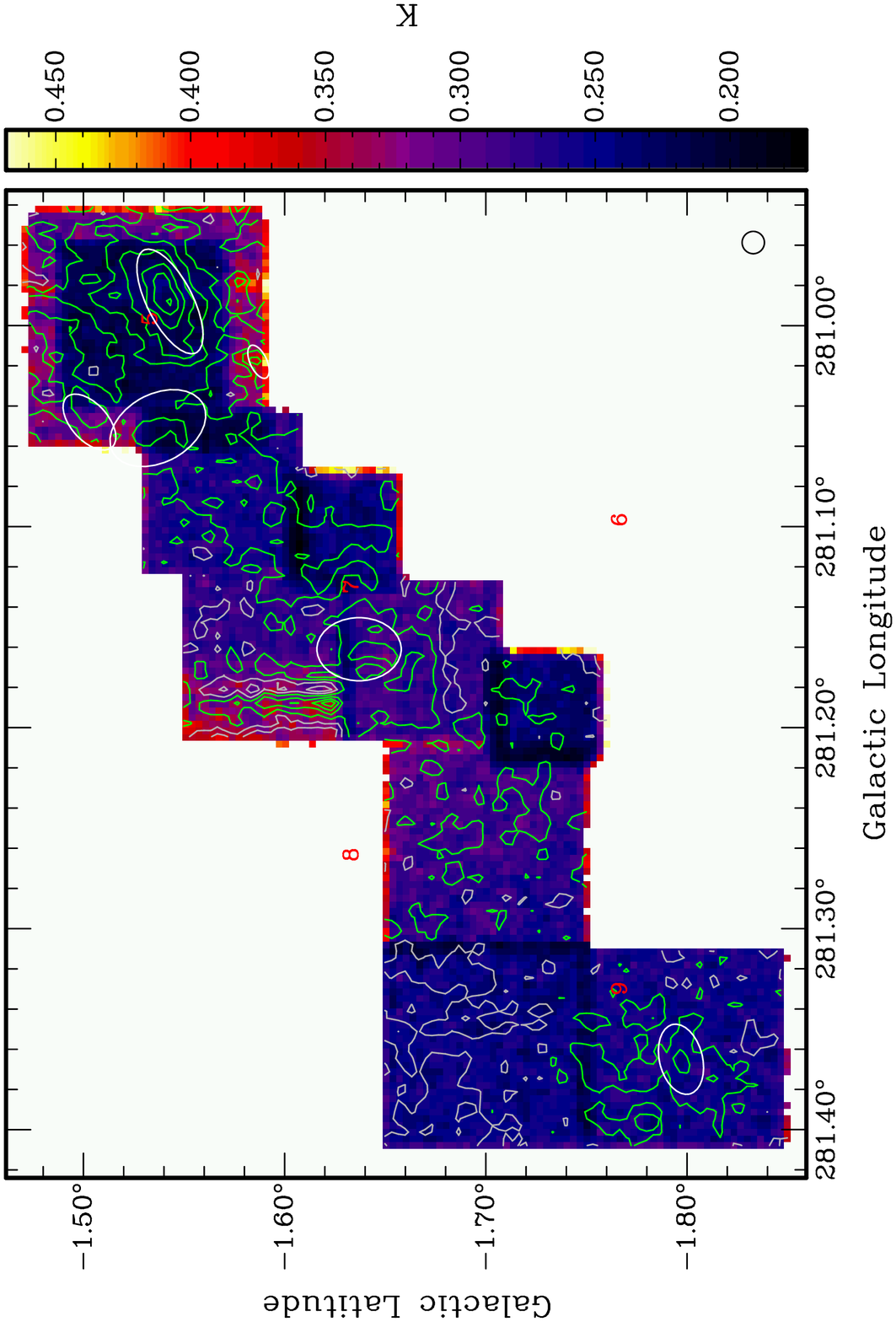}}}
\centerline{(c){\includegraphics[angle=-90,scale=0.30]{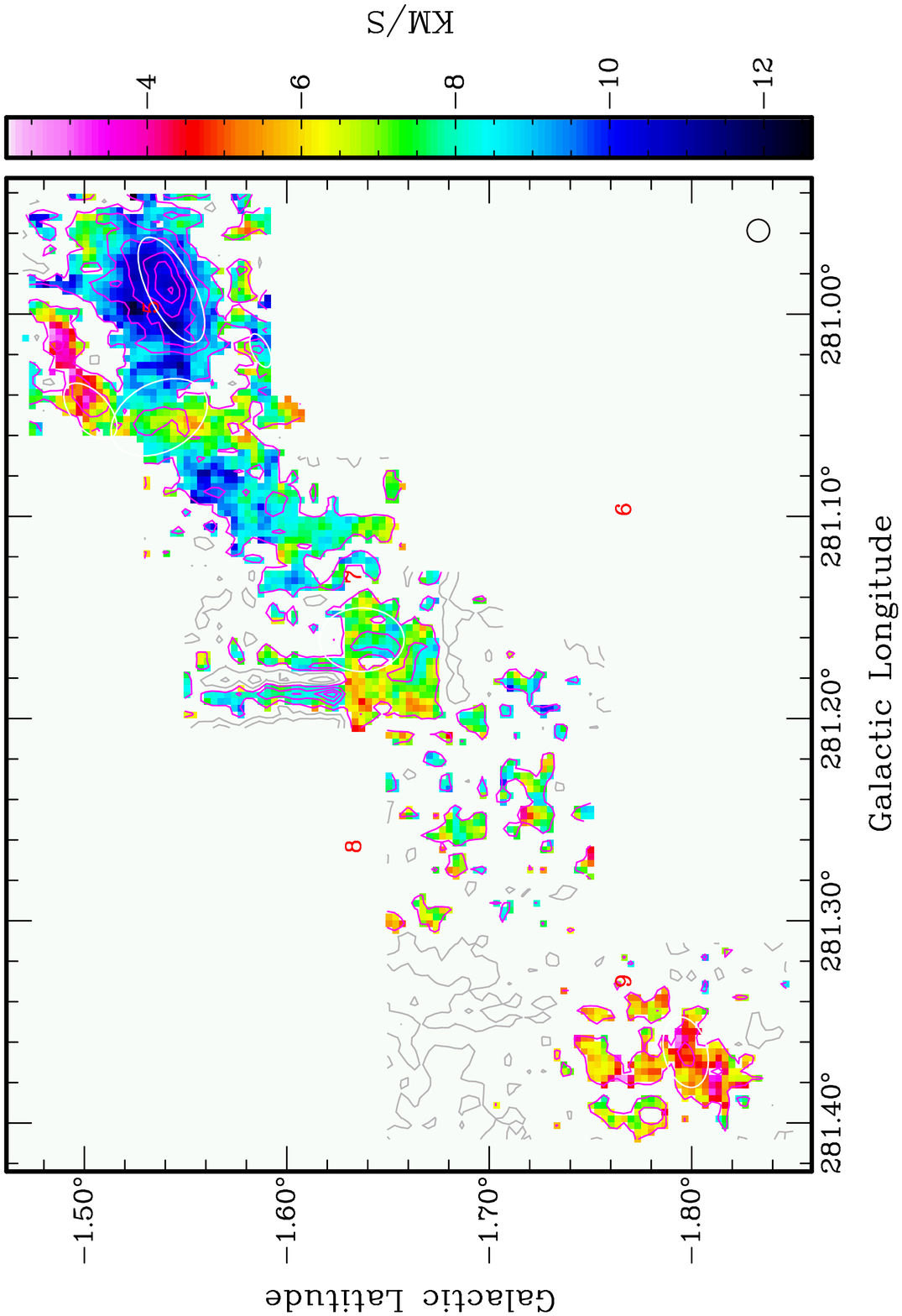}}
		(d){\includegraphics[angle=-90,scale=0.30]{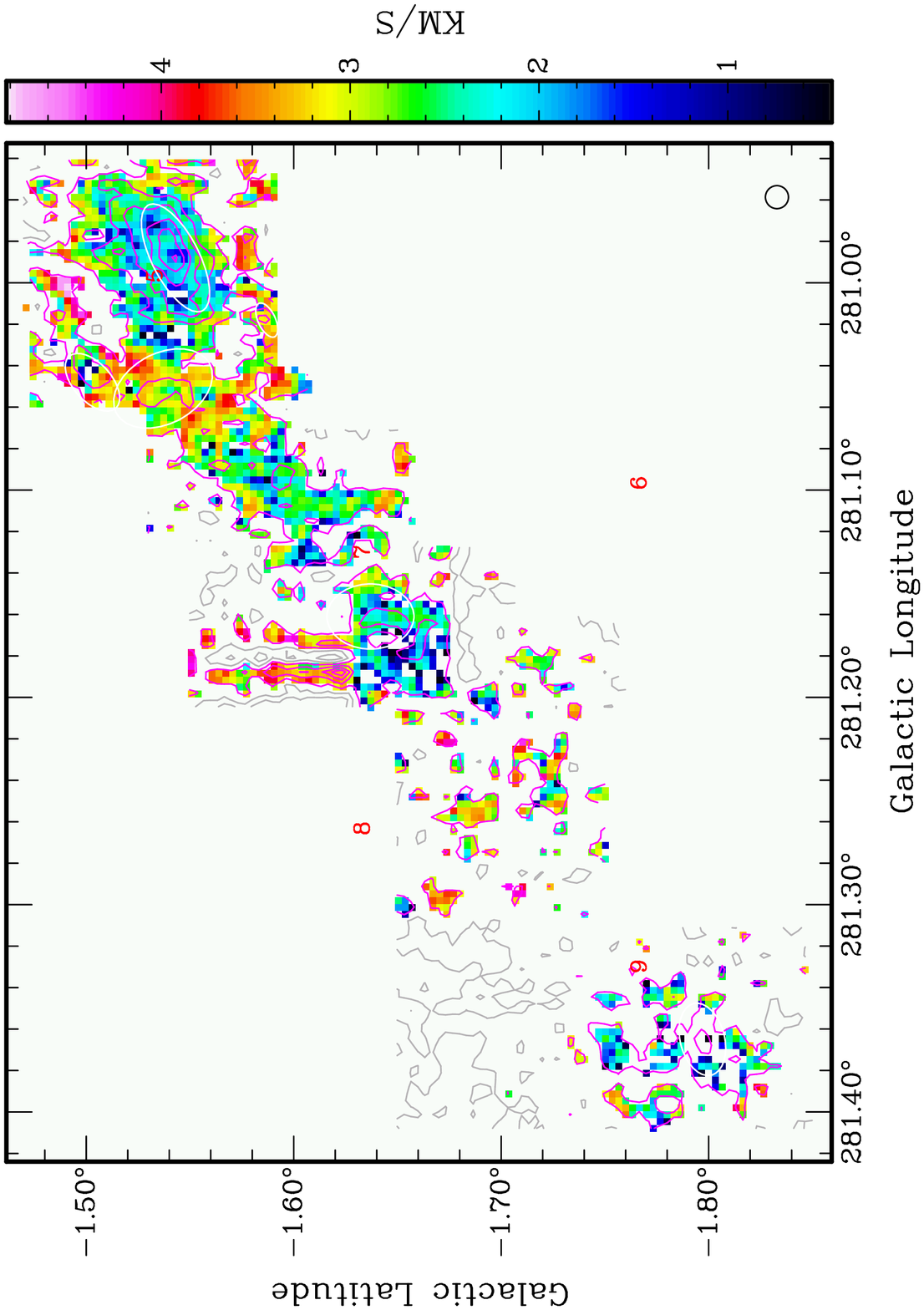}}}
\caption{\small Higher-moment Mopra \hcop \joz\ images for Region 1 sources BYF\,5--9, with contours of integrated intensity from Fig.\,\ref{reg1} at 3$\sigma$ (= 0.954\,K\kms) intervals.  At a distance of 3.2\,kpc, the 40$''$ Mopra beam (lower right corner) scales to 0.621\,pc.  ($a$) $T_p$,  ($b$) rms,  ($c$)  $V_{\rm LSR}$,  ($d$) $\sigma_{V}$.
\label{momR1}}
\end{figure*}

\clearpage

\begin{figure*}[htp]
(a){\includegraphics[angle=-90,scale=0.4]{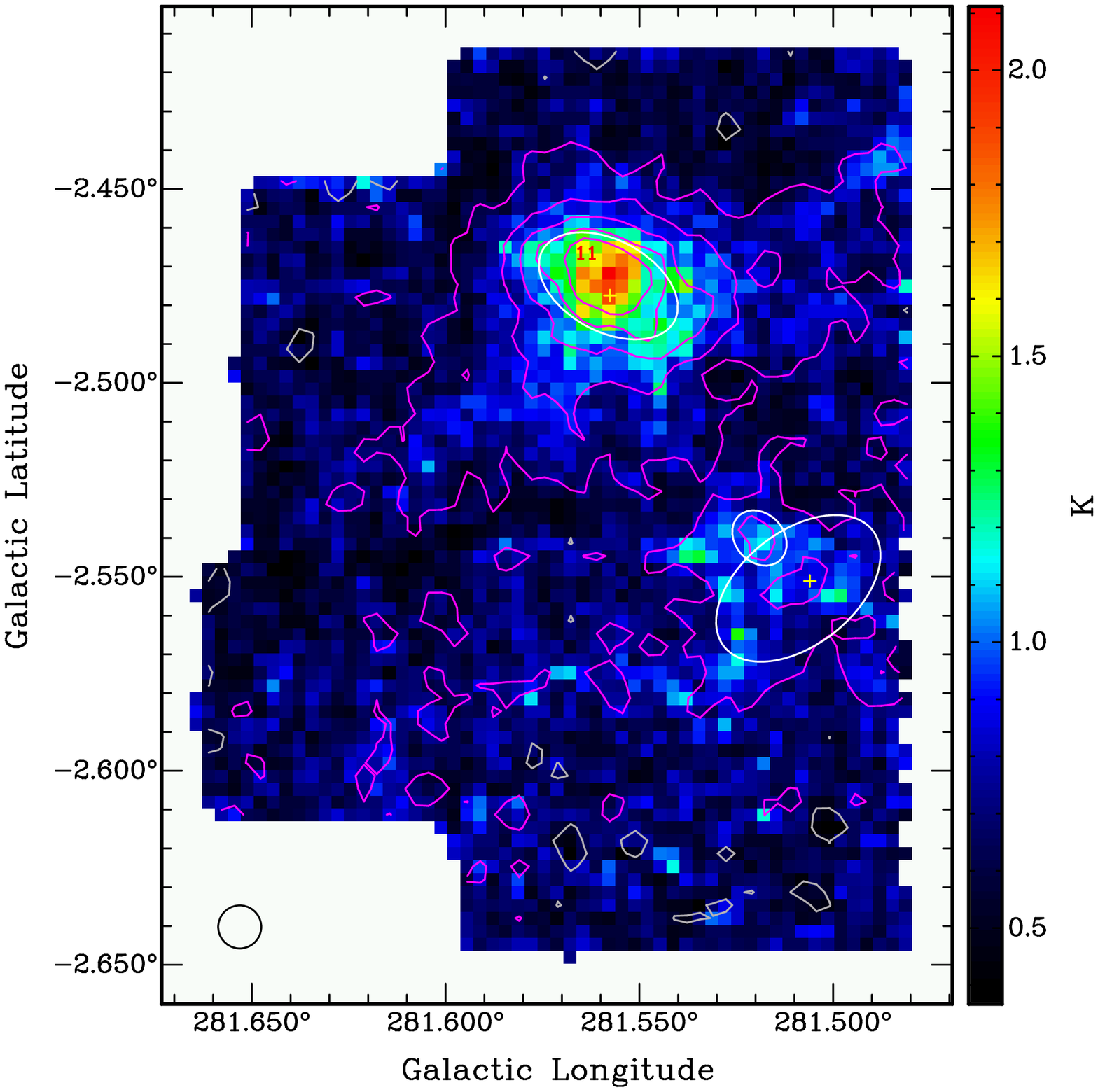}}
(b){\includegraphics[angle=-90,scale=0.4]{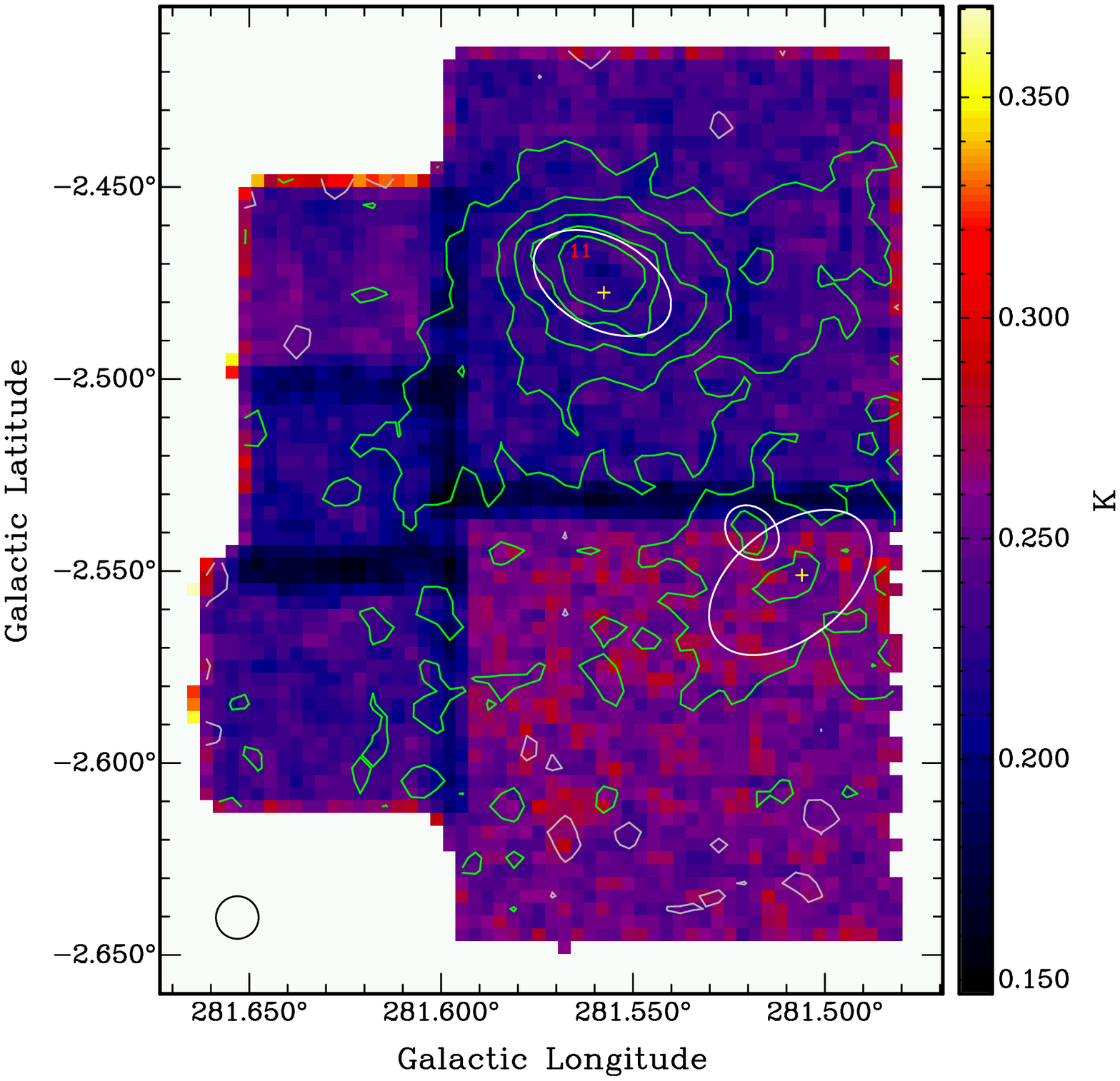}}
(c){\includegraphics[angle=0,scale=0.4]{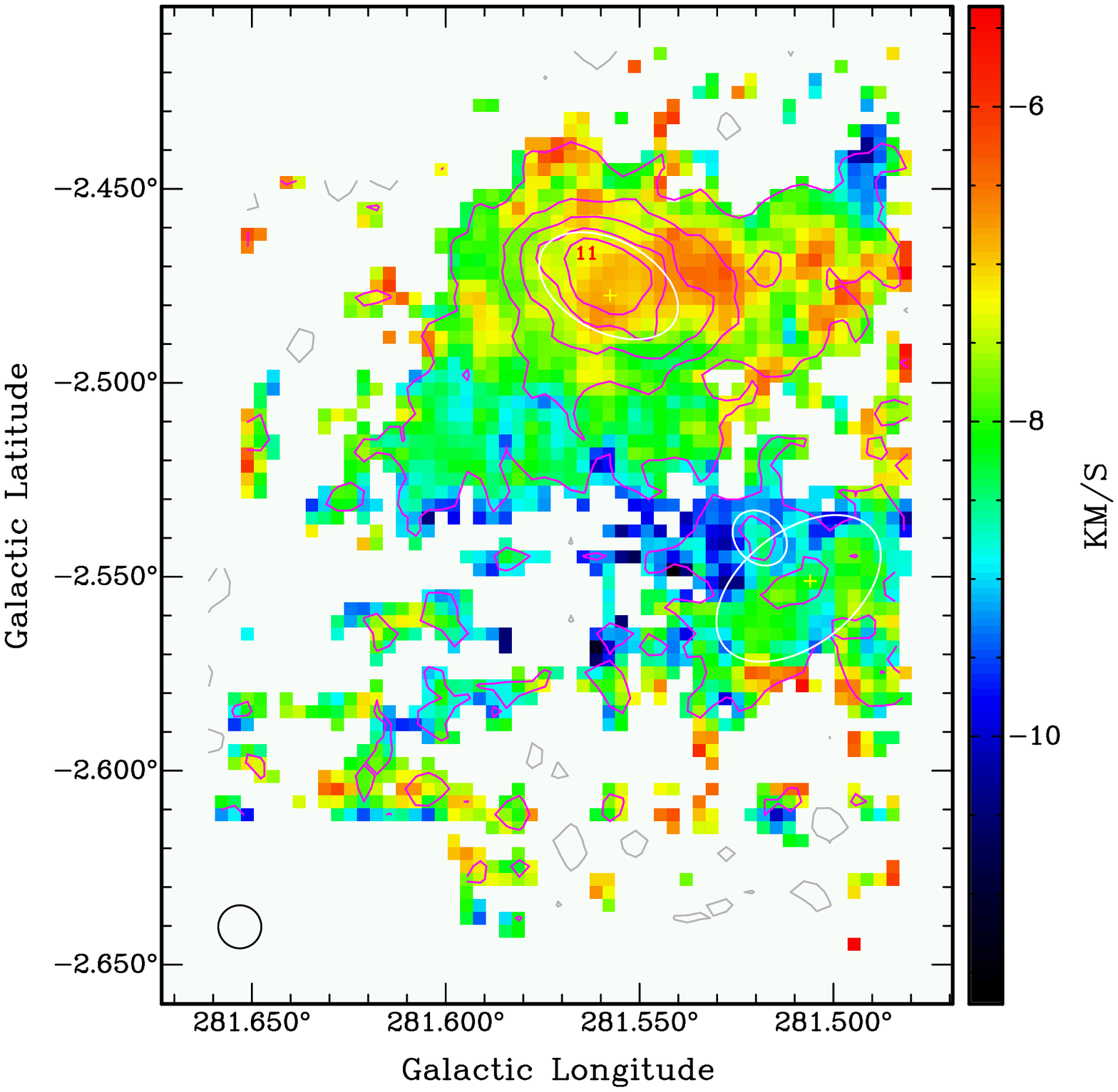}}
(d){\includegraphics[angle=0,scale=0.4]{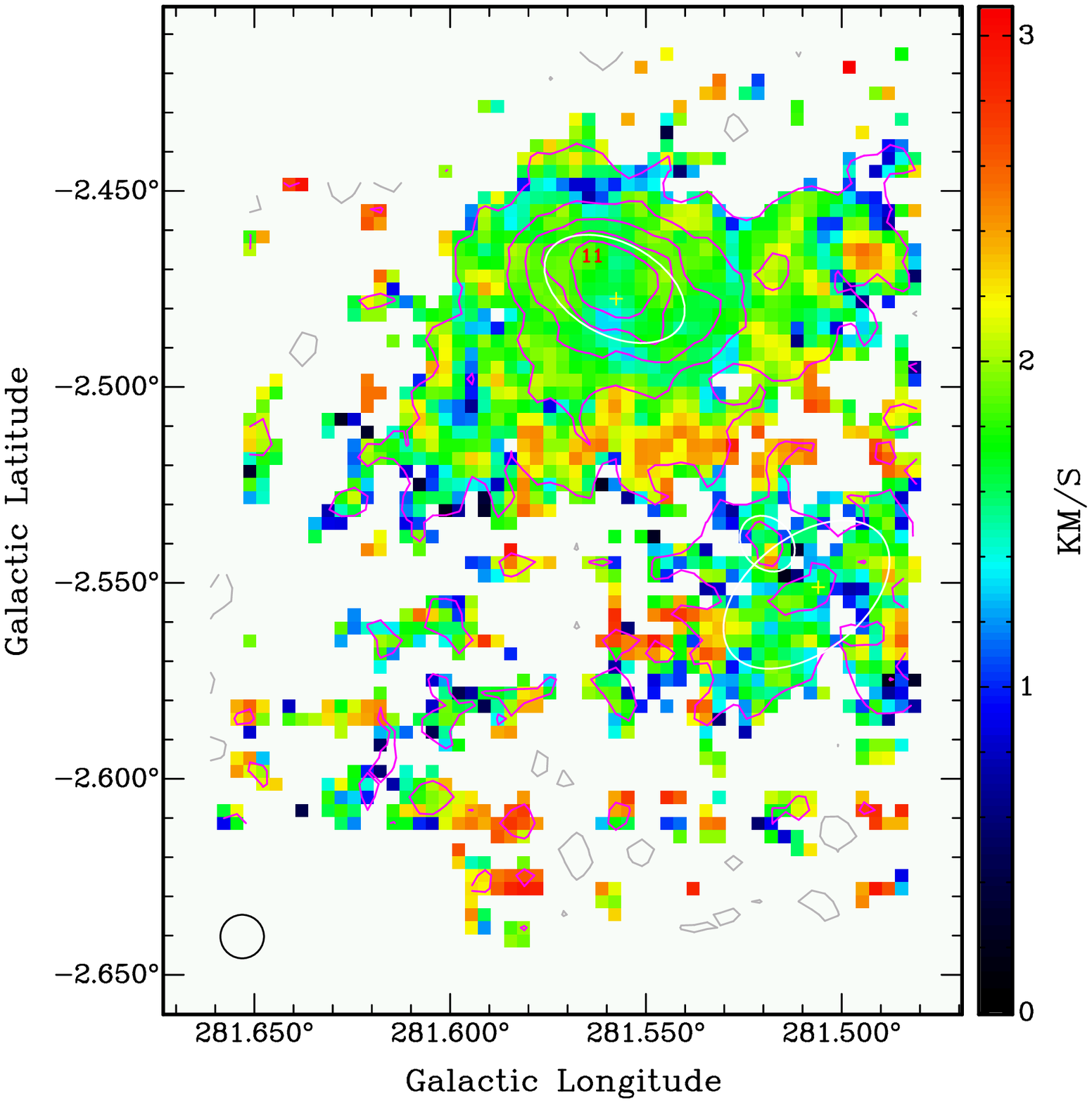}}
\caption{\small Same as Fig.\,\ref{momR1}, but for isolated source BYF\,11.  Contours are every 4$\sigma$ = 0.952\,K\kms, and at 3.2\,kpc the 40$''$ Mopra beam (lower left corner) scales to 0.621\,pc.  ($a$) $T_p$,  ($b$) rms,  ($c$) $V_{\rm LSR}$,  ($d$) $\sigma_{V}$.
\label{momBYF11}}
\end{figure*}

\clearpage

\begin{figure*}[htp]
\centerline{(a){\includegraphics[angle=-90,scale=0.35]{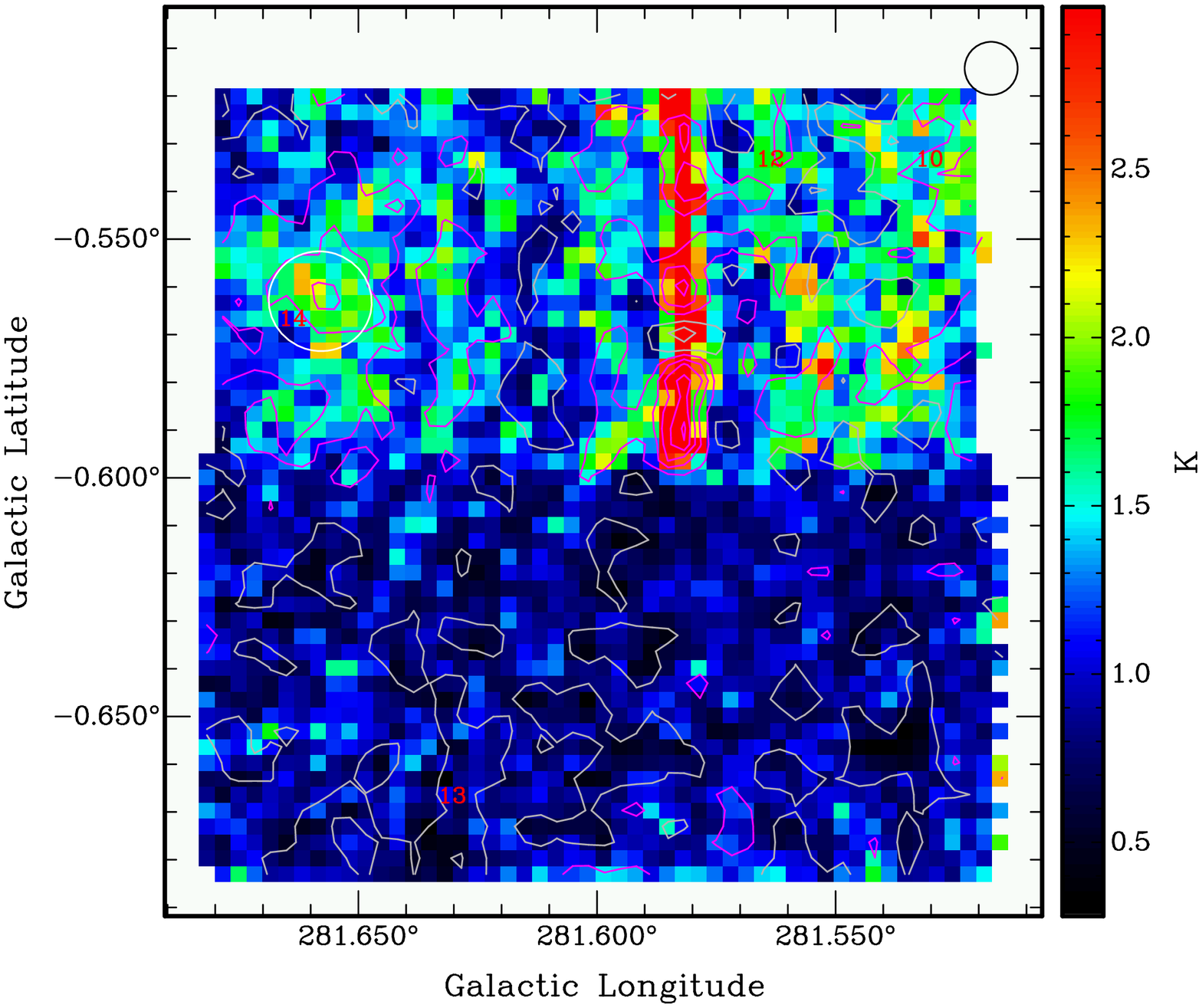}}
		(b){\includegraphics[angle=-90,scale=0.35]{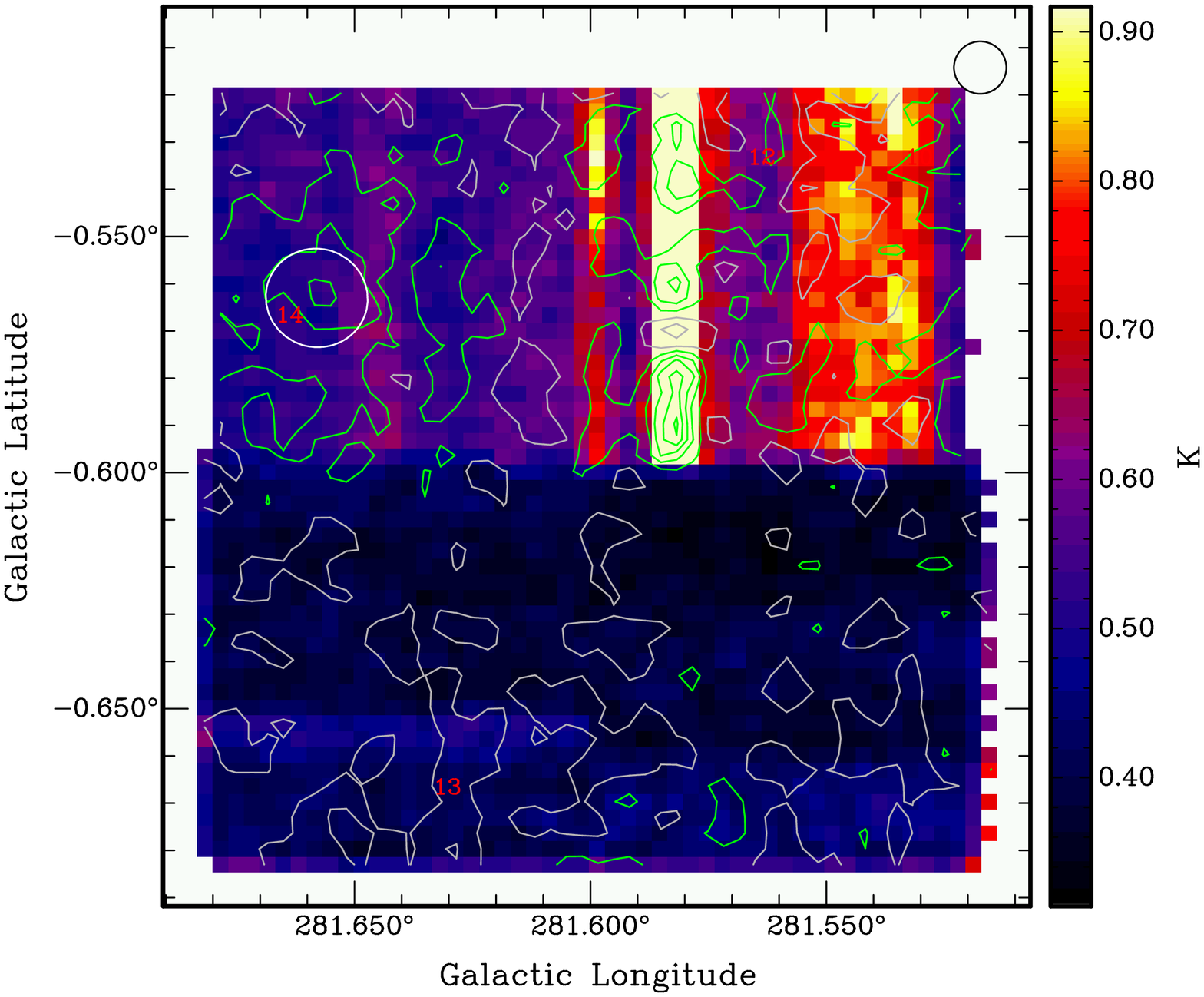}}}
\centerline{(c){\includegraphics[angle=-90,scale=0.35]{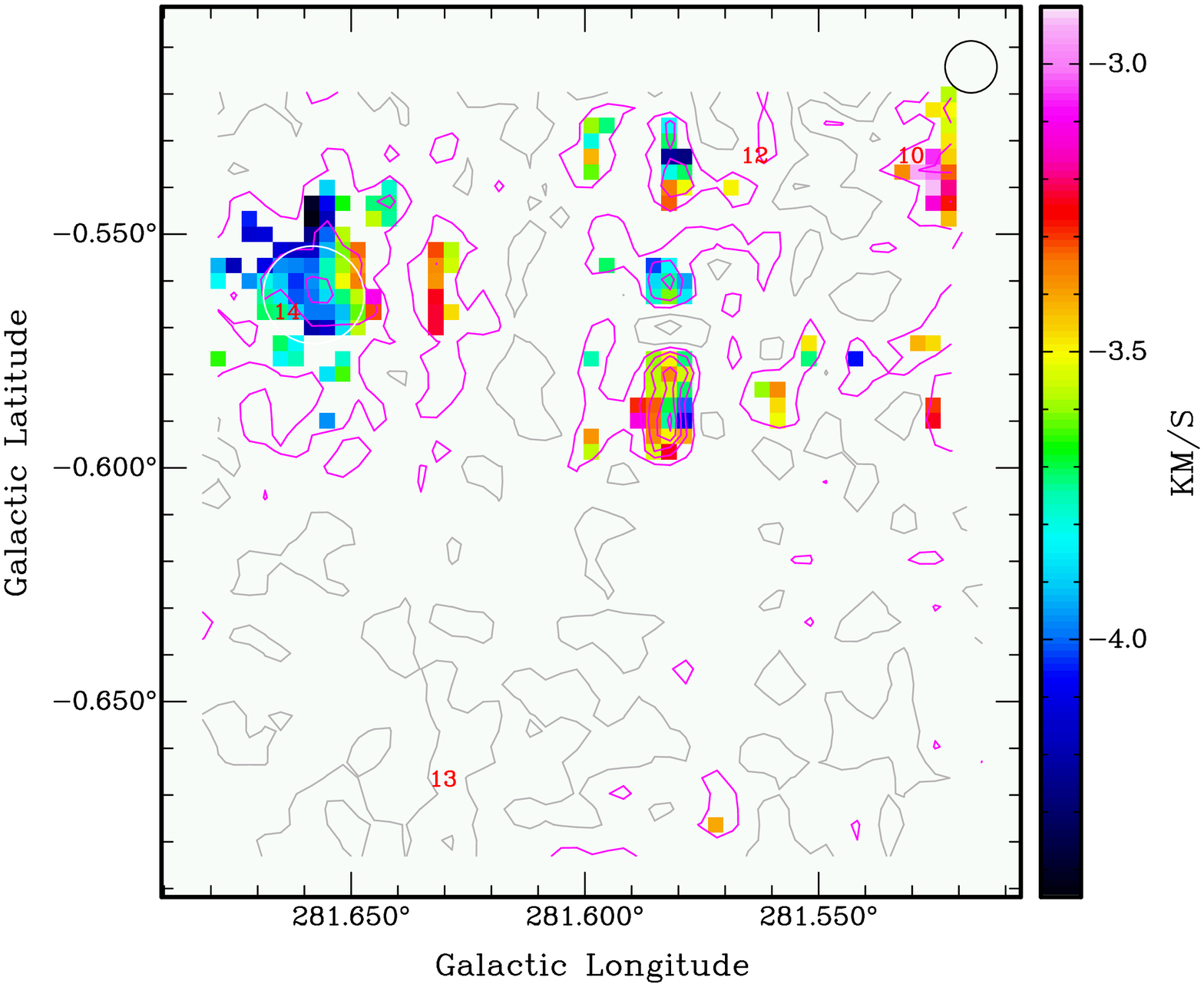}}
		(d){\includegraphics[angle=-90,scale=0.35]{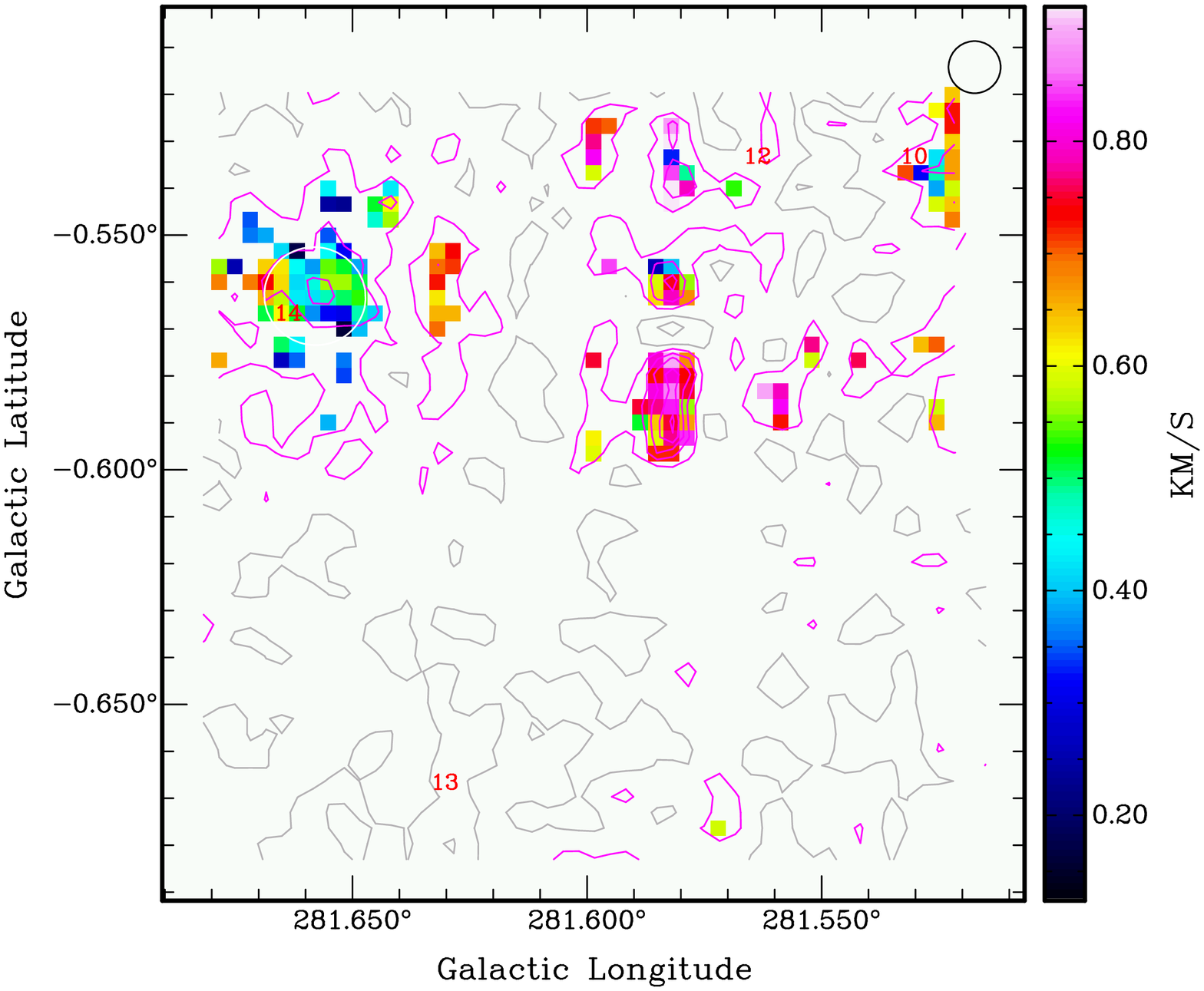}}}
\caption{\small Same as Fig.\,\ref{momR1}, but for Region 2a sources BYF\,10 and 12--14.  Contours are every 2$\sigma$ = 0.558\,K\kms\, and at 3.2\,kpc the 40$''$ Mopra beam (upper right corner) scales to 0.621\,pc.  ($a$) $T_p$,  ($b$) rms,  ($c$) $V_{\rm LSR}$,  ($d$) $\sigma_{V}$.
\label{momR2a}}
\end{figure*}

\clearpage

\begin{figure*}[htp]
\centerline{(a){\includegraphics[angle=0,scale=0.35]{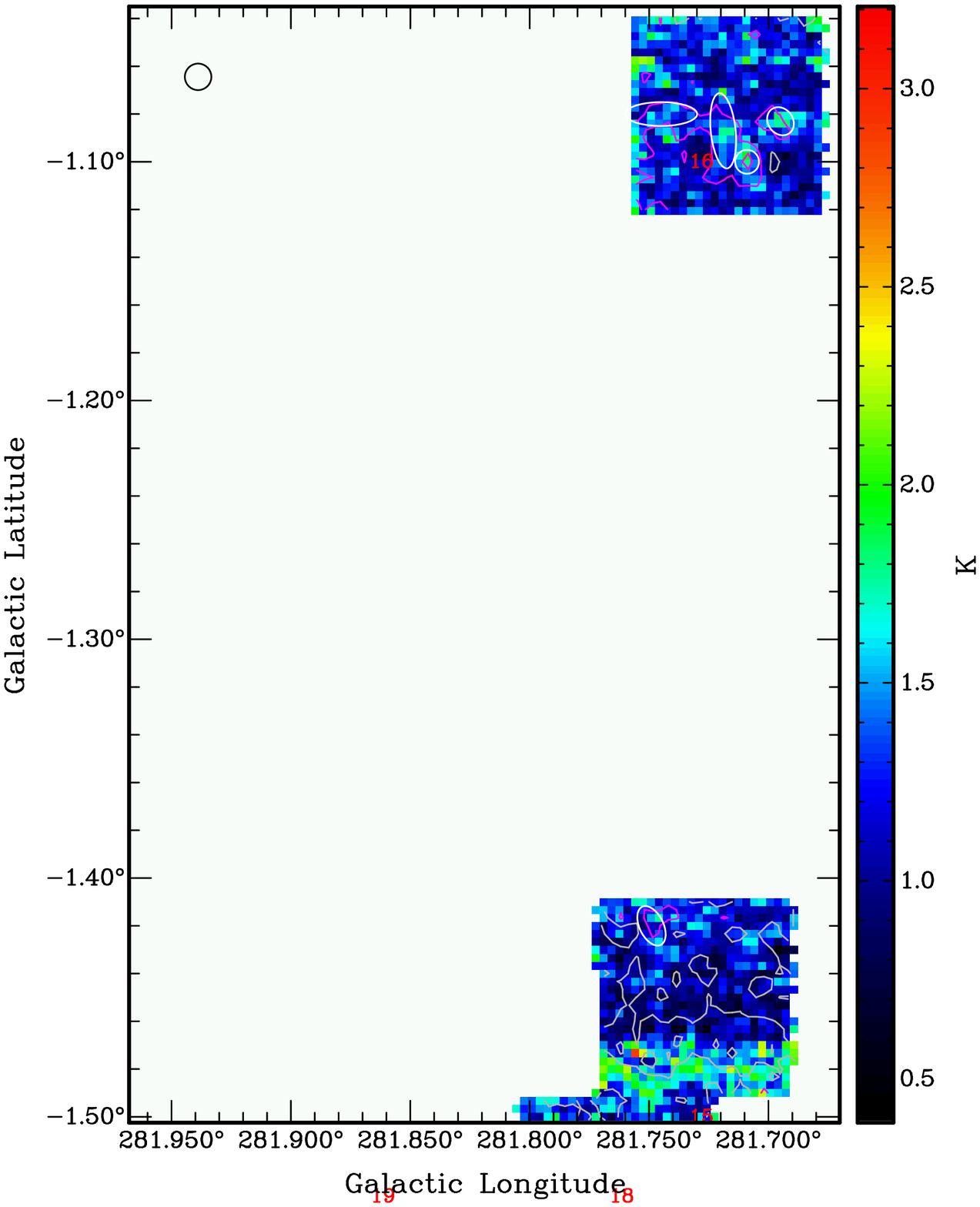}}
		(b){\includegraphics[angle=0,scale=0.35]{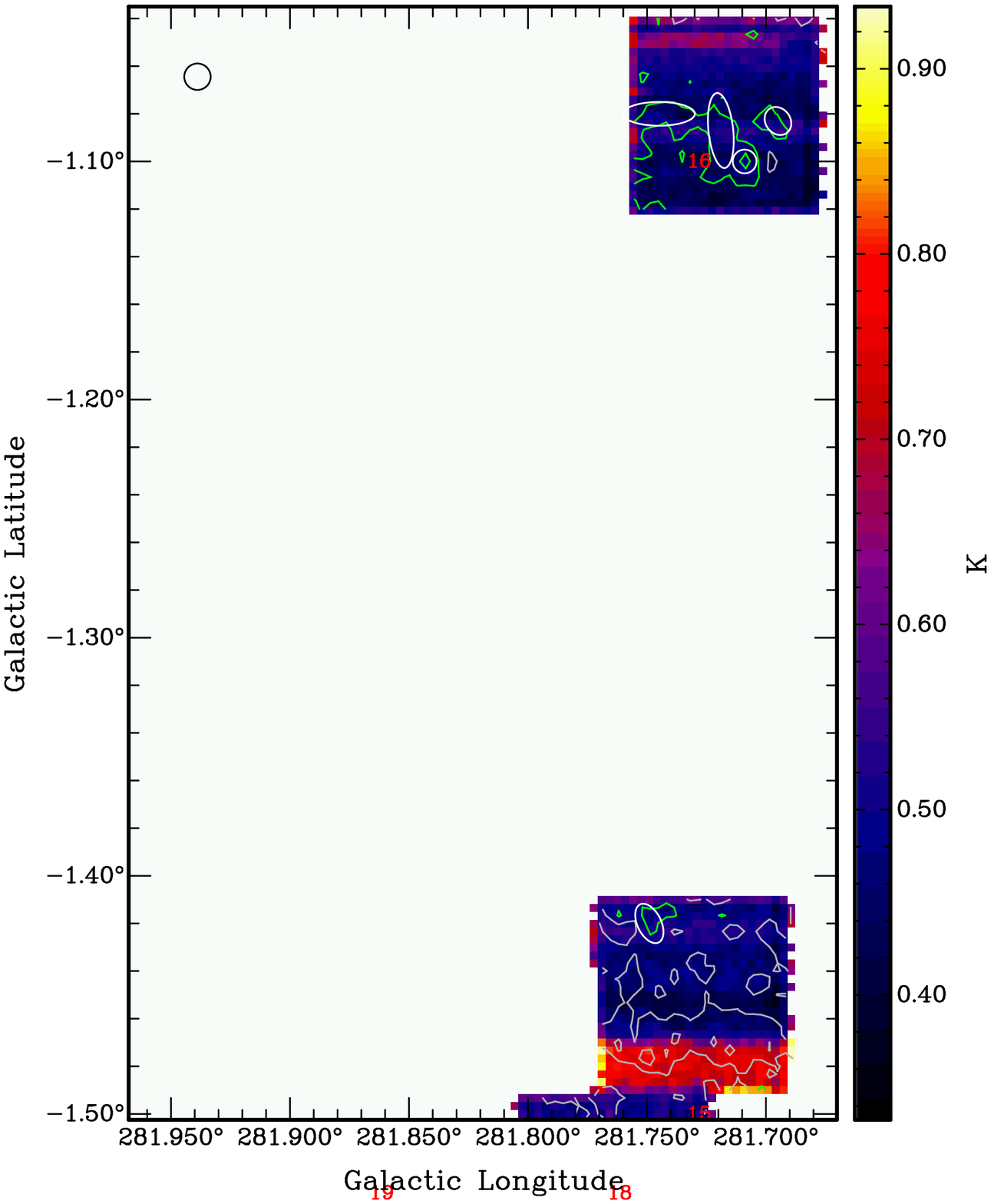}}}
\centerline{(c){\includegraphics[angle=0,scale=0.35]{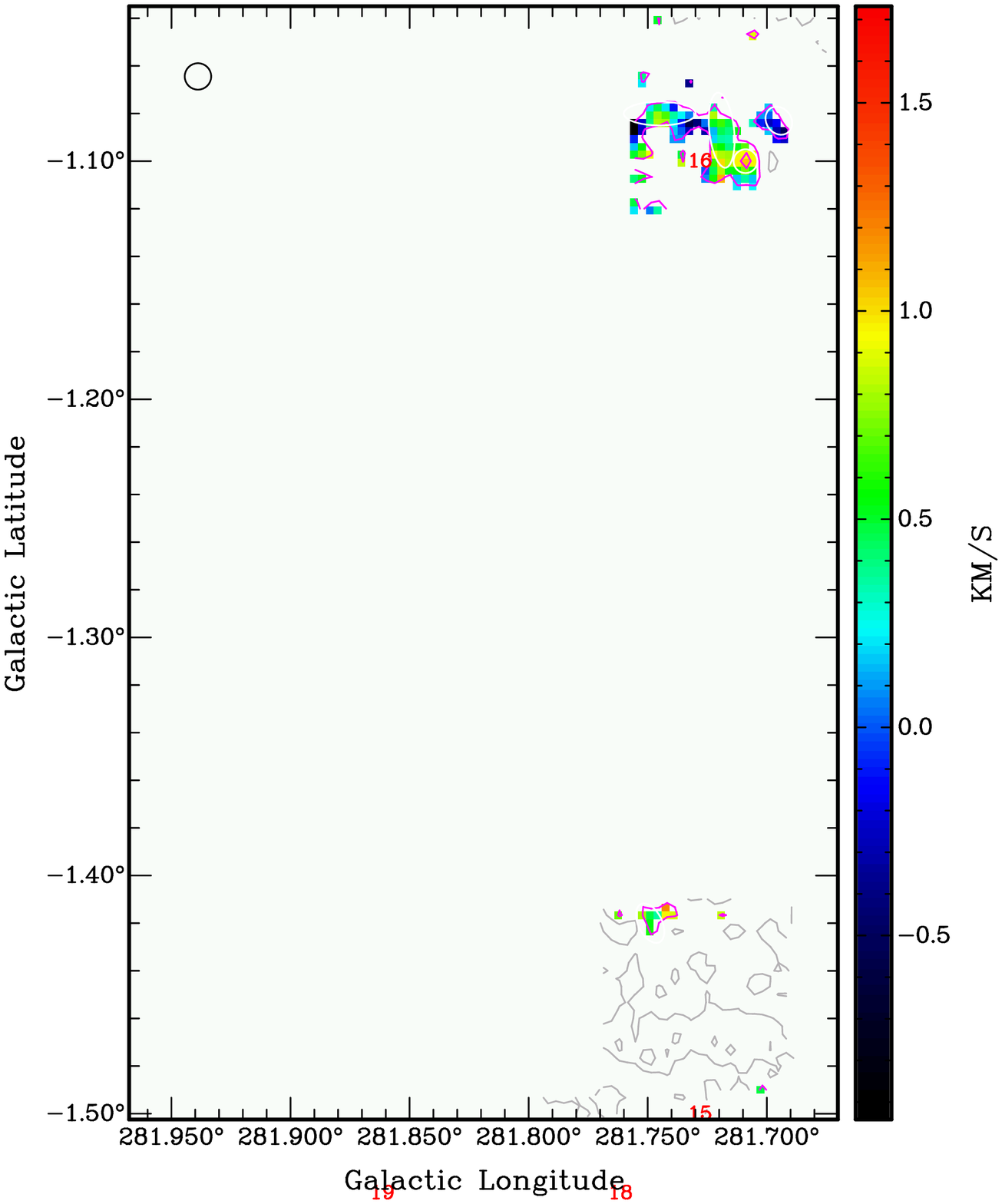}}
		(d){\includegraphics[angle=0,scale=0.35]{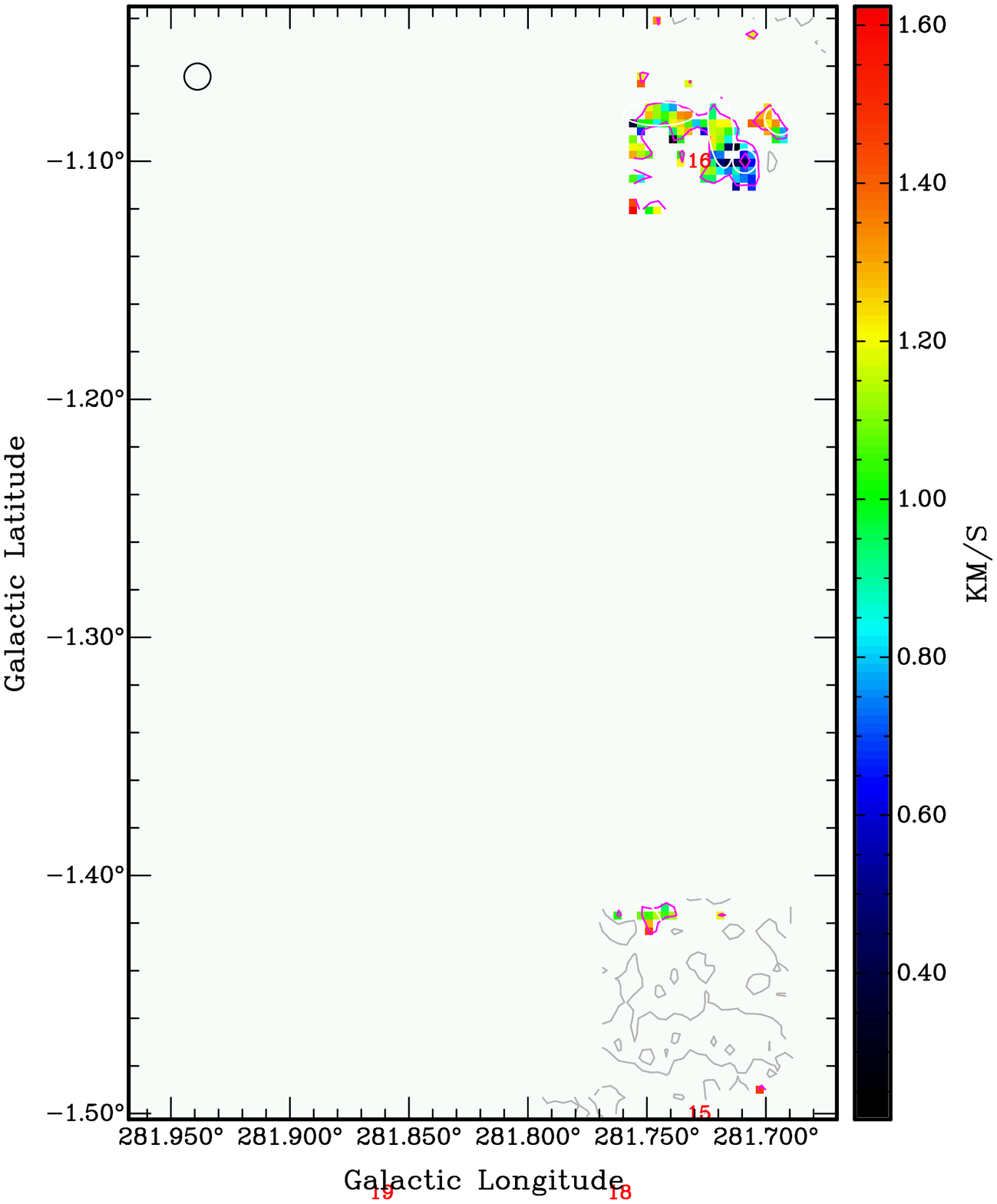}}}
\caption{\small Same as Fig.\,\ref{momR1}, but for Region 2b sources BYF\,15 and 16.  Contours are every 3$\sigma$ = 1.041\,K\kms\, and at 3.2\,kpc the 40$''$ Mopra beam (upper left corner) scales to 0.621\,pc.  ($a$) $T_p$,  ($b$) rms,  ($c$) $V_{\rm LSR}$,  ($d$) $\sigma_{V}$.
\label{momR2b}}
\end{figure*}

\clearpage

\begin{figure*}[htp]
\centerline{(a){\includegraphics[angle=0,scale=0.35]{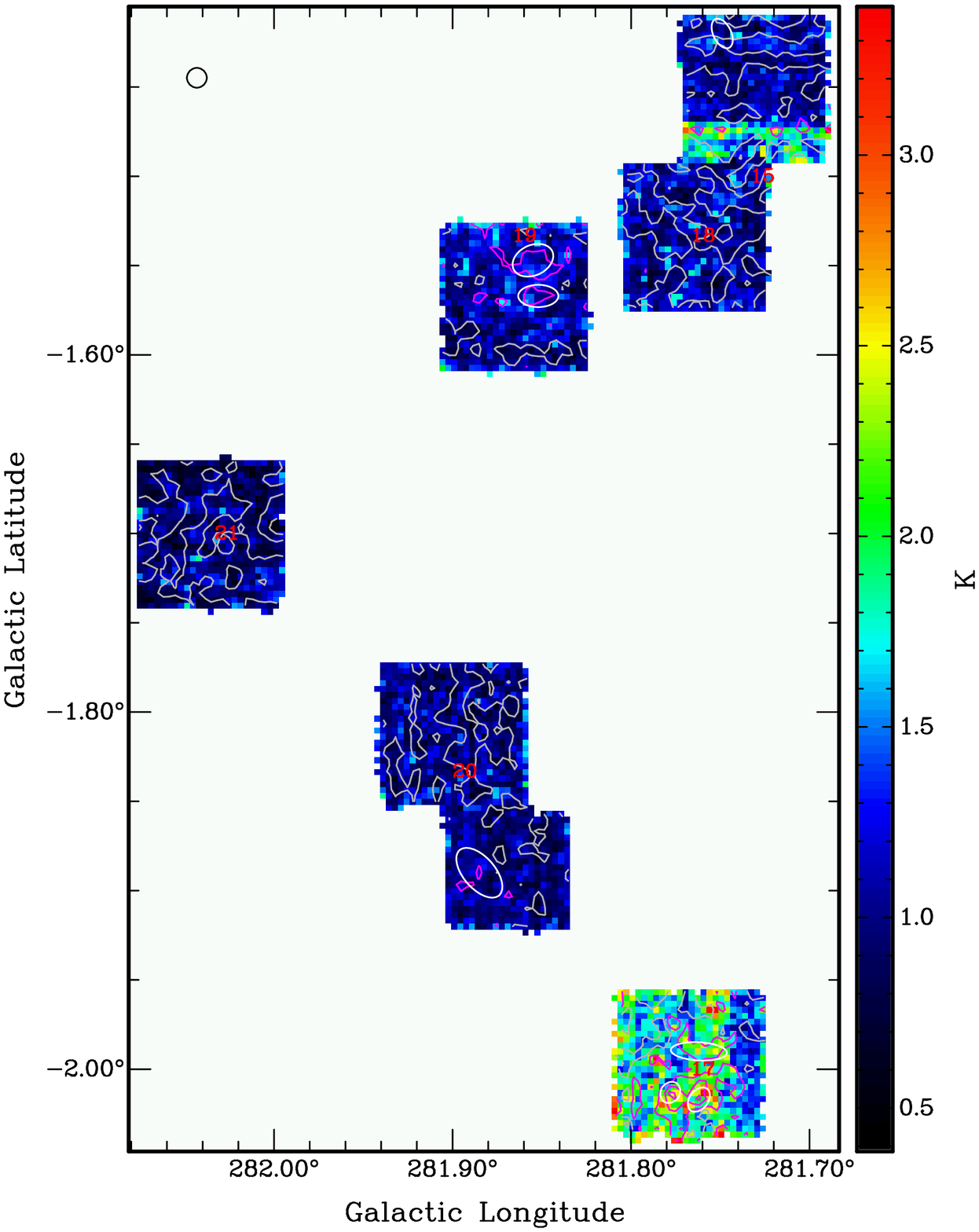}}
		(b){\includegraphics[angle=0,scale=0.35]{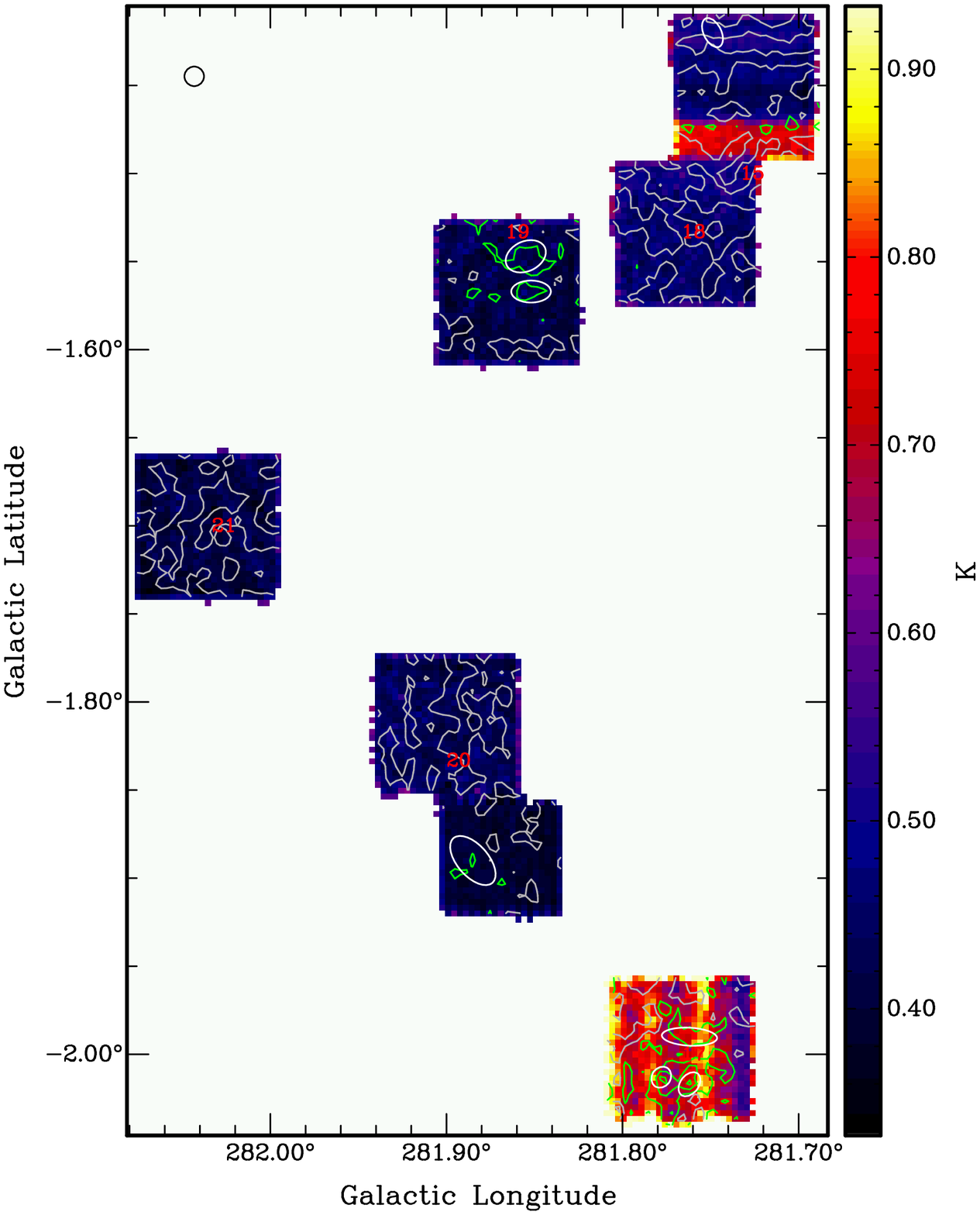}}}
\centerline{(c){\includegraphics[angle=0,scale=0.35]{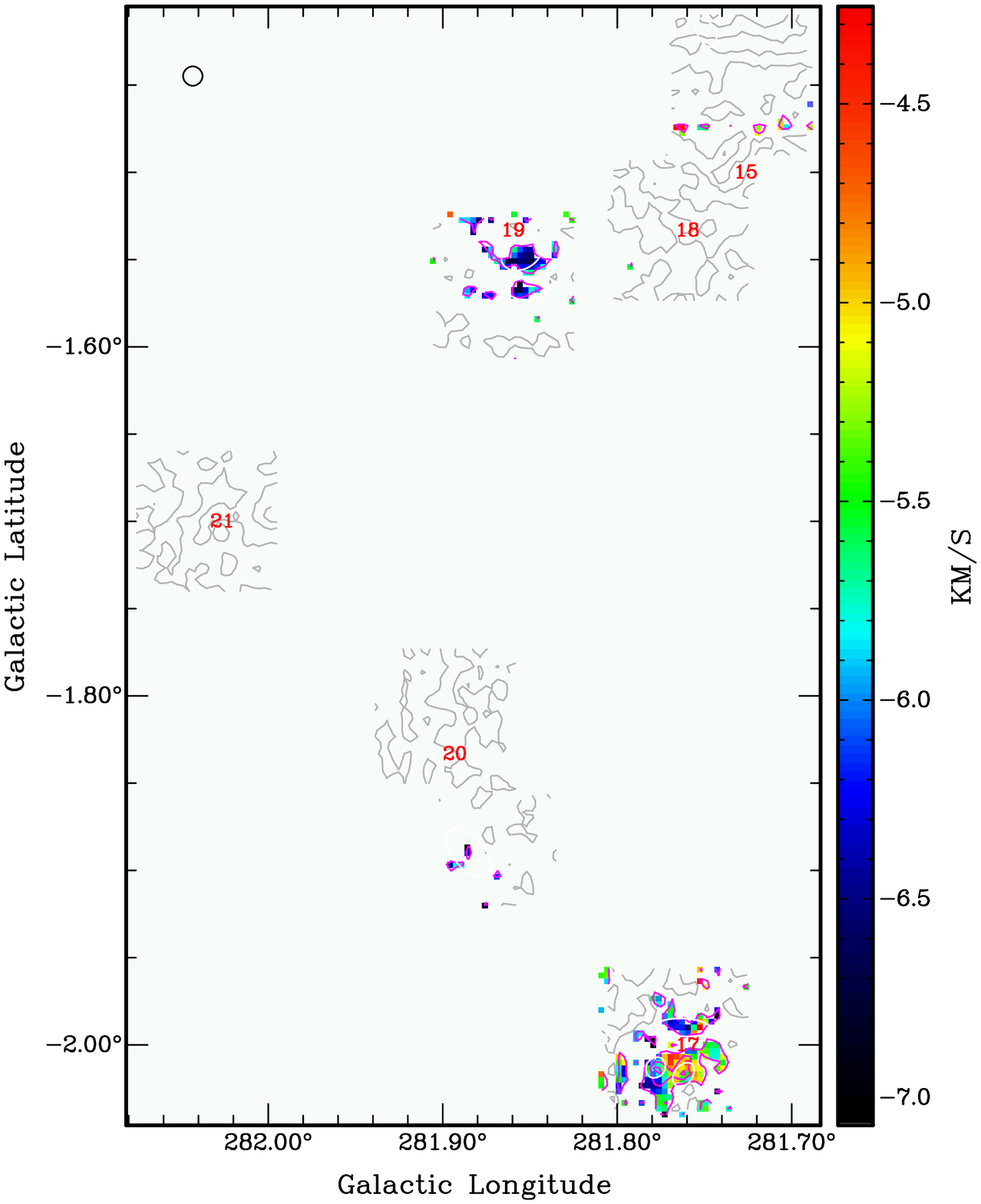}}
		(d){\includegraphics[angle=0,scale=0.35]{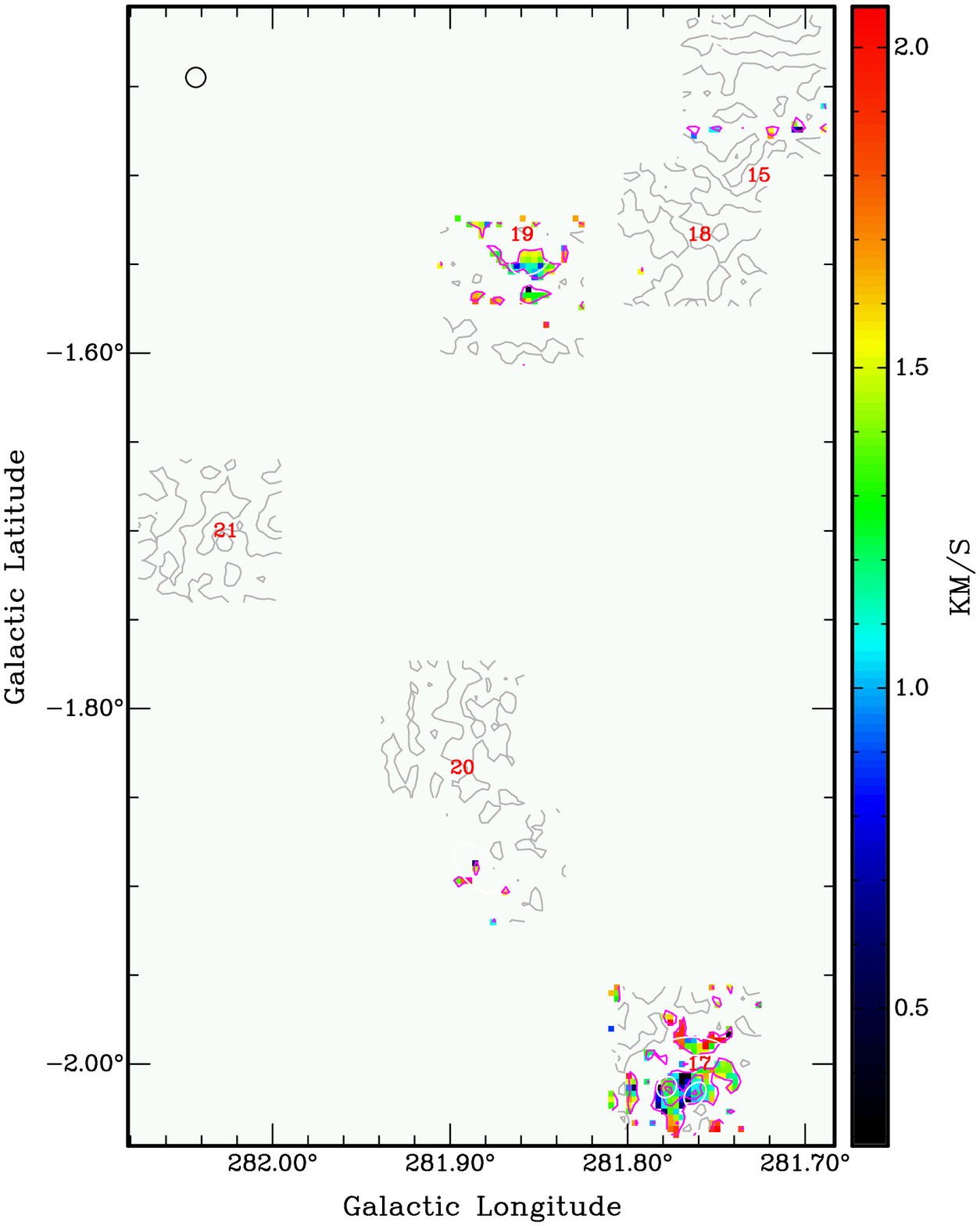}}}
\caption{\small Same as Fig.\,\ref{momR1}, but for Region 2b \& 3 sources BYF\,15 and 17--21.  Contours are every 3$\sigma$ = 1.122\,K\kms\, and at 3.2\,kpc the 40$''$ Mopra beam (upper left corner) scales to 0.621\,pc.  ($a$) $T_p$,  ($b$) rms,  ($c$) $V_{\rm LSR}$,  ($d$) $\sigma_{V}$.
\label{momR3a}}
\end{figure*}

\clearpage

\begin{figure*}[htp]
\centerline{(a){\includegraphics[angle=-90,scale=0.25]{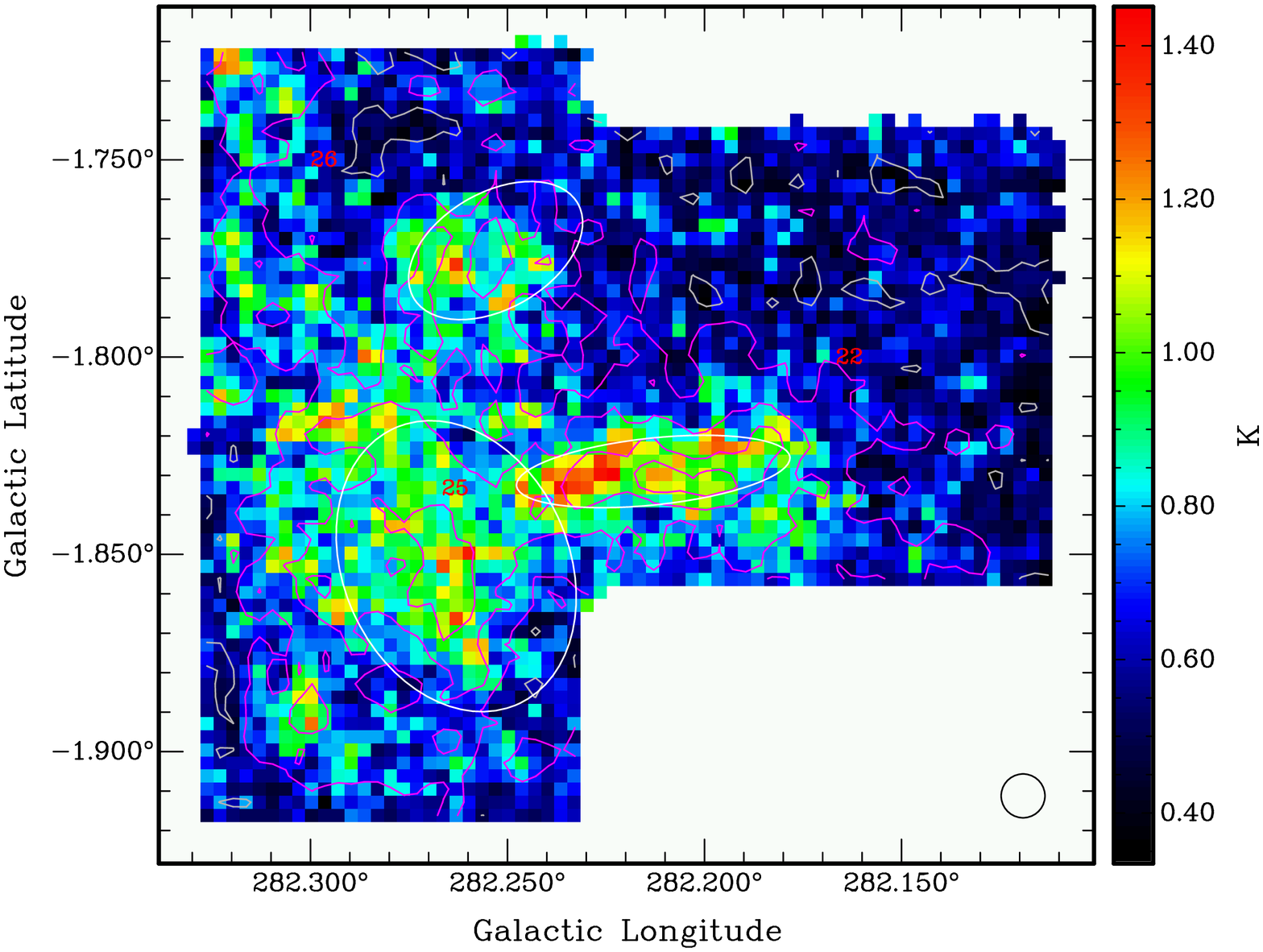}}
		(b){\includegraphics[angle=-90,scale=0.25]{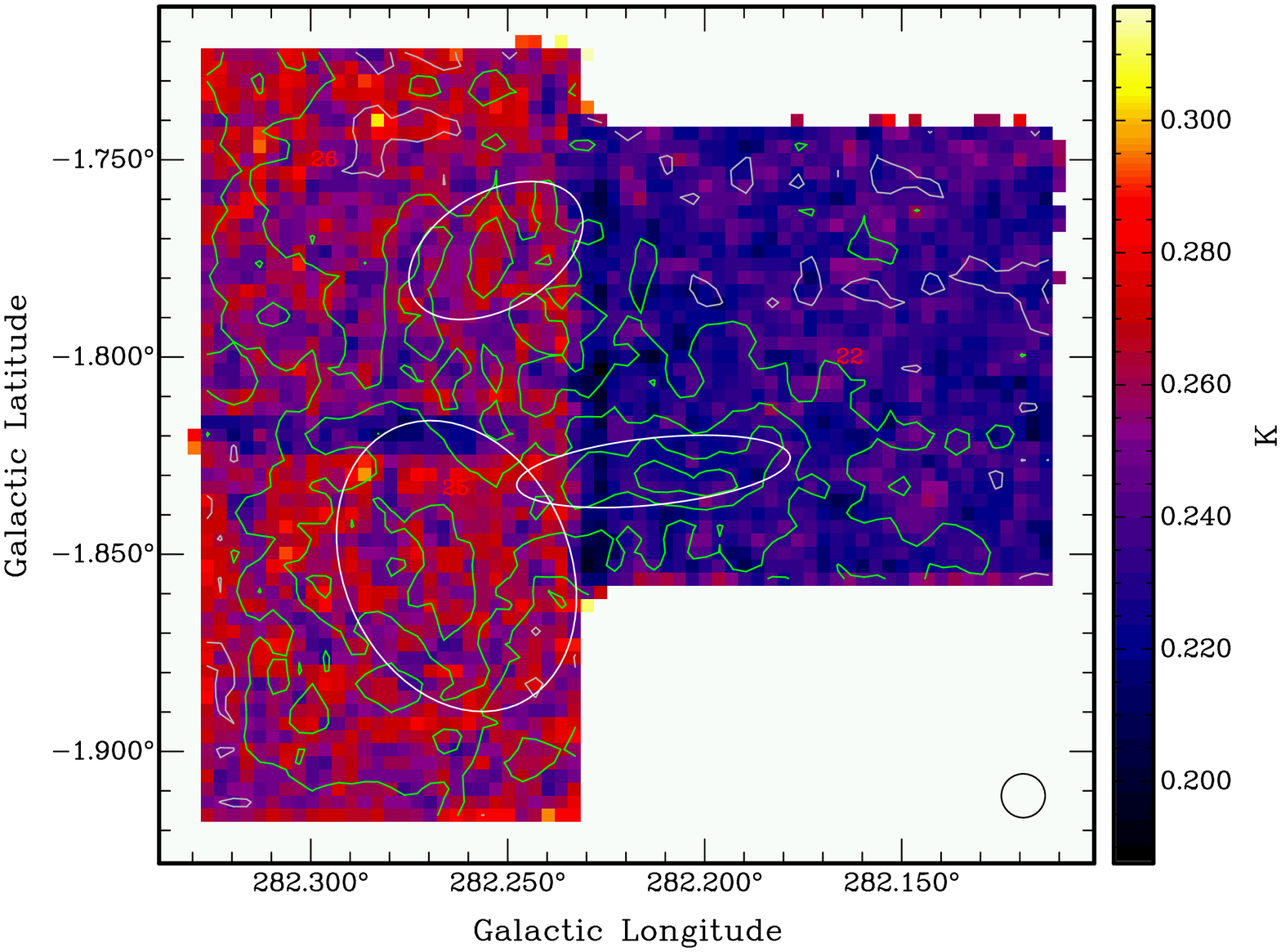}}}
\centerline{(c){\includegraphics[angle=-90,scale=0.25]{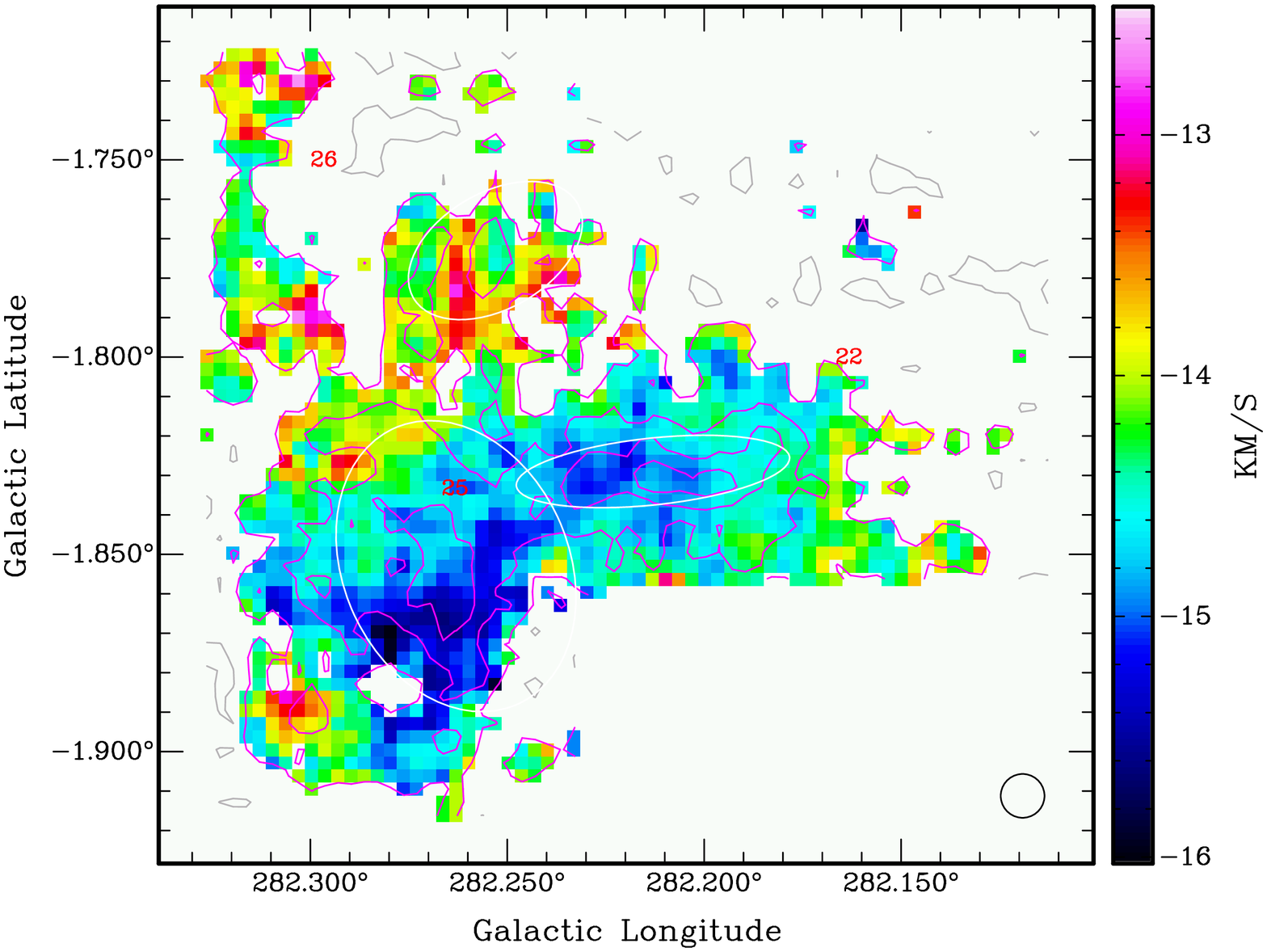}}
		(d){\includegraphics[angle=-90,scale=0.25]{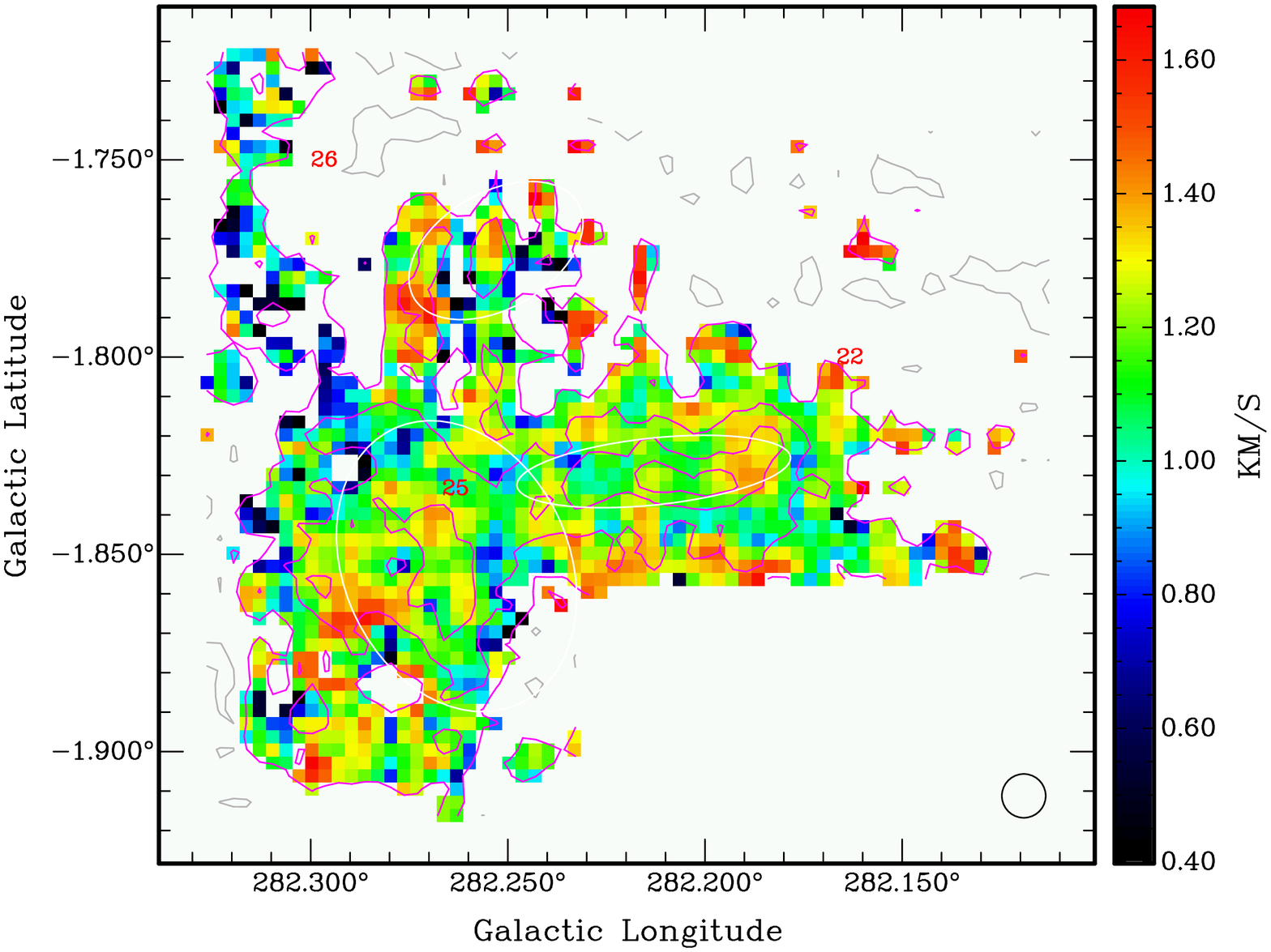}}}
\caption{\small Same as Fig.\,\ref{momR1}, but for Region 3 sources BYF\,22, 25, and 26.  Contours are every 3$\sigma$ = 0.558\,K\kms\, and at 3.2\,kpc the 40$''$ Mopra beam (lower right corner) scales to 0.621\,pc.  ($a$) $T_p$,  ($b$) rms,  ($c$) $V_{\rm LSR}$,  ($d$) $\sigma_{V}$.
\label{momR3b}}
\end{figure*}

\clearpage

\begin{figure*}[htp]
\centerline{(a){\includegraphics[angle=0,scale=0.3]{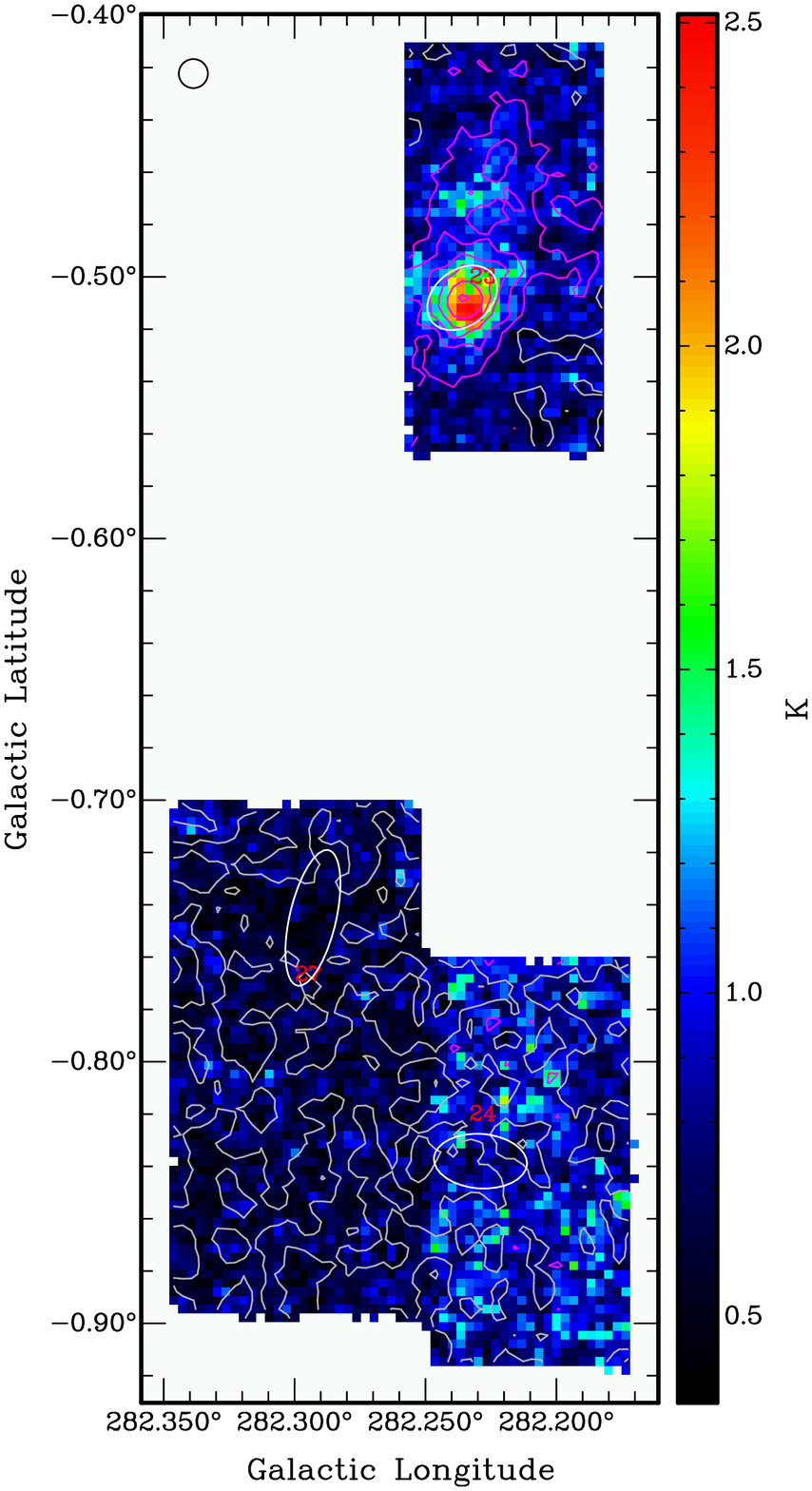}}
		(b){\includegraphics[angle=0,scale=0.3]{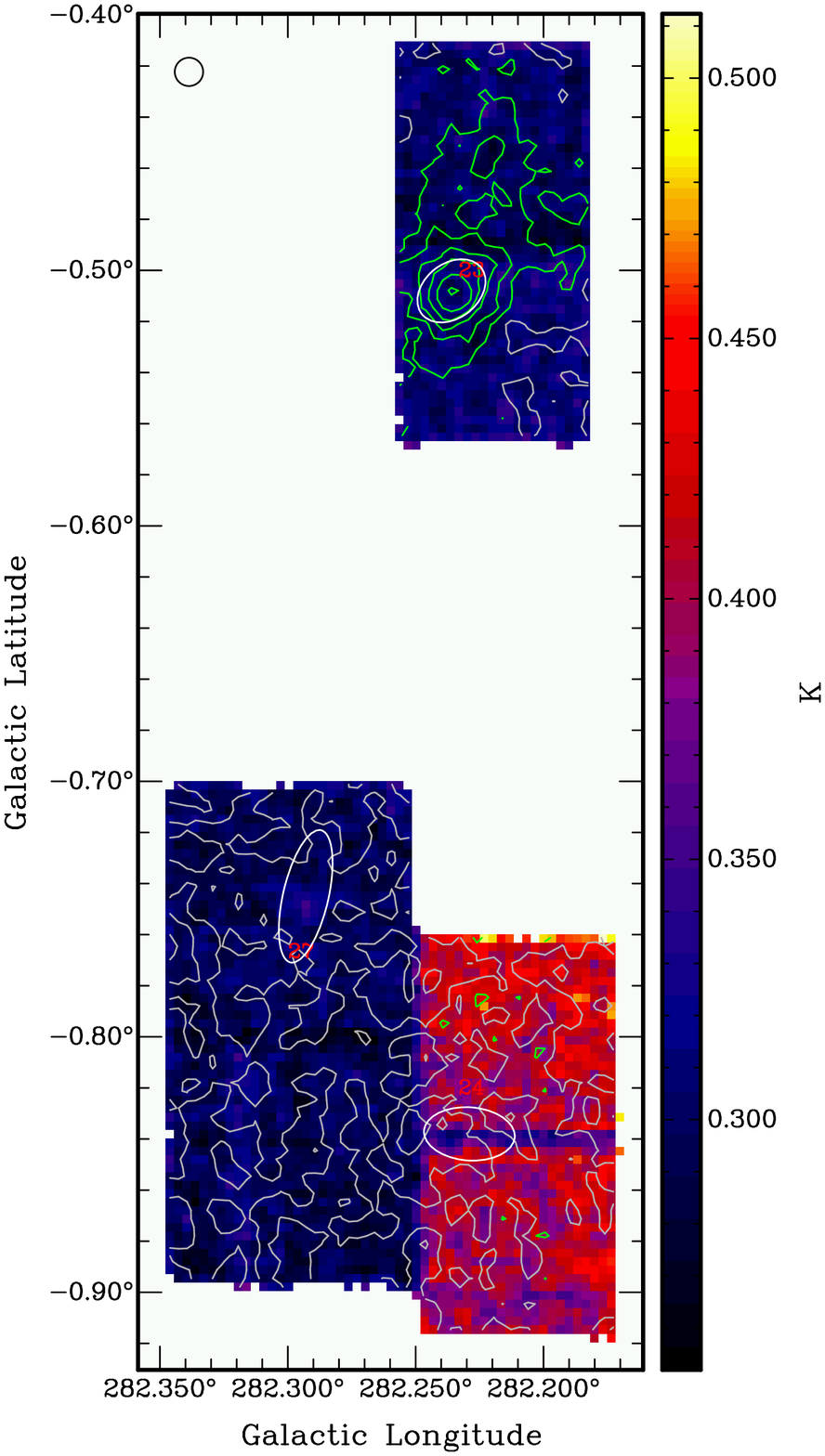}}}
\centerline{(c){\includegraphics[angle=0,scale=0.3]{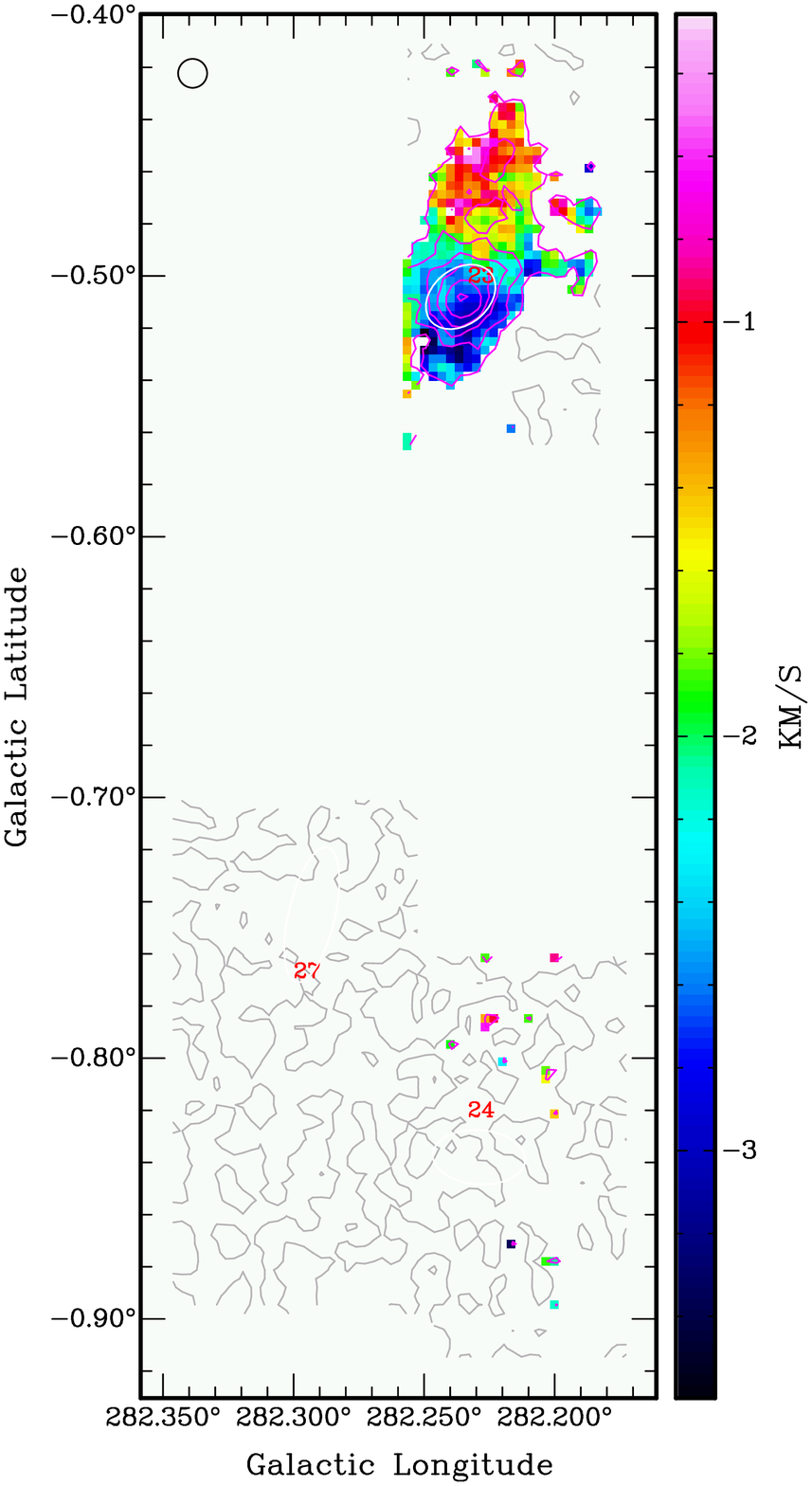}}
		(d){\includegraphics[angle=0,scale=0.3]{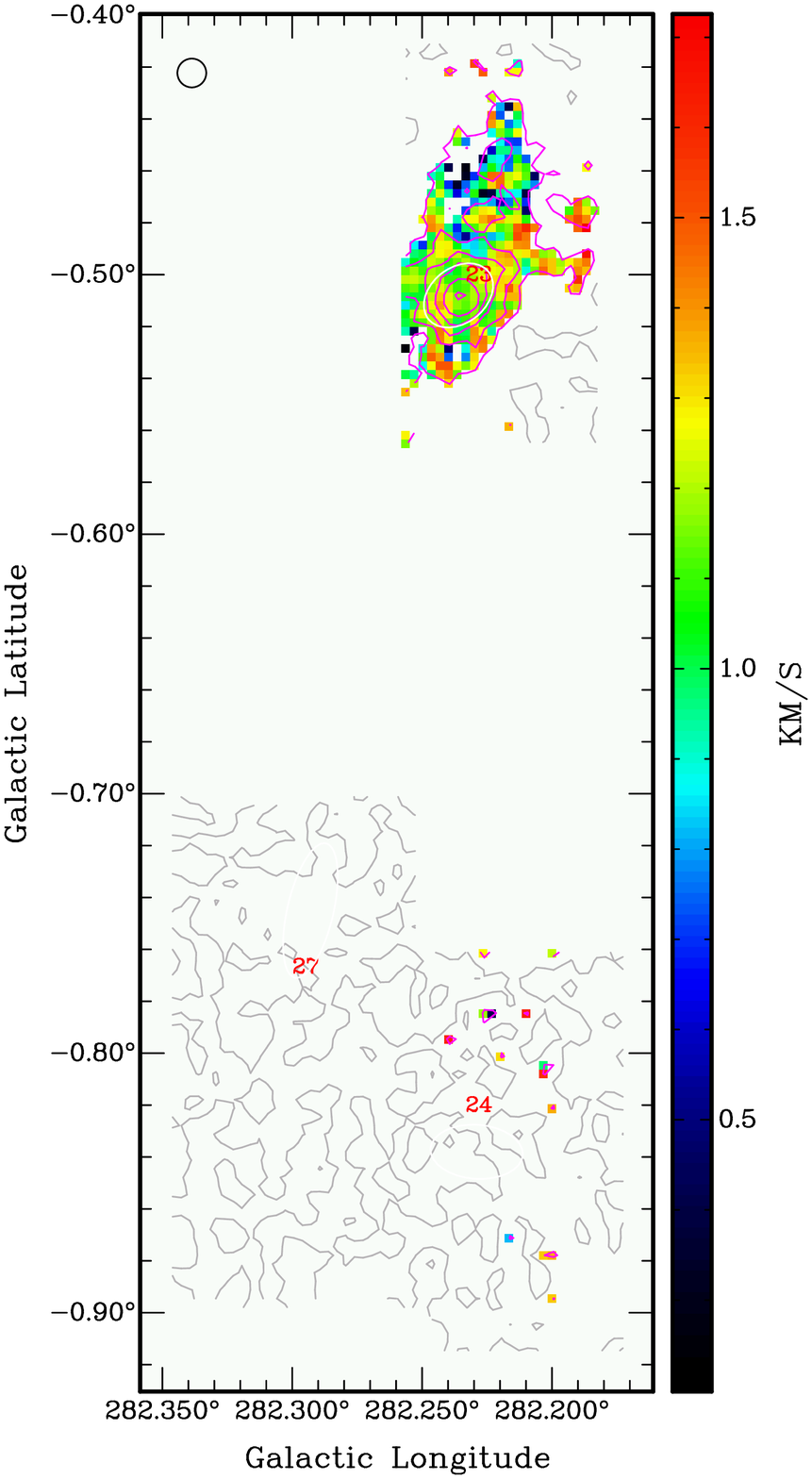}}}
\caption{\small Same as Fig.\,\ref{momR1}, but for Region 2c source BYF\,23.  Contours are every 3$\sigma$ = 0.780\,K\kms\, and at 3.2\,kpc the 40$''$ Mopra beam (upper left corner) scales to 0.621\,pc.  ($a$) $T_p$,  ($b$) rms,  ($c$) $V_{\rm LSR}$,  ($d$) $\sigma_{V}$.
\label{momR2c23}}
\end{figure*}

\clearpage

\begin{figure*}[htp]
\centerline{(a){\includegraphics[angle=0,scale=0.3]{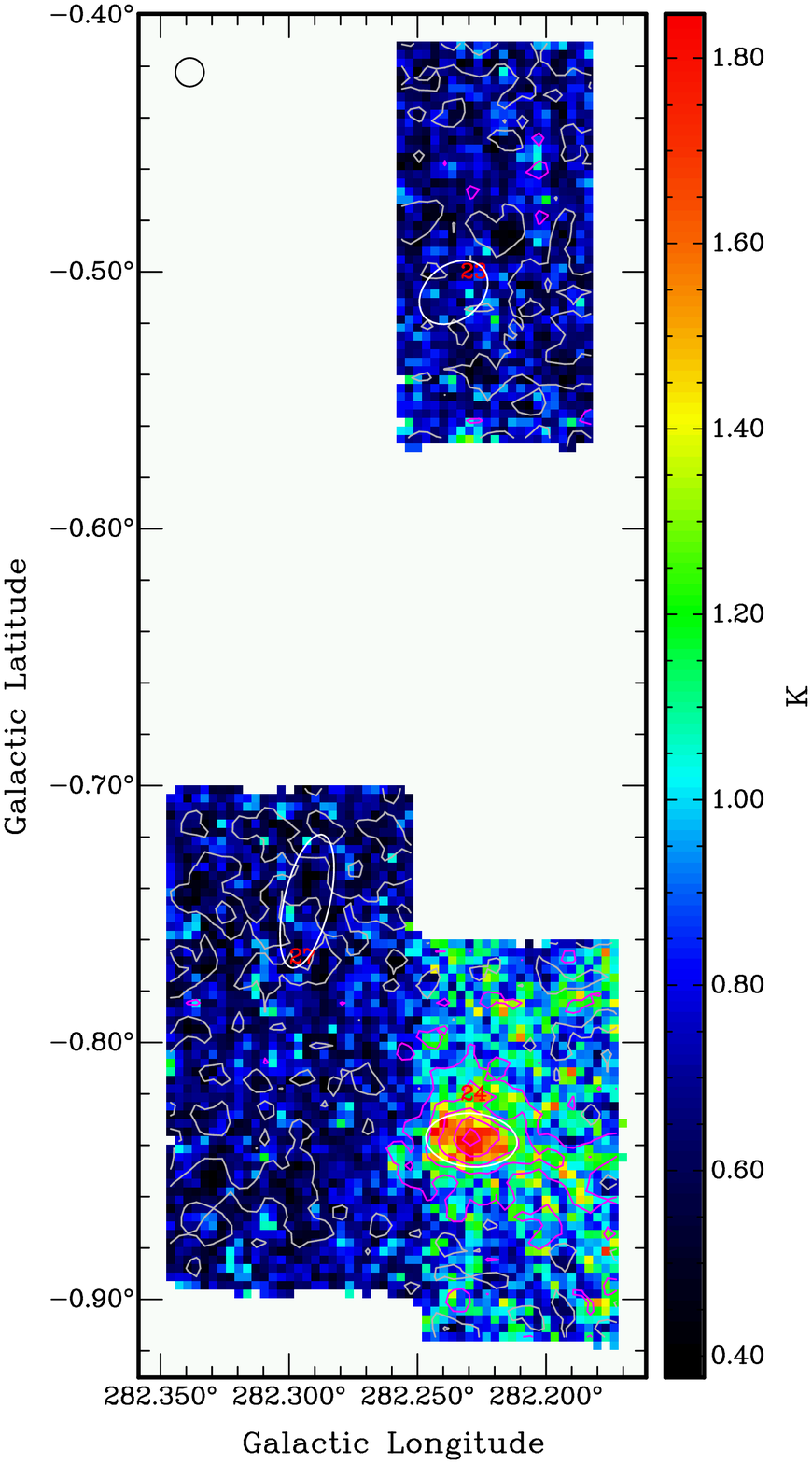}}
		(b){\includegraphics[angle=0,scale=0.3]{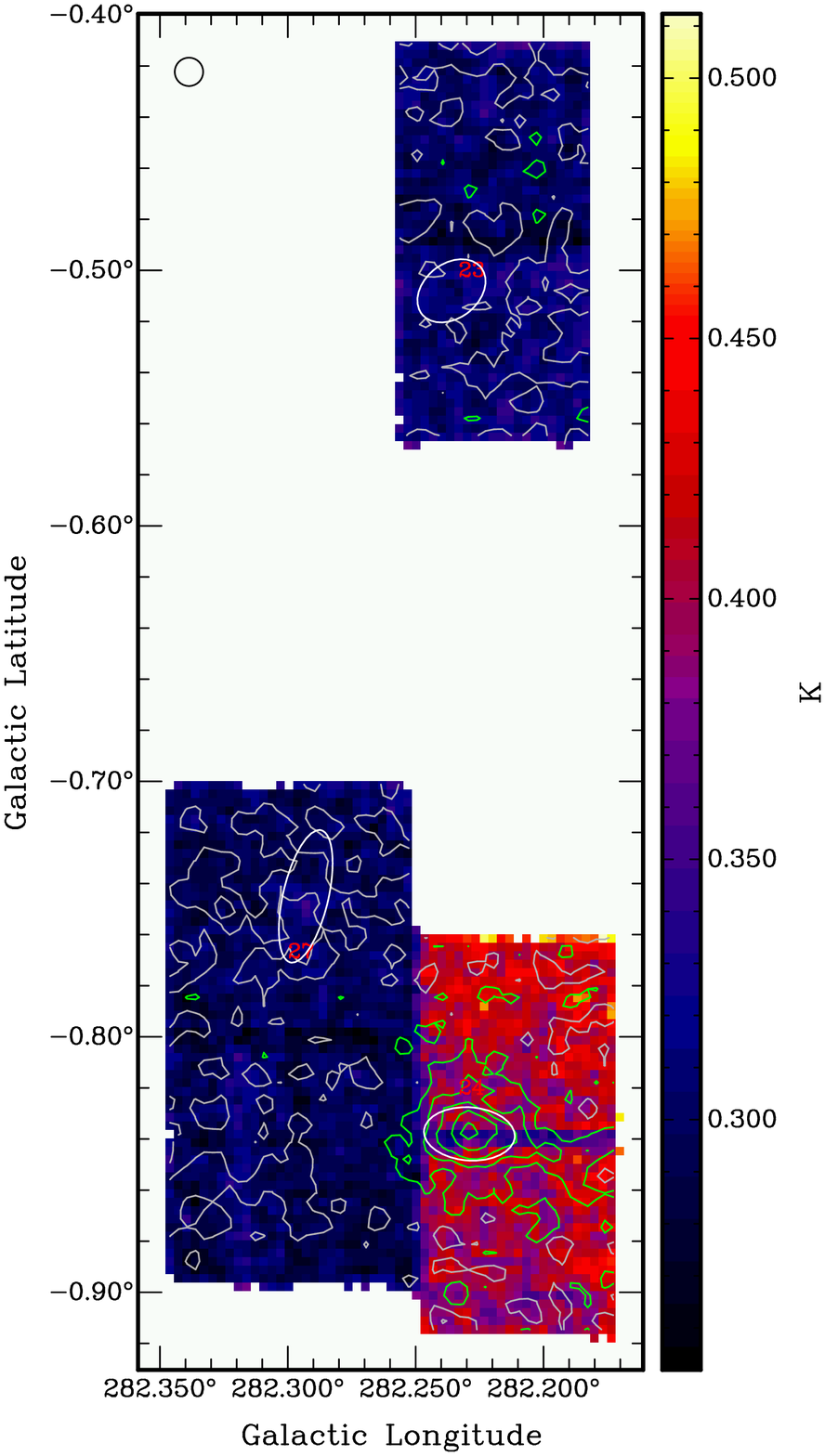}}}
\centerline{(c){\includegraphics[angle=0,scale=0.3]{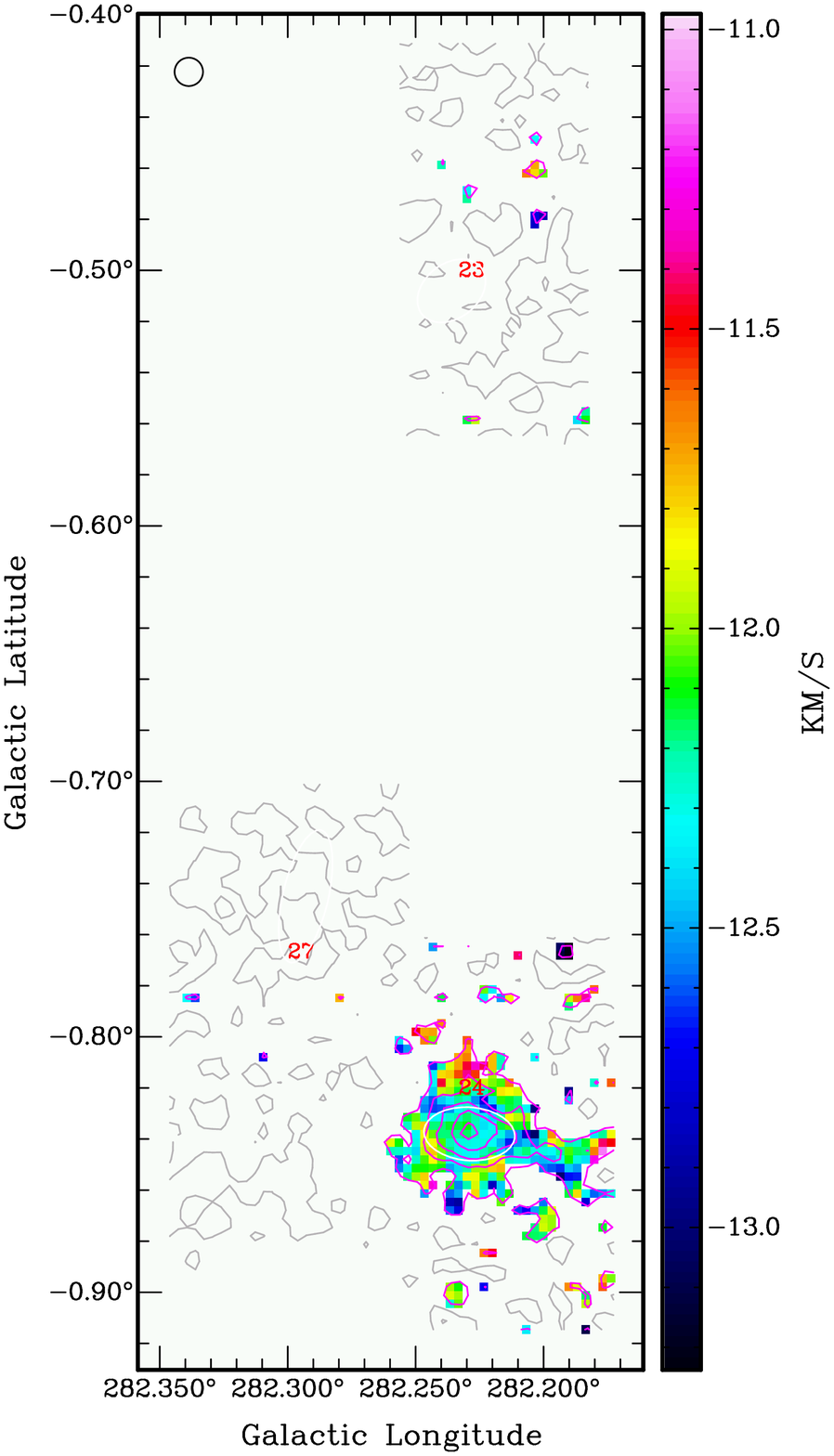}}
		(d){\includegraphics[angle=0,scale=0.3]{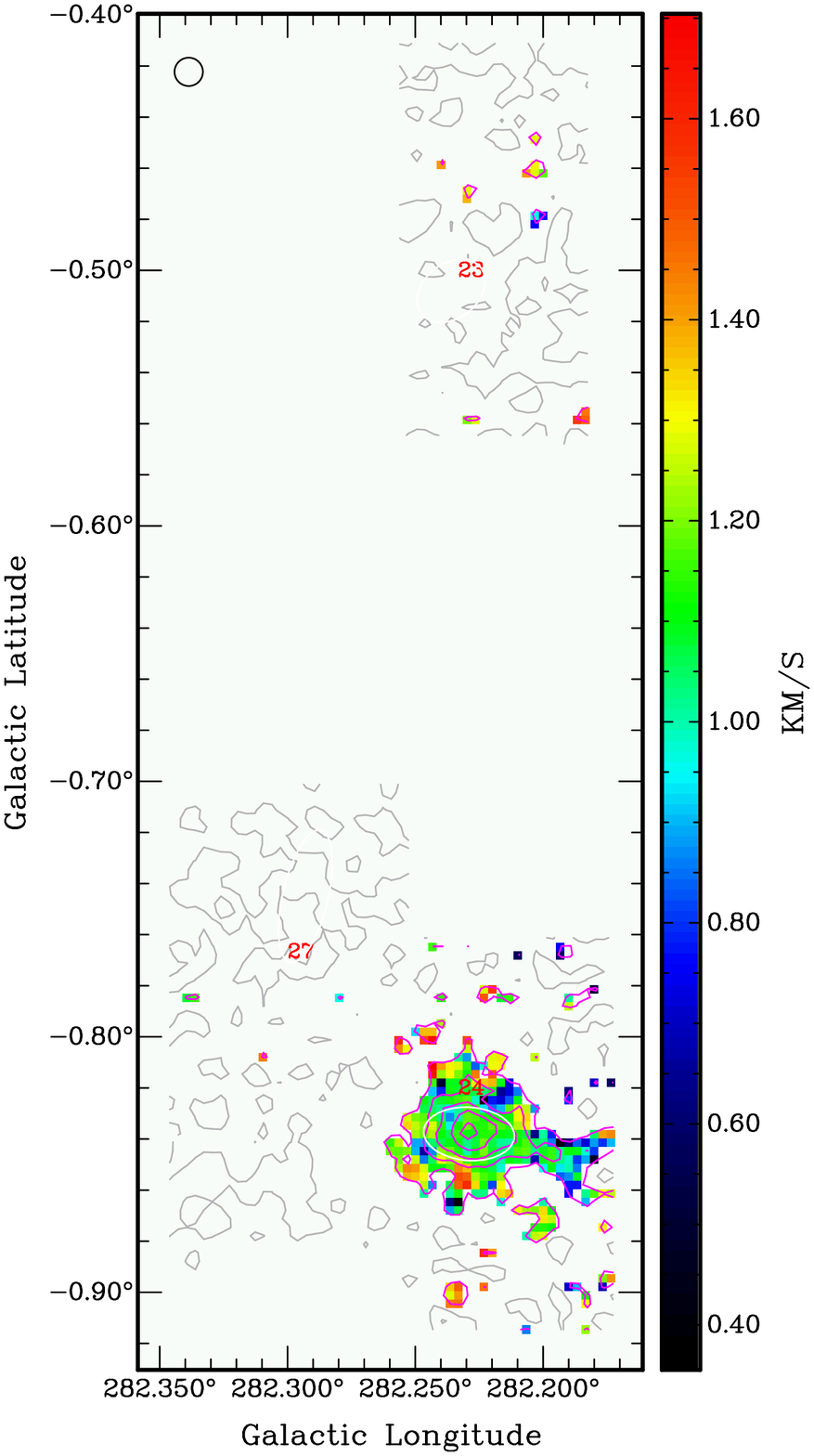}}}
\caption{\small Same as Fig.\,\ref{momR1}, but for Region 2c source BYF\,24.  Contours are every 3$\sigma$ = 0.729\,K\kms\, and at 3.2\,kpc the 40$''$ Mopra beam (upper left corner) scales to 0.621\,pc.  ($a$) $T_p$,  ($b$) rms,  ($c$) $V_{\rm LSR}$,  ($d$) $\sigma_{V}$.
\label{momR2c24}}
\end{figure*}

\clearpage

\begin{figure*}[htp]
\centerline{(a){\includegraphics[angle=0,scale=0.3]{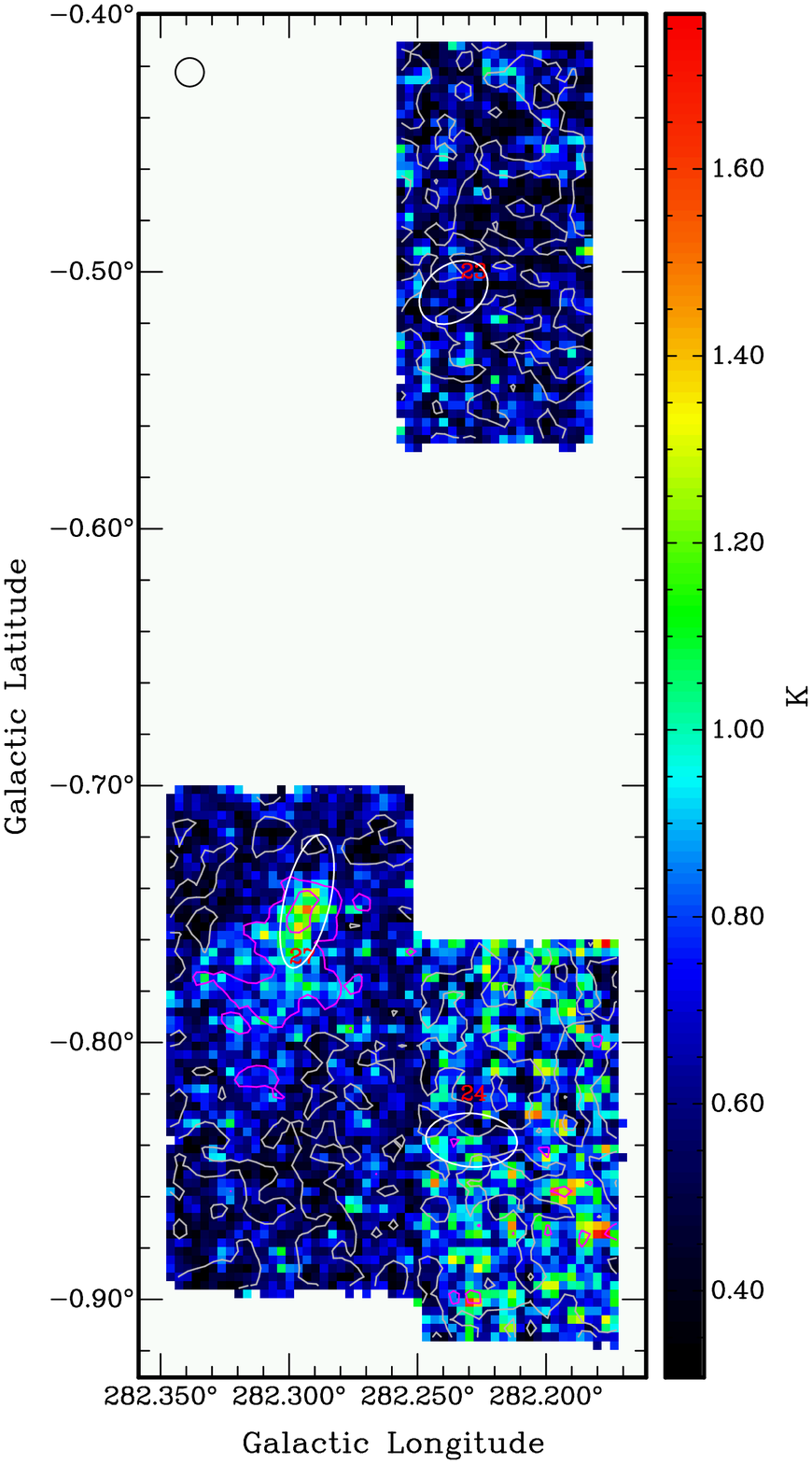}}
		(b){\includegraphics[angle=0,scale=0.3]{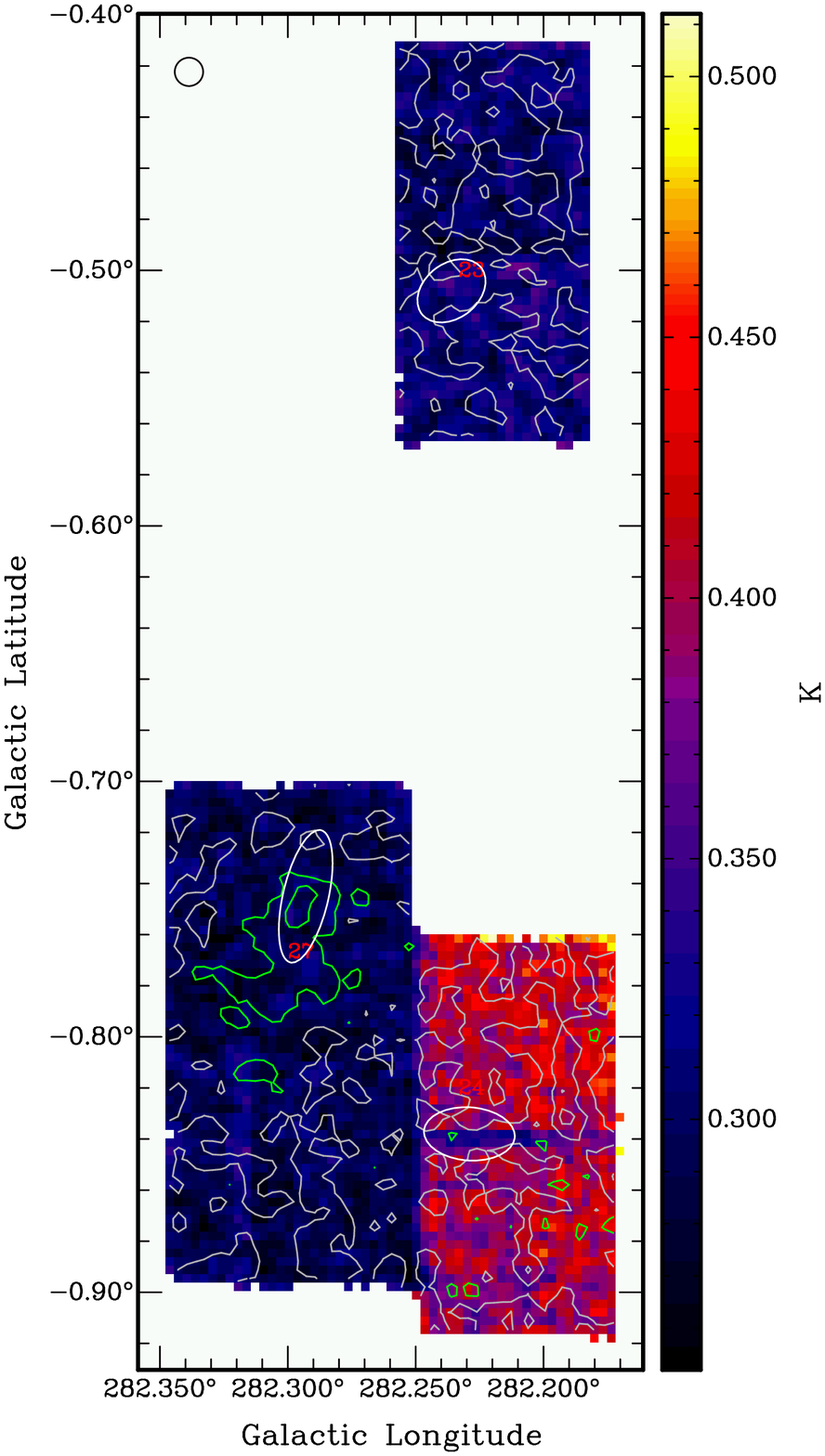}}}
\centerline{(c){\includegraphics[angle=0,scale=0.3]{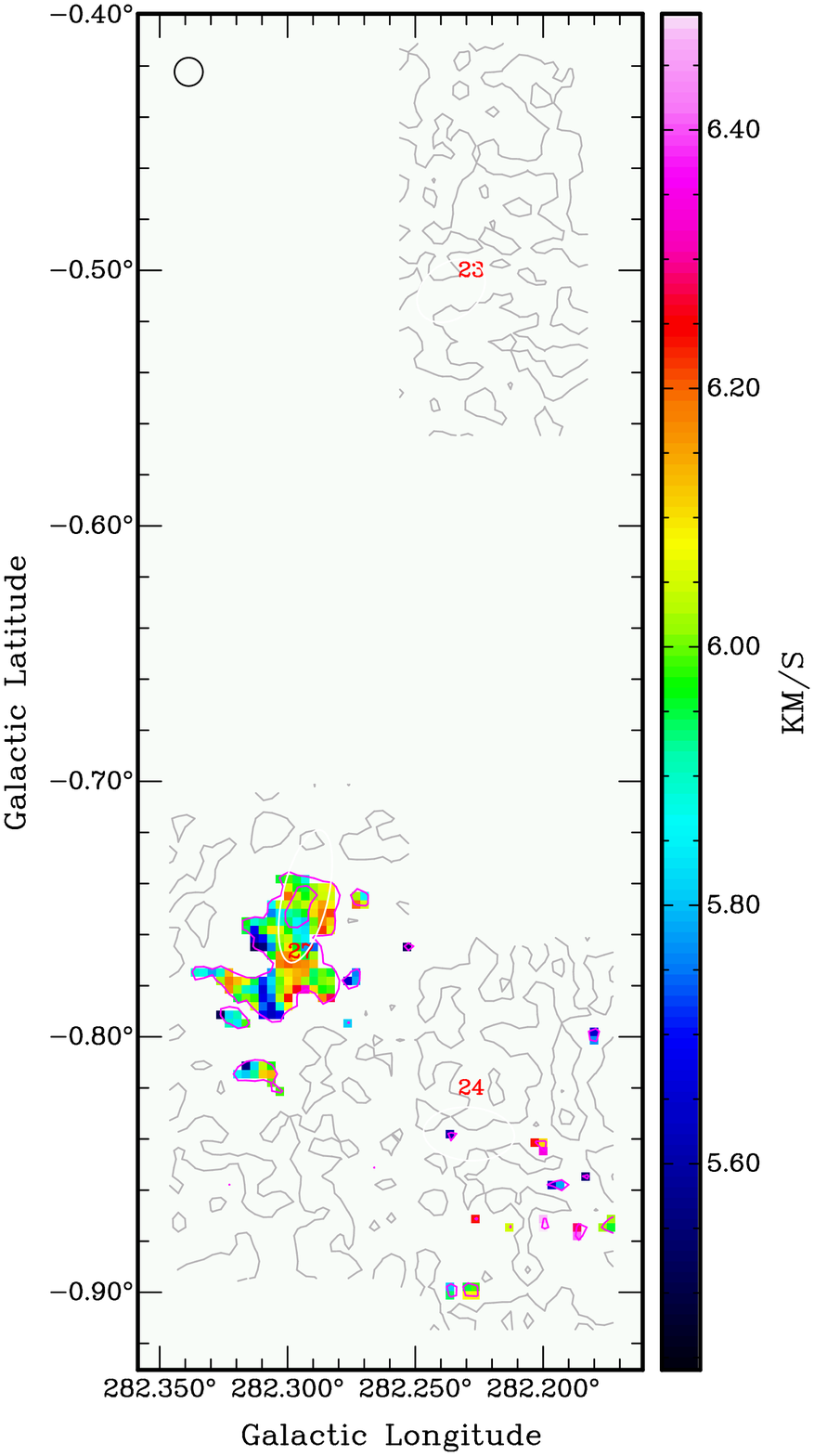}}
		(d){\includegraphics[angle=0,scale=0.3]{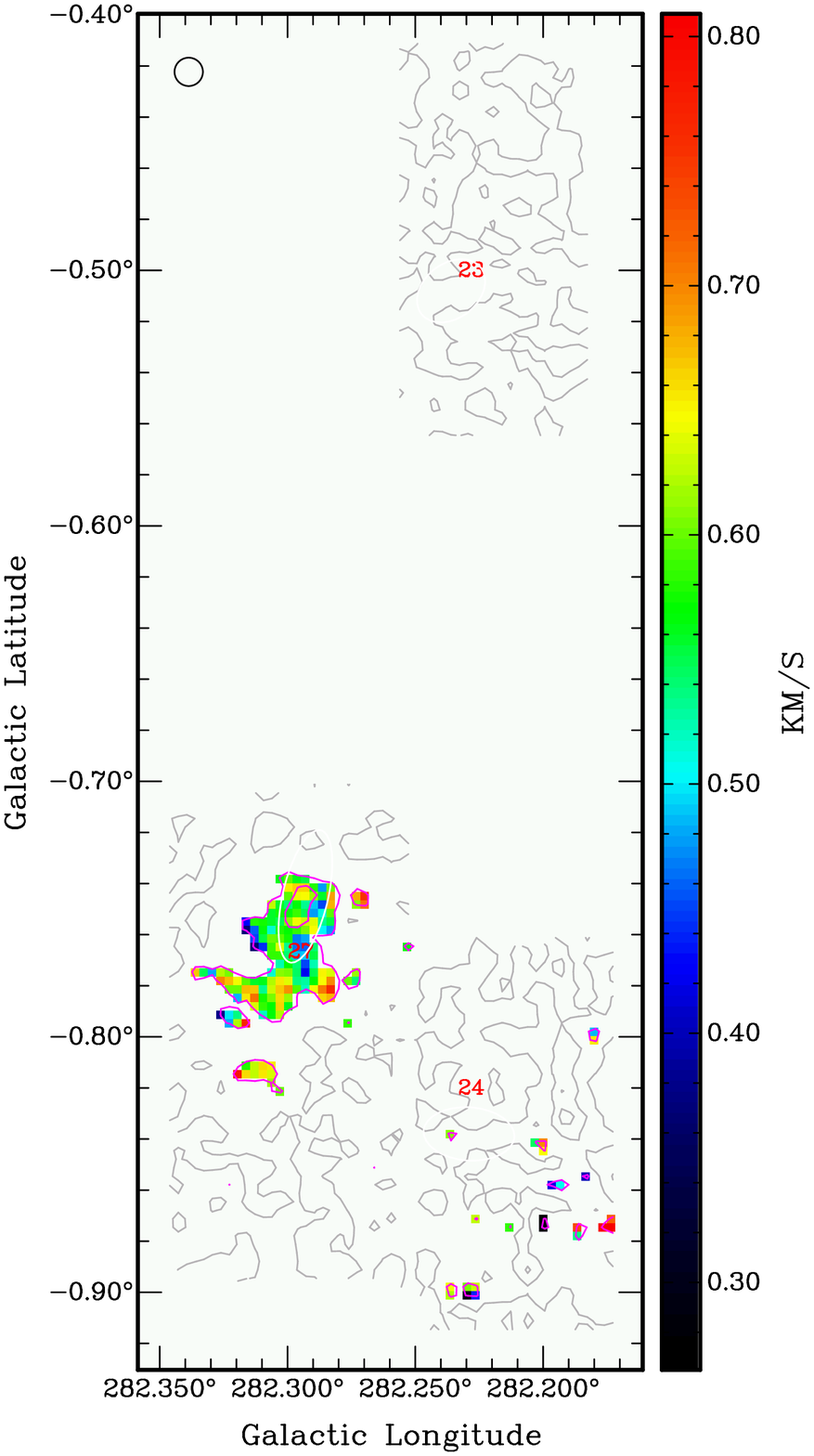}}}
\caption{\small Same as Fig.\,\ref{momR1}, but for Region 2c source BYF\,27.  Contours are every 3$\sigma$ = 0.507\,K\kms\, and at 3.2\,kpc the 40$''$ Mopra beam (upper left corner) scales to 0.621\,pc.  ($a$) $T_p$,  ($b$) rms,  ($c$) $V_{\rm LSR}$,  ($d$) $\sigma_{V}$.
\label{momR2c27}}
\end{figure*}

\clearpage

\begin{figure*}[htp]
\centerline{(a){\includegraphics[angle=-90,scale=0.3]{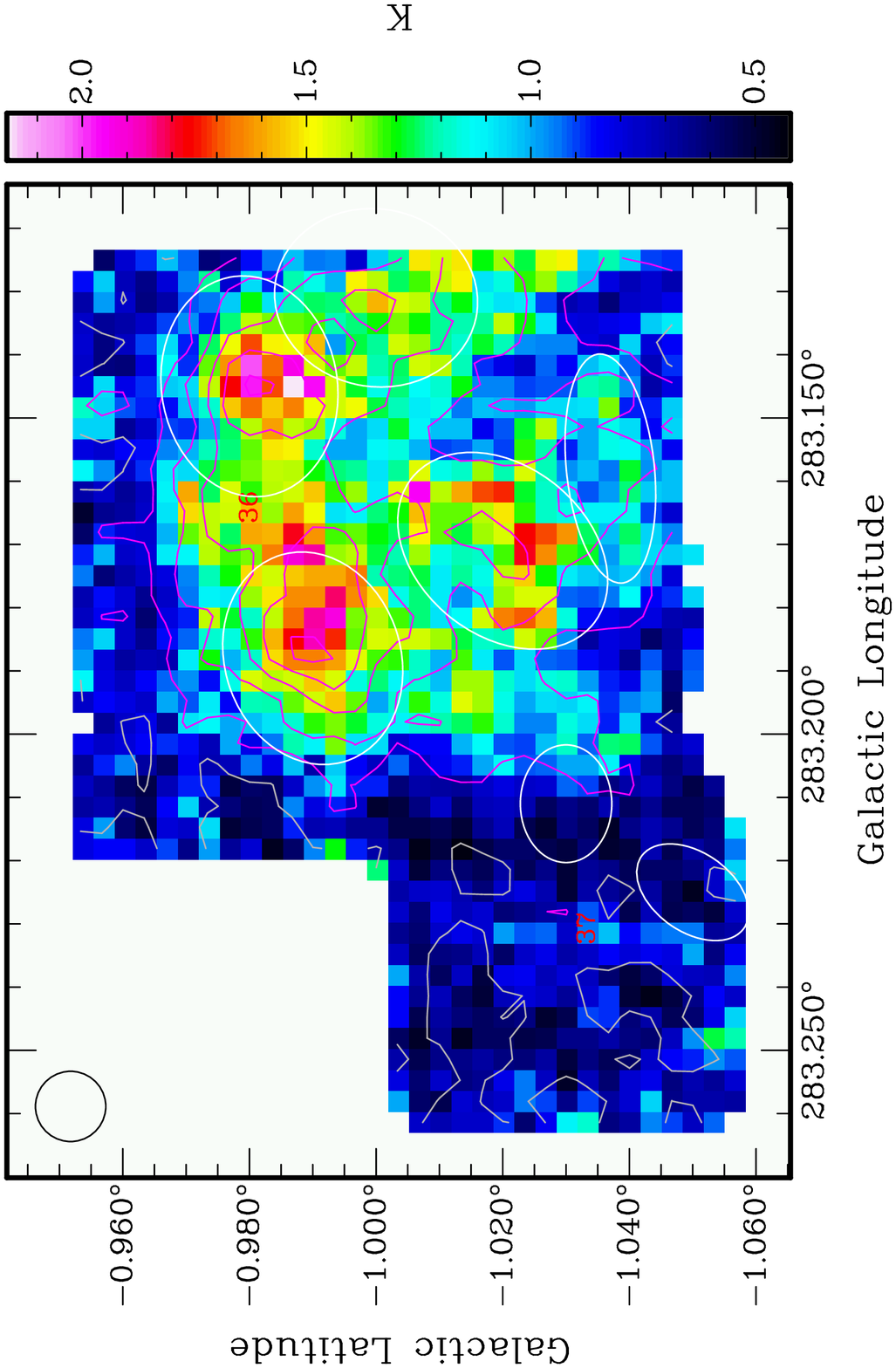}}
		(b){\includegraphics[angle=-90,scale=0.3]{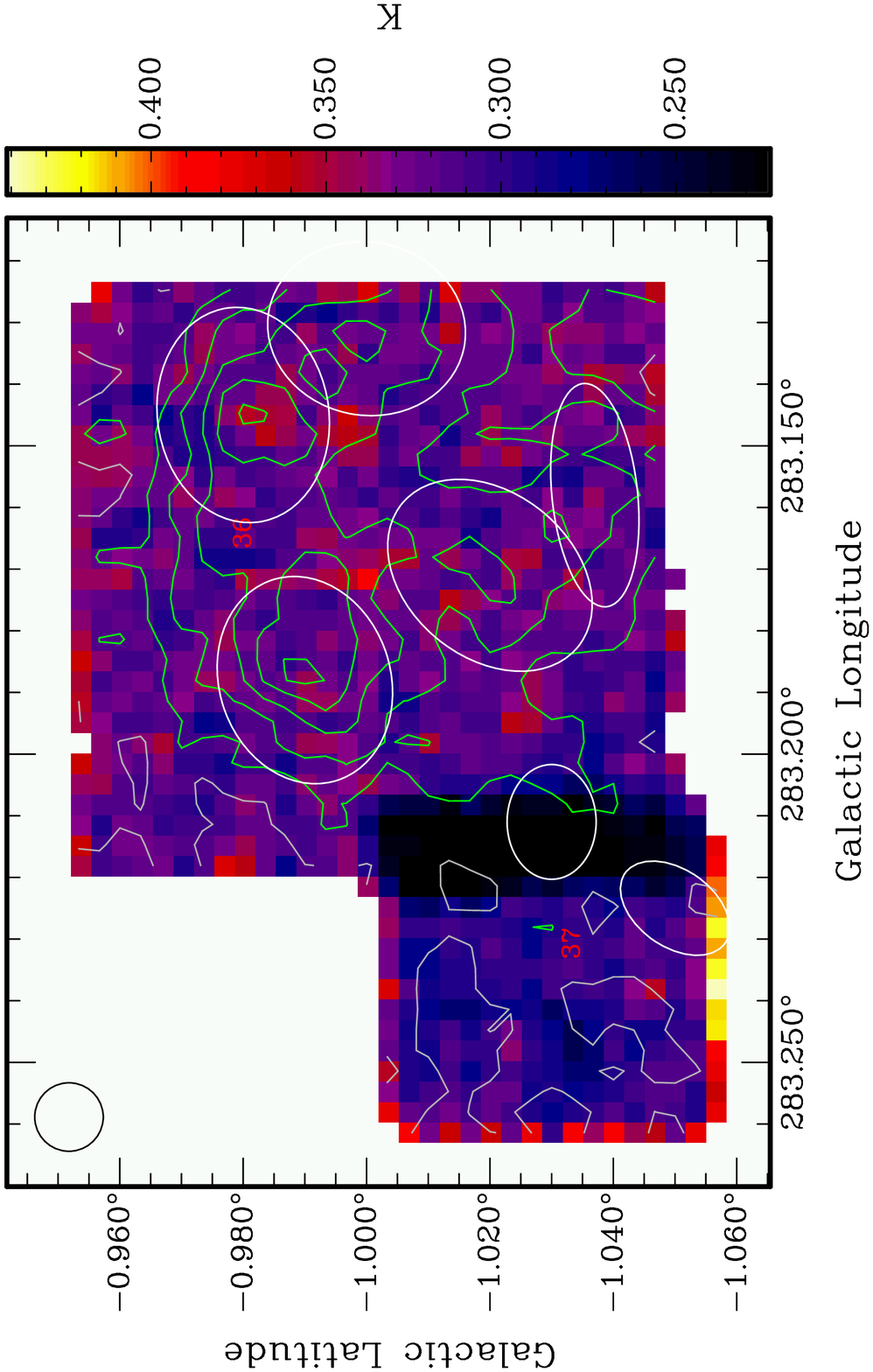}}}
\centerline{(c){\includegraphics[angle=-90,scale=0.3]{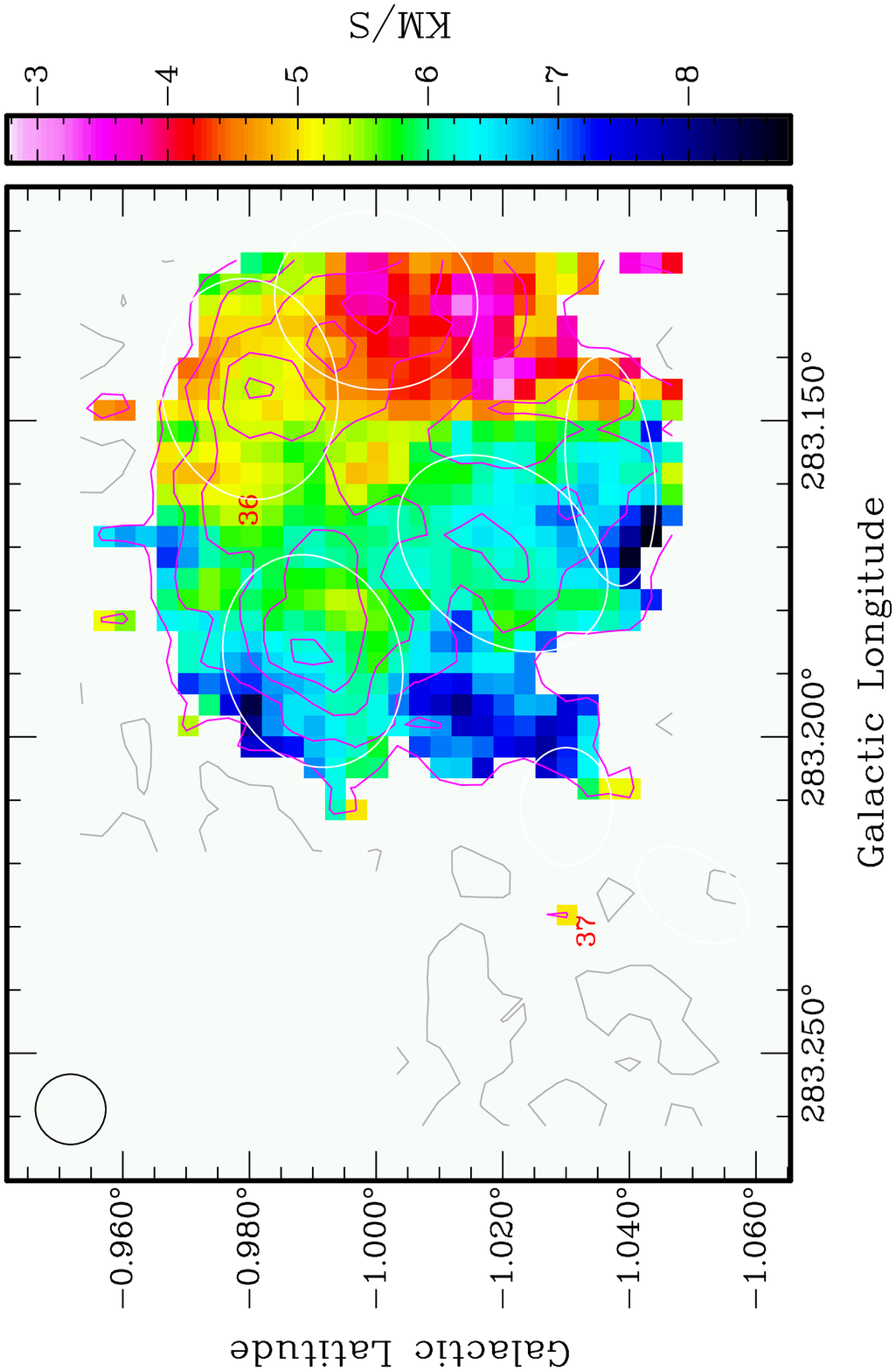}}
		(d){\includegraphics[angle=-90,scale=0.3]{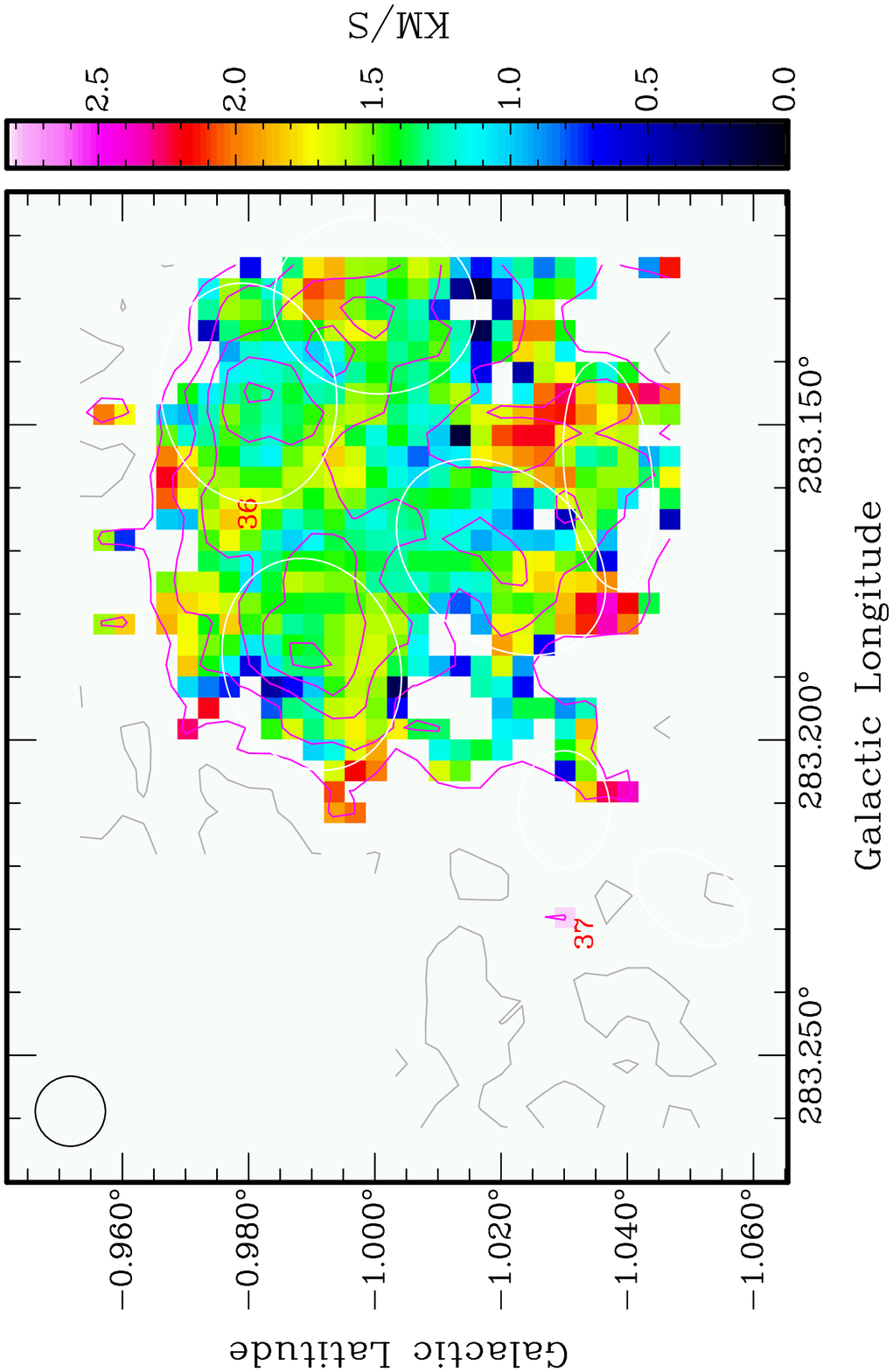}}}
\caption{\small Same as Fig.\,\ref{momR1}, but for Region 5 source BYF\,36.  Contours are every 3$\sigma$ = 0.891\,K\kms\, and at 3.2\,kpc the 40$''$ Mopra beam (upper left corner) scales to 0.621\,pc.  ($a$) $T_p$,  ($b$) rms,  ($c$) $V_{\rm LSR}$,  ($d$) $\sigma_{V}$.
\label{momR5a}}
\end{figure*}

\clearpage

\begin{figure*}[htp]
\centerline{(a){\includegraphics[angle=-90,scale=0.3]{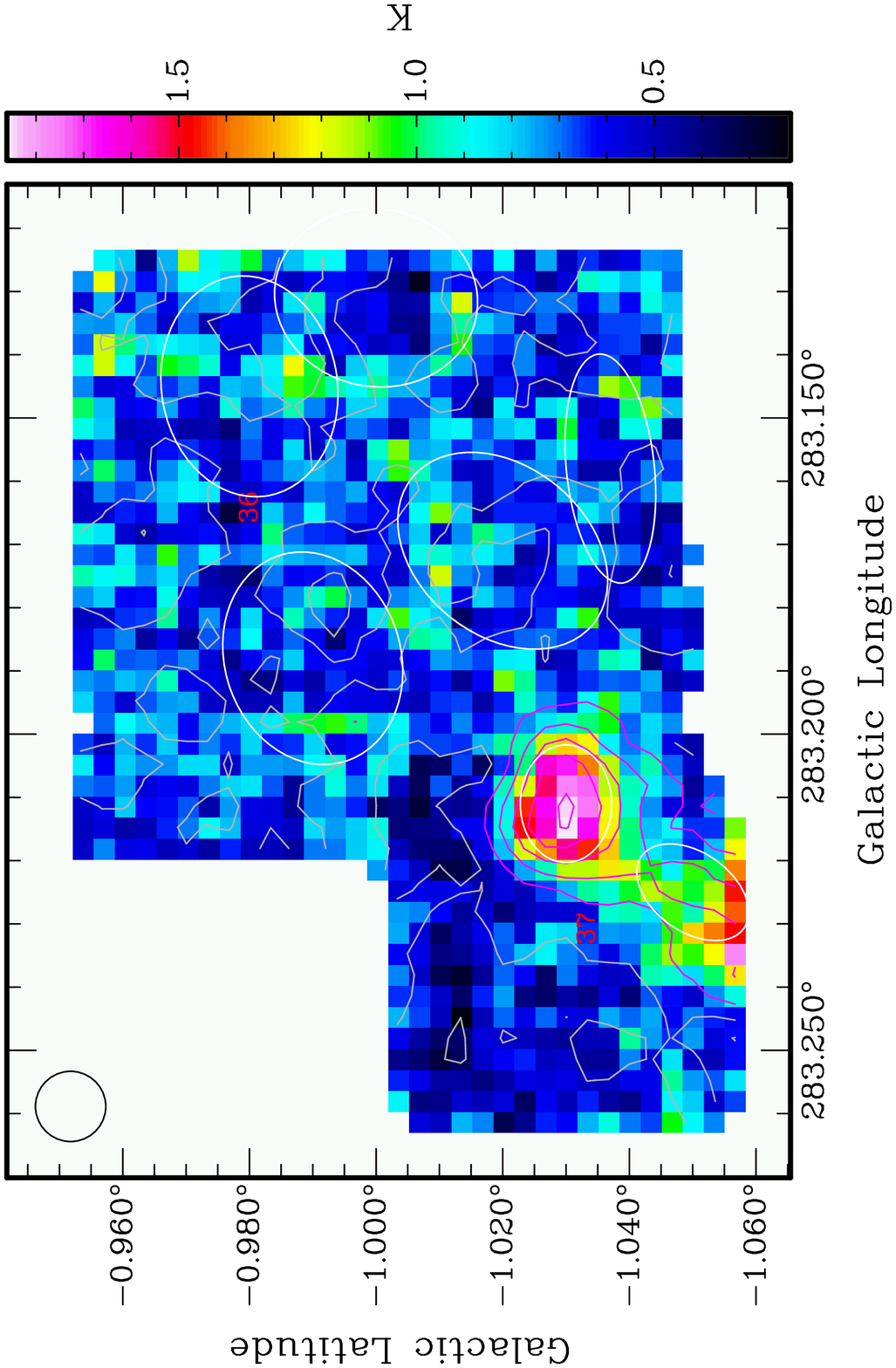}}
		(b){\includegraphics[angle=-90,scale=0.3]{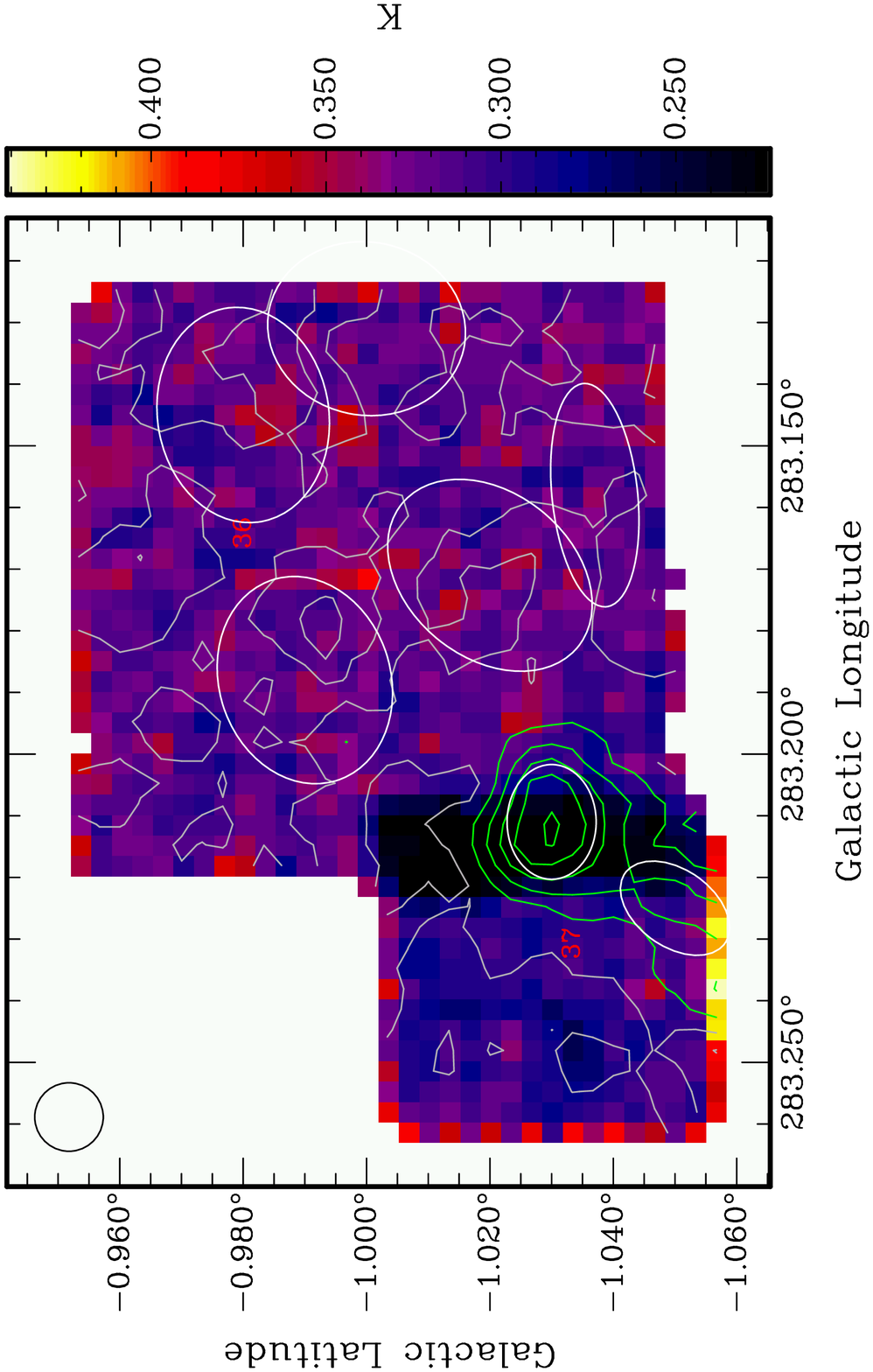}}}
\centerline{(c){\includegraphics[angle=-90,scale=0.3]{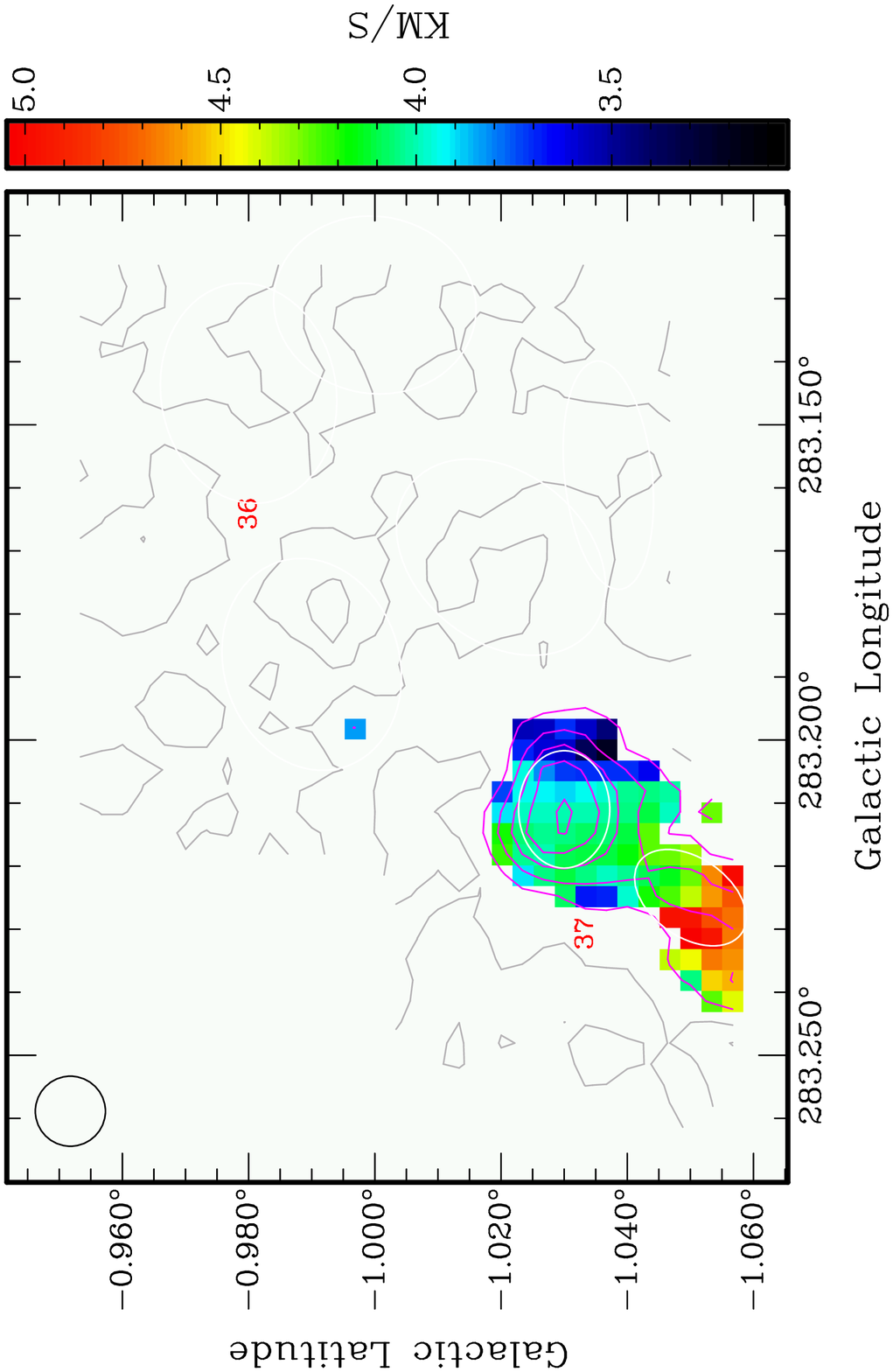}}
		(d){\includegraphics[angle=-90,scale=0.3]{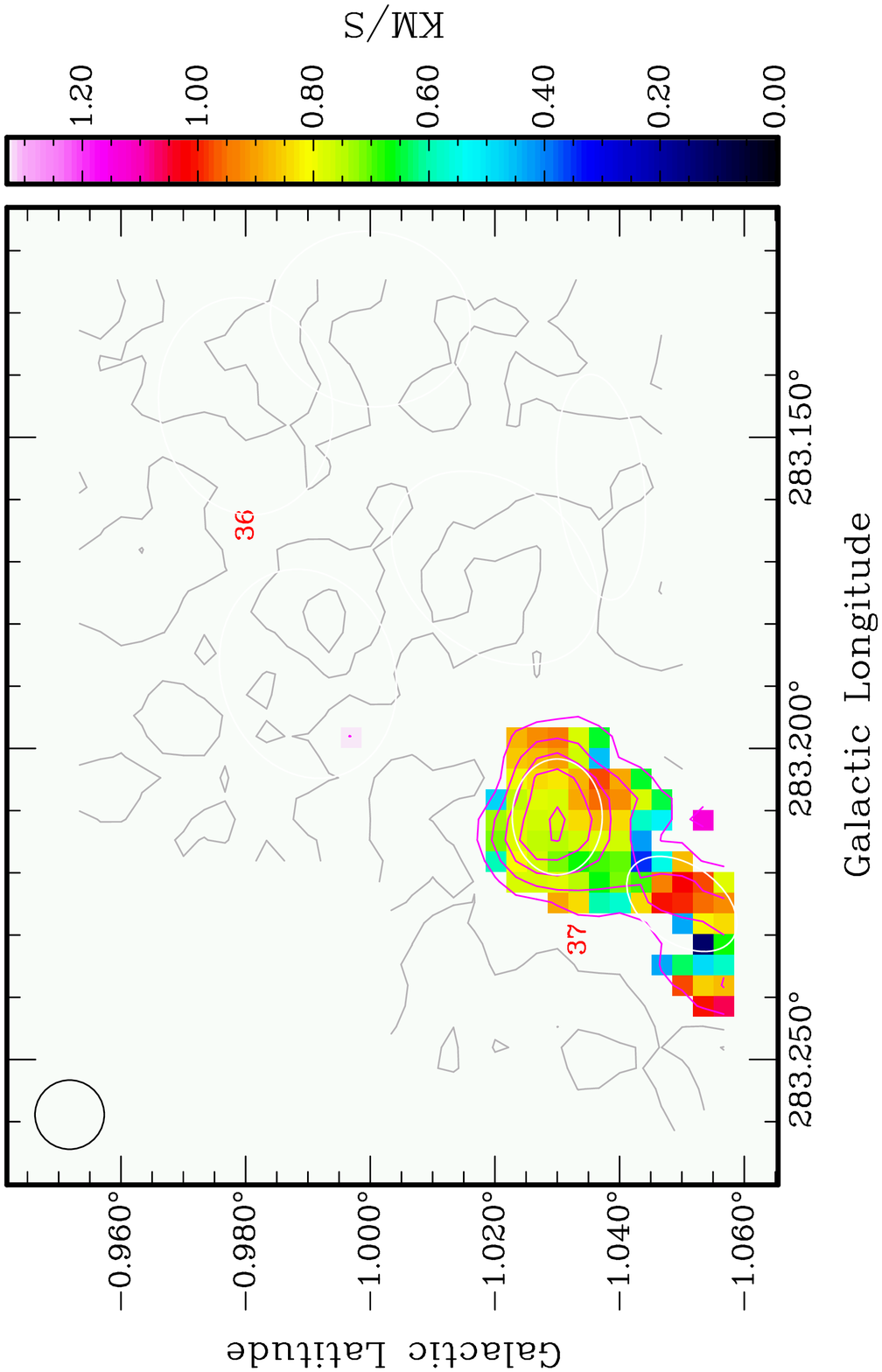}}}
\caption{\small Same as Fig.\,\ref{momR1}, but for Region 5 source BYF\,37.  Contours are every 3$\sigma$ = 0.624\,K\kms\, and at 3.2\,kpc the 40$''$ Mopra beam (upper left corner) scales to 0.621\,pc.  ($a$) $T_p$,  ($b$) rms,  ($c$) $V_{\rm LSR}$,  ($d$) $\sigma_{V}$.
\label{momR5b}}
\end{figure*}

\clearpage

\clearpage

\begin{figure*}[htp]
\centerline{(a){\includegraphics[angle=-90,scale=0.3]{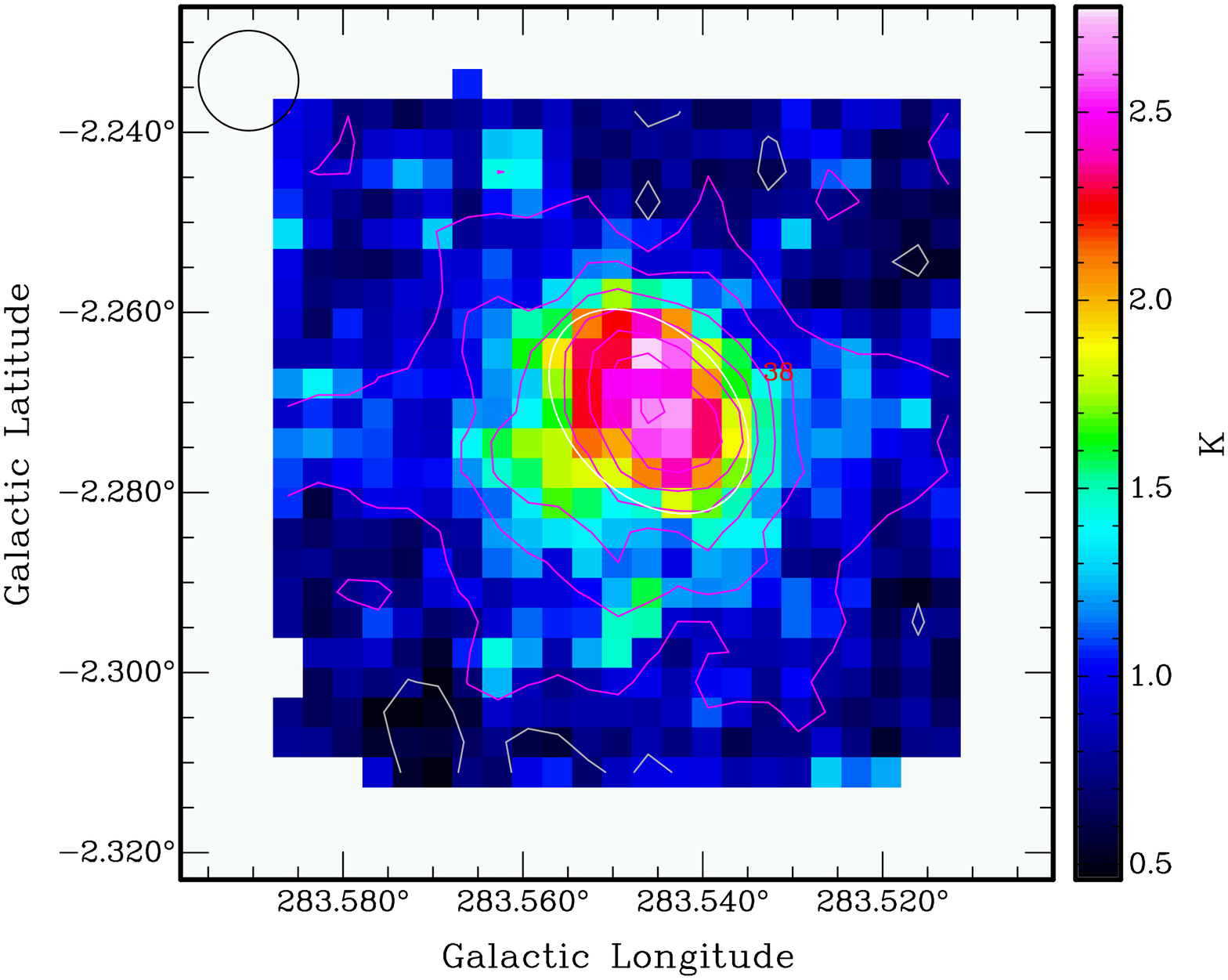}}
		(b){\includegraphics[angle=-90,scale=0.3]{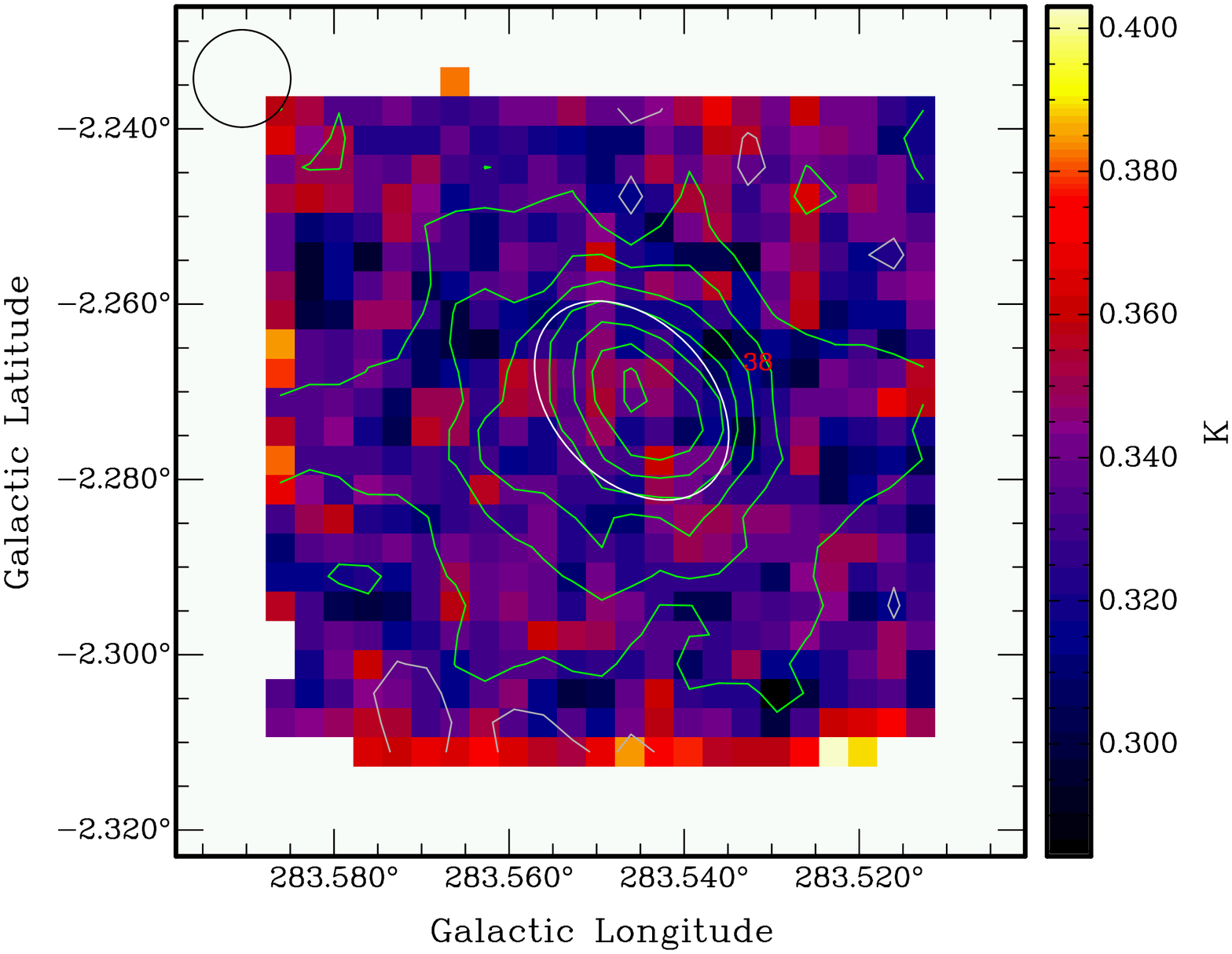}}}
\centerline{(c){\includegraphics[angle=-90,scale=0.3]{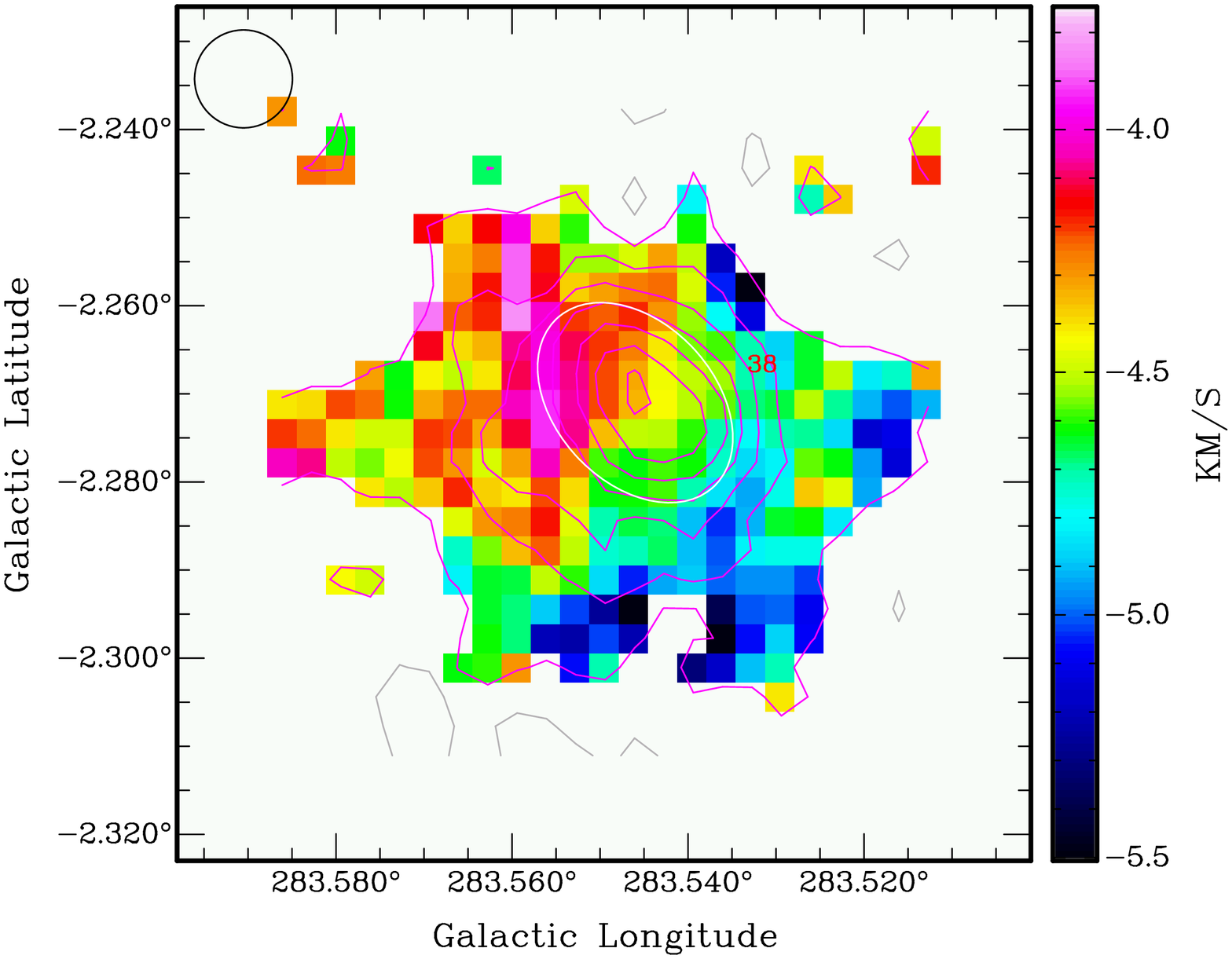}}
		(d){\includegraphics[angle=-90,scale=0.3]{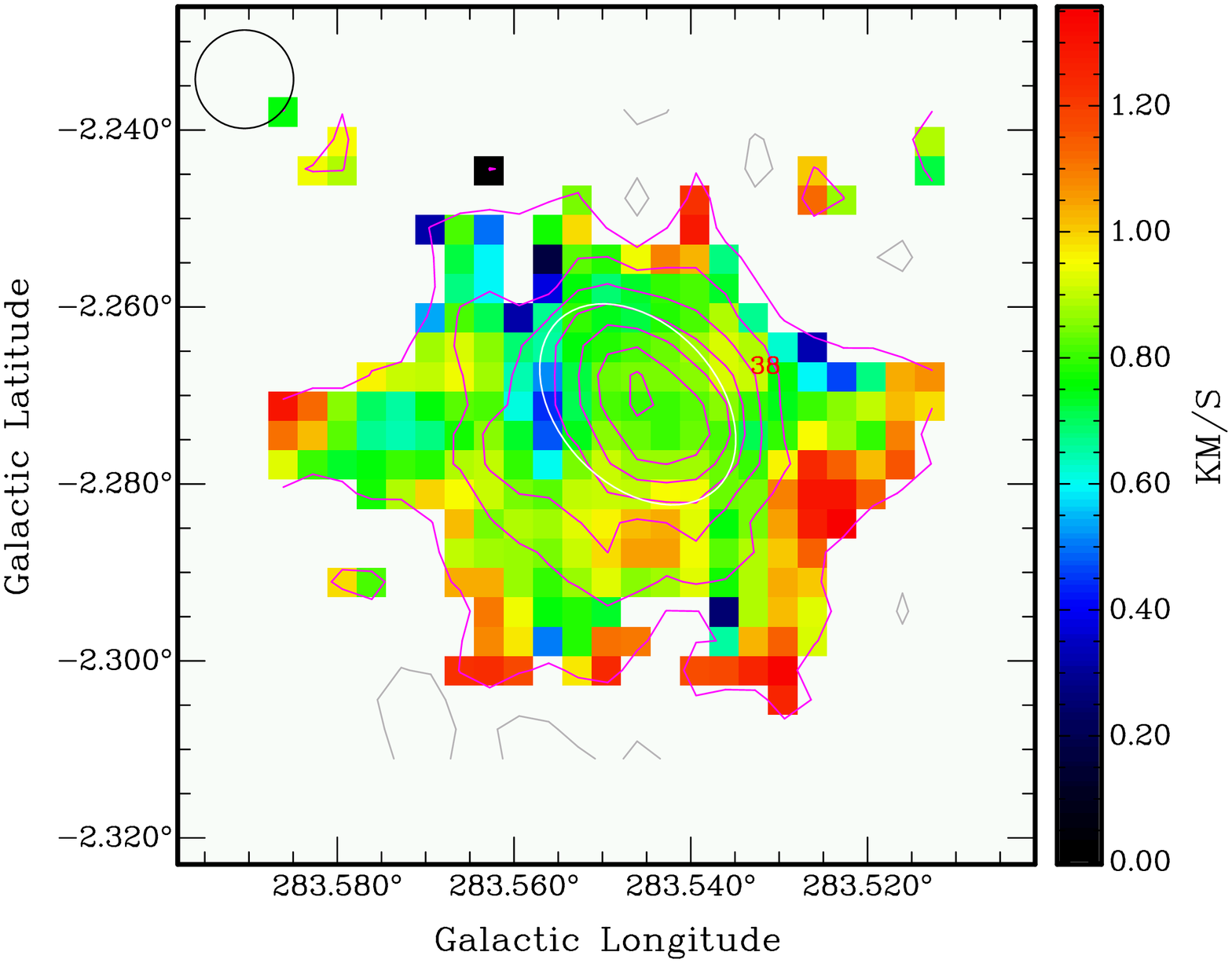}}}
\caption{\small Same as Fig.\,\ref{momR1}, but for isolated source BYF\,38.  Contours are every 3$\sigma$ = 0.657\,K\kms, and at 2.0\,kpc the 40$''$ Mopra beam (upper left corner) scales to 0.388\,pc.  ($a$) $T_p$,  ($b$) rms,  ($c$) $V_{\rm LSR}$,  ($d$) $\sigma_{V}$.
\label{momBYF38}}
\end{figure*}

\clearpage

\begin{figure*}[htp]
\centerline{(a){\includegraphics[angle=0,scale=0.35]{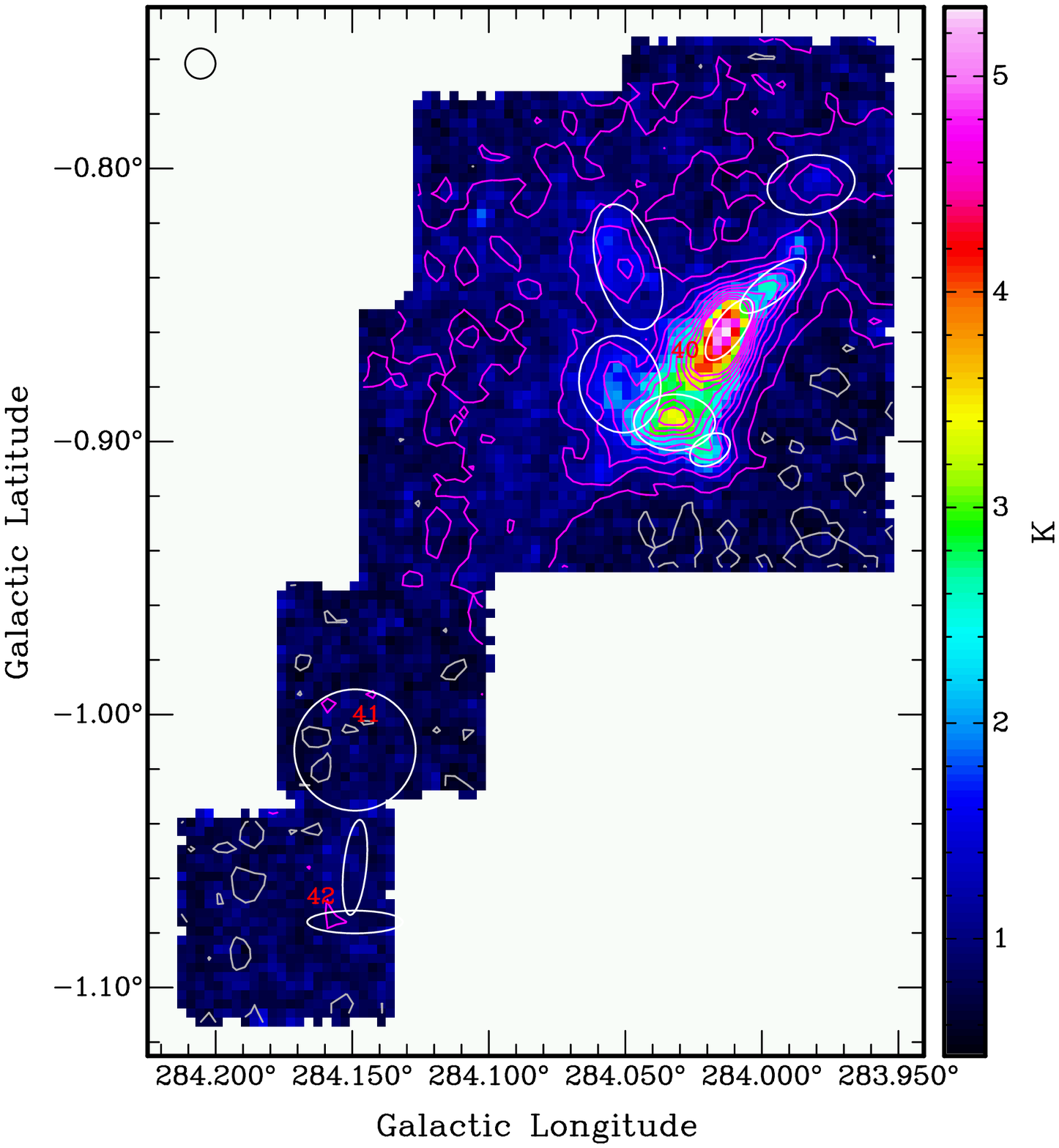}}
		(b){\includegraphics[angle=0,scale=0.37]{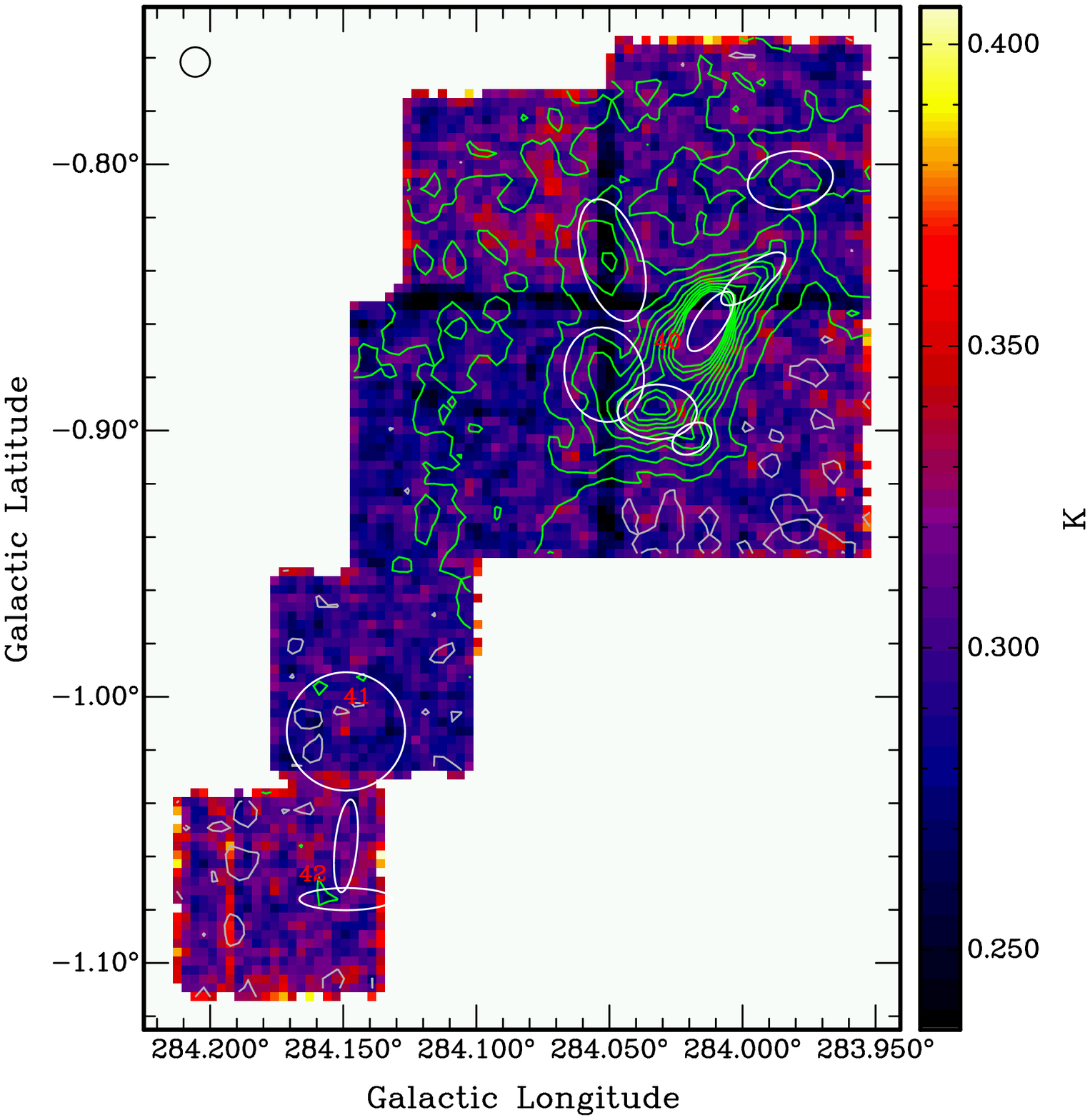}}}
\centerline{(c){\includegraphics[angle=0,scale=0.35]{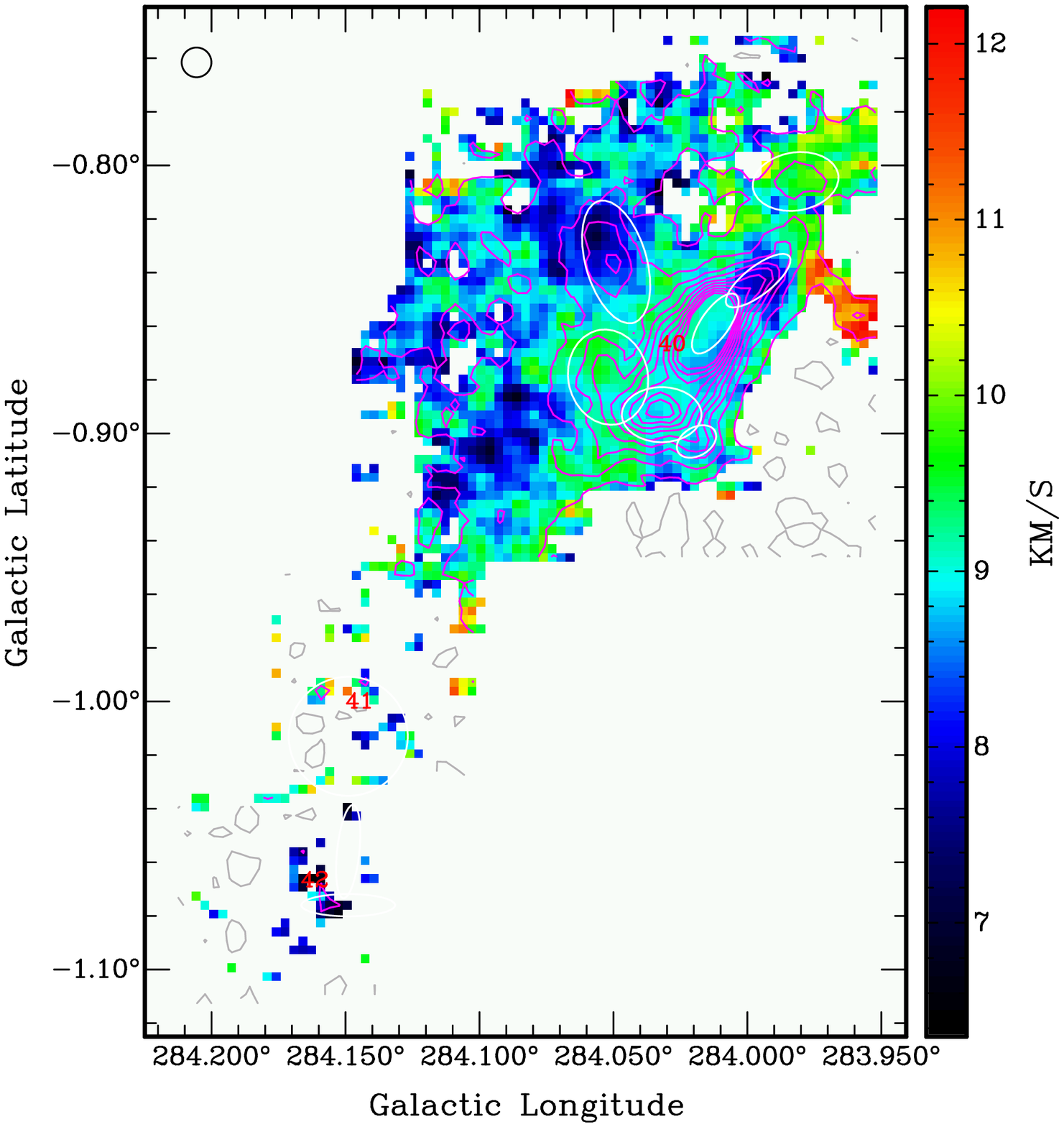}}
		(d){\includegraphics[angle=0,scale=0.35]{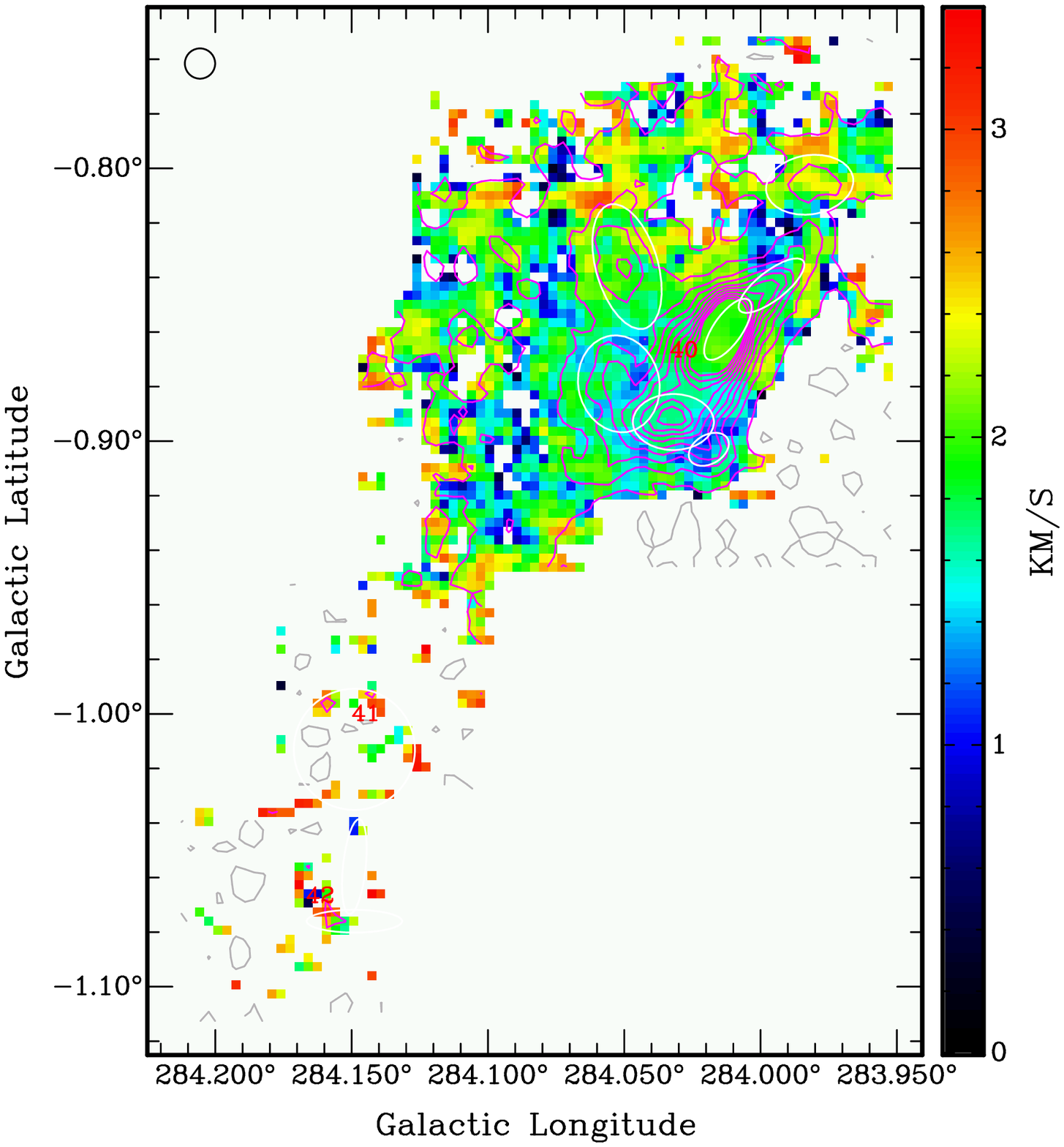}}}
\caption{\small Same as Fig.\,\ref{momR1}, but for Region 6 source BYF\,40.  Contours are every 4$\sigma$ = 1.248\,K\kms\, and at 6.6\,kpc the 40$''$ Mopra beam (upper left corner) scales to 1.280\,pc.  ($a$) $T_p$,  ($b$) rms,  ($c$) $V_{\rm LSR}$,  ($d$) $\sigma_{V}$.
\label{momR6a}}
\end{figure*}

\clearpage

\begin{figure*}[htp]
\centerline{(a){\includegraphics[angle=0,scale=0.35]{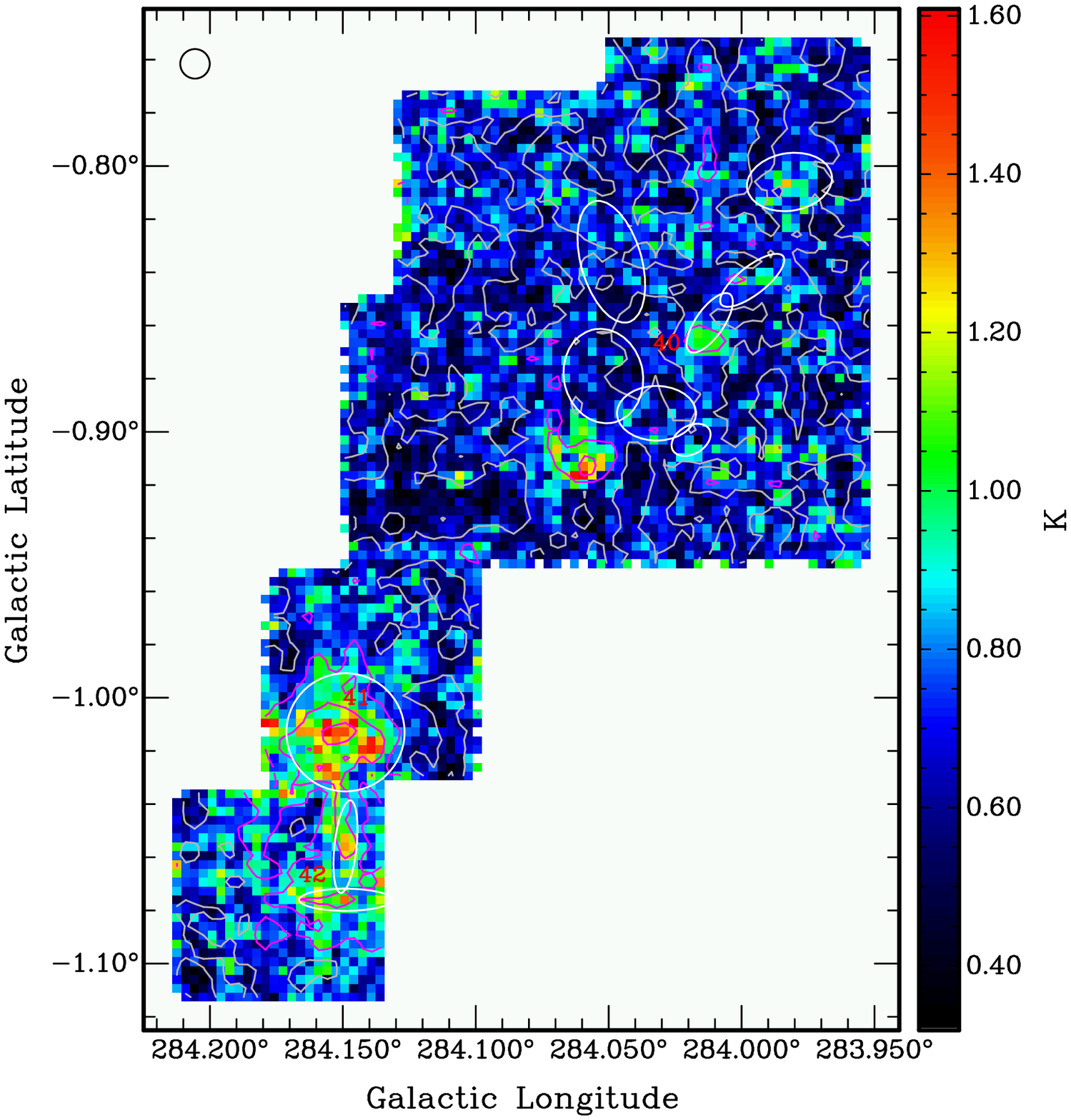}}
		(b){\includegraphics[angle=0,scale=0.36]{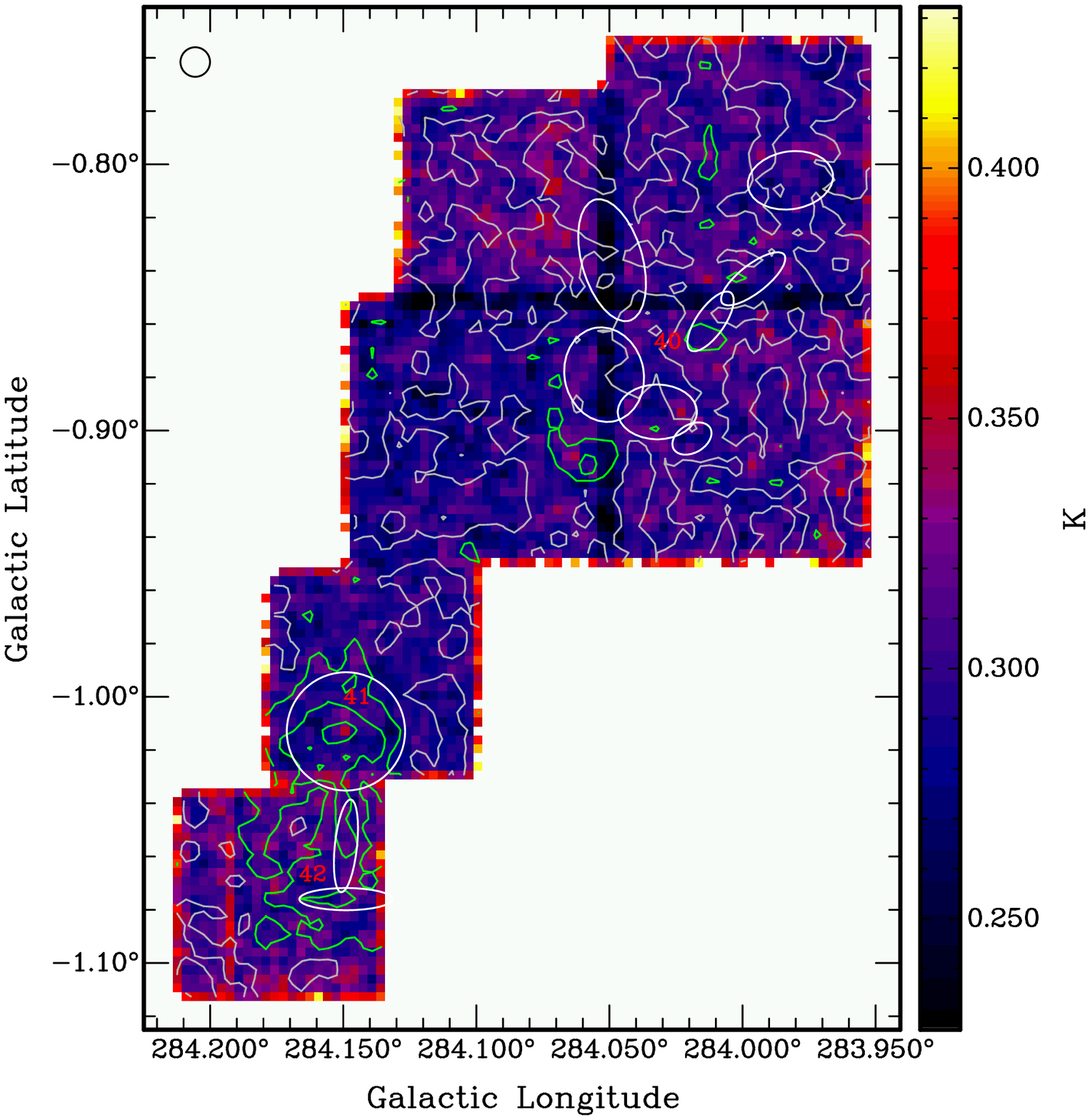}}}
\centerline{(c){\includegraphics[angle=0,scale=0.35]{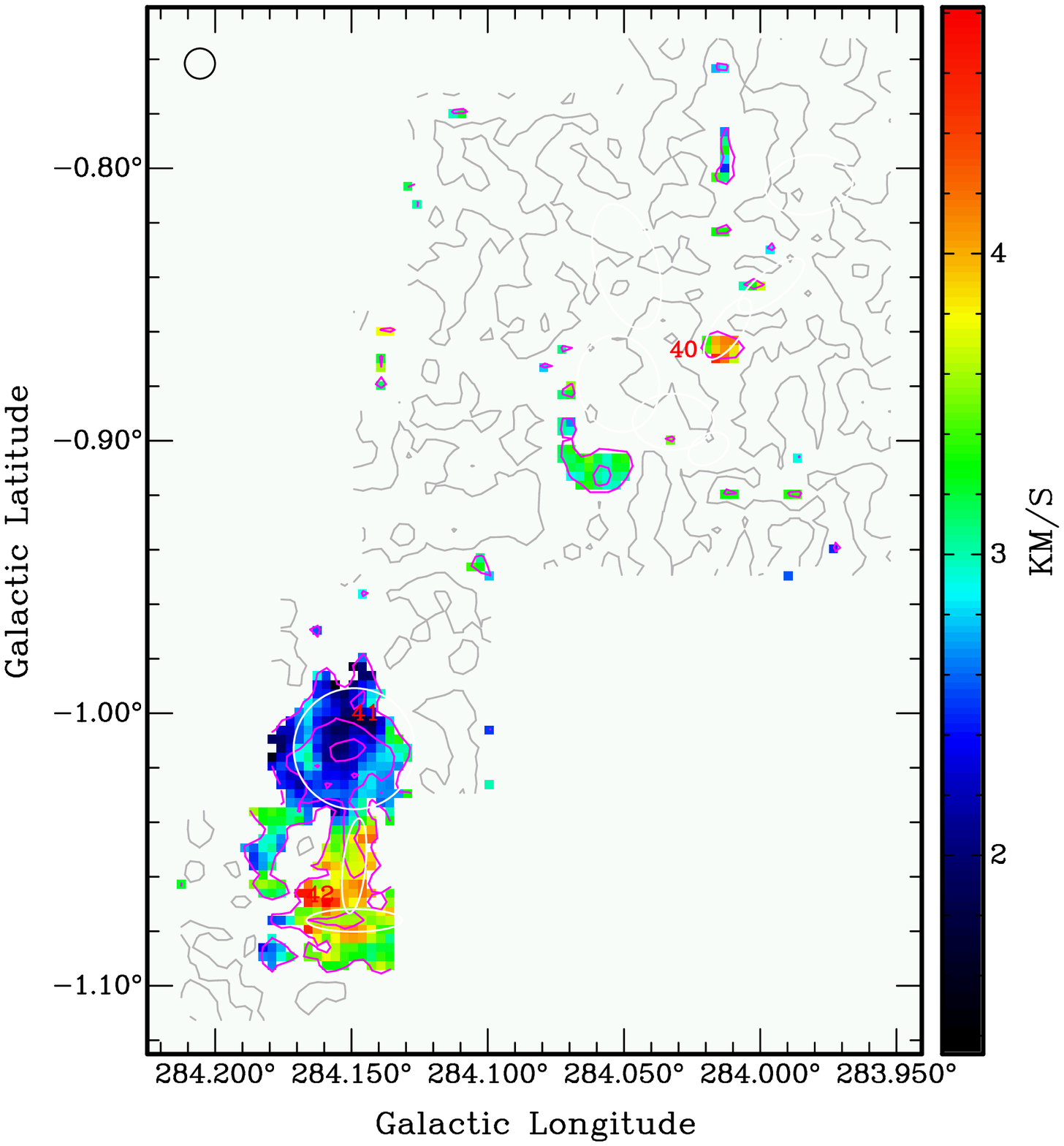}}
		(d){\includegraphics[angle=0,scale=0.35]{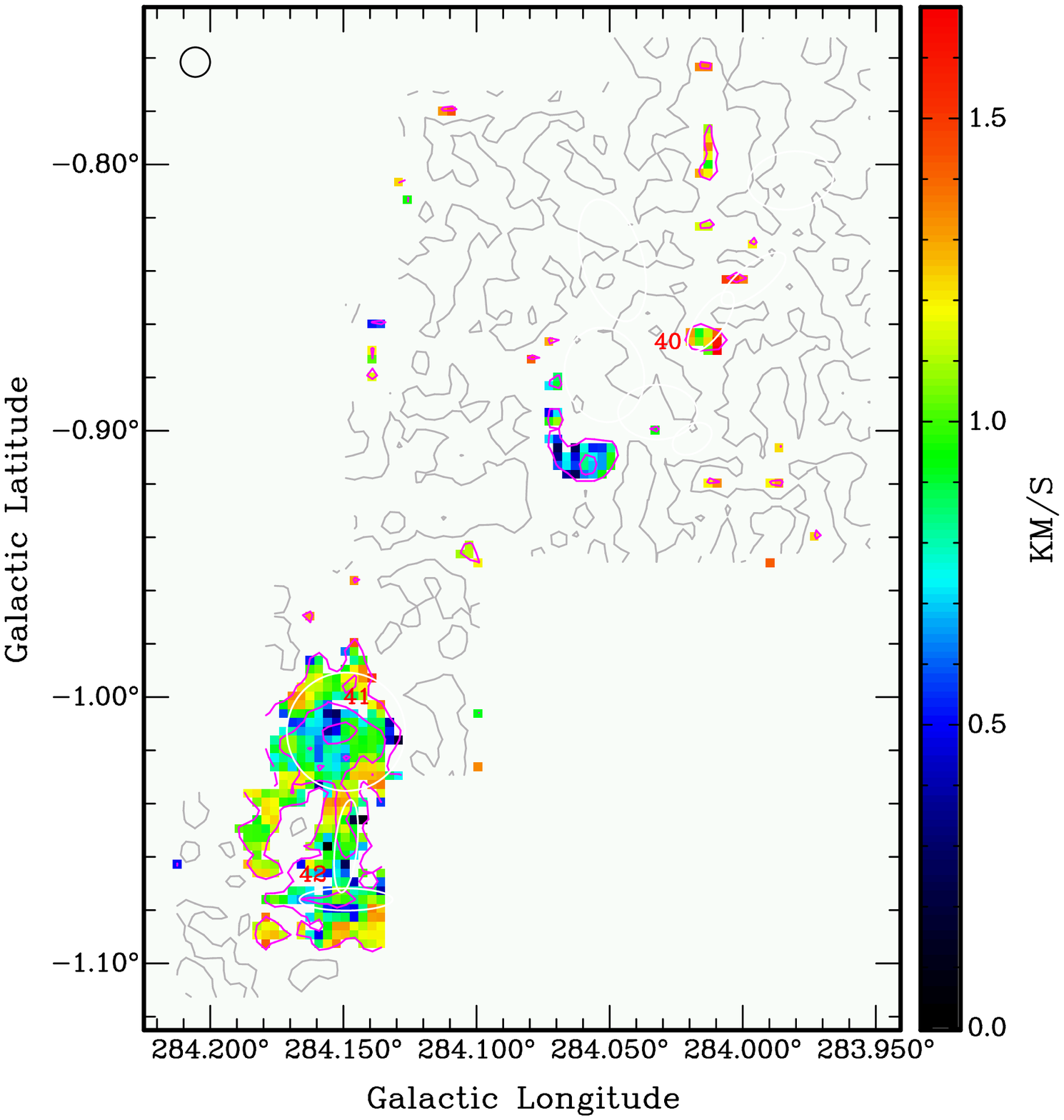}}}
\caption{\small Same as Fig.\,\ref{momR1}, but for Region 6 sources BYF\,41 and 42.  Contours are every 3$\sigma$ = 0.639\,K\kms\, and at 6.6\,kpc the 40$''$ Mopra beam (upper left corner) scales to 1.280\,pc.  ($a$) $T_p$,  ($b$) rms,  ($c$) $V_{\rm LSR}$,  ($d$) $\sigma_{V}$.
\label{momR6b}}
\end{figure*}

\clearpage

\begin{figure*}[htp]
\centerline{(a){\includegraphics[angle=-90,scale=0.25]{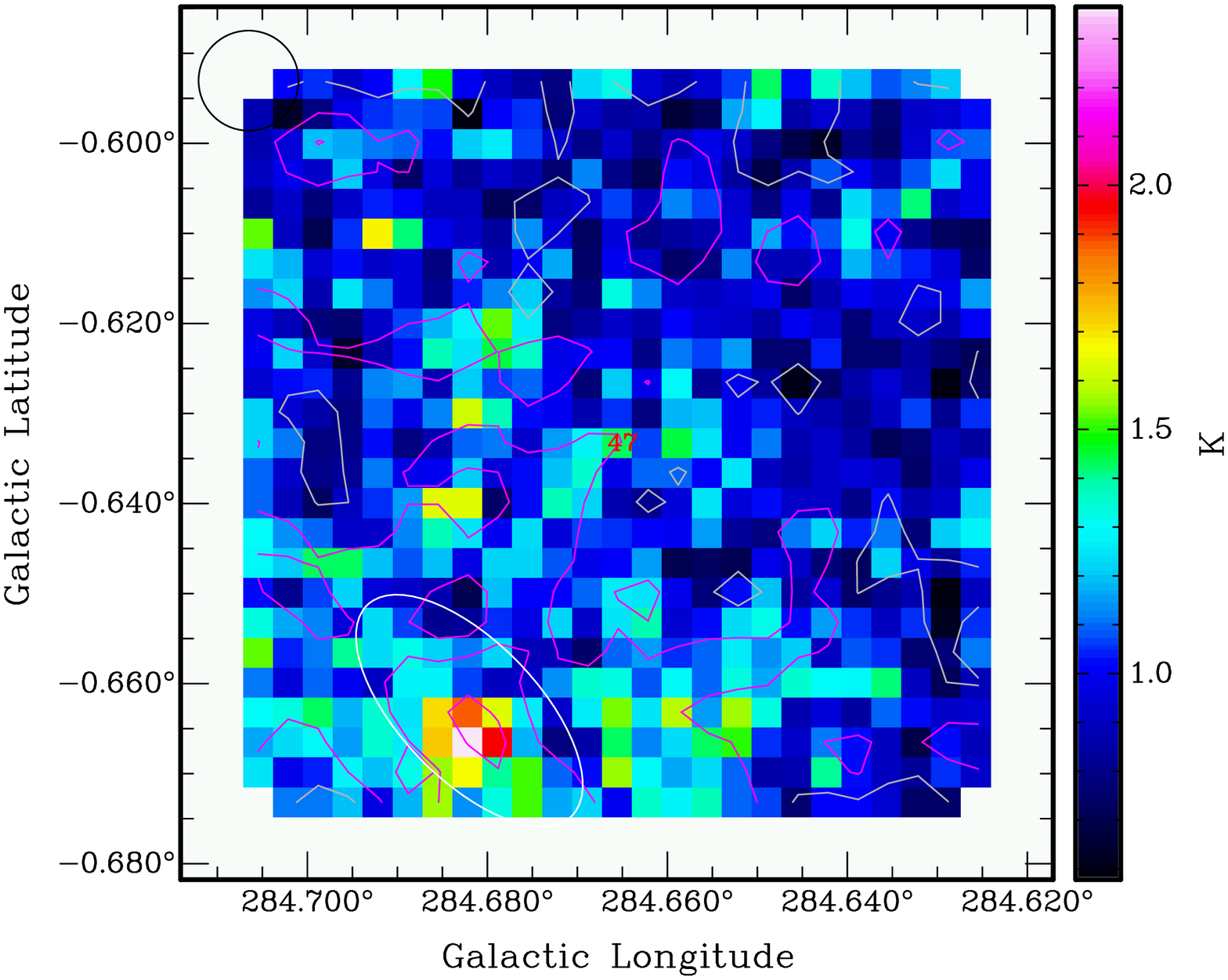}}
		(b){\includegraphics[angle=-90,scale=0.25]{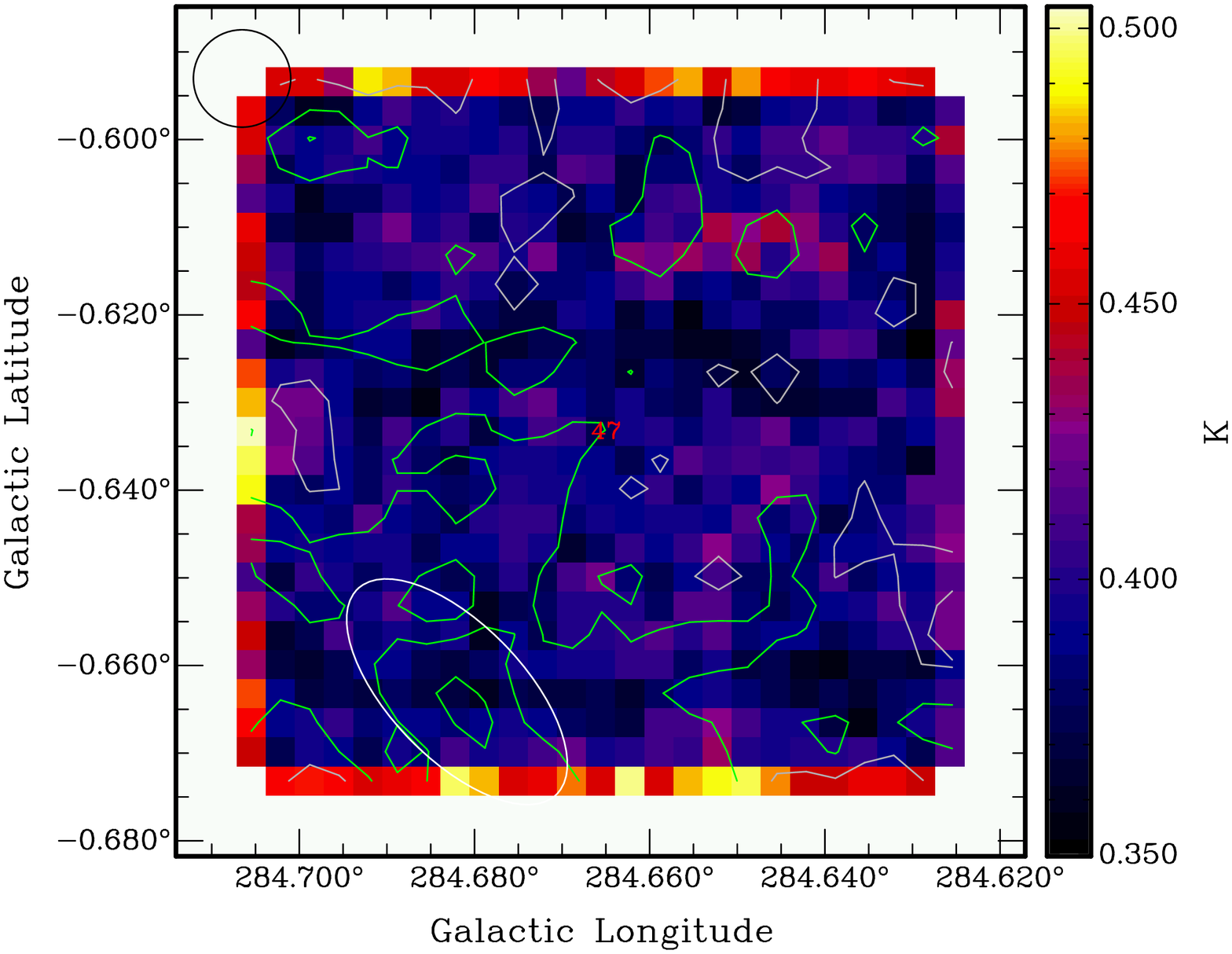}}}
\centerline{(c){\includegraphics[angle=-90,scale=0.25]{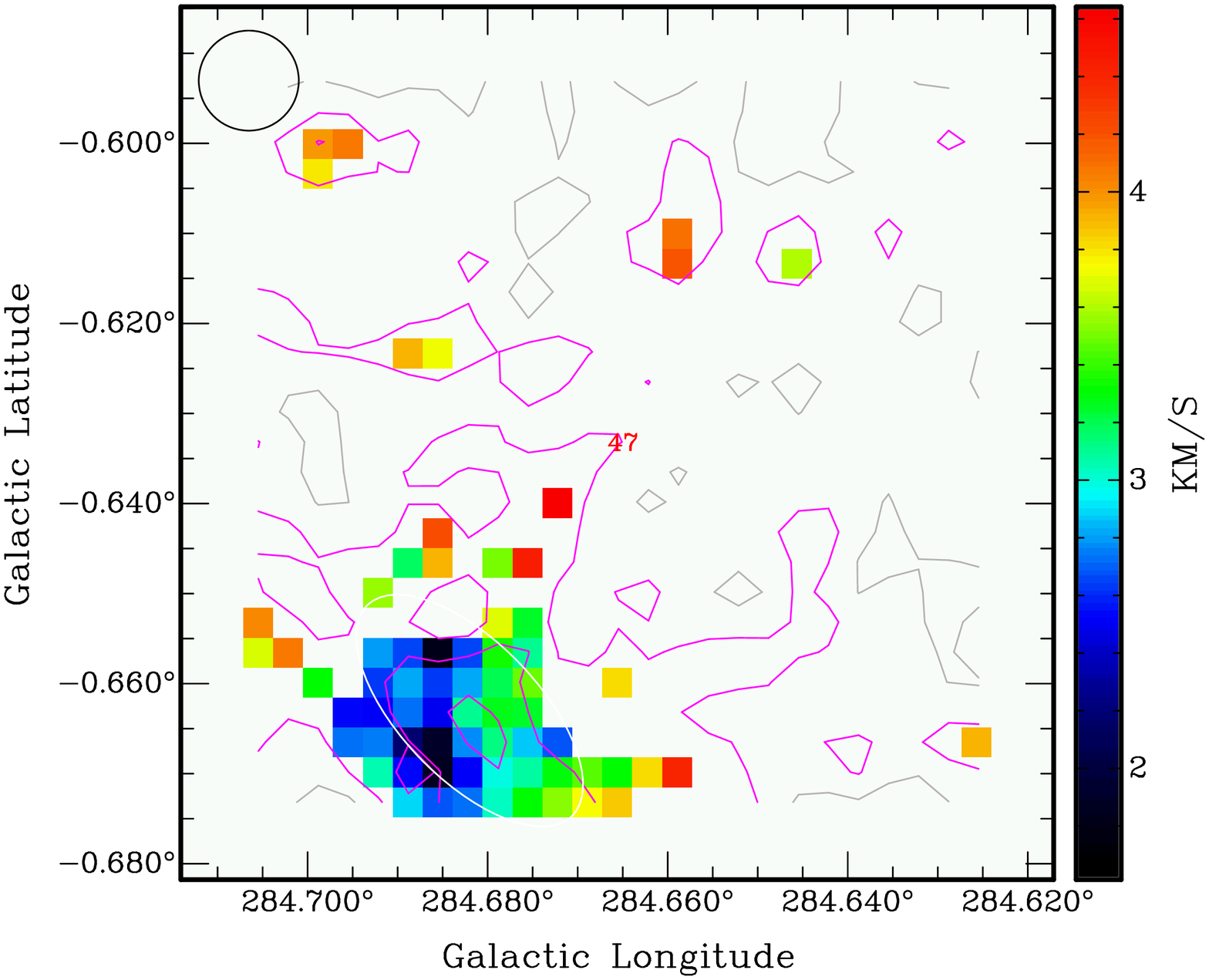}}
		(d){\includegraphics[angle=-90,scale=0.25]{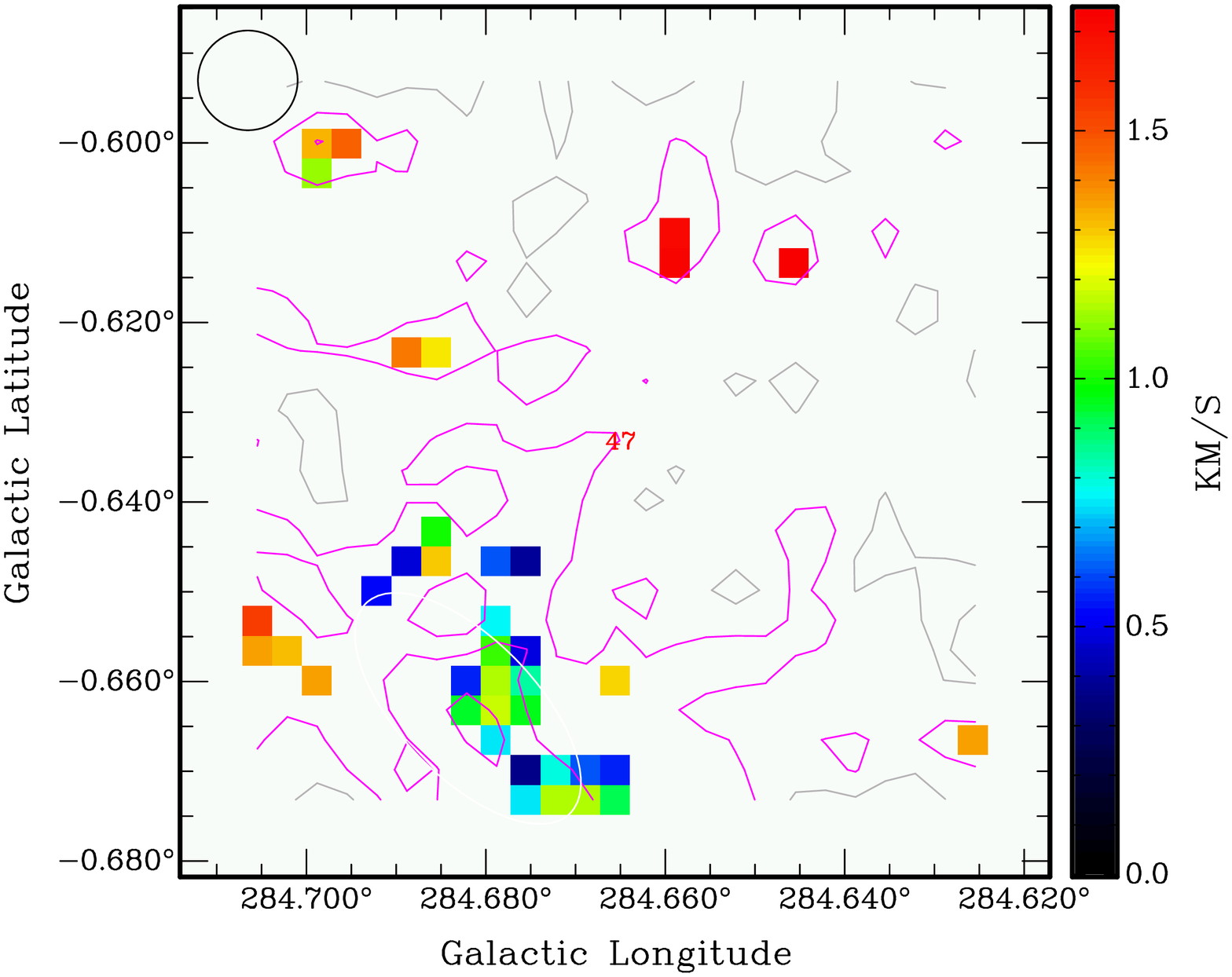}}}
\caption{\small Same as Fig.\,\ref{momR1}, but for Region 7 source BYF\,47.  Contours are every 2$\sigma$ = 0.636\,K\kms\, and at 5.3\,kpc the 40$''$ Mopra beam (upper left corner) scales to 1.028\,pc.  ($a$) $T_p$,  ($b$) rms,  ($c$) $V_{\rm LSR}$,  ($d$) $\sigma_{V}$.
\label{momR7}}
\end{figure*}

\clearpage

\begin{figure*}[htp]
(a){\includegraphics[angle=-90,scale=0.37]{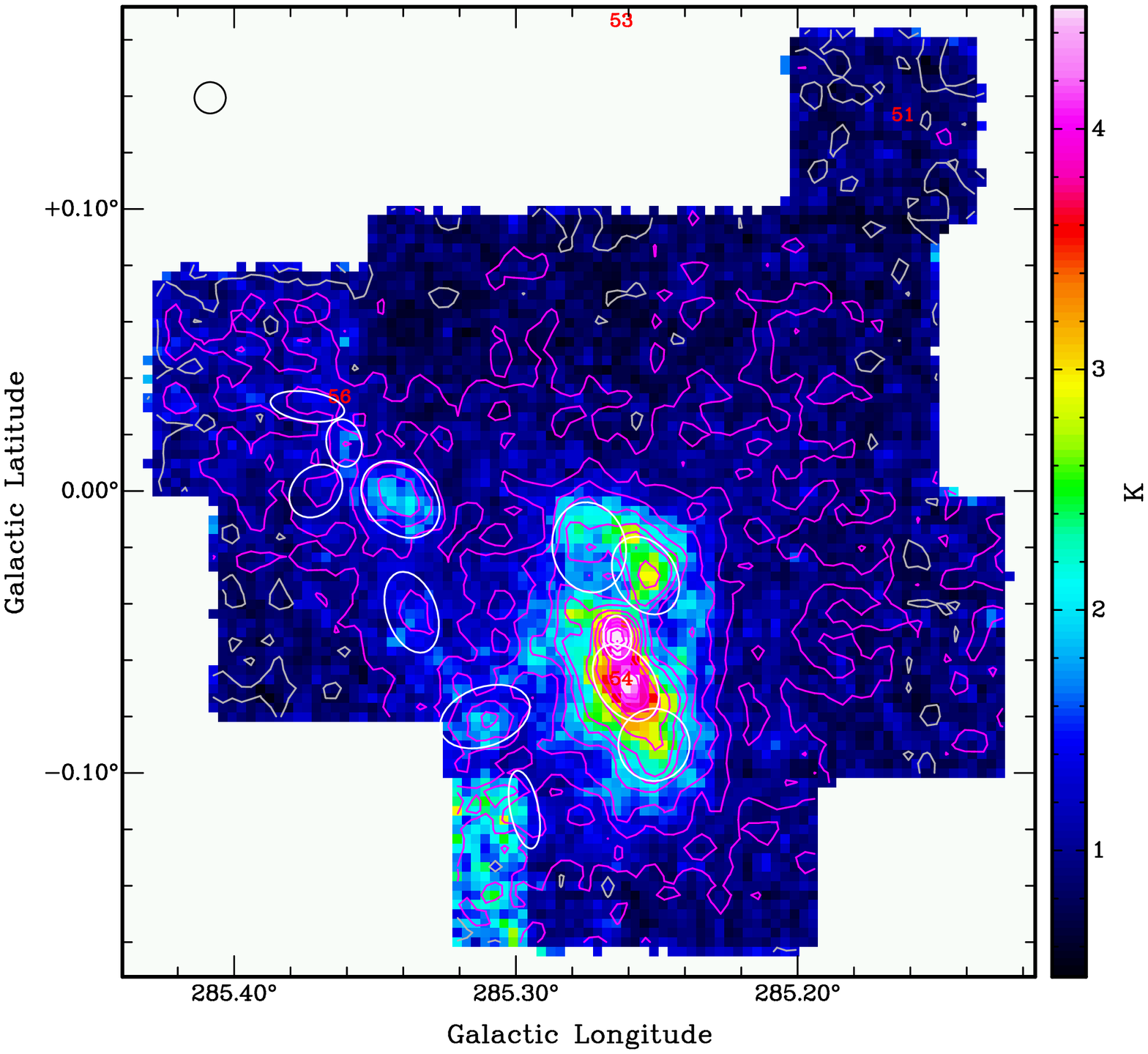}}
(b){\includegraphics[angle=-90,scale=0.37]{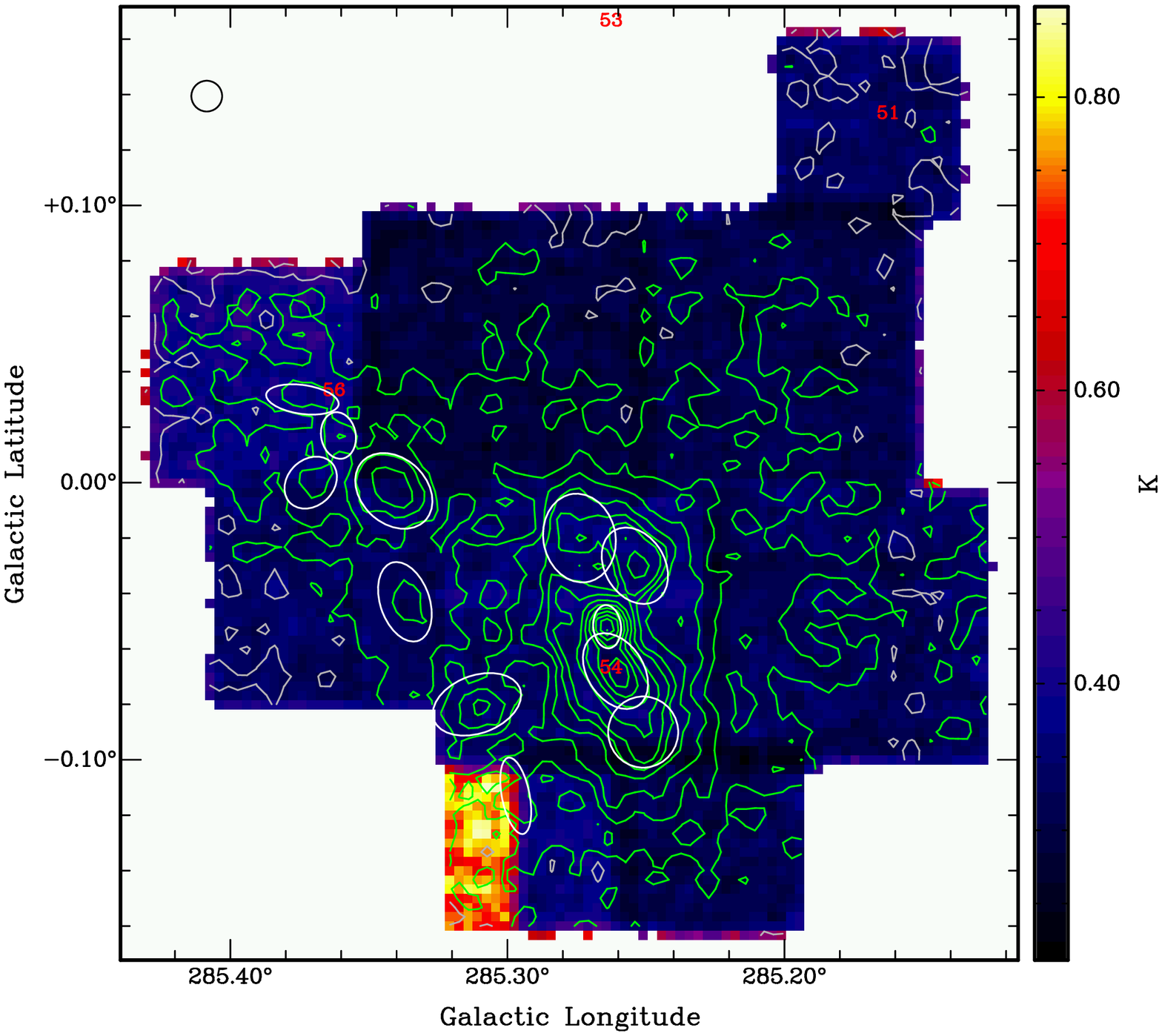}}
(c){\includegraphics[angle=-90,scale=0.37]{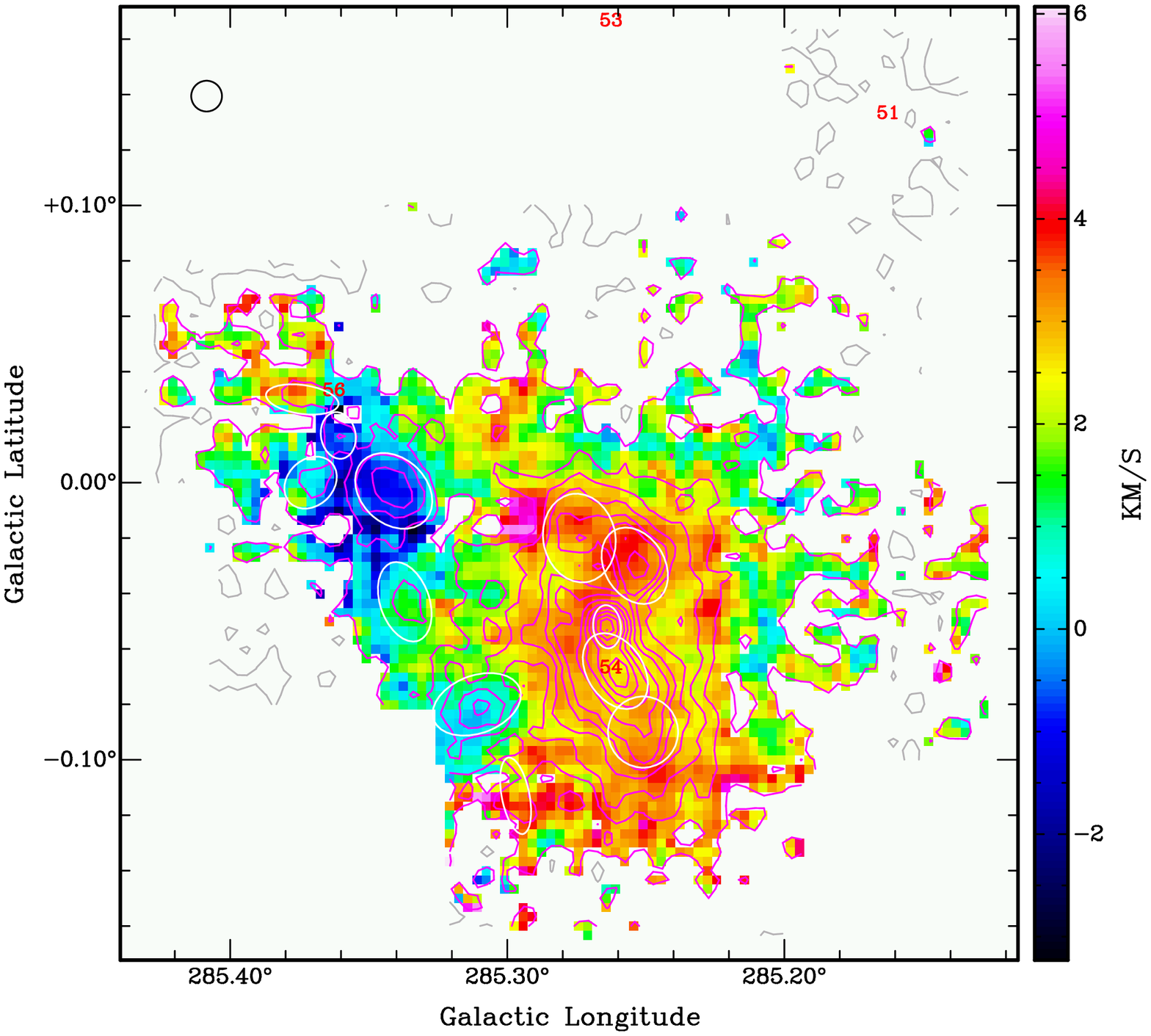}}
(d){\includegraphics[angle=-90,scale=0.37]{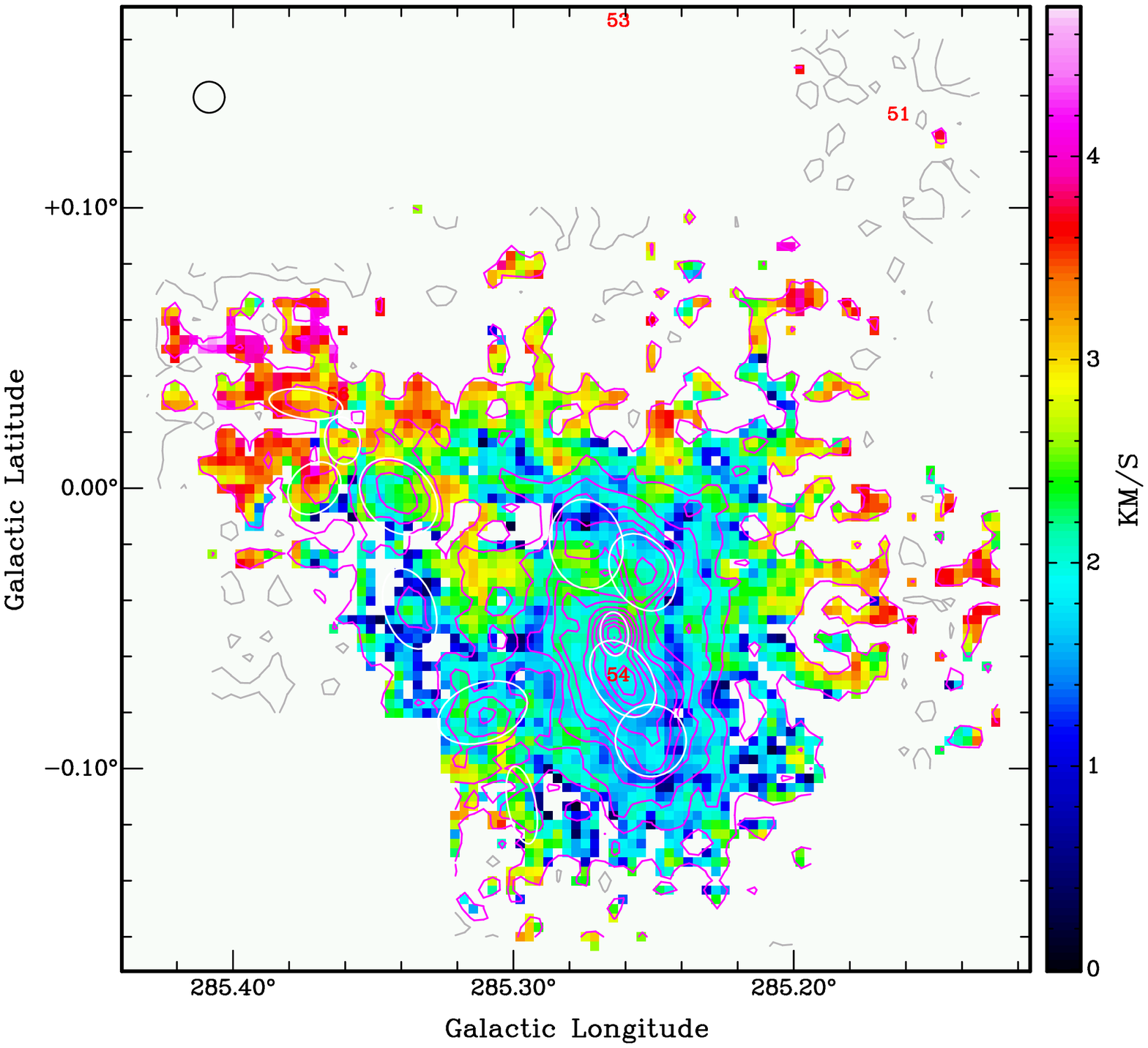}}
\caption{\small Same as Fig.\,\ref{momR1}, but for Region 8 sources BYF\,51--56.  Contours are every 3$\sigma$ = 1.203\,K\kms\ up to 18$\sigma$, and then every 6$\sigma$; at 5.3\,kpc the 40$''$ Mopra beam (upper left corner) scales to 1.028\,pc.  ($a$) $T_p$,  ($b$) rms,  ($c$) $V_{\rm LSR}$,  ($d$) $\sigma_{V}$.
\label{momR8}}
\end{figure*}

\clearpage

\begin{figure*}[htp]
(a){\includegraphics[angle=-90,scale=0.40]{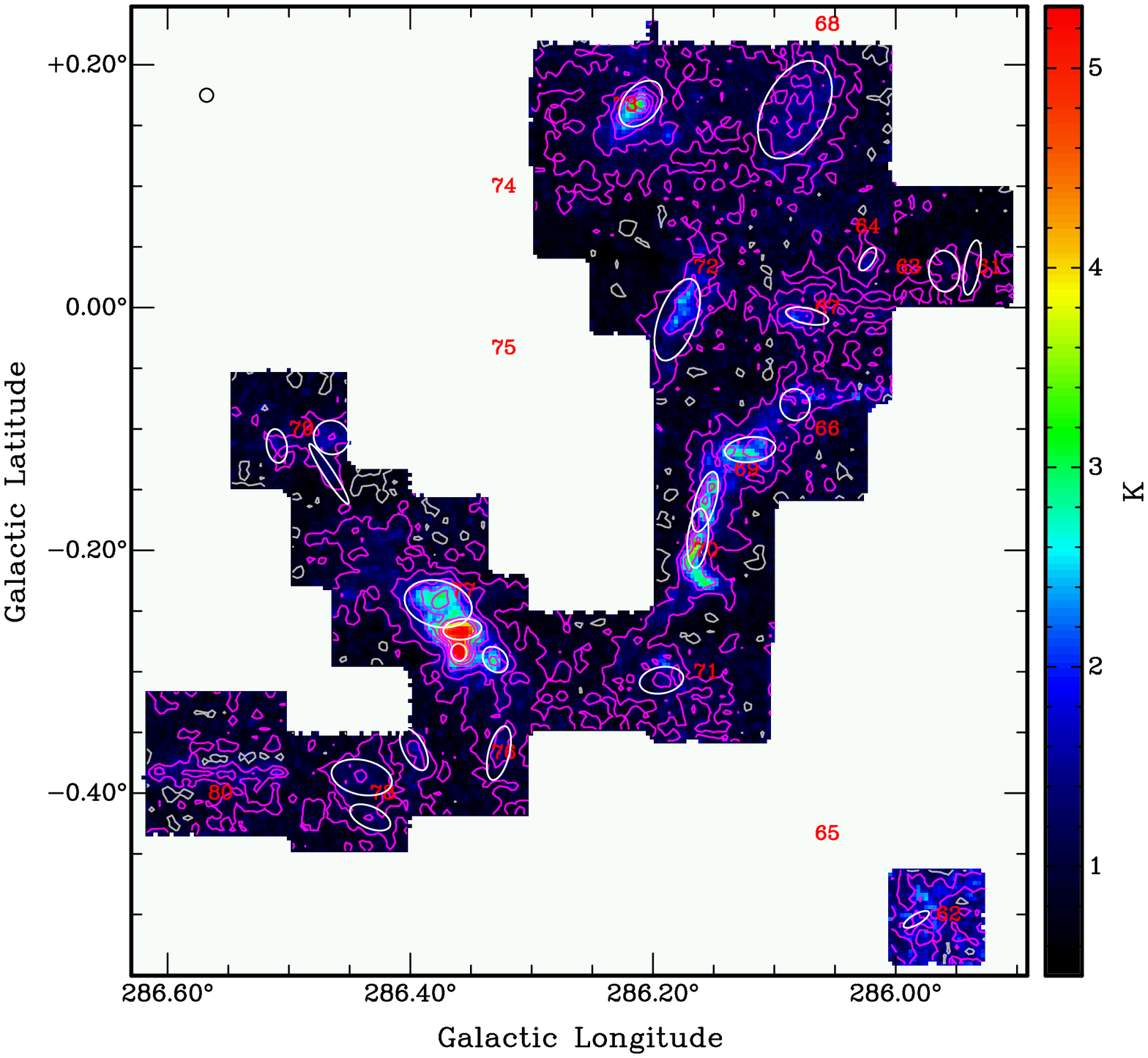}}
(b){\includegraphics[angle=-90,scale=0.40]{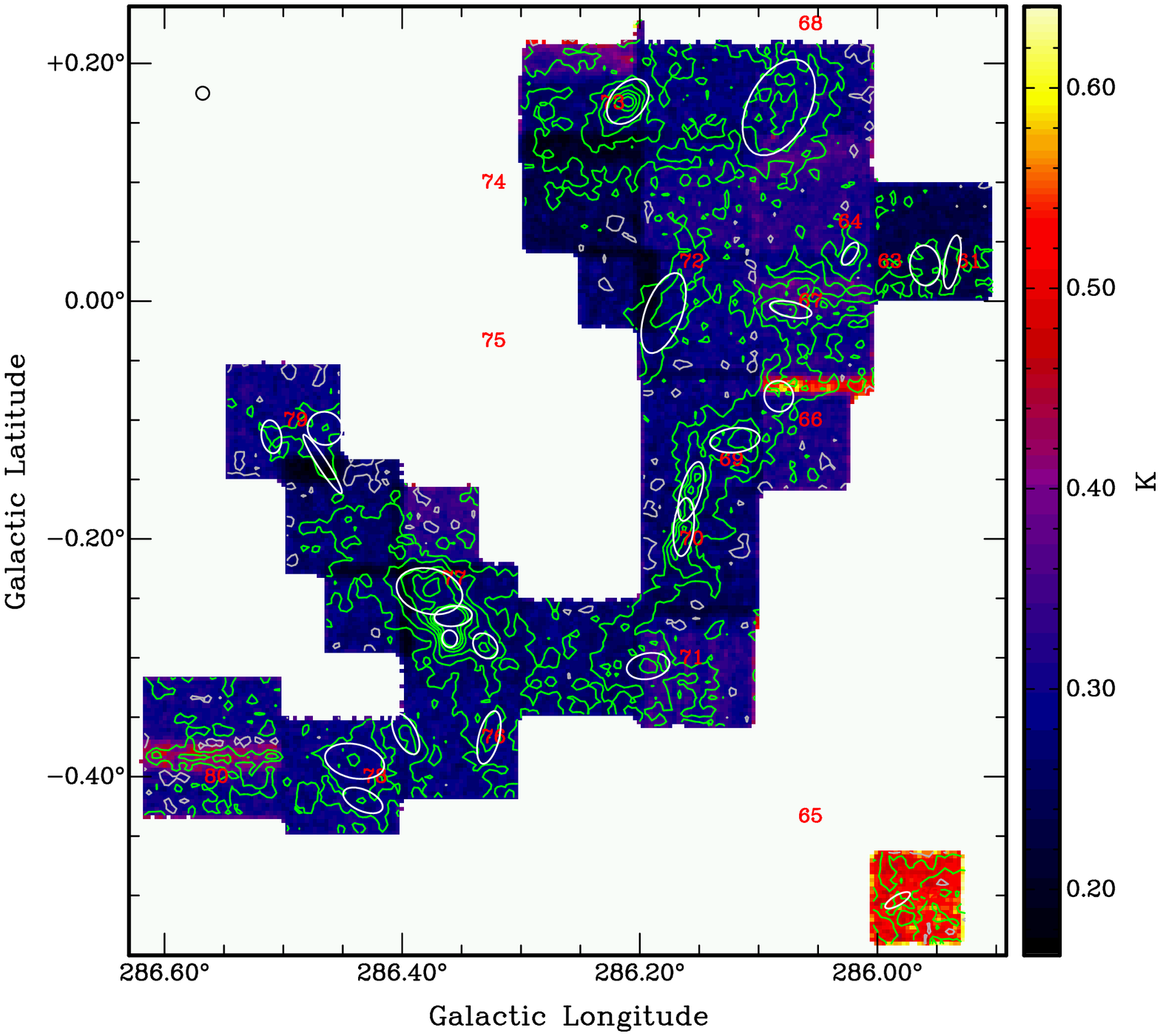}}
(c){\includegraphics[angle=-90,scale=0.40]{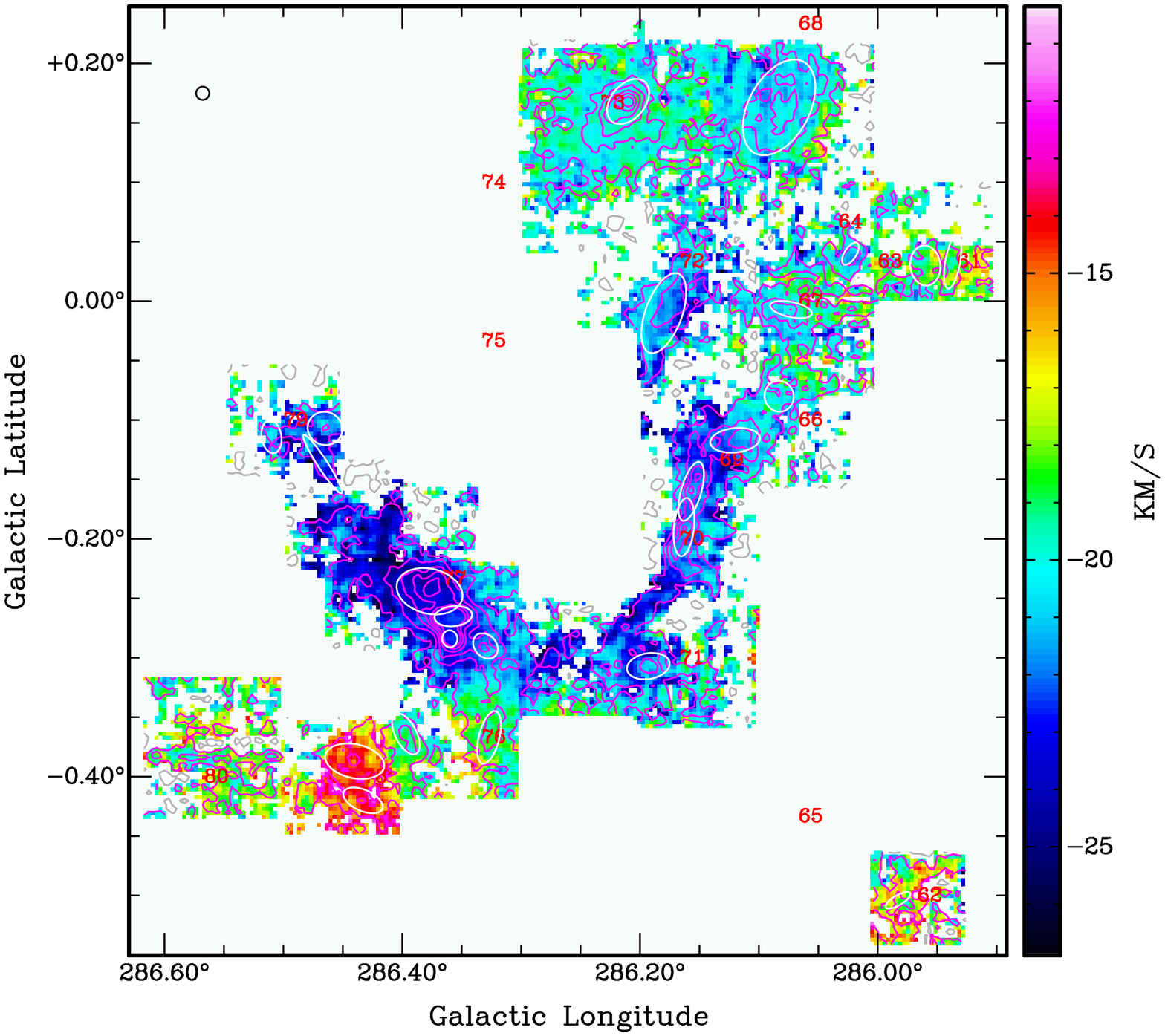}}
(d){\includegraphics[angle=-90,scale=0.40]{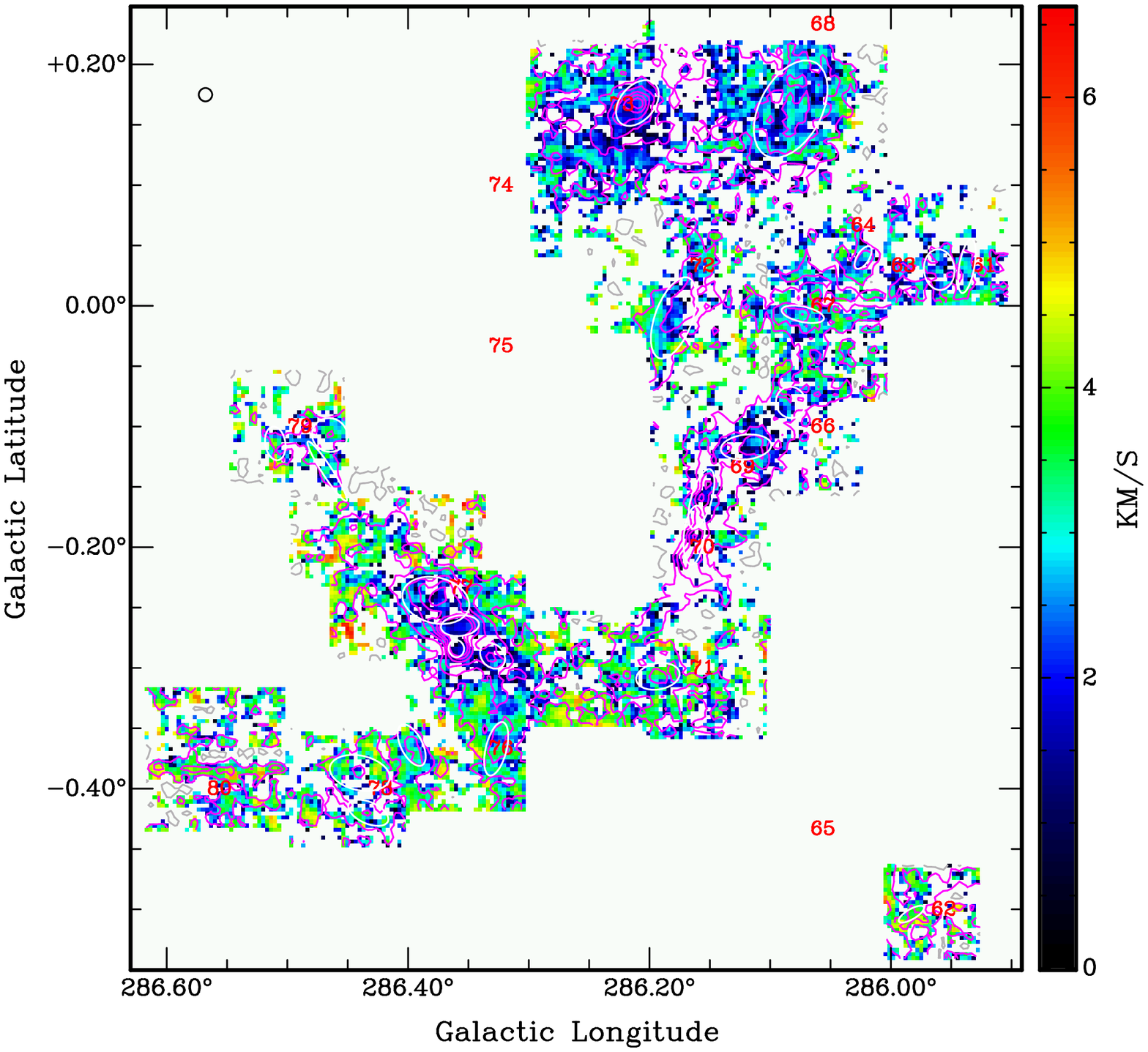}}
\caption{\small Same as Fig.\,\ref{momR1}, but for Region 9 sources BYF\,61--80.  Contours are every 4$\sigma$ = 1.616\,K\kms, and at 2.5\,kpc the 40$''$ Mopra beam (upper left corner) scales to 0.485\,pc.  ($a$) $T_p$,  ($b$) rms,  ($c$) $V_{\rm LSR}$,  ($d$) $\sigma_{V}$.
\label{momR9}}
\end{figure*}

\clearpage

\begin{figure*}[htp]
(a){\includegraphics[angle=0,scale=0.40]{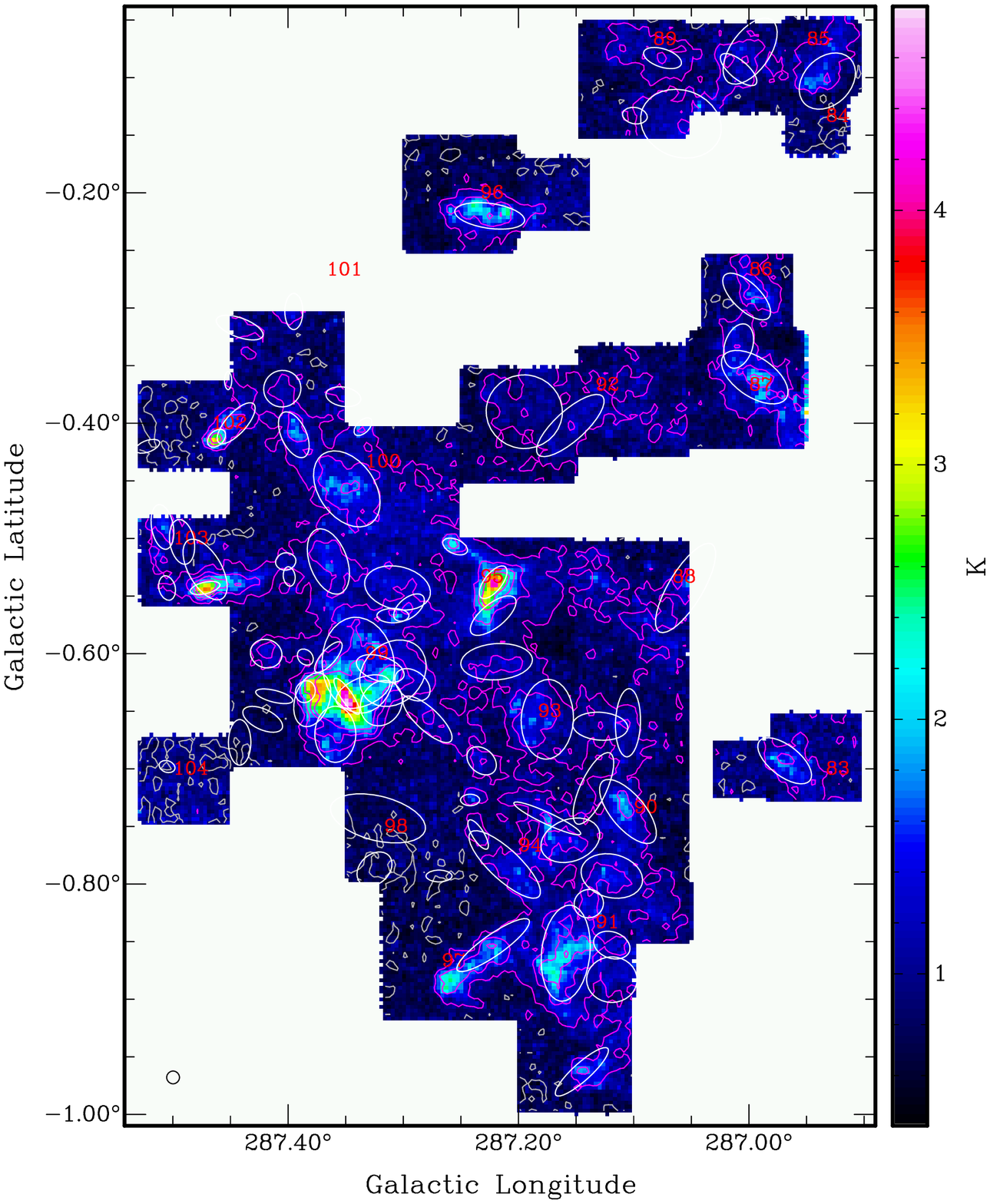}}
(b){\includegraphics[angle=0,scale=0.40]{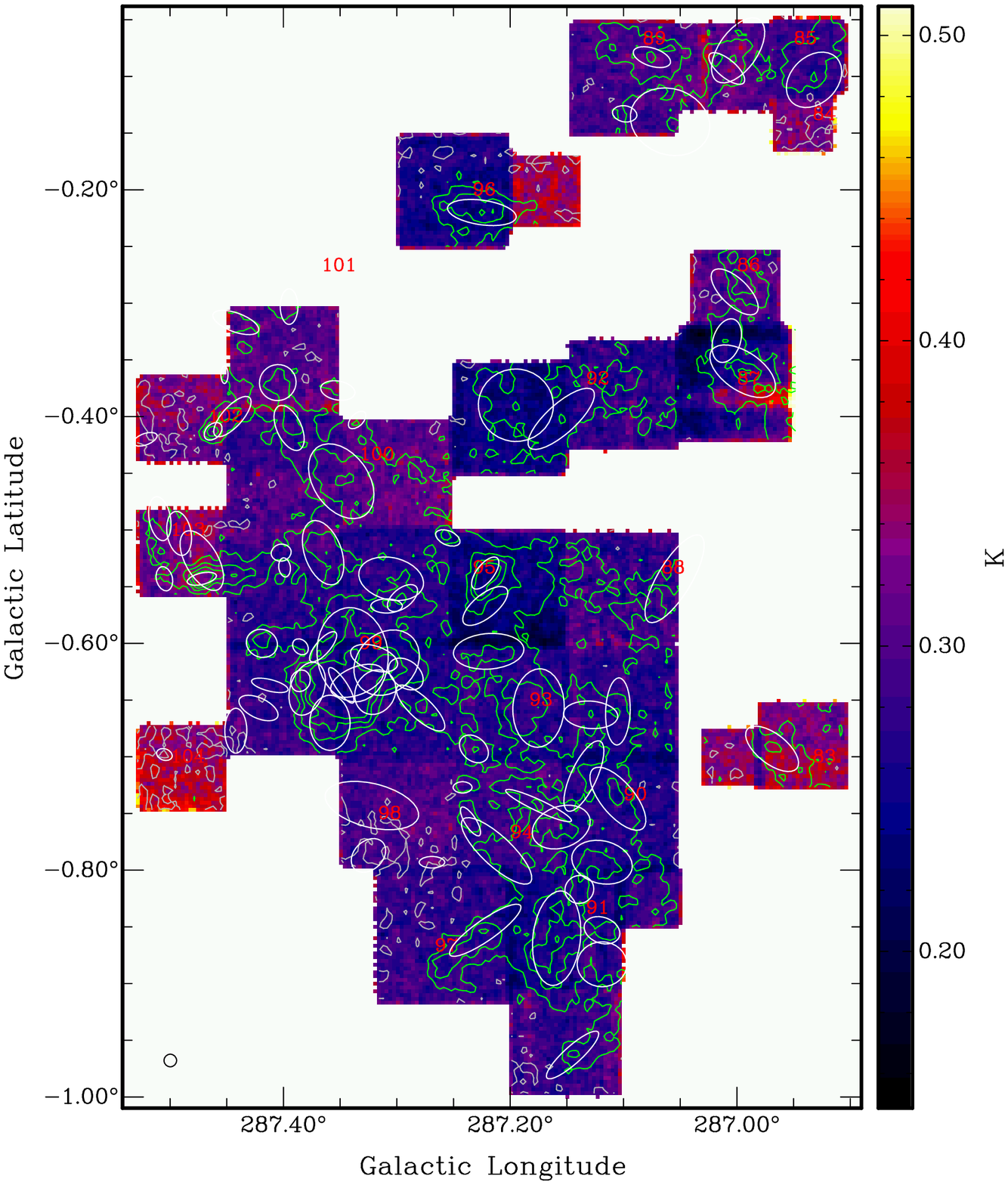}}
(c){\includegraphics[angle=0,scale=0.40]{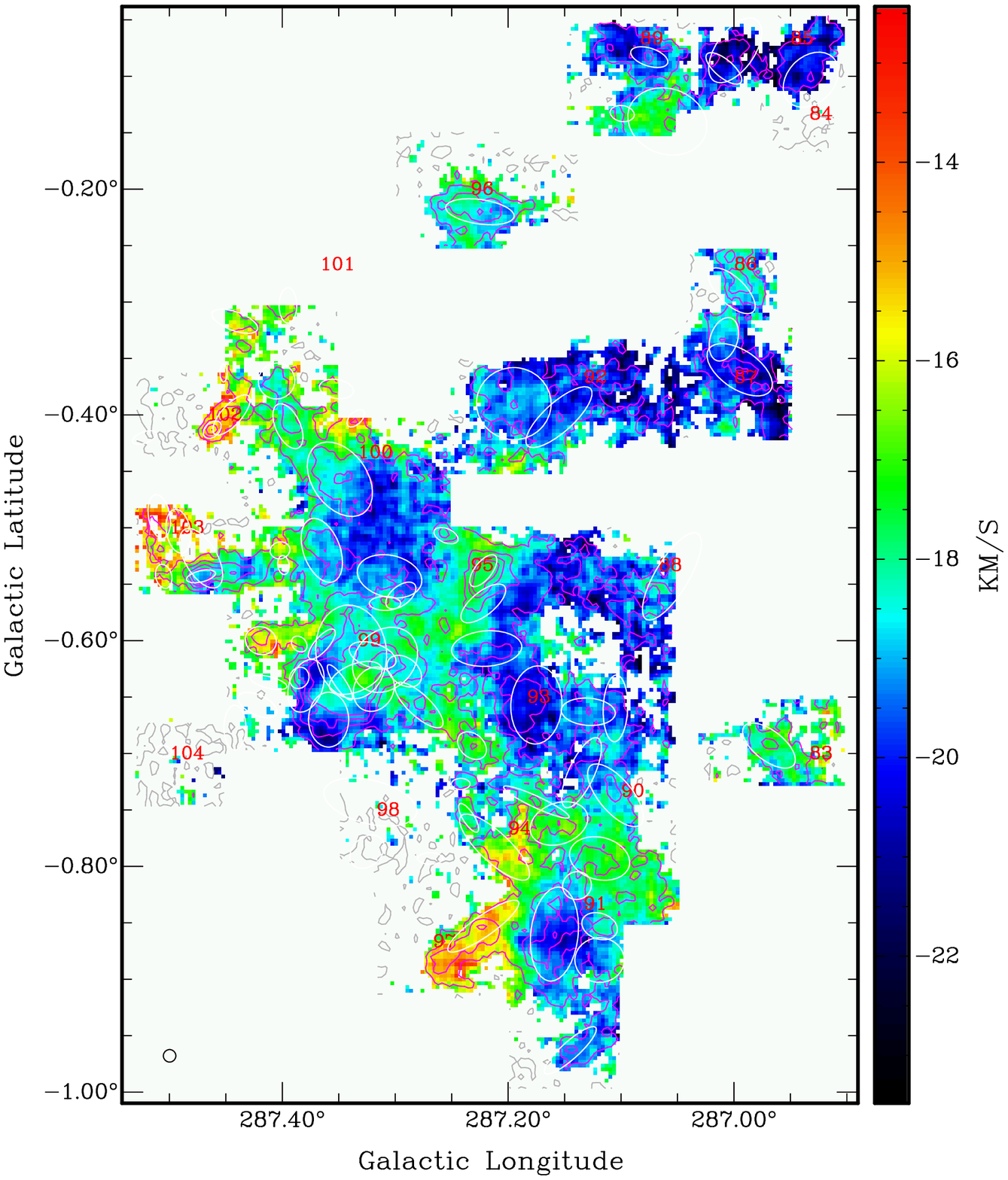}}
(d){\includegraphics[angle=0,scale=0.40]{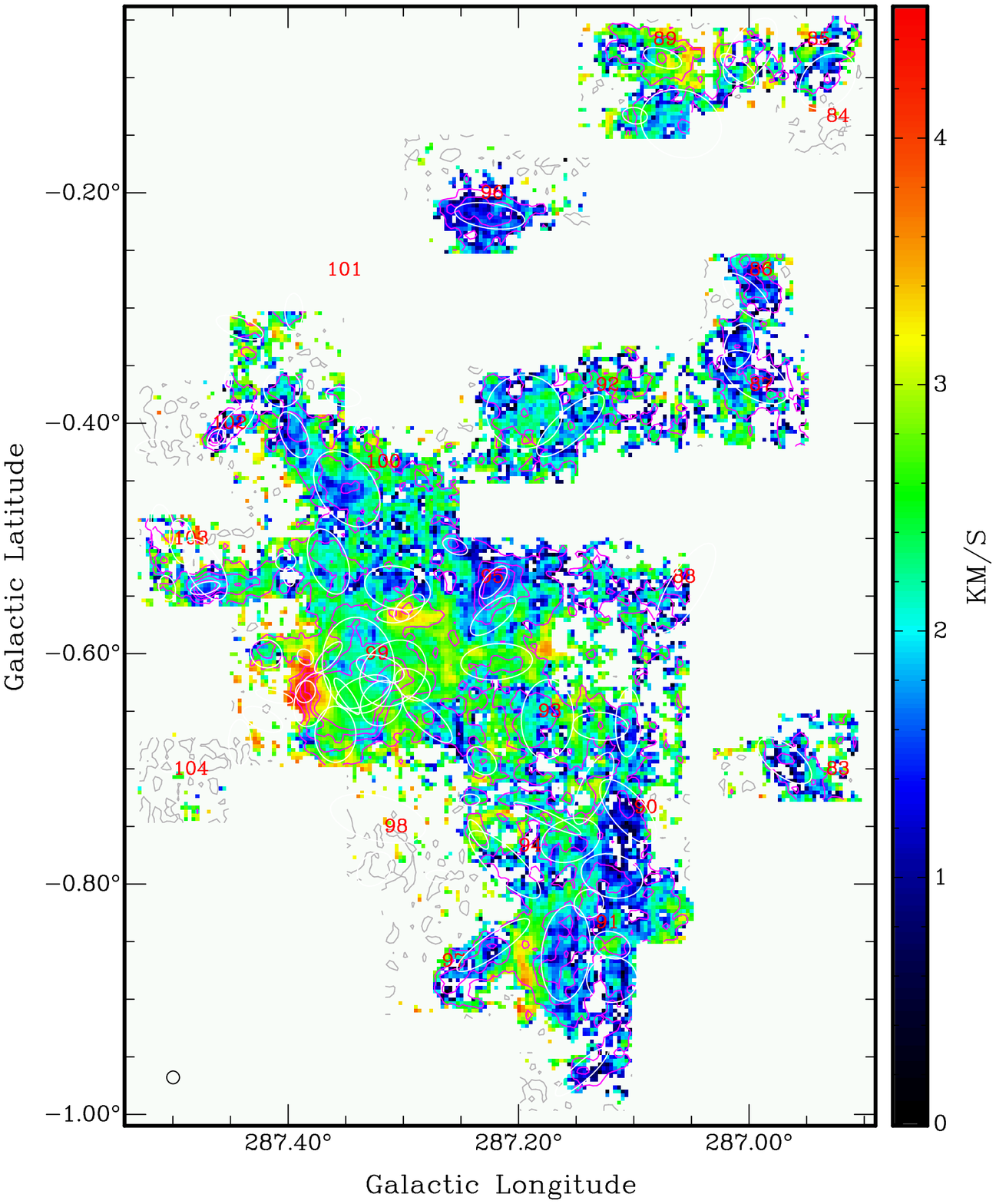}}
\caption{\small Same as Fig.\,\ref{momR1}, but for Region 10 sources BYF\,83--104.  Contours are every 5$\sigma$ = 1.60\,K\kms, and at 2.5\,kpc the 40$''$ Mopra beam (lower left corner) scales to 0.485\,pc.  ($a$) $T_p$,  ($b$) rms,  ($c$) $V_{\rm LSR}$,  ($d$) $\sigma_{V}$.
\label{momR10}}
\end{figure*}

\clearpage

\begin{figure*}[htp]
(a){\includegraphics[angle=-90,scale=0.40]{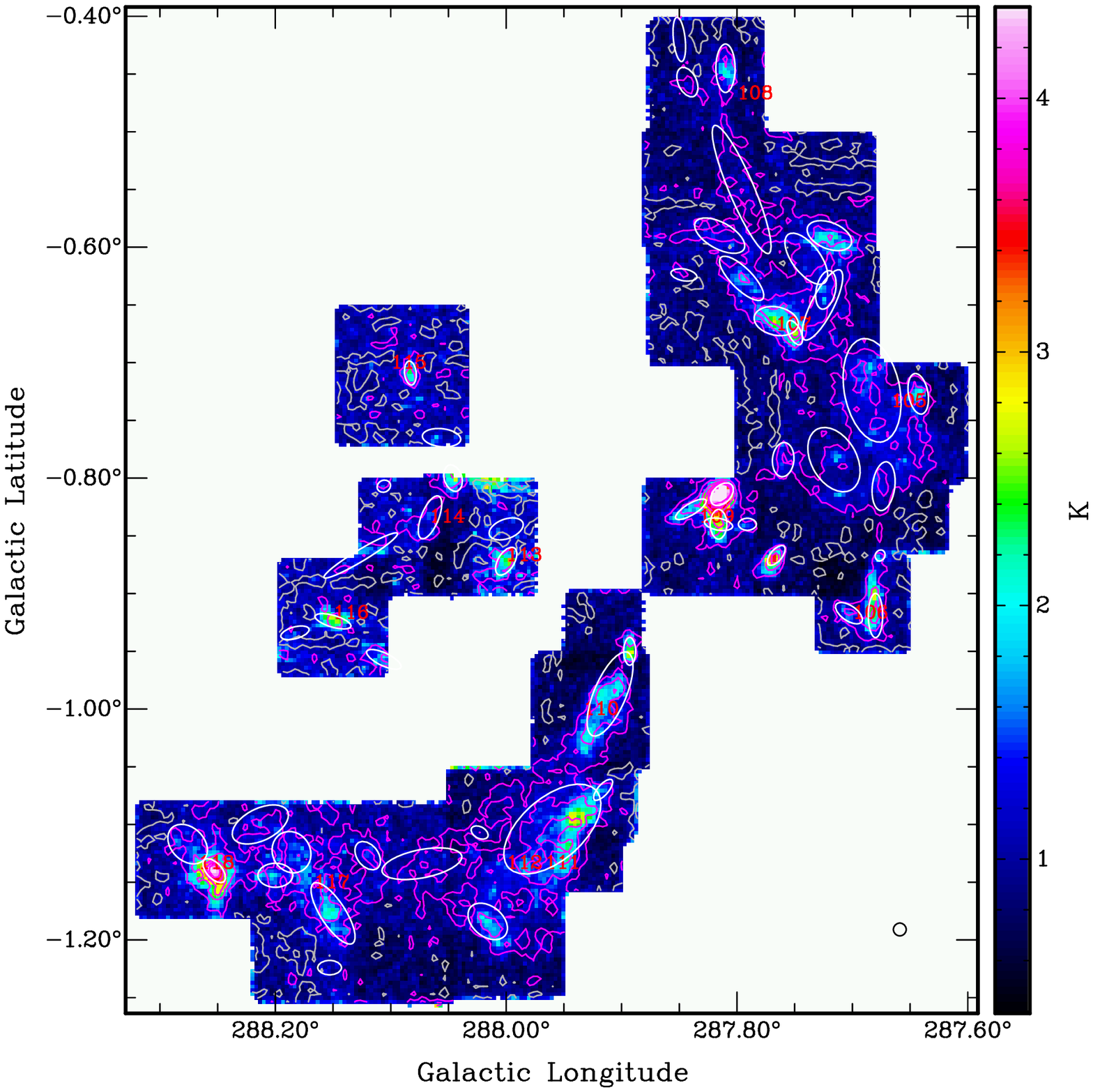}}
(b){\includegraphics[angle=-90,scale=0.40]{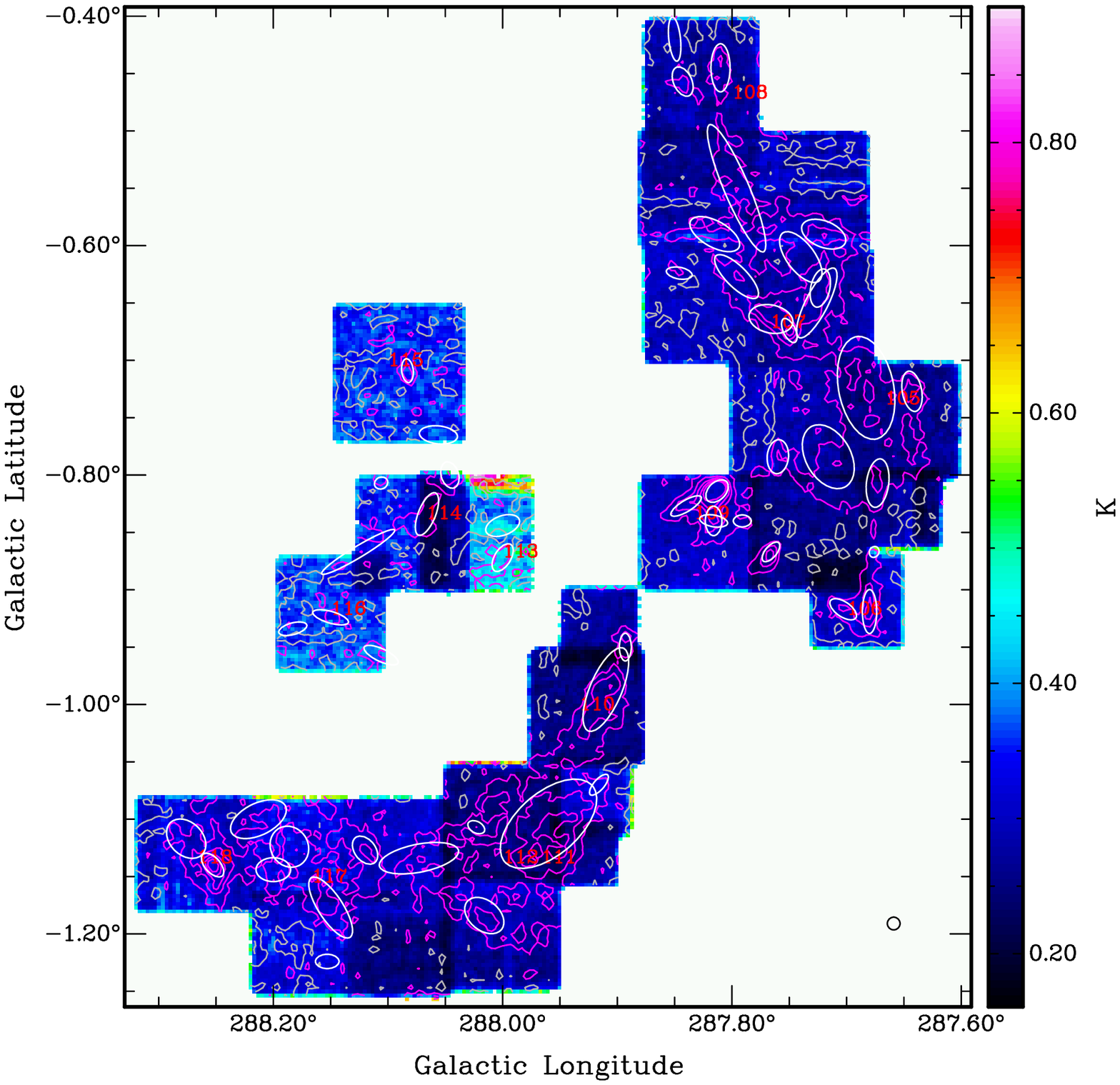}}
(c){\includegraphics[angle=-90,scale=0.40]{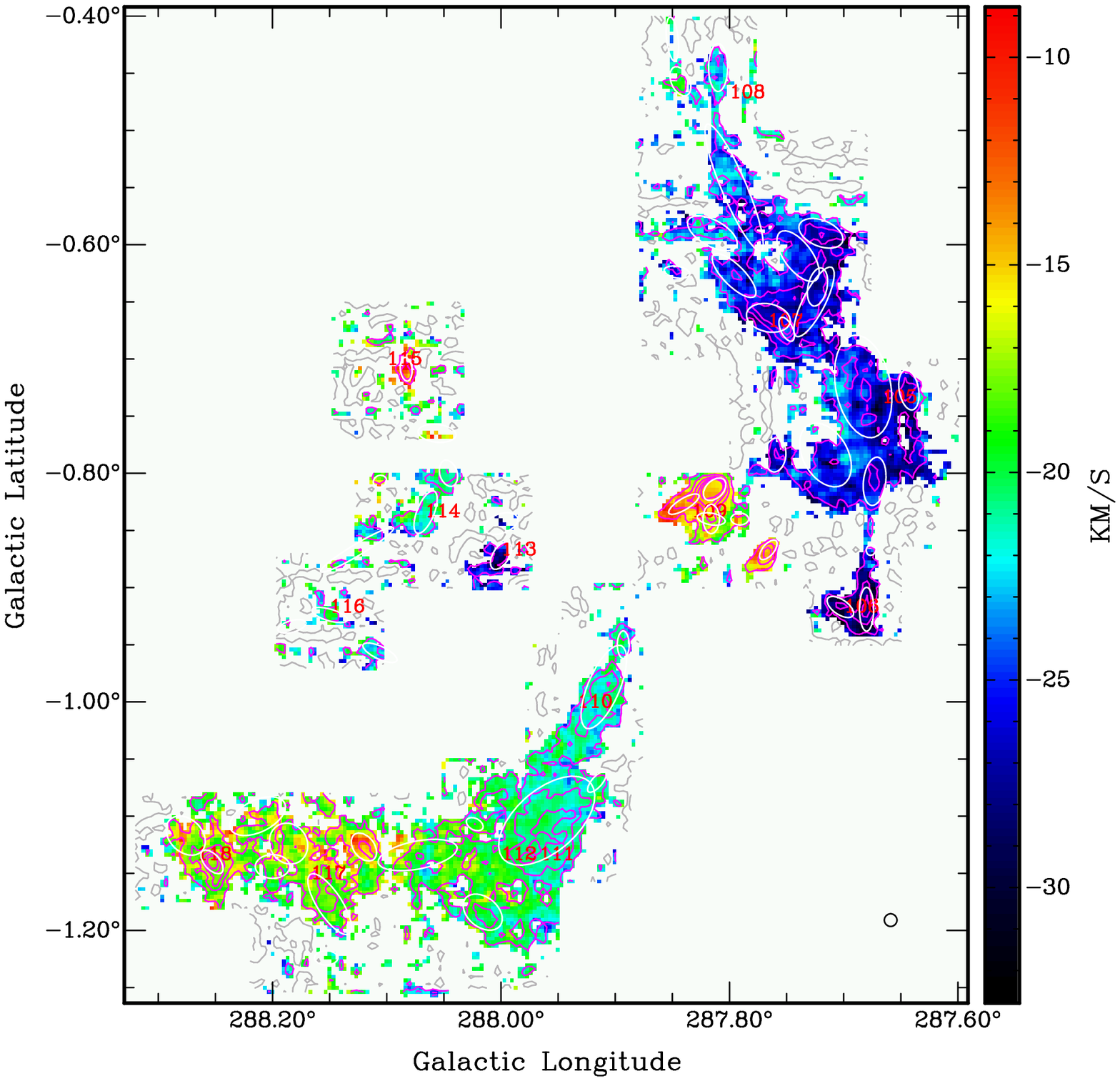}}
(d){\includegraphics[angle=-90,scale=0.40]{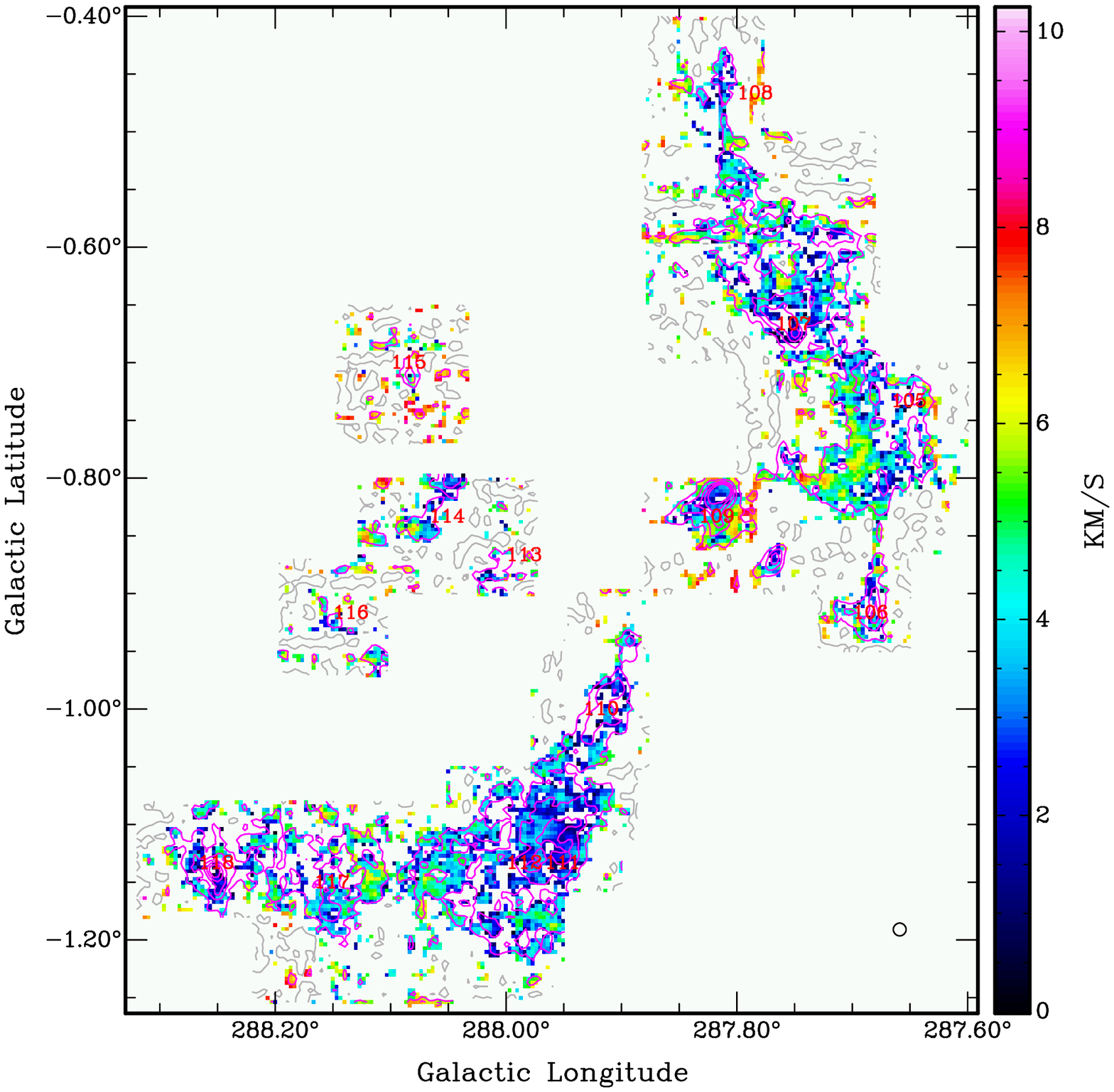}}
\caption{\small Same as Fig.\,\ref{momR1}, but for Region 11 sources BYF\,105--118.  Contours are every 4$\sigma$ = 2.04\,K\kms, and at 2.5\,kpc the 40$''$ Mopra beam (lower right corner) scales to 0.485\,pc.  ($a$) $T_p$,  ($b$) rms,  ($c$) $V_{\rm LSR}$,  ($d$) $\sigma_{V}$.
\label{momR11}}
\end{figure*}

\clearpage

\begin{figure*}[ht]
(a){\includegraphics[angle=-90,scale=0.35]{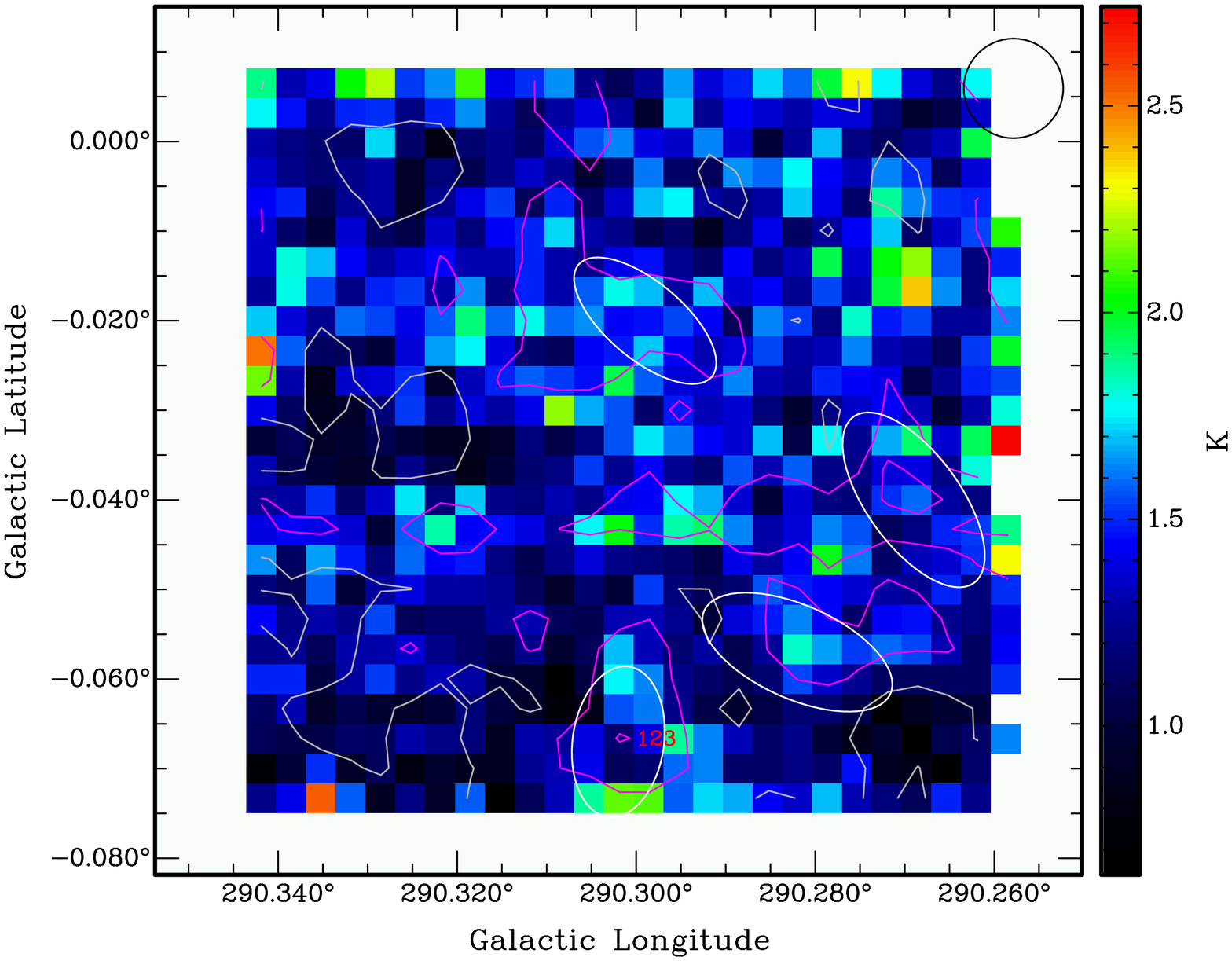}}
(b){\includegraphics[angle=-90,scale=0.35]{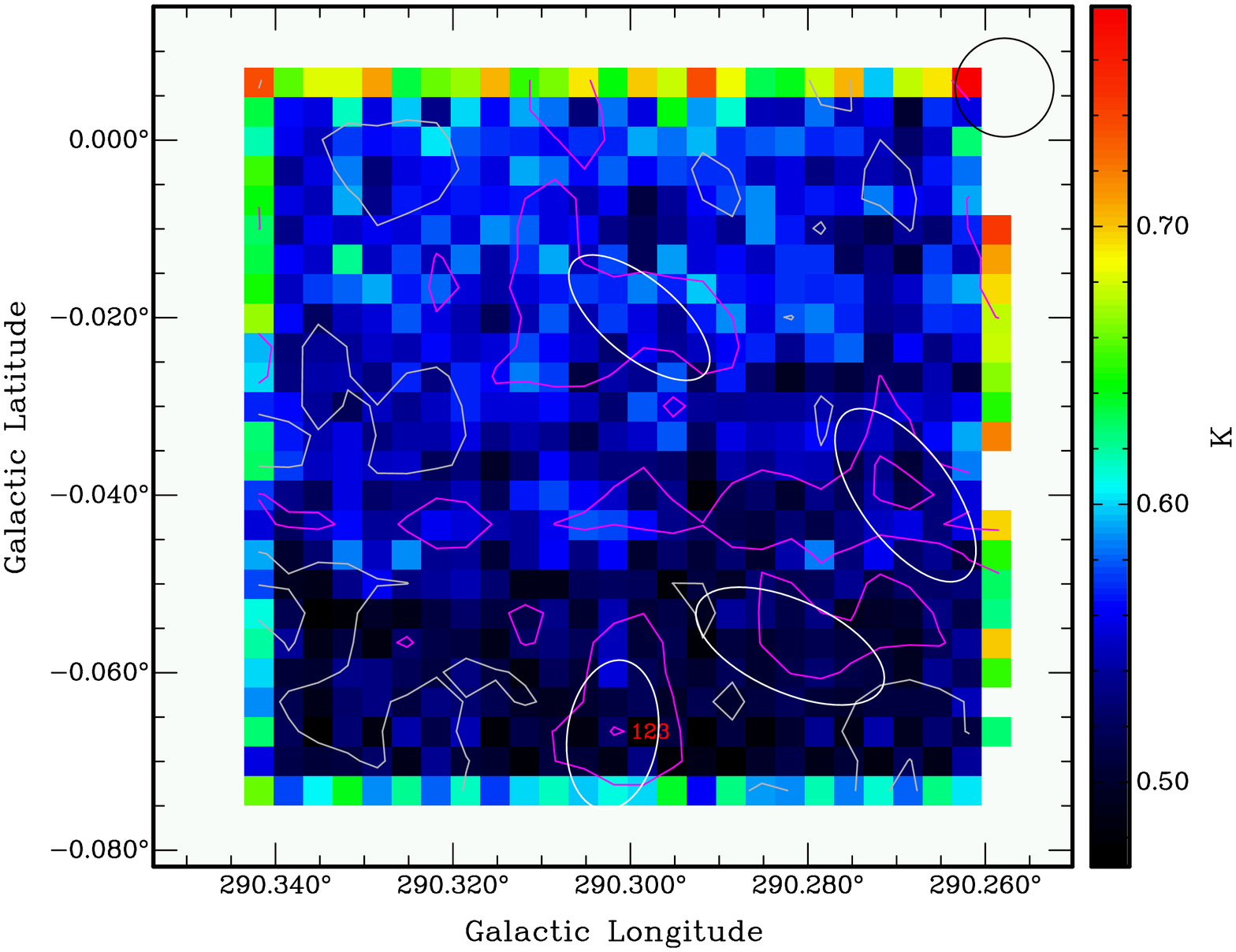}} \\
(c){\includegraphics[angle=-90,scale=0.35]{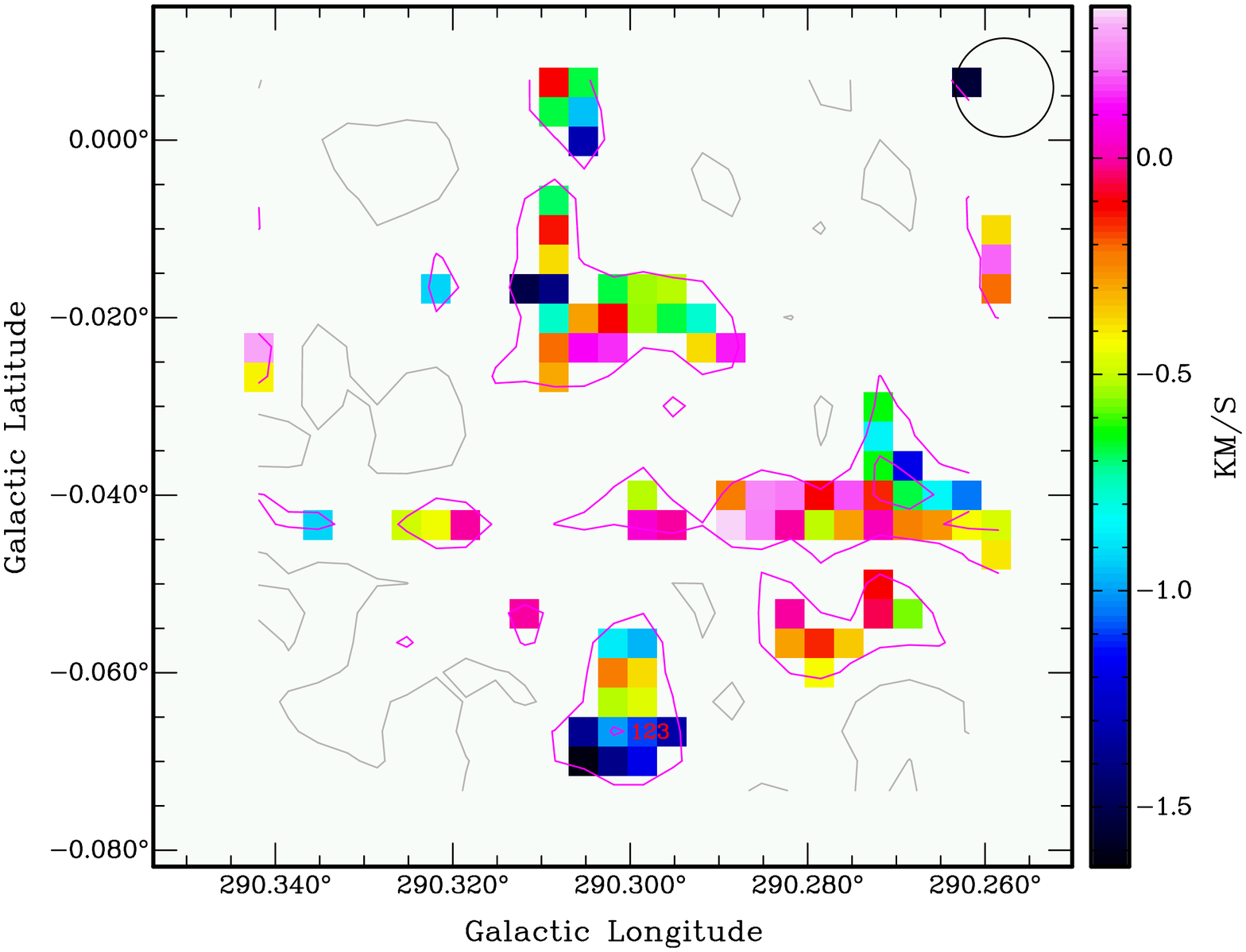}}
(d){\includegraphics[angle=-90,scale=0.35]{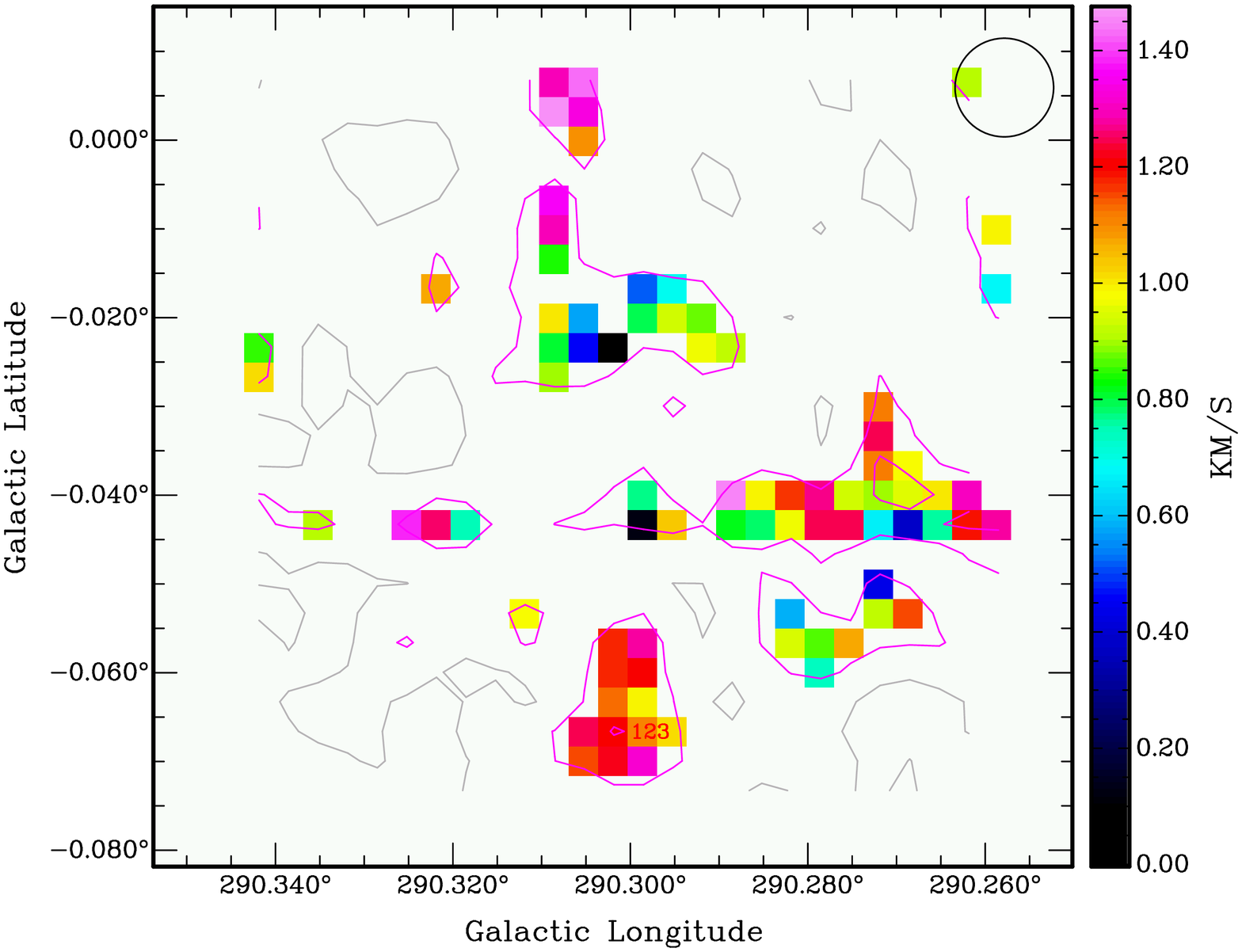}}
\caption{\small Same as Fig.\,\ref{momR1}, but for isolated source BYF\,123.  Contours are every 2$\sigma$ = 0.836\,K\kms, and at 6.8\,kpc the 40$''$ Mopra beam (upper right corner) scales to 1.32\,pc.  ($a$) $T_p$,  ($b$) rms,  ($c$) $V_{\rm LSR}$,  ($d$) $\sigma_{V}$.
\label{momBYF123}}
\end{figure*}

\clearpage

\begin{figure*}[ht]
(a){\includegraphics[angle=-90,scale=0.35]{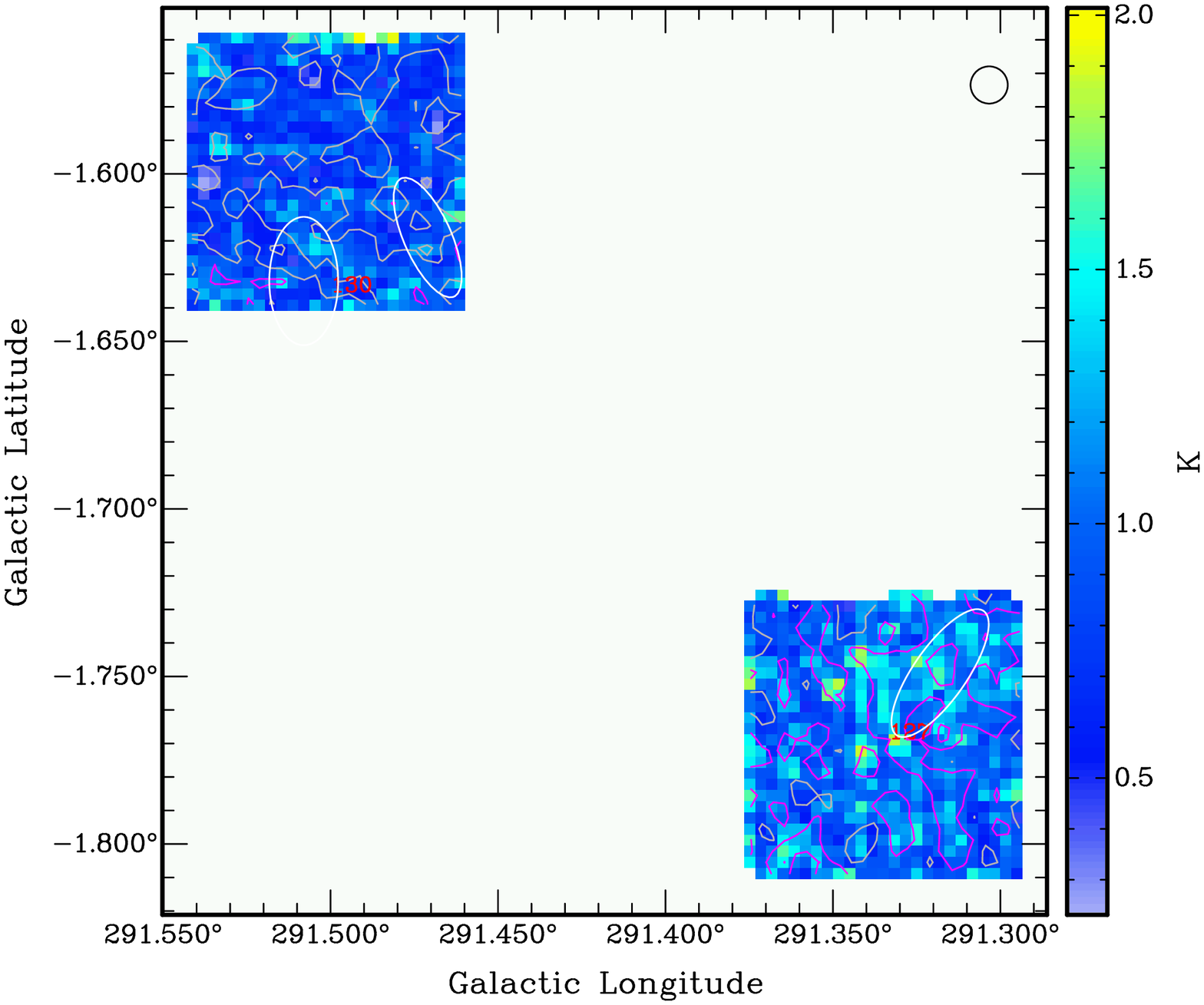}}
(b){\includegraphics[angle=-90,scale=0.35]{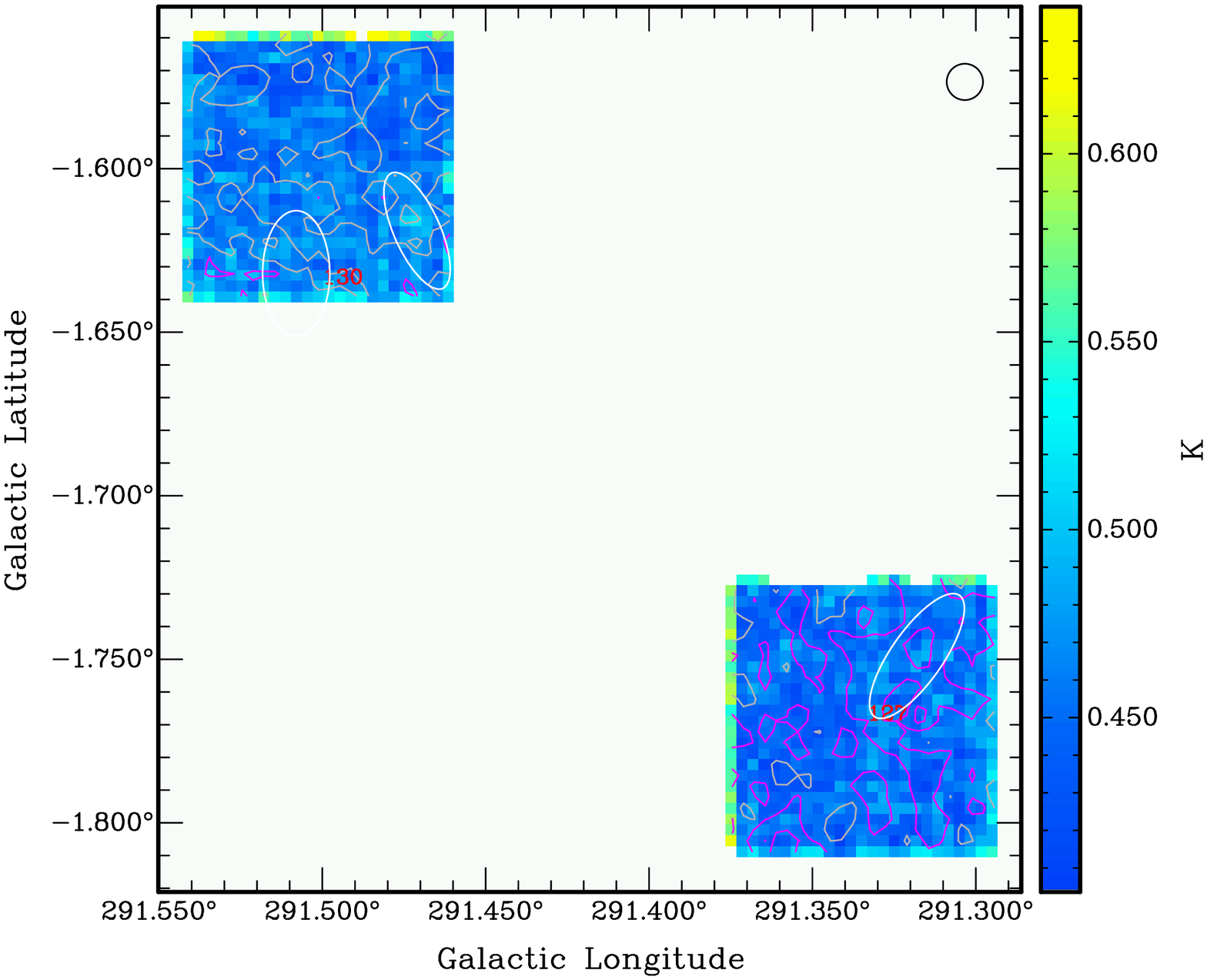}}
(c){\includegraphics[angle=-90,scale=0.35]{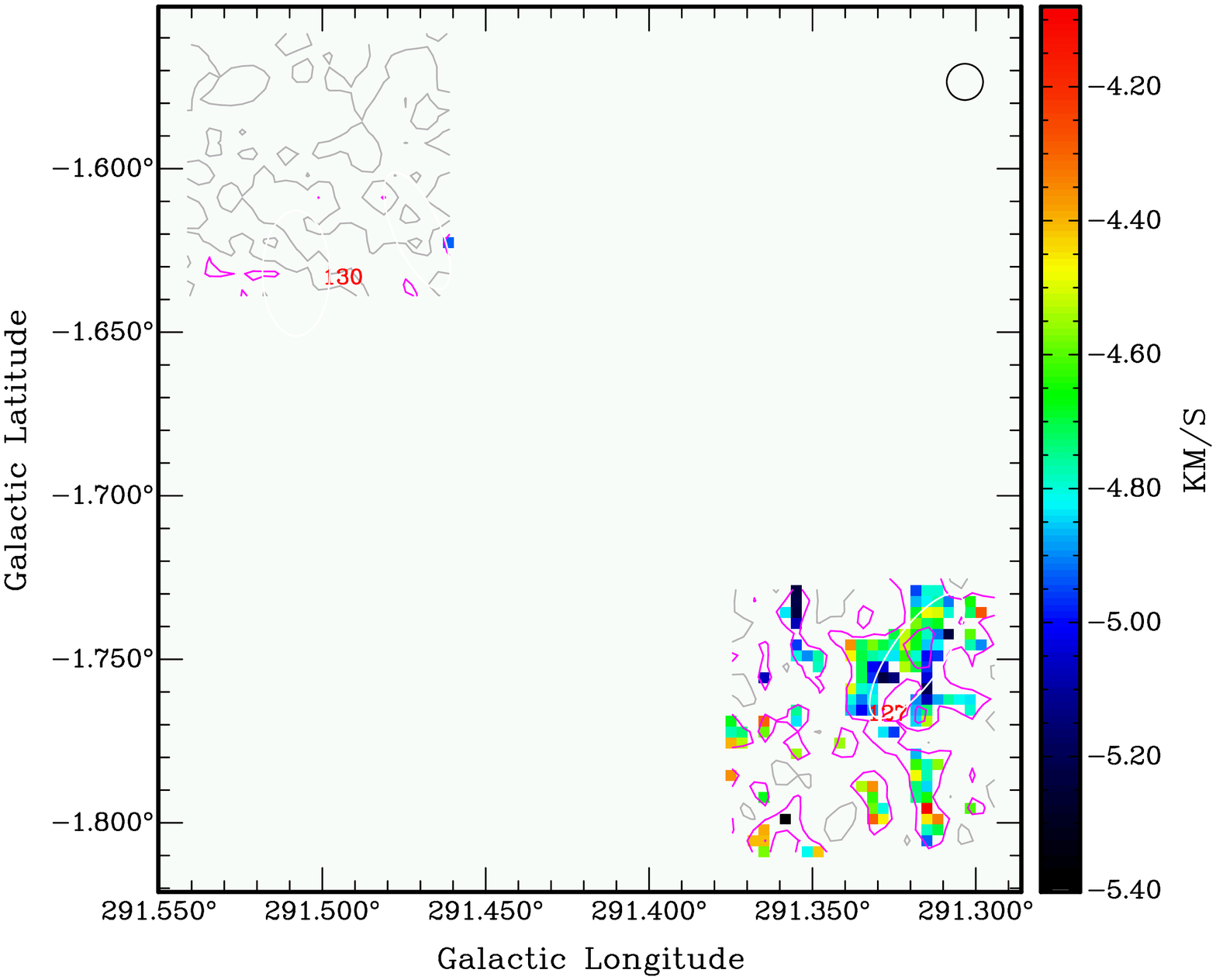}}
(d){\includegraphics[angle=-90,scale=0.35]{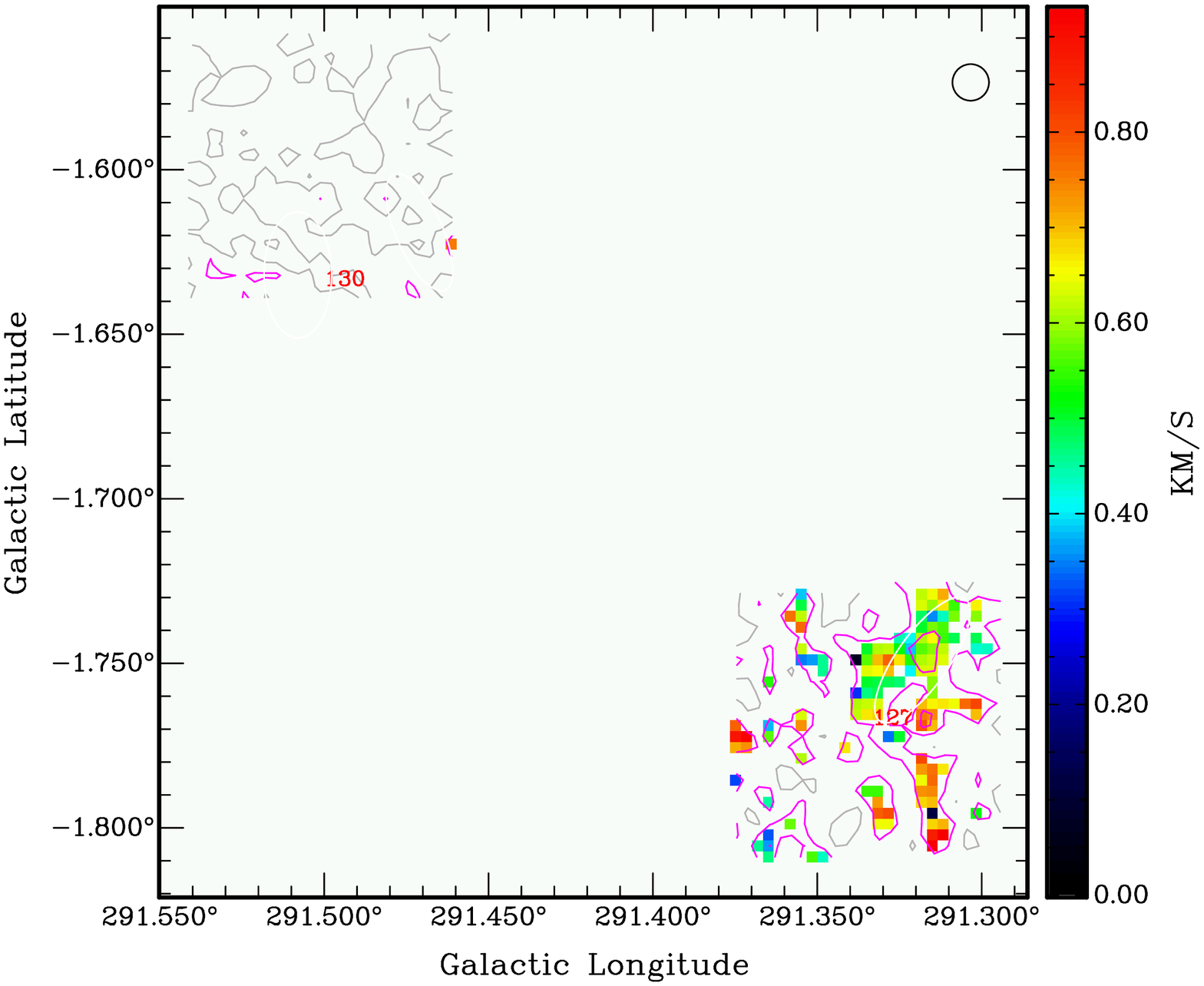}}
\caption{\small Same as Fig.\,\ref{momR1}, but for Region 12 source BYF\,127.  Contours are every 2$\sigma$ = 0.460\,K\kms, and at 1.1\,kpc the 40$''$ Mopra beam (lower left corner) scales to 0.213\,pc.  ($a$) $T_p$,  ($b$) rms,  ($c$) $V_{\rm LSR}$,  ($d$) $\sigma_{V}$.
\label{momR12a}}
\end{figure*}

\clearpage

\begin{figure*}[ht]
(a){\includegraphics[angle=-90,scale=0.35]{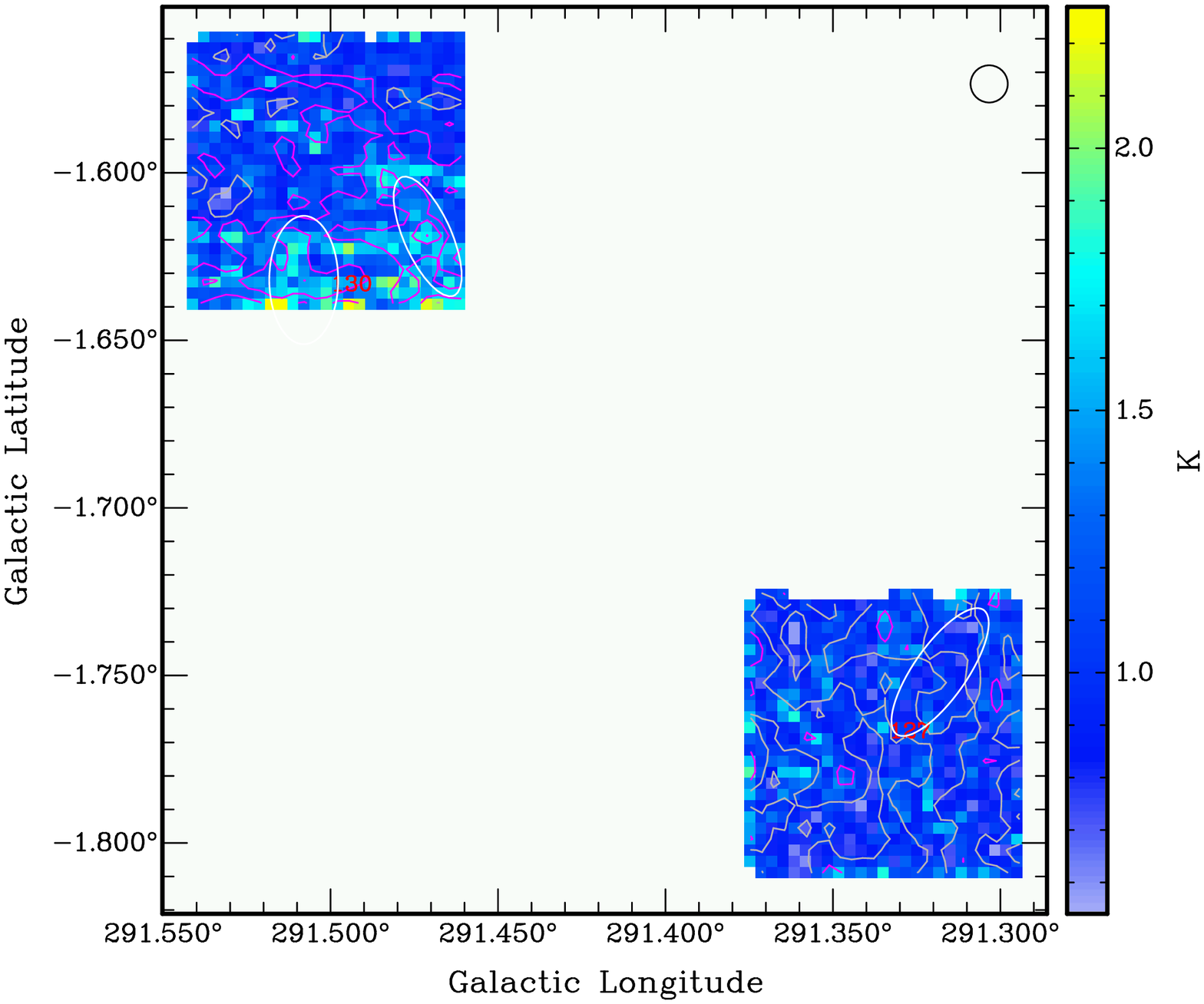}}
(b){\includegraphics[angle=-90,scale=0.35]{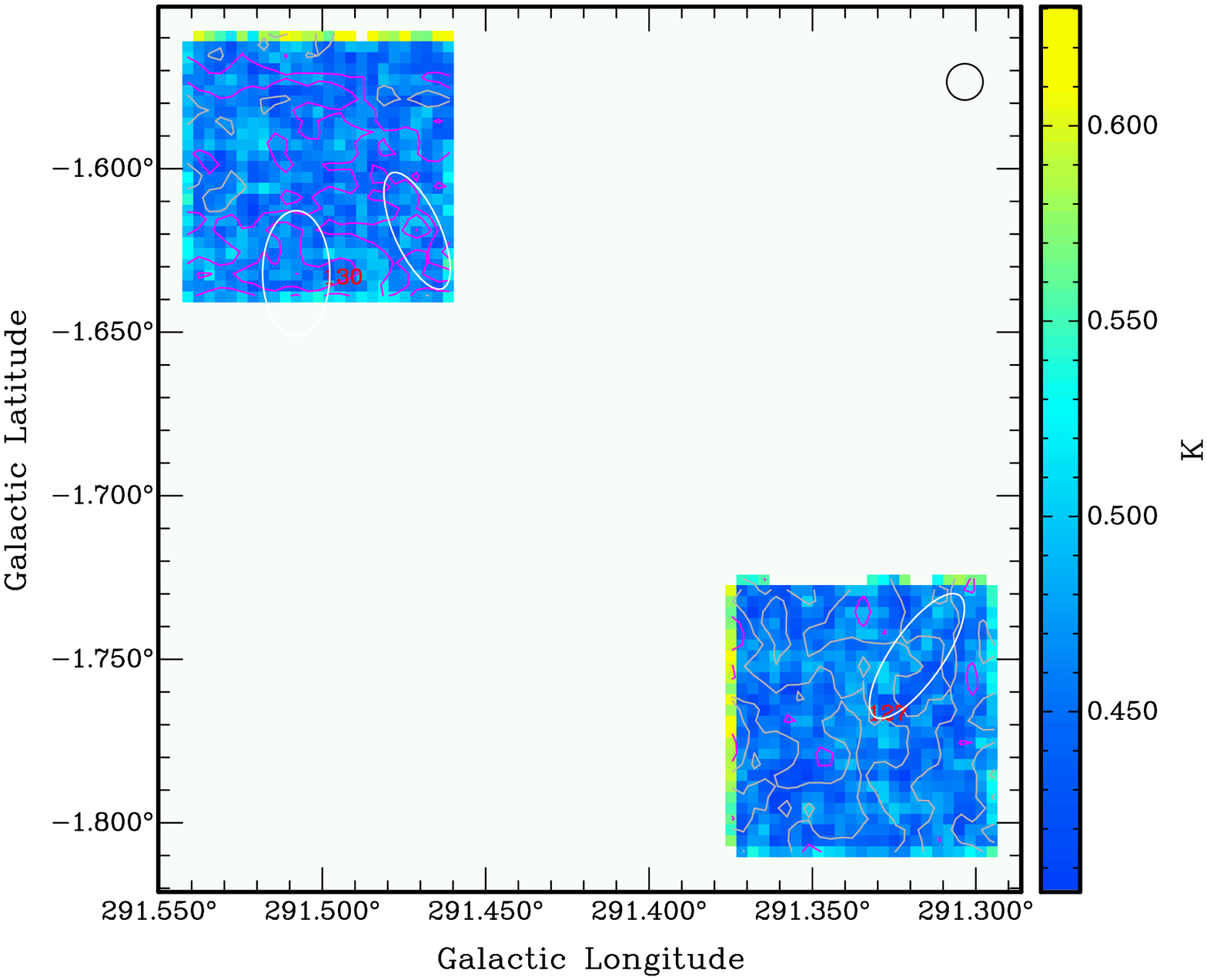}}
(c){\includegraphics[angle=-90,scale=0.35]{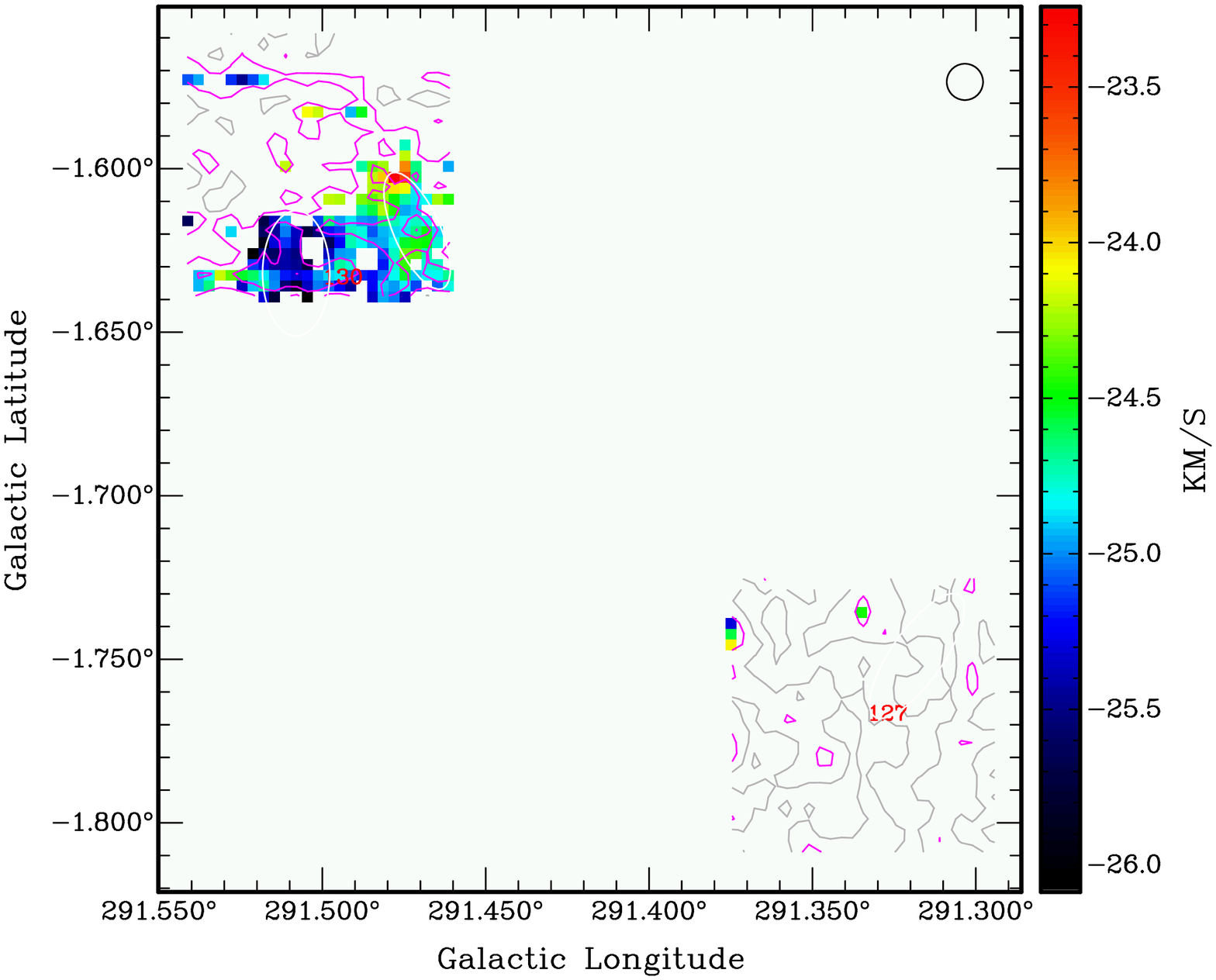}}
(d){\includegraphics[angle=-90,scale=0.35]{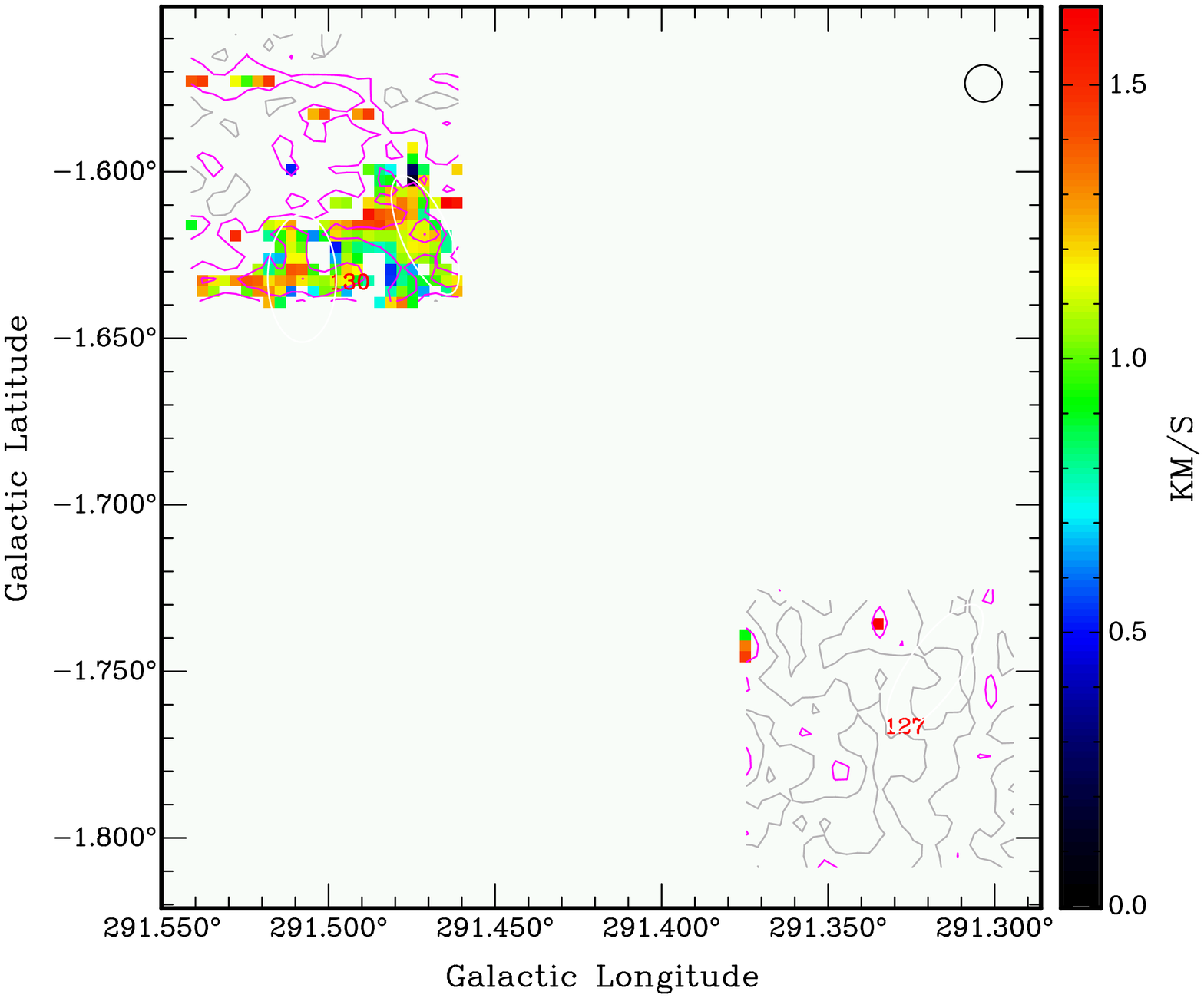}}
\caption{\small Same as Fig.\,\ref{momR1}, but for Region 12 source BYF\,130.  Contours are every 2$\sigma$ = 0.676\,K\kms, and at 2.4\,kpc the 40$''$ Mopra beam (lower left corner) scales to 0.465\,pc.  ($a$) $T_p$,  ($b$) rms,  ($c$) $V_{\rm LSR}$,  ($d$) $\sigma_{V}$.
\label{momR12b}}
\end{figure*}

\clearpage

\begin{figure*}[ht]
(a){\includegraphics[angle=-90,scale=0.30]{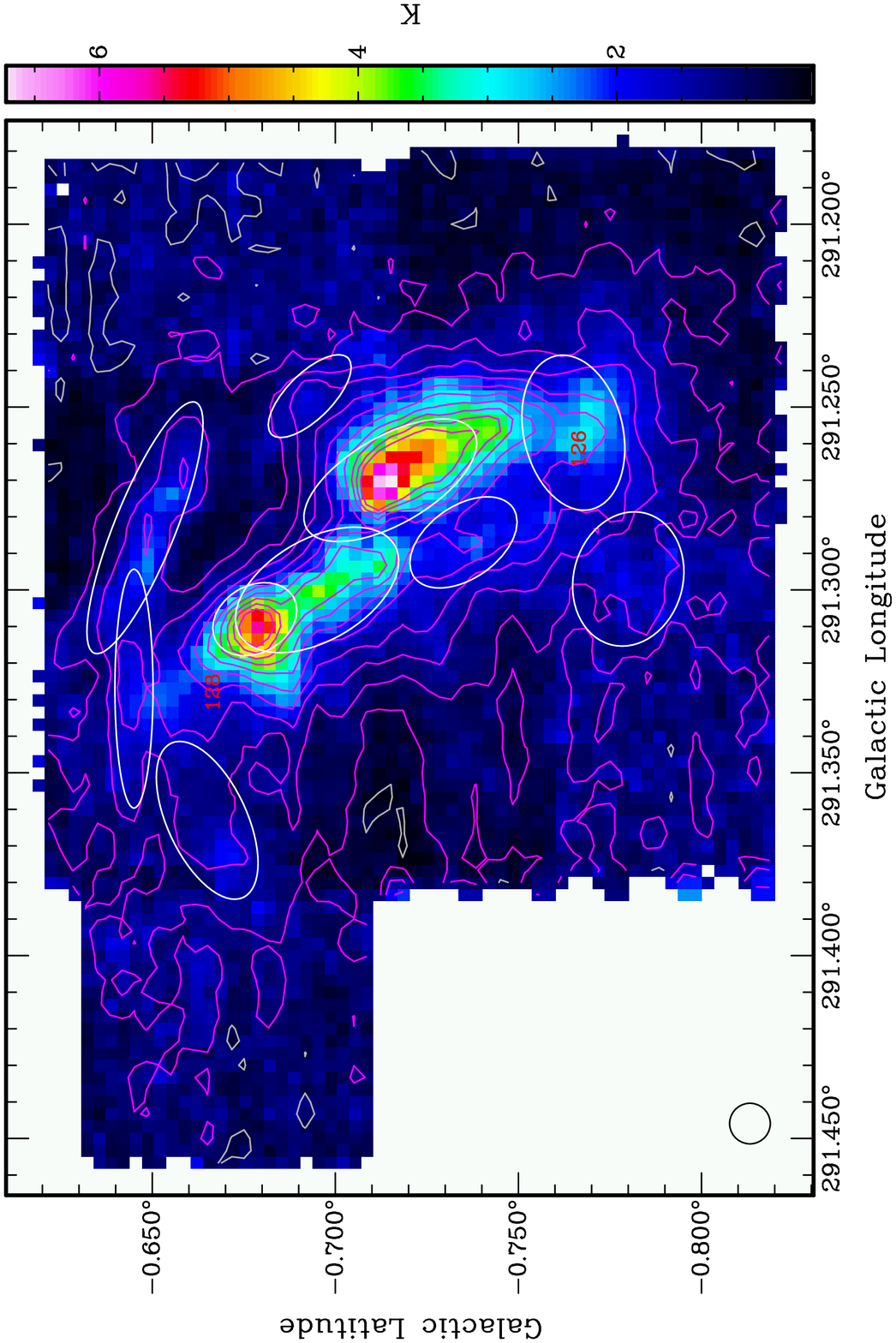}}
(b){\includegraphics[angle=-90,scale=0.30]{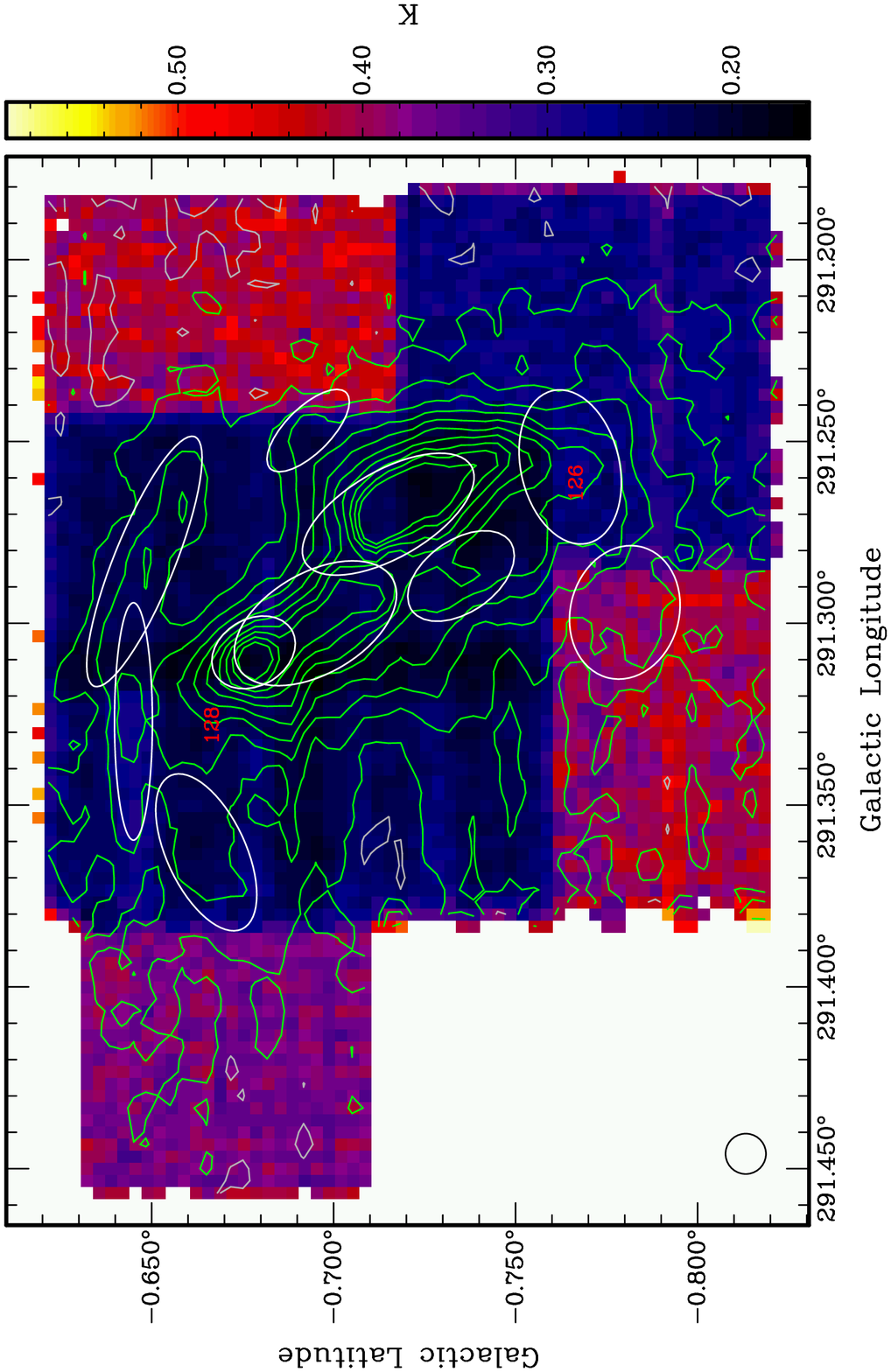}}
(c){\includegraphics[angle=-90,scale=0.30]{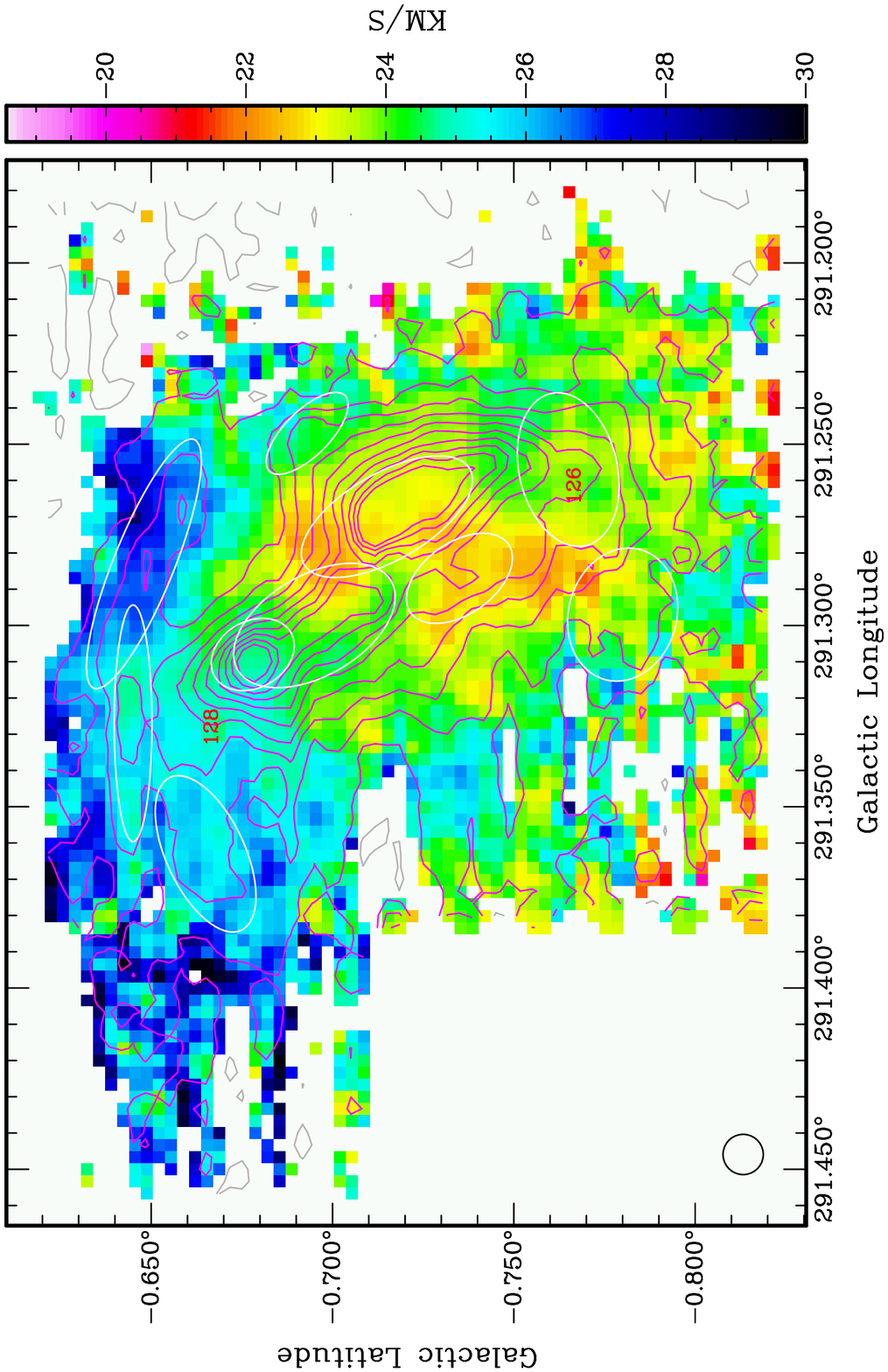}}
(d){\includegraphics[angle=-90,scale=0.30]{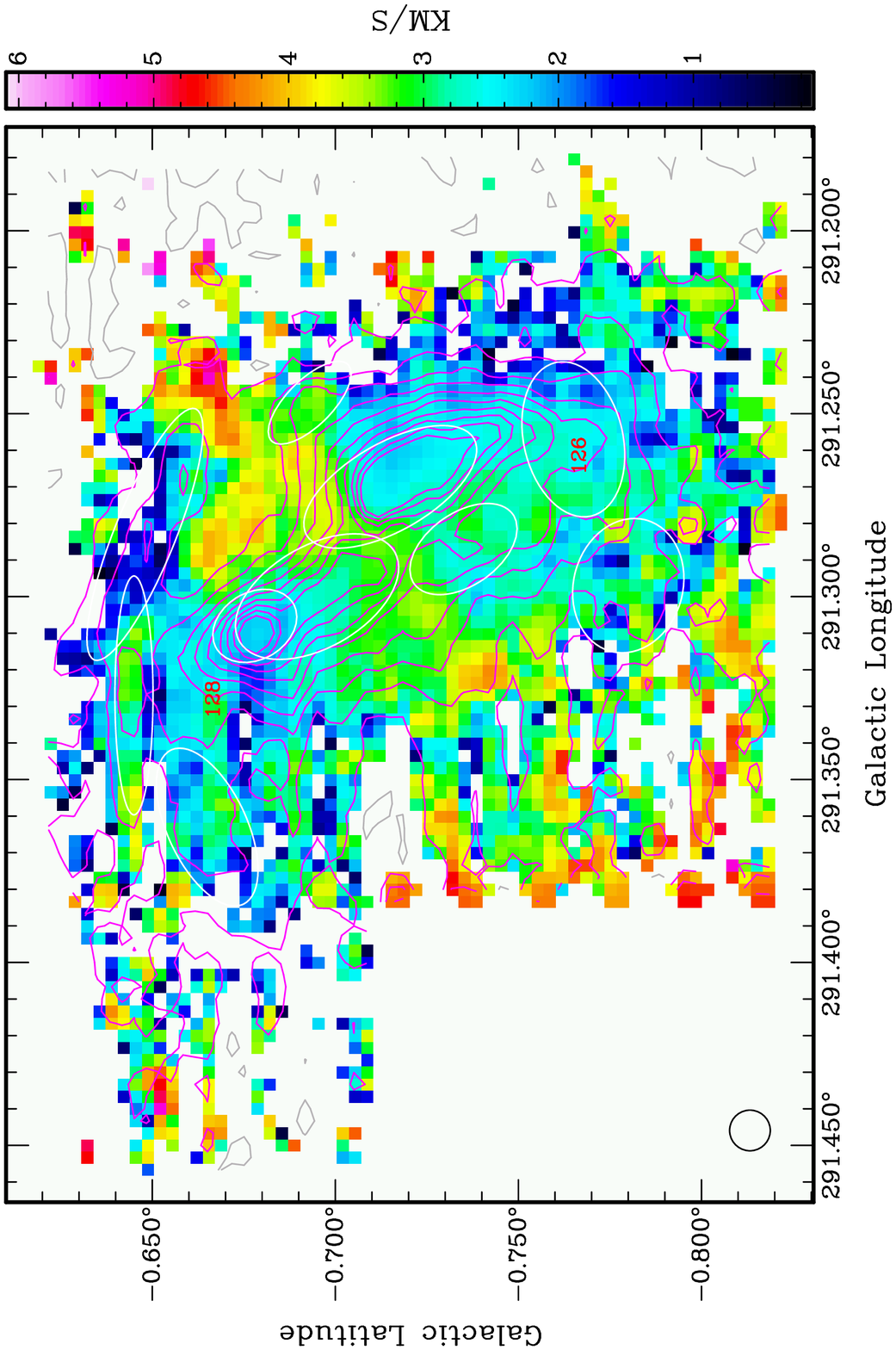}}
\caption{\small Same as Fig.\,\ref{momR1}, but for Region 13 sources BYF\,126 and 128.  Contours are every 5$\sigma$ = 2.12\,K\kms, and at 2.4\,kpc the 40$''$ Mopra beam (lower left corner) scales to 0.465\,pc.  ($a$) $T_p$,  ($b$) rms,  ($c$) $V_{\rm LSR}$,  ($d$) $\sigma_{V}$.
\label{momR13a}}
\end{figure*}

\clearpage

\begin{figure*}[ht]
\centerline{(a){\includegraphics[angle=-90,scale=0.30]{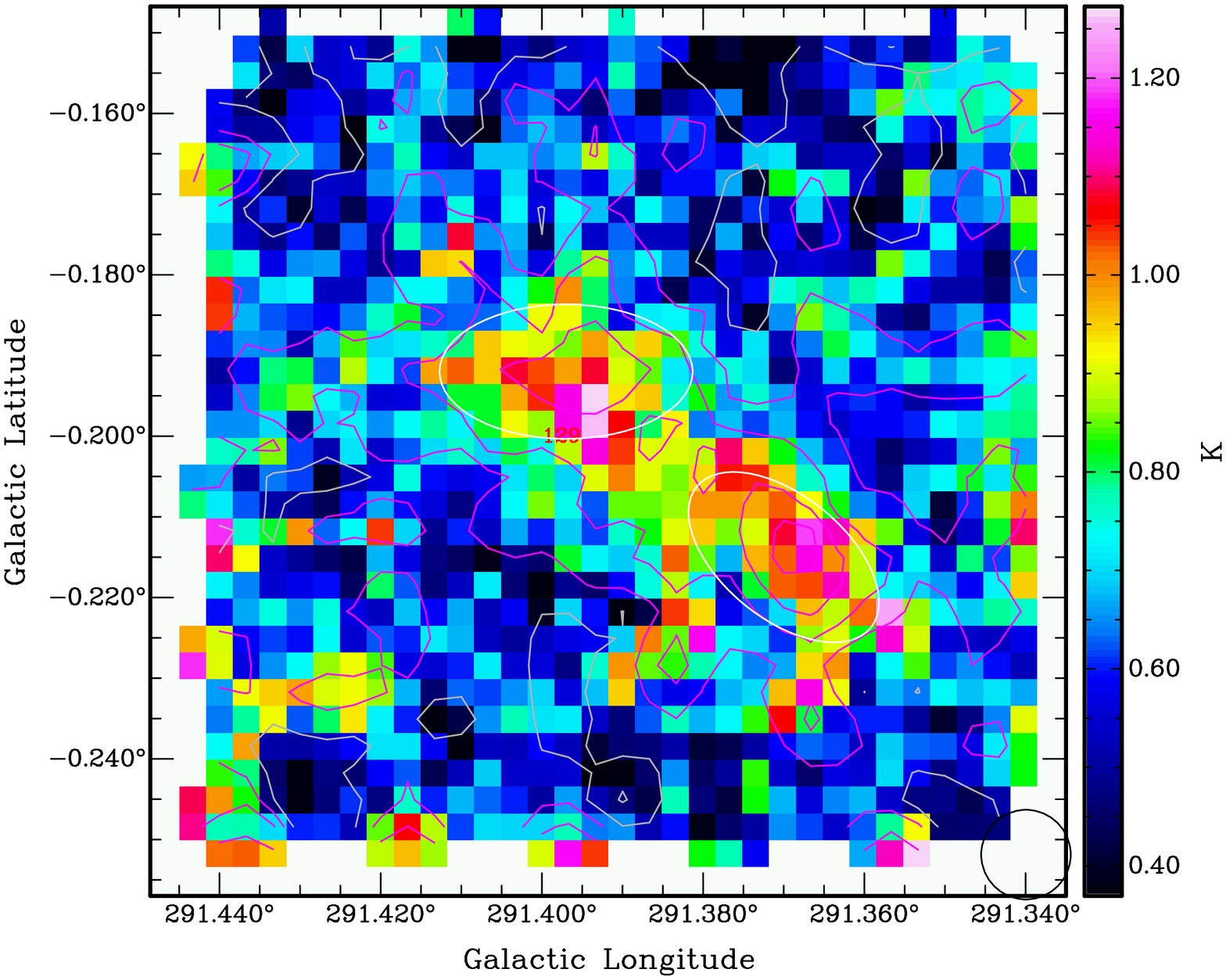}}
		(b){\includegraphics[angle=-90,scale=0.30]{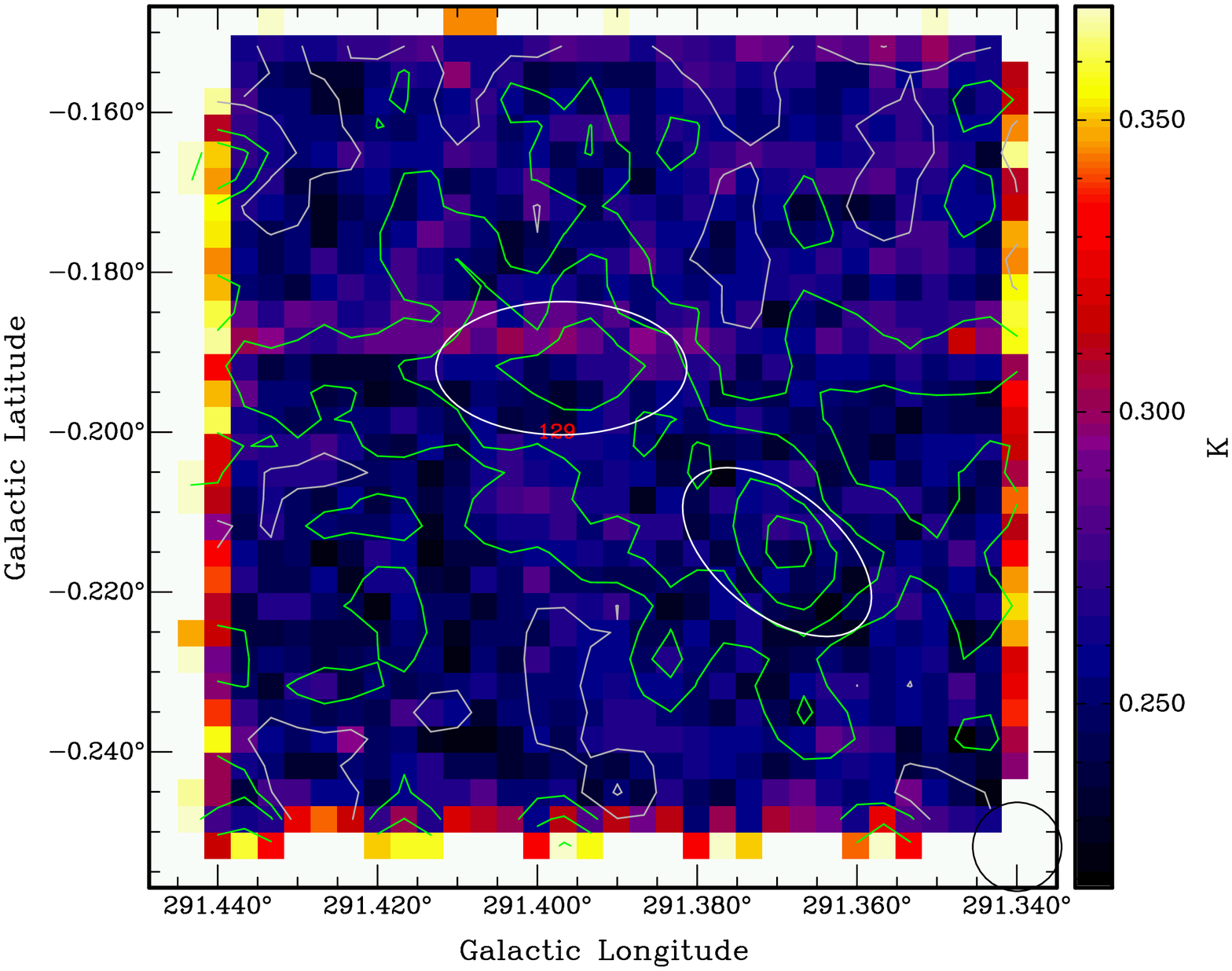}}}
\centerline{(c){\includegraphics[angle=-90,scale=0.30]{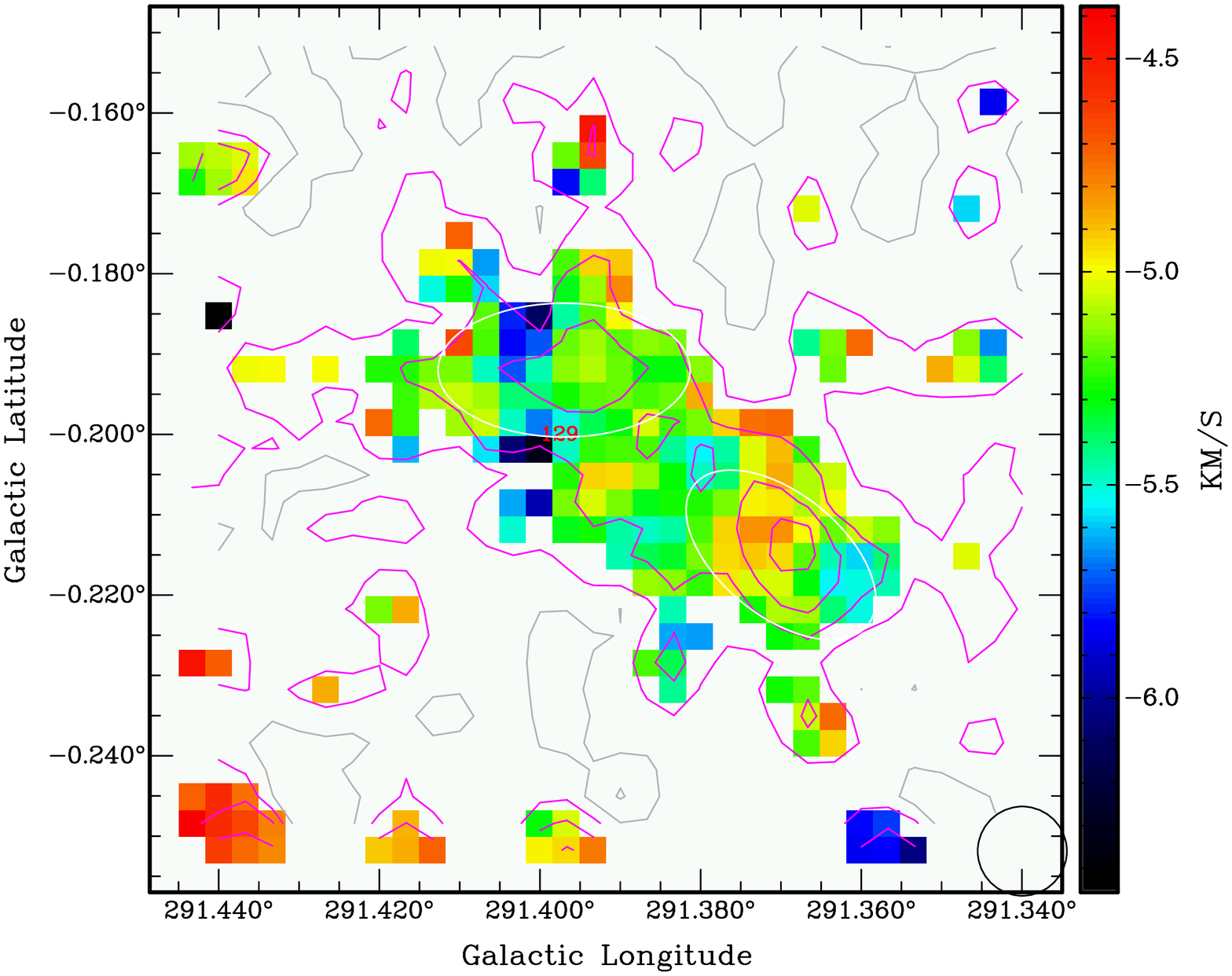}}
		(d){\includegraphics[angle=-90,scale=0.30]{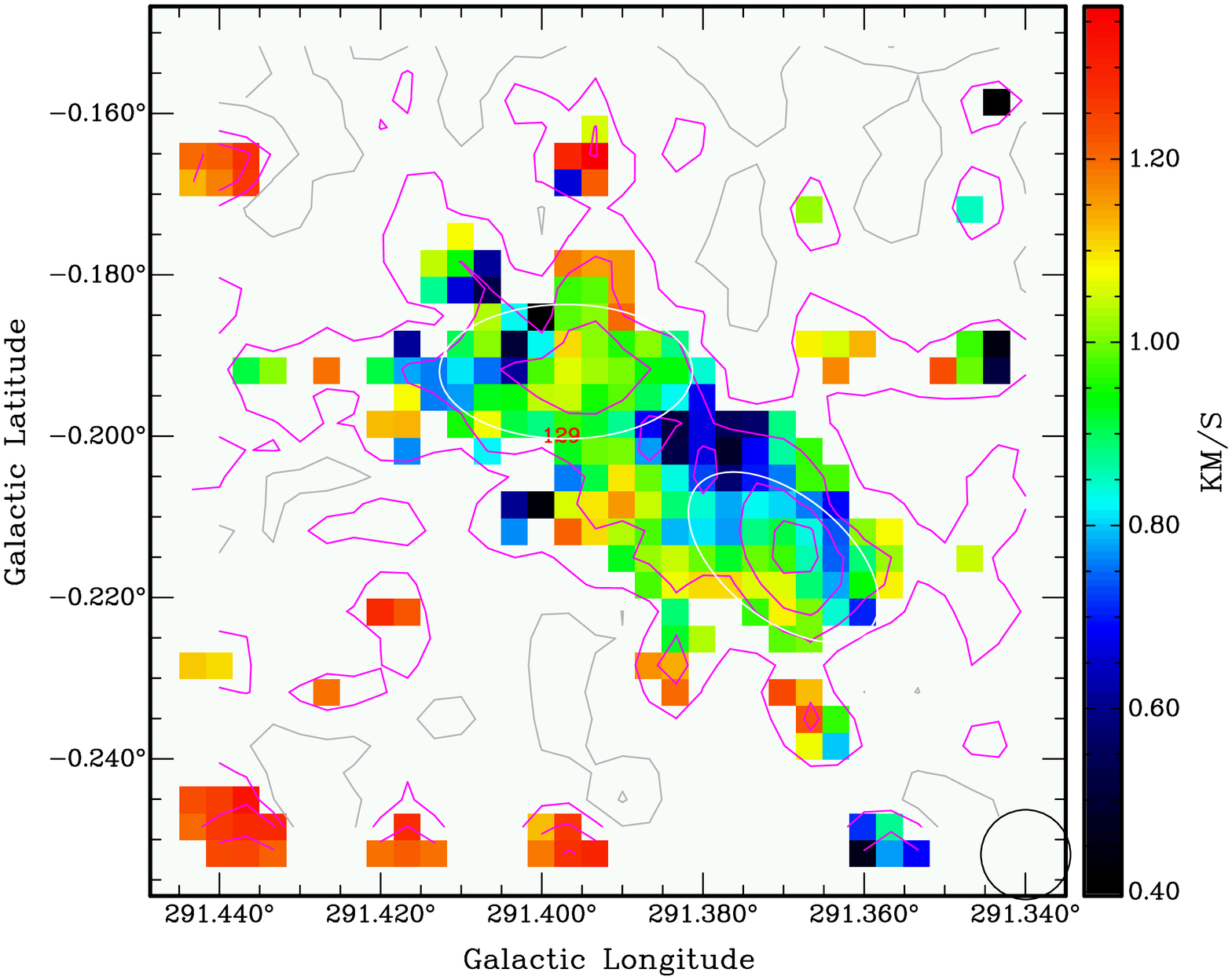}}}
\caption{\small Same as Fig.\,\ref{momR1}, but for Region 13 source BYF\,129.  Contours are every 2$\sigma$ = 0.358\,K\kms, and at 1.2\,kpc the 40$''$ Mopra beam (lower right corner) scales to 0.233\,pc.  ($a$) $T_p$,  ($b$) rms,  ($c$) $V_{\rm LSR}$,  ($d$) $\sigma_{V}$.
\label{momR13b}}
\end{figure*}

\clearpage

\begin{figure*}[ht]
(a){\includegraphics[angle=-90,scale=0.40]{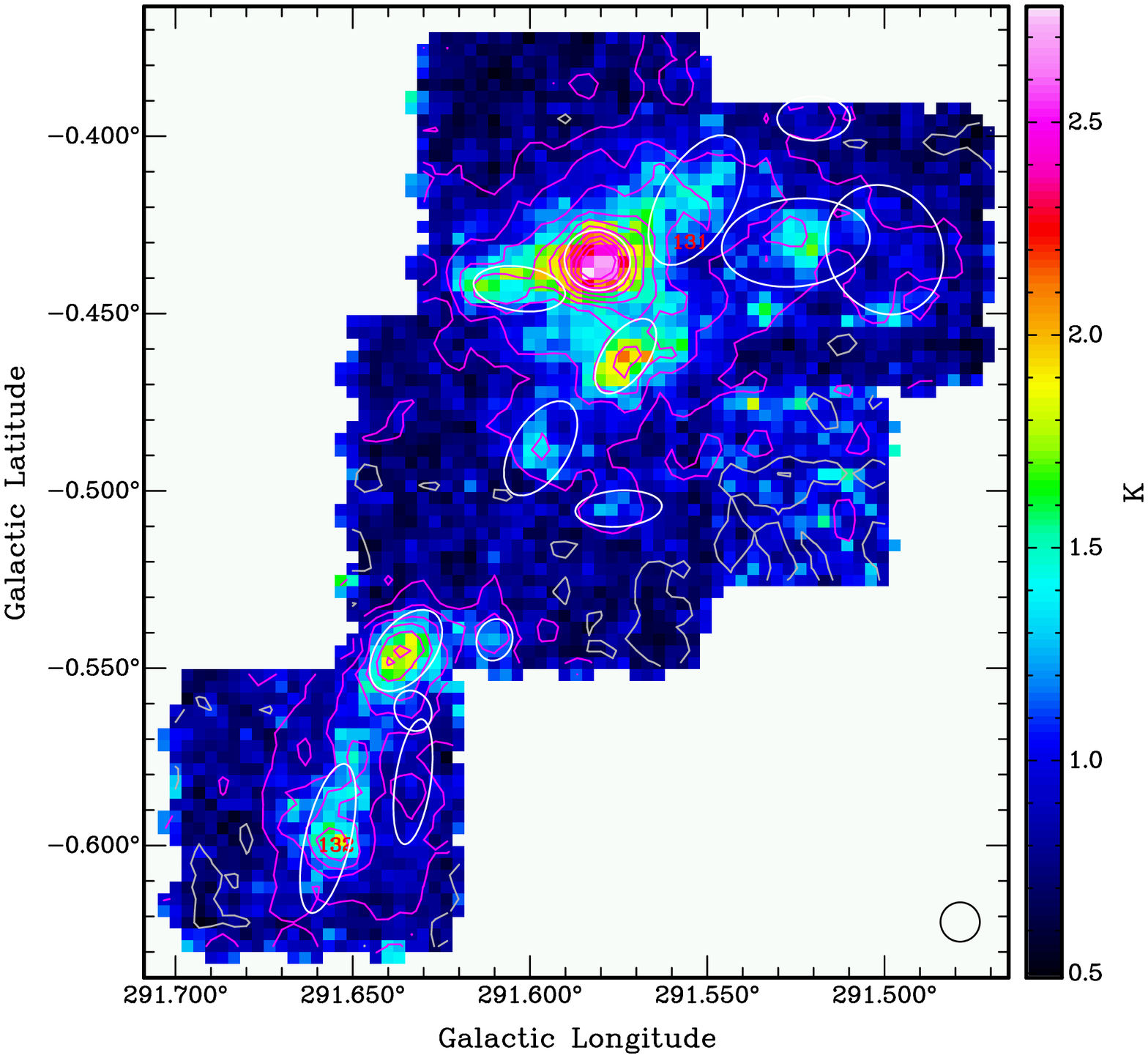}}
(b){\includegraphics[angle=-90,scale=0.40]{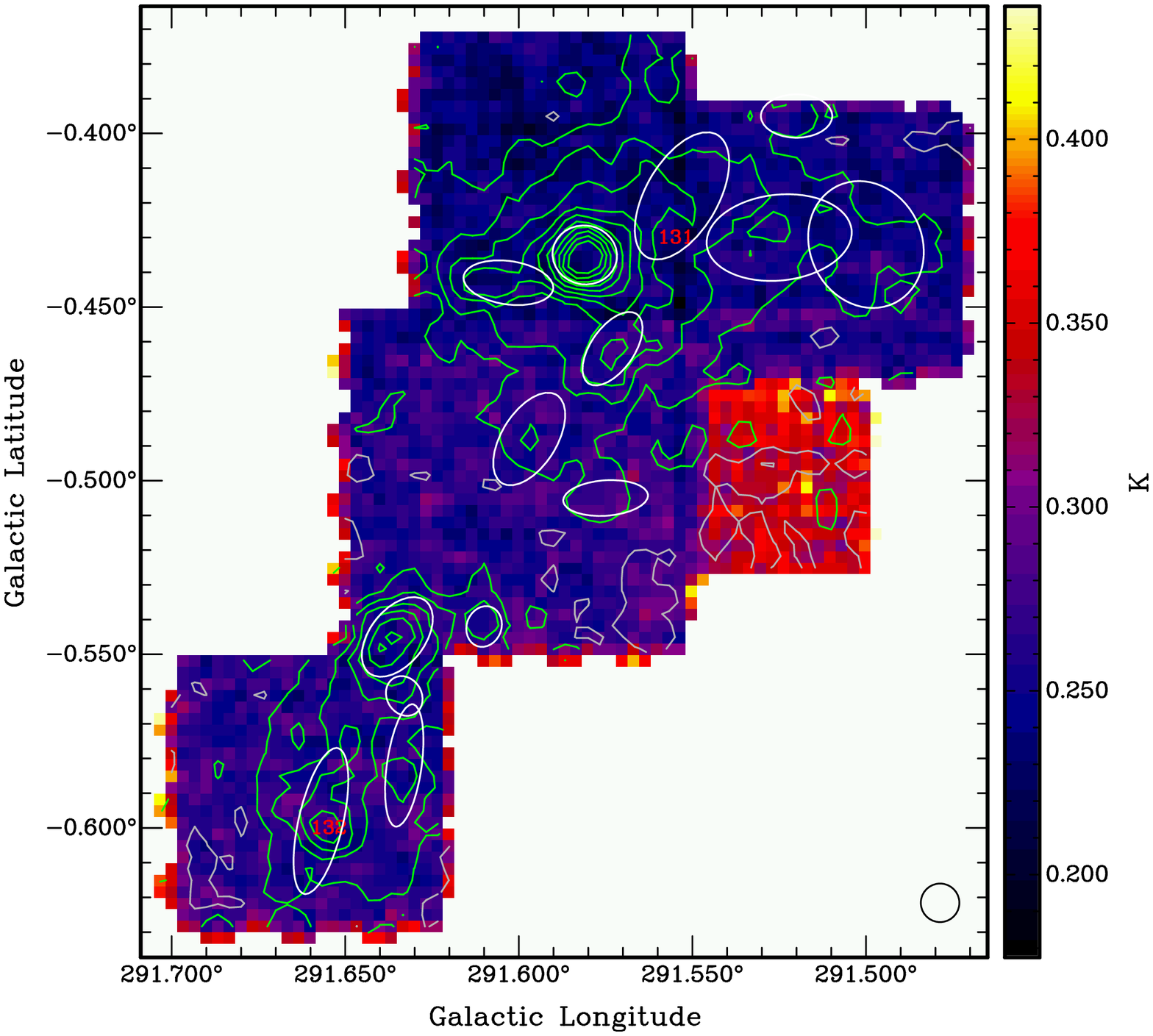}}
(c){\includegraphics[angle=-90,scale=0.40]{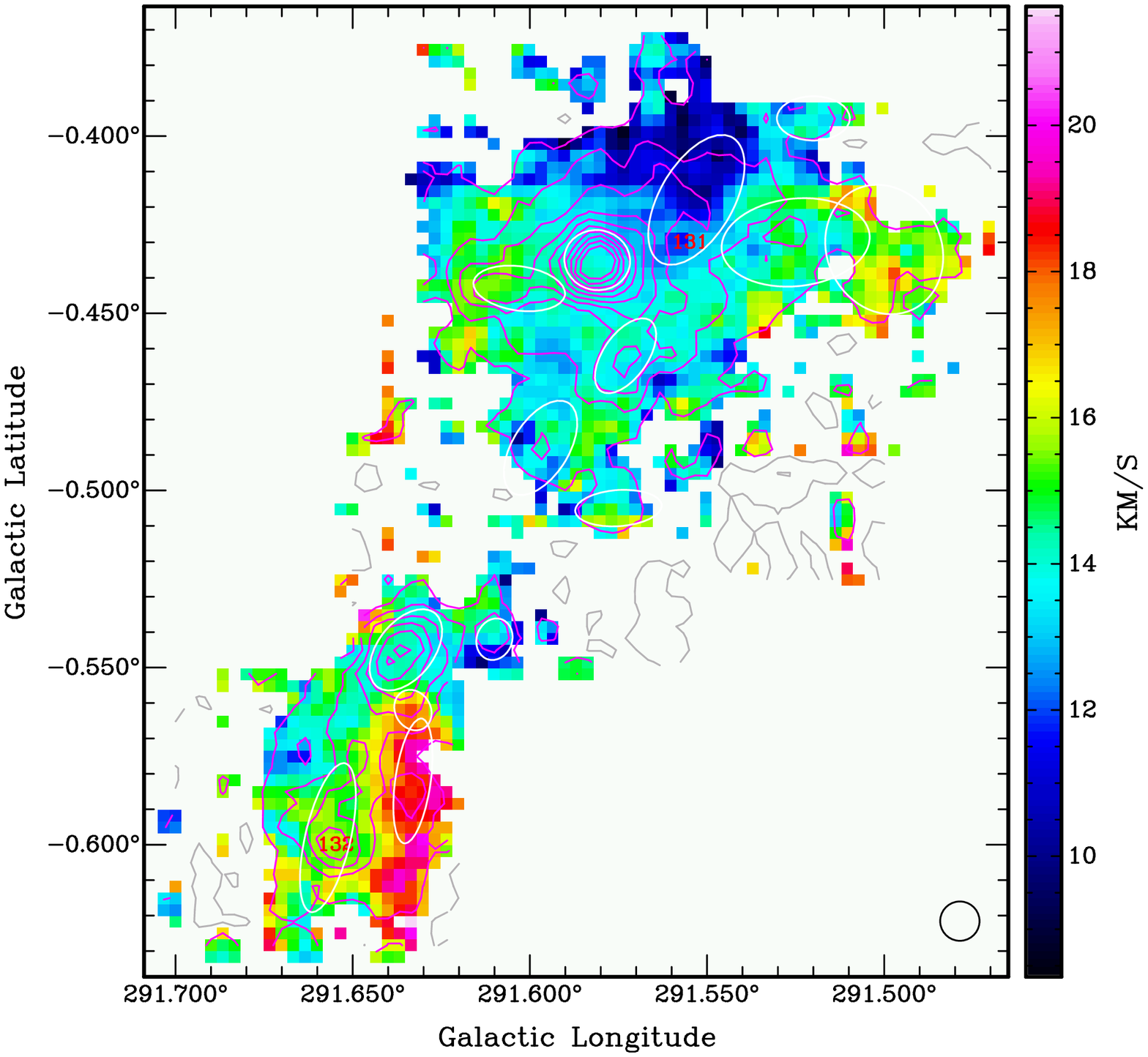}}
(d){\includegraphics[angle=-90,scale=0.40]{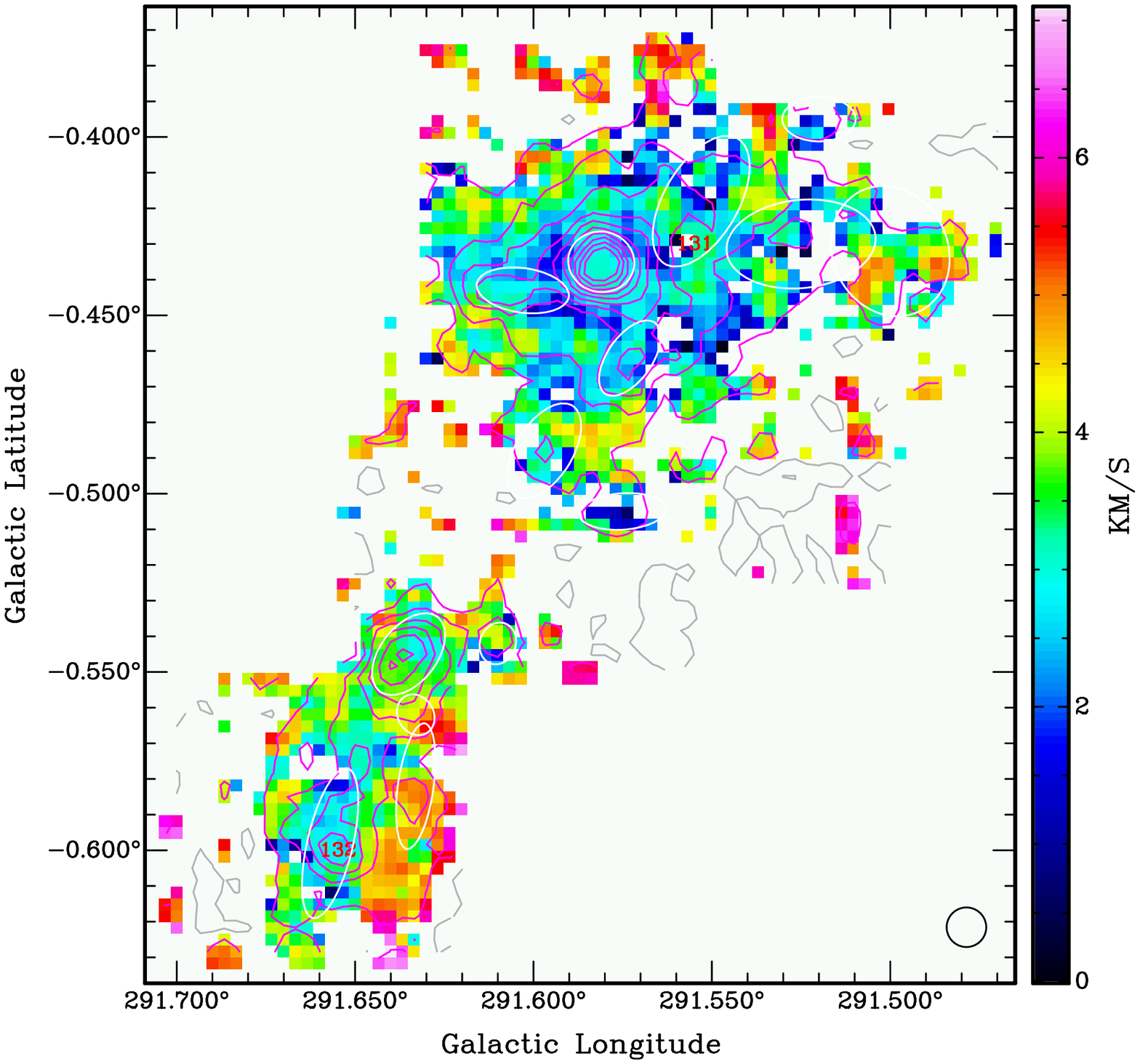}}
\caption{\small Same as Fig.\,\ref{momR1}, but for Region 13 sources BYF\,131 and 132.  Contours are every 4$\sigma$ = 1.60\,K\kms, and at 6.0\,kpc the 40$''$ Mopra beam (lower right corner) scales to 1.16\,pc.  ($a$) $T_p$,  ($b$) rms,  ($c$) $V_{\rm LSR}$,  ($d$) $\sigma_{V}$.
\label{momR13c}}
\end{figure*}

\clearpage

\begin{figure*}[ht]
\centerline{(a){\includegraphics[angle=-90,scale=0.30]{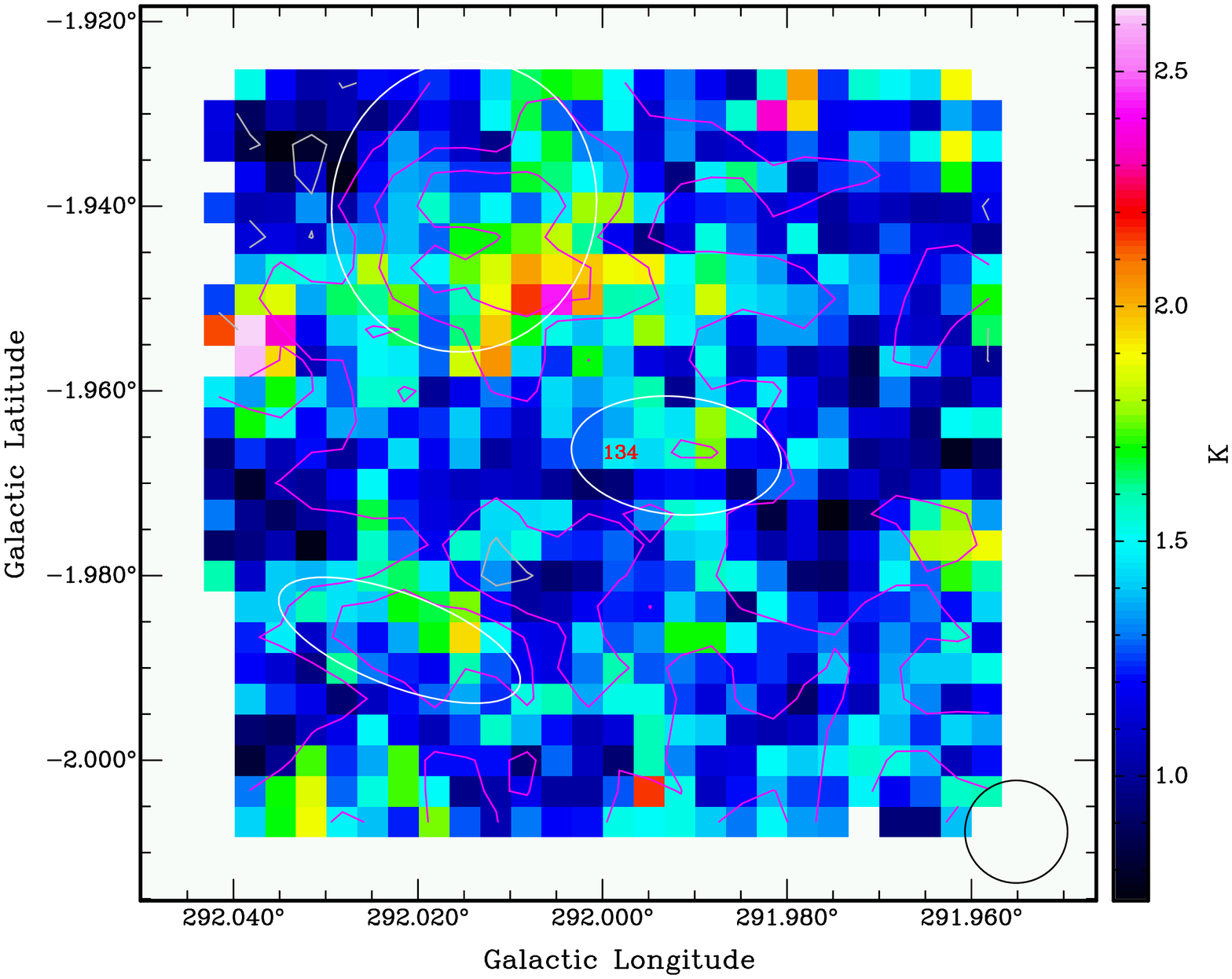}}
		(b){\includegraphics[angle=-90,scale=0.30]{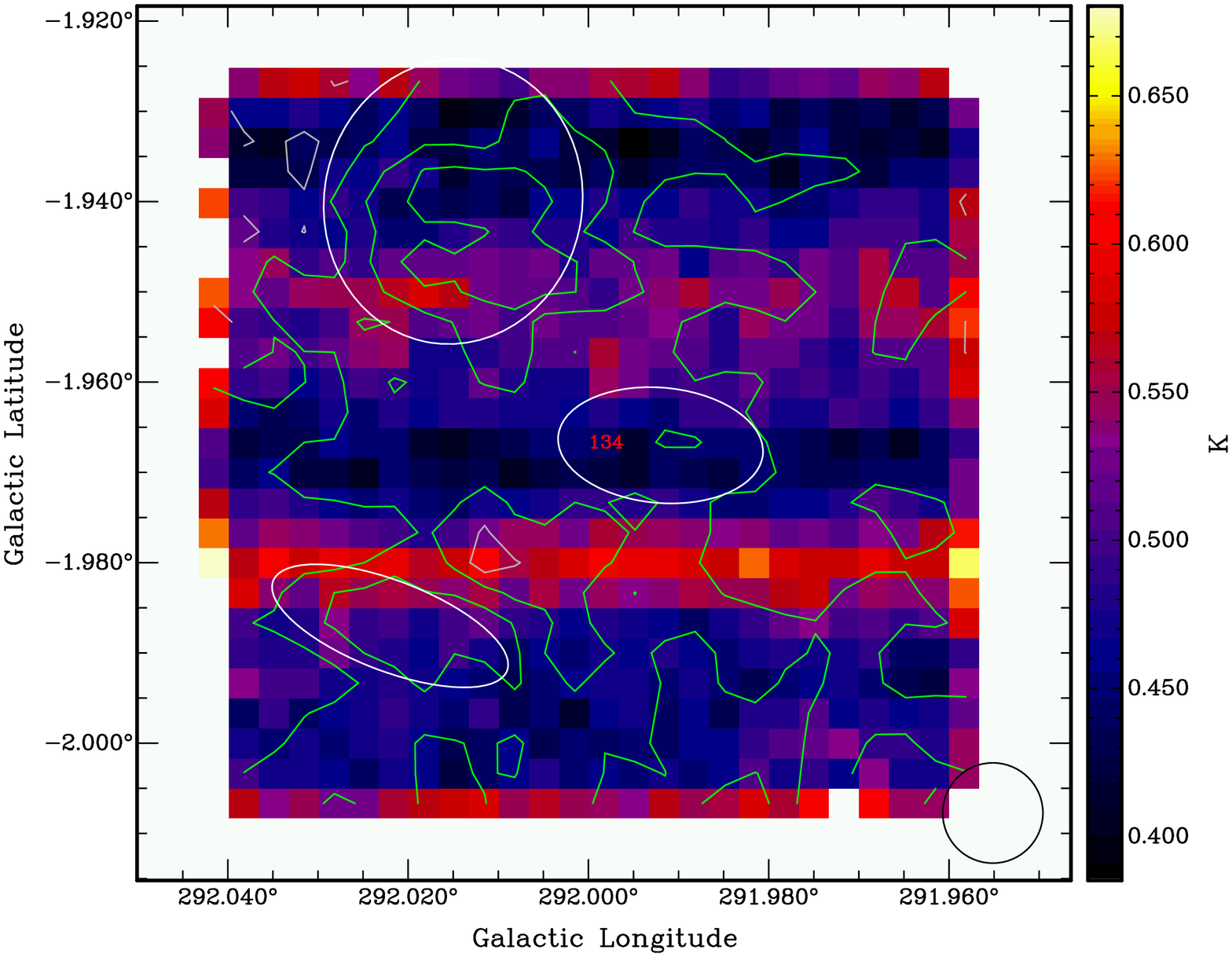}}}
\centerline{(c){\includegraphics[angle=-90,scale=0.30]{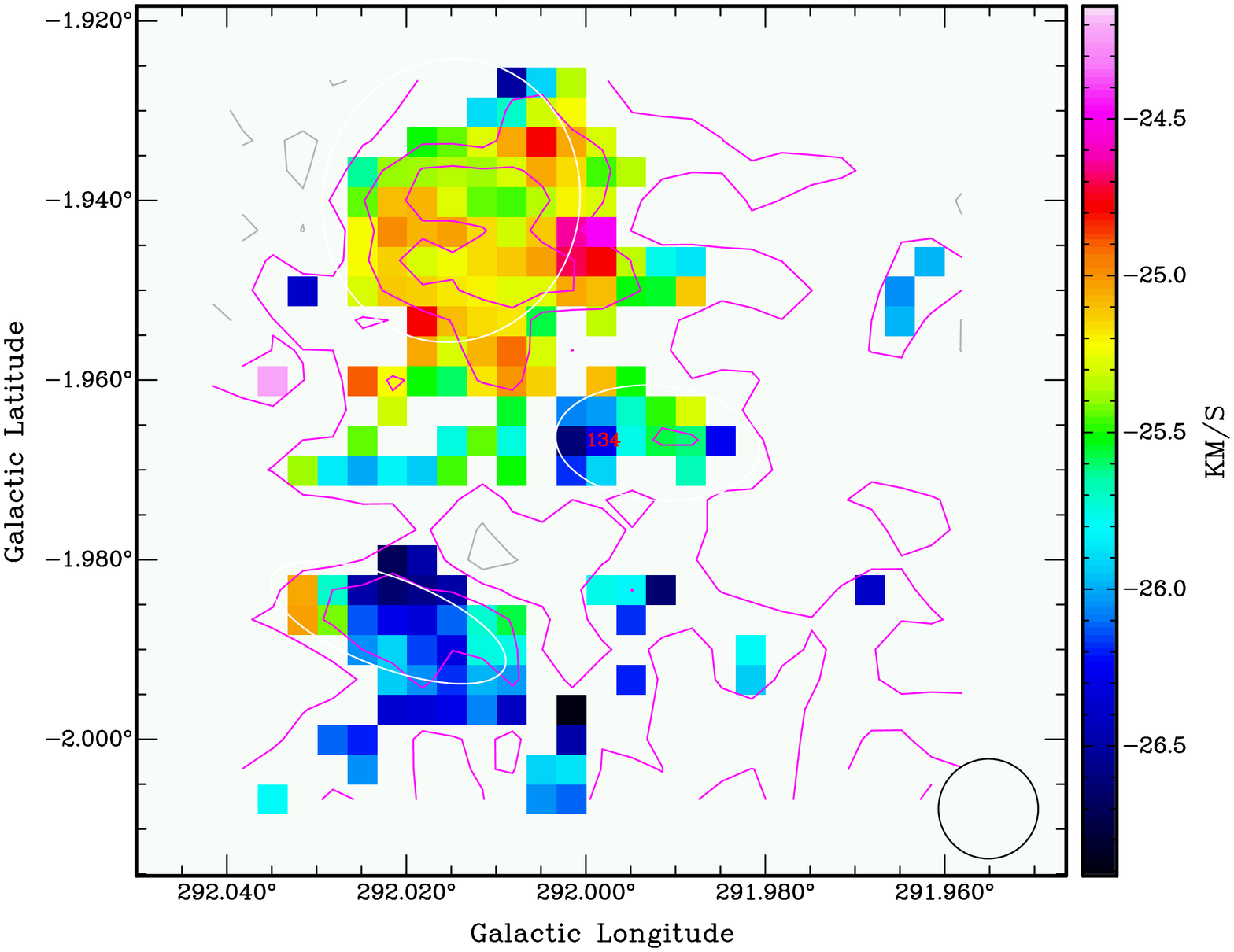}}
		(d){\includegraphics[angle=-90,scale=0.30]{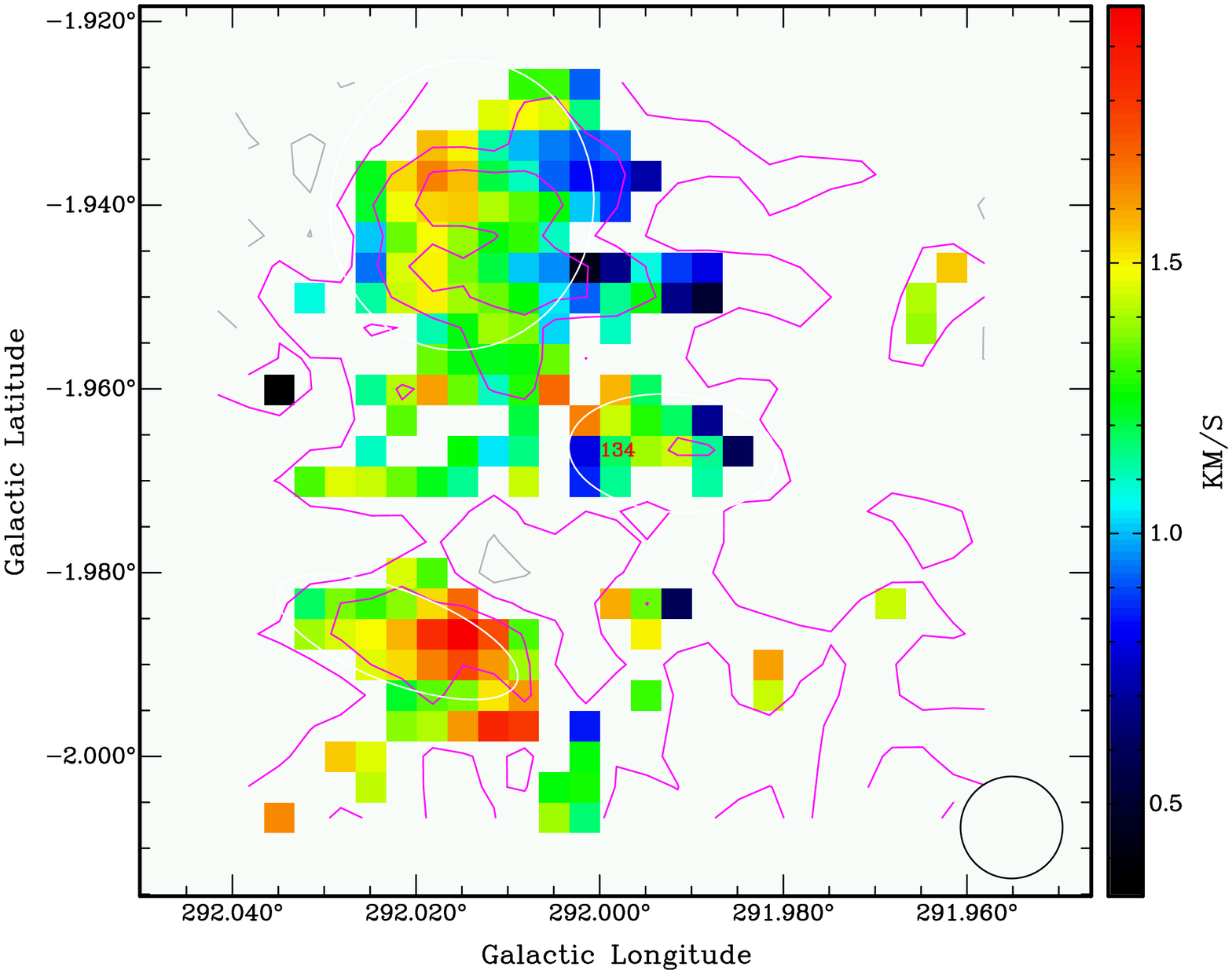}}}
\caption{\small Same as Fig.\,\ref{momR1}, but for Region 15 source BYF\,134.  Contours are every 2$\sigma$ = 0.766\,K\kms, and at 2.4\,kpc the 40$''$ Mopra beam (lower right corner) scales to 0.465\,pc.  ($a$) $T_p$,  ($b$) rms,  ($c$) $V_{\rm LSR}$,  ($d$) $\sigma_{V}$.
\label{momR15}}
\end{figure*}

\clearpage

\begin{figure*}[ht]
(a){\includegraphics[angle=-90,scale=0.30]{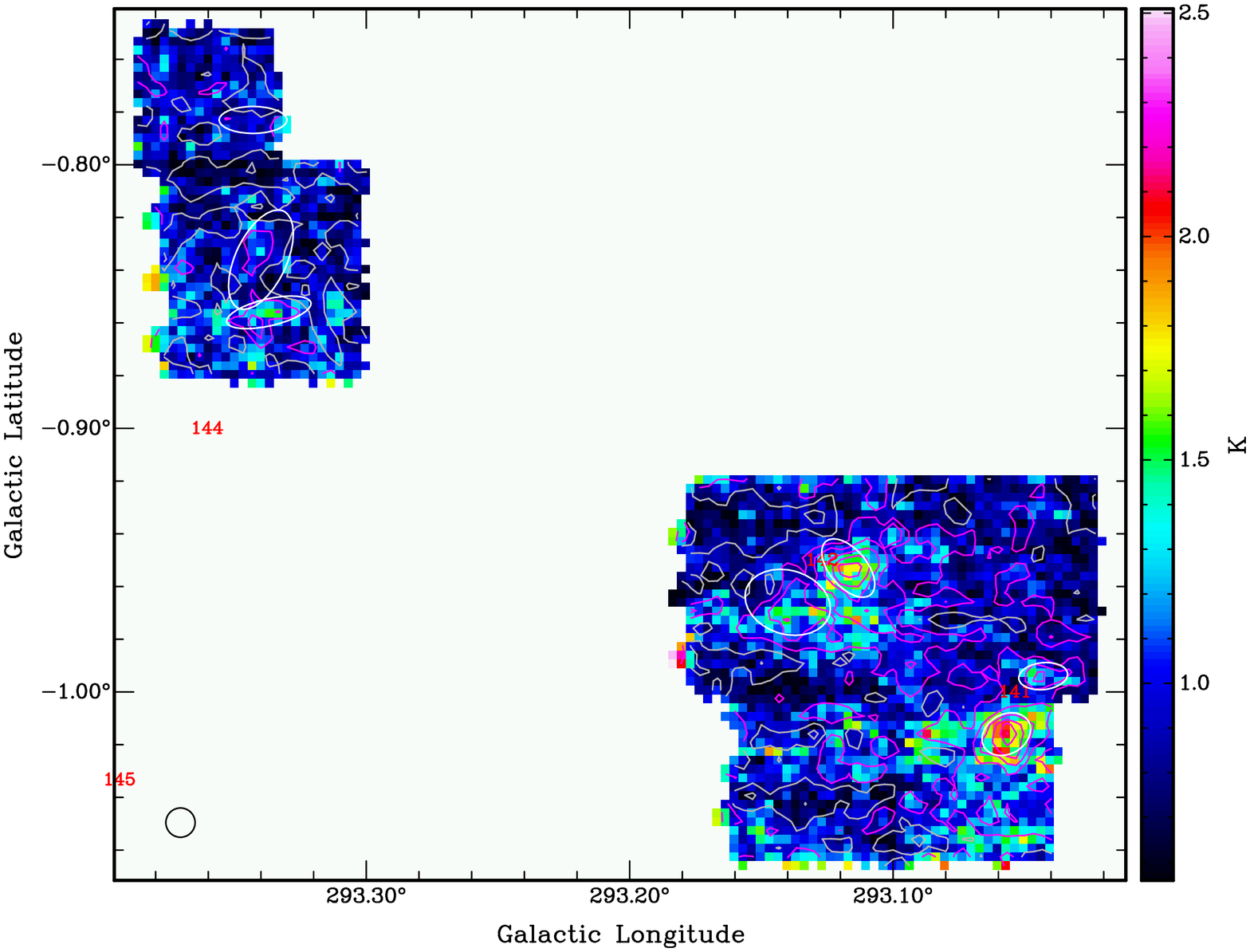}}
(b){\includegraphics[angle=-90,scale=0.30]{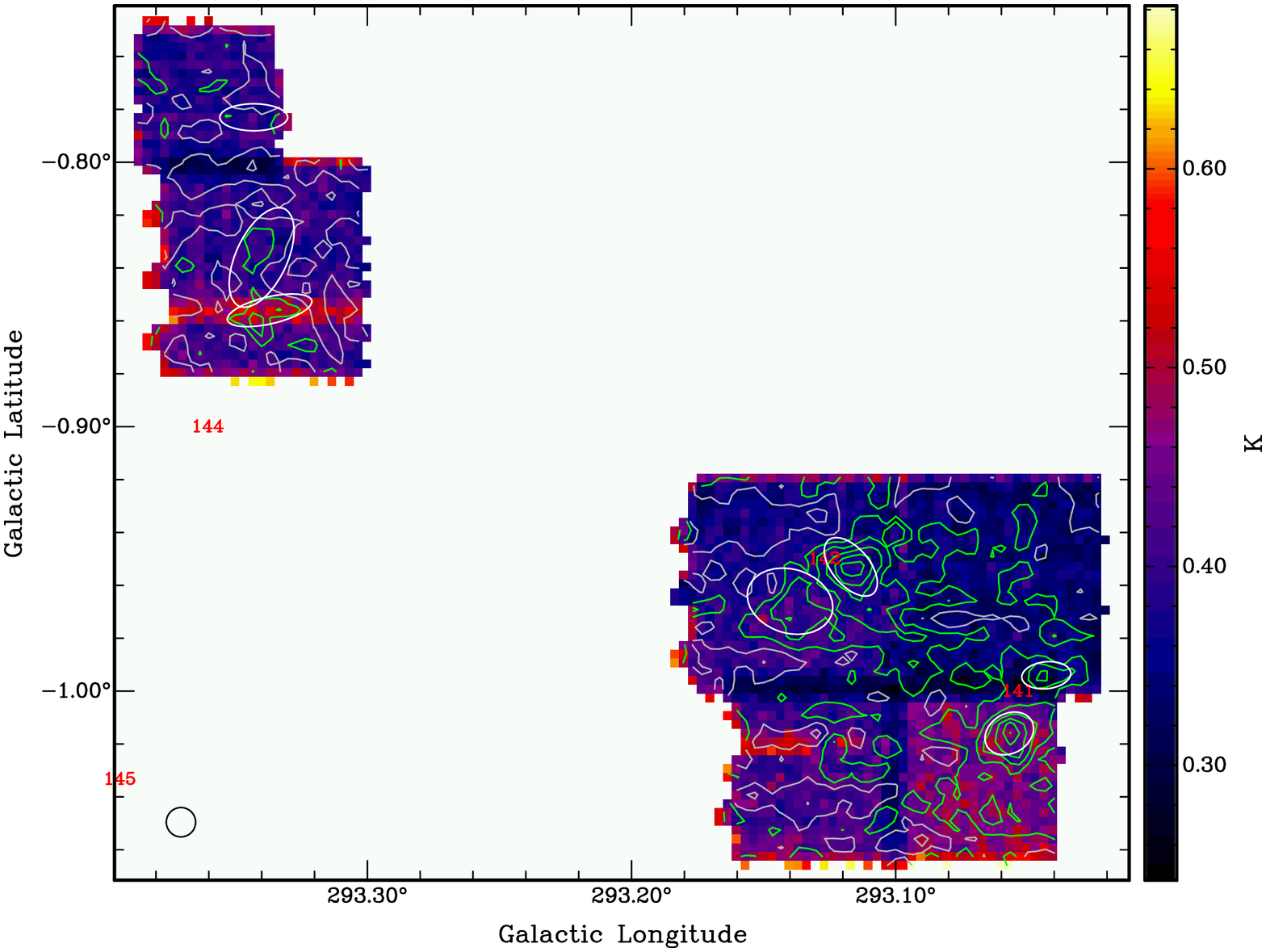}}
(c){\includegraphics[angle=-90,scale=0.30]{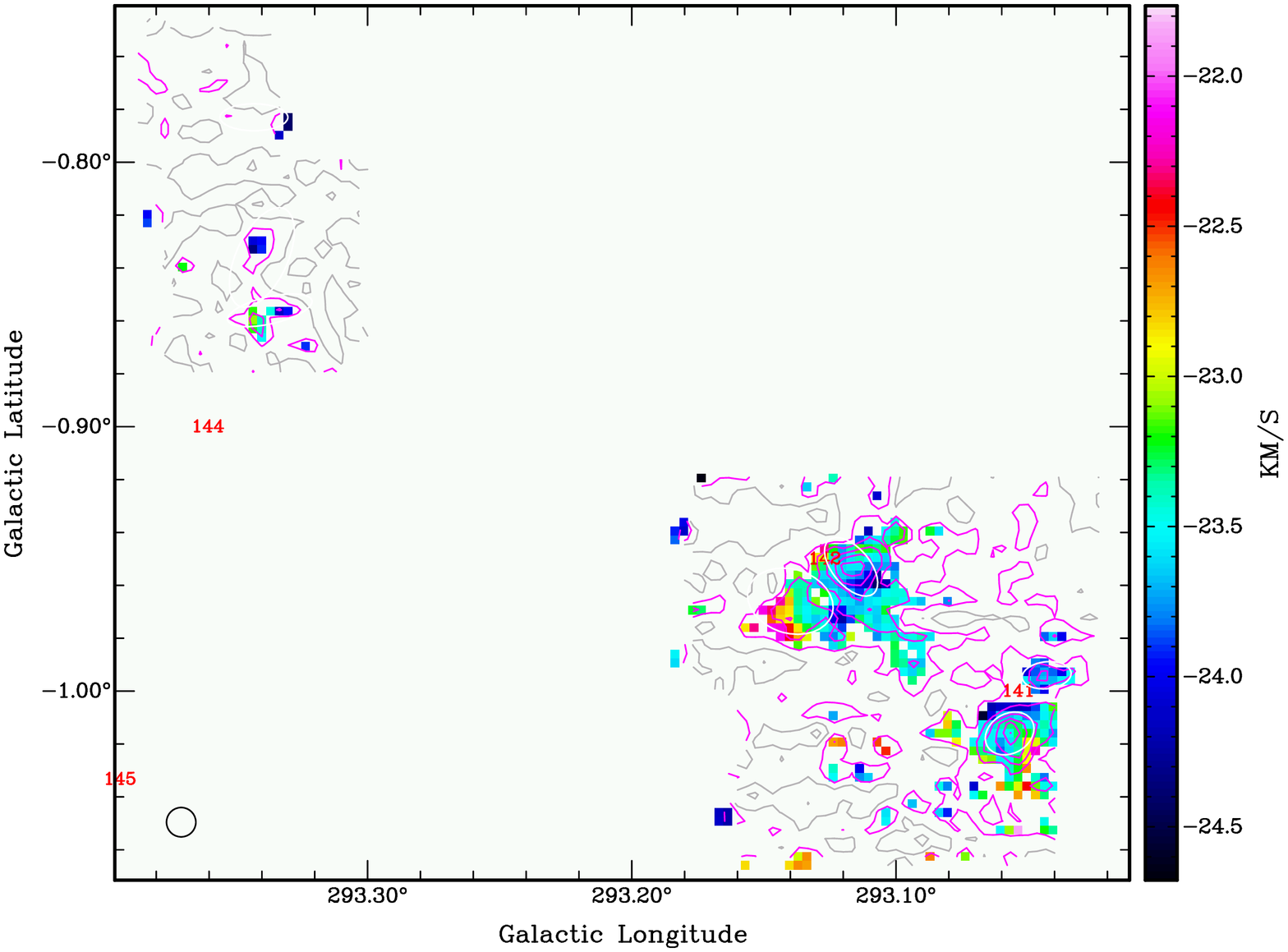}}
(d){\includegraphics[angle=-90,scale=0.30]{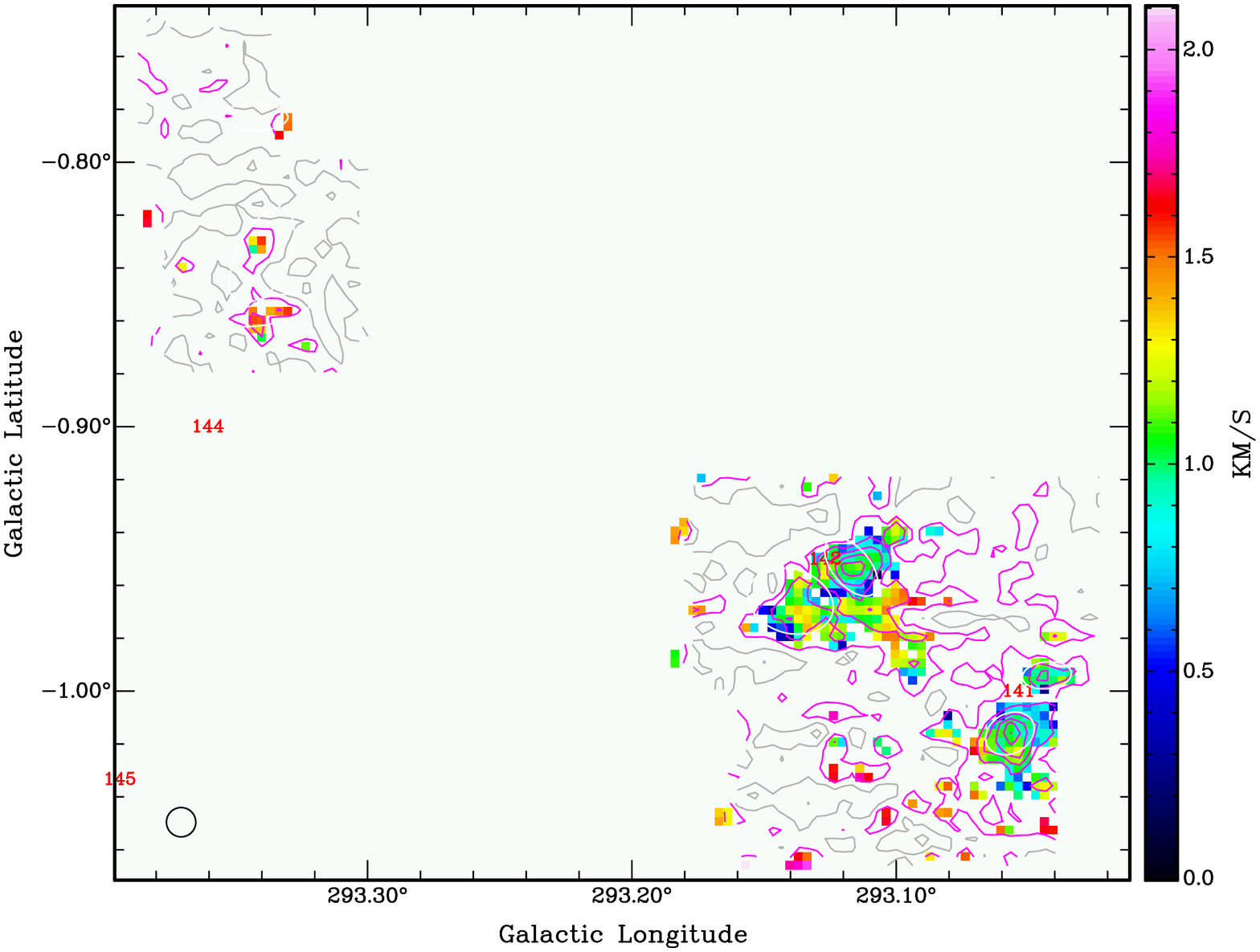}}
\caption{\small Same as Fig.\,\ref{momR1}, but for Region 16 sources BYF\,141 and 142.  Contours are every 2$\sigma$ = 0.606\,K\kms, and at 2.4\,kpc the 40$''$ Mopra beam (lower left corner) scales to 0.465\,pc.  ($a$) $T_p$,  ($b$) rms,  ($c$) $V_{\rm LSR}$,  ($d$) $\sigma_{V}$.
\label{momR16a}}
\end{figure*}

\clearpage

\begin{figure*}[ht]
(a){\includegraphics[angle=-90,scale=0.30]{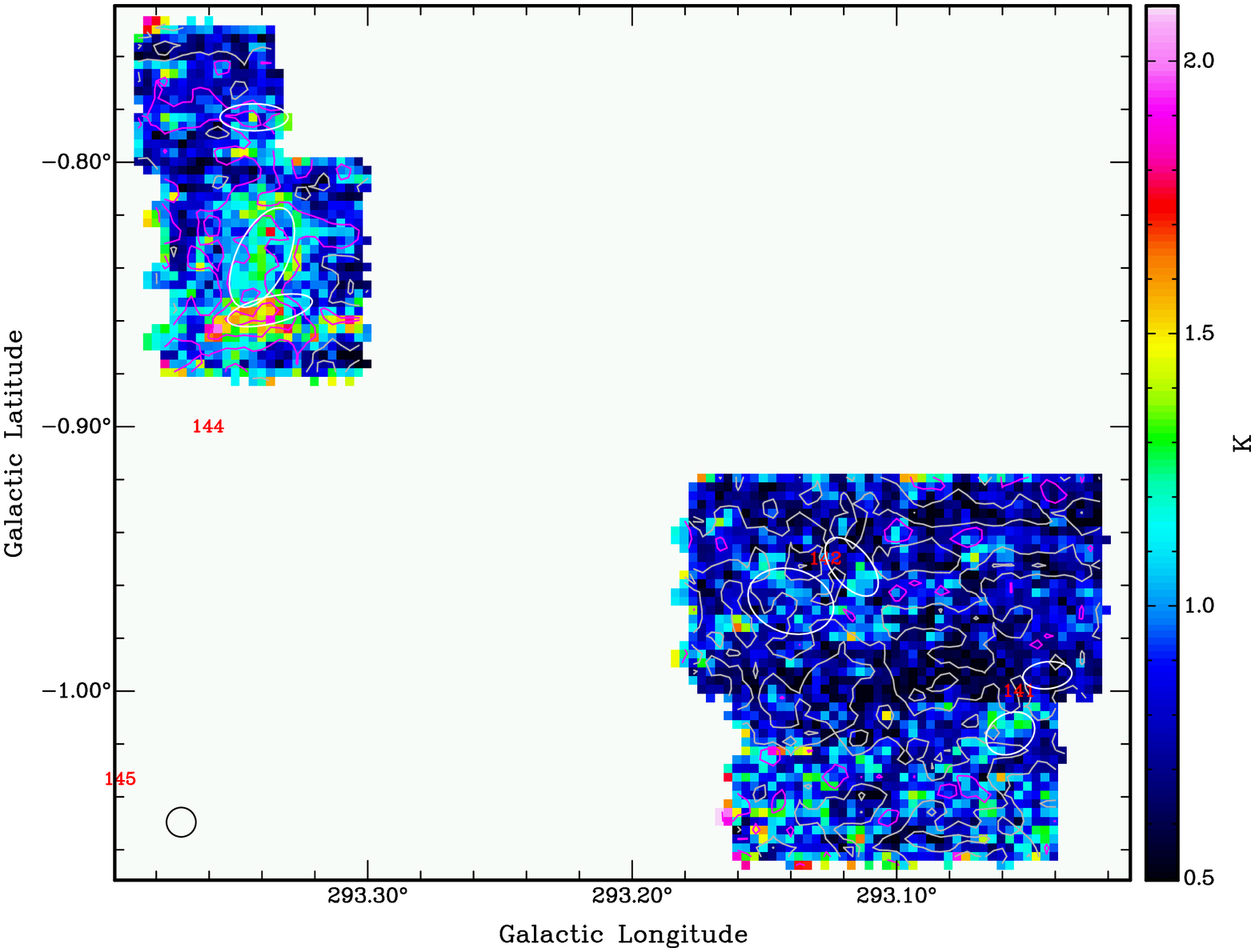}}
(b){\includegraphics[angle=-90,scale=0.30]{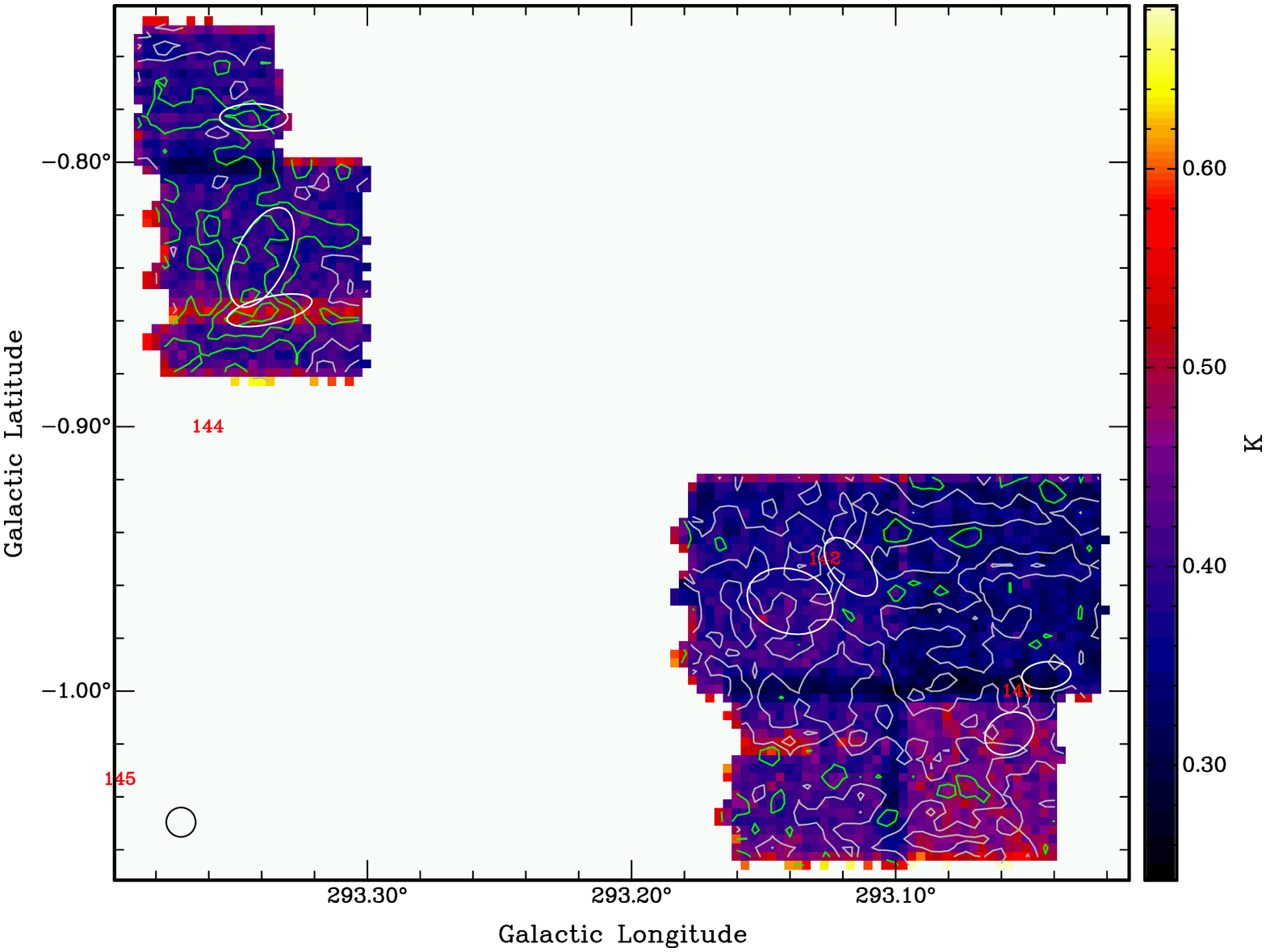}}
(c){\includegraphics[angle=-90,scale=0.30]{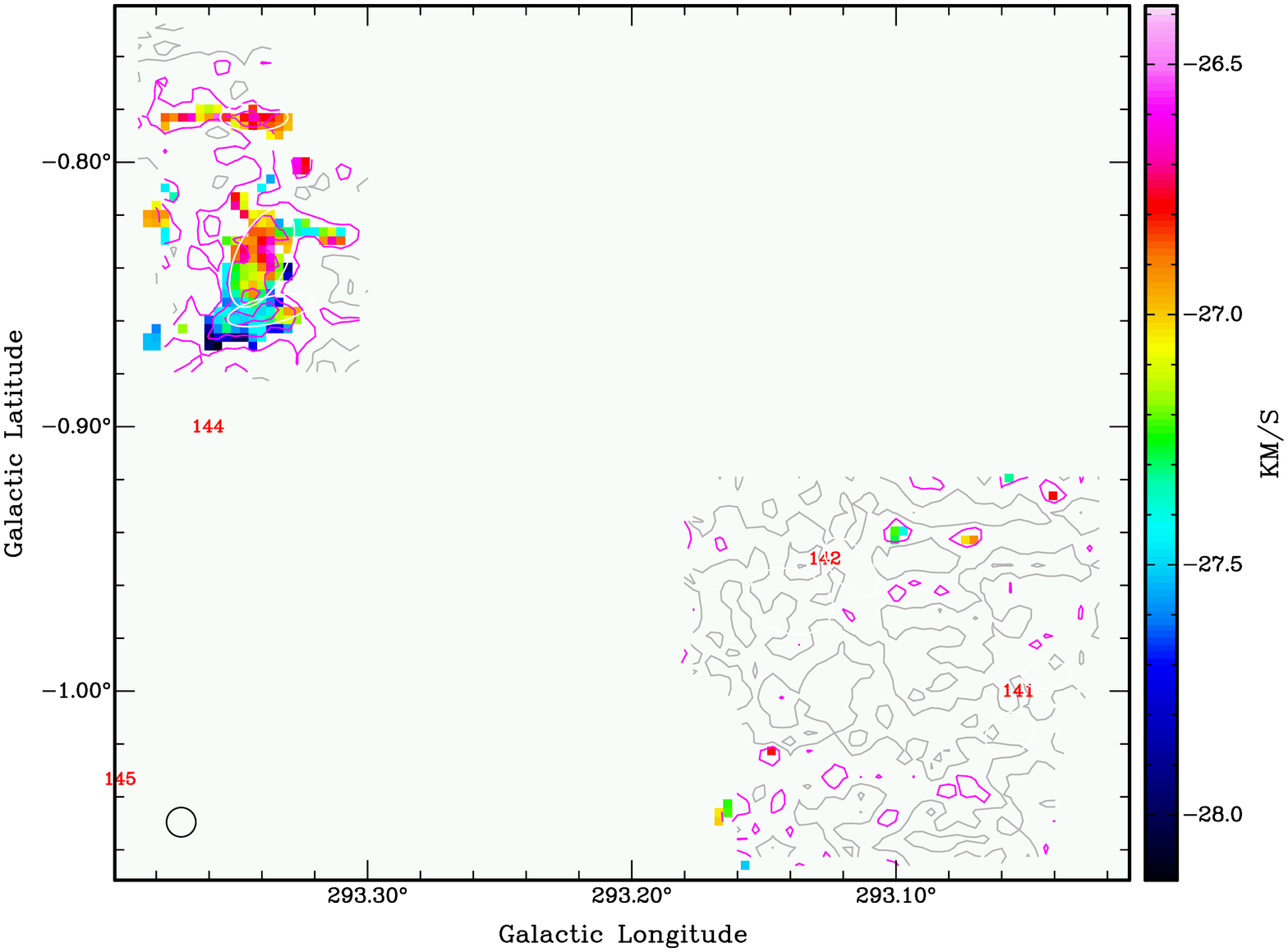}}
(d){\includegraphics[angle=-90,scale=0.30]{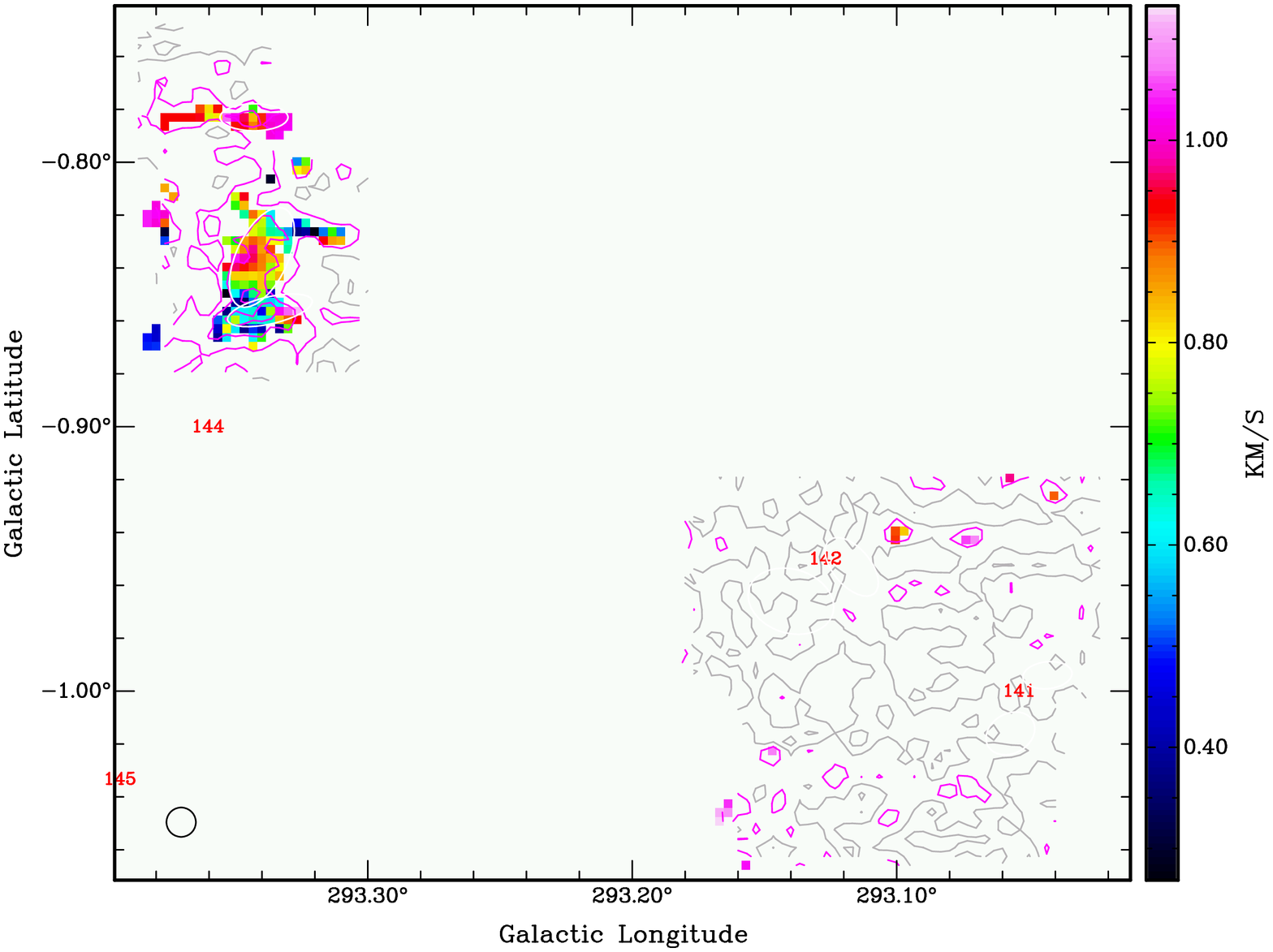}}
\caption{\small Same as Fig.\,\ref{momR1}, but for Region 16 source BYF\,144.  Contours are every 2$\sigma$ = 0.500\,K\kms, and at 2.4\,kpc the 40$''$ Mopra beam (lower left corner) scales to 0.465\,pc.  ($a$) $T_p$,  ($b$) rms,  ($c$) $V_{\rm LSR}$,  ($d$) $\sigma_{V}$.
\label{momR16b}}
\end{figure*}

\newpage

\begin{figure*}[ht]
(a){\includegraphics[angle=-90,scale=0.30]{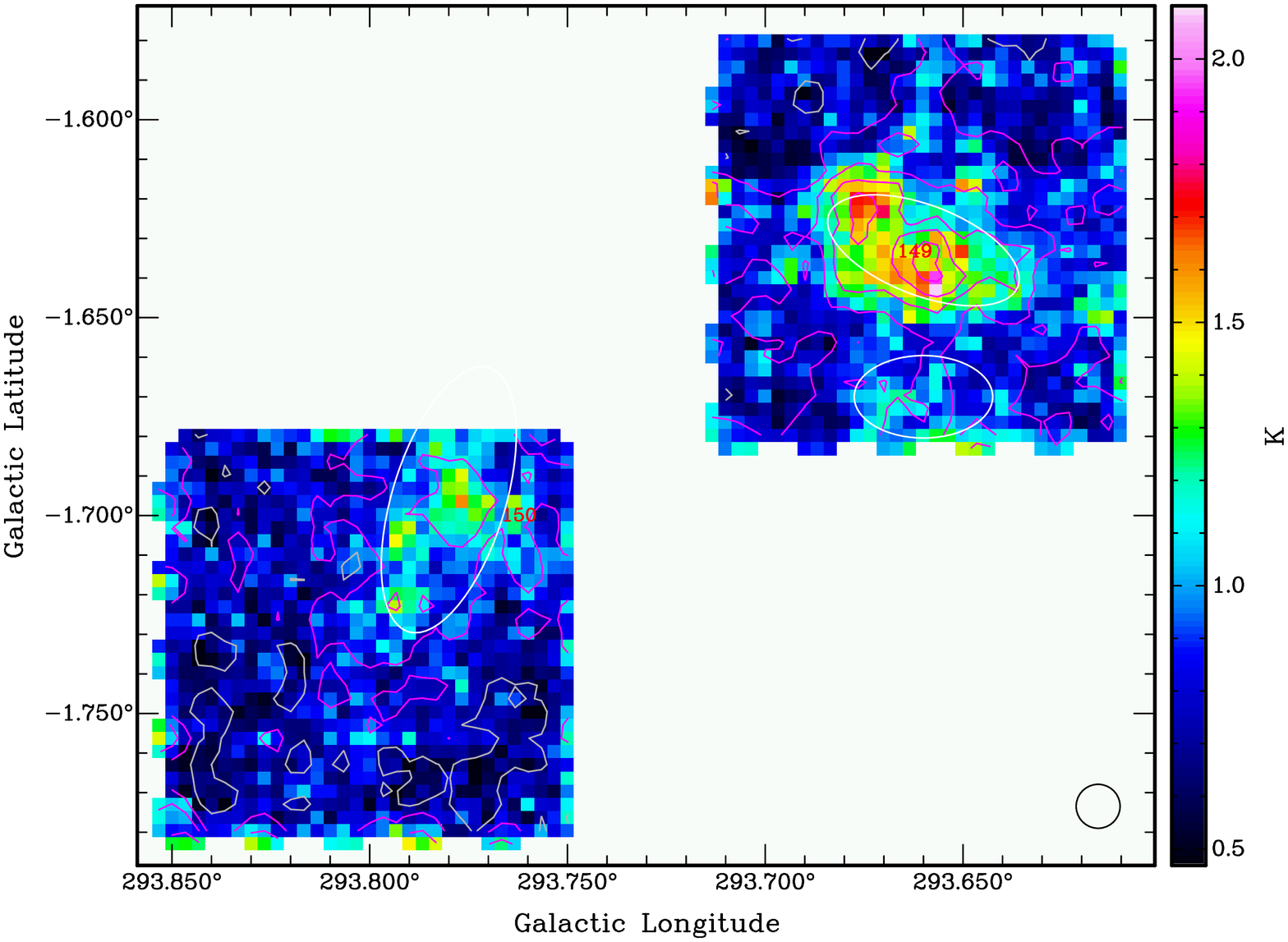}}
(b){\includegraphics[angle=-90,scale=0.30]{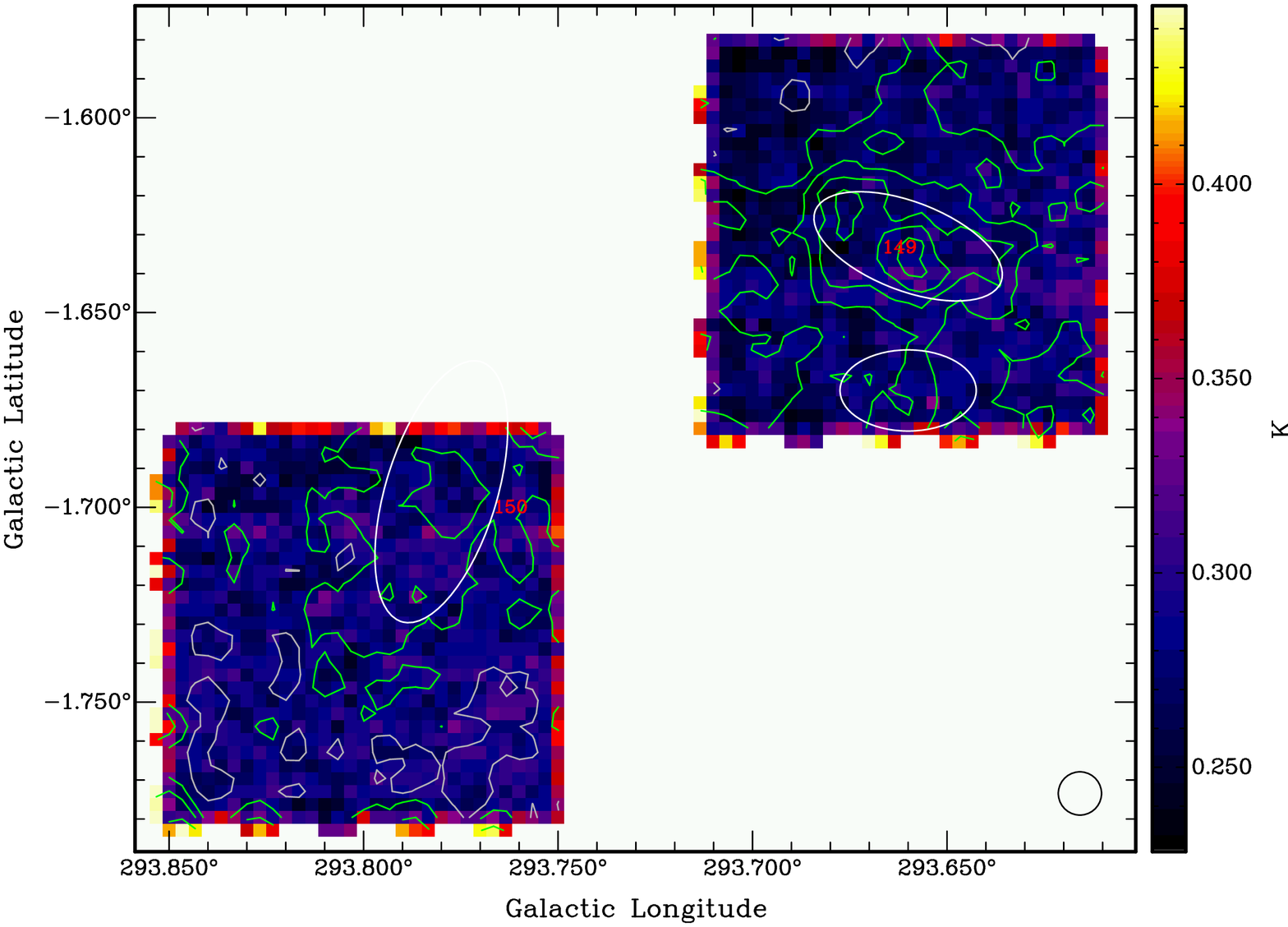}}
(c){\includegraphics[angle=-90,scale=0.30]{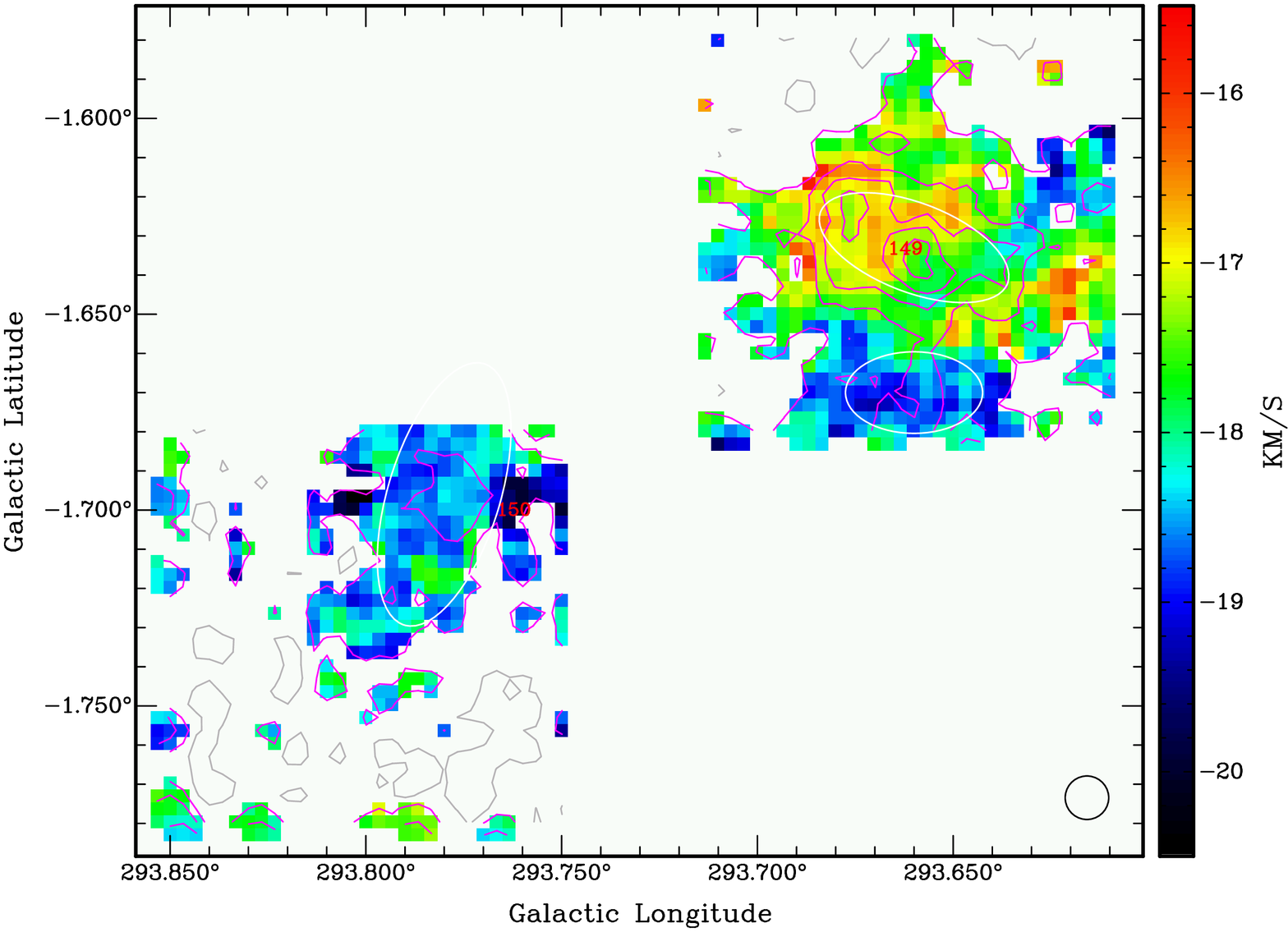}}
(d){\includegraphics[angle=-90,scale=0.30]{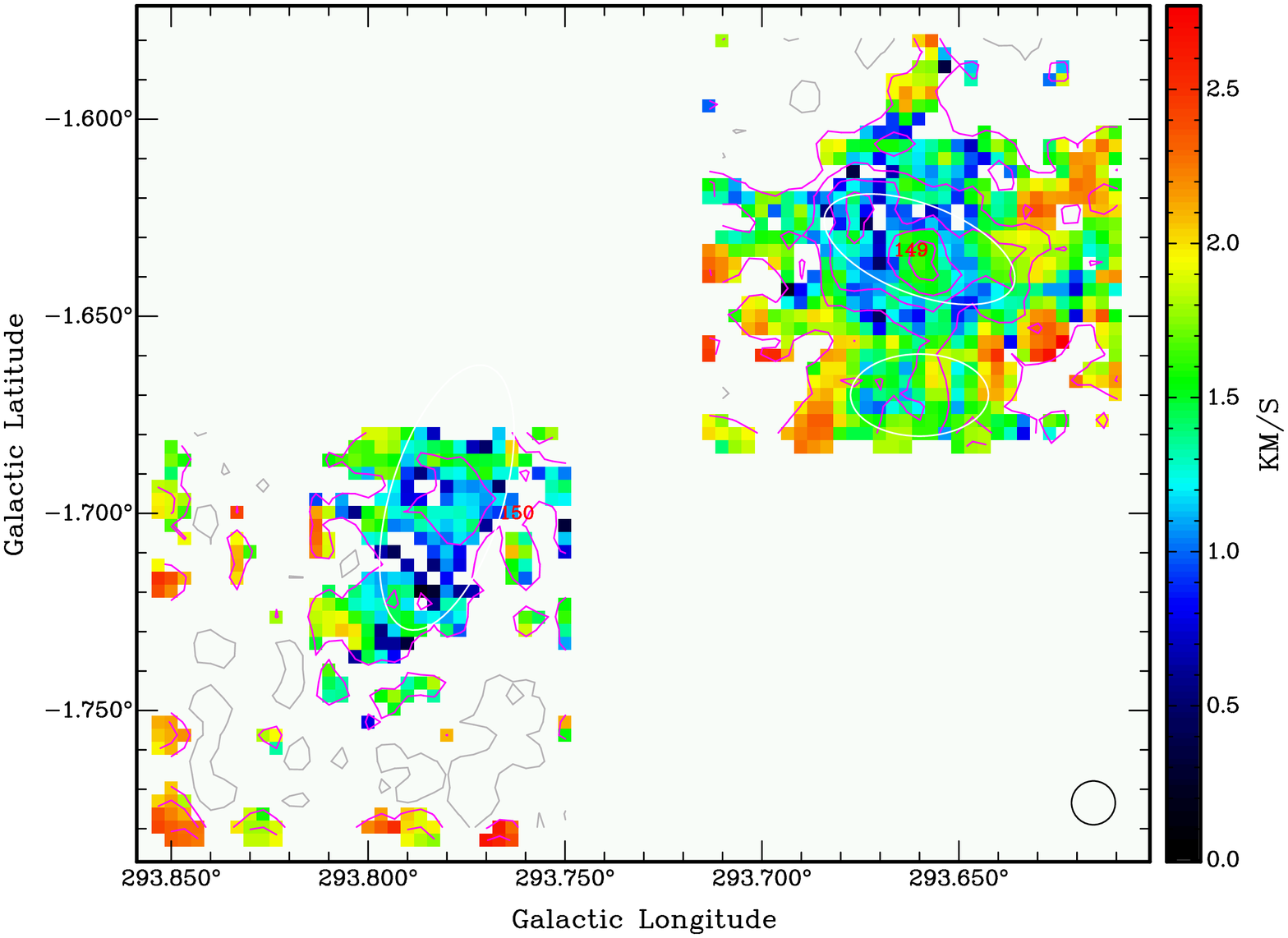}}
\caption{\small Same as Fig.\,\ref{momR1}, but for Region 18 sources BYF\,149 and 150.  Contours are every 3$\sigma$ = 0.810\,K\kms, and at 2.4\,kpc the 40$''$ Mopra beam (lower right corner) scales to 0.465\,pc.  ($a$) $T_p$,  ($b$) rms,  ($c$) $V_{\rm LSR}$,  ($d$) $\sigma_{V}$.
\label{momR18}}
\end{figure*}

\newpage

\begin{figure*}[ht]
\centerline{(a){\includegraphics[angle=-90,scale=0.31]{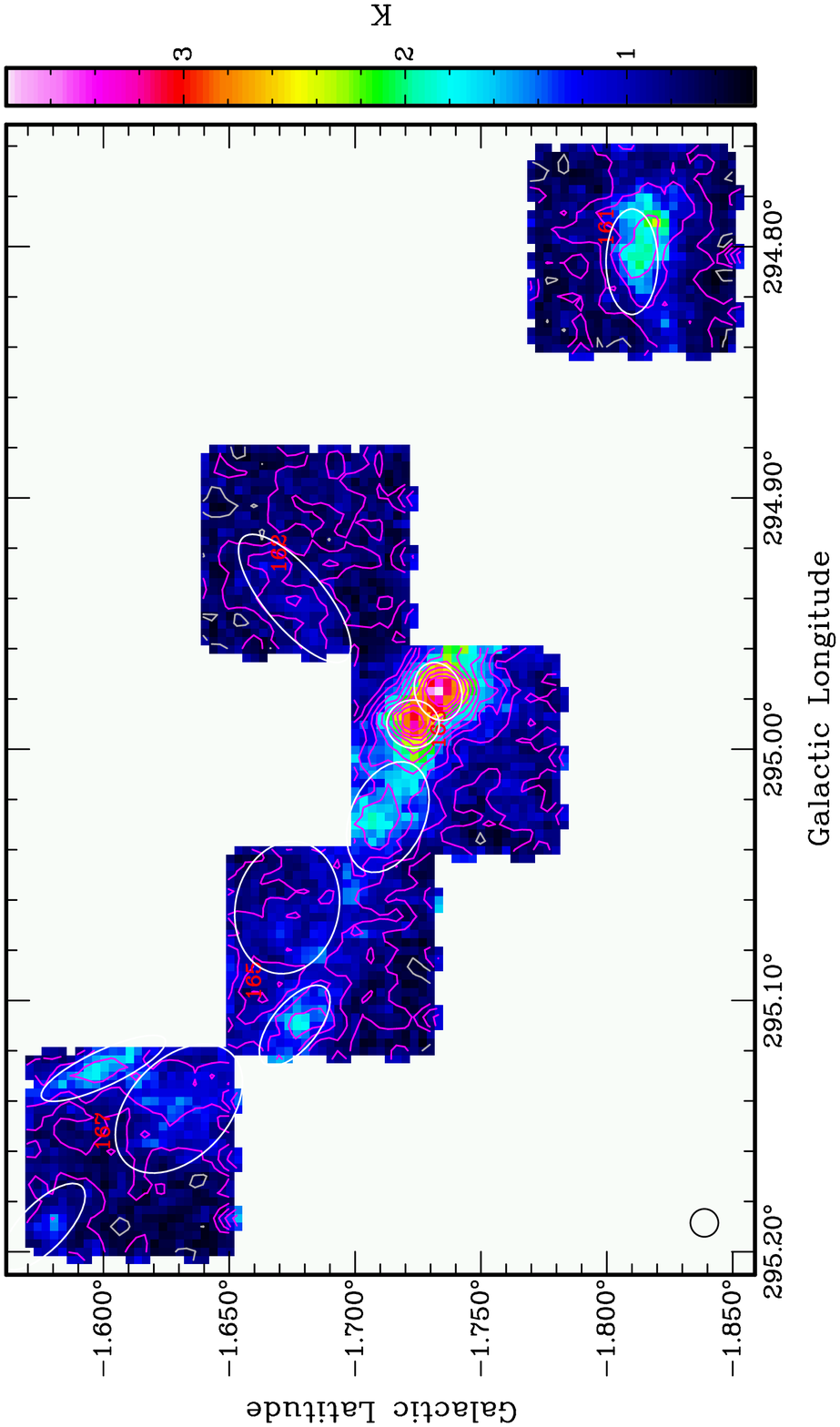}}}
\centerline{(b){\includegraphics[angle=-90,scale=0.31]{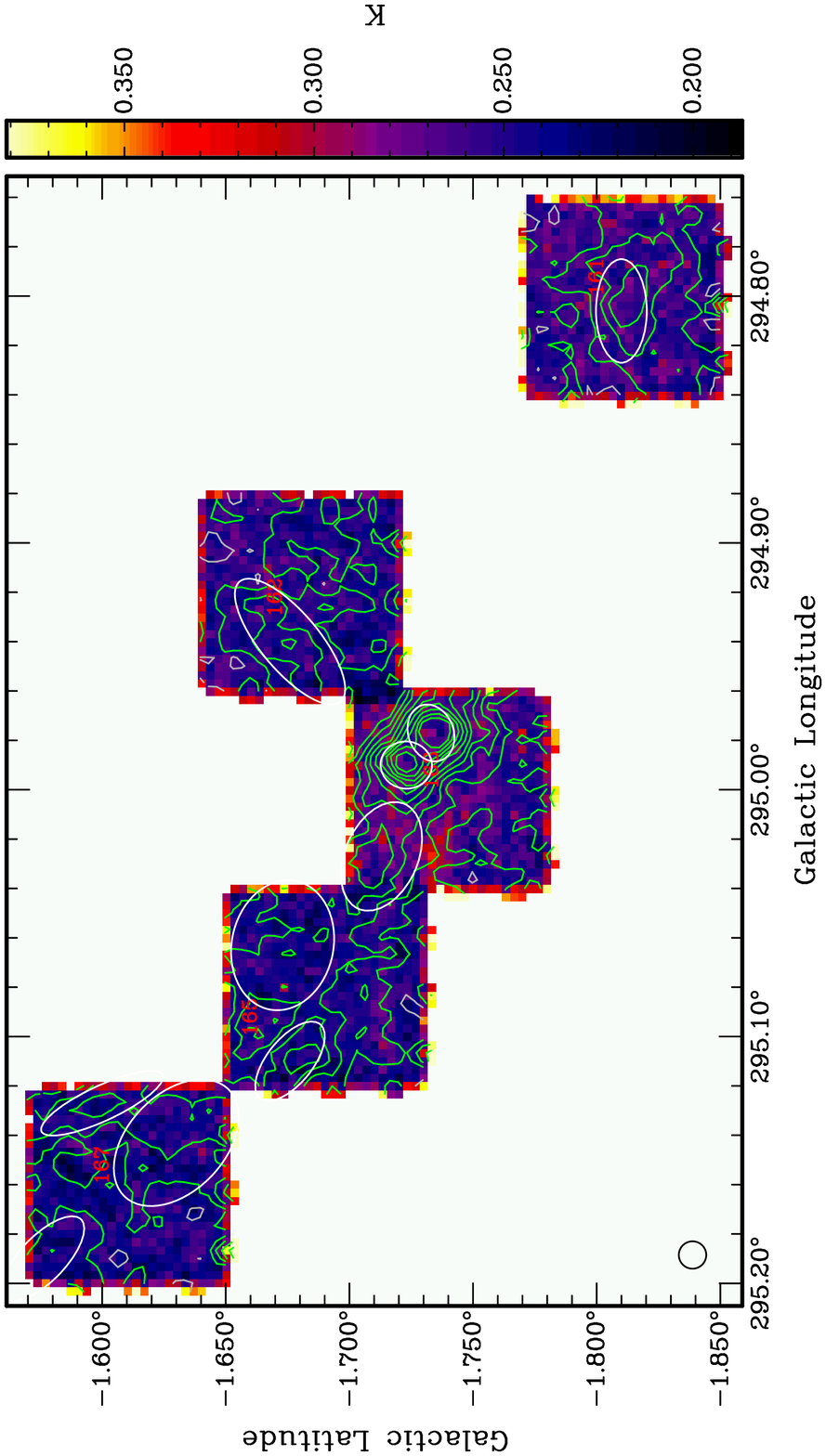}}}
\centerline{(c){\includegraphics[angle=-90,scale=0.31]{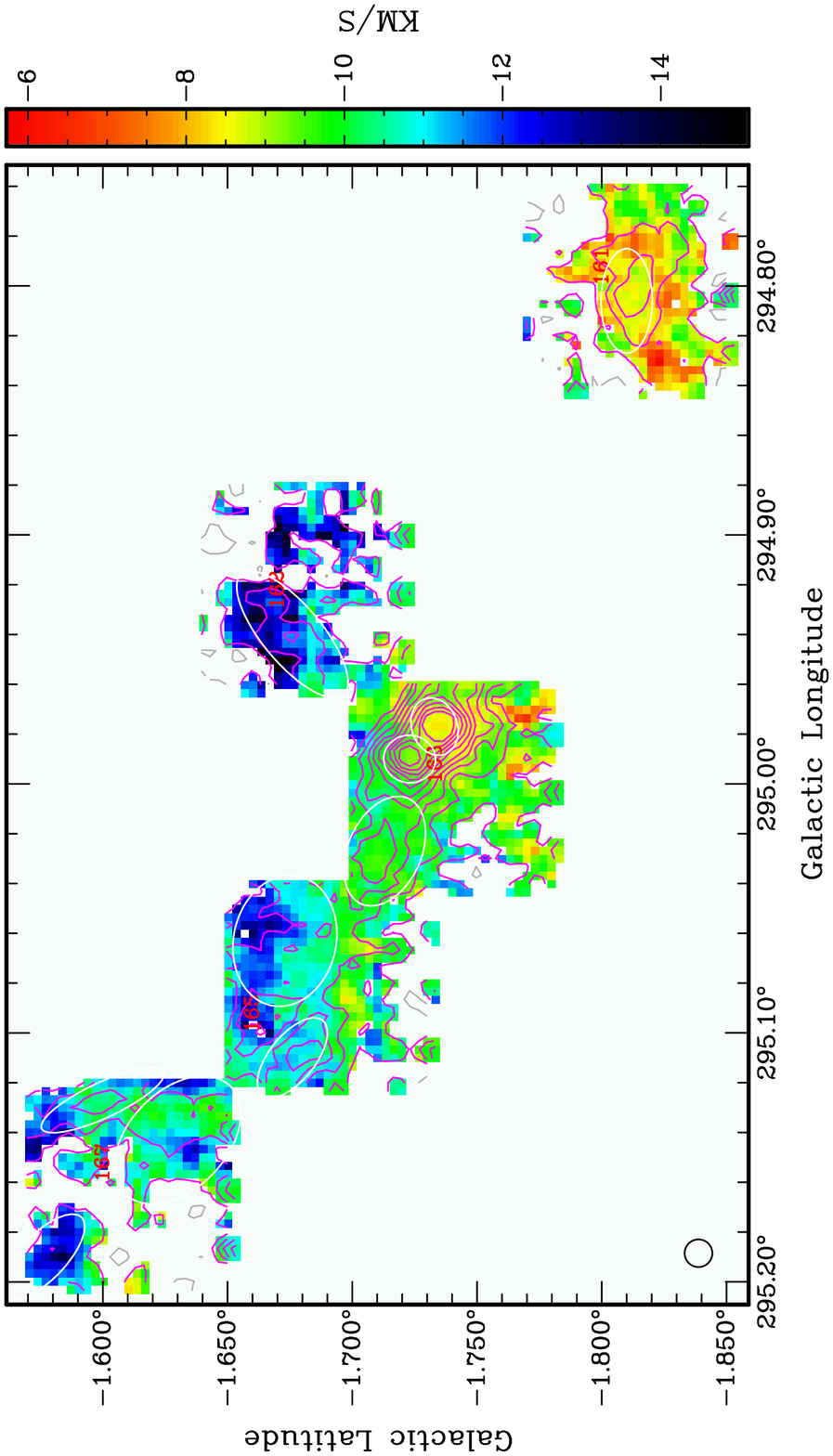}}}
\centerline{(d){\includegraphics[angle=-90,scale=0.31]{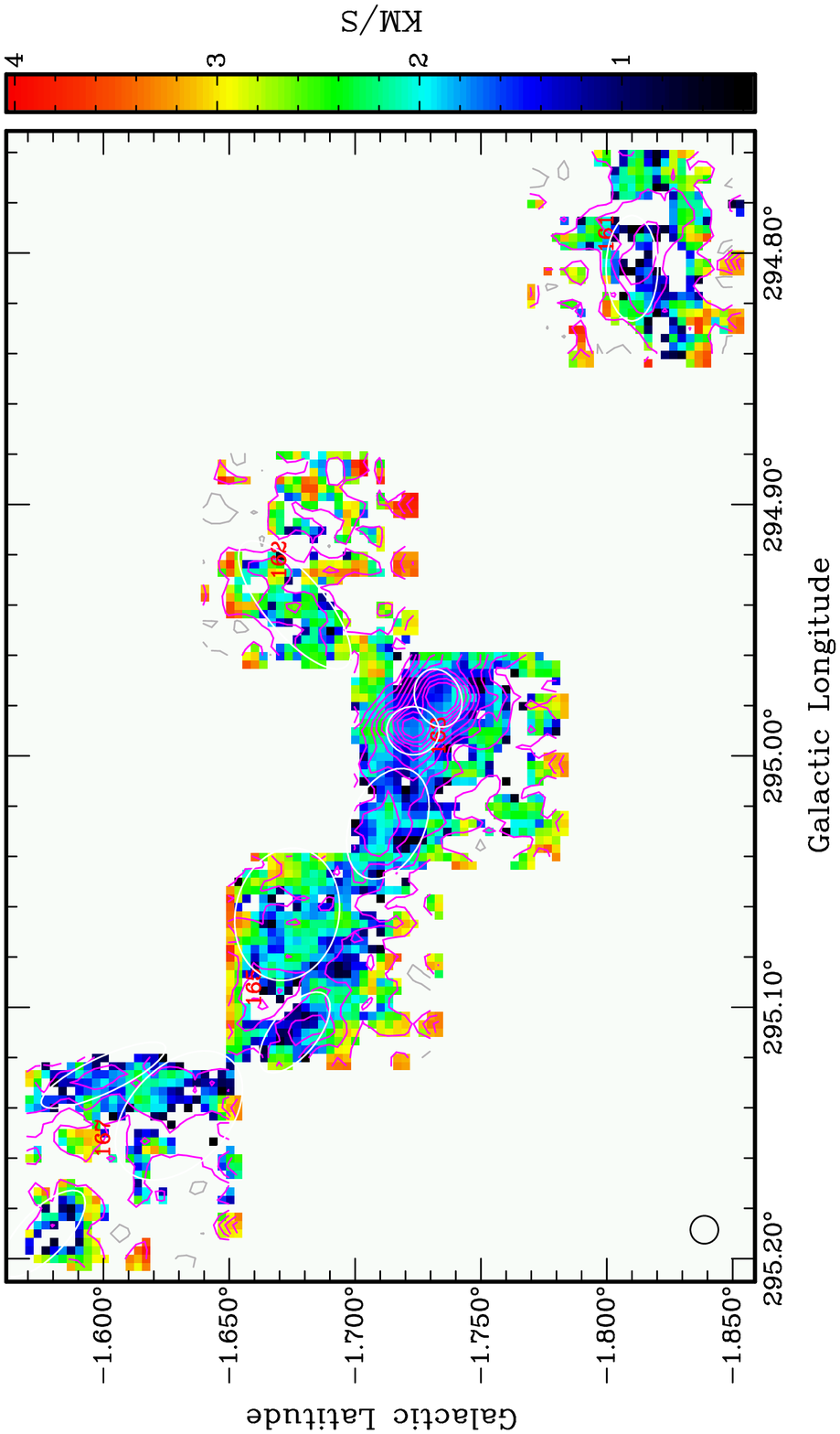}}}
\caption{\small Same as Fig.\,\ref{momR1}, but for Region 21 sources BYF\,161--167.  Contours are every 3$\sigma$ = 0.924\,K\kms, and at 2.4\,kpc the 40$''$ Mopra beam (lower left corner) scales to 0.465\,pc.  ($a$) $T_p$,  ($b$) rms,  ($c$) $V_{\rm LSR}$,  ($d$) $\sigma_{V}$.
\label{momR21}}
\end{figure*}

\newpage

\begin{figure*}[ht]
\centerline{(a){\includegraphics[angle=-90,scale=0.40]{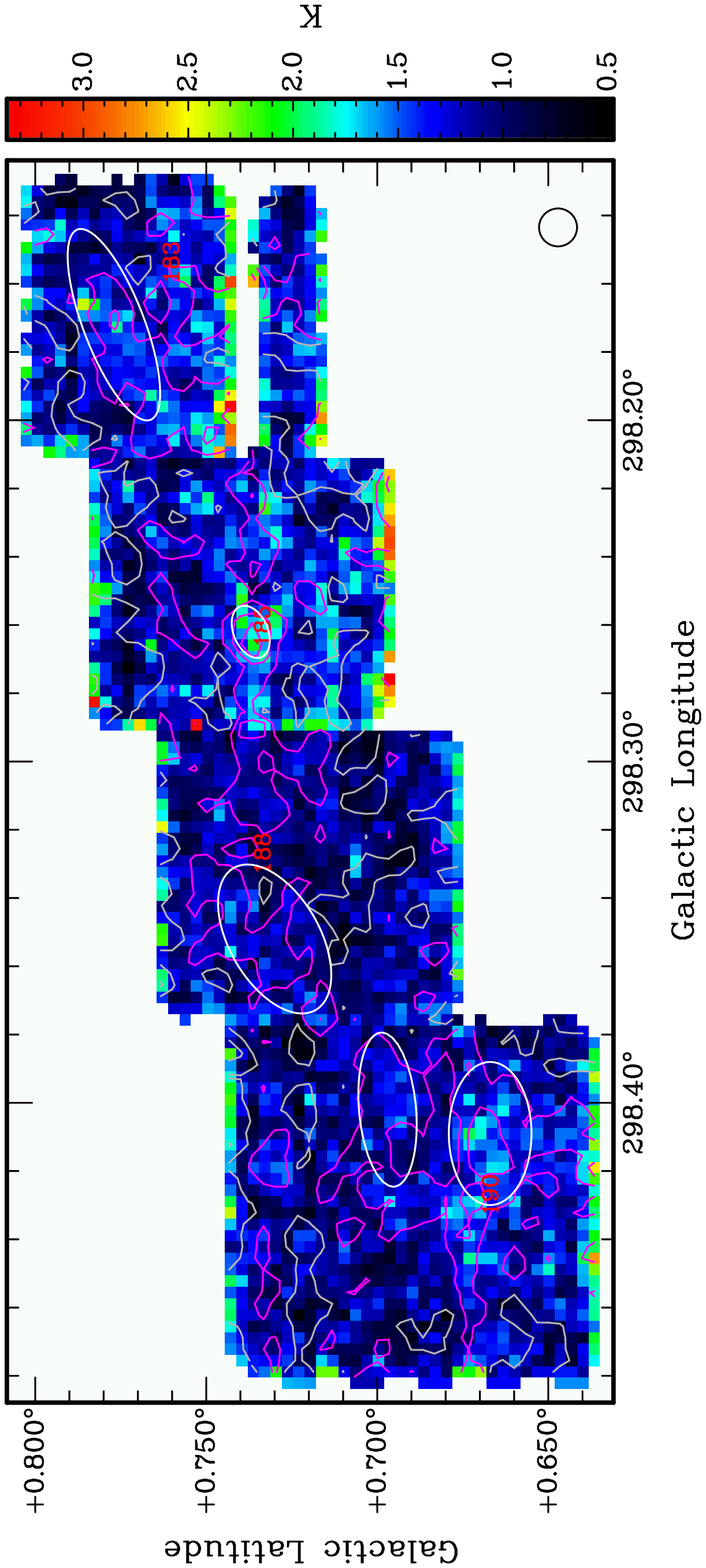}}}
\centerline{(b){\includegraphics[angle=-90,scale=0.40]{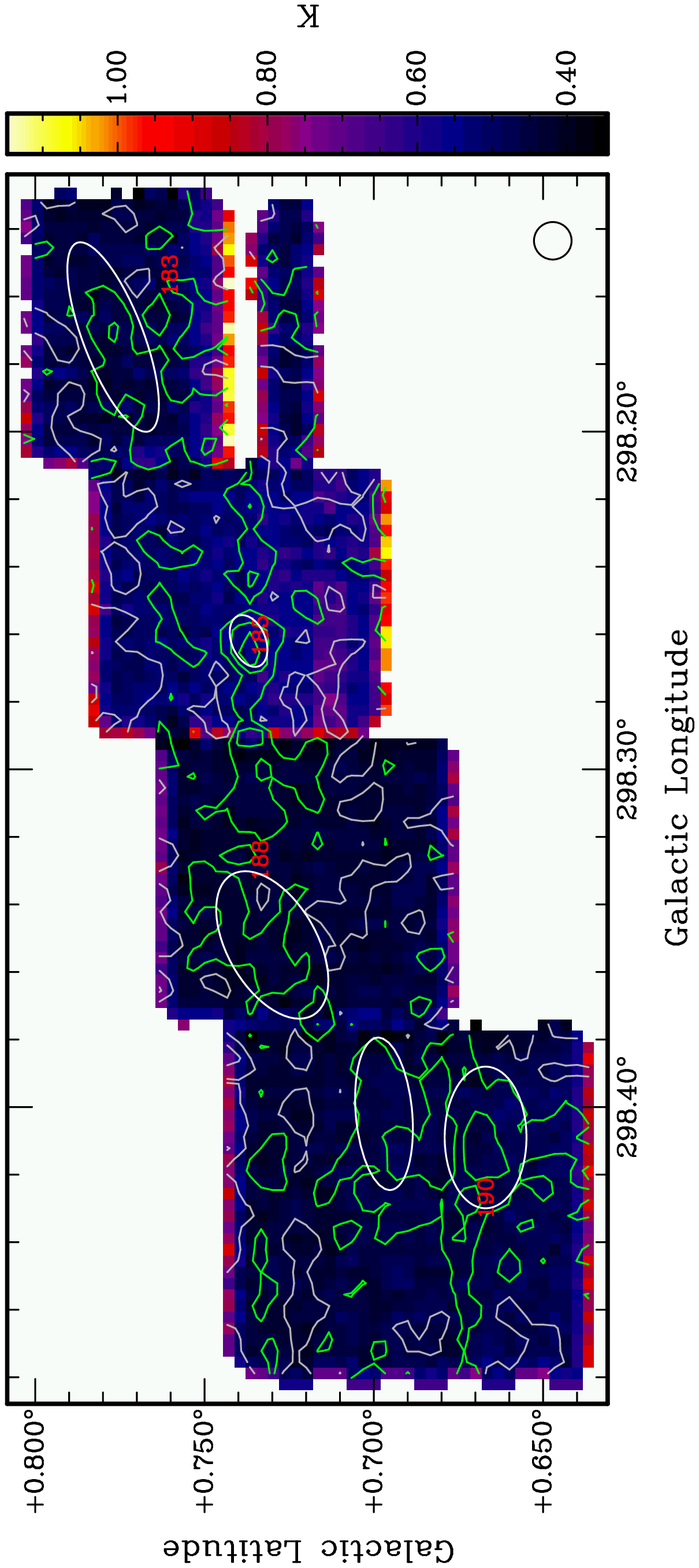}}}
\centerline{(c){\includegraphics[angle=-90,scale=0.40]{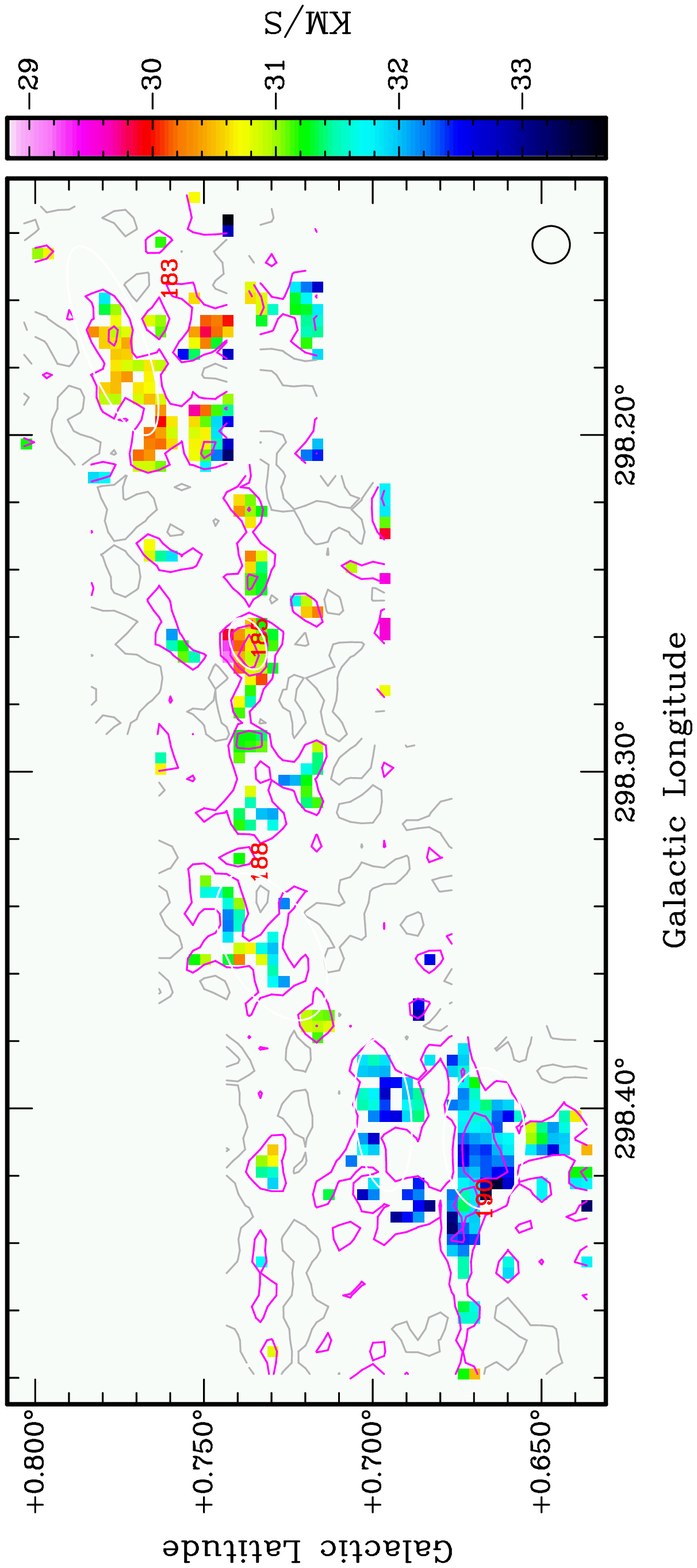}}}
\centerline{(d){\includegraphics[angle=-90,scale=0.40]{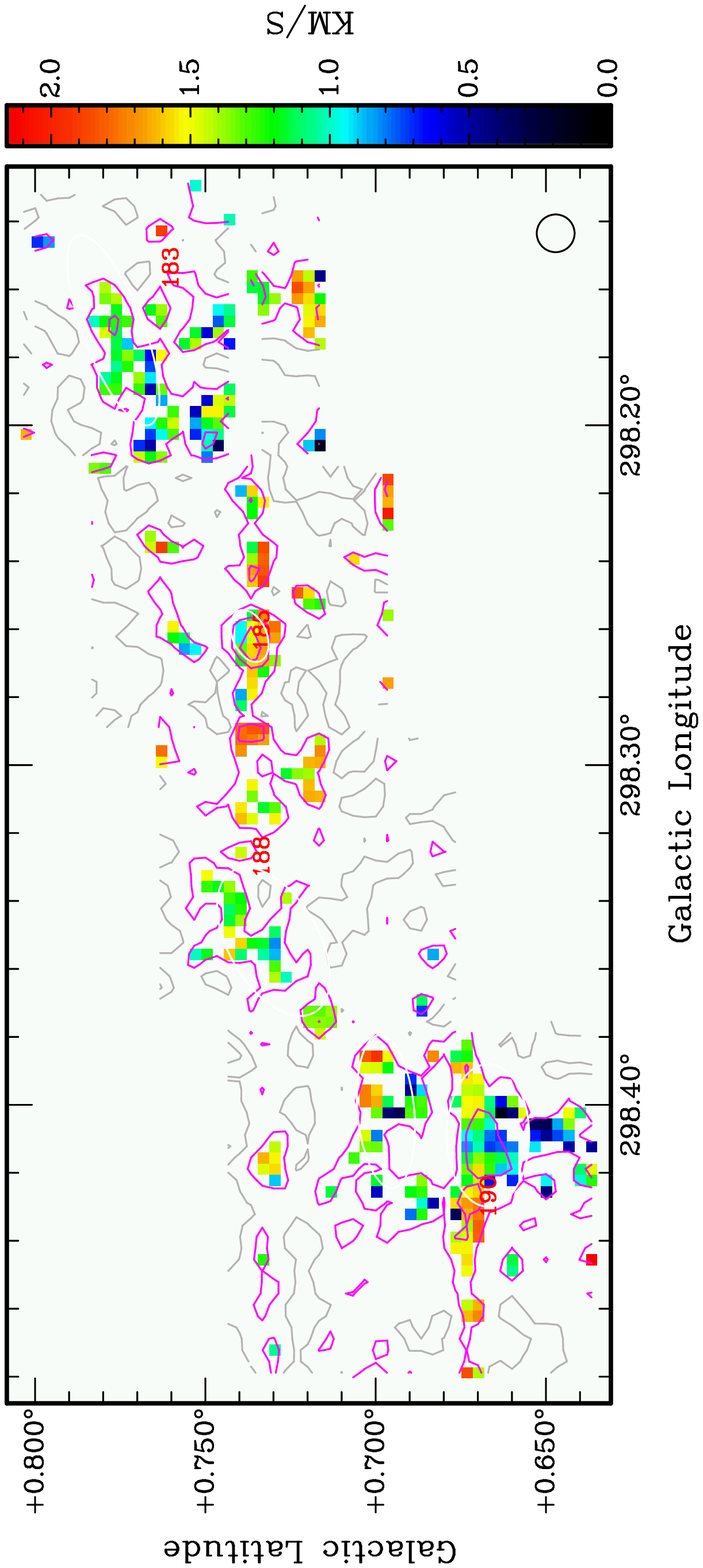}}}
\caption{\small Same as Fig.\,\ref{momR1}, but for Region 23 sources BYF\,183--190.  Contours are every 2$\sigma$ = 0.828\,K\kms, and at 4.7\,kpc the 40$''$ Mopra beam (lower right corner) scales to 0.911\,pc.  ($a$) $T_p$,  ($b$) rms,  ($c$) $V_{\rm LSR}$,  ($d$) $\sigma_{V}$.
\label{momR23}}
\end{figure*}

\clearpage

\begin{figure*}[ht]
\centerline{(a){\includegraphics[angle=0,scale=0.30]{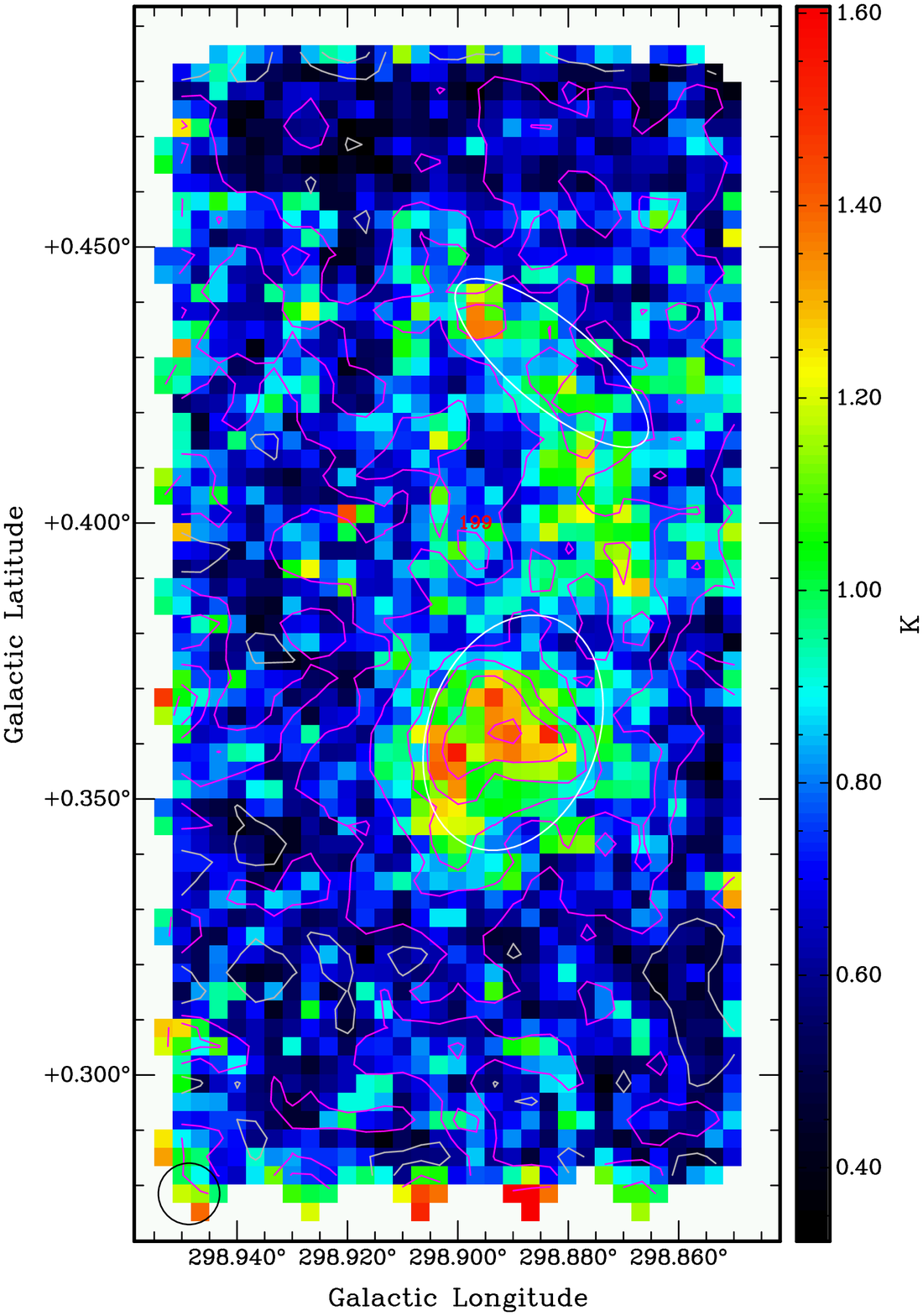}}
		(b){\includegraphics[angle=0,scale=0.30]{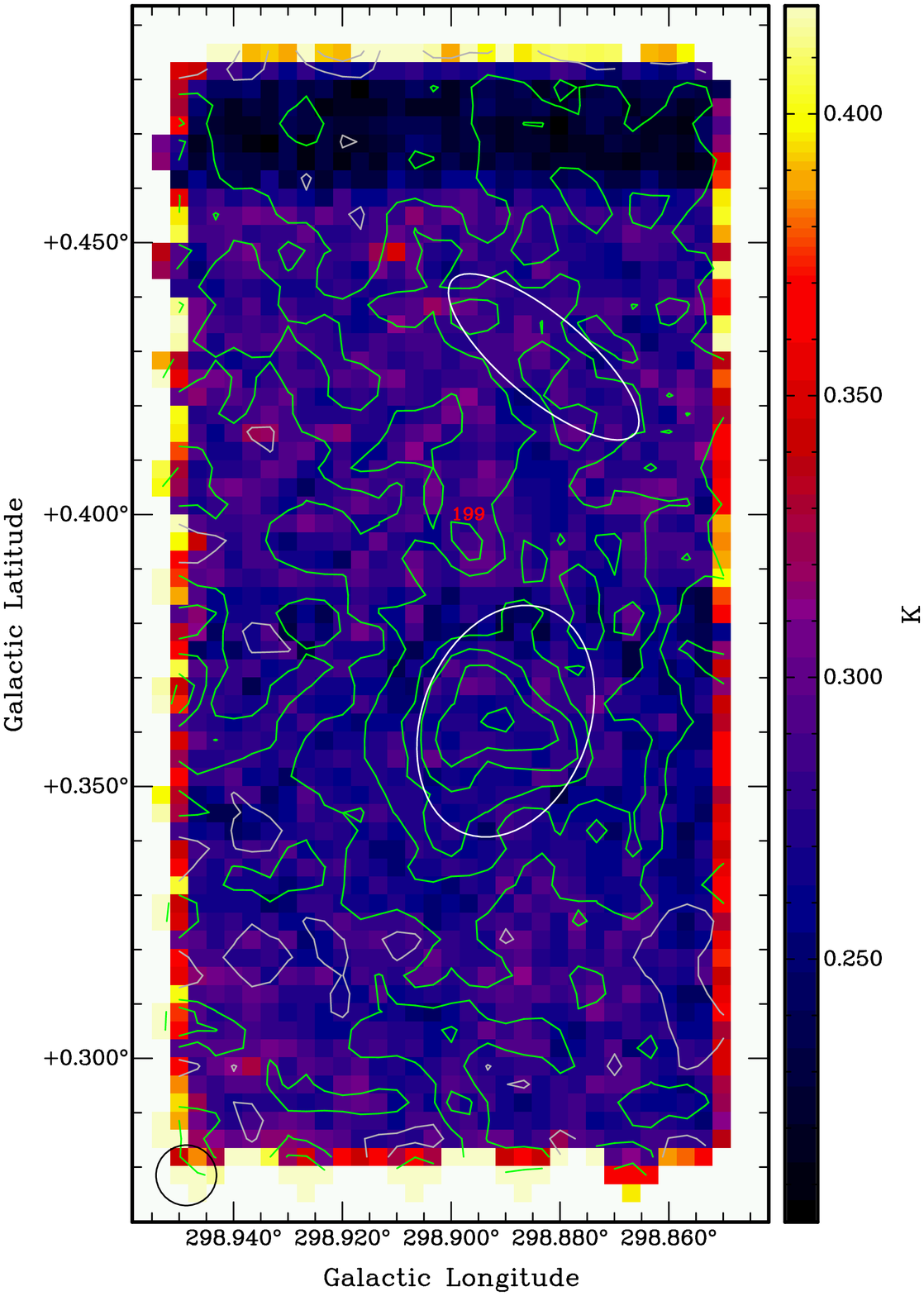}}}
\centerline{(c){\includegraphics[angle=0,scale=0.30]{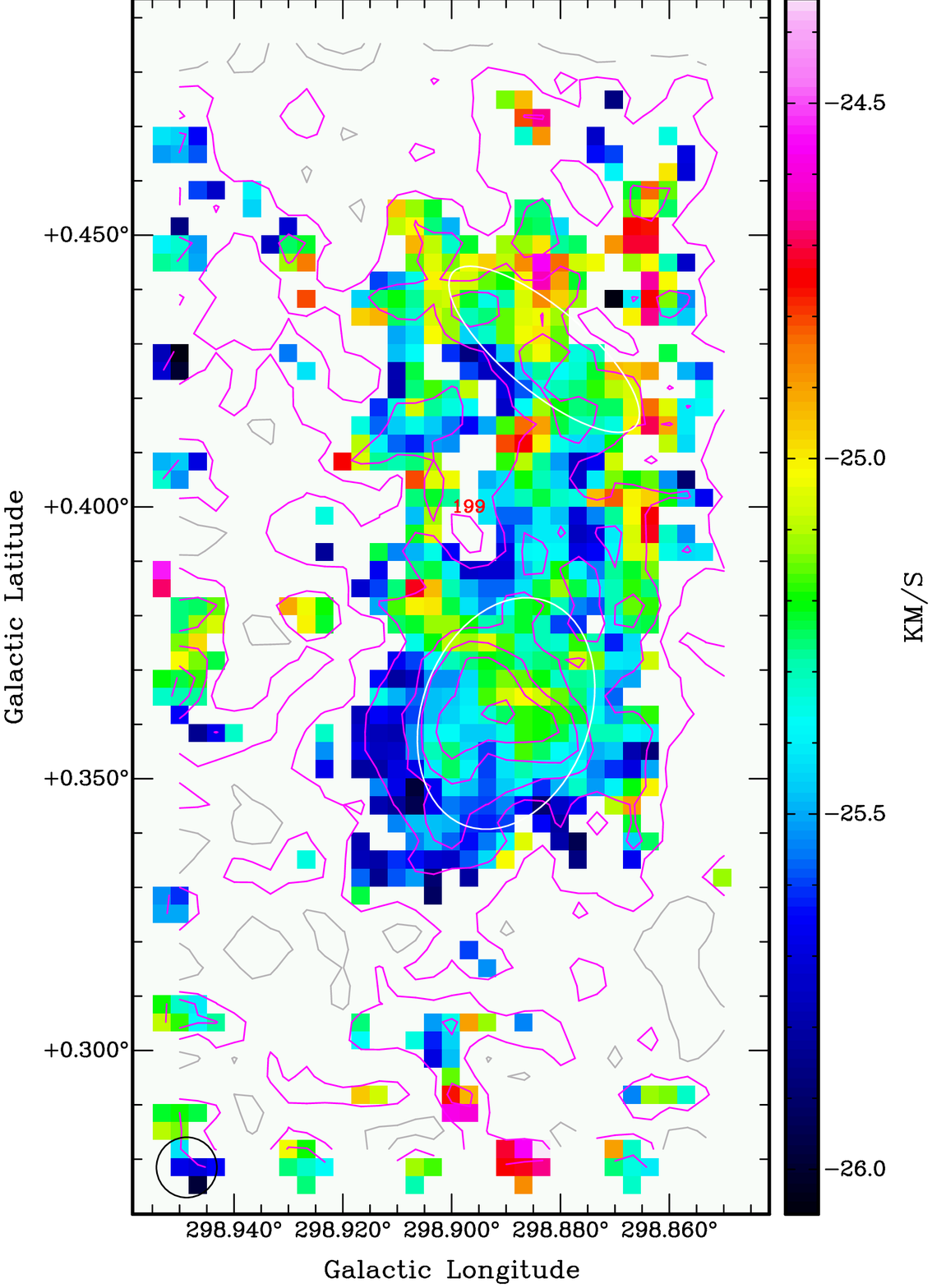}}
		(d){\includegraphics[angle=0,scale=0.30]{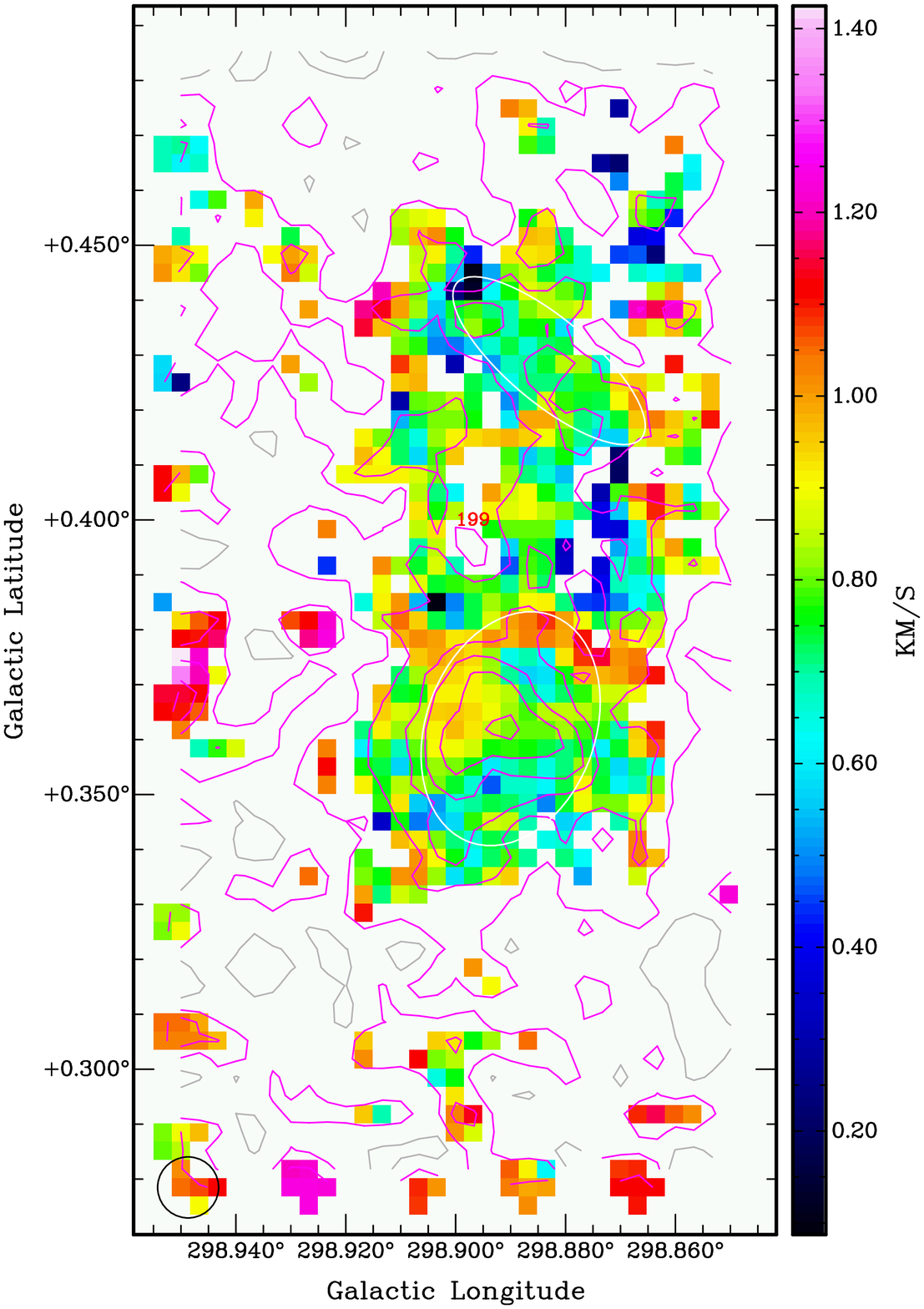}}}
\caption{\small Same as Fig.\,\ref{momR1}, but for Region 26 source BYF\,199.  Contours are every 2$\sigma$ = 0.368\,K\kms, and at 4.7\,kpc the 40$''$ Mopra beam (lower left corner) scales to 0.911\,pc.  ($a$) $T_p$,  ($b$) rms,  ($c$) $V_{\rm LSR}$,  ($d$) $\sigma_{V}$.
\label{momR26a}}
\end{figure*}

\clearpage

\begin{figure*}[ht]
\centerline{(a){\includegraphics[angle=-90,scale=0.42]{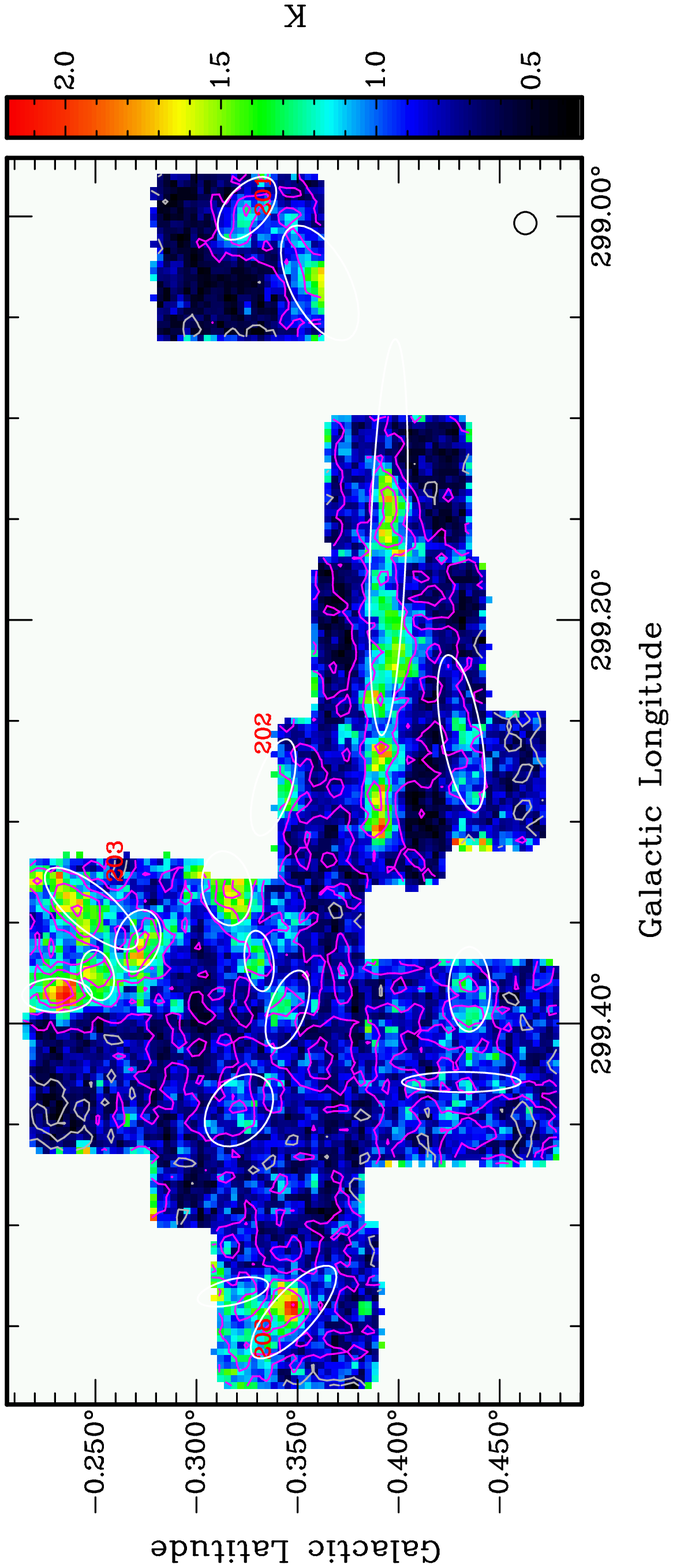}}} 
\centerline{(b){\includegraphics[angle=-90,scale=0.42]{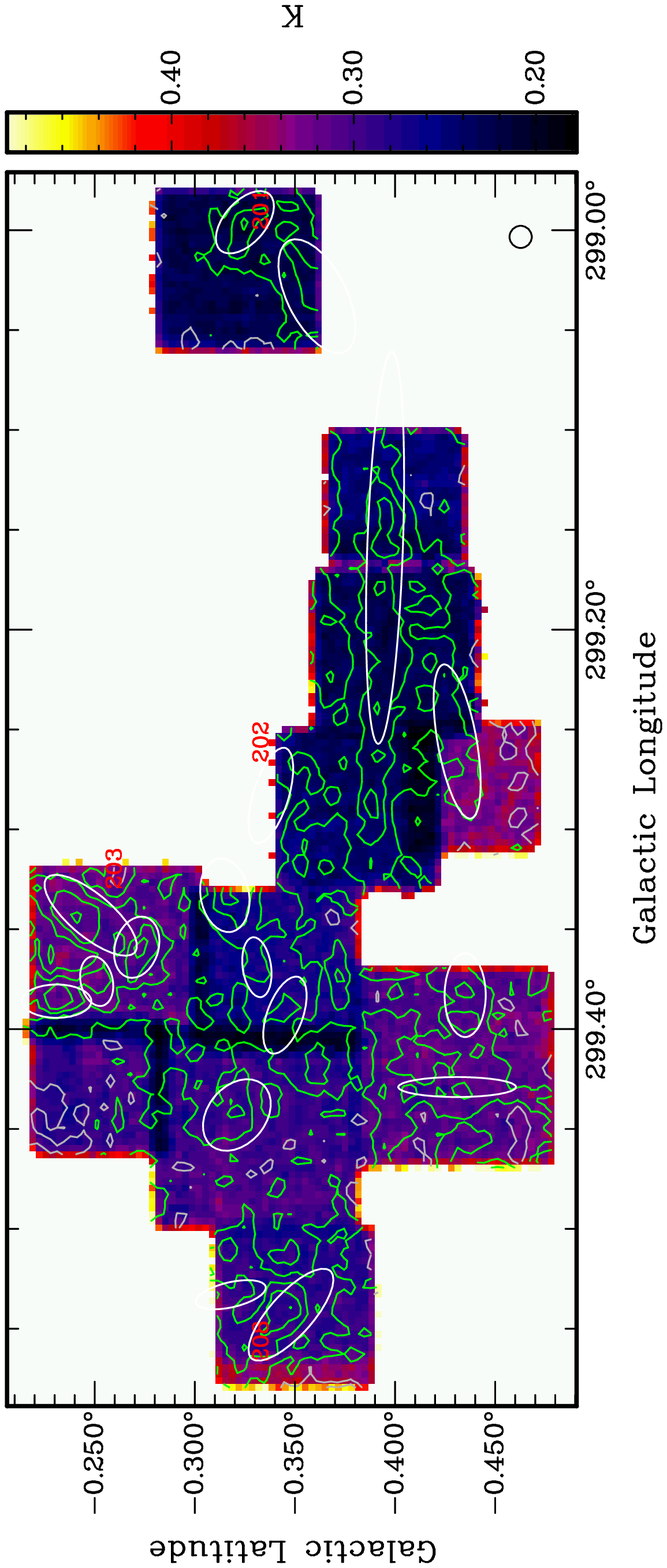}}}
\centerline{(c){\includegraphics[angle=-90,scale=0.42]{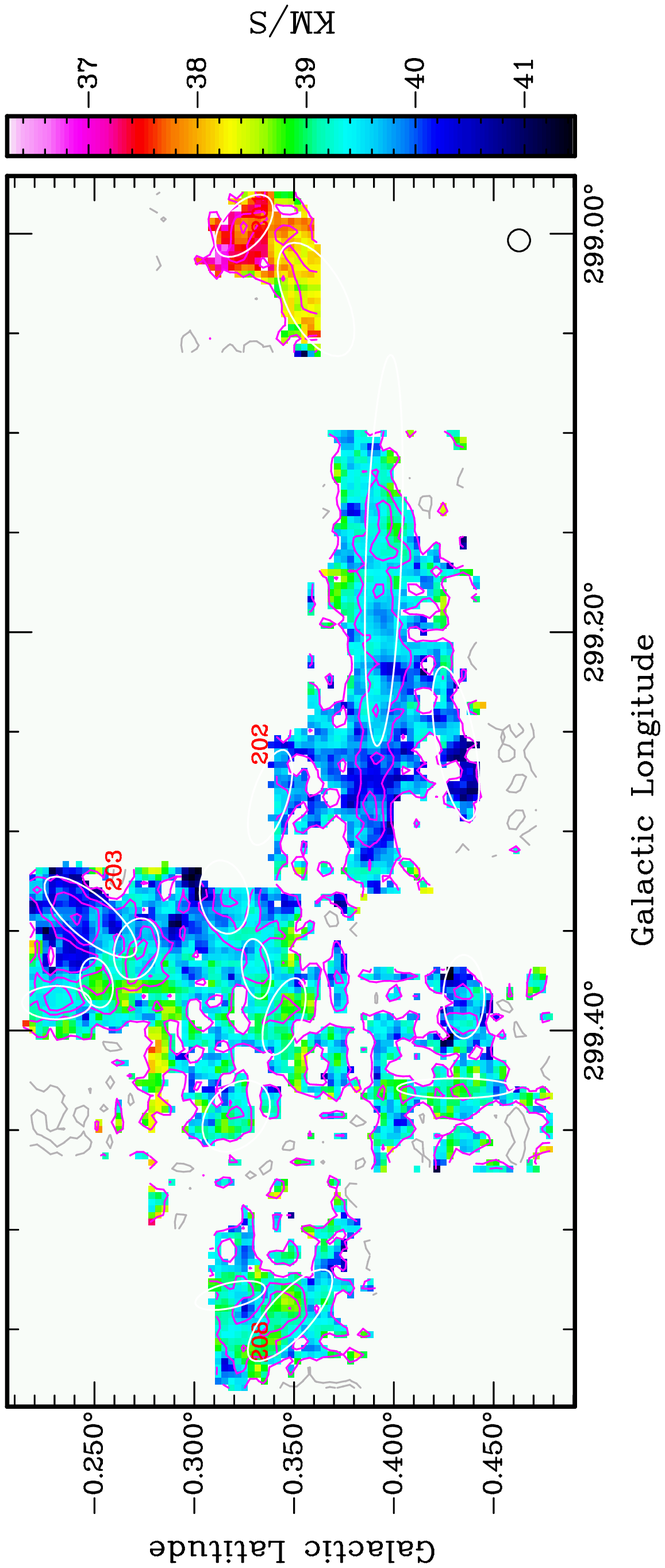}}}
\centerline{(d){\includegraphics[angle=-90,scale=0.42]{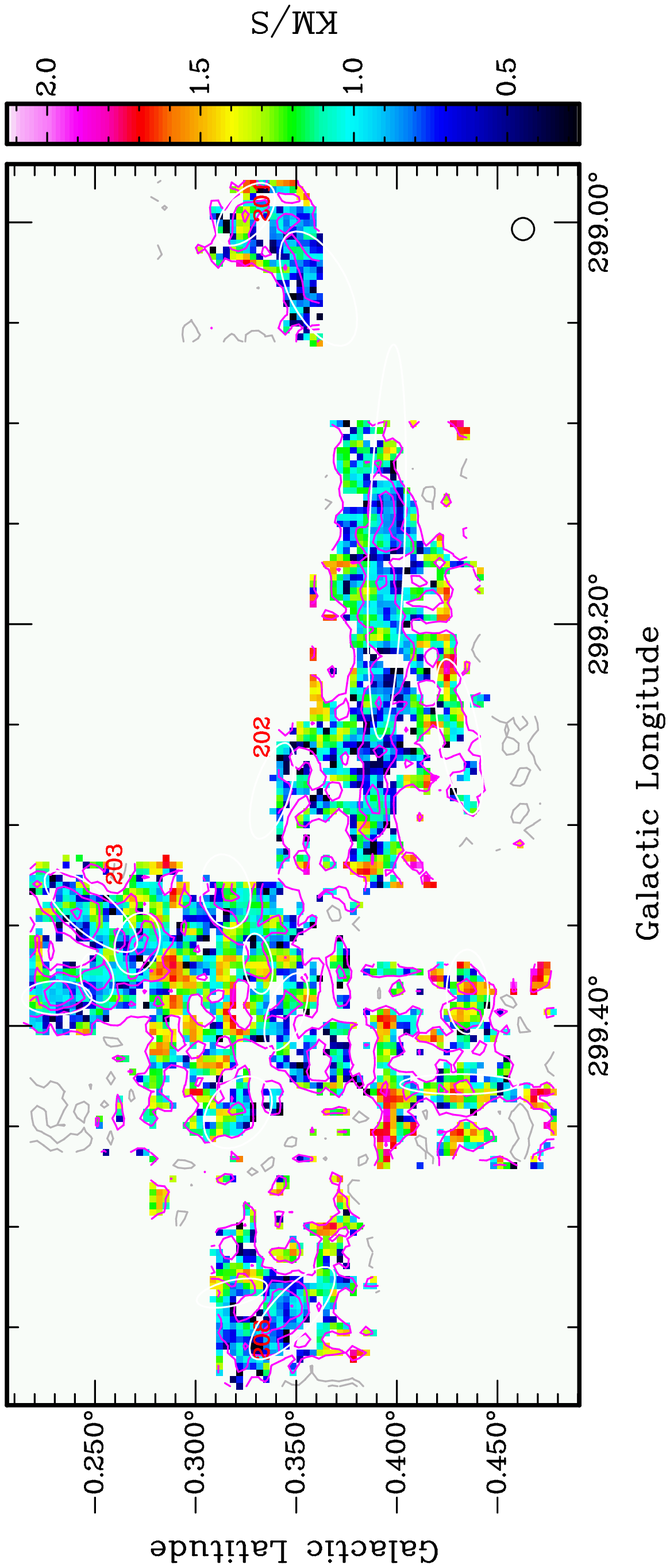}}}
\caption{\small Same as Fig.\,\ref{momR1}, but for Region 26 sources BYF\,201--208.  Contours are every 3$\sigma$ = 0.690\,K\kms, and at 4.7\,kpc the 40$''$ Mopra beam (lower right corner) scales to 0.911\,pc.  ($a$) $T_p$,  ($b$) rms,  ($c$) $V_{\rm LSR}$,  ($d$) $\sigma_{V}$.
\label{momR26b}}
\end{figure*}

\clearpage




\clearpage


\end{document}